\documentclass[11pt]{report}

\usepackage{PhysRep}
\usepackage{Lep2Rep}
\usepackage{gg}
\usepackage{ff}
\usepackage{4f_s05}
\usepackage{smat}
\usepackage{gc}
\usepackage{mw}

\usepackage[english]{babel}
\usepackage{graphicx,rotating}
\usepackage{a4p,here}
\usepackage{cite,mcite,epsfig}

\renewcommand{\Huge}{\huge}
\newcommand{\updates}[1]%
{\fbox{\parbox{\linewidth}{\textbf{Updates with respect to summer 2004:}\\#1}}}

\input{rotate}

\begin{document}
\flushbottom
\begin{titlepage}
\begin{center}
\Large {EUROPEAN ORGANIZATION FOR NUCLEAR RESEARCH}
\end{center}

\begin{flushright}
       CERN-PH-EP/2005-051 \\
       LEPEWWG/2005-01  \\
       ALEPH 2005-004 PHYSICS 2005-004 \\
       DELPHI 2005-027-PHYS-947 \\
       L3 Note 2832   \\
       OPAL PR 413    \\
       hep-ex/0511027 \\
       {\bf 9 November 2005}
\end{flushright}

\begin{center}
\boldmath
\Huge {\bf A Combination of Preliminary \\
               Electroweak Measurements and \\
            Constraints on the Standard Model\\[.5cm]

}
\unboldmath

\vspace*{0.5cm}
\Large {\bf
The LEP Collaborations\footnote{ The LEP Collaborations each take
responsibility for the preliminary results of their own experiment.}
ALEPH, DELPHI, L3, OPAL, and \\ the LEP Electroweak Working
Group\footnote{ WWW access at {\tt http://www.cern.ch/LEPEWWG}

The members of the 
LEP Electroweak Working Group 
who contributed significantly to this
note are: \\
J.~Alcaraz,         %
A.~Bajo-Vaquero,    %
P.~Bambade,         %
E.~Barberio,        %
A.~Blondel,         %
D.~Bourilkov,       %
P.~Checchia,        %
R.~Chierici,        %
R.~Clare,           %
J.~D'Hondt,         %
B.~de~la~Cruz,      %
P.~de~Jong,         %
G.~Della~Ricca,     %
M.~Dierckxsens,     %
D.~Duchesneau,      %
G.~Duckeck,         %
M.~Elsing,          %
M.W.~Gr\"unewald,   %
A.~Gurtu,           %
J.B.~Hansen,        %
R.~Hawkings,        %
St.~Jezequel,       %
R.W.L.~Jones,       %
T.~Kawamoto,        %
E.~Lan{\c c}on,     %
W.~Liebig,          %
L.~Malgeri,         %
S.~Mele,            %
E.~Migliore,        %
M.N.~Minard,        %
K.~M\"onig,         %
C.~Parkes,          %
U.~Parzefall,       %
B.~Pietrzyk,        %
G.~Quast,           %
P.~Renton,          %
S.~Riemann,         %
K.~Sachs,           %
A.~Straessner,      %
D.~Strom,           %
R.~Tenchini,        %
F.~Teubert,         %
M.A.~Thomson,       %
S.~Todorova-Nova,   %
A.~Valassi,         %
A.~Venturi,         %
H.~Voss,            %
C.P.~Ward,          %
N.K.~Watson,        %
P.S.~Wells,         %
St.~Wynhoff.        %
}\\
}
\vskip 0.5cm
\large\textbf{Prepared from Contributions of the LEP Experiments \\
to the 2005 Summer Conferences.}\\
\end{center}
\vfill
\begin{abstract}
  This note presents a combination of published and preliminary
  electroweak results from the four LEP collaborations ALEPH, DELPHI,
  L3 and OPAL based on electron-positron collision data taken at
  centre-of-mass energies above the Z-pole, $130~\GeV$ to $209~\GeV$
  (\LEPII), as prepared for the 2005 summer conferences.  Averages are
  derived for di-fermion cross sections and forward-backward
  asymmetries, photon-pair, W-pair, Z-pair, single-W and single-Z
  cross sections, electroweak gauge boson couplings, W mass and width
  and W decay branching ratios.  An investigation of the interference
  of photon and Z-boson exchange is presented, and colour reconnection
  and Bose-Einstein correlation analyses in W-pair production are
  combined.  The main changes with respect to the experimental results
  presented in 2004 are updates to some 4-fermion cross sections,
  final results on BE correlations and a new preliminary combination
  of the mass and width of the W boson.
  
  Including the precision electroweak measurements performed at the Z
  pole published recently, the results are compared with precise
  electroweak measurements from other experiments, notably CDF and
  D\O\ at the Tevatron.  Constraints on the input parameters of the
  Standard Model are derived from the results obtained in high-$Q^2$
  interactions, and used to predict results in low-$Q^2$ experiments,
  such as atomic parity violation, Moller scattering, and
  neutrino-nucleon scattering.

\end{abstract}
\end{titlepage}
\setcounter{page}{3}
\renewcommand{\thefootnote}{\arabic{footnote}}
\setcounter{footnote}{0}

\chapter{Introduction}
\label{sec-Intro}

This article presents an updated summary of combined results on
electroweak observables measured in high-energy electron-positron
collisions.  The results of the LEP experiments ALEPH, DELPHI, L3,
OPAL and the SLD experiment at SLC based on data collected at the Z
resonance and their combinations reported in previous
summaries~\cite{bib-EWEP-04} have since then been finalised and are
published~\cite{bib-Z-pole}.  All Z-pole results and observables
derived thereof, in particular the various effective couplings of the
neutral weak current, are reported in Reference~\citen{bib-Z-pole} and
are no longer described in this yearly update.

Since 1996 the electron-positron collider LEP has run at
centre-of-mass energies above the Z pole, $\sqrt{s}\ge130~\GeV$
(\LEPII), and mainly above the W-pair production threshold.  In 2000,
the final year of data taking at LEP, the maximum centre-of-mass
energy of close to $209~\GeV$ was attained, although most of the data
taken in 2000 was collected at 205 and 207~\GeV.  By the end of
$\LEPII$ operations, a total integrated luminosity of approximately
$700~\pb$ per experiment was recorded above the Z resonance.

The electroweak $\LEPII$ measurements discussed here consist of
di-fermion cross sections and forward-backward asymmetries; di-photon
production, W-pair, Z-pair, single-W and single-Z production cross
sections, and electroweak gauge boson self couplings.  W boson
properties, like mass, width and decay branching ratios are also
measured. Studies on photon/Z interference in fermion-pair production
as well as on colour reconnection and Bose-Einstein correlations in
W-pair production are presented.  The $\LEPII$ combinations presented
here supersede the previous analyses~\cite{bib-EWEP-04}.  Most
measurements are still preliminary.  Note that in some cases some
experiments have already published final results which are not yet
included in the combinations presented here.

This note is organised as follows:
\begin{description}
\item [Chapter~\ref{sec-GG}]   Photon-pair production at energies above the Z;
\item [Chapter~\ref{sec-FF}]   Fermion-pair production at energies above the Z;
\item [Chapter~\ref{sec-smat}] Photon/Z-boson interference;
\item [Chapter~\ref{sec-4F}]   W and four-fermion production;
\item [Chapter~\ref{sec-GC}]   Electroweak gauge boson self couplings;
\item [Chapter~\ref{sec-CR}]   Colour reconnection in W-pair events;
\item [Chapter~\ref{sec-BE}]   Bose-Einstein correlations in W-pair events;
\item [Chapter~\ref{sec-MW}]   W-boson mass and width;
\item [Chapter~\ref{sec-MSM}] Interpretation of all results, including
  final published Z-pole results~\cite{bib-Z-pole} from {\LEPI} and
  SLD, as well as results from CDF and D\O, in terms of constraints on
  the Standard Model (SM);
\item [Chapter~\ref{sec-Conc}] Conclusions including prospects for the future.
\end{description}
To allow a quick assessment, a box highlighting the updates is given
at the beginning of each chapter.

\boldmath
\chapter{Photon-Pair Production at \LEPII}
\label{sec-GG}
\unboldmath

\updates{ Unchanged w.r.t. summer 2002: ALEPH, L3 and OPAL have
  provided final results for the complete {\LEPII} dataset, DELPHI up
  to 1999 data and preliminary results for the 2000 data.  

  Note that some recent publications~\cite{Abdallah:2004rc} are not
  yet included in this combination.  }

\section{Introduction}
The reaction $\eeggga$ provides a clean test of QED at LEP energies
and is well suited to detect the presence of non-standard physics.
The differential QED cross-section at the Born level in the
relativistic limit is given by \cite{ref:QED,ref:radcor}:
\begin{equation}
\xb = \frac{\alpha^2}{s} 
\frac{1+\cos^2\theta}{1 -\cos^2\theta} \; .
\end{equation}
Since the two final state particles are identical the polar angle
$\theta$ is defined such that $\ct > 0$. Various models with 
deviations from this cross-section will be discussed in section \ref{gg:sec:fit}.
Results on the $\ge$2-photon  final state using the high energy data 
collected by the four LEP collaborations are reported by the individual
experiments \cite{gg:ref:LEPGG}.
Here the results of the LEP working group %
dedicated to the combination of the $\eeggga$ measurements
are reported.  Results are given for the averaged total cross-section
and for global fits to the differential cross-sections.

\section{Event Selection}
This channel is very clean and the event selection, which is similar
for all experiments, is based on the presence of at least two
energetic clusters in the electromagnetic calorimeters.  A minimum
energy is required, typically $(E_1+ E_2)/\sqrt{s}$ larger than 0.3 to
0.6, where $E_1$ and $E_2$ are the energies of the two most energetic
photons.  In order to remove $\ee$ events, charged tracks are in
general not allowed except when they can be associated to a photon
conversion in one hemisphere.

The polar angle is defined in order to minimise effects due to 
initial state radiation as
\[
\ct =\left.\left| \sin (\frac{\theta_1 - \theta_2}{2}) \right| 
        \right/ \sin (\frac{\theta_1 + \theta_2}{2}) \; ,   \] 
where $\theta_1$ and $\theta_2$ are the polar angles of the two most energetic photons.
The acceptance in polar angle is in the range of 0.90 to 0.96 on 
$|\ct|$, depending on the experiment.

With these criteria, the selection efficiencies are in the range of
68\% to 98\% and the residual background (from $\ee$ events and
from $\eetautau$ with $\tau^{\pm} \rightarrow\rm e^{\pm}\nu
\bar{\nu}$) is very small, 0.1\% to 1\%.  Detailed descriptions of
the event selections performed by the four collaborations can be found
in \cite{gg:ref:LEPGG}.

\section{Total cross-section}

\begin{figure}[t]
   \begin{center} \mbox{
          \epsfxsize=16.0cm
           \epsffile{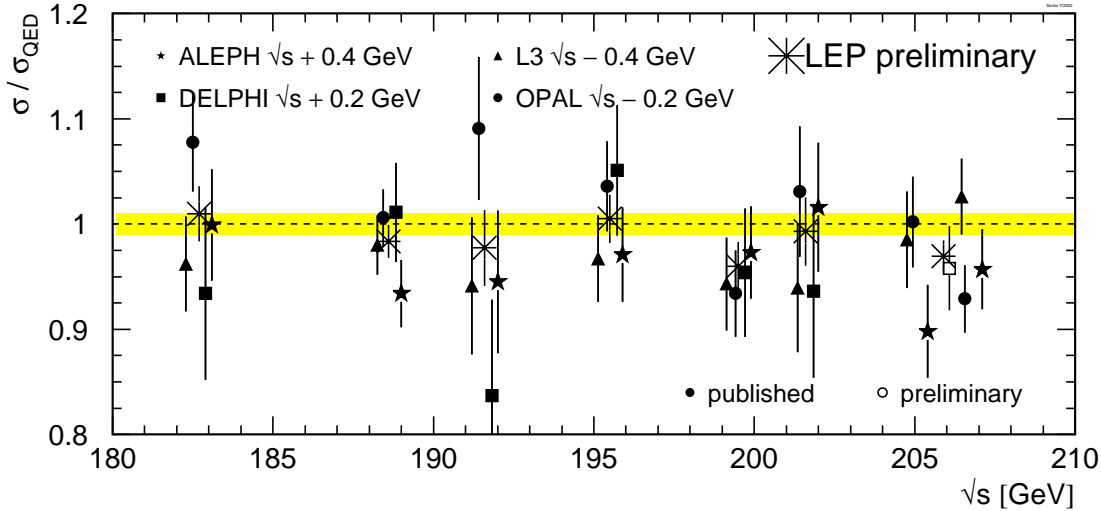}
           } \end{center}
\vspace{-1cm}
\caption{Cross-section ratios 
$r = \sigma_{\rm meas} / \sigma_{\rm QED}$ at different energies.
The measurements of the single experiments are displaced by 
$\pm$ 200 or 400 \MeV\ from the actual energy for clarity. Filled symbols
indicate published results, open symbols stand for preliminary numbers.
The average over the experiments at each energy is shown as a star. 
Measurements between 203 and 209 \GeV\ are averaged to one energy point. 
The theoretical error is not included in the experimental errors 
but is represented as the shaded band.
}
\label{gg:fig:xsn}
\end{figure}

The total cross-sections are combined using a $\chi^2$ minimisation.
For simplicity, given the different angular acceptances,
the ratios of the measured cross-sections relative to the QED 
expectation, \mbox{$r = \sigma_{\rm meas} / \sigma_{\rm QED}$},
are averaged. Figure \ref{gg:fig:xsn} shows the measured ratios $r_{i,k}$ 
of the experiments $i$ at energies $k$ with their statistical
and systematic errors %
added in quadrature. There are no significant sources of experimental 
systematic errors that are correlated between experiments. The theoretical error on 
the QED prediction, which is fully correlated between energies and experiments
is taken into account after the combination.

Denoting with $\Delta$ the vector of residuals between the measurements
and the expected ratios, three different averages are performed:
\begin{enumerate}
\item per energy $k=1,\ldots,7$: $\Delta_{i,k} = r_{i,k} - x_k$ 
\item per experiment $i=1,\ldots,4$: $\Delta_{i,k} = r_{i,k} - y_i$ 
\item global value:  $\Delta_{i,k} = r_{i,k} - z$ 
\end{enumerate}
The seven fit parameters per energy $x_k$ are shown in Figure 
\ref{gg:fig:xsn} as LEP combined cross-sections. They are correlated
with correlation coefficients ranging from 5\% to 20\%. 
The four fit-parameters per experiment $y_i$ are uncorrelated
between each other, the results are given in Table \ref{gg:tab:xsn}
together with the single global fit parameter $z$.

No significant deviations from the QED expectations are found.
The global ratio is below unity by 1.8 standard deviations not 
accounting for the error on the radiative corrections.
This theory error can be assumed to be about 10\% of the applied
radiative correction and hence depends on the selection. 
For this combination it is assumed to be 1\% which is of 
same size as the experimental error (1.0\%).

\begin{table}[hbt]
\begin{center}
\begin{tabular}{|l|r@{$\pm$}l|}\hline
Experiment & \multicolumn{2}{c|}{cross-section ratio} \\\hline\hline
ALEPH  & 0.953 & 0.024 \\
DELPHI & 0.976 & 0.032 \\
L3     & 0.978 & 0.018 \\
OPAL   & 0.999 & 0.016 \\ \hline
global & 0.982 & 0.010 \\ \hline
\end{tabular}
\caption[]{Cross-section ratios 
$r = \sigma_{\rm meas} / \sigma_{\rm QED}$ for the four LEP experiments
averaged over all energies and the global average over all experiments
and energies. The error includes the statistical and experimental
systematic error but no error from theory.
}
\label{gg:tab:xsn}
\end{center}
\end{table}

\begin{figure}[btp]
   \begin{center} 
   \mbox{\epsfxsize=8.0cm\epsffile{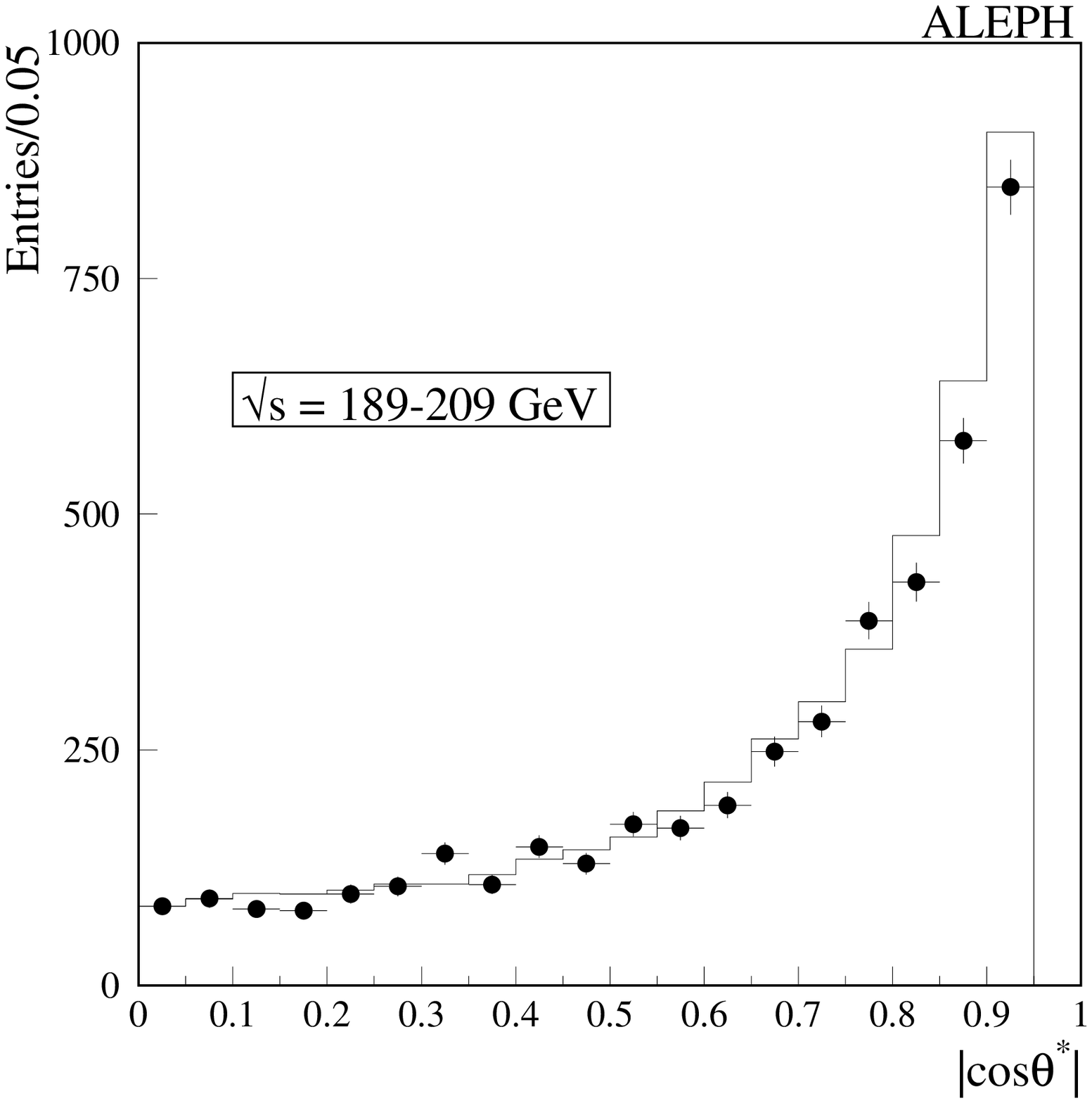}}
   \raisebox{1.5cm}{\epsfxsize=8.0cm\epsffile{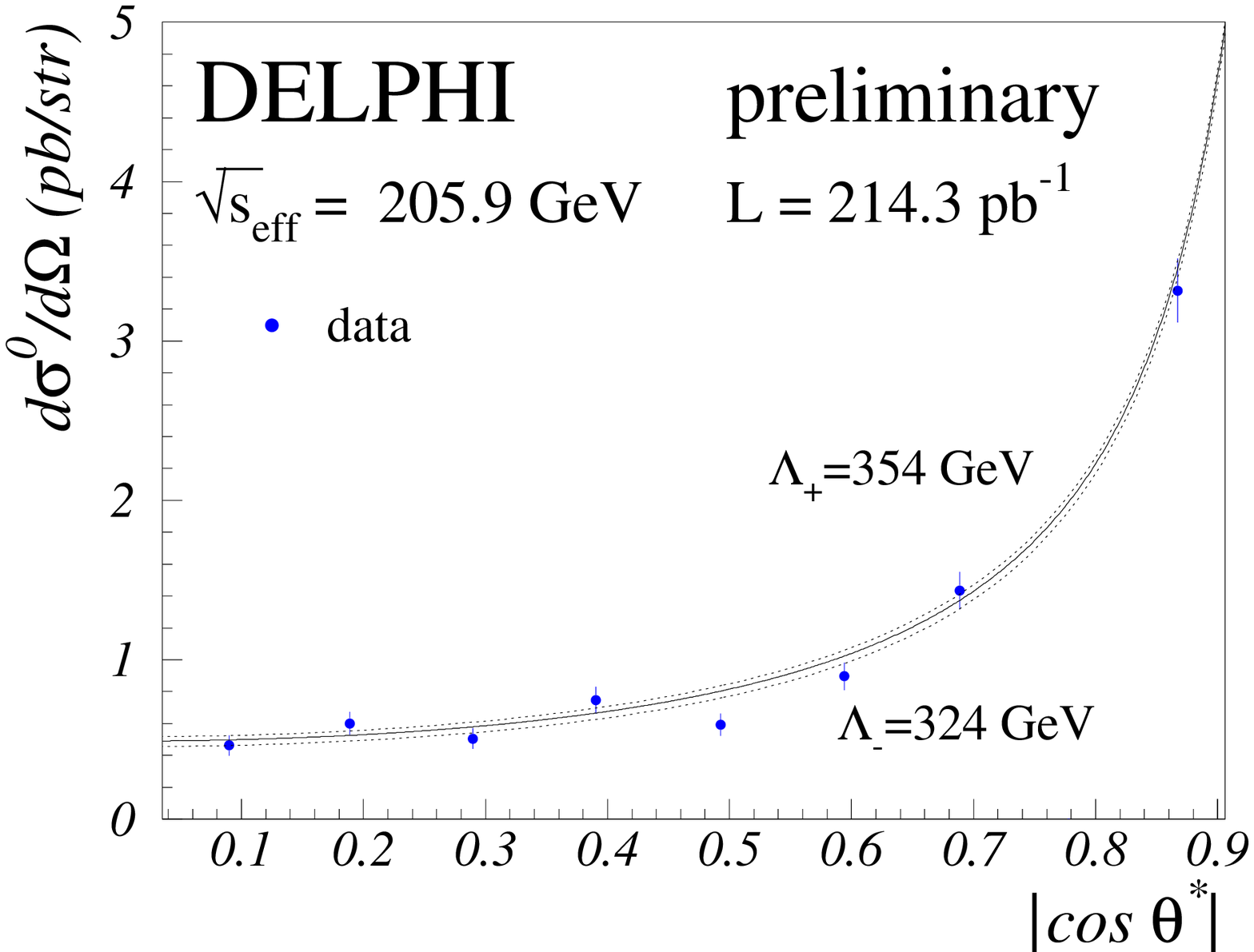}}\\
   \mbox{\epsfxsize=8.0cm\epsffile{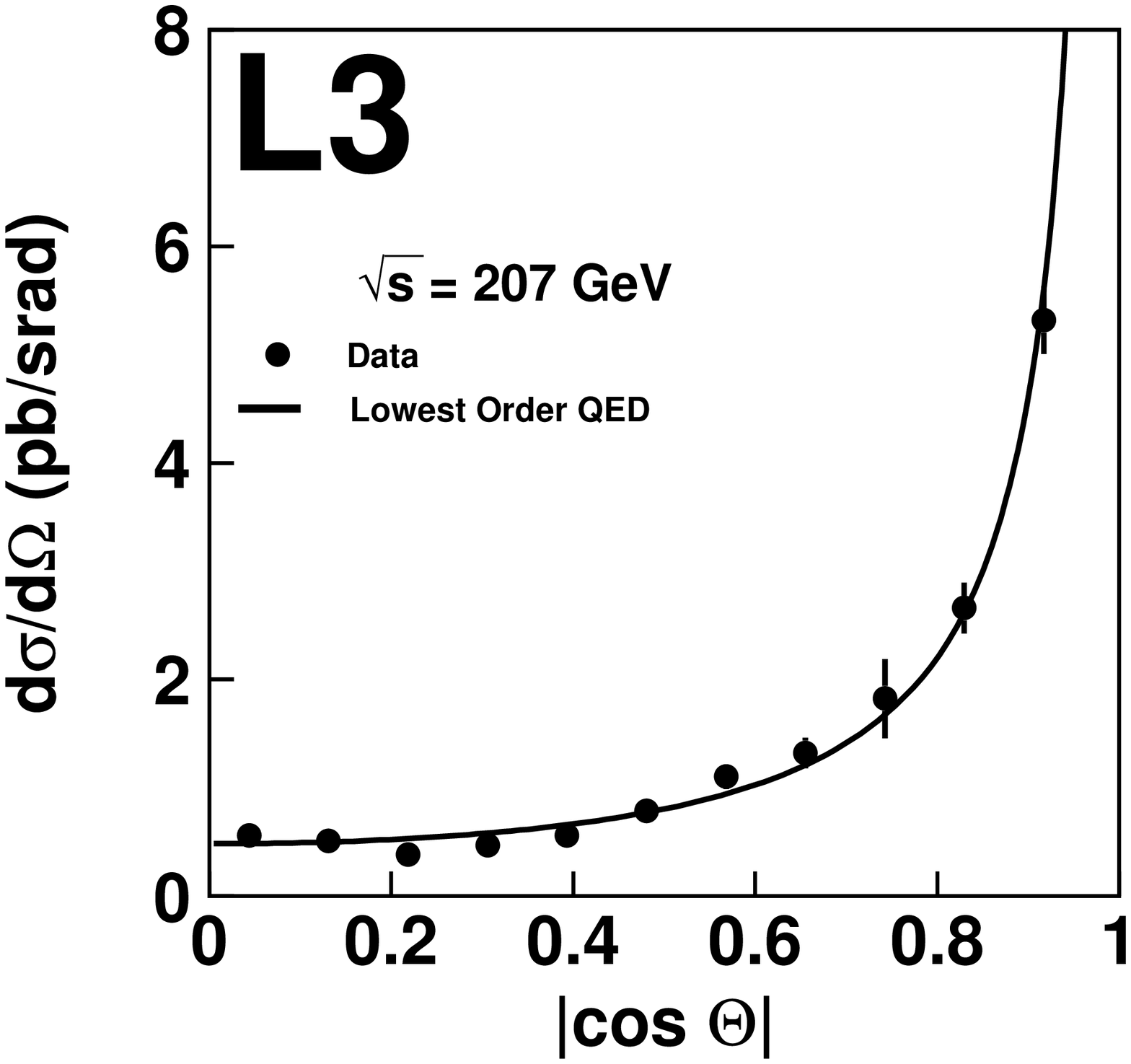}}
   \mbox{\epsfxsize=8.0cm\epsffile{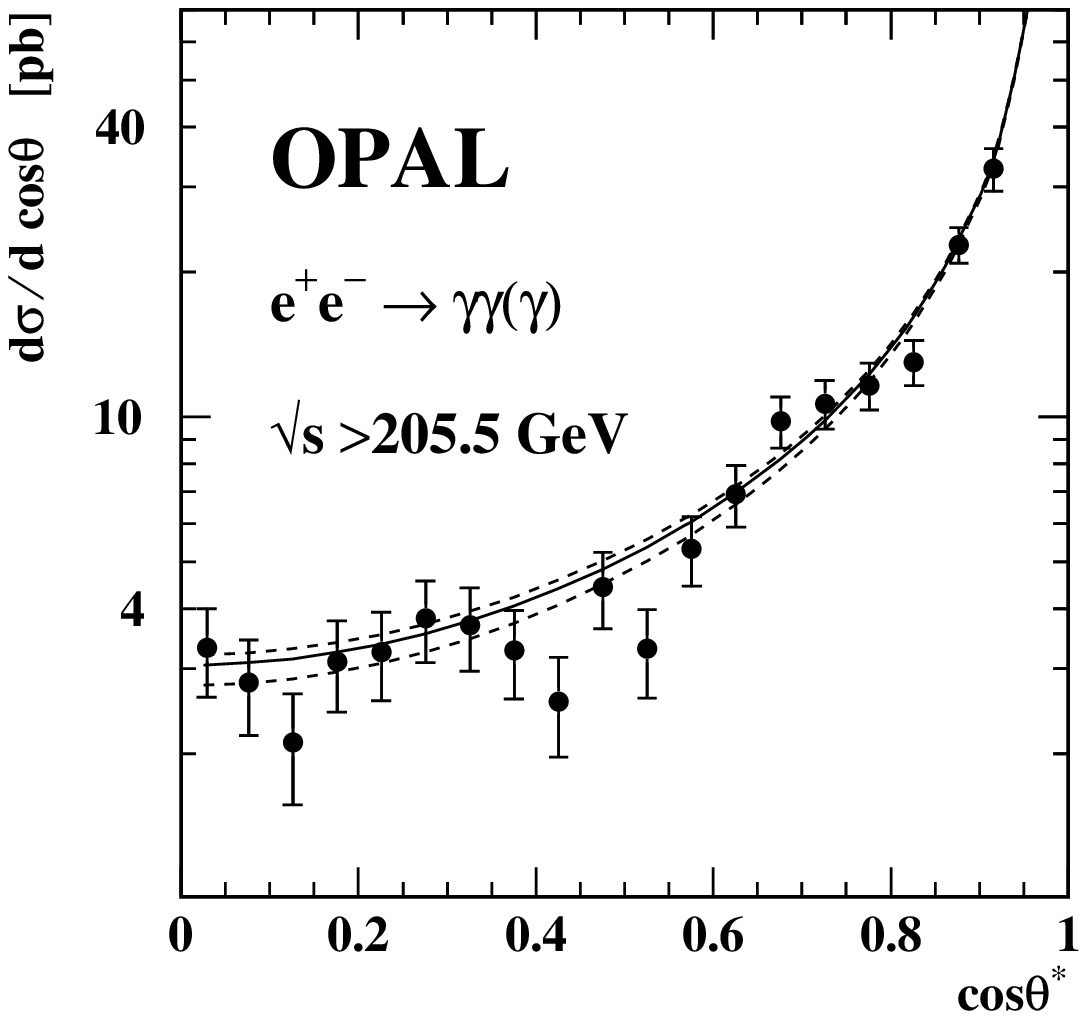}}
   \end{center}
\caption{Examples for angular distributions of the four LEP experiments.
Points are the data and the curves are the QED prediction (solid) and
the individual fit results for $\Lpm$ (dashed). ALEPH shows the
uncorrected number of observed events, the expectation is presented as
histogram. 
}
\label{gg:fig:ADLO}
\end{figure}

\section{Global fit to the differential cross-sections}
\label{gg:sec:fit}

\begin{table}[tb]
\begin{center}
\begin{tabular}{|l|c|c|c|c|c|} \hline
  & \multicolumn{2}{c|}{data used} & \multicolumn{2}{c|}{sys. error $[ \% ]$}&$\left | \rm{cos} \theta \right |$ \\ 
  & published & preliminary & experimental & theory & \\\hline
ALEPH  & 189 -- 207 &  --  & 2 & 1& 0.95 \\
DELPHI & 189 -- 202 & 206 & 2.5 & 1 & 0.90 \\
L3     & 183 -- 207 &  --  & 2.1 & 1 & 0.96 \\
OPAL   & 183 -- 207 &  --  & 0.6 -- 2.9 & 1 & 0.93 \\\hline
\end{tabular}
\caption[]{ The data samples used for the global fit to the
  differential cross-sections, the systematic errors, the assumed
  error on the theory and the polar angle acceptance for the LEP
  experiments.}
\label{gg:tab:stat} 
\end{center} 
\end{table}

The global fit is based on angular distributions at energies between
183 and 207 \GeV\ from the individual experiments. As an example,
angular distributions from each experiment are shown in
Figure~\ref{gg:fig:ADLO}. Combined differential cross-sections are not
available yet, since they need a common binning of the distributions.
All four experiments give results including the whole year 2000 
data-taking. Apart from the 2000 DELPHI data all inputs are final, 
as shown in Table~\ref{gg:tab:stat}.  
The systematic errors arise from the luminosity
evaluation (including theory uncertainty on the small-angle Bhabha
cross-section computation), from the selection efficiency and the
background evaluations and from radiative corrections. The last
contribution, owing to the fact that the available $\eeggga$
cross-section calculation is based on $\cal O$$(\alpha^3)$ code,
is assumed to be 1\% and is considered correlated among energies and experiments.

Various model predictions
are fitted to these angular distributions taking into account the
experimental systematic error correlated between energies for each
experiment and the error on the theory.
A binned log likelihood fit is performed
with one free parameter for the model and five fit parameters
used to keep the normalisation free within the systematic errors
of the theory and the four experiments. Additional fit parameters are
needed to accommodate the angular dependent systematic errors of OPAL.

The following models of new physics are considered. 
The simplest ansatz is a short-range exponential deviation from the 
Coulomb field parameterised by cut-off 
parameters $\Lpm$~\cite{gg:ref:drell,gg:ref:low}. 
This leads to a differential cross-section of the form
\begin{equation}
\xl   =  \xb \pm \frac{\alpha^2 \pi s}{\Lambda_\pm^4}(1+\cos^2{\theta}) \; .
\label{gg:lambda}
\end{equation}

New effects can also be introduced in effective Lagrangian theory
\cite{gg:ref:eboli}. Here dimension-6 terms lead to anomalous 
$\rm ee\gamma$ couplings. The resulting deviations in the differential 
cross-section are similar in form to those given in 
Equation~\ref{gg:lambda}, but with a slightly different definition of the
parameter: $\Lambda_6^4 = \frac{2}{\alpha}\Lambda_+^4$.
While for the ad hoc included cut-off parameters $\Lpm$ both signs are
allowed the physics motivated parameter $\Lambda_6$ occurs only with the
positive sign.
Dimension 7 and 8 Lagrangians introduce $\rm ee\gamma\gamma$ contact
interactions and result in an angle-independent term added to the Born
cross-section:
\begin{equation}
\xq  =  \xb + \frac{s^2}{16}\frac{1}{\Lambda'{}^6} \; .
\end{equation}
The associated parameters are given by 
$\Lambda_7 = \Lambda'$ and $\Lambda_8^4 = m_{\rm e} {\Lambda'}^3$ for
dimension~7 and dimension~8 couplings, respectively.
The subscript refers to the dimension of the Lagrangian.

Instead of an ordinary electron, an excited electron $\rm e^\ast$
with mass $\mestar$
could be exchanged in the $t$-channel \cite{gg:ref:low,gg:ref:estar}. 
In the most general case $\rm \rm e^\ast e \gamma$ couplings would lead
to a large anomalous magnetic moment of the electron 
\cite{gg:ref:g2_brodsky}. 
This effect can be avoided by a chiral magnetic coupling of the form~\cite{gg:ref:boudjema:1993}:
\begin{equation}
{\cal L}_{\rm e^\ast e \gamma} = 
\frac{1}{2\Lambda} \bar{e^\ast} \sigma^{\mu\nu}
\left[ g f \frac{\tau}{2}W_{\mu\nu} + g' f' \frac{Y}{2} B_{\mu\nu}
\right] e_L + \mbox{h.c.} \; ,
\end{equation}
where $\tau$ are the Pauli matrices and $Y$ is the hypercharge.
The parameters of the model are the compositeness scale $\Lambda$
and the weight factors $f$ and $f'$ associated to the gauge fields 
$W$ and $B$ with Standard Model couplings $g$ and $g'$.
For the process $\eeggga$, 
the following cross-section results~\cite{gg:ref:vachon}: 
\begin{eqnarray}
\xe & = & \xb  \\
 & + & \frac{\alpha^2 \pi}{2}\frac{f_\gamma^4}{\Lambda^4}\mestar^2 \left[
\frac{p^4}{(p^2-\mestar^2)^2} + \frac{q^4}{(q^2-\mestar^2)^2} +
\frac{\frac{1}{2} s^2 \sin^2\theta}{(p^2-\mestar^2)(q^2-\mestar^2)} \right]
\; , \nonumber \end{eqnarray}
with $f_\gamma = -\frac{1}{2}(f+f')$, $p^2=-\frac{s}{2}(1-\ct)$ and 
$q^2=-\frac{s}{2}(1+\ct)$. 
Effects vanish in the case of $f = -f'$. The cross-section does not
depend on the sign of $f_\gamma$.

Theories of quantum gravity in extra spatial dimensions could solve the 
hierarchy problem because gravitons would be allowed to travel in 
more than 3+1 space-time dimensions \cite{gg:ref:ad}. 
While in these models the Planck mass $M_D$
in $D=n+4$ dimensions is chosen to be of electroweak scale the usual
Planck mass $M_{\rm Pl}$ in four dimensions would be
\begin{equation} M_{\rm Pl}^2 = R^n M_D^{n+2} \; ,\end{equation}
where $R$ is the compactification radius of the additional dimensions.
Since gravitons couple to the energy-momentum tensor, their
interaction with photons is as weak as with fermions. However, the huge
number of Kaluza-Klein excitation modes in the extra dimensions may 
give rise to
observable effects. These effects depend on the scale $M_s (\sim M_D)$ 
which may be as low as ${\cal O}(\rm TeV)$. Model dependencies
are absorbed in the parameter $\lambda$ which cannot be explicitly
calculated without knowledge of the full theory, the sign is undetermined. 
The parameter $\lambda$ is expected to be 
of ${\cal O}(1)$ and for this analysis it is assumed that $\lambda = \pm 1$. 
The expected differential cross-section is given by \cite{gg:ref:ad}:
\begin{equation}
\xg = \xb - {\alpha s} \; \frac{\lambda}{M_s^4}\;(1+\cos^2{\theta})
    + \frac{s^3}{8 \pi} \;  \frac{\lambda^2}{M_s^8} \;(1-\cos^4{\theta})
    \; .
\end{equation}

\section{Fit Results}

Where possible the fit parameters are chosen such that the likelihood
function is approximately Gaussian. The preliminary results of the
fits to the differential cross-sections are given in
Table~\ref{gg:tab:results}.  No significant deviations with respect to
the QED expectations are found (all the parameters are compatible with
zero) and therefore 95\% confidence level limits are obtained by
renormalising the probability distribution of the fit parameter to the
physically allowed region. 
The asymmetric limits $x_{95}^{\pm}$ 
on the fitting parameter are obtained by:
\begin{equation} \frac{\int^{x_{95}^+}_0 \Gamma(x,\mu ,\sigma ) dx}
         {\int^{\infty   }_0 \Gamma(x,\mu ,\sigma ) dx} = 0.95 \; 
    \hspace{7mm}\mbox{and}\hspace{7mm}
    \frac{\int_{x_{95}^-}^0 \Gamma(x,\mu ,\sigma ) dx}
         {\int_{-\infty  }^0 \Gamma(x,\mu ,\sigma ) dx} = 0.95 \; , 
         \label{limeq}\end{equation}
where $\Gamma$ is a Gaussian with the central value and error of the fit
result denoted by $\mu$ and $\sigma$, respectively. This is equivalent
to the integration of a Gaussian probability function as a
function of the fit parameter. The 95 \% CL limits on the model parameters 
are derived from the limits on the 
fit parameters, e.g. the limit on $\Lambda_+$ is obtained as 
$[x_{95}^+(\Lambda^{-4}_{\pm})]^{-1/4}$.

\begin{figure}[hbtp]
\vspace*{-1cm}
   \begin{center}\mbox{
          \epsfxsize=15.0cm
           \epsffile{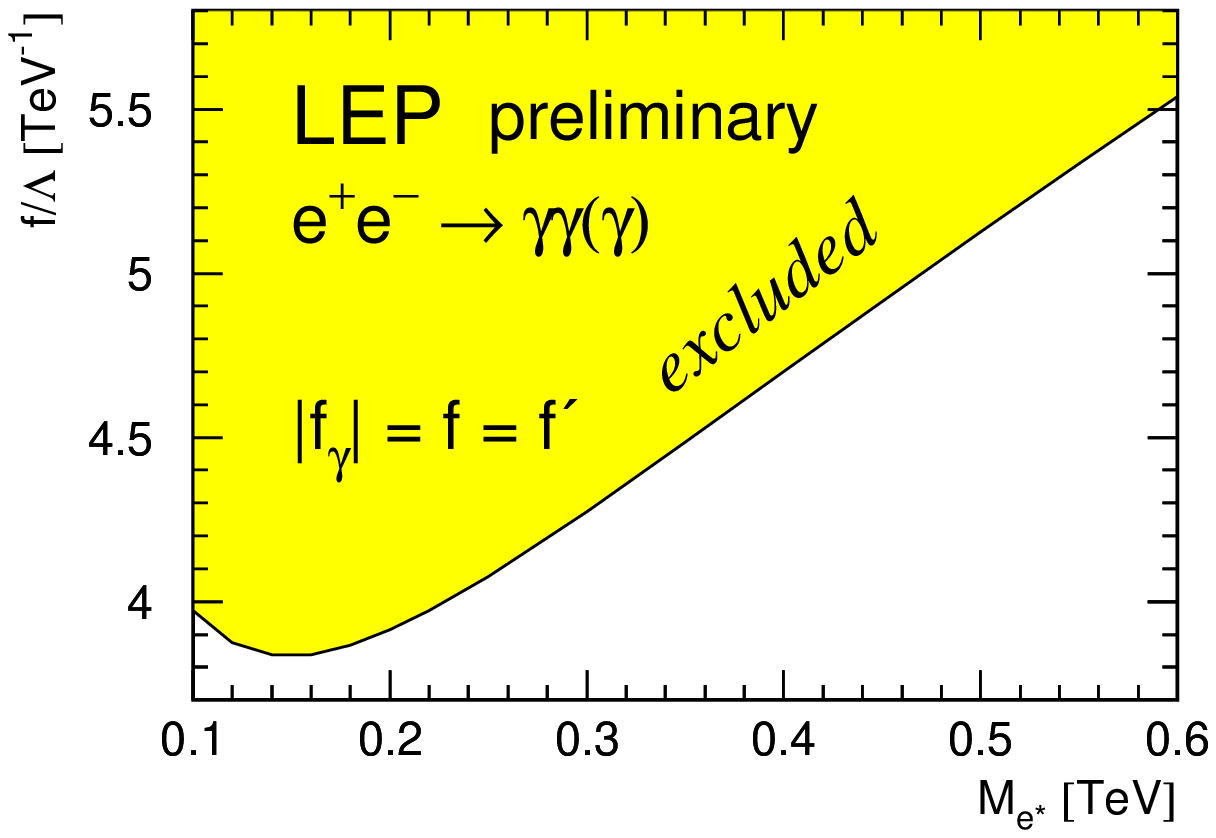}
           } \end{center}
\vspace{-1cm}
\caption{95\% CL limits on the coupling $f_\gamma/\Lambda$ of an excited
electron as a function of $\mestar$.
In the case of $f=f'$ it follows that $|f_\gamma| = f$.
It is assumed that $\Lambda=\mestar$.}
\label{gg:fig:estar}
\vspace*{-2cm}
   \begin{center}\mbox{
          \epsfxsize=15.0cm
           \epsffile{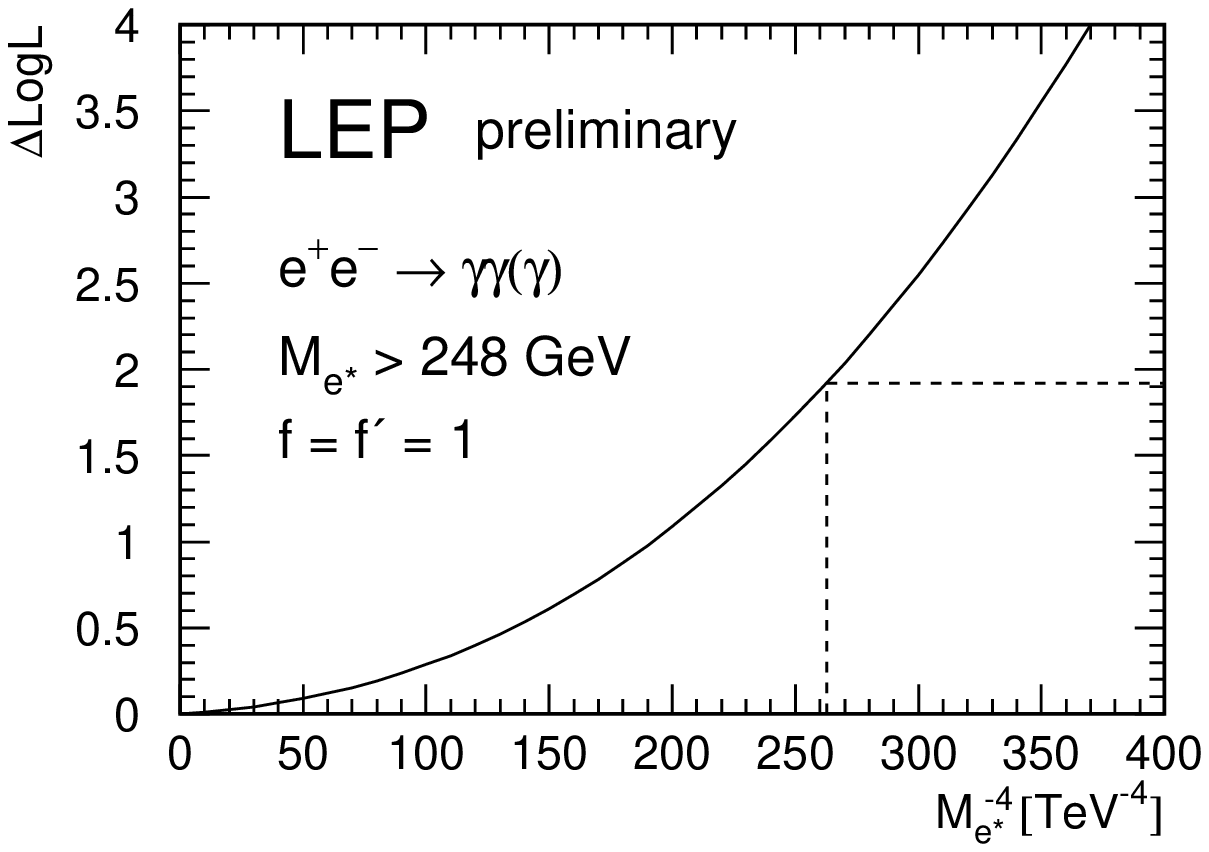}
           } \end{center}
\vspace{-1cm}
\caption{Log likelihood difference 
$\Delta\mbox{LogL} = -\ln{\cal L}+\ln{\cal L}_{\rm max}$
as a function of $\mestar^{-4}$. The coupling is fixed at $f = f' = 1$. 
The value corresponding to $\Delta\mbox{LogL} = 1.92$ is $\mestar$ = 248 \GeV.}
\label{gg:fig:ll}
\end{figure}

The only model with more than one free model parameter is the search 
for excited electrons. In this case only one out of the two parameters 
$f_\gamma$ and $\mestar$ is determined while the other is fixed.
It is assumed that $\Lambda=\mestar$. For limits on the coupling
$f_\gamma/\Lambda$ 
a scan over $\mestar$ is performed. The fit result at 
$\mestar = 200 \mbox{GeV}$ is included in Table~\ref{gg:tab:results},
limits for all masses are presented in Figure~\ref{gg:fig:estar}. 
For the determination of the excited electron mass 
the fit cannot be expressed in terms of a linear fit parameter.
For $|f_\gamma| =1$ the curve of the negative log likelihood, 
$\Delta\mbox{LogL}$, as a function of $\mestar$ is shown
in Figure \ref{gg:fig:ll}. The value corresponding to 
$\Delta\mbox{LogL} = 1.92$ is \mbox{$\mestar$ = 248 \GeV}.

\begin{table}[htb]
\begin{center}
\renewcommand{\arraystretch}{1.5}
\begin{tabular}{|c|c|r@{ }l|}\hline
Fit parameter & Fit result &
\multicolumn{2}{c|}{95\%\ CL limit [\GeV]}\\  \hline
 &  & $\Lambda_+ >$ &  392 \\
 \raisebox{2.2ex}[-2.2ex]{$\Lpm^{-4}$} &
 \raisebox{2.2ex}[-2.2ex]{$
 \left(-12.5{+25.1 \atop -24.7}\right)\cdot 10^{-12}$ \GeV$^{-4}$}
 & $\Lambda_- > $&  364  \\\hline
$\Lambda_7^{-6}$ & $ \left(-0.91{+1.81 \atop -1.78}\right)\cdot
                                  10^{-18}$ \GeV$^{-6}$
& $\Lambda_7 > $&  831   \\ \hline
\multicolumn{2}{|c|}{derived from $\Lambda_+$}& $\Lambda_6 > $&  1595   \\
\multicolumn{2}{|c|}{derived from $\Lambda_7$}& $\Lambda_8 > $&  23.3   \\\hline
 &  & $\lambda = +1$: $M_s >$ &  933  \\
 \raisebox{2.2ex}[-2.2ex]{$\lambda/M_s^4$} &
 \raisebox{2.2ex}[-2.2ex]{ $ \left(0.29{+0.57 \atop -0.58}\right)\cdot
                                         10^{-12}$ \GeV$^{-4} $ }
 & $\lambda = -1$: $M_s >$&  1010 \\ \hline
 $f_\gamma^4 (\mestar=200 \rm \GeV)$ & 
  $0.037{+0.202 \atop -0.198 }$ & 
 \multicolumn{2}{c|}{$f_\gamma/\Lambda <  3.9 \mbox{ \TeV}^{-1}$} \\ \hline
\end{tabular}
\caption[]{ The preliminary combined fit parameters 
and the 95$\%$  confidence level limits for the four LEP experiments.}
\label{gg:tab:results} 
\end{center} 
\end{table}

\section{Conclusion}
The LEP collaborations study the $\eeggga$ channel up to the highest
available centre-of-mass energies. The total cross-section results are
combined in terms of the ratios with respect to the QED expectations.
No deviations are found. The differential cross-sections are fit
following different parametrisations from models predicting deviations
from QED. No evidence for deviations is found and therefore combined
95\% confidence level limits are given.

\boldmath
\chapter{Fermion-Pair Production at \LEPII}
\label{sec-FF}
\unboldmath

\updates{ Unchanged w.r.t. summer 2003: Results are preliminary.

  Note that some recent publications~\cite{Abbiendi:2003dh,
  Abbiendi:2004vw, LEP2L3ffbar} are not yet included in this
  combination.  }

\section{Introduction}

\begin{table}[p]
 \begin{center}
 \begin{tabular}{|c|c|c|c|}
  \hline
   Year & Nominal Energy & Actual Energy & Luminosity \\
        &     $\GeV$     &    $\GeV$     &  pb$^{-1}$ \\
  \hline
  \hline
   1995 &      130       &    130.2      & $\sim 3  $ \\
        &      136       &    136.2      & $\sim 3  $ \\
  \cline{2-4}
        &  $133^{\ast}$ &     133.2      & $\sim 6  $ \\
  \hline
   1996 &      161       &    161.3      & $\sim 10 $ \\
        &      172       &    172.1      & $\sim 10 $ \\
  \cline{2-4}
        &  $167^{\ast}$ &     166.6      & $\sim 20 $ \\
  \hline
   1997 &      130       &    130.2      & $\sim 2  $ \\
        &      136       &    136.2      & $\sim 2  $ \\
        &      183       &    182.7      & $\sim 50 $ \\
  \hline
   1998 &      189       &    188.6      & $\sim 170$ \\
  \hline
   1999 &      192       &    191.6      & $\sim 30 $ \\
        &      196       &    195.5      & $\sim 80 $ \\
        &      200       &    199.5      & $\sim 80 $ \\
        &      202       &    201.6      & $\sim 40 $ \\
  \hline
   2000 &      205       &    204.9      & $\sim 80 $ \\
        &      207       &    206.7      & $\sim140 $ \\
  \hline
 \end{tabular}
 \end{center}
 \caption{The nominal and actual centre-of-mass energies for data
          collected during $\LEPII$ operation in each year. The approximate
          average luminosity analysed per experiment at each energy is also
          shown. Values marked with 
          a $^{\ast}$ are average energies for 1995 and 1996 used 
          for heavy flavour results. The data taken at nominal energies of
          130 GeV and 136 GeV in 1995 and 1997 are combined by most 
          experiments.}
 \label{ff:tab:ecms}
\end{table}

During the $\LEPII$ program LEP delivered collisions
at energies from $\sim 130$ $\GeV$ to $\sim 209$ $\GeV$. The 4 LEP experiments
have made measurements on the $\eeff$ process over this range of energies,
and a preliminary combination of these data is discussed in this note.
 
In the years 1995 through 1999 LEP delivered luminosity at a number of 
distinct centre-of-mass energy points. In 2000 most of the luminosity
was delivered close to 2 distinct energies, but there was also
a significant fraction of the luminosity delivered in, more-or-less, a 
continuum of energies. To facilitate the combination of the data,
the 4 LEP experiments all divided the data they collected in 2000 
into two energy bins: from 202.5 to 205.5 $\GeV$; and 205.5 $\GeV$ and above.
The nominal and actual centre-of-mass energies to which the LEP data are
averaged for each year are given in Table~\ref{ff:tab:ecms}.

A number of measurements on the process $\eeff$ exist and are combined.
The preliminary averages of cross-section and forward-backward asymmetry
measurements are discussed in Section \ref{ff:sec-ave-xsc-afb}.
The results presented in this section update those presented 
in~\cite{bib-EWEP-02}.
Complete results of the combinations are available on the 
web page~\cite{ff:ref:ffbar_web}.
In Section~\ref{ff:sec-dsdc} a preliminary average of the differential
cross-sections measurements, $\dsdc$, for the channels $\eeee$,
$\eemumu$ and $\eetautau$ is presented. 
In Section~\ref{ff:sec-hvflv} a preliminary combination of the
heavy flavour results $\Rb$, $\Rc$, $\Abb$ and $\Acc$ from $\LEPII$ is 
presented. In Section~\ref{ff:sec-interp} the combined results are interpreted
in terms of contact interactions and the exchange of $\Zprime$ bosons, the 
exchange of leptoquarks or squarks and the exchange of gravitons in large
extra dimensions. The results are summarised in section~\ref{ff:sec-conc}.

\section{Averages for Cross-sections and Asymmetries}
\label{ff:sec-ave-xsc-afb}

In this section the results of the preliminary combination of
cross-sections and asymmetries are given.
The individual experiments' analyses of cross-sections and forward-backward
asymmetries are discussed in~\cite{ff:ref:expts}. 

Cross-section results are combined for the $\eeqq$, $\eemumu$ and $\eetautau$ 
channels, forward-backward asymmetry measurements are combined for
the $\mumu$ and $\tautau$ final states. The averages are made for the
samples of events with high effective centre-of-mass energies, $\sqrt{\spr}$. 
\begin{figure}[tp]
 \begin{center}
 \mbox{\epsfig{file=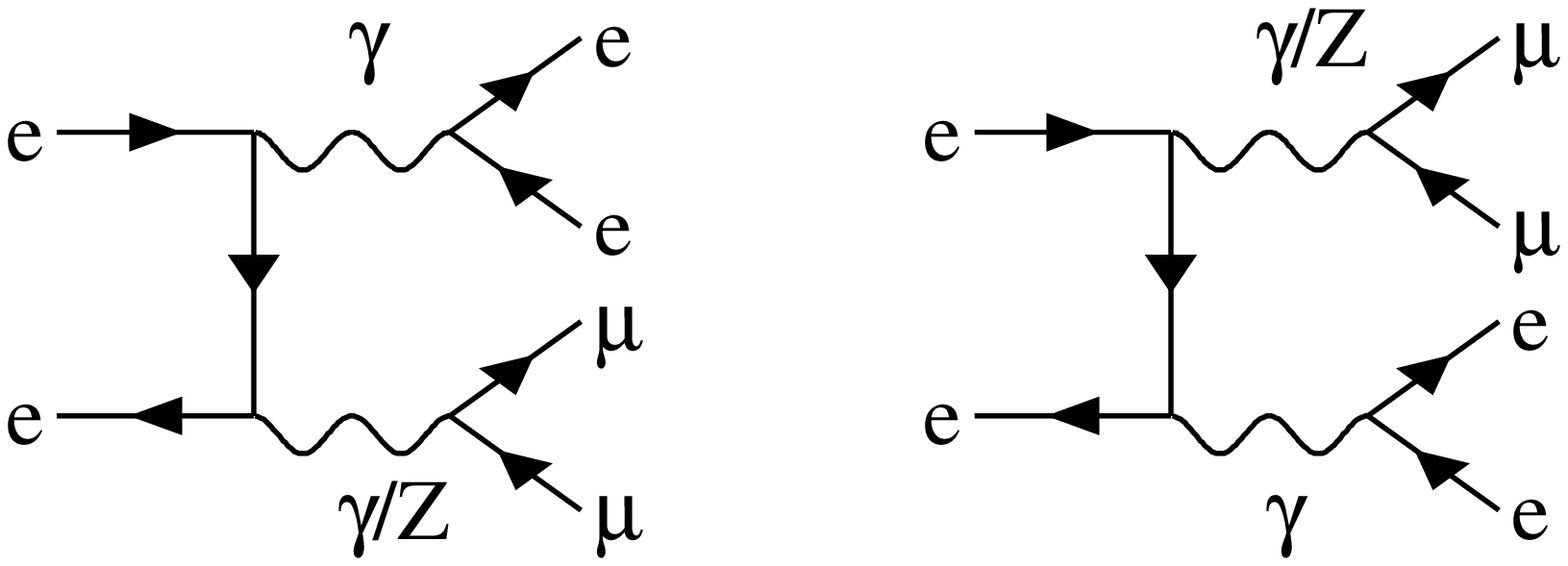,width=14cm}}
 \end{center}
 \caption{Diagrams leading to the the production of initial state non-singlet
          electron-positron pairs in $\eemumu$, which are considered as signal
          in the common signal definition.}
\label{ff:fig:isnspairs}
\end{figure}
Individual experiments have their own \ff\ signal definitions; corrections are
applied to bring the measurements to a common signal definitions:
\begin{itemize}
\item $\sqrt{\spr}$ is taken to be the mass of the
      $s$-channel propagator, with the $\ff$ signal being defined by the cut 
      $\sqrt{\spr/s} > 0.85$. 

\item ISR-FSR photon interference is subtracted to 
      render the propagator mass unambiguous.

\item Results are given for the full $4\pi$ angular acceptance. 

\item Initial state non-singlet diagrams \cite{ff:ref:lepffwrkshp}, 
      see for example Figure~\ref{ff:fig:isnspairs},
      which lead to events containing additional fermions pairs are considered
      as part of the two fermion signal. In such events, the additional 
      fermion pairs are typically lost down the beampipe of the experiments, 
      such that the visible event topologies are usually similar to a 
      difermion events with photons radiated from the initial state.
\end{itemize}
The corrected measurement of a cross-section or a forward backward asymmetry, 
$\mathrm{M_{LEP}}$, corresponding to the
common signal definition,  is computed from the
experimental measurement $\mathrm{M_{exp}}$,
\begin{eqnarray}
{\mathrm{M_{LEP}}} = {\mathrm{M_{exp}}} + ({\mathrm{P_{LEP}}} -
                                              {\mathrm{P_{exp}}}),
\end{eqnarray}
\noindent
where $\mathrm{P_{exp}}$ is the prediction for the measurement obtained 
for the experiments signal definition and $\mathrm{P_{LEP}}$ is the
prediction for the common signal definition. The predictions are computed with
ZFITTER~\cite{ff:ref:ZFITTER}.

In choosing a common signal definition there is a tension between the need to
have a definition which is practical to implement in event generators and
semi-analytical calculations, one which comes close to describing the 
underlying hard processes and one which most closely matches what is actually
measured in experiments. Different signal definitions represent different 
balances between these needs. To illustrate how different choices 
would effect the quoted results a second signal definition is 
studied by calculating different predictions using ZFITTER:
\begin{itemize}
 \item For dilepton events, $\sqrt{\spr}$ is taken to be 
       the bare invariant mass of the outgoing difermion pair (\ie,
       the invariant mass excluding all radiated photons). 
       
 \item For hadronic events, it is taken to be the mass of the $s$-channel 
       propagator. 

 \item In both cases, ISR-FSR photon interference is included and the signal
       is defined by the cut $\sqrt{\spr/s} > 0.85$. When calculating the 
       contribution to the hadronic cross-section due to ISR-FSR interference,
       since the propagator mass is ill-defined, it is replaced by the bare 
       $\qq$ mass.
\end{itemize}
The definition of the hadronic cross-section is close to that used to
define the signal for the heavy quark measurements given in 
Section~\ref{ff:sec-hvflv}.

Theoretical uncertainties associated with the Standard Model predictions 
for each of the measurements are not included during the averaging procedure,
but must be included when assessing the compatibility of the data with 
theoretical predictions.
The theoretical uncertainties on the Standard Model predictions amount to
$0.26\%$ on $\sigma(\qq)$, $0.4\%$ on $\sigma(\mumu)$ and
$\sigma(\tautau)$, $2\%$ on $\sigma(\ee)$, 
and 0.004 on the leptonic forward-backward 
asymmetries~\cite{ff:ref:lepffwrkshp}.

The average is performed using the best linear unbiased estimator
technique (BLUE)~\cite{common_bib:BLUE}, which is equivalent to a $\chi^{2}$ 
minimisation. All data from nominal centre-of-mass 
energies of 130--207 GeV are averaged at the same time.

Particular care is taken to ensure that the correlations between the 
hadronic cross-sections are reasonably estimated. 
The errors are broken down into 5 
categories, with the ensuing correlations accounted for in the combinations:
\begin{itemize}

\item[1)] The statistical uncertainty plus uncorrelated systematic 
uncertainties, combined in quadrature.

\item[2)] The systematic uncertainty for the final state X which is 
fully correlated between energy points for that experiment.

\item[3)] The systematic uncertainty for experiment Y which is fully 
correlated  between different final states for this energy point.

\item[4)] The systematic uncertainty for the final state X which is 
fully correlated between energy points  and between different experiments.

\item[5)] The systematic uncertainty which is fully correlated between 
energy points and between different experiments for all final states.
\end{itemize}
Uncertainties in the hadronic cross-sections arising from fragmentation
models and modelling of ISR are treated as fully correlated between 
experiments. Despite some differences between the models used and the 
methods of evaluating the errors in the different experiments, 
there are significant common elements in the estimation of these sources 
of uncertainty. 

New, preliminary, results from ALEPH are included in the average. The 
updated ALEPH measurements use a lower cut on the effective
centre-of-mass energy, which makes the signal definition of 
ALEPH closer to the combined LEP signal definition.

Table~\ref{ff:tab:xsafbres} gives the averaged cross-sections
and forward-backward asymmetries for all energies.
The differences in the results obtained when using predictions
of ZFITTER for the second signal definition are also given.
The differences are significant when compared to the precision obtained
from averaging together the measurements at all energies.
The $\chi^{2}$ per degree of freedom for the average of the $\LEPII$ $\ff$ 
data is $160/180$. Most correlations are rather small, with the largest 
components at any given pair of energies being between the hadronic 
cross-sections. The other off-diagonal terms in the correlation 
matrix are smaller than $10\%$. The correlation matrix between the 
averaged hadronic cross-sections at different centre-of-mass energies 
is given in Table~\ref{ff:tab:hadcorrel}.

\begin{figure}[p]
 \begin{center}
 \mbox{\epsfig{file=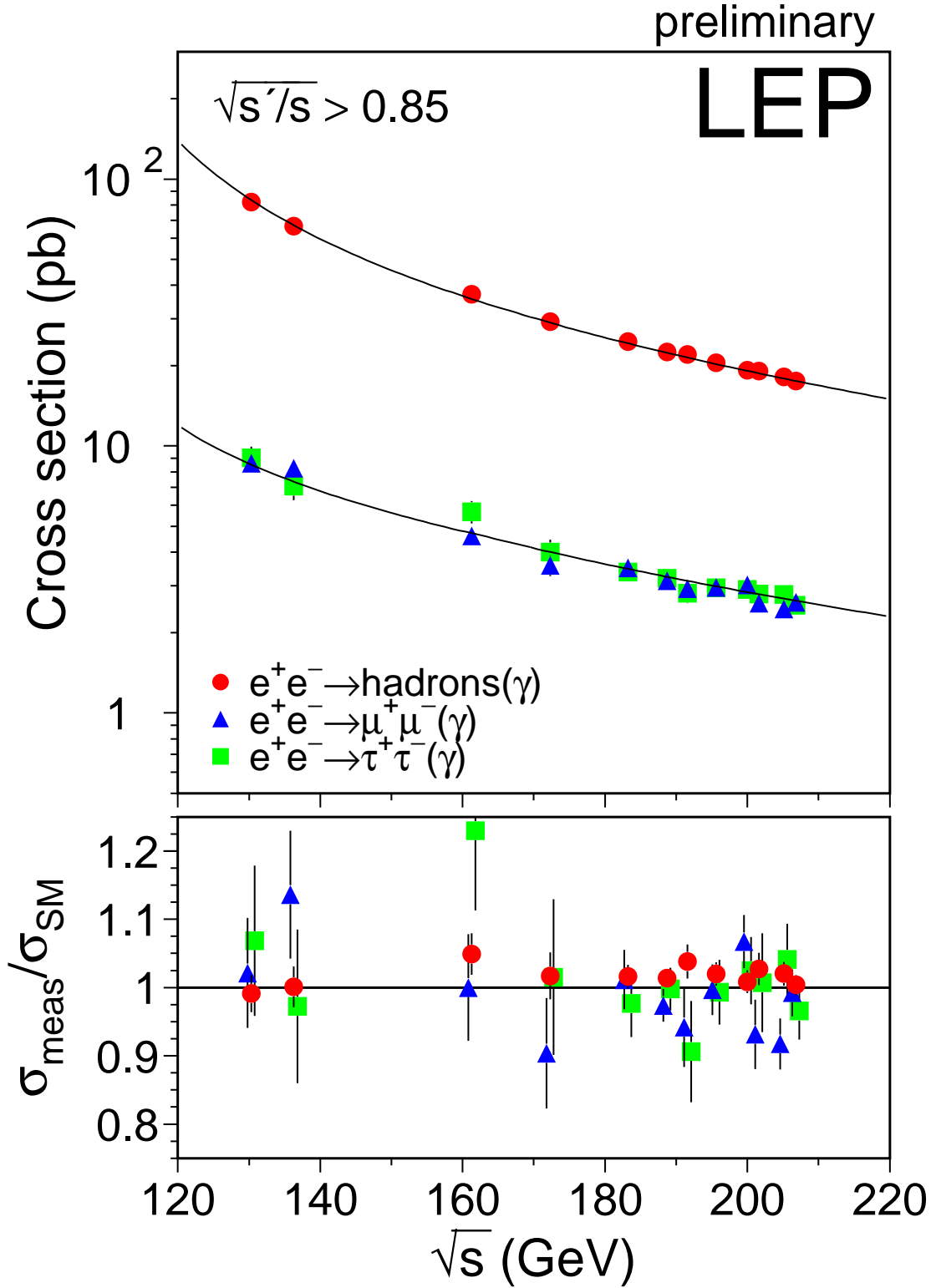,width=15cm}}
 \end{center}
 \caption{Preliminary combined LEP results on the cross-sections for 
          $\qq$, $\mumu$ and $\tautau$ final states, as a function of 
          centre-of-mass energy. The expectations of the SM, 
          computed with ZFITTER~\capcite{ff:ref:ZFITTER}, are shown as curves.
          The lower plot shows the ratio of the data divided by the SM.}
\label{ff:fig-xs_lep}
\end{figure}

\begin{figure}[p]
 \begin{center}
 \mbox{\epsfig{file=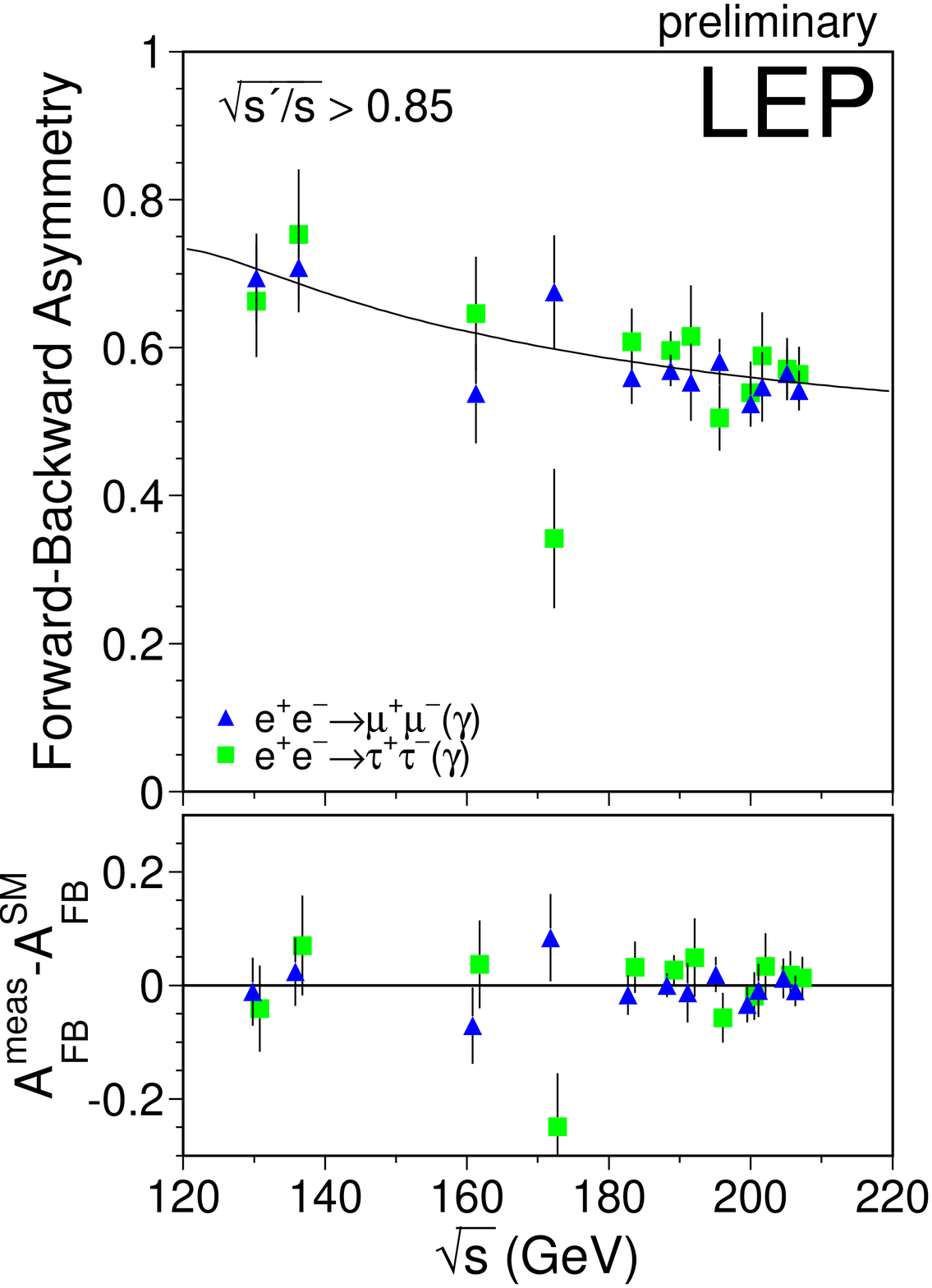,width=15cm}}
 \end{center}
 \caption{Preliminary combined LEP results on the forward-backward 
          asymmetry for $\mumu$  and $\tautau$ final states as a function of 
          centre-of-mass energy. The expectations of the SM 
          computed  with ZFITTER~\capcite{ff:ref:ZFITTER}, are shown as 
          curves. The lower plot shows differences between the data 
          and the SM.}
 \label{ff:fig-afb_lep}
\end{figure}

Figures~\ref{ff:fig-xs_lep} and~\ref{ff:fig-afb_lep} show the LEP 
averaged cross-sections and asymmetries, respectively, as a 
function of the centre-of-mass energy, together with the SM predictions. 
There is good agreement between the SM expectations and the measurements of the
individual experiments and the combined averages.
The cross-sections for hadronic final states at most of the energy points 
are somewhat above the SM expectations. Taking into account the correlations
between the data points and also taking into account the theoretical error
on the SM predictions, 
the ratio of the measured cross-sections to the SM expectations, averaged over 
all energies, is approximately a $1.7$ standard deviation excess. 
It is concluded that there is no significant evidence in the results of the
combinations for physics beyond the SM in the process $\eeff$.

\begin{table}[p]
 \begin{center}
 \begin{turn}{90}
 \begin{tabular}{cc}
 \begin{tabular}{|c|c|r@{$\pm$}l|c|c|}
 \hline
 $\sqrt{s}$ &          &
 \multicolumn{2}{c|}{Average} &
                    &
                               \\
 ($\GeV$) & Quantity   &
 \multicolumn{2}{|c|}{value}  &
 \multicolumn{1}{|c|}{SM} &
 \multicolumn{1}{|c|}{$\Delta$}  \\
 \hline\hline
  130 & $\sigma(q\overline{q})$                      & 82.1   &  2.2   & 82.8   & -0.3   \\
  130 & $\sigma(\mu^{+}\mu^{-})$                     &  8.62  &  0.68  &  8.44  & -0.33  \\
  130 & $\sigma(\tau^{+}\tau^{-})$                   &  9.02  &  0.93  &  8.44  & -0.11  \\
  130 & $\mathrm{A_{FB}}(\mu^{+}\mu^{-})$            &  0.694 &  0.060 &  0.705 &  0.012 \\
  130 & $\mathrm{A_{FB}}(\tau^{+}\tau^{-})$          &  0.663 &  0.076 &  0.704 &  0.012 \\
 \hline
  136 & $\sigma(q\overline{q})$                      & 66.7   &  2.0   & 66.6   & -0.2   \\
  136 & $\sigma(\mu^{+}\mu^{-})$                     &  8.27  &  0.67  &  7.28  & -0.28  \\
  136 & $\sigma(\tau^{+}\tau^{-})$                   &  7.078 &  0.820 &  7.279 & -0.091 \\
  136 & $\mathrm{A_{FB}}(\mu^{+}\mu^{-})$            &  0.708 &  0.060 &  0.684 &  0.013 \\
  136 & $\mathrm{A_{FB}}(\tau^{+}\tau^{-})$          &  0.753 &  0.088 &  0.683 &  0.014 \\
 \hline
  161 & $\sigma(q\overline{q})$                      & 37.0   &  1.1   & 35.2   & -0.1   \\
  161 & $\sigma(\mu^{+}\mu^{-})$                     &  4.61  &  0.36  &  4.61  & -0.18  \\
  161 & $\sigma(\tau^{+}\tau^{-})$                   &  5.67  &  0.54  &  4.61  & -0.06  \\
  161 & $\mathrm{A_{FB}}(\mu^{+}\mu^{-})$            &  0.538 &  0.067 &  0.609 &  0.017 \\
  161 & $\mathrm{A_{FB}}(\tau^{+}\tau^{-})$          &  0.646 &  0.077 &  0.609 &  0.016 \\
 \hline
  172 & $\sigma(q\overline{q})$                      & 29.23  &  0.99  & 28.74  & -0.12  \\
  172 & $\sigma(\mu^{+}\mu^{-})$                     &  3.57  &  0.32  &  3.95  & -0.16  \\
  172 & $\sigma(\tau^{+}\tau^{-})$                   &  4.01  &  0.45  &  3.95  & -0.05  \\
  172 & $\mathrm{A_{FB}}(\mu^{+}\mu^{-})$            &  0.675 &  0.077 &  0.591 &  0.018 \\
  172 & $\mathrm{A_{FB}}(\tau^{+}\tau^{-})$          &  0.342 &  0.094 &  0.591 &  0.017 \\
 \hline
  183 & $\sigma(q\overline{q})$                      & 24.59  &  0.42  & 24.20  & -0.11  \\
  183 & $\sigma(\mu^{+}\mu^{-})$                     &  3.49  &  0.15  &  3.45  & -0.14  \\
  183 & $\sigma(\tau^{+}\tau^{-})$                   &  3.37  &  0.17  &  3.45  & -0.05  \\
  183 & $\mathrm{A_{FB}}(\mu^{+}\mu^{-})$            &  0.559 &  0.035 &  0.576 &  0.018 \\
  183 & $\mathrm{A_{FB}}(\tau^{+}\tau^{-})$          &  0.608 &  0.045 &  0.576 &  0.018 \\
 \hline
  189 & $\sigma(q\overline{q})$                      & 22.47  &  0.24  & 22.156 & -0.101 \\
  189 & $\sigma(\mu^{+}\mu^{-})$                     &  3.123 &  0.076 &  3.207 & -0.131 \\
  189 & $\sigma(\tau^{+}\tau^{-})$                   &  3.20  &  0.10  &  3.20  & -0.048 \\
  189 & $\mathrm{A_{FB}}(\mu^{+}\mu^{-})$            &  0.569 &  0.021 &  0.569 &  0.019 \\
  189 & $\mathrm{A_{FB}}(\tau^{+}\tau^{-})$          &  0.596 &  0.026 &  0.569 &  0.018 \\
 \hline
 \end{tabular}
 &
 \begin{tabular}{|c|c|r@{$\pm$}l|c|c|}
 \hline
 $\sqrt{s}$ &          &
 \multicolumn{2}{c|}{Average} &
                    &
                               \\
 ($\GeV$) & Quantity   &
 \multicolumn{2}{|c|}{value}  &
 \multicolumn{1}{|c|}{SM} &
 \multicolumn{1}{|c|}{$\Delta$}  \\
 \hline\hline
  192 & $\sigma(q\overline{q})$                      & 22.05  &  0.53  & 21.24  & -0.10  \\
  192 & $\sigma(\mu^{+}\mu^{-})$                     &  2.92  &  0.18  &  3.10  & -0.13  \\
  192 & $\sigma(\tau^{+}\tau^{-})$                   &  2.81  &  0.23  &  3.10  & -0.05  \\
  192 & $\mathrm{A_{FB}}(\mu^{+}\mu^{-})$            &  0.553 &  0.051 &  0.566 &  0.019 \\
  192 & $\mathrm{A_{FB}}(\tau^{+}\tau^{-})$          &  0.615 &  0.069 &  0.566 &  0.019 \\
 \hline
  196 & $\sigma(q\overline{q})$                      & 20.53  &  0.34  & 20.13  & -0.09  \\
  196 & $\sigma(\mu^{+}\mu^{-})$                     &  2.94  &  0.11  &  2.96  & -0.12  \\
  196 & $\sigma(\tau^{+}\tau^{-})$                   &  2.94  &  0.14  &  2.96  & -0.05  \\
  196 & $\mathrm{A_{FB}}(\mu^{+}\mu^{-})$            &  0.581 &  0.031 &  0.562 &  0.019 \\
  196 & $\mathrm{A_{FB}}(\tau^{+}\tau^{-})$          &  0.505 &  0.044 &  0.562 &  0.019 \\
 \hline
  200 & $\sigma(q\overline{q})$                      & 19.25  &  0.32  & 19.09  & -0.09  \\
  200 & $\sigma(\mu^{+}\mu^{-})$                     &  3.02  &  0.11  &  2.83  & -0.12  \\
  200 & $\sigma(\tau^{+}\tau^{-})$                   &  2.90  &  0.14  &  2.83  & -0.04  \\
  200 & $\mathrm{A_{FB}}(\mu^{+}\mu^{-})$            &  0.524 &  0.031 &  0.558 &  0.019 \\
  200 & $\mathrm{A_{FB}}(\tau^{+}\tau^{-})$          &  0.539 &  0.042 &  0.558 &  0.019 \\
 \hline
  202 & $\sigma(q\overline{q})$                      & 19.07  &  0.44  & 18.57  & -0.09  \\
  202 & $\sigma(\mu^{+}\mu^{-})$                     &  2.58  &  0.14  &  2.77  & -0.12  \\
  202 & $\sigma(\tau^{+}\tau^{-})$                   &  2.79  &  0.20  &  2.77  & -0.04  \\
  202 & $\mathrm{A_{FB}}(\mu^{+}\mu^{-})$            &  0.547 &  0.047 &  0.556 &  0.020 \\
  202 & $\mathrm{A_{FB}}(\tau^{+}\tau^{-})$          &  0.589 &  0.059 &  0.556 &  0.019 \\
 \hline
  205 & $\sigma(q\overline{q})$                      & 18.17  &  0.31  & 17.81  & -0.09  \\
  205 & $\sigma(\mu^{+}\mu^{-})$                     &  2.45  &  0.10  &  2.67  & -0.11  \\
  205 & $\sigma(\tau^{+}\tau^{-})$                   &  2.78  &  0.14  &  2.67  & -0.042 \\
  205 & $\mathrm{A_{FB}}(\mu^{+}\mu^{-})$            &  0.565 &  0.035 &  0.553 &  0.020 \\
  205 & $\mathrm{A_{FB}}(\tau^{+}\tau^{-})$          &  0.571 &  0.042 &  0.553 &  0.019 \\
 \hline
  207 & $\sigma(q\overline{q})$                      & 17.49  &  0.26  & 17.42  & -0.08  \\
  207 & $\sigma(\mu^{+}\mu^{-})$                     &  2.595 &  0.088 &  2.623 & -0.111 \\
  207 & $\sigma(\tau^{+}\tau^{-})$                   &  2.53  &  0.11  &  2.62  & -0.04  \\
  207 & $\mathrm{A_{FB}}(\mu^{+}\mu^{-})$            &  0.542 &  0.027 &  0.552 &  0.020 \\
  207 & $\mathrm{A_{FB}}(\tau^{+}\tau^{-})$          &  0.564 &  0.037 &  0.551 &  0.019 \\
 \hline
 \end{tabular}
 \end{tabular}
 \end{turn}
 \end{center}
\caption{Preliminary combined LEP results for $\eeff$, with cross
 section quoted in units of picobarn.
 All the results correspond to the first signal definition. The Standard Model
 predictions are from ZFITTER \capcite{ff:ref:ZFITTER}.
 The difference, $\Delta$, in the predictions of ZFITTER for 
 second definition relative to the first are given in the final column.
 The quoted uncertainties do not include the theoretical 
 uncertainties on the corrections discussed in the text.}
\label{ff:tab:xsafbres}
\end{table}
\begin{table}[p]
 \vskip 3cm
 \begin{center}
 \begin{turn}{90}
 \begin{tabular}{|c|c|c|c|c|c|c|c|c|c|c|c|c|}
 \hline
 $\begin{array}[b]{c}\roots \\ (\GeV) \end{array}$
       & 130    & 136    & 161    & 172    & 183    & 189    & 192    & 196    & 200    & 202    & 205    & 207    \\
 \hline\hline
 130 &  1.000 &  0.071 &  0.080 &  0.072 &  0.114 &  0.146 &  0.077 &  0.105 &  0.120 &  0.086 &  0.117 &  0.138 \\
 136 &  0.071 &  1.000 &  0.075 &  0.067 &  0.106 &  0.135 &  0.071 &  0.097 &  0.110 &  0.079 &  0.109 &  0.128 \\
 161 &  0.080 &  0.075 &  1.000 &  0.077 &  0.120 &  0.153 &  0.080 &  0.110 &  0.125 &  0.090 &  0.124 &  0.145 \\
 172 &  0.072 &  0.067 &  0.077 &  1.000 &  0.108 &  0.137 &  0.072 &  0.099 &  0.112 &  0.081 &  0.111 &  0.130 \\
 183 &  0.114 &  0.106 &  0.120 &  0.108 &  1.000 &  0.223 &  0.117 &  0.158 &  0.182 &  0.129 &  0.176 &  0.208 \\
 189 &  0.146 &  0.135 &  0.153 &  0.137 &  0.223 &  1.000 &  0.151 &  0.206 &  0.235 &  0.168 &  0.226 &  0.268 \\
 192 &  0.077 &  0.071 &  0.080 &  0.072 &  0.117 &  0.151 &  1.000 &  0.109 &  0.126 &  0.090 &  0.118 &  0.138 \\
 196 &  0.105 &  0.097 &  0.110 &  0.099 &  0.158 &  0.206 &  0.109 &  1.000 &  0.169 &  0.122 &  0.162 &  0.190 \\
 200 &  0.120 &  0.110 &  0.125 &  0.112 &  0.182 &  0.235 &  0.126 &  0.169 &  1.000 &  0.140 &  0.184 &  0.215 \\
 202 &  0.086 &  0.079 &  0.090 &  0.081 &  0.129 &  0.168 &  0.090 &  0.122 &  0.140 &  1.000 &  0.132 &  0.153 \\
 205 &  0.117 &  0.109 &  0.124 &  0.111 &  0.176 &  0.226 &  0.118 &  0.162 &  0.184 &  0.132 &  1.000 &  0.213 \\
 207 &  0.138 &  0.128 &  0.145 &  0.130 &  0.208 &  0.268 &  0.138 &  0.190 &  0.215 &  0.153 &  0.213 &  1.000 \\
 \hline
 \end{tabular}
 \end{turn}
 \end{center}
\caption{The correlation coefficients between averaged hadronic cross-sections
         at different energies.}
\label{ff:tab:hadcorrel}
 \vskip 5cm
\end{table}

\clearpage

\section{Averages for Differential Cross-sections}
\label{ff:sec-dsdc}

\subsection{{\boldmath{\ee}} final state}
\label{ff:sec-dsdc-ee}

The LEP experiments have measured the differential cross-section, $\dsdc$, 
for the $\eeee$ channel.%
A preliminary combination of these results is made by performing 
a $\chi^{2}$ fit to the measured differential cross-sections,
using the statistical errors as given by the experiments. In contrast
to the muon and tau  channels (Section~\ref{ff:sec-dsdc-mm-tt})
the higher statistics makes the use of expected statistical errors unnecessary.
The combination includes data from 189 $\GeV$ to 207 $\GeV$ from
all experiments but DELPHI. 
The data used in the 
combination are summarised in Table~\ref{ff:tab:ee_inputs}. 

Each experiment's data are binned according to an agreed common definition,
which takes into account the large forward peak of Bhabha scattering:
\begin{itemize}
 \item 10 bins for $\cos\theta$ between $0.0$  and $0.90$ and
 \item  5 bins for $\cos\theta$ between $-0.90$ and $0.0$
\end{itemize}
at each energy. The scattering angle, $\theta$, is
the angle of the negative lepton with respect to the incoming electron 
direction in the lab coordinate system. 
The outer acceptances of the most 
forward and most backward bins for which the experiments present
their data are different. The ranges in $\cos\theta$ of 
the individual experiments and the average are given in 
Table~\ref{ff:tab:acptee}. Except for the binning, each experiment uses their 
own signal definition, for example different experiments have different
acollinearity cuts to select events.
The signal definition used for the LEP average 
corresponds to an acollinearity cut of $\rm 10^{\circ}$. The experimental
measurements are corrected to the common signal definition following the
procedure described in Section~\ref{ff:sec-ave-xsc-afb}. The theoretical 
predictions are taken from the Monte Carlo event generator 
BHWIDE~\cite{ff:ref:BHWIDE}.

Correlated systematic errors between different experiments, energies and bins
at the same energy, arising from uncertainties on the overall normalisation,
and from migration of events between forward and backward bins with the same
absolute value of $\cos\theta$ due to uncertainties in the corrections for
charge confusion, were considered in the averaging procedure.

\begin{figure}[p]
 \begin{center}
  \epsfig{file=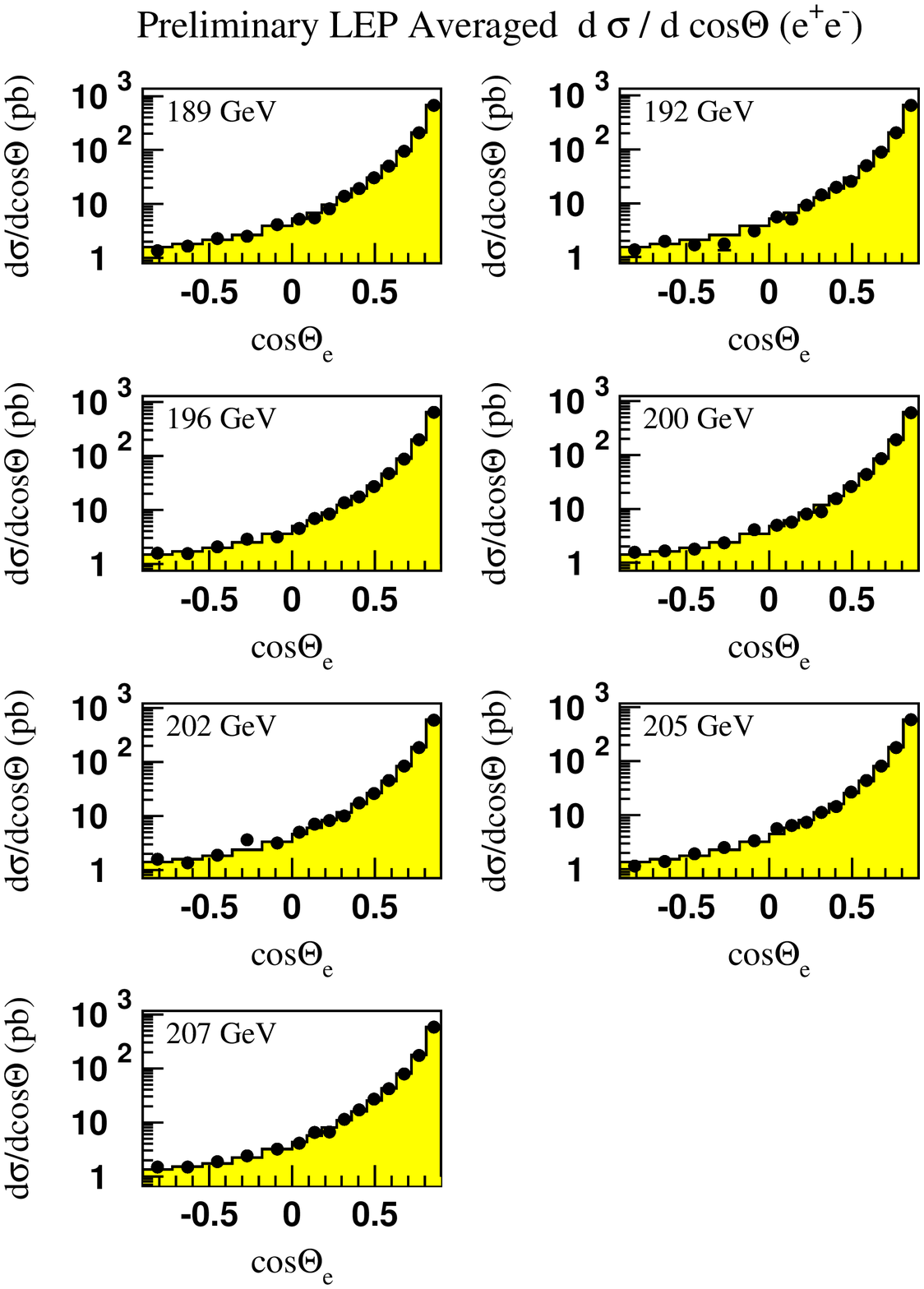,width=0.98\textwidth}
 \end{center}
 \caption{LEP averaged differential cross-sections for $\eeee$ at
          energies of 189--207 $\GeV$. The SM
          predictions, shown as solid histograms, are computed with
          BHWIDE~\capcite{ff:ref:BHWIDE}.}
 \label{ff:fig:dsdc-res-ee}
 \vskip 2cm 
\end{figure}

An average for all energies between 189--207 $\GeV$ is performed.
The results of the averages are shown in Figure~\ref{ff:fig:dsdc-res-ee}.
The $\chi^{2}$ per degree of freedom for the average is $190.8/189$.

The correlations between bins in the average are well below $5\%$ of the total
error on the averages in each bin for most of the cases, and exceed $10\%$ for
the most forward bin for the energy points with the highest accumulated
statistics.
The agreement between the averaged data and the predictions from the Monte
Carlo generator BHWIDE is good.
\subsection{{\boldmath{\mumu}}and {\boldmath{\tautau}} final states}
\label{ff:sec-dsdc-mm-tt}

The LEP experiments have measured the differential cross-section, $\dsdc$, 
for the $\eemumu$ and $\eetautau$ channels for samples of events 
with high effective centre-of-mass energy, $\sqrt{s'/s}>0.85$. 
A preliminary combination of these results is
made using the BLUE technique. The statistical error associated with
each measurement is taken as the expected statistical error on the 
differential cross-section, computed from the expected number of events 
in each bin for each experiment. Using a Monte Carlo simulation it has 
been shown that this method provides a good approximation to the exact 
likelihood method based on Poisson statistics~\cite{ff:ref:lepff-osaka}.

The combination includes data from 183 $\GeV$ to 207 $\GeV$, but not all 
experiments
provided data at all energies. 
The data used in the combination are summarised in
Table~\ref{ff:tab:inputs}.

Each experiment's data are binned in 10 bins of $\cos\theta$ at each 
energy, using their own signal definition. The scattering angle, $\theta$, is
the angle of the negative lepton with respect to the incoming electron 
direction in the lab coordinate system. The outer acceptances of the most 
forward and most backward bins for which the four experiments present 
their data are different. This was accounted for as part of the correction to 
a common signal definition. The ranges in $\cos\theta$ for the measurements of 
the individual experiments and the average are given in 
Table~\ref{ff:tab:acpt}. The signal definition used corresponded 
to the first definition given in Section~\ref{ff:sec-ave-xsc-afb}.

Correlated systematic errors between different experiments, channels and 
energies, arising from uncertainties on the overall normalisation are 
considered in the averaging procedure.
All data from all energies are combined in a single fit to obtain
averages at each centre-of-mass energy yielding the full covariance matrix 
between the different measurements at all energies.

\begin{figure}[p]
 \begin{center}
  \epsfig{file=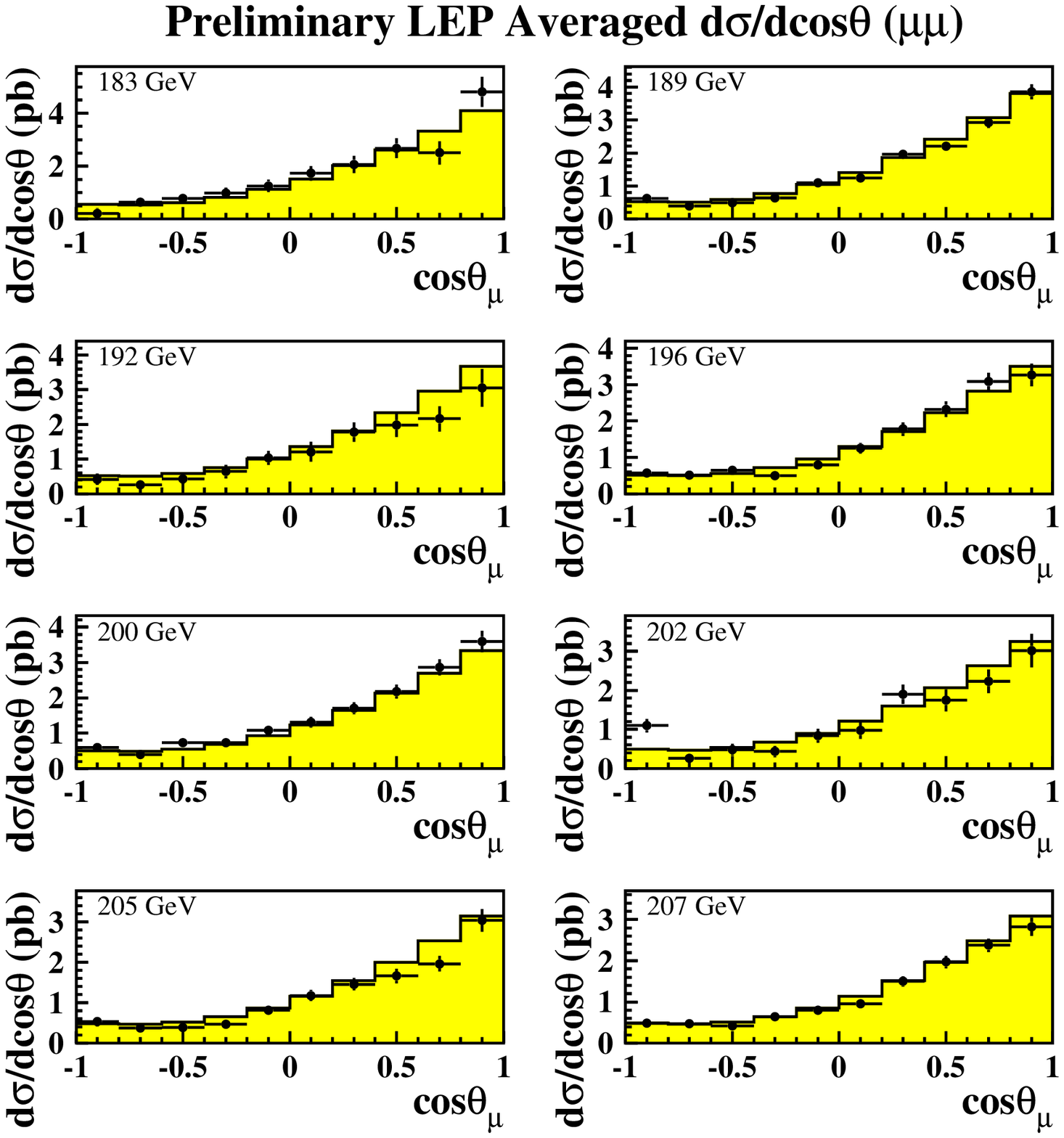,width=0.98\textwidth}
 \end{center}
 \caption{LEP averaged differential cross-sections for $\eemumu$ at
          energies of 183--207 $\GeV$. The SM
          predictions, shown as solid histograms, are computed with
          ZFITTER~\capcite{ff:ref:ZFITTER}.}
 \label{ff:fig:dsdc-res-mm}
 \vskip 2cm 
\end{figure}

\begin{figure}[p]
 \begin{center}
  \epsfig{file=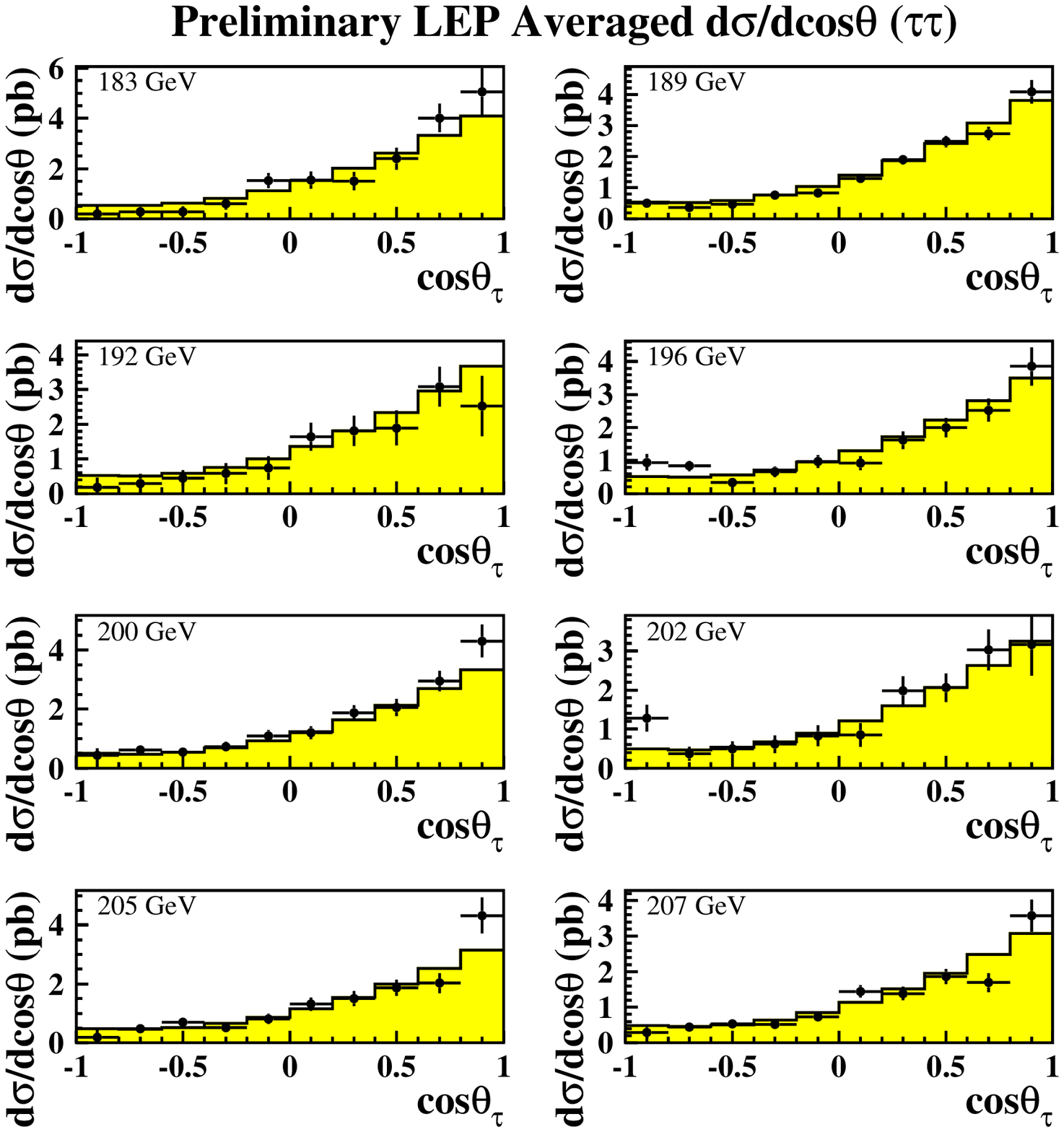,width=0.98\textwidth}
 \end{center}
 \caption{LEP averaged differential cross-sections for $\eetautau$ at 
          energies of 183--207 $\GeV$. The SM
          predictions, shown as solid histograms, are computed with
          ZFITTER~\capcite{ff:ref:ZFITTER}.}
 \label{ff:fig:dsdc-res-tt}
 \vskip 2cm 
\end{figure}

The results of the averages are shown in Figures~\ref{ff:fig:dsdc-res-mm} 
and~\ref{ff:fig:dsdc-res-tt}. 
The correlations between bins in the average are less that 
$2\%$ of the total error on the averages in each bin.
Overall the agreement between the averaged data and the predictions
is reasonable, with a $\chi^{2}$ of $200$ for $160$ degrees of freedom. 
At 202 $\GeV$ the measured differential cross-sections in the most backward 
bins, $-1.00 < \cos\theta < 0.8$, for both muon and tau final states are 
above the predictions. The data at 202 $\GeV$ suffer
from rather low delivered luminosity, with less than 4 events
expected in each experiment in each channel in this backward 
$\cos\theta$ bin. The agreement between the data
and the predictions in the same $\cos\theta$ bin is more consistent at 
higher energies. 

\begin{table}[htbp]
 \begin{center}
 \begin{tabular}{|l|cccc|}
 \hline
                     & \multicolumn{4}{|c|}{$\eeee$}             \\
 \cline{2-5}
  $\sqrt{s}$($\GeV$) &      A   &      D   &      L   &      O   \\
 \hline 
  189                & {\sc{P}} & {\sc{-}} & {\sc{P}} & {\sc{F}} \\
 \hline
  192--202           & {\sc{P}} & {\sc{-}} & {\sc{P}} & {\sc{P}} \\
 \hline
  205--207           & {\sc{P}} & {\sc{-}} & {\sc{P}} & {\sc{P}} \\
 \hline
 \end{tabular}
 \end{center}
 \caption{Differential cross-section data provided by the LEP 
          collaborations (ALEPH, DELPHI, L3 and OPAL) for $\eeee$.
          Data indicated with {\sc{F}} are final, published data.
          Data marked with {\sc{P}} are preliminary. 
          Data marked with a {\sc{-}} were not available for combination.}
 \label{ff:tab:ee_inputs}
\end{table}
\begin{table}[htbp]
 \begin{center}
 \begin{tabular}{|l|c|c|}
  \hline
  Experiment                       & $\cos\theta_{min}$ & $\cos\theta_{max}$ \\
  \hline
  \hline 
   ALEPH  ($\sqrt{s'/s}>0.85$)     &    $-0.90$         &     $0.90$         \\
   L3     (acol. $<\ 25^{\circ}$)  &    $-0.72$         &     $0.72$         \\
   OPAL   (acol. $<\ 10^{\circ}$)  &    $-0.90$         &     $0.90$         \\
  \hline
  \hline
   Average (acol. $<\ 10^{\circ}$) &    $-0.90$         &     $0.90$         \\
  \hline
 \end{tabular}
 \end{center}
 \caption{The acceptances for which experimental data are presented 
          for the $\eeee$ channel
          and the acceptance for the LEP average.}
 \label{ff:tab:acptee}
\end{table}
\begin{table}[htbp]
 \begin{center}
 \begin{tabular}{|l|cccc|cccc|}
 \hline
                     & \multicolumn{4}{|c|}{$\eemumu$}           
                     & \multicolumn{4}{|c|}{$\eetautau$}         \\
 \cline{2-9}
  $\sqrt{s}$($\GeV$) &      A   &      D   &      L   &      O   
                     &      A   &      D   &      L   &      O   \\
 \hline 
 \hline
  183                & {\sc{-}} & {\sc{F}} & {\sc{-}} & {\sc{F}} 
                     & {\sc{-}} & {\sc{F}} & {\sc{-}} & {\sc{F}} \\
 \hline
  189                & {\sc{P}} & {\sc{F}} & {\sc{F}} & {\sc{F}}  
                     & {\sc{P}} & {\sc{F}} & {\sc{F}} & {\sc{F}} \\
 \hline
  192--202           & {\sc{P}} & {\sc{P}} & {\sc{P}} & {\sc{P}} 
                     & {\sc{P}} & {\sc{P}} & {\sc{-}} & {\sc{P}} \\
 \hline
  205--207           & {\sc{P}} & {\sc{P}} & {\sc{P}} & {\sc{P}} 
                     & {\sc{P}} & {\sc{P}} & {\sc{-}} & {\sc{P}} \\
 \hline
 \end{tabular}
 \end{center}
 \caption{Differential cross-section data provided by the LEP 
          collaborations (ALEPH, DELPHI, L3 and OPAL) for $\eemumu$ and 
          $\eetautau$ combination at different centre-of-mass energies. 
          Data indicated with {\sc{F}} are final, published data. 
          Data marked with {\sc{P}} are preliminary. 
          Data marked with a {\sc{-}} were not available for combination.}
 \label{ff:tab:inputs}
\end{table}
\begin{table}[htbp]
 \begin{center}
 \begin{tabular}{|l|c|c|}
  \hline
  Experiment                   & $\cos\theta_{min}$ & $\cos\theta_{max}$ \\
  \hline
  \hline 
   ALEPH                       &    $-0.95$         &     $0.95$         \\
   DELPHI ($\eemumu$ 183)      &    $-0.94$         &     $0.94$         \\
   DELPHI ($\eemumu$ 189--207) &    $-0.97$         &     $0.97$         \\
   DELPHI ($\eetautau$)        &    $-0.96$         &     $0.96$         \\
   L3                          &    $-0.90$         &     $0.90$         \\
   OPAL                        &    $-1.00$         &     $1.00$         \\
  \hline
  \hline
   Average                     &    $-1.00$         &     $1.00$         \\
  \hline
 \end{tabular}
 \end{center}
 \caption{The acceptances for which experimental data are presented 
          and the acceptance for the LEP average.
          For DELPHI the acceptance is shown for the different channels and 
          for the muons for different centre of mass energies. For all other
          experiments the acceptance is the same for muon and tau-lepton 
          channels and for all energies provided.}
 \label{ff:tab:acpt}
\end{table}

\clearpage

\section{Averages for Heavy Flavour Measurements}
\label{ff:sec-hvflv}

This section presents a preliminary combination of both 
published~\cite{ff:ref:hfpublished}
and preliminary~\cite{ff:ref:hfpreliminary} measurements of the 
ratios cross section ratios $R_{\mathrm{q}}$ defined as  
${\mathrm{\frac{\sigma_{q \overline {q} } }{\sigma_{had}}}}$
for b and c production, $\Rb$ and $\Rc$,  
and the forward-backward asymmetries, $\Abb$ and 
$\Acc$, from the LEP collaborations at centre-of-mass 
energies in the range of 130 $\GeV$ to 207 $\GeV$. 
Table~\ref{ff:tab:hfinput} summarises all the inputs that have been combined so far.

A common signal definition is defined for all the measurements, requiring:
\begin{list}{$\bullet$}{\setlength{\itemsep}{0ex}
                        \setlength{\parsep}{0ex}
                        \setlength{\topsep}{0ex}}
 \item{an effective centre-of-mass energy $\sqrt{s^{\prime}} > 0.85 \sqrt{s}$}
 \item{no subtraction of ISR and FSR photon interference contribution and}
 \item{extrapolation to full angular acceptance.}
\end{list}
Systematic errors are divided into three categories: uncorrelated errors, 
errors correlated between the measurements of each experiment, and 
errors common to all experiments. 

Due to the fact that $\Rc$ measurements are only provided by a single 
experiment and are strongly correlated with $\Rb$ measurements, 
it was decided to fit the b sector and c sector separately, 
the other flavour's measurements being fixed to their Standard Model 
predictions. 
In addition, these fitted values are used to set limits upon physics beyond 
the Standard Model, such as contact term interactions, in which only one 
quark flavour is assumed to be effected by the new physics during each fit,
therefore this averaging method is consistent with the interpretations.

Full details concerning the combination procedure
can be found in~\cite{ff:ref:hfconfnote}. 

\begin{figure}[p]
\begin{center}
\mbox{\epsfig{file=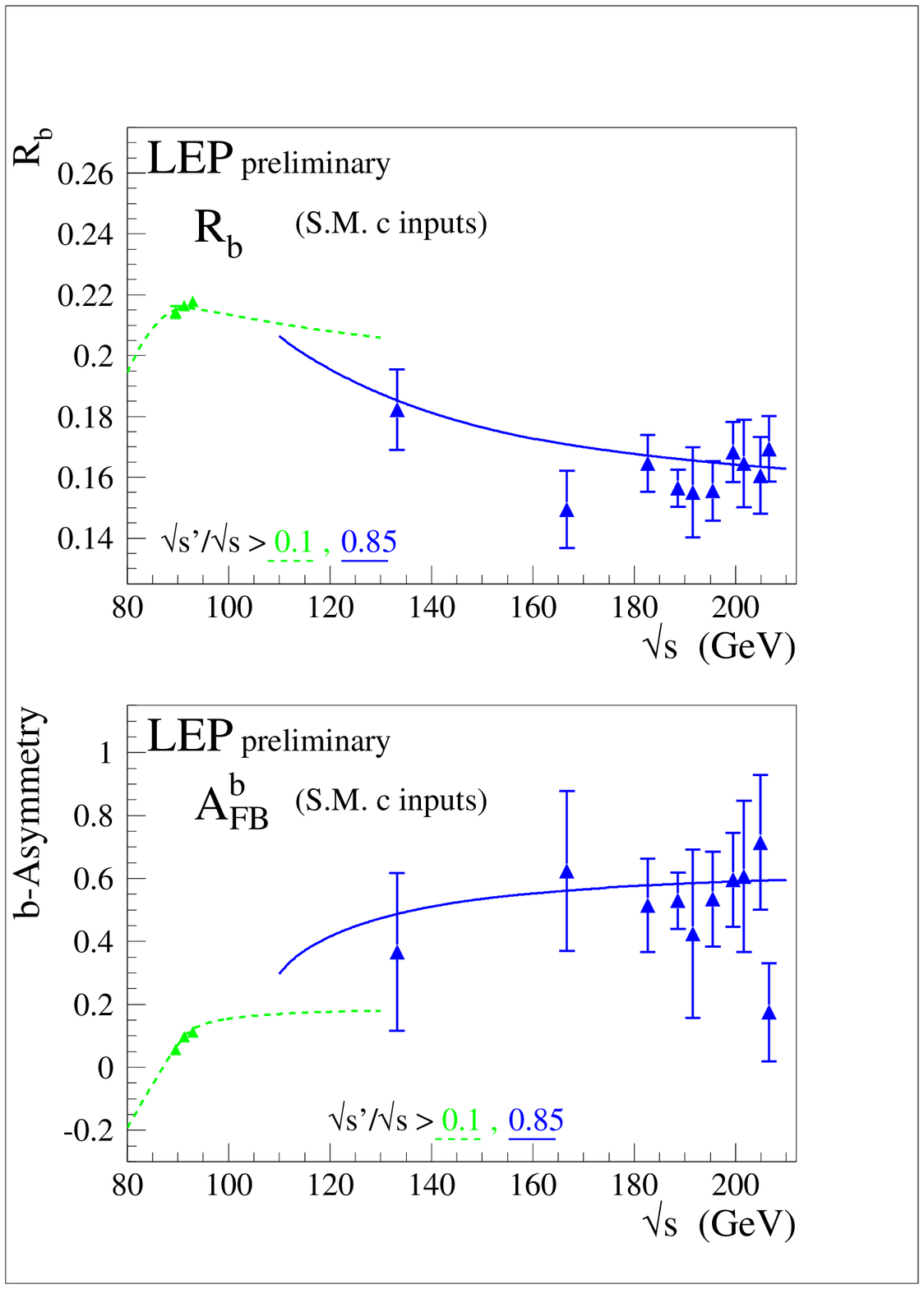,height=20cm}}
\end{center}
\caption{Preliminary combined LEP measurements of $\Rb$ and $\Abb$. 
  Solid lines represent the Standard Model prediction for the high
  $\sqrt{s'}$ selection used at $\LEPII$ and dotted lines the inclusive
  prediction used at $\LEPI$. Both are computed with
  ZFITTER\capcite{ff:ref:hfzfit}. The $\LEPI$ measurements have been
  taken from \capcite{ff:ref:hflep1-99}.}
\label{ff:fig:hfbres}
\end{figure}

\begin{figure}[p]
\begin{center}
\mbox{\epsfig{file=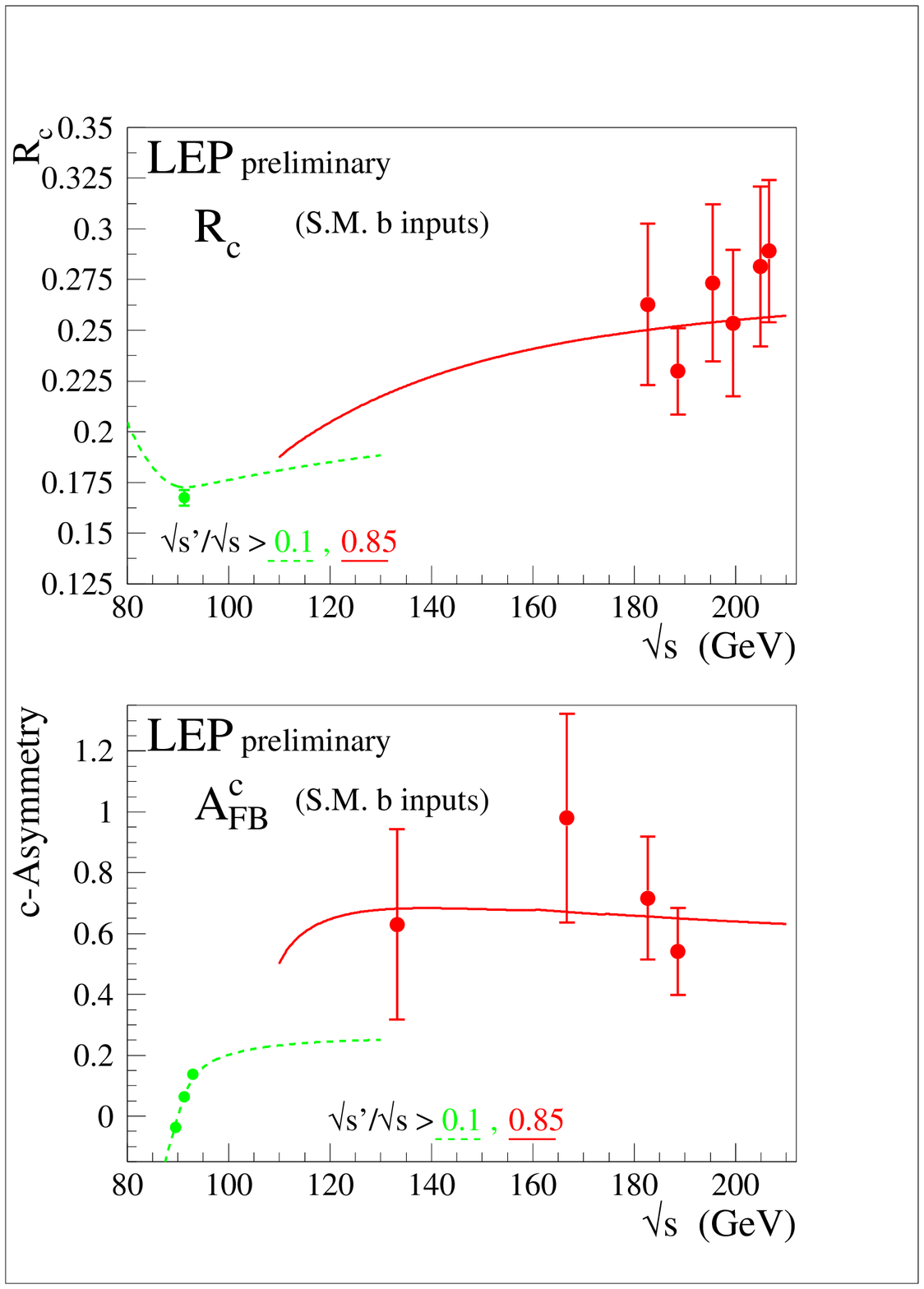,height=20cm}}
\end{center}
\caption{Preliminary combined LEP measurements of $\Rc$ and $\Acc$. 
  Solid lines represent the Standard Model prediction for the high
  $\sqrt{s'}$ selection used at $\LEPII$ and dotted lines the
  inclusive prediction used at $\LEPI$.  Both are computed with
  ZFITTER~\capcite{ff:ref:hfzfit}. The $\LEPI$ measurements have been
  taken from~\capcite{ff:ref:hflep1-99}.}
\label{ff:fig:hfcres} 
\end{figure}

The results of the combination are presented in Table~\ref{ff:tab:hfbresults} 
and Table~\ref{ff:tab:hfcresults} 
and in Figures~\ref{ff:fig:hfbres} and~\ref{ff:fig:hfcres}. 
The results for both b and c sector are in agreement with the Standard 
Model predictions of ZFITTER. 
The averaged discrepancies with respect to the Standard Model predictions 
is -2.08 $\sigma$ for $\Rb$, +0.30 $\sigma$ for $\Rc$, -1.56  $\sigma$ for $\Abb$ and -0.24 $\sigma $ for $\Acc$. 
A list of the error contributions from the combination at 189~$\GeV$ is shown 
in Table~\ref{ff:tab:hferror}.

\begin{table}[htbp]
\begin{center}
\begin{tabular}{|l|cccc|cccc|cccc|cccc|}
\hline 
 $\sqrt{s}$ ($\GeV$)
            & \multicolumn{4}{|c|}{$\Rb$}
            & \multicolumn{4}{|c|}{$\Rc$}
            & \multicolumn{4}{|c|}{$\Abb$}
            & \multicolumn{4}{|c|}{$\Acc$} \\
\cline{2-17}
            & A & D & L & O & A & D & L & O & A & D & L & O & A & D & L & O \\
\hline\hline
133         & {\sc{F}} & {\sc{F}} & {\sc{F}} & {\sc{F}}
            & {\sc{-}} & {\sc{-}} & {\sc{-}} & {\sc{-}}  
            & {\sc{-}} & {\sc{F}} & {\sc{-}} & {\sc{F}}  
            & {\sc{-}} & {\sc{F}} & {\sc{-}} & {\sc{F}} \\
\hline
167         & {\sc{F}} & {\sc{F}} & {\sc{F}} & {\sc{F}}
            & {\sc{-}} & {\sc{-}} & {\sc{-}} & {\sc{-}} 
            & {\sc{-}} & {\sc{F}} & {\sc{-}} & {\sc{F}} 
            & {\sc{-}} & {\sc{F}} & {\sc{-}} & {\sc{F}}   \\
\hline 
183         & {\sc{F}} & {\sc{P}} & {\sc{F}} & {\sc{F}}
            & {\sc{F}} & {\sc{-}} & {\sc{-}} & {\sc{-}}  
            & {\sc{F}} & {\sc{-}} & {\sc{-}} & {\sc{F}} 
            & {\sc{P}} & {\sc{-}} & {\sc{-}} & {\sc{F}}  \\
\hline 
189         & {\sc{P}} & {\sc{P}} & {\sc{F}} & {\sc{F}}
            & {\sc{P}} & {\sc{-}} & {\sc{-}} & {\sc{-}}  
            & {\sc{P}} & {\sc{P}} & {\sc{F}} & {\sc{F}} 
            & {\sc{P}} & {\sc{-}} & {\sc{-}} & {\sc{F}} \\
\hline 
192 to 202  & {\sc{P}} & {\sc{P}} & {\sc{P}} & {\sc{-}} 
            & {\sc{P*}} & {\sc{-}} & {\sc{-}} & {\sc{-}} 
            & {\sc{P}} & {\sc{P}} & {\sc{-}} & {\sc{-}} 
            & {\sc{-}} & {\sc{-}} & {\sc{-}} & {\sc{-}} \\ 
\hline 
205 and 207 & {\sc{-}} & {\sc{P}} & {\sc{P}} & {\sc{-}} 
            & {\sc{P}} & {\sc{-}} & {\sc{-}} & {\sc{-}} 
            & {\sc{P}} & {\sc{P}} & {\sc{-}} & {\sc{-}} 
            & {\sc{-}} & {\sc{-}} & {\sc{-}} & {\sc{-}} \\
\hline
\end{tabular}
\end{center}
\caption{Data provided by the ALEPH, DELPHI, L3, OPAL collaborations 
         for combination at different centre-of-mass energies. 
         Data indicated with {\sc{F}} are final, published data. 
         Data marked with {\sc{P}} are preliminary and for data marked 
         with {\sc{P*}}, not all energies are supplied.
         Data marked with a {\sc{-}} were not supplied for combination.}
\label{ff:tab:hfinput} 
\end{table}
\begin{table}[htbp]
\begin{center}
\begin{tabular}{|l|c|c|}
\hline 
$\sqrt{s}$ ($\GeV$) & $\Rb$
                    & $\Abb$ \\
\hline\hline
133      & 0.1822 $\pm$ 0.0132 & 0.367 $\pm$ 0.251 \\
         & (0.1867)            & (0.504)           \\
\hline
167      & 0.1494 $\pm$ 0.0127 & 0.624 $\pm$ 0.254 \\
         & (0.1727)            & (0.572)           \\
\hline 
183      & 0.1646 $\pm$ 0.0094 & 0.515 $\pm$ 0.149 \\
         & (0.1692)            & (0.588)           \\
\hline 
189      & 0.1565 $\pm$ 0.0061 & 0.529 $\pm$ 0.089 \\
         & (0.1681)            &  (0.593)          \\
\hline
192      & 0.1551 $\pm$ 0.0149 & 0.424 $\pm$ 0.267 \\
         & (0.1676)            & (0.595)           \\
\hline 
196      & 0.1556 $\pm$ 0.0097 & 0.535 $\pm$ 0.151 \\
         & (0.1670)            & (0.598)           \\
\hline
200      & 0.1683 $\pm$ 0.0099 & 0.596 $\pm$ 0.149 \\
         & (0.1664)            & (0.600)           \\
\hline
202      & 0.1646 $\pm$ 0.0144 & 0.607 $\pm$ 0.241 \\
         & (0.1661)            & (0.601)           \\
\hline
205      & 0.1606 $\pm$ 0.0126 & 0.715 $\pm$ 0.214 \\
         & (0.1657)            & (0.603)           \\
\hline
207      & 0.1694 $\pm$ 0.0107 & 0.175 $\pm$ 0.156 \\
         & (0.1654)            & (0.604)           \\
\hline
\end{tabular}
\end{center}
\caption[]{Combined results on $\Rb $ and $\Abb$. Quoted errors 
represent the statistical and systematic errors added in quadrature. 
For comparison, the Standard Model predictions computed with 
ZFITTER~\capcite{ff:ref:hfzfit} are given in parentheses. }
\label{ff:tab:hfbresults}
\end{table}
\begin{table}[htbp]
\begin{center}
\begin{tabular}{|l|c|c|}
\hline 
$\sqrt{s}$ ($\GeV$) & $\Rc$
                    & $\Acc$ \\
\hline\hline
133      & -   & 0.630 $\pm$ 0.313 \\
         &     & (0.684)           \\
\hline
167      & -   & 0.980 $\pm$ 0.343 \\
         &     & (0.677)           \\
\hline 
183      & 0.2628 $\pm$ 0.0397  & 0.717 $\pm$ 0.201 \\
         & (0.2472)    & (0.663)           \\
\hline 
189      & 0.2298 $\pm$ 0.0213  & 0.542 $\pm$ 0.143 \\
         & (0.2490)    & (0.656)           \\
\hline
196      & 0.2734 $\pm$ 0.0387 & - \\
         & (0.2508)            &   \\
\hline
200      & 0.2535 $\pm$ 0.0360 & - \\
         & (0.2518)            &   \\
\hline
205      & 0.2816 $\pm$ 0.0394 & - \\
         & (0.2530)            &   \\
\hline
207      & 0.2890 $\pm$ 0.0350 & - \\
         & (0.2533)            &   \\
\hline
\end{tabular}
\end{center}
\caption{Combined results on $\Rc$ and $\Acc$. Quoted errors 
represent the statistical and systematic errors added in quadrature. 
For comparison, the Standard Model predictions computed with 
ZFITTER~\capcite{ff:ref:hfzfit} are given in parentheses. }
\label{ff:tab:hfcresults}
\end{table}
\begin{table}[htbp]
\begin{center}
\begin{tabular}{|l|c|c||c|c|}
\hline 
Error list & $\Rb$ (189 $\GeV$) 
           & $\Abb$ (189 $\GeV$) 
           & $\Rc$ (189 $\GeV$) 
           & $\Acc$ (189 $\GeV$)  \\

\hline\hline
statistics    & 0.0057  & 0.084 & 0.0169 & 0.119 \\ 
\hline 
internal syst & 0.0020  & 0.025 & 0.0109 & 0.042 \\
common syst   & 0.0007  & 0.011 & 0.0072 & 0.069 \\
total syst    & 0.0021  & 0.027 & 0.0130 & 0.081 \\ 
\hline 
total error   & 0.0061  & 0.089 & 0.0213 & 0.143 \\ 
\hline 
\end{tabular}
\end{center}
\caption{Error breakdown at 189 $\GeV$.}
\label{ff:tab:hferror} 
\end{table}
\section{Interpretation}
\label{ff:sec-interp}

The combined measurements presented above are interpreted in a variety
of models.
The cross-section and asymmetry results are used to place limits 
on contact interactions between leptons and quarks and, using 
the results on heavy flavour production, on contact interaction between 
electrons and $b$ and $c$ quarks specifically.
Limits on the mass of a possible additional heavy neutral boson, $\Zprime$,
are obtained for a variety of models.
Using the combined differential cross-sections for \ee\ final states, 
limits on contact interactions in the $\eeee$ channel and limits on the 
scale of gravity in models with large extra-dimensions are presented.
Limits are also derived on the masses of leptoquarks - assuming
a coupling of electromagnetic strength. 
In all cases the Born level predictions for the physics beyond the Standard 
Model have been corrected to take into account QED radiation.

\subsection{Contact Interactions}
\label{ff:sec-cntc}

The averages of cross-sections and forward-backward asymmetries for 
muon-pair and tau-lepton pair and the cross-sections for $\qq$ 
final states are used to search for 
contact interactions between fermions. 

Following~\cite{ff:ref:ELPthr}, contact interactions are parameterised 
by an effective Lagrangian, $\cal{L}_{\mathrm{eff}}$, which is added to the 
Standard Model Lagrangian and has the form:
\begin{eqnarray}
 \mbox{$\cal{L}$}_{\mathrm{eff}} = 
                        \frac{g^{2}}{(1+\delta)\Lambda^{2}} 
                          \sum_{i,j=L,R} \eta_{ij} 
                           \overline{e}_{i} \gamma_{\mu} e_{i}
                            \overline{f}_{j} \gamma^{\mu} f_{j},
\end{eqnarray}
where $g^{2}/{4\pi}$ is taken to be 1 by convention, $\delta=1 (0)$ for 
$f=e ~(f \neq e)$, $\eta_{ij}=\pm 1$ or $0$ for different interaction types,
$\Lambda$ is the scale of the contact interactions,
$e_{i}$ and $f_{j}$ are left or right-handed spinors. 
By assuming different helicity coupling between the initial 
state and final state currents, a set of different models can be defined
from this Lagrangian~\cite{ff:ref:Kroha}, with either
constructive ($+$) or destructive ($-$) interference between the 
Standard Model process and the contact interactions. The models and 
corresponding choices of $\eta_{ij}$ are given in Table~\ref{ff:tab:cntcdef}.
The models LL$^{\pm}$, RR$^{\pm}$, VV$^{\pm}$, AA$^{\pm}$, LR$^{\pm}$, 
RL$^{\pm}$, V0$^{\pm}$, A0$^{\pm}$ are considered here since 
these models lead to large deviations in $\eeff$ at $\LEPII$.
The corresponding energies scales for the models with constructive
or destructive interference are denoted by $\Lambda^{+}$ and $\Lambda^{-}$
respectively.

For leptonic final states 4 different fits are made
\begin{itemize}
 \item individual fits to contact interactions in $\eemumu$ and $\eetautau$
       using the measured cross-sections and asymmetries,
 \item fits to $\eell$ (simultaneous fits to $\eemumu$ and $\eetautau$)
       again using the measured cross-sections and asymmetries,
 \item fits to $\eeee$, using the measured differential cross-sections. 
\end{itemize}
For the inclusive hadronic final states three different model 
assumptions are used to fit the total hadronic cross-section
\begin{itemize}
\item the contact interactions affect only one quark flavour of up-type 
      using the measured hadronic cross-sections,
\item the contact interactions affect only one quark flavour of down-type 
      using the measured hadronic cross-sections,
\item the contact interactions contribute to all quark final states with 
      the same strength. 
\end{itemize}

Limits on contact interactions between electrons and $b$ and $c$ quarks
are obtained using all the heavy flavour $\LEPII$ combined results 
from 133 $\GeV$ to 207 $\GeV$ given in Tables~\ref{ff:tab:hfbresults} 
and~\ref{ff:tab:hfcresults}.
For the purpose of fitting contact interaction models to the data, 
$\Rb$ and $\Rc$ are converted to cross-sections 
$\sigma_{\bb}$ and $\sigma_{\cc}$ using the averaged ${\qq}$ cross-section of 
section \ref{ff:sec-ave-xsc-afb} corresponding to the second signal 
definition.  
In the calculation of errors, the correlations between $\Rb$, $\Rc$ and 
$\sigma_{\qq}$ are assumed to be negligible.
These results are of particular interest since they are inaccessible
to ${\mathrm{p\bar{p}}}$ or ep colliders.

For the purpose of fitting contact interaction models to the data, 
the parameter $\epsilon=1/\Lambda^{2}$ is used, with
$\epsilon=0$ in the limit that there are no contact interactions. 
This parameter is allowed to take both positive and negative values in 
the fits. 
Theoretical uncertainties on the Standard Model predictions are taken 
from~\cite{ff:ref:lepffwrkshp}.

The values of $\epsilon$ extracted for each model are all compatible 
with the Standard Model expectation $\epsilon=0$, at the two standard 
deviation level. As expected, 
the errors on $\epsilon$ are typically a factor of two 
smaller than those obtained from a single LEP experiment with the same data
set. The fitted values of $\epsilon$ are converted into  
$95\%$ confidence level lower limits on $\Lambda$. 
The limits are obtained by integrating the likelihood function in 
$\epsilon$ over the physically allowed values\footnote{To be able to obtain 
confidence limits from the likelihood function in $\epsilon$ 
it is necessary to convert the likelihood to a probability density function 
for $\epsilon$; this is done by 
multiplying by a prior probability function. Simply integrating the 
likelihood over $\epsilon$ is equivalent to multiplying by a uniform 
prior probability function in $\epsilon$.}, 
$\epsilon \ge 0$ for each $\Lambda^{+}$ limit and $\epsilon \le 0$ for 
$\Lambda^{-}$ limits.

\begin{figure}[p]
 \begin{center}
  \begin{tabular}{cc}
  \epsfig{file=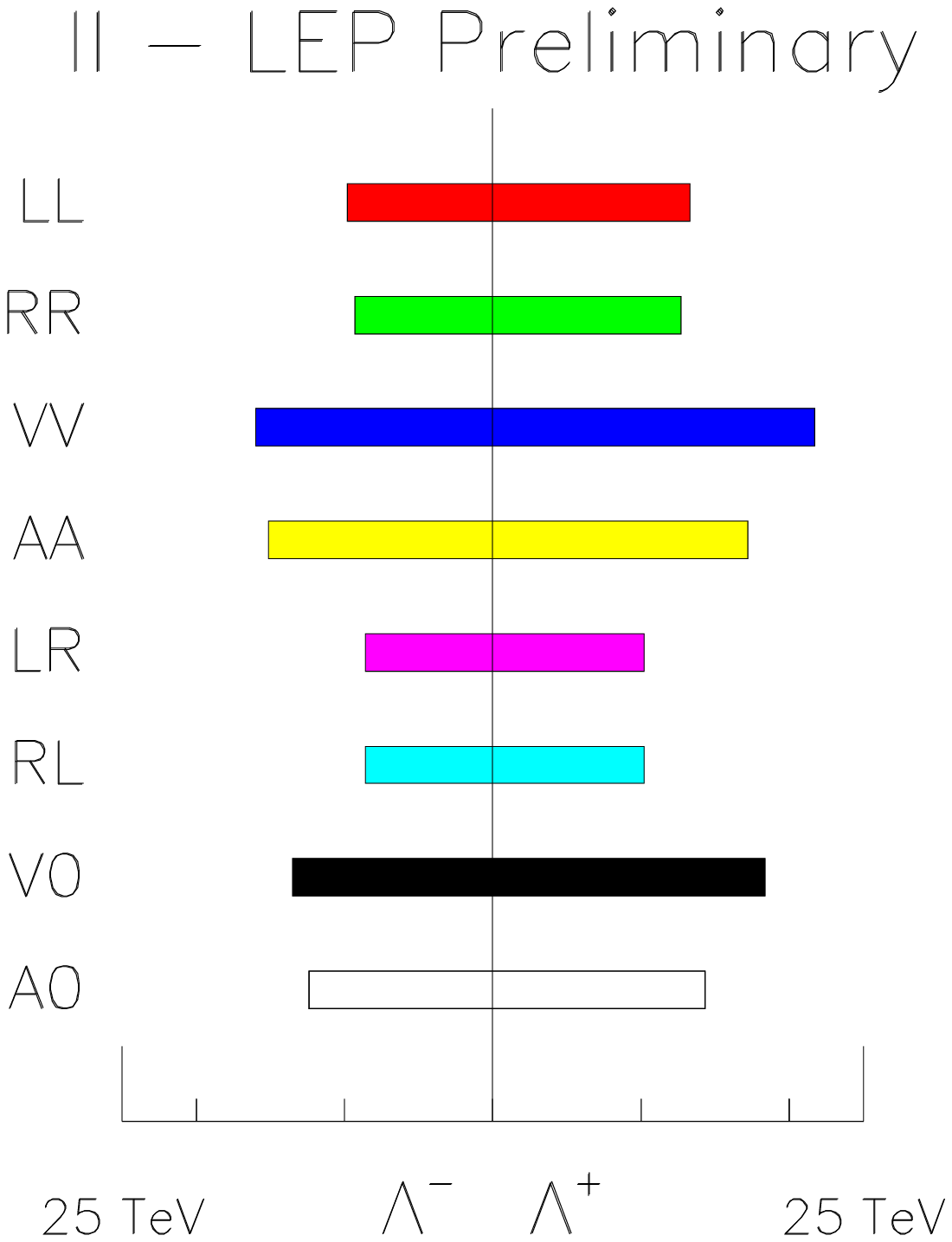,width=0.36\textwidth} &
  \epsfig{file=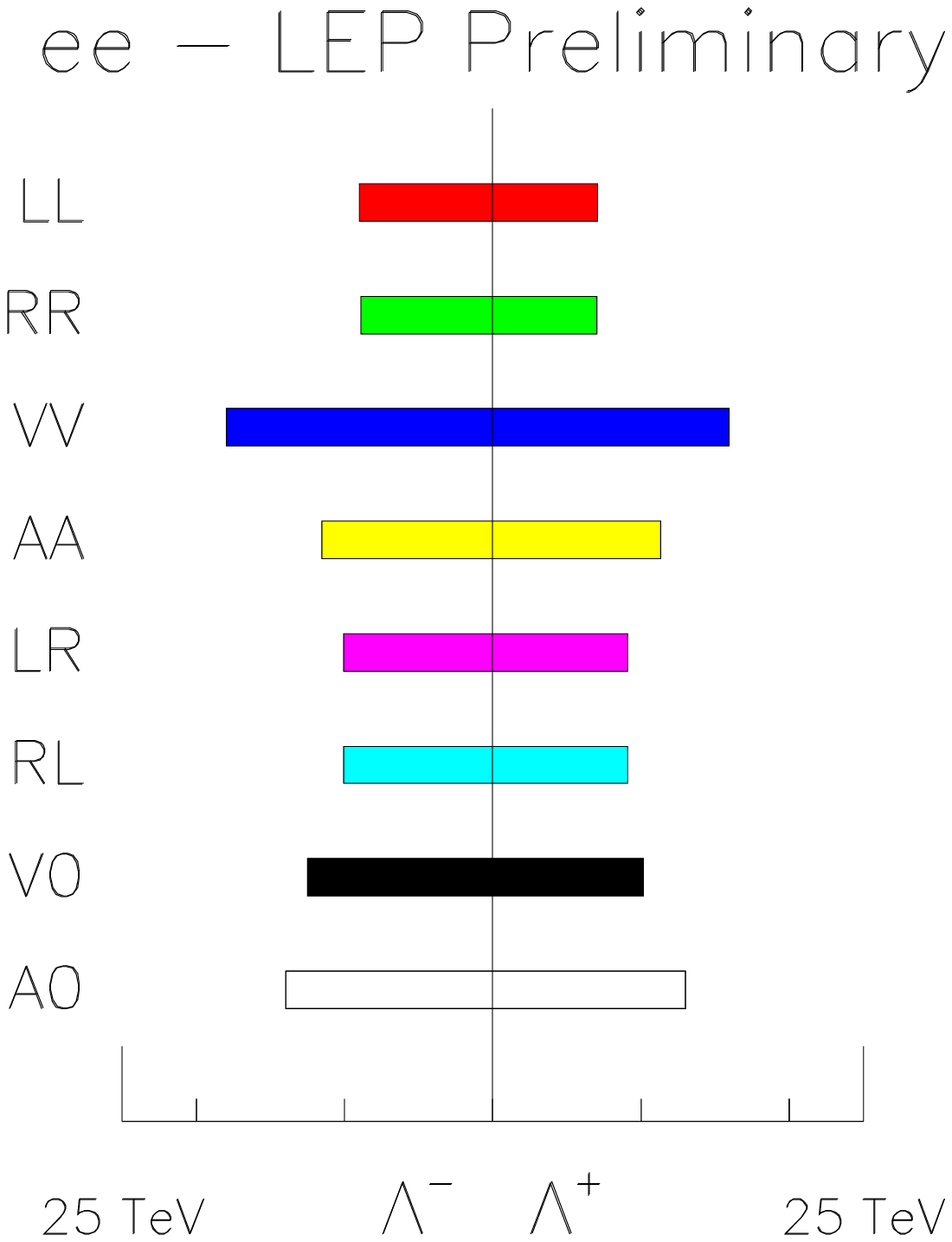,width=0.36\textwidth} \\
  \multicolumn{2}{c}
   {\epsfig{file=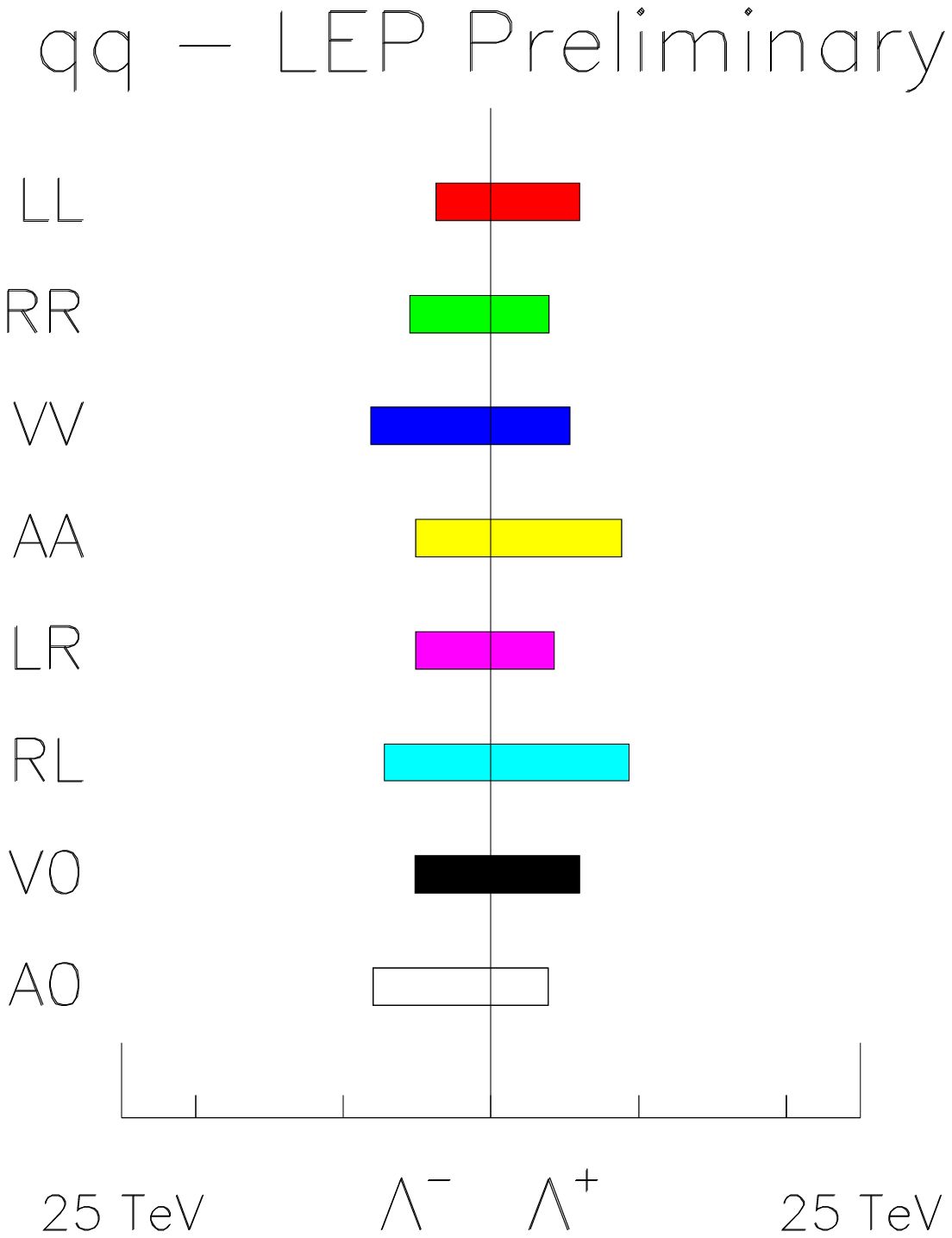,width=0.36\textwidth}} \\ 
  \epsfig{file=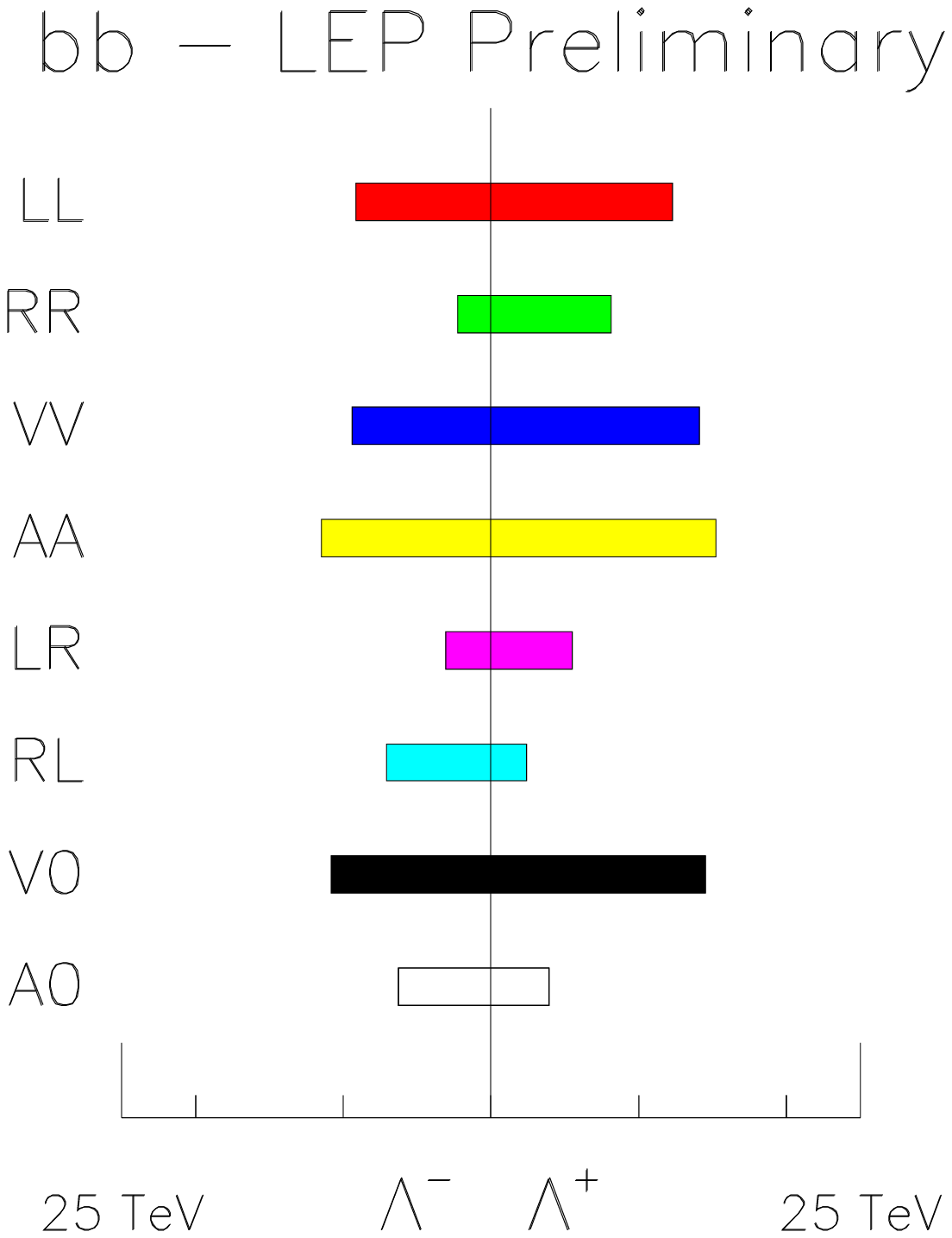,width=0.36\textwidth} &
  \epsfig{file=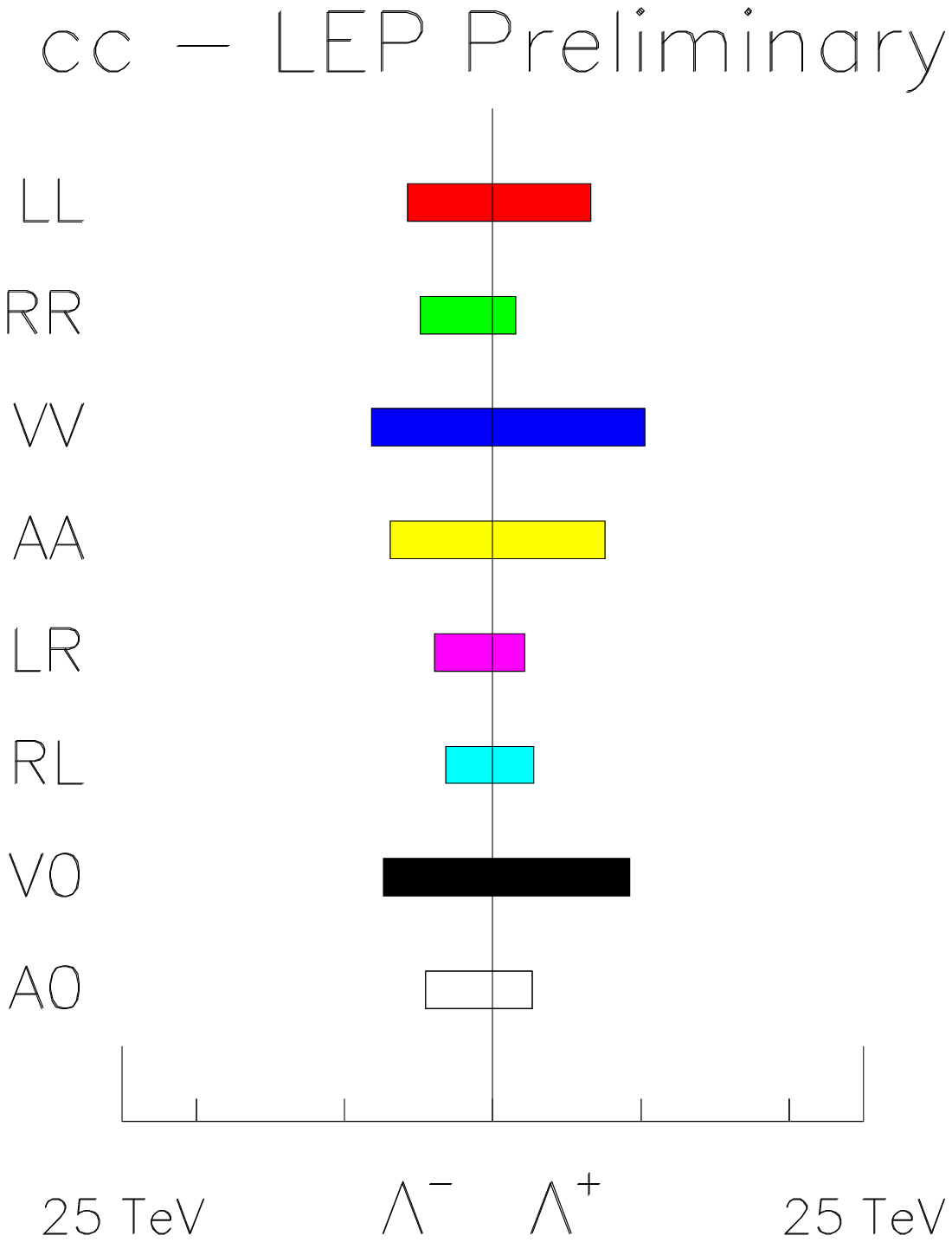,width=0.36\textwidth} \\
  \end{tabular}
 \end{center}
 \caption{The limits on $\Lambda$ for $\eell$ assuming
          universality in the contact interactions between 
          $\eell$ ($\ell \neq e$), for $\eeee$, for $\eeqq$ assuming 
          equal strength contact interactions for quarks and for
          $\eebb$ and $\eecc$.}
 \label{ff:fig:cntc}
\end{figure}

The fitted values of $\epsilon$ and their 68$\%$ confidence level 
uncertainties together with the 95$\%$ confidence level lower limit 
on ${\mathrm{\Lambda}}$ are shown in Table \ref{ff:tab:cntceps} for
the fits to $\eell$ ($\ell \neq e$), $\eeee$ , inclusive $\eeqq$, $\eebb$ 
and $\eecc$. Table \ref{ff:tab:cntclmb} shows only the limits 
obtained on the scale $\Lambda$ for other fits. The limits are shown 
graphically in Figure \ref{ff:fig:cntc}.

For the VV model with positive interference and assuming
electromagnetic coupling strength instead of $g^{2}/{4\pi} = 1$,
the scale $\Lambda$ obtained in the $\eeee$ channel is converted to 
an upper limit on the electron size:
\begin{eqnarray}
\mathrm{r_e < 1.4 \times 10^{-19} m}
\end{eqnarray}
Models with stronger couplings will make this upper limit even tighter.

\begin{table}[tp]
 \begin{center}
  \begin{tabular}{|c|c|c|c|c|}
   \hline
   Model      & $\eta_{LL}$ & $\eta_{RR}$ & $\eta_{LR}$ & $\eta_{RL}$ \\
   \hline\hline
   LL$^{\pm}$ &   $\pm 1$   &      0      &      0      &      0      \\
   \hline
   RR$^{\pm}$ &      0      &   $\pm 1$   &      0      &      0      \\
   \hline
   VV$^{\pm}$ &   $\pm 1$   &   $\pm 1$   &   $\pm 1$   &   $\pm 1$   \\
   \hline
   AA$^{\pm}$ &   $\pm 1$   &   $\pm 1$   &   $\mp 1$   &   $\mp 1$   \\
   \hline
   LR$^{\pm}$ &      0      &      0      &   $\pm 1$   &      0      \\
   \hline
   RL$^{\pm}$ &      0      &      0      &      0      &   $\pm 1$   \\
   \hline
   V0$^{\pm}$ &   $\pm 1$   &   $\pm 1$   &      0      &      0      \\
   \hline
   A0$^{\pm}$ &      0      &      0      &  $\pm 1$    &   $\pm 1$   \\
   \hline
  \end{tabular}
 \end{center}
 \caption{Choices of $\eta_{ij}$ for different contact interaction models}
 \label{ff:tab:cntcdef}.
\end{table}
\begin{table} 
 \renewcommand{\arraystretch}{1.2}
  \begin{center}
  \begin{tabular}{cc}

\begin{tabular}{|c||c|cc|}
\hline
\multicolumn{4}{|c|}{$\eell$} \\
\hline 
\hline
       & $\epsilon$   & $\Lambda^{-}$ & $\Lambda^{+}$ \\
 Model & (TeV$^{-2}$) &     (TeV)     &     (TeV)     \\
\hline 
\hline    
~~~~LL~~~~ &  -0.0044$^{+0.0035}_{-0.0035}$ &  9.8  & 13.3 \\
\hline
    RR     &  -0.0049$^{+0.0039}_{-0.0039}$ &  9.3  & 12.7 \\
\hline                                                     
    VV     &  -0.0016$^{+0.0013}_{-0.0014}$ & 16.0  & 21.7  \\
\hline
    AA     &  -0.0013$^{+0.0017}_{-0.0017}$ & 15.1  & 17.2  \\
\hline                                                     
    LR     &  -0.0036$^{+0.0052}_{-0.0054}$ &  8.6  & 10.2 \\
\hline
    RL     &  -0.0036$^{+0.0052}_{-0.0054}$ &  8.6  & 10.2 \\
\hline                                                     
    V0     &  -0.0023$^{+0.0018}_{-0.0018}$ & 13.5  & 18.4  \\
\hline
    A0     &  -0.0018$^{+0.0026}_{-0.0026}$ & 12.4  & 14.3 \\
\hline
\end{tabular}
&
\begin{tabular}{|c||c|cc|}
\hline
\multicolumn{4}{|c|}{$\eeee$} \\
\hline 
\hline
       & $\epsilon$   & $\Lambda^{-}$ & $\Lambda^{+}$ \\
 Model & (TeV$^{-2}$) &      (TeV)    &     (TeV)     \\
\hline    
\hline    
~~~~LL~~~~ &  ~0.0049$^{+0.0084}_{-0.0084}$ & 9.0  & 7.1  \\
\hline
    RR     &  ~0.0056$^{+0.0082}_{-0.0092}$ & 8.9  & 7.0  \\
\hline                                                     
    VV     &  ~0.0004$^{+0.0022}_{-0.0016}$ &18.0  &15.9  \\
\hline
    AA     &  ~0.0009$^{+0.0041}_{-0.0039}$ &11.5  &11.3  \\
\hline                                                     
    LR     &  ~0.0008$^{+0.0064}_{-0.0052}$ &10.0  & 9.1  \\
\hline
    RL     &  ~0.0008$^{+0.0064}_{-0.0052}$ &10.0  & 9.1  \\ 
\hline                                                     
    V0     &  ~0.0028$^{+0.0038}_{-0.0045}$ &12.5  &10.2  \\
\hline
    A0     &  -0.0008$^{+0.0028}_{-0.0030}$ &14.0  &13.0  \\
\hline
\end{tabular}
\\

\\
\multicolumn{2}{c}{
\begin{tabular}{|c||c|cc|}
\hline
\multicolumn{4}{|c|}{$\eeqq$} \\
\hline 
\hline
       & $\epsilon$   & $\Lambda^{-}$ & $\Lambda^{+}$ \\
 Model & (TeV$^{-2}$) &      (TeV)    &     (TeV)     \\
\hline    
\hline    
~~~~LL~~~~ &  ~0.0152$^{+0.0064}_{-0.0076}$ & 3.7  & 6.0  \\
\hline
    RR     &  -0.0208$^{+0.0103}_{-0.0082}$ & 5.5  & 3.9  \\
\hline                                                     
    VV     &  -0.0096$^{+0.0051}_{-0.0037}$ & 8.1  & 5.3  \\
\hline
    AA     &  ~0.0068$^{+0.0033}_{-0.0034}$ & 5.1  & 8.8  \\
\hline                                                     
    LR     &  -0.0308$^{+0.0172}_{-0.0055}$ & 5.1  & 4.3  \\
\hline
    RL     &  -0.0108$^{+0.0057}_{-0.0054}$ & 7.2  & 9.3  \\ 
\hline                                                     
    V0     &  ~0.0174$^{+0.0057}_{-0.0074}$ & 5.1  & 6.0  \\
\hline
    A0     &  -0.0092$^{+0.0049}_{-0.0041}$ & 8.0  & 3.9  \\
\hline
\end{tabular}
}
\\

\\
\begin{tabular}{|c|c|cc|}
\hline
\multicolumn{4}{|c|}{$\eebb$} \\
\hline
\hline
       & $\epsilon$   & $\Lambda^{-}$ & $\Lambda^{+}$ \\
 Model & (TeV$^{-2}$) &      (TeV)    &     (TeV)     \\
\hline
\hline
~~~~LL~~~~ & -0.0038$^{+ 0.0044}_{- 0.0047}$ &    9.1 &   12.3 \\
\hline
    RR     & -0.1729$^{+ 0.1584}_{- 0.0162}$ &    2.2 &    8.1 \\
\hline
    VV     & -0.0040$^{+ 0.0039}_{- 0.0041}$ &    9.4 &   14.1 \\
\hline
    AA     & -0.0022$^{+ 0.0029}_{- 0.0031}$ &   11.5 &   15.3 \\
\hline
    LR     & -0.0620$^{+ 0.0692}_{- 0.0313}$ &    3.1 &    5.5 \\
\hline
    RL     &  0.0180$^{+ 0.1442}_{- 0.0249}$ &    7.0 &    2.4 \\
\hline
    V0     & -0.0028$^{+ 0.0032}_{- 0.0033}$ &   10.8 &   14.5 \\
\hline
    A0     &  0.0375$^{+ 0.0193}_{- 0.0379}$ &    6.3 &    3.9 \\
\hline
\end{tabular}
&
\begin{tabular}{|c|c|cc|}
\hline
\multicolumn{4}{|c|}{$\eecc$} \\
\hline
\hline
       & $\epsilon$   & $\Lambda^{-}$ & $\Lambda^{+}$ \\
 Model & (TeV$^{-2}$) &      (TeV)    &     (TeV)     \\
\hline
\hline
~~~~LL~~~~ & -0.0091$^{+ 0.0126}_{- 0.0126}$ &    5.7 &    6.6 \\
\hline
    RR     &  0.3544$^{+ 0.0476}_{- 0.3746}$ &    4.9 &    1.5 \\
\hline
    VV     & -0.0047$^{+ 0.0057}_{- 0.0060}$ &    8.2 &   10.3 \\
\hline
    AA     & -0.0059$^{+ 0.0095}_{- 0.0090}$ &    6.9 &    7.6 \\
\hline
    LR     &  0.1386$^{+ 0.0555}_{- 0.1649}$ &    3.9 &    2.1 \\
\hline
    RL     &  0.0106$^{+ 0.0848}_{- 0.0757}$ &    3.1 &    2.8 \\
\hline
    V0     & -0.0058$^{+ 0.0075}_{- 0.0071}$ &    7.4 &    9.2 \\
\hline
    A0     &  0.0662$^{+ 0.0564}_{- 0.0905}$ &    4.5 &    2.7 \\
\hline
\end{tabular}

\end{tabular}
   \caption{The fitted values of $\epsilon$ and the derived 95\% confidence 
            level lower limits on the parameter $\Lambda$
            of contact interaction derived from fits to lepton-pair
            cross-sections and asymmetries and from fits to hadronic 
            cross-sections. The limits $\Lambda_+$ and $\Lambda_-$
            given in TeV correspond to the upper and lower signs of the 
            parameters $\eta_{ij}$ in Table \ref{ff:tab:cntcdef}.
            For $\leptlept$ ($\ell \neq e$) the couplings to $\mumu$ and 
            $\tautau$ are a assumed to be universal and for inclusive 
            $\qq$ final states 
            all quarks are assumed to experience contact interactions 
            with the same strength.}
  \label{ff:tab:cntceps}
  \end{center} 
\end{table}
\begin{table}[tp]
 \renewcommand{\arraystretch}{1.1}
  \begin{center}
  \begin{tabular}{c}
    \begin{tabular}{|c||cc|cc|}
\hline
\multicolumn{5}{|c|}{leptons} \\
\hline
\hline       
           &\multicolumn{2}{|c|}{$\mu^+\mu^-$}
           &\multicolumn{2}{|c|}{$\tau^+ \tau^-$} \\
 Model     &~~$\Lambda_-$~~
                 &~~$\Lambda_+$~~
                       &~~$\Lambda_-$~~
                            &~~$\Lambda_+$~~ \\
\hline \hline  
~~~~LL~~~~ & 8.5  & 12.5 & 9.1  & 8.6  \\
    RR     & 8.1  & 11.9 & 8.7  & 8.2  \\
\hline                                                     
    VV     & 14.3 & 19.7 & 14.2 & 14.5 \\
    AA     & 12.7 & 16.4 & 14.0 & 11.3 \\
\hline                                                     
    LR     & 7.9  & 8.9  & 2.2  & 7.9  \\ 
    RL     & 7.9  & 8.9  & 2.2  & 7.9  \\
\hline                                                     
    V0     & 11.7 & 17.2 & 12.7 & 11.8 \\
    A0     & 11.5 & 12.4 & 9.8  & 10.8 \\
\hline
    \end{tabular}
\\

\\
    \begin{tabular}{|c||cc|cc|}
\hline
\multicolumn{5}{|c|}{hadrons} \\
\hline
\hline
           &\multicolumn{2}{|c|}{up-type}
           &\multicolumn{2}{|c|}{down-type} \\
 Model     &~~$\Lambda_-$~~
                 &~~$\Lambda_+$~~
                       &~~$\Lambda_-$~~
                             &~~$\Lambda_+$~~ \\
\hline \hline  
~~~~LL~~~~ & 6.7  & 10.2 & 10.6 & 6.0  \\
    RR     & 5.7  & 8.3  & 2.2  & 4.3  \\
\hline                   
    VV     & 9.6  & 14.3 & 11.4 & 7.0  \\
    AA     & 8.0  & 11.5 & 13.3 & 7.7  \\
\hline
    LR     & 4.2  & 2.3  & 2.7  & 3.5  \\
    RL     & 3.5  & 2.8  & 4.2  & 2.4  \\
\hline                                                     
    V0     & 8.7  & 13.4 & 12.5 & 7.1  \\
    A0     & 4.9  & 2.8  & 4.2  & 3.3  \\
\hline
    \end{tabular}
   \end{tabular}
   \caption{The 95\% confidence level lower limits on the parameter 
            $\Lambda$
            of contact interaction derived from fits to lepton-pair 
            cross-sections and asymmetries and from fits to hadronic 
            cross-sections. The limits $\Lambda_+$ and $\Lambda_-$
            given in TeV correspond to the upper and lower signs of the 
            parameters $\eta_{ij}$ in Table \ref{ff:tab:cntcdef}.
            For hadrons the limits for up-type and down-type quarks
            are derived assuming a single up or down type quark undergoes
            contact interactions.}
   \label{ff:tab:cntclmb}
  \end{center}
\end{table}
\subsection{Models with $\mathbf{\Zprime}$ Bosons}
\label{ff:sec-zprime}

The combined hadronic and leptonic cross-sections and the leptonic 
forward-backward asymmetries are used to fit the data to models including 
an additional, heavy, neutral boson, $\Zprime$.

Fits are made to $\MZp$, the mass of a $\Zprime$ for models 
resulting from an E$_6$ GUT and L-R symmetric models~\cite{ff:ref:zprime-thry}
and for the Sequential Standard Model (SSM)~\cite{ff:ref:sqsm}, which proposes the 
existence of a $\Zprime$ with exactly the same coupling to fermions as 
the standard Z. $\LEPII$ data alone does not significantly constrain
the mixing angle between the Z and $\Zprime$ fields, $\thtzzp$.
However results from a single experiment, in which $\LEPI$ data is used in the 
fit, show that the mixing is consistent with zero (see for 
example~\cite{ff:ref:lep1zprime}). So for these fits $\thtzzp$ was fixed to 
zero.

\begin{figure}[tp]
  \begin{flushleft}
   \begin{tabular}{ll}  
    \mbox{\epsfig{file=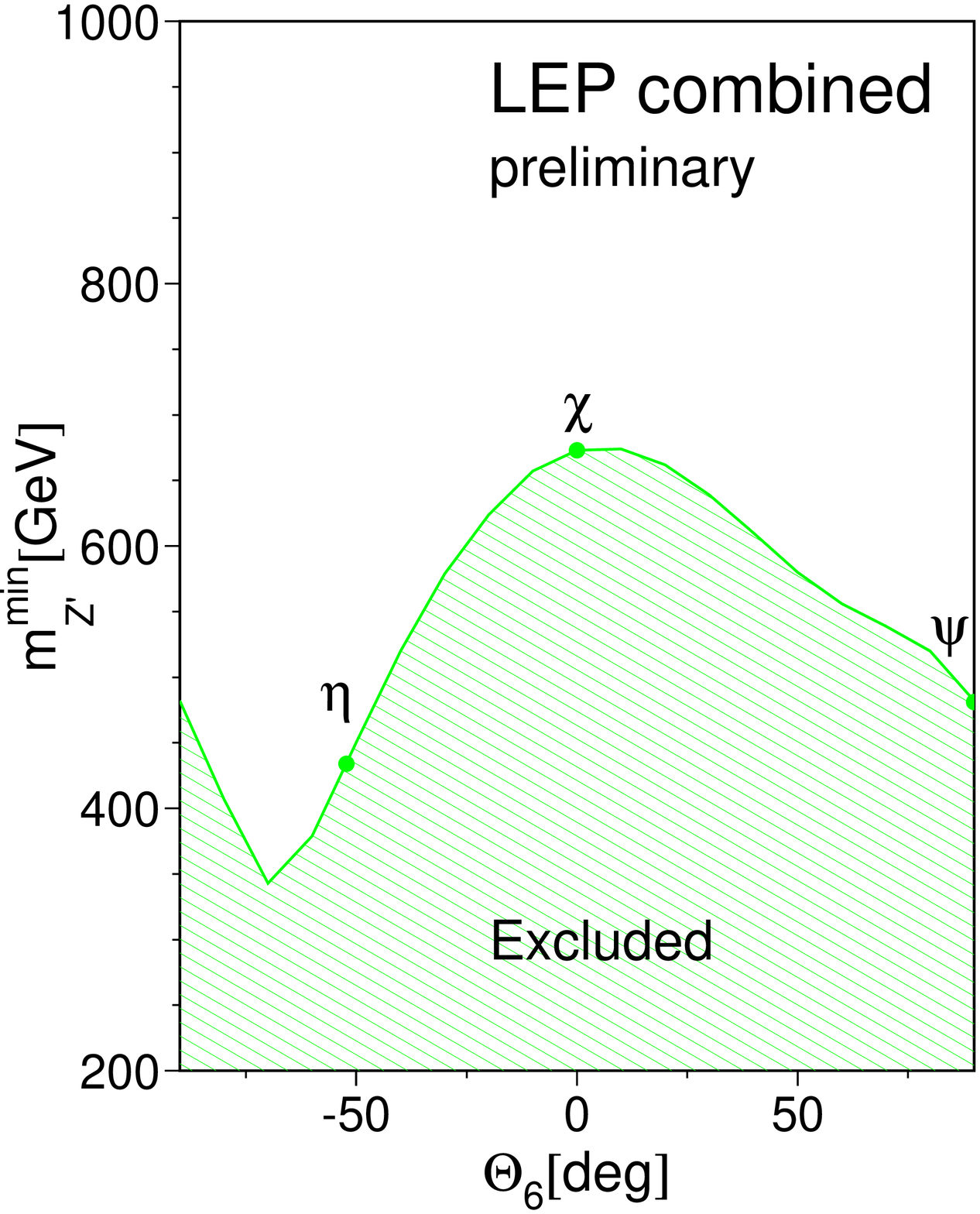,width=0.50\textwidth}}
    \mbox{\epsfig{file=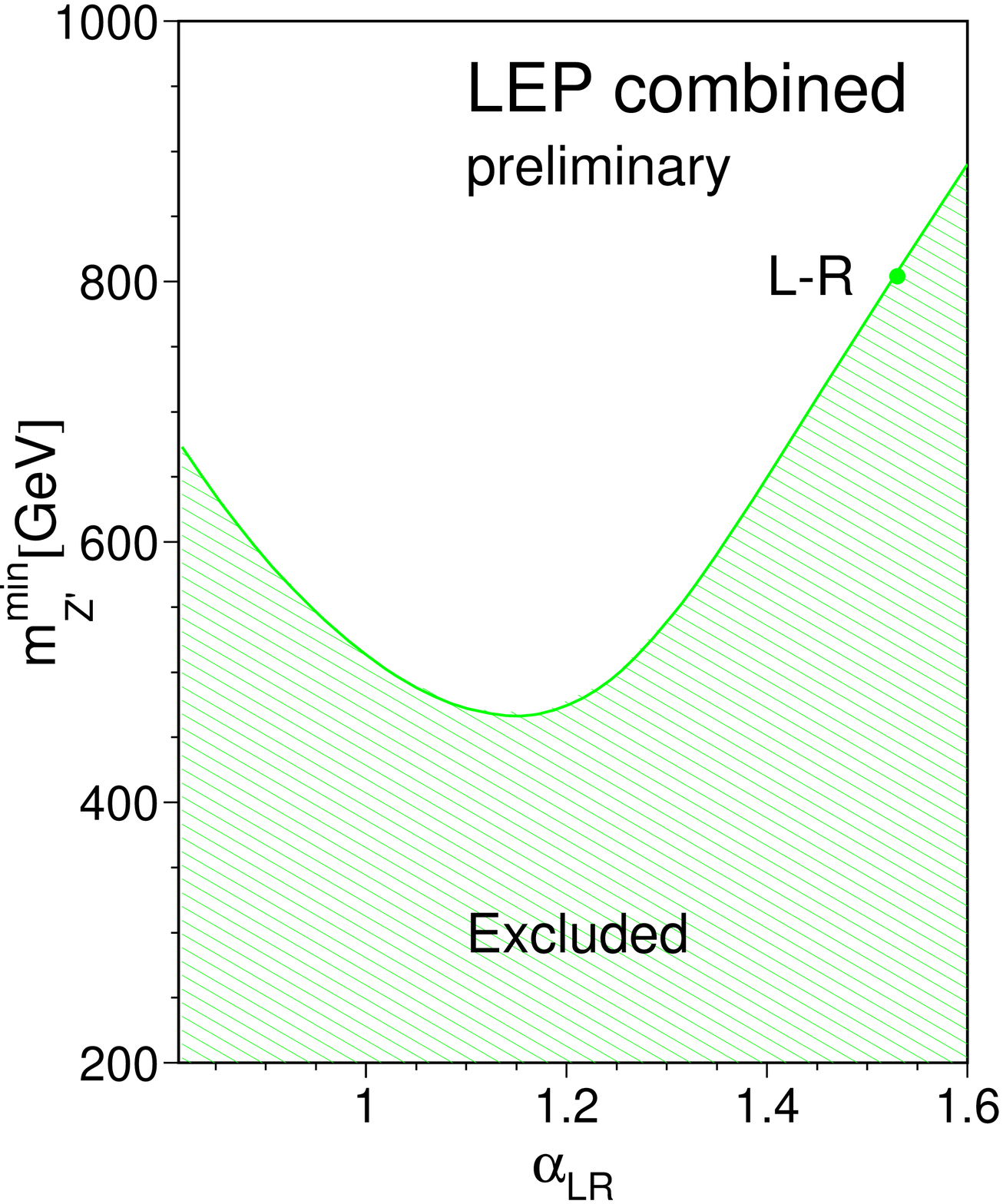,width=0.50\textwidth}}
   \end{tabular}
  \caption{The 95\% confidence level limits on $\MZp$ as a function of 
           the model parameter $\theta_6$ for E$_6$ models and 
           $\alpha_{\mathrm{LR}}$ for left-right models. 
           The Z-$\Zprime$ mixing is fixed, $\thtzzp=0$.}  
   \label{ff:fig:zp_e6-lr}
  \end{flushleft}
\end{figure}

No significant evidence is found for the existence of a $\Zprime$ boson
in any of the models. 
The procedure to find limits on the Z$'$ mass corresponds to that in case 
of  contact interactions: for large masses the exchange of a Z$'$ can be 
approximated by contact terms, $\Lambda \propto \MZp$.
The lower limits on the Z$'$ mass are shown in Figure \ref{ff:fig:zp_e6-lr} 
varying the parameters $\theta_6$ for the E$_6$ models and  
$\alpha_{\mathrm{LR}}$ for the left-right models. 
The results for the specific models 
$\chi,~\psi~,\eta$ ($\theta_6=0,~\pi/2,~- \arctan \sqrt{5/3}$), 
L-R ($\alpha_{\mathrm{LR}}$=1.53) and SSM are shown in 
Table~\ref{ff:tab:zprime_mass_lim}.

\begin{table}[tp]
  \renewcommand{\arraystretch}{1.15}
  \begin{center}
    \begin{tabular}{|c||c|c|c|c|c|}
\hline
Z$'$ model                   & $\chi$ & $\psi$ & $\eta$ & L-R   & SSM  \\
  \hline  \hline 
M$_{Z'}^{limit}$ (GeV/c$^2$) & 673    & 481    & 434    & 804   & 1787 \\
 \hline
 \end{tabular}
 \caption{The 95\% confidence level lower limits on the $\Zprime$ mass for
 $\chi,~\psi,~\eta$, L-R and SSM models.}
 \label{ff:tab:zprime_mass_lim}
 \end{center} 
\end{table}
\subsection{Leptoquarks and R-parity violating squarks }
\label{ff:sec-lq-sq}

Leptoquarks (LQ) would mediate quark-lepton transitions. 
Following the notations in  Reference~\cite{ff:ref:lq-thry,ff:ref:lq-squ}, 
scalar leptoquarks, $S_I$, and vector leptoquarks,$V_I$ are indicated 
based on spin and isospin $I$. Leptoquarks with the same Isospin but 
with different hypercharges are distinguished by an additional tilde. 
See Reference~\citen{ff:ref:lq-squ} for further details.
They carry fermion numbers, $F=L+3B$. 
It is assumed that leptoquark couplings to quark-lepton 
pairs preserve baryon- and lepton-number. 
The couplings $g_L,~g_R$, are labelled according to the chirality 
of the lepton.        

\SBL{1/2} and \SL{0} leptoquarks are equivalent to up-type anti-squarks and 
down-type squarks, respectively. Limits in terms of the leptoquark coupling 
are  then exactly equivalent to limits on $\mathrm \lambda_{1jk}$ in the 
Lagrangian ${\mathrm  \lambda_{1jk}L_{1}Q_{j}\bar D_{k}}$. 

At LEP, the exchange of a leptoquark can modify the hadronic
cross-sections and asymmetries, as described at the Born level by the equations
given in Reference~\citen{ff:ref:lq-squ}. Using the LEP combined measurements 
of hadronic cross-sections, and the measurements of heavy quark production, 
$\Rb$, $\Rc$, $\Abb$ and $\Acc$, upper limits can be set on the leptoquark's 
coupling $g$ as a function of its mass \MLQ\ for leptoquarks coupling electrons to first, second and third generation quarks.
For convenience, one type of leptoquark is assumed to be much lighter than 
the others. Furthermore, experimental constraints on the product $g_L  g_R$ 
allow the study leptoquarks assuming either only $g_L \neq 0$
or $g_R \neq 0$. Limits are then denoted by either (L) for leptoquarks coupling
to left handed leptons or (R) for leptoquarks coupling to right handed leptons.

In the processes $\eeuu$ and $\eedd$ first generation leptoquarks could be 
exchanged in $u$- or $t$-channel (F=2 or  F=0) which would lead to a change 
of the hadronic cross-section.
In the processes $\eecc$ and $\eebb$ the exchange of leptoquarks with 
cross-generational couplings can alter the \qq\ angular distribution, 
especially at low polar angle. 
The reported measurements on heavy quark production have been extrapolated 
to $4\pi$ acceptance, using SM predictions, from the measurements performed 
in restricted angular ranges, corresponding to the acceptance of the
vertex-detector in each experiment.
Therefore, when fitting limits on leptoquarks' coupling to the 2nd or 3rd 
generation of quarks, the LEP combined results for b and c sector are 
extrapolated back to an angular range of $\left| \cos \theta \right| < 0.85$
using ZFITTER predictions.  

The following measurements are used to constrain different types of leptoquarks
\begin{itemize}  
\item For leptoquarks coupling electrons to 1$^{\mathrm{st}}$ generation 
      quarks, all LEP combined hadronic cross-sections at centre-of-mass 
      energies from 130 GeV to 207 GeV are used

\item For leptoquarks coupling electrons to 2$^{\mathrm{nd}}$ generation 
      quarks, $\sigma_{\cc}$ is calculated from $\Rc$ and the hadronic 
      cross-section at the energy points where $\Rc$ is 
      measured. The measurements of $\sigma_{\cc}$ and $\Acc$ are then 
      extrapolated back to $\left| \cos \theta \right| < 0.85$.
      Since measurements in the c-sector are scarce and originate from, 
      at most, 2 experiments, hadronic cross-sections, extrapolated down to 
      $\left| \cos \theta \right| < 0.85$ are also used in the fit, with an 
      average $10\%$ correlated errors. 

\item For leptoquarks coupling electrons to 3$^{\mathrm{rd}}$ generation 
      quarks, only ${\mathrm \sigma_{b\bar b}}$ and \Abb, extrapolated 
      back to a $\left| \cos \theta \right| < 0.85$ are used.
\end{itemize}

\begin{figure}[tp]
\begin{center}
\mbox{\epsfig{file=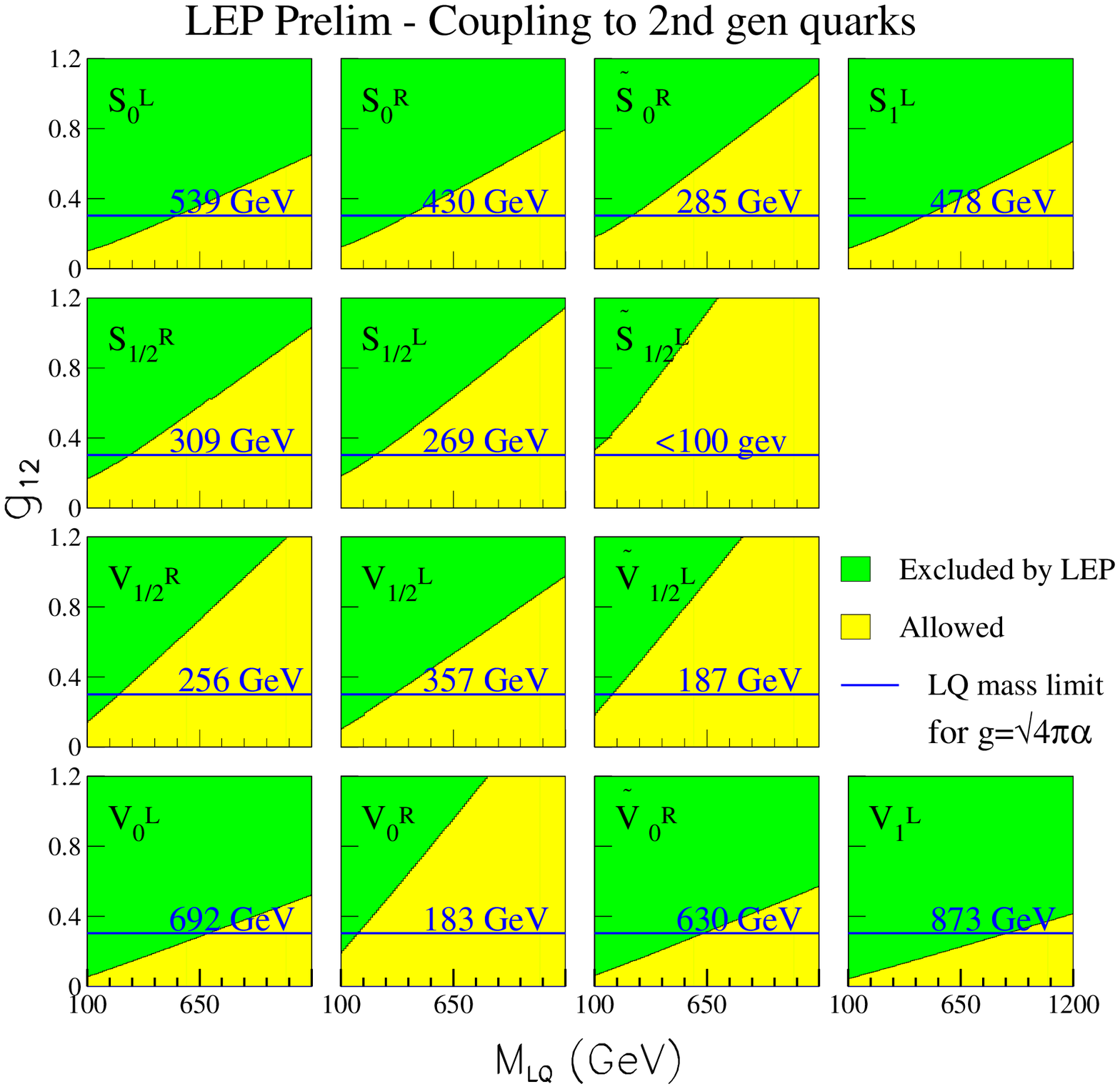,width=15cm}}
\caption{95\% confidence level limit on the coupling 
         of leptoquarks to 2nd generation of quarks.}  
\label{ff:fig:lq-2nd} 
\end{center}
\end{figure}

\begin{figure}[tp]
\begin{center}
\begin{tabular}{c}
\mbox{\epsfig{file=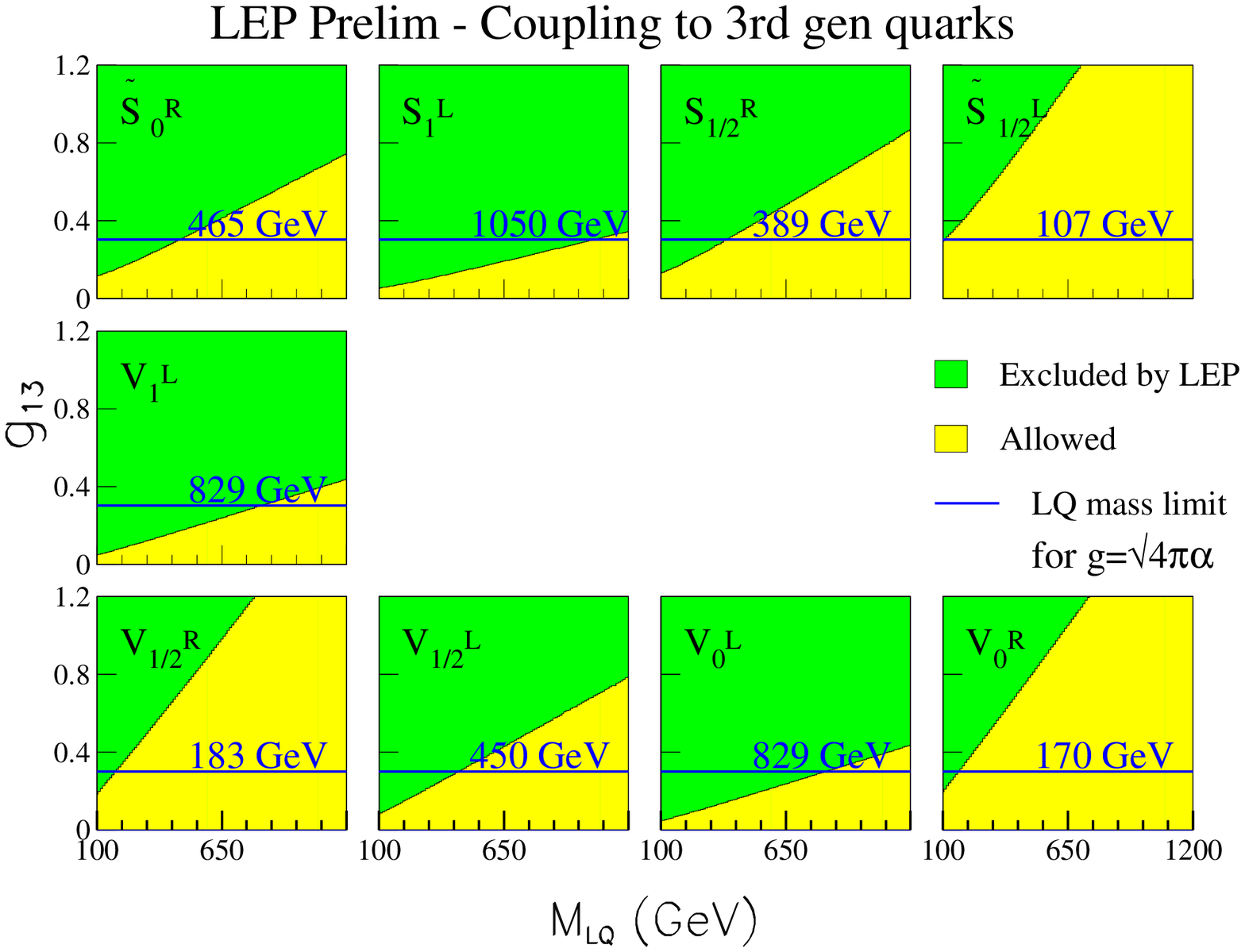,width=15cm}}
\end{tabular}
\caption{95\% confidence level limit on the coupling 
         of leptoquarks to 3rd generation of quarks.}  
\label{ff:fig:lq-3rd} 
\end{center}
\end{figure}

The 95$\%$ confidence level lower limits on masses $\MLQ$ are derived 
assuming a coupling of electromagnetic strength, 
$g = \sqrt{4\pi \alpha_{em}}$, where $\alpha_{em}$ is the fine structure 
constant. The results are summarised in    
Table~\ref{ff:tab:lq-mass}. These results complement the leptoquark searches 
at HERA~\cite{ff:ref:lq-h1,ff:ref:lq-zeus} and the 
Tevatron~\cite{ff:ref:lq-tevatron}.
Figures~\ref{ff:fig:lq-2nd} and \ref{ff:fig:lq-3rd} give the 95\% confidence 
level limits on the 
coupling as a function of the leptoquark mass for leptoquarks coupling 
electrons to the second and third generations of quarks.

\begin{table}[tp]                                                              
  \renewcommand{\arraystretch}{1.35}
 \begin{center}
\begin{tabular}{|c|c c|c|c c|c|c|}
\hline
 \multicolumn{8}{|c|}{Limit on scalar LQ mass (GeV/$c^{2}$)} \\
\hline\hline
 & \SL{0} & \SR{0} & \SBR{0} & \SL{\lqhalf} & \SR{\lqhalf} & \SBL{\lqhalf} & \SL{1} \\
\hline\hline
$LQ_{1st}$ &  655  &  520  &   202   &   178  &   232  &   -  &  361 \\
\hline
$LQ_{2nd}$ &  539  &  430  &   285   &   269  &   309  &   -  &  478 \\
\hline
$LQ_{3rd}$ &  NA  &  NA   &   465   &   NA  &   389  &   107  &  1050 \\
\hline
\multicolumn{8}{c}{\null}\\

\hline
 \multicolumn{8}{|c|}{Limit on vector LQ mass (GeV/$c^{2}$)} \\
\hline\hline
  & \VL{0} & \VR{0} & \VBR{0} & \VL{\lqhalf} & \VR{\lqhalf} & \VBL{\lqhalf} & \VL{1} \\
\hline\hline
$ LQ_{1st}$ &  917  &   165  &   489   &   303  &   227  &   176   &   659  \\
\hline
$ LQ_{2nd}$ &  692  &   183  &   630   &   357  &   256  &   187   &   873  \\
\hline
$ LQ_{3rd}$ &  829   &   170  &   NA   &   451  &   183  &   NA   &   829

  \\
\hline
\end{tabular}
\caption{$95\%$ confidence level lower limits on the LQ mass for leptoquarks 
         coupling between electrons and
         the first, second and third generation of quarks.
         A dash indicates that no limit can be  set and N.A denotes 
         leptoquarks coupling only to top quarks and hence not visible at LEP.}
\label{ff:tab:lq-mass}
\end{center} 
\end{table}                                                                  
\subsection{Low Scale Gravity in Large Extra Dimensions}
\label{ff:sec-grav}

The averaged differential cross-sections for $\eeee$ are used to search for 
the effects of graviton exchange in large extra dimensions.

A new approach to the solution of the hierarchy problem has
been proposed in~\cite{ff:ref:ADD,ff:ref:ADD2,ff:ref:ADD3}, which brings close
the electroweak scale $\rm m_{EW} \sim 1\; TeV$ and the
Planck scale $\rm M_{Pl} = \frac{1}{\sqrt{G_N}} \sim 10^{15}\; TeV$.
In this framework the effective 4 dimensional $\rm M_{Pl}$ is
connected to a new $\rm M_{Pl(4+n)}$ scale in a (4+n) dimensional
theory:
\begin{eqnarray}
\mathrm{M_{Pl}^2 \sim M_{Pl(4+n)}^{2+n} R^n},
\end{eqnarray}
where there are n extra compact spatial dimensions of radius
$\rm \sim R$.

In the production of fermion- or boson-pairs in $\ee$ collisions this class of
models can be manifested through virtual effects due to the exchange of
gravitons (Kaluza-Klein excitations). As discussed in~\cite{ff:ref:Hewett,ff:ref:Rizzo,ff:ref:Giudice,ff:ref:Lykken,ff:ref:Shrock},
the exchange of spin-2 gravitons modifies in a unique way the differential 
cross-sections for fermion pairs, providing clear signatures. These models 
introduce an effective scale (ultraviolet cut-off).
Adopting the notation from~\cite{ff:ref:Hewett} 
the gravitational mass scale is called $\mathrm{M_H}$. 
The cut-off scale is supposed to be of the order of the
fundamental gravity scale in 4+n dimensions.

The parameter $\varepsilon_{H}$ is defined as
\begin{eqnarray}
 \varepsilon_{H} = \frac{\lambda}{\mathrm{M_H^4}},
\end{eqnarray}
where the coefficient $\rm \lambda$ is of $\rm \mathcal{O}(1)$ and can not be
calculated explicitly without knowledge of the full quantum gravity
theory. In the following analysis we will assume that
$\rm \lambda = \pm 1$ in order to study both the cases of positive
and negative interference.
To compute the deviations from the Standard Model due to virtual graviton
exchange the calculations~\cite{ff:ref:Giudice,ff:ref:Rizzo} were used.

Theoretical uncertainties on the Standard Model predictions are taken 
from~\cite{ff:ref:lepffwrkshp}. The full correlation matrix of the 
differential cross-sections, obtained in our averaging procedure, is
used in the fits. This is an improvement compared to previous combined analyses
of published or preliminary LEP data on Bhabha scattering, performed before
this detailed information was available (see 
e.g.~\cite{ff:ref:Bourilkov:1999,ff:ref:Bourilkov:2000,ff:ref:Bourilkov:2001}).

\begin{figure}[p]
 \begin{center}
  \epsfig{file=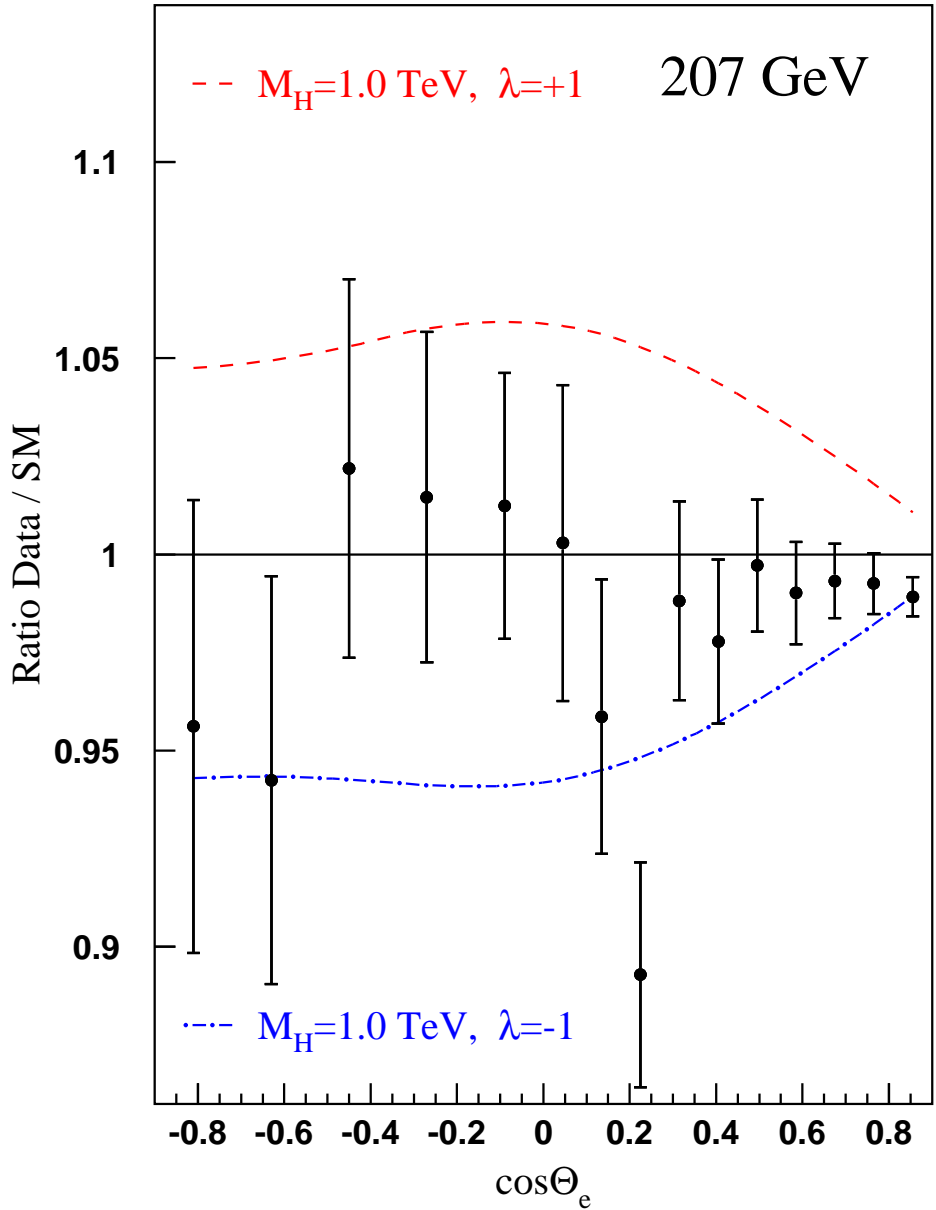,width=0.8\textwidth}
 \end{center}
 \caption{Ratio of the LEP averaged differential cross-section for $\eeee$
          at energy of 207 $\GeV$ compared to the SM prediction. The effects
          expected from virtual graviton exchange are also shown.}
 \label{ff:fig:dsdc-ee-207-lsg}
 \vskip 2cm 
\end{figure}

The extracted value of $\varepsilon_{H}$ is compatible 
with the Standard Model expectation $\varepsilon_{H}=0$.
The errors on $\varepsilon_{H}$ are $\sim 1.5$
smaller than those obtained from a single LEP experiment with the same data
set. The fitted value of $\varepsilon_{H}$ is converted into  
$95\%$ confidence level lower limits on $\mathrm{M_H}$
by integrating the likelihood function over the 
physically allowed values, $\varepsilon_{H} \ge 0$ for $\lambda = +1$ and 
$\varepsilon_{H} \le 0$ for $\lambda = -1$ giving:
\begin{eqnarray}
\mathrm{M_H} & > & 1.20~\TeV\qquad\mathrm{for}~\lambda = +1\,, \\
\mathrm{M_H} & > & 1.09~\TeV\qquad\mathrm{for}~\lambda = -1\,.
\end{eqnarray}
An example of our analysis for the highest energy point is
shown in Figure~\ref{ff:fig:dsdc-ee-207-lsg}.

The interference of virtual graviton exchange amplitudes with both 
t-channel and s-channel Bhabha scattering amplitudes makes this the
most sensitive search channel at LEP. The results obtained here would not
be strictly valid if the luminosity measurements of the LEP experiments,
based on the very same process, are also significantly affected by graviton
exchange.
As shown in~\cite{ff:ref:Bourilkov:1999}, the effect on the cross-section
in the luminosity angular range is so small that it can safely be neglected
in this analysis.

\section{Summary}
\label{ff:sec-conc}

A preliminary combination of the $\LEPII$ $\eeff$ cross-sections (for hadron, 
muon and tau-lepton final states) and
forward-backward asymmetries (for muon and tau final states) 
from LEP running at energies from 130~$\GeV$  to 207~$\GeV$ has been made. 
The results from the four LEP experiments are in good 
agreement with each other. 
The averages for all energies are shown given in Table~\ref{ff:tab:xsafbres}.
Overall the data agree 
with the Standard Model predictions of ZFITTER, although the combined hadronic 
cross-sections are on average $1.7$ standard deviations above the predictions.
Further information is available at~\cite{ff:ref:ffbar_web}.

Preliminary differential cross-sections, $\dsdc$, for $\eeee$, $\eemumu$ and
$\eetautau$ were combined. Results are shown in 
Figures~\ref{ff:fig:dsdc-res-ee}, \ref{ff:fig:dsdc-res-mm} 
and~\ref{ff:fig:dsdc-res-tt}.

A preliminary average of results on heavy flavour production at $\LEPII$ 
has also been made for measurements of $\Rb$, $\Rc$, $\Abb$
and $\Acc$, using results from LEP centre-of-mass energies
from 130 to 207 $\GeV$. Results are given in Tables~\ref{ff:tab:hfbresults}
and~\ref{ff:tab:hfcresults}  and 
shown graphically in Figures~\ref{ff:fig:hfbres} and~\ref{ff:fig:hfcres}.
The results are in good agreement with the predictions of the SM. 

The preliminary averaged cross-section and forward-backward asymmetry results 
together with the combined results on heavy flavour production have been
interpreted in a  variety of models. 
Limits on the scale of contact interactions between leptons and quarks 
and in $\eeee$ and also 
between electrons and specifically $\bb$ and $\cc$ final states have been 
determined.
A full set of limits are given in Tables~\ref{ff:tab:cntceps} and
\ref{ff:tab:cntclmb}.
The $\LEPII$ averaged cross-sections have been used to obtain lower limits 
on the mass of a possible $\Zprime$ boson 
in different models. Limits range from $340$ to $1787$ $\GeV/c^2$ depending
on the model. 
Limits on the masses of leptoquarks have been derived from the hadronic 
cross-sections. The limits range from $101$ to $1036$ $\GeV/c^2$ depending on 
the type of leptoquark.
Limits on the scale of gravity in models with large extra dimensions have been 
obtained from combined differential cross-sections for $\eeee$; for
positive interference between the new physics and the Standard model the limit
is $1.20$ TeV and for negative interference $1.09$ TeV.

\boldmath
\chapter{Investigation of the Photon/Z-Boson Interference}
\label{sec-smat}
\unboldmath

\updates{ Unchanged w.r.t. summer 2002: Results are preliminary. 

  Note that some recent publications~\cite{Abbiendi:2003dh,
  LEP2L3ffbar} are not yet included in this combination.  }

\section{Introduction}

The S-Matrix ansatz provides a coherent way of describing
LEP measurements of the cross-section and forward-backward 
asymmetries in $s$-channel $\eeff$ processes at centre-of-mass energies 
around the $\Zzero$ resonance, from the
{\LEPI} program, and the measurements at centre-of-mass 
energies from  130 -- 207~GeV from the {\LEPII} program.

Compared with the standard 5 and 9 parameter descriptions of the
measurements at the $\Zzero$~\cite{smat:ref:5and9}, the S-Matrix formalism
includes an extra 3 parameters (assuming lepton universality) or 7 parameters
(without lepton universality) which explicitly determine the
contributions to the cross-sections and forward-backward asymmetries
of the interference between the exchange of a $\Zzero$ and a photon.
The {\LEPI} data alone cannot tightly constrain these interference 
terms, in particular the interference term for hadronic cross-sections, 
since their contributions are small around the $\Zzero$ resonance and change 
sign at the pole.
Due to strong correlations between the size of the hadronic interference 
term and the mass of the $\Zzero$, this leads to a larger error on the 
fitted mass of the $\Zzero$ compared to the standard 5 and 9 parameter fits, 
where the hadronic interference term is fixed to the value predicted in the
Standard Model. Including the {\LEPII} data 
leads to a significant improvement in the constraints on the interference terms
and a corresponding reduction in the uncertainty on the mass of the 
$\Zzero$. This results in a measurement of {$\MZ$} which is almost as 
sensitive as the standard results, 
but without constraining the interference to the Standard Model prediction.
This chapter describes the first, preliminary, combination of data from the
full data sets of the 4 LEP experiments, to obtain a LEP combined results 
on the parameters of the S-Matrix ansatz. These results update those
of a previous combination~\cite{smat:ref:smat1997} which was based on
preliminary {\LEPI} data and only partial statistics from the full {\LEPII} 
data set.

Different strategies are used to combine the {\LEPI} and {\LEPII} data.
For {\LEPI} data, an average of the individual experiment's results
on the S-Matrix parameters is made. This approach is rather similar to
the method used to combine the results of the 5 and 9 parameter fits.
To include {\LEPII} data, a fit is made to LEP combined 
measurements of cross-sections and asymmetries above the $\Zzero$, 
taking into account the results of the {\LEPI} combination of S-Matrix
parameters.

In Section~\ref{smat:sec:ansatz} the parameters of the
S-Matrix ansatz are explained. In Sections~\ref{smat:sec:lep1} 
and~\ref{smat:sec:lep2} the average of the {\LEPI} data and the 
inclusion of the {\LEPII} data are described. The results are discussed in 
Section~\ref{smat:sec:discuss} and conclusions are drawn in 
Section~\ref{smat:sec:conc}.

\section{The S-Matrix Ansatz}
\label{smat:sec:ansatz}

The S-matrix
ansatz~\cite{smat:ref:bcms,*smat:ref:rgs,*smat:ref:arr,*smat:ref:tr}
is a rigorous approach to describe the cross-sections and
forward-backward asymmetries in the $s$-channel $\ee$ annihilations
under the assumption that the processes can be parameterised as the
exchange of a massless and a massive vector boson, in which the
couplings of the bosons including their interference are treated as
free parameters.

In this model, the cross-sections can be parametrised
 as follows:
\begin{equation}
\sigma^0_{tot, f}(s)=\frac{4}{3}\pi\alpha^2
\left[
      \frac{\gf^{tot}}{s}
     +\frac{\jf^{tot} (s-\MZbar^2) + \rf^{tot} \, s}
           {(s-\MZbar^2)^2 + \MZbar^2 \GZbar^2}
\right]
\,\,\,\mathrm{with}\,\,\mathrm{f=had,e,\mu,\tau}\,,
\label{smat:eqn:eq1}
\end{equation}
while the forward-backward asymmetries are given by:
\begin{equation}
A^0_\mathrm{fb, f}(s)=\pi\alpha^2
\left[
 \frac{\gf^{fb}}{s} +
 \frac{\jf^{fb} (s-\MZbar^2) + \rf^{fb} \, s}
           {(s-\MZbar^2)^2 + \MZbar^2 \GZbar^2}
\right]
                       / {\sigma^0_{\mathrm{tot, f}}(s)}\,,
\label{smat:eqn:eq2}
\end{equation}
where $\roots$ is the centre-of-mass energy.
The parameters $\rf$ and $\jf$ scale the $\Zzero$ exchange and the 
\mbox{$\Zzero-\gamma$} 
interference contributions to the total cross-section and forward-backward
asymmetries. The contribution $\gf$ of the pure $\gamma$ exchange was fixed to 
the value predicted by QED in all fits. Neither the hadronic charge
asymmetry, nor the flavour tagged quark forward-backward asymmetries are 
considered here, 
which leaves 16 free parameters to described the LEP data: 14 $\rf$ and
$\jf$ parameters and the mass and width of the massive $\Zzero$ resonance.
Applying the constraint of lepton universality reduces this to 8 
parameters.

In the Standard Model the $\Zzero$ exchange term, the $\Zzero-\gamma$ 
interference term and the photon exchange term are given in terms of the
fermion charges and their effective vector and axial couplings to the
$\Zzero$ by:
\begin{equation}
\begin{array}{l@{}l@{}l@{}l}
\displaystyle
\rtotf & = & \kappa^2
             \left[\gae^2+\gve^2\rule{0mm}{4mm}\right]
             \left[\gaf^2+\gvf^2\right]
            -2\kappa\,\gve\,\gvf C_{Im}         \\[5mm]
\jtotf & = & 2\kappa\,\gve\,\gvf \left(C_{Re}+C_{Im}\right) \\[5mm]
\gtotf & = & Q^2_{\mathrm{e}}Q^2_{\mathrm{f}}\left|F_A(\MZ)\right|^2 \\[5mm]
\rfbf  & = & 4\kappa^2\gae\,\gve\,\gaf\,\gvf
            -2\kappa\,\gae\,\gaf C_{Im}         \\[5mm]
\jfbf  & = & 2\kappa\,\gae\,\gaf \left(C_{Re}+C_{Im}\right) \\[5mm]
\gfbf  & = & 0 \,,
\end{array}
\end{equation}
with the following definitions:
\begin{equation}
\begin{array}{l}
\displaystyle
\kappa    = \dfrac{G_F\MZ^2}{2\sqrt{2\,}\pi\alpha} \approx 1.50\\[5mm]
C_{Im}    = \dfrac{\GZ}{\MZ}  \left.Q_{\mathrm{e}}Q_{\mathrm{f}}\right.
                                \mathrm{Im} \left\{F_A(\MZ)\right\} \\[5mm]
C_{Re}    =                   \left.Q_{\mathrm{e}}Q_{\mathrm{f}}\right.
                                \mathrm{Re} \left\{F_A(\MZ)\right\} \\[5mm]
F_A(\MZ)  = \dfrac{\alpha(\MZ)}{\alpha} \,,
\end{array}
\end{equation}
where $\alpha(\MZ)$ is the complex fine-structure constant, and 
$\alpha\equiv\alpha(0)$.
The photonic virtual and bremsstrahlung corrections are included through 
the convolution of Equations~\ref{smat:eqn:eq1} and~\ref{smat:eqn:eq2}
with radiator functions as in the 5 and 9 parameter fits.
The expressions of the S-Matrix parameters in terms of the
effective vector and axial-vector couplings given above 
neglect the imaginary parts of the effective couplings.

The usual definitions of the mass $\MZ$ and width $\GZ$ of a Breit-Wigner 
resonance are used, the width being $s$-dependent, such that:
\begin{equation}
\begin{array}{lllll@{}c@{}r}
  \MZ & \equiv  & \MZbar\sqrt{1+\GZbar^2/\MZbar^2} & \approx & \MZbar & + & 34.20~\MeV/c^2\phantom{\,,} \\[3mm]
  \GZ & \equiv  & \GZbar\sqrt{1+\GZbar^2/\MZbar^2} & \approx & \GZbar & + &  0.94~\MeV\,.\phantom{/c^2}
\end{array}
\label{smat:eqn:smat-ls}
\end{equation}

In the following fits, the predictions from the S-Matrix ansatz and
the QED convolution for cross-sections and asymmetries are made using
SMATASY~\cite{smat:ref:smatasy}, which in turn uses
ZFITTER~\cite{smat:ref:lep2xsafbave} to calculate the QED convolution
of the electroweak kernel.  In case of the $\ee$ final state,
$t$-channel and $s/t$ interference contributions are added to the
$s$-channel ansatz.

\section{LEP combination}
\label{smat:sec:comb}

In the following sections the combinations of the results from the
individual LEP experiments are described: firstly the {\LEPI}
combination, then the combination of both {\LEPI} and {\LEPII} data.
The results from these combinations are compared in
Section~\ref{smat:sec:discuss}.  Although all 16 parameters are
averaged during the combination, only results for the
parameters $\MZ$ and $\jtoth$ are reported here. Systematic studies
specific to the other parameters are ongoing.

\subsection{\LEPI\ combination}
\label{smat:sec:lep1}

Individual LEP experiments have their own determinations of the
16 S-Matrix parameters~\cite{smat:ref:aleph,smat:ref:delphi,smat:ref:l3lep1,smat:ref:opallep1} from {\LEPI} data alone, using the full {\LEPI} data sets.

These results are averaged using a multi-parameter BLUE technique
based on an extension of Reference~\citen{common_bib:BLUE}. Sources of 
systematic uncertainty correlated between the experiments have been 
investigated, using techniques described in~\cite{smat:ref:5and9} and are
accounted for in the averaging procedure and benefiting from the experience
gained in those combinations.

The parameters $\MZ$ and $\jtoth$ are the most sensitive of all 16
S-matrix parameters to the inclusion of the {\LEPII} data, and are
also the
most interesting ones in the context of the 5 and 9 parameter fits. For
these parameters the most significant source of systematic error which
is correlated between experiments comes from the uncertainty on the
$\ee$ collision energy as determined by models of the LEP RF system
and calibrations using the resonant depolarisation technique. These
errors amount to $\pm 3$ MeV on $\MZ$ and $\pm 0.16$ on $\jtoth$ with
a correlation coefficient of $-0.86$.  The LEP averaged values of
$\MZ$ and $\jtoth$ are given in Table~\ref{smat:tab:lepres}, together
with their correlation coefficient.  The $\chi^2/$D.O.F. for the
average of all 16 parameters is 62.0/48, corresponding to a
probability of $8\%$, which is acceptable.

\begin{table}[tpb]
\begin{center}
\renewcommand{\arraystretch}{1.2}
\begin{tabular}{|c|r@{$\pm$}l|r@{$\pm$}l|c|}
\hline
 & \multicolumn{2}{c|}{ $\MZ$ [GeV] }& \multicolumn{2}{c|}{ $\jtoth$}
 & correlation \\ \hline\hline
{\LEPI} only  & 91.1925 & 0.0059 &  -0.084 & 0.324  & -0.935 \\
{\LEPI} \& {\LEPII} & 91.1869 & 0.0023 &   0.277 & 0.065  & -0.461 \\ \hline
\end{tabular}
\caption{Averaged {\LEPI} and {\LEPII} S-Matrix results for $\MZ$ and $\jtoth$.}
\label{smat:tab:lepres}
\end{center}
\end{table}
\subsection{{\LEPI} and {\LEPII} combination}
\label{smat:sec:lep2}

Some experiments have determined S-Matrix parameters using 
their {\LEPI} and {\LEPII} measured cross-sections and forward-backward
asymmetries~\cite{smat:ref:aleph,smat:ref:delphi,smat:ref:l3lep2,smat:ref:opallep2}. To do a full LEP combination
would require each experiment to provide S-Matrix results and would require 
an analysis of the correlated systematic errors on each measured parameter.

However, preliminary combinations of the measurements of forward-backward 
asymmetries and cross-sections from all 4 LEP experiments, for the 
full~{\LEPII} period, have already been made~\cite{smat:ref:lep2xsafbave} and 
correlations between these measurements have been estimated. The combination
procedure averages measurements of cross-sections and asymmetry for those
events with reduced centre-of-mass energies, $\rootsp$, close to the actual 
centre-of-mass energy of the $\ee$ beams, $\roots$, removing those
events which are less sensitive to the $\Zzero-\gamma$ interference where, 
predominantly, initial state radiation reduces the centre-of-mass
energy to close to the mass of the $\Zzero$. The only significant correlations
are those between hadronic cross-section measurements at different energies,
which are around 20--40\%, depending on energies.

The predictions from SMATASY are fitted to the combined 
{\LEPII} cross-section and forward-backward asymmetry 
measurements~\cite{smat:ref:lep2xsafbave}.
The signal definition 1 of Reference~\citen{smat:ref:lep2xsafbave} is 
used for the data and for the predictions of SMATASY. 
Theoretical uncertainties on the S-Matrix predictions for the {\LEPII}
results and on the corrections of the $\LEPII$ data to the common signal 
defintion are taken to be the same as for the 
Standard Model predictions of ZFITTER~\cite{smat:ref:lep2xsafbave} which
are dominated by uncertainties in the QED convolution. These 
amount to a relative uncertainty of $0.26\%$ on the hadronic cross-sections, 
fully correlated between all {\LEPII} energies.

The fit also uses as inputs the averaged {\LEPI} S-Matrix
parameters and covariance matrix. These inputs effectively constrain those
parameters, such as $\MZ$, which are not accurately determined by 
{\LEPII} data.
There are no significant correlations between the {\LEPI} and {\LEPII} inputs.

The LEP averaged values of $\MZ$ and $\jtoth$ for both {\LEPI} and {\LEPII}
data are given in Table~\ref{smat:tab:lepres}, together with their
correlation coefficient. 
The $\chi^2/$D.O.F. for the average of all 16 parameters 
is 64.4/60, corresponding to a probability of $33\%$, which is good.

\subsection{Discussion}
\label{smat:sec:discuss}

In the {\LEPI} combination the measured values of the Z boson mass
$\MZ = 91.1925 \pm 0.0059$~GeV agrees well with the results of the standard 
9 parameter fit ($91.1876 \pm 0.0021$~GeV) albeit with a significantly
larger error, resulting from the correlation with the large uncertainty 
on $\jtoth$ which is then the dominant source of uncertainty on $\MZ$ in the
S-Matrix fits.
The measured value of $\jtoth = -0.084 \pm 0.324 $ , also agrees with the 
prediction of the Standard Model ($0.2201^{+0.0032}_{-0.0137}$). 

\begin{figure}[p]
 \begin{center}
  \mbox{\epsfig{file=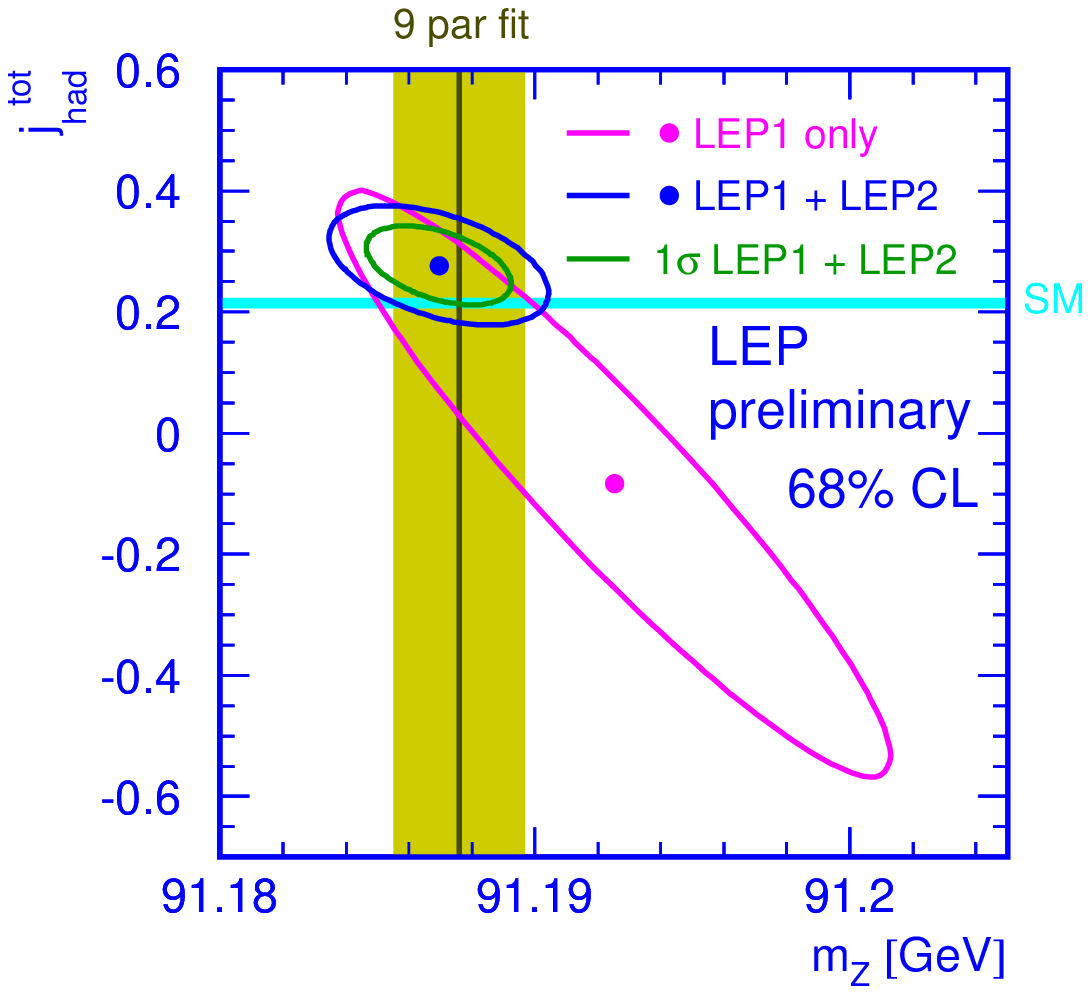,width=\textwidth}}
  \caption{Error ellipses for $\MZ$ and $\jtoth$ for {\LEPI} (at 39\% and
    68\%) and the combination of {\LEPI} and {\LEPII} (at 68\%).}
  \label{smat:fig:mzjtoth}
 \end{center}
\end{figure}

Including the {\LEPII} data brings a significant improvement in the
uncertainty on the size of the interference between $\Zzero$ and photon
exchange compared to {\LEPI} data alone.  The measured value 
$\jtoth = 0.277 \pm 0.065$, agrees well with the values predicted from the 
Standard Model. Correspondingly, the uncertainty
on the the mass of the $\Zzero$ in this ansatz, $2.3$ MeV, is close to 
the precision obtained from {\LEPI} data alone using the standard 9 parameter
fit, $2.1$ MeV. The slightly larger error is due to the uncertainty on
$\jtoth$ which amounts to 0.9~MeV. 
The measured value, $\MZ = 91.1869 \pm 0.0023$~GeV, agrees with that 
obtained from the standard 9 parameter fits.
The results are summarised in Figure~\ref{smat:fig:mzjtoth}.

The good agreement found between the values of $\MZ$ and $\jtoth$ and their
expectations provide a validation of the approach taken in the standard 5 
and 9 parameter fits, in which the size of the interference between $\Zzero$
boson and photon exchange in the hadronic cross-sections was fixed to the
Standard Model expectation.

The precision on $\jtoth$ is slightly better than that obtained by the VENUS 
collaboration~\cite{smat:ref:venus} of $\pm 0.08$, which was obtained using
preliminary results from {\LEPI} and their own 
measurements of the hadronic cross-section below the $\Zzero$ resonance.
The measurement of the hadronic cross-sections from VENUS~\cite{smat:ref:venus}
and TOPAZ~\cite{smat:ref:topaz} could be included in the future to give a 
further reduction in the uncertainty on $\jtoth$.

Work is in progress to understand those sources of systematic error,
correlated between experiments, which are significant
for the remaining S-Matrix parameter that have not been presented here. 
In particular, for $\jtote$ and $\jfbe$, it is important to understand the 
errors resulting from $t$-channel contributions to the $\eeee$ process.
These errors have only limited impact on the standard 5 and 9 parameter fits.

\section{Conclusion}
\label{smat:sec:conc}

Results for the S-Matrix parameter $\MZ$ and $\jtoth$ have been presented 
for {\LEPI} data alone and for a fit using the full data sets for
{\LEPI}  and {\LEPII} from all 4 LEP experiments. 
Inclusion of {\LEPII} data brings a significant improvement 
in the determination of $\jtoth$, the fitted value $0.277 \pm 0.065$, agrees 
well with the values predicted from the Standard Model. As a result
in the improvement of the uncertainty in $\jtoth$, the uncertainty on the 
fitted value of $\MZ$ approaches that of the standard 5 and 9 parameter
fits and the measured value $\MZ = 91.1869 \pm 0.0023$~GeV is compatible
with that from the standard fits.

\boldmath
\chapter{W and Four-Fermion Production at \LEPII}
\label{sec-4F}
\unboldmath

\updates{ The WW cross-section, $\rww$ and the W branching ratios
  combinations are updated including the final reviewed \Aleph\
  results, with very minor changes with respect to the last
  combination.  The determination of $|\mathrm{V}_{\mathrm{cs}}|$ is
  also updated. An update of the combination of the W polar angle
  differential cross-sections using final \Aleph\, \Delphi\ and \Ltre\
  results is presented. The single boson We$\nu$ and Zee
  cross-sections are updated with the final DELPHI numbers.  A first
  combination of Z$\gamma^*$ final states using the \Delphi\ and
  \Ltre\ inputs is also made. All combinations are preliminary. }

\section{Introduction}
\label{4f_sec:introduction}

This chapter summarises the present status of the combination of published 
and preliminary
results of the four LEP experiments on four-fermion cross-sections 
for the Summer 2005 Conferences.
If not stated otherwise, all presented results use the full $\LEPII$ data 
sample at \CoM\ energies up to 209 GeV, supersede the results presented
at the Summer 2004 Conferences~\cite{4f_bib:4f_s04} and have to be 
considered as preliminary.

The \CoM\ energies and the corresponding integrated luminosities are provided
by the experiments and are the same used for previous 
conferences. The LEP energy value in each point (or group
of points) is the luminosity-weighted average of those values. 

Cross-section results from different experiments are combined 
by $\chi^2$ minimisation using the Best Linear Unbiased Estimate method 
described in Ref.~\cite{common_bib:BLUE}, properly taking into account the correlations 
between the systematic uncertainties.

The detailed inputs from the experiments and the resulting LEP
combined values, with the full breakdown of systematic errors 
is described in Appendix~\ref{4f_sec:appendix}.
Experimental results are compared with recent theoretical predictions,
many of which were developed in the framework 
of the $\LEPII$ \MC\ workshop~\cite{4f_bib:fourfrep}. 

\section{W-pair production cross-section}
\label{4f_sec:WWxsec}

ALEPH, DELPHI and L3 have presented final results 
on the W-pair ({\sc CC03}~\cite{4f_bib:fourfrep}) 
production cross-section and W branching ratios for all $\LEPII$ \CoM\ 
energies~\cite{4f_bib:adloww161,4f_bib:adloww172,4f_bib:aleww,4f_bib:delww,4f_bib:ltrww}.
OPAL has final results from 161 to 189~GeV~\cite{4f_bib:adloww161,4f_bib:adloww172,4f_bib:opaww189}
and preliminary measurements at $\roots=192$--207~GeV~\cite{4f_bib:opawwsc01}.

With respect to the Summer 2004 Conferences, new final results from ALEPH are now included 
in the LEP averages. The difference in the combined results is marginal with respect to
last summer. 
The same grouping of the systematic errors consolidated in previous 
combinations~\cite{4f_bib:4f_s04} was used.

The detailed inputs used for the combinations are given in Appendix~\ref{4f_sec:appendix}.

The measured statistical errors are used for the combination; 
after building the full 32$\times$32 covariance matrix for the measurements,
the $\chi^2$ minimisation fit is performed by matrix algebra,
as described in Ref.~\cite{4f_bib:valassi},
and is cross-checked using Minuit~\cite{MINUIT}.

The results from each experiment 
for the W-pair production cross-section
are shown in Table~\ref{4f_tab:wwxsec}, 
together with the LEP combination at each energy. 
All measurements assume Standard Model values for the W decay branching fractions.
The results for \CoM\ energies between 183 and 207 GeV,
for which new LEP averages have been computed,
supersede the ones presented in~\cite{4f_bib:4f_s04}.
For completeness, 
the measurements at 161 and 172~GeV are also listed in the table.

\renewcommand{\arraystretch}{1.2}
\begin{table}[hbtp]
\begin{center}
\hspace*{-0.5cm}
\begin{tabular}{|c|c|c|c|c|c|r|} 
\hline
\roots & \multicolumn{5}{|c|}{WW cross-section (pb)} 
       & \multicolumn{1}{|c|}{$\chi^2/\textrm{d.o.f.}$} \\
\cline{2-6} 
(GeV)      & \Aleph\                & \Delphi\               &
             \Ltre\                 & \Opal\                 &
             LEP                    &                        \\
\hline
161.3      & $\phz4.23\pm0.75^*$    & 
             $\phz3.67^{\phz+\phz0.99\phz*}_{\phz-\phz0.87}$& 
             $\phz2.89^{\phz+\phz0.82\phz*}_{\phz-\phz0.71}$& 
             $\phz3.62^{\phz+\phz0.94\phz*}_{\phz-\phz0.84}$& 
             $\phz3.69\pm0.45\phs^*$  & 
             $\left\} \hspace*{2mm} \phz1.3\phz/\phz3 \right.$ \\
172.1      & $11.7\phz\pm1.3\phz^*$ & $11.6\phz\pm1.4\phz^*$ &
             $12.3\phz\pm1.4\phz^*$ & $12.3\phz\pm1.3\phz^*$ &
             $12.0\phz\pm0.7\phz\phs^*$ & 
             $\left\} \hspace*{2mm} \phz0.22/\phz3 \right.$ \\
182.7      & $15.86\pm0.63^*$       & $16.07\pm0.70^*$       &
             $16.53\pm0.72^*$       & $15.43\pm0.66^*$       &
             $15.88\pm0.35\phs^*$     & 
             \multirow{8}{20.3mm}{$
               \hspace*{-0.3mm}
               \left\}
                 \begin{array}[h]{rr}
                   &\multirow{8}{8mm}{\hspace*{-4.2mm}26.6/24}\\
                   &\\ &\\ &\\ &\\ &\\ &\\ &\\  
                 \end{array}
               \right.
               $}\\
188.6      & $15.78\pm0.36^*$       & $16.09\pm0.42^*$       &
             $16.17\pm0.41^*$       & $16.30\pm0.39^*$       &
             $16.03\pm0.21\phs^*$   & \\
191.6      & $17.10\pm0.90\phs^*$   & $16.64\pm1.00^*$     &
             $16.11\pm0.92\phs^*$   & $16.60\pm0.99\phs$     &
             $16.56\pm0.48\phs$     & \\
195.5      & $16.60\pm0.54\phs^*$   & $17.04\pm0.60^*$     &
             $16.22\pm0.57\phs^*$   & $18.59\pm0.75\phs$     &
             $16.90\pm0.31\phs$     & \\
199.5      & $16.93\pm0.52\phs^*$   & $17.39\pm0.57^*$     &
             $16.49\pm0.58\phs^*$   & $16.32\pm0.67\phs$     &
             $16.76\pm0.30\phs$     & \\
201.6      & $16.63\pm0.71\phs^*$   & $17.37\pm0.82^*$     &
             $16.01\pm0.84\phs^*$   & $18.48\pm0.92\phs$     &
             $16.99\pm0.41\phs$     & \\
204.9      & $16.84\pm0.54\phs^*$   & $17.56\pm0.59^*$     &
             $17.00\pm0.60\phs^*$   & $15.97\pm0.64\phs$     &
             $16.79\pm0.31\phs$     & \\
206.6      & $17.42\pm0.43\phs^*$   & $16.35\pm0.47^*$     &
             $17.33\pm0.47\phs^*$   & $17.77\pm0.57\phs$     &
             $17.15\pm0.25\phs$     & \\
\hline
\end{tabular}
\caption{%
W-pair production cross-section from the four LEP
experiments and combined values at all recorded \CoM\ energies.
All results are preliminary, with the exception of those indicated by $^*$. 
The measurements between 183 and 207 GeV
have been combined in one global fit, taking into account 
inter-experiment as well as inter-energy correlations of systematic errors.
The results for the combined LEP W-pair production cross-section 
at 161 and 172~GeV are taken 
from~\protect\cite{4f_bib:lepewwg97,4f_bib:lepewwg98} respectively.}
\label{4f_tab:wwxsec}
\end{center}
\vspace*{-4mm}
\end{table}
\renewcommand{\arraystretch}{1.}

\begin{figure}[p]
\centering
\epsfig{figure=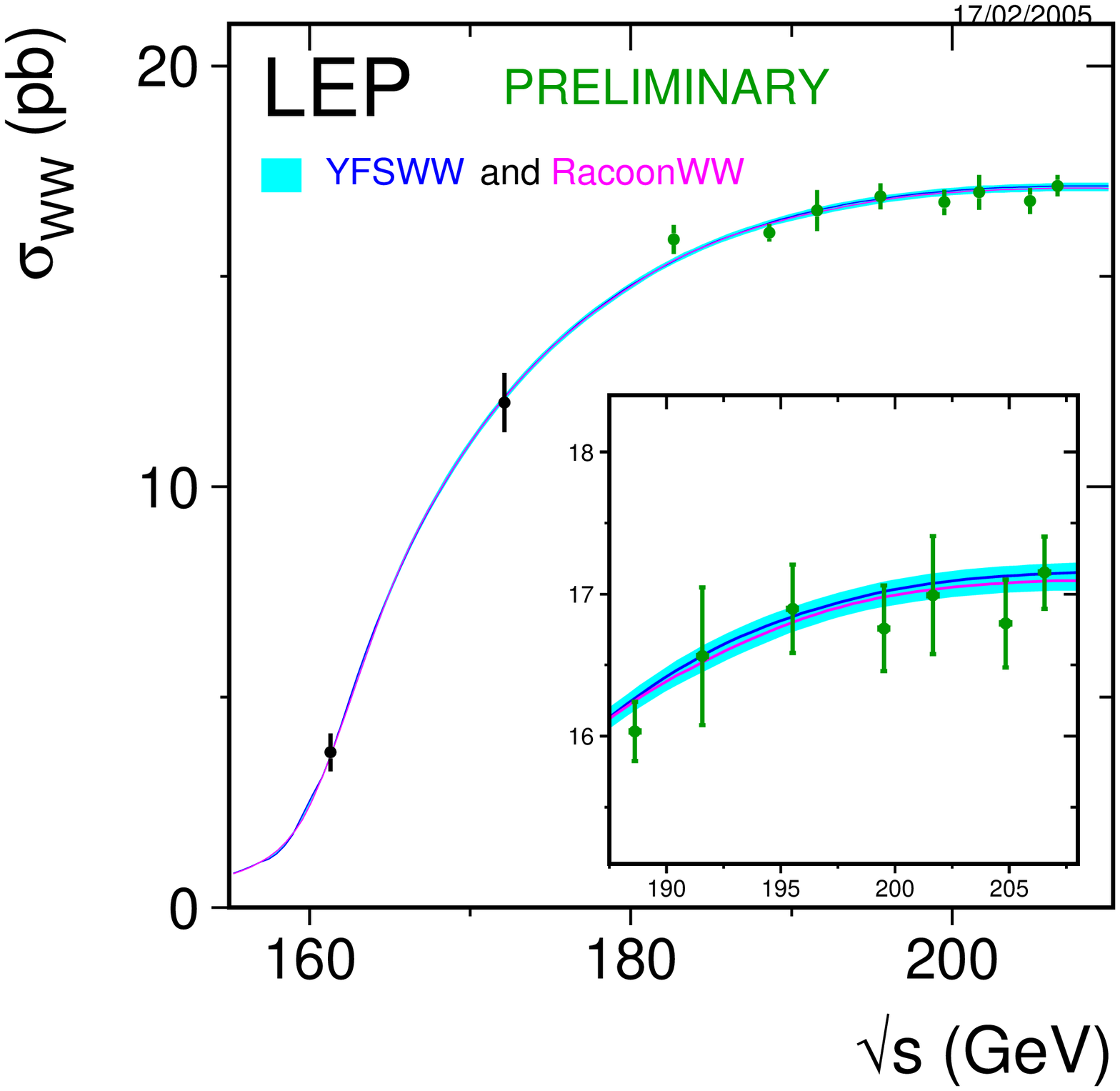,width=0.9\textwidth}
\caption{%
  Measurements of the W-pair production cross-section,
  compared to the predictions 
  of \RacoonWW~\protect\cite{4f_bib:racoonww} 
  and \YFSWW~\protect\cite{4f_bib:yfsww}. 
  The shaded area represents the uncertainty on the theoretical predictions,
  estimated in $\pm$2\% for $\roots\!<\!170$ GeV
  and ranging from 0.7 to 0.4\% above 170~GeV.
}
\label{4f_fig:sww_vs_sqrts}
\end{figure}

Figure~\ref{4f_fig:sww_vs_sqrts} shows the combined LEP W-pair cross-section 
measured as a function of the \CoM\ energy.
The experimental points are compared with the theoretical calculations 
from \YFSWW~\cite{4f_bib:yfsww} and \RacoonWW~\cite{4f_bib:racoonww} 
between 155 and 215 GeV for $\Mw=80.35$~GeV.
The two codes have been extensively compared and 
agree at a level better than 0.5\% 
at the $\LEPII$ energies~\cite{4f_bib:fourfrep}.
The calculations above 170 GeV, 
based for the two programs on the so-called leading pole~(LPA) 
or double pole approximations~(DPA)~\cite{4f_bib:dpa}, 
have theoretical uncertainties 
decreasing from 0.7\% at 170 GeV
to about 0.4\% at \CoM\ energies larger than 200 GeV,
while in the threshold region, where the codes are run in Improved Born
Approximation, a larger theoretical uncertainty of 2\% is 
assigned~\cite{4f_bib:dpaerr}.
This theoretical uncertainty is represented by
the blue band in Figure~\ref{4f_fig:sww_vs_sqrts}. 
An error of 50 MeV on the W mass would translate 
into additional errors of 0.1\% (3.0\%) 
on the cross-section predictions at 200 GeV (161 GeV, respectively).
All results, up to the highest \CoM\ energies, 
are in agreement with the considered theoretical predictions.

The agreement between the measured W-pair cross-section,
$\sww^\mathrm{meas}$,
and its expectation according to a given theoretical model,
$\sww^\mathrm{theo}$,
can be expressed quantitatively in terms of their ratio
\begin{equation}
\rww = \frac{\sww^\mathrm{meas}}{\sww^\mathrm{theo}} ,
\end{equation}
averaged over the measurements performed by the four experiments 
at different energies in the $\LEPII$ region.
The above procedure has been used to compare the measurements
at the eight energies between 183 and 207 GeV to the predictions 
of \Gentle~\cite{4f_bib:gentle}, \KoralW~\cite{4f_bib:koralw}, 
\YFSWW~\cite{4f_bib:yfsww} and \RacoonWW~\cite{4f_bib:racoonww}.
The measurements at 161 and 172 GeV
have not been used in the combination 
because they were performed using data samples of low statistics
and because of the high sensitivity of the cross-section 
to the value of the W mass at these energies.

The combination of the ratio $\rww$ is performed
using as input from the four experiments the 32 cross-sections
measured at each of the eight energies.
These are then converted into 32 ratios by dividing them
by the considered theoretical predictions,
listed in Appendix~\ref{4f_sec:appendix}.
The full 32$\times$32 covariance matrix for the ratios
is built taking into account the same sources
of systematic errors used for the combination 
of the W-pair cross-sections at these energies.

The small statistical errors 
on the theoretical predictions at the various energies,
taken as fully correlated for the four experiments
and uncorrelated between different energies, are also translated into errors 
on the individual measurements of $\rww$.
The theoretical errors on the predictions,
due to the physical and technical precision of the generators used, 
are not propagated to the individual ratios but are used when comparing 
the combined values of $\rww$ to unity.
For each of the four models considered,
two fits are performed:
in the first, eight values of $\rww$ at the different energies are extracted,
averaged over the four experiments;
in the second, only one value of $\rww$ is determined,
representing the global agreement of measured and predicted cross-sections
over the whole energy range.

\renewcommand{\arraystretch}{1.2}
\begin{table}[bhtp]
\vspace*{-0mm}
\begin{center}
\hspace*{-0.3cm}
\begin{tabular}{|c|c|c|} 
\hline
\roots (GeV) & $\rww^{\footnotesize\YFSWW}$ & $\rww^{\footnotesize\RacoonWW}$ \\
\hline
182.7      & $1.034\pm0.023$ & $1.033\pm0.023$ \\
188.6      & $0.986\pm0.013$ & $0.987\pm0.013$ \\
191.6      & $1.000\pm0.029$ & $1.003\pm0.029$ \\
195.5      & $1.003\pm0.019$ & $1.006\pm0.019$ \\
199.5      & $0.985\pm0.018$ & $0.987\pm0.018$ \\
201.6      & $0.995\pm0.024$ & $0.998\pm0.024$ \\
204.9      & $0.980\pm0.018$ & $0.983\pm0.018$ \\
206.6      & $1.000\pm0.015$ & $1.004\pm0.015$ \\
\hline
$\chi^2$/d.o.f & 26.6/24         & 26.6/24        \\
\hline
\hline
Average        & $0.994\pm0.009$ & $0.996\pm0.009$ \\
\hline
$\chi^2$/d.o.f & 32.2/31         & 32.0/31        \\
\hline
\end{tabular}
\caption{%
Ratios of LEP combined W-pair cross-section measurements
to the expectations according to 
\YFSWW~\protect\cite{4f_bib:yfsww} and 
\RacoonWW~\protect\cite{4f_bib:racoonww}.
For each of the two models,
two fits are performed,
one to the LEP combined values 
of $\rww$ at the eight energies between 183 and 207~GeV,
and another to the LEP combined average of $\rww$ over all energies.
The results of the fits are given in the table
together with the resulting $\chi^2$.
Both fits take into account inter-experiment 
as well as inter-energy correlations of systematic errors.
}
\label{4f_tab:wwratio}
\end{center}
\vspace*{-6mm}
\end{table}
\renewcommand{\arraystretch}{1.}

\begin{figure}[tp]
\centering
\vspace*{-0.5truecm}
\mbox{
  \fbox{\epsfig{figure=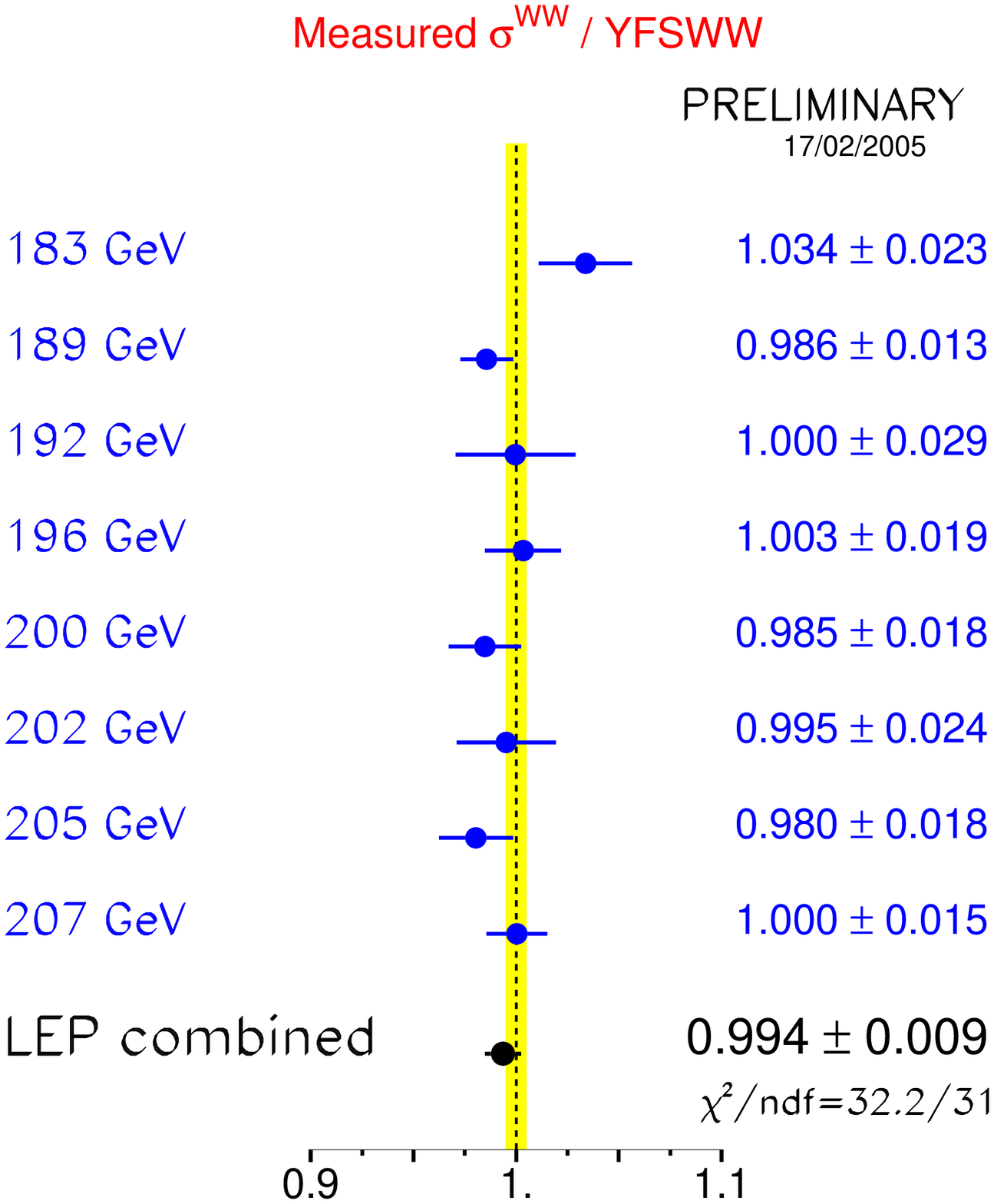,width=0.45\textwidth}}
  \hspace*{0.04\textwidth}
  \fbox{\epsfig{figure=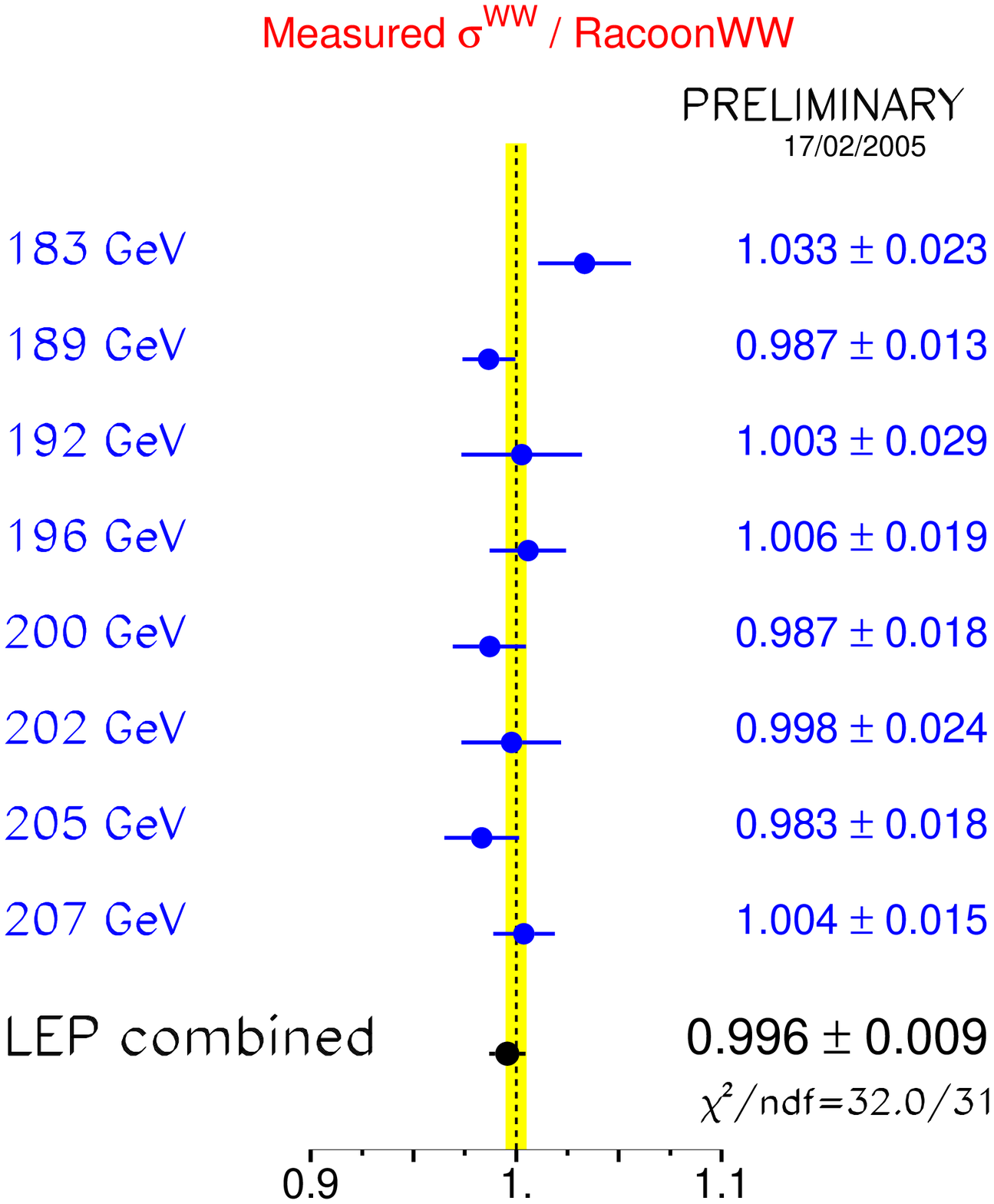,width=0.45\textwidth}}
  }
\vspace*{-0.5truecm}
\caption{%
  Ratios of LEP combined W-pair cross-section measurements
  to the expectations according to 
  \YFSWW~\protect\cite{4f_bib:yfsww} and 
  \RacoonWW~\protect\cite{4f_bib:racoonww}
  The yellow bands represent constant relative errors 
  of 0.5\% on the two 
  cross-section predictions.
}
\label{4f_fig:rww}
\end{figure} 

The results of the two fits to $\rww$
for \YFSWW\ and \RacoonWW\ are given in Table~\ref{4f_tab:wwratio}.
As already qualitatively noted from Figure~\ref{4f_fig:sww_vs_sqrts},
the LEP measurements of the W-pair cross-section above threshold
are in very good agreement to the predictions and can test the theory
at the level of better than 1\%.
In contrast, the predictions from \Gentle\ and \KoralW\ are
about 3\% too high with respect to the measurements; the equivalent
values of $\rww$ in those cases are, respectively, $0.969\pm0.009$
and $0.975\pm0.009$.

The main differences between these two sets of predictions
come from non-leading $\oa$ electroweak radiative corrections
to the W-pair production process and non-factorisable corrections,
which are included 
(in the LPA/DPA approximation~\cite{4f_bib:dpa})
in both \YFSWW\ and \RacoonWW, but not in \Gentle\ and \KoralW.
The data clearly prefer the computations which
more precisely include $\oa$ radiative corrections.

The results of the fits for \YFSWW\ and \RacoonWW are also shown 
in Figure~\ref{4f_fig:rww}, where relative errors of 0.5\% on the 
cross-section predictions have been assumed.
For simplicity in the figure the energy dependence of the theory error
on the W-pair cross-section has been neglected.

\section{W branching ratios and  $|\mathrm{V}_{\mathrm{cs}}|$ }
\label{4f_sec:WWBR}

From the partial cross-sections WW$\rightarrow\mathrm{4f}$ measured by the 
four experiments at all energies above 161~GeV, the W decay branching fractions 
\mbox{$\mathcal{B}(\mathrm{W}\rightarrow\textrm{f}\overline{\textrm{f}}')$}
are determined, with and without the assumption of lepton universality.

The two combinations use as inputs from the experiments the three leptonic branching 
fractions, with their systematic and observed statistical errors
and their correlation matrices.
In the fit with lepton universality, the branching fraction to hadrons is 
determined from that to leptons by constraining the sum to unity.
The part of the systematic error correlated between experiments is properly
accounted for when building the full covariance matrix.

\begin{figure}[tp]
\centering
\vspace*{-0.5truecm}
\mbox{
  \fbox{\epsfig{figure=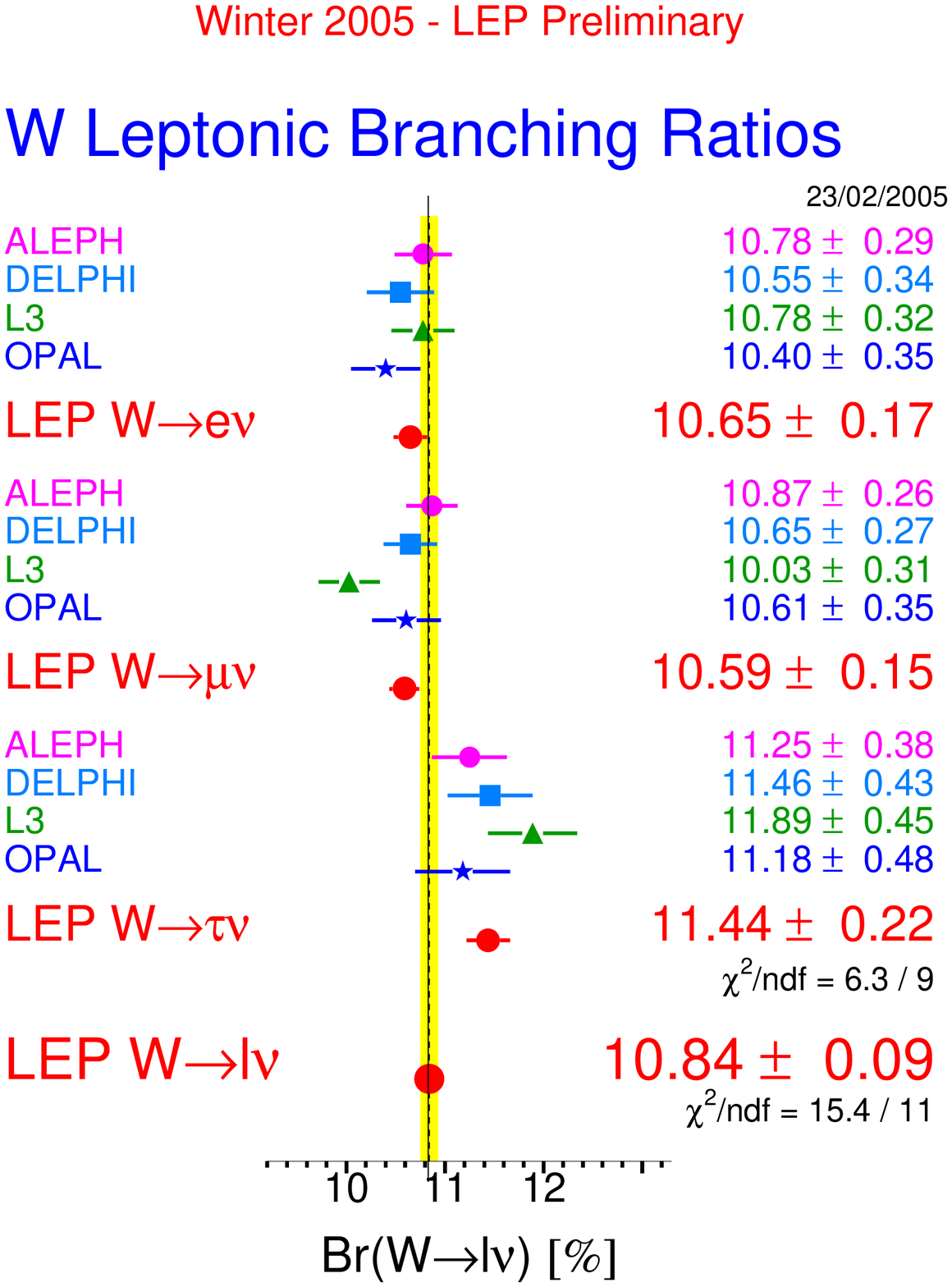,width=0.45\textwidth}}
  \hspace*{0.04\textwidth}
  \fbox{\epsfig{figure=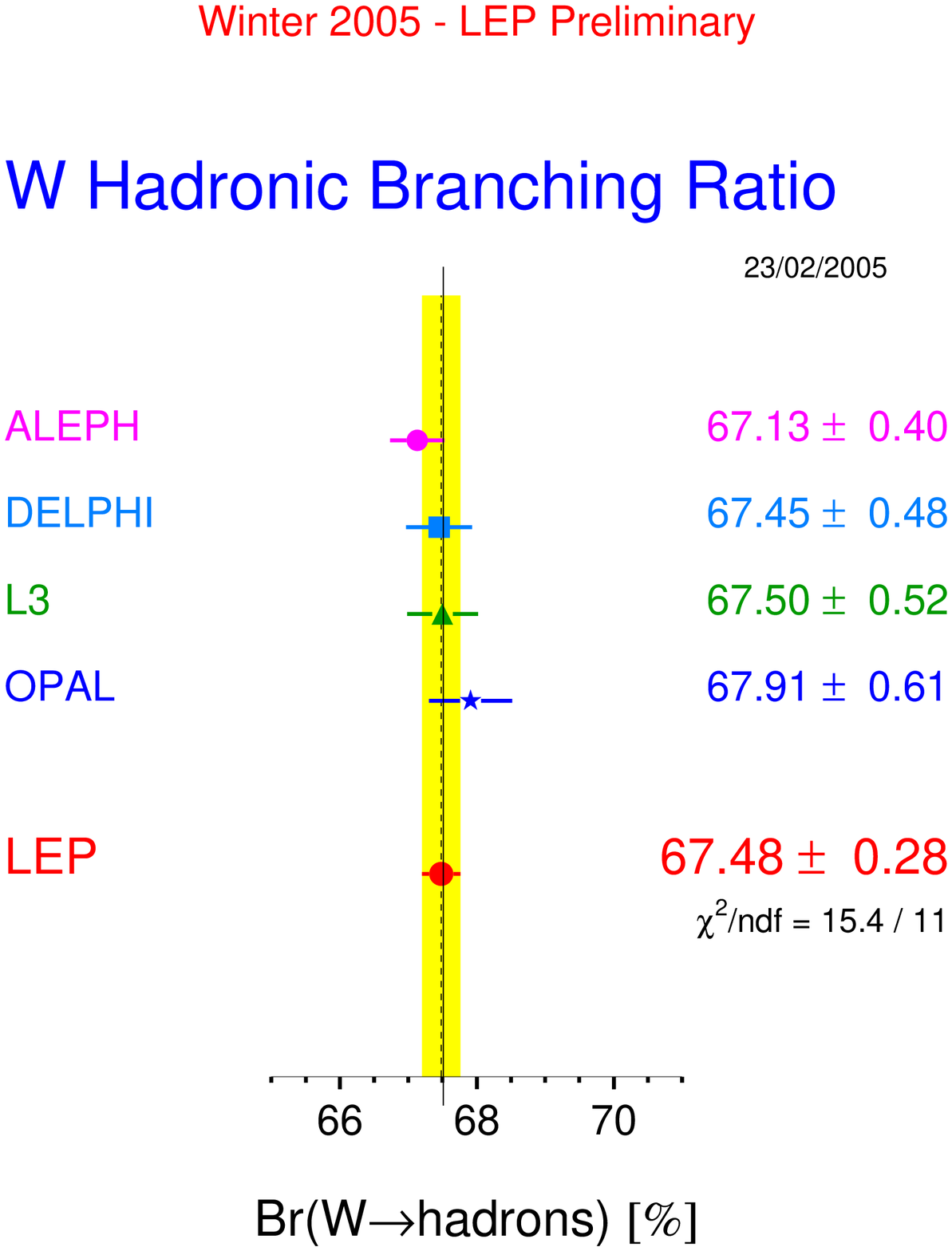,width=0.45\textwidth}}
  }
\vspace*{-0.5truecm}
\caption{%
  Leptonic and hadronic W branching fractions, as measured by 
  the experiments, and the LEP combined values according to the
  procedures described in the text.}
\label{4f_fig:brw}
\end{figure}

The detailed inputs used for the combinations are given in 
Appendix~\ref{4f_sec:appendix}.
The results from each experiment are given in Table~\ref{tab:wwbra} 
together with the result of the LEP combination. The same results are 
shown in Figure~\ref{4f_fig:brw}.

\renewcommand{\arraystretch}{1.2}
\begin{table}[hbtp]
\begin{center}
\begin{tabular}{|c|c|c|c|c|} 
\cline{2-5}
\multicolumn{1}{c|}{$\quad$} & \multicolumn{3}{|c|}{Lepton} & Lepton \\
\multicolumn{1}{c|}{$\quad$} & \multicolumn{3}{|c|}{non--universality} & universality \\
\hline
Experiment 
         & \wwbr(\Wtoenu) & \wwbr(\Wtomnu) & \wwbr(\Wtotnu)  
         & \wwbr({\mbox{$\mathrm{W}\rightarrow\mathrm{hadrons}$}}) \\
         & [\%] & [\%] & [\%] & [\%]  \\
\hline
\Aleph\  & $10.78\pm0.29^*$ & $10.87\pm0.26^*$ & $11.25\pm0.38^*$ & $67.13\pm0.40^*$ \\
\Delphi\ & $10.55\pm0.34^*$ & $10.65\pm0.27^*$ & $11.46\pm0.43^*$ & $67.45\pm0.48^*$ \\
\Ltre\   & $10.78\pm0.32^*$ & $10.03\pm0.31^*$ & $11.89\pm0.45^*$ & $67.50\pm0.52^*$ \\
\Opal\   & $10.40\pm0.35$ & $10.61\pm0.35$ & $11.18\pm0.48$ & $67.91\pm0.61$ \\
\hline
LEP      & $10.65\pm0.17$ & $10.59\pm0.15$ & $11.44\pm0.22$ & $67.48\pm0.28$ \\
\hline
$\chi^2/\textrm{d.o.f.}$ & \multicolumn{3}{|c|}{6.3/9} & 15.4/11 \\
\hline
\end{tabular}
\caption{
  Summary of W branching fractions derived from W-pair production 
  cross sections measurements up to 207 GeV \CoM\ energy. All results
  are preliminary with the exception of those indicated by $^*$. } 
\label{tab:wwbra} 
\end{center}
\end{table}
\renewcommand{\arraystretch}{1.}

The results of the fit which does not make use of the lepton universality
assumption show a negative correlation of 19.5\% (13.2\%) between the 
\Wtotnu\  and \Wtoenu\  (\Wtomnu)  branching fractions, while between the
electron and muon decay channels there is a positive correlation of 11.0\%.

From the results on the leptonic branching ratios an excess of the branching 
ratio \Wtotnu\ with respect to the other leptons is evident. 
The excess can be quantified with the two-by-two comparison of these 
branching fractions, which represents a test of lepton universality 
in the decay of on--shell W bosons at the level of 2.9\%:
\begin{eqnarray*}
\wwbr\mathrm{(\Wtomnu)} \, / \, \wwbr\mathrm{(\Wtoenu)} \,
& = & 0.994 \pm 0.020 \, ,\\
\wwbr\mathrm{(\Wtotnu)} \; / \, \wwbr\mathrm{(\Wtoenu)} \,
& = & 1.074 \pm 0.029 \, ,\\
\wwbr\mathrm{(\Wtotnu)} \, / \, \wwbr\mathrm{(\Wtomnu)} 
& = & 1.080 \pm 0.028 \, .
\end{eqnarray*}
The branching fractions in taus with respect to electrons and muons 
differ by more than two standard deviations, where the correlations have
been taken into account. The branching fractions of W into electrons and into
muons perfectly agree.

Assuming only partial lepton universality the ratio between the tau fractions
and the average of electrons and muons can also be computed:
\begin{eqnarray*}
2\wwbr\mathrm{(\Wtotnu)} \, / \, (\wwbr\mathrm{(\Wtoenu)}+\wwbr\mathrm{(\Wtomnu)}) \,
& = & 1.077 \pm 0.026 \,
\end{eqnarray*}
resulting in a poor agreement at the level of 2.8 standard deviations, with all
correlations included.

If complete lepton universality is assumed,
the measured hadronic branching fraction can be determined, yielding 
$67.48\pm0.18\mathrm{(stat.)}\pm0.21\mathrm{(syst.)}\%$,
whereas for the leptonic one gets 
$10.84\pm0.06\mathrm{(stat.)}\pm0.07\mathrm{(syst.)}\%$.
These results are consistent with their Standard Model expectations,
of 67.51\% and 10.83\% respectively.
The systematic error receives equal contributions 
from the correlated and uncorrelated sources.

Within the Standard Model, 
the branching fractions of the W boson depend on the six matrix elements 
$|\mathrm{V}_{\mathrm{qq'}}|$ of the Cabibbo--Kobayashi--Maskawa (CKM) 
quark mixing matrix not involving the top quark. 
In terms of these matrix elements, 
the leptonic branching fraction of the W boson 
$\mathcal{B}(\Wtolnu)$ is given by
\begin{equation*}
  \frac{1}{\mathcal{B}(\Wtolnu)}\quad = \quad 3 
  \Bigg\{ 1 + 
          \bigg[ 1 + \frac{\alpha_{\mathrm{s}}(\mathrm{M}^2_{\mathrm{W}})}{\pi} 
          \bigg] 
          \sum_{\tiny\begin{array}{c}i=(u,c),\\j=(d,s,b)\\\end{array}}
          |\mathrm{V}_{ij}|^2 
  \Bigg\},
\end{equation*} 
where $\alpha_{\mathrm{s}}(\mathrm{M}^2_{\mathrm{W}})$ is the strong coupling
constant. 
Taking $\alpha_{\mathrm{s}}(\mathrm{M}^2_{\mathrm{W}})=0.119\pm0.002$~\cite{4f_bib:pdg02},
and using the experimental knowledge of the sum
$|\mathrm{V}_{\mathrm{ud}}|^2+|\mathrm{V}_{\mathrm{us}}|^2+|\mathrm{V}_{\mathrm{ub}}|^2+
 |\mathrm{V}_{\mathrm{cd}}|^2+|\mathrm{V}_{\mathrm{cb}}|^2=1.0476\pm0.0074$~\cite{4f_bib:pdg02}, 
the above result can be interpreted as a measurement of $|\mathrm{V}_{\mathrm{cs}}|$ 
which is the least well determined of these matrix elements:
\begin{equation*}
  |\mathrm{V}_{\mathrm{cs}}|\quad=\quad0.976\,\pm\,0.014.
\end{equation*}
The error includes 
a $\pm0.0006$ contribution from the uncertainty on $\alpha_{\mathrm{s}}$
and a $\pm0.004$ contribution from the uncertainties 
on the other CKM matrix elements,
the largest of which is that on $|\mathrm{V}_{\mathrm{cd}}|$.
These contributions are negligible
in the error on this determination of $|\mathrm{V}_{\mathrm{cs}}|$,
which is dominated by the $\pm0.013$ experimental error 
from the measurement of the W branching fractions. The value of $|\mathrm{V}_{\mathrm{cs}}|$
is in agreement with unity.

\clearpage

\section{Combination of the cos$\theta_{\mathrm{W}^-}$ distribution}
\label{4f_sec:dsdcost}

\subsection{Introduction and definitions}

In addition to measuring the total $\WW$ cross-section, the LEP
experiments produce results for the differential cross-section, 
$\mathrm{d}(\sigma_{\mathrm{WW}})/\mathrm{d}(\costw)$ 
($\costw$ is the polar angle of the produced $\mathrm{W}^-$ 
with respect to the $\mathrm{e}^-$ beam direction). 
The LEP combination of these measurements will allow future theoretical 
models which predict deviations in this distribution 
to be tested against the LEP data in a direct and as much as possible
model independent manner. To reconstruct the 
$\costw$ distribution it is necessary to identify the charges of the
decaying W bosons. This can only be performed without significant 
ambiguity when one of W-boson decays via $\mathrm{W}\rightarrow e \nu$
or $\mathrm{W}\rightarrow\mu\nu$ (in which case the lepton provides the
charge tag). Consequently, the 
combination of the differential cross-section measurements is
performed for the $\qqen$ and $\qqmn$ channels combined.
Selected $\qqtn$ events are not considered due to the larger 
backgrounds and difficulties in determining the tau lepton charge.

The measured $\qqen$ and $\qqmn$ differential cross-sections are corrected 
to correspond to the {\sc CC03} set of diagrams with the additional
constraint that the charged lepton is more than $20^\circ$ away from the $\epem$ 
beam direction, $|\theta_{\ell^\pm}|>20^\circ$. This angular
requirement corresponds closely to the experimental acceptance of the 
four LEP experiments and also greatly reduces the difference between the 
full $4f$ cross-section and the {\sc CC03} cross-section by reducing the 
contribution of $t$-channel diagrams in the $\qqen$ final 
state\footnote{With this requirement the difference between the total 
{\sc CC20} and {\sc CC03} $\qqen$ cross-sections is approximately 3.5\,\%,
as opposed to 24.0\,\% without the lepton angle requirement. For the $\qqmn$
channel the differences between the {\sc CC10} and {\sc CC03} cross-sections
are less than 1\,\% in both cases.}. The anlge $\costw$ is reconstructed 
from the four-momenta of the fermions from the ${\mathrm{W}^-}$ decay
using the {\sc ECALO5} photon recombination scheme\cite{4f_bib:fourfrep}.

\subsection{LEP combination method} 

The LEP combination is performed in ten bins of $\costw$. Because
the differential cross-section distribution evolves with 
$\roots$, reflecting the changing relative $s-$ and $t-$ channel 
contributions, the LEP data are
divided into four $\roots$ ranges: 
$180.0 < \roots \le 184.0$; $184.0 < \roots \le 194.0$;
$194.0 < \roots \le 204.0$; and $204.0 < \roots \le 210.0$.
It has been verified for each $\roots$ range that 
the differences in the differential cross-sections at the mean 
value of $\roots$  compared to the luminosity weighted
sum of the differential cross-sections reflecting the actual distribution
of the data across $\roots$ are negligible compared to the statistical errors.

The experimental resolution in LEP on the reconstructed minus generated 
value of $\costw$ is typically 0.15-0.2 and, as a result,
there is a significant migration between generated and reconstructed
bins of $\costw$. The effects of bin-to-bin migration are not explicitely
unfolded, instead each experiment obtains the cross-section in 
$i^{th}$ bin of the differential distribution, $\sigma_i$, from
\begin{eqnarray}
      \sigma_i & = & {{N_i-b_i}\over{\epsilon_i\cal{L}}}, 
\end{eqnarray}
where: 
\begin{itemize}
   \item[ $N_i$ ] is the observed number of $\qqen$/$\qqmn$ events 
                  reconstructed in the $i$th bin of the $\costw$ distribution.
   \item[ $b_i$ ] is the expected number of background
                  events in bin $i$. The contribution from four-fermion
                  background is treated as in each of the experiments 
                  $\WW$ cross-section analyses.
   \item[ $\epsilon_i$ ] 
                  is the Monte Carlo efficiency in bin $i$, 
                  defined as $\epsilon_i=S_i/G_i$ where $S_i$ is  
                  the number of selected {\sc CC03} MC $\qqln$ events  
                  reconstructed in bin $i$ and $G_i$ is the number of
                  MC {\sc CC03} $\qqen$/$\qqmn$ events with generated
                  \costw\
                  (calculated using the ECALO5 recombination scheme)
                  lying in the $i$th bin ($|\theta_{\ell^\pm}|>20^\circ$). 
                  Selected $\qqtn$
                  events are included in the
                  numerator of the efficiency.
\end{itemize}

This bin-by-bin efficiency correction method has the advantages 
of simplicity and that the resulting $\sigma_i$ are uncorrelated.
The main disadvantage of this procedure is that bin-by-bin migrations
between generated and reconstructed $\costw$ are corrected purely on the 
basis of the Standard Model expectation. If the data deviate
from it the resulting differential cross-section may be therefore biased 
toward the Standard Model expectation. 
However, the validity of the simple correction
procedure has been tested by considering a range of deviations from
the SM. Specifically the SM $\costw$ distribution was reweighted by 
$1+0.10 \,( \costw-1.0 ) $,  
$1-0.20 \,\cosstw  $  ,
$1+0.20 \,\cosstw  $  and
$1-0.40 \,\cosetw  $ and data samples generated corresponding to
the combined LEP luminosity. These reweighting functions represent deviations
which are large compared to the statistics of the combined LEP 
measurements. The bin-by-bin correction method was found to result in
good $\chi^2$ distributions when the extracted $\costw$ distributions were
compared with the underlying generated distribution ({\em e.g.} the worst
case gave a mean $\chi^2$ of 11.3 for the 10 degrees of freedom 
corresponding to the ten $\costw$ bins).

For the LEP combination the systematic uncertainties on 
measured differential cross-sections are broken down into two terms:
errors which are 100~\% correlated between bins and experiments
and errors which are correlated between bins but uncorrelated 
between experiments. This procedure reflects the the fact that the dominant
systematic errors affect the overall normalisation of the measured 
distributions rather than the shape.

\subsection{Results}

For the Winter Conferences 2005 the combination of the W angular distribution 
has been performed using final numbers by \Aleph\ ~\cite{4f_bib:aleww},
\Delphi\ ~\cite{4f_bib:delww} and \Ltre\ ~\cite{4f_bib:ltrww}. 
The detailed inputs by the experiments are reported in the 
appendix~\ref{4f_sec:appendix}, whereas Table~\ref{4f_tab:dsdcost} presents
the combined LEP results according to the above described procedure. In the 
table the error breakdown bin by bin is also reported.
 
\begin{table}[hbtp]
\begin{center}
\begin{small}
\begin{tabular}{|c|c|c|}
\hline
$\sqrt{s}$ interval (GeV) & Total luminosity (pb$^{-1}$) & Lumi weighted $\sqrt{s}$ (GeV) \\
180-184 & 163.90 & 182.66 \\
\hline
\end{tabular}
\begin{tabular}{|c|c|c|c|c|c|c|c|c|c|c|}
\hline
cos$\theta_{\mathrm{W}-}$ bin $i$ & 1 & 2 & 3 & 4 & 5 & 6 & 7 & 8 & 9 & 10 \\
$\sigma_i$  (pb)            & 0.515 & 0.633 & 0.772 & 1.295 & 1.370 & 2.090 & 2.659 & 2.489 & 4.406 & 5.619 \\
$\delta\sigma_i$  (pb)      & 0.131 & 0.139 & 0.155 & 0.250 & 0.217 & 0.288 & 0.328 & 0.287 & 0.451 & 0.512 \\
$\delta\sigma_i$(stat) (pb) & 0.129 & 0.137 & 0.153 & 0.249 & 0.215 & 0.285 & 0.325 & 0.283 & 0.447 & 0.508 \\
$\delta\sigma_i$(syst) (pb) & 0.019 & 0.018 & 0.020 & 0.024 & 0.027 & 0.045 & 0.043 & 0.050 & 0.062 & 0.067 \\
\hline
\end{tabular}

\begin{tabular}{|c|c|c|}
\hline
$\sqrt{s}$ interval (GeV) & Total luminosity (pb$^{-1}$) & Lumi weighted $\sqrt{s}$ (GeV) \\
184-194 & 587.95 & 189.09 \\
\hline
\end{tabular}
\begin{tabular}{|c|c|c|c|c|c|c|c|c|c|c|}
\hline
cos$\theta_{\mathrm{W}-}$ bin $i$ & 1 & 2 & 3 & 4 & 5 & 6 & 7 & 8 & 9 & 10 \\
$\sigma_i$  (pb)            & 0.748 & 0.811 & 1.012 & 1.091 & 1.314 & 1.735 & 2.225 & 2.877 & 4.161 & 5.748 \\
$\delta\sigma_i$  (pb)      & 0.087 & 0.090 & 0.101 & 0.102 & 0.112 & 0.149 & 0.171 & 0.180 & 0.213 & 0.256 \\
$\delta\sigma_i$(stat) (pb) & 0.085 & 0.089 & 0.099 & 0.099 & 0.109 & 0.144 & 0.167 & 0.174 & 0.203 & 0.243 \\
$\delta\sigma_i$(syst) (pb) & 0.017 & 0.017 & 0.021 & 0.024 & 0.026 & 0.037 & 0.036 & 0.047 & 0.063 & 0.078 \\
\hline
\end{tabular}

\begin{tabular}{|c|c|c|}
\hline
$\sqrt{s}$ interval (GeV) & Total luminosity (pb$^{-1}$) & Lumi weighted $\sqrt{s}$ (GeV) \\
194-204 & 605.05 & 198.38 \\
\hline
\end{tabular}
\begin{tabular}{|c|c|c|c|c|c|c|c|c|c|c|}
\hline
$\sigma_i$  (pb)            & 0.685 & 0.623 & 1.043 & 0.982 & 1.178 & 1.598 & 2.155 & 3.026 & 4.080 & 6.379 \\
$\delta\sigma_i$  (pb)      & 0.091 & 0.072 & 0.099 & 0.096 & 0.108 & 0.134 & 0.149 & 0.190 & 0.213 & 0.262 \\
$\delta\sigma_i$(stat) (pb) & 0.090 & 0.071 & 0.098 & 0.094 & 0.106 & 0.129 & 0.145 & 0.185 & 0.203 & 0.248 \\
$\delta\sigma_i$(syst) (pb) & 0.013 & 0.014 & 0.017 & 0.021 & 0.022 & 0.034 & 0.034 & 0.042 & 0.062 & 0.084 \\
\hline
cos$\theta_{\mathrm{W}-}$ bin $i$ & 1 & 2 & 3 & 4 & 5 & 6 & 7 & 8 & 9 & 10 \\
\hline
\end{tabular}

\begin{tabular}{|c|c|c|}
\hline
$\sqrt{s}$ interval (GeV) & Total luminosity (pb$^{-1}$) & Lumi weighted $\sqrt{s}$ (GeV) \\
204-210 & 630.51 & 205.92 \\
\hline
\end{tabular}
\begin{tabular}{|c|c|c|c|c|c|c|c|c|c|c|}
\hline
cos$\theta_{\mathrm{W}-}$ bin $i$ & 1 & 2 & 3 & 4 & 5 & 6 & 7 & 8 & 9 & 10 \\
$\sigma_i$  (pb)            & 0.495 & 0.601 & 0.726 & 1.084 & 1.293 & 1.603 & 2.291 & 2.764 & 4.443 & 7.760 \\
$\delta\sigma_i$  (pb)      & 0.065 & 0.076 & 0.084 & 0.115 & 0.112 & 0.136 & 0.172 & 0.166 & 0.217 & 0.307 \\
$\delta\sigma_i$(stat) (pb) & 0.064 & 0.075 & 0.083 & 0.113 & 0.109 & 0.130 & 0.169 & 0.160 & 0.206 & 0.291 \\
$\delta\sigma_i$(syst) (pb) & 0.013 & 0.013 & 0.015 & 0.022 & 0.024 & 0.037 & 0.033 & 0.045 & 0.070 & 0.098 \\
\hline
\end{tabular}
\end{small}
\caption[]{Combined W$^{-}$ differential angular cross-section in the 10 angular bins for the four chosen energy 
intervals. 
For each energy range, the sum of the measured integrated luminosities and the luminosity weighted centre-of-mass 
energy is reported.
The results per angular bin in each of the energy interval are then presented: $\sigma_{i}$ indicates 
the average of d[$\sigma_{\mathrm{WW}}$(BR$_{e\nu}$+BR$_{\mu\nu}$)]/dcos$\theta_{\mathrm{W}^-}$ 
in the $i$-th bin of cos$\theta_{\mathrm{W}^-}$ with width 0.2.
The values, in each bin, of the total, statistical and systematic errors are reported as well. 
All values are expressed in pb
}
\label{4f_tab:dsdcost} 
\end{center}
\end{table}

\begin{figure}[p]
\centering
\epsfig{figure=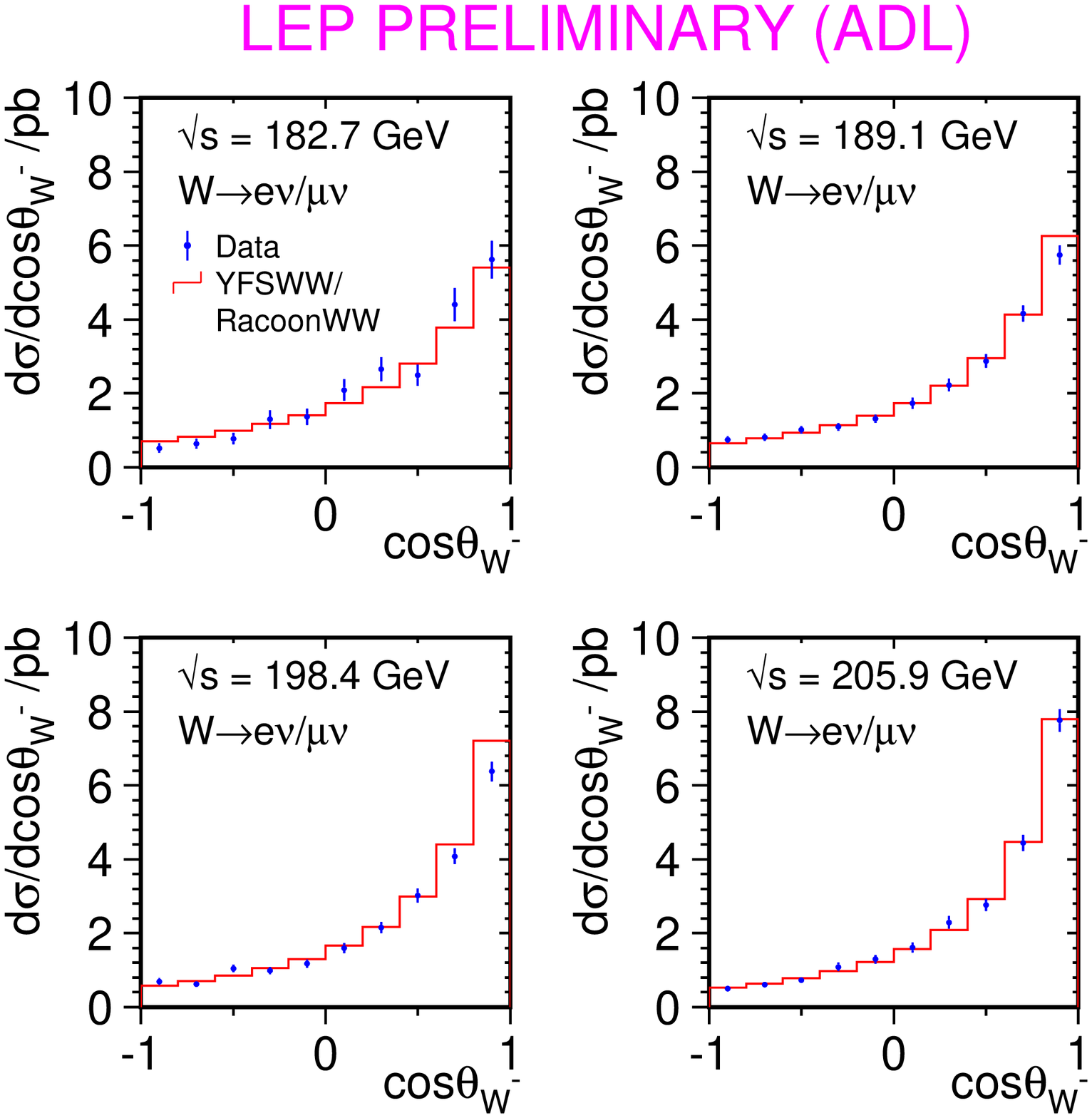,width=0.95\textwidth}
\caption{%
LEP combined d[$\sigma_{\mathrm{WW}}$(BR$_{e\nu}$+BR$_{\mu\nu}$)]/dcos$\theta_{\mathrm{W}^-}$ 
distributions for the four chosen energy intervals. The combined values (points) are superimposed
with the four-fermion predictions from \KandY~\protect\cite{4f_bib:kandy}. 
}
\label{4f_fig:dsdcost}
\end{figure}

The result is also presented in Figure~\ref{4f_fig:dsdcost}, where the combined data are superimposed
to the four-fermion theory predictions from \YFSWW\ and from \RacoonWW\, that provide 
undistinguishable results on the plot scale.

\clearpage

\section{Single-W production cross-section}
\label{4f_sec:wenxsec}

The LEP combination of the single-W production cross-section has been updated using 
the final \Delphi~\cite{4f_bib:delswsc05} results, and supersede the last combination 
presented at the 2004 Summer Conferences~\cite{4f_bib:4f_s04}.

Single-W production at $\LEPII$ is defined as the complete $t$-channel 
subset of Feynman diagrams
contributing to e$\nu_\mathrm{e}$f$\bar{\mathrm{f}}'$ final states,
with additional cuts on kinematic variables
to exclude the regions of phase space dominated by multiperipheral diagrams,
where the cross-section calculation is affected by large uncertainties.
The kinematic cuts used in the signal definitions are:
\mbox{m$_{\qq}>45$~GeV/c$^2$} for the $\enu\qq$ final states,
\mbox{E$_\ell>20$~GeV} for the $\enu\lnu$ final states 
with $\ell=\mu$ or $\tau$,
and finally \mbox{$|\cos\theta_\mathrm{e^-}|>0.95$}, 
\mbox{$|\cos\theta_\mathrm{e^+}|<0.95$} and 
\mbox{E$_\mathrm{e^+}>20$~GeV} (or the charge conjugate cuts)
for the $\enu\enu$ final states.

In the LEP combination the correlation of the systematic errors in energy and among
experiments is properly taken into account.
The expected statistical errors have been used for all measurements,
given the limited statistical precision of the single-W cross-section measurements.

The total and the hadronic single-W cross-sections, less contamined by
$\gamma\gamma$ interaction contributions, are combined independently;
the inputs by the four LEP experiments between 183 and 207~GeV are 
listed in Tables~\ref{4f_tab:swxsechad} and~\ref{4f_tab:swxsectot},
and the corresponding LEP combined values presented.

\renewcommand{\arraystretch}{1.2}
\begin{table}[htbp]
\begin{center}
\hspace*{-0.0cm}
\begin{tabular}{|c|c|c|c|c|c|c|} 
\hline
\roots & \multicolumn{5}{|c|}{Single-W hadronic cross-section (pb)} &  \\
\cline{2-6} 
(GeV) & \Aleph\ & \Delphi\ & \Ltre\ & \Opal\ & LEP & $\chi^2/\textrm{d.o.f.}$ \\
\hline
182.7 & $0.44^{\phz+\phz0.29*}_{\phz-\phz0.24}\phs$ & $0.11^{\phz+\phz0.31*}_{\phz-\phz0.14}\phs$ & 
$0.58^{\phz+\phz0.23\phz*}_{\phz-\phz0.20}$ & --- & 
$0.42\pm0.15\phs$ &
             \multirow{8}{20.3mm}{$
               \hspace*{-0.3mm}
               \left\}
                 \begin{array}[h]{rr}
                   &\multirow{8}{8mm}{\hspace*{-4.2mm}13.3/16}\\
                   &\\ &\\ &\\ &\\ &\\ &\\ &\\  
                 \end{array}
               \right.
               $}\\
188.6 & $0.33^{\phz+\phz0.16*}_{\phz-\phz0.15}\phs$ & $0.57^{\phz+\phz0.21*}_{\phz-\phz0.20}\phs$ &
$0.52^{\phz+\phz0.14\phz*}_{\phz-\phz0.13}$ & $0.53^{\phz+\phz0.14}_{\phz-\phz0.13}\phs$ & $0.48\pm0.08\phs$ &  \\
191.6 & $0.52^{\phz+\phz0.52*}_{\phz-\phz0.40}\phs$ & $0.30^{\phz+\phz0.48*}_{\phz-\phz0.31}\phs$ &
$0.84^{\phz+\phz0.44\phz*}_{\phz-\phz0.37}\phs$ & --- &
$0.56\pm0.25\phs$ & \\
195.5 & $0.61^{\phz+\phz0.28*}_{\phz-\phz0.25}\phs$ & $0.50^{\phz+\phz0.30*}_{\phz-\phz0.27}\phs$ &
$0.66^{\phz+\phz0.25\phz*}_{\phz-\phz0.23}\phs$ & --- &
$0.60\pm0.14\phs$ & \\
199.5 & $1.06^{\phz+\phz0.30*}_{\phz-\phz0.27}\phs$ & $0.57^{\phz+\phz0.28*}_{\phz-\phz0.26}\phs$ &
$0.37^{\phz+\phz0.22\phz*}_{\phz-\phz0.20}\phs$ & --- &
$0.65\pm0.14\phs$ & \\
201.6 & $0.72^{\phz+\phz0.39*}_{\phz-\phz0.33}\phs$ & $0.67^{\phz+\phz0.40*}_{\phz-\phz0.36}\phs$ &
$1.10^{\phz+\phz0.40\phz*}_{\phz-\phz0.35}\phs$ & --- &
$0.82\pm0.20\phs$ & \\
204.9 & $0.34^{\phz+\phz0.24*}_{\phz-\phz0.21}\phs$ & $0.99^{\phz+\phz0.33*}_{\phz-\phz0.31}\phs$ & 
$0.42^{\phz+\phz0.25\phz*}_{\phz-\phz0.21}\phs$ & --- &
$0.54\pm0.15\phs$ & \\
206.6 & $0.64^{\phz+\phz0.21*}_{\phz-\phz0.19}\phs$ & $0.81^{\phz+\phz0.23*}_{\phz-\phz0.22}\phs$ & 
$0.66^{\phz+\phz0.20\phz*}_{\phz-\phz0.18}\phs$ & --- &
$0.69\pm0.12\phs$ & \\
\hline
\end{tabular}
\end{center}
\vspace*{-0.3cm}
\caption{%
  Single-W production cross-section from the four LEP
  experiments and combined values 
  for the eight energies between 183 and 207~GeV,
  in the hadronic decay channel of the W boson.
  All results are preliminary with the exception of those indicated by $^*$.}
\label{4f_tab:swxsechad}
\end{table}
\renewcommand{\arraystretch}{1.}

\renewcommand{\arraystretch}{1.2}
\begin{table}[ht]
\begin{center}
\hspace*{-0.0cm}
\begin{tabular}{|c|c|c|c|c|c|c|} 
\hline
\roots & \multicolumn{5}{|c|}{Single-W total cross-section (pb)} & \\
\cline{2-6} 
(GeV) & \Aleph\ & \Delphi\ & \Ltre\ & \Opal\ & LEP & $\chi^2/\textrm{d.o.f.}$ \\
\hline
182.7 & $0.60^{\phz+\phz0.32*}_{\phz-\phz0.26}\phs$ & $0.69^{\phz+\phz0.42*}_{\phz-\phz0.25}\phs$ &
$0.80^{\phz+\phz0.28\phz*}_{\phz-\phz0.25}$ & --- &
$0.70\pm0.17\phs$ & 
             \multirow{8}{20.3mm}{$
               \hspace*{-0.3mm}
               \left\}
                 \begin{array}[h]{rr}
                   &\multirow{8}{8mm}{\hspace*{-4.2mm}8.1/16}\\
                   &\\ &\\ &\\ &\\ &\\ &\\ &\\  
                 \end{array}
               \right.
               $}\\

188.6 & $0.55^{\phz+\phz0.18*}_{\phz-\phz0.16}\phs$ & $0.75^{\phz+\phz0.23*}_{\phz-\phz0.22}\phs$ &
$0.69^{\phz+\phz0.16\phz*}_{\phz-\phz0.15}$ & $0.67^{\phz+\phz0.17}_{\phz-\phz0.15}\phs$ &
$0.66\pm0.09\phs$ & \\

191.6 & $0.89^{\phz+\phz0.58*}_{\phz-\phz0.44}\phs$ & $0.40^{\phz+\phz0.55*}_{\phz-\phz0.33}\phs$ &
$1.11^{\phz+\phz0.48\phz*}_{\phz-\phz0.41}\phs$ & --- &
$0.81\pm0.28\phs$ & \\

195.5 & $0.87^{\phz+\phz0.31*}_{\phz-\phz0.27}\phs$ & $0.68^{\phz+\phz0.34*}_{\phz-\phz0.38}\phs$ &
$0.97^{\phz+\phz0.27\phz*}_{\phz-\phz0.25}\phs$ & --- &
$0.85\pm0.16\phs$ & \\

199.5 & $1.31^{\phz+\phz0.32*}_{\phz-\phz0.29}\phs$ & $0.95^{\phz+\phz0.34*}_{\phz-\phz0.30}\phs$ &
$0.88^{\phz+\phz0.26\phz*}_{\phz-\phz0.24}\phs$ & --- &
$1.05\pm0.16\phs$ & \\

201.6 & $0.80^{\phz+\phz0.42*}_{\phz-\phz0.35}\phs$ & $1.24^{\phz+\phz0.52*}_{\phz-\phz0.43}\phs$ &
$1.50^{\phz+\phz0.45\phz*}_{\phz-\phz0.40}\phs$ & --- &
$1.17\pm0.23\phs$ & \\

204.9 & $0.65^{\phz+\phz0.27*}_{\phz-\phz0.23}\phs$ & $1.06^{\phz+\phz0.37*}_{\phz-\phz0.32}\phs$ & 
$0.78^{\phz+\phz0.29\phz*}_{\phz-\phz0.25}\phs$ & --- &
$0.80\pm0.17\phs$ & \\

206.6 & $0.81^{\phz+\phz0.22*}_{\phz-\phz0.20}\phs$ & $1.14^{\phz+\phz0.28*}_{\phz-\phz0.25}\phs$ & 
$1.08^{\phz+\phz0.21\phz*}_{\phz-\phz0.20}\phs$ & --- &
$1.00\pm0.14\phs$ &  \\
\hline
\end{tabular}
\end{center}
\vspace*{-0.3cm}
\caption{%
  Single-W total production cross-section from the four LEP
  experiments and combined values 
  for the eight energies between 183 and 207~GeV.
  All results are preliminary with the exception of those indicated by $^*$.}
\label{4f_tab:swxsectot}
\vspace*{-0.3cm}
\end{table}
\renewcommand{\arraystretch}{1.}

\begin{figure}[p]
\centering
\epsfig{figure=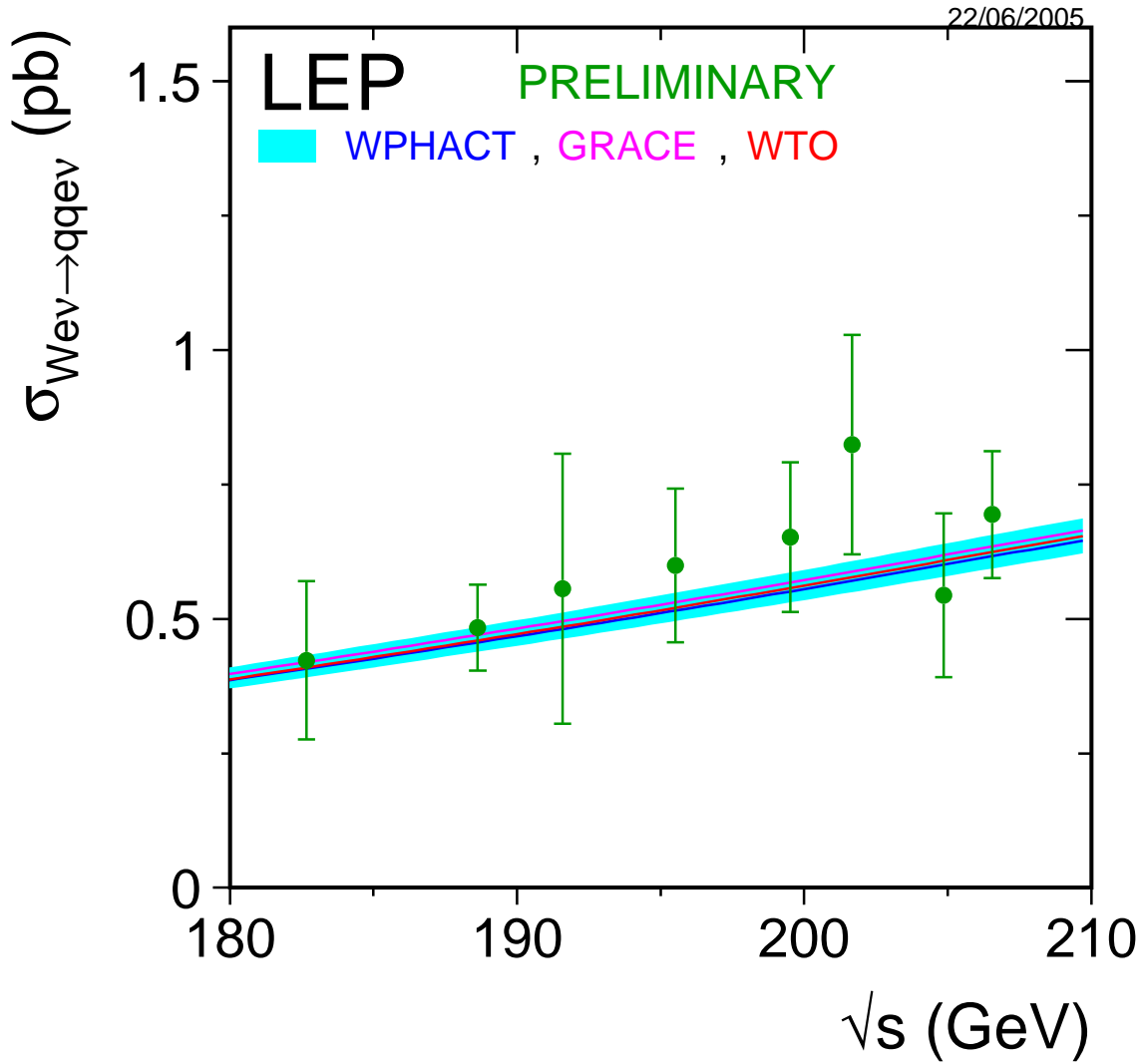,width=0.9\textwidth}
\caption{%
  Measurements of the single-W production cross-section 
  in the hadronic decay channel of the W boson, 
  compared to the predictions of 
  \WTO~\protect\cite{4f_bib:wto}, 
  \WPHACT~\protect\cite{4f_bib:wphact} 
  and \Grace~\protect\cite{4f_bib:grace} . 
  The shaded area represents the $\pm5$\% uncertainty 
  on the predictions.
}
\label{4f_fig:swen_had}
\end{figure}

\begin{figure}[p]
\centering
\epsfig{figure=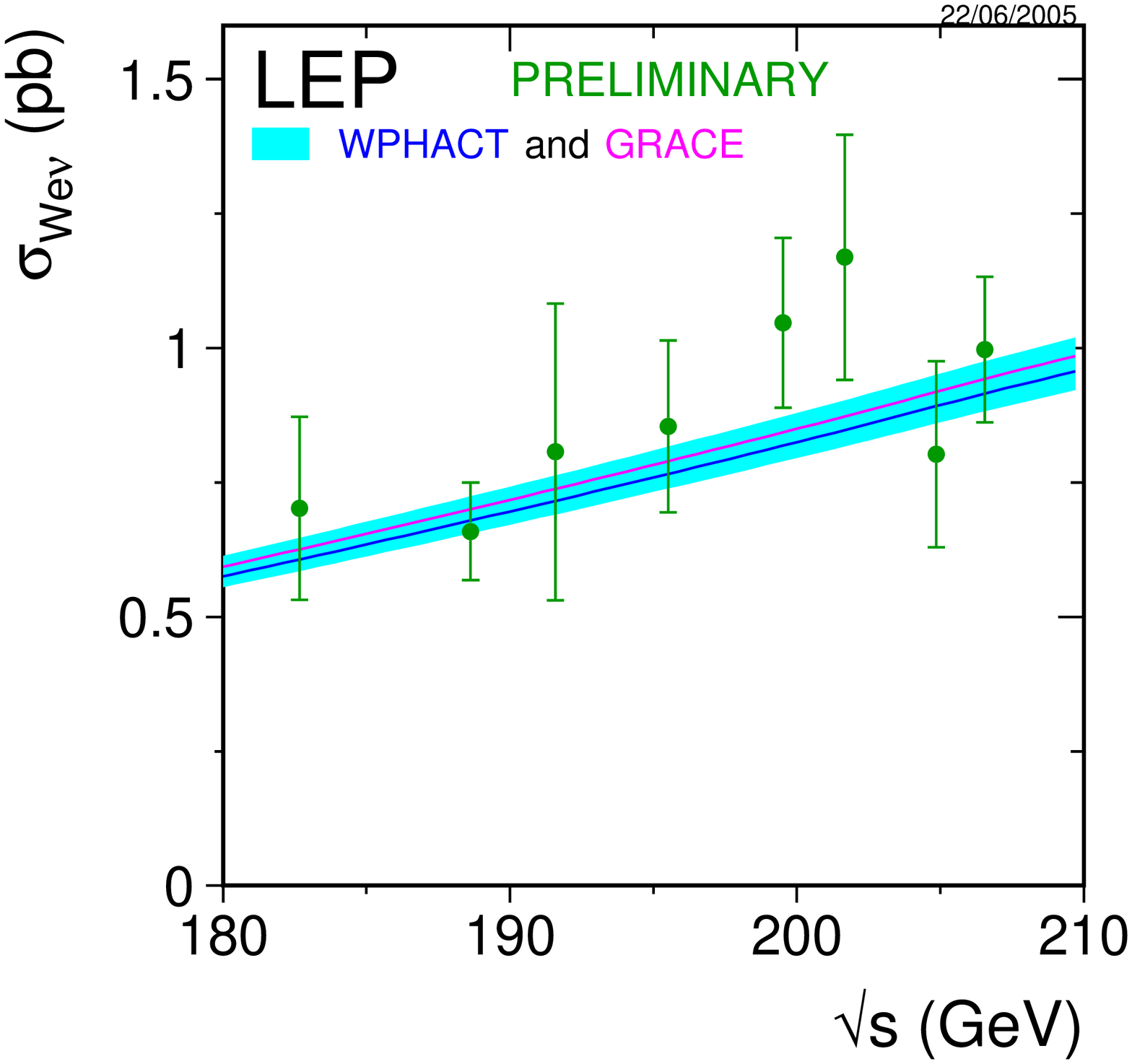,width=0.9\textwidth}
\caption{%
  Measurements of the single-W total production cross-section,
  compared to the predictions of \WPHACT\  and \Grace. 
  The shaded area represents the $\pm5$\% uncertainty 
  on the predictions.
}
\label{4f_fig:swen_all}
\end{figure}

The LEP measurements of the single-W cross-section are shown,
as a function of the LEP \CoM\ energy, 
in Figure~\ref{4f_fig:swen_had} for the hadronic decays
and in Figure~\ref{4f_fig:swen_all} for all decays of the W boson.
In the two figures, 
the measurements are compared with the expected values 
from \WPHACT~\cite{4f_bib:wphact} and \Grace~\cite{4f_bib:grace}.
\WTO~\cite{4f_bib:wto}, which includes fermion-loop corrections for the
hadronic final states, is also used in Figure~\ref{4f_fig:swen_had}.
As discussed more in detail 
in~\cite{4f_bib:wwichep00} and~\cite{4f_bib:fourfrep},
the theoretical predictions are scaled upward 
to correct for the implementation of QED radiative corrections 
at the wrong energy scale {\it s}.
The full correction factor of 4\%,
derived~\cite{4f_bib:fourfrep} by the comparison 
to the theoretical predictions from \SWAP~\cite{4f_bib:swap},
is conservatively taken as a systematic error.
This uncertainty dominates the $\pm$5\% theoretical error 
currently assigned to these 
predictions~\cite{4f_bib:wwichep00,4f_bib:fourfrep},
represented by the shaded area 
in Figures~\ref{4f_fig:swen_had} and~\ref{4f_fig:swen_all}.
All results, up to the highest \CoM\ energies, 
are in agreement with the theoretical predictions.

\begin{figure}[h]
\centering
\vspace*{-0.2truecm}
\mbox{
  \fbox{\epsfig{figure=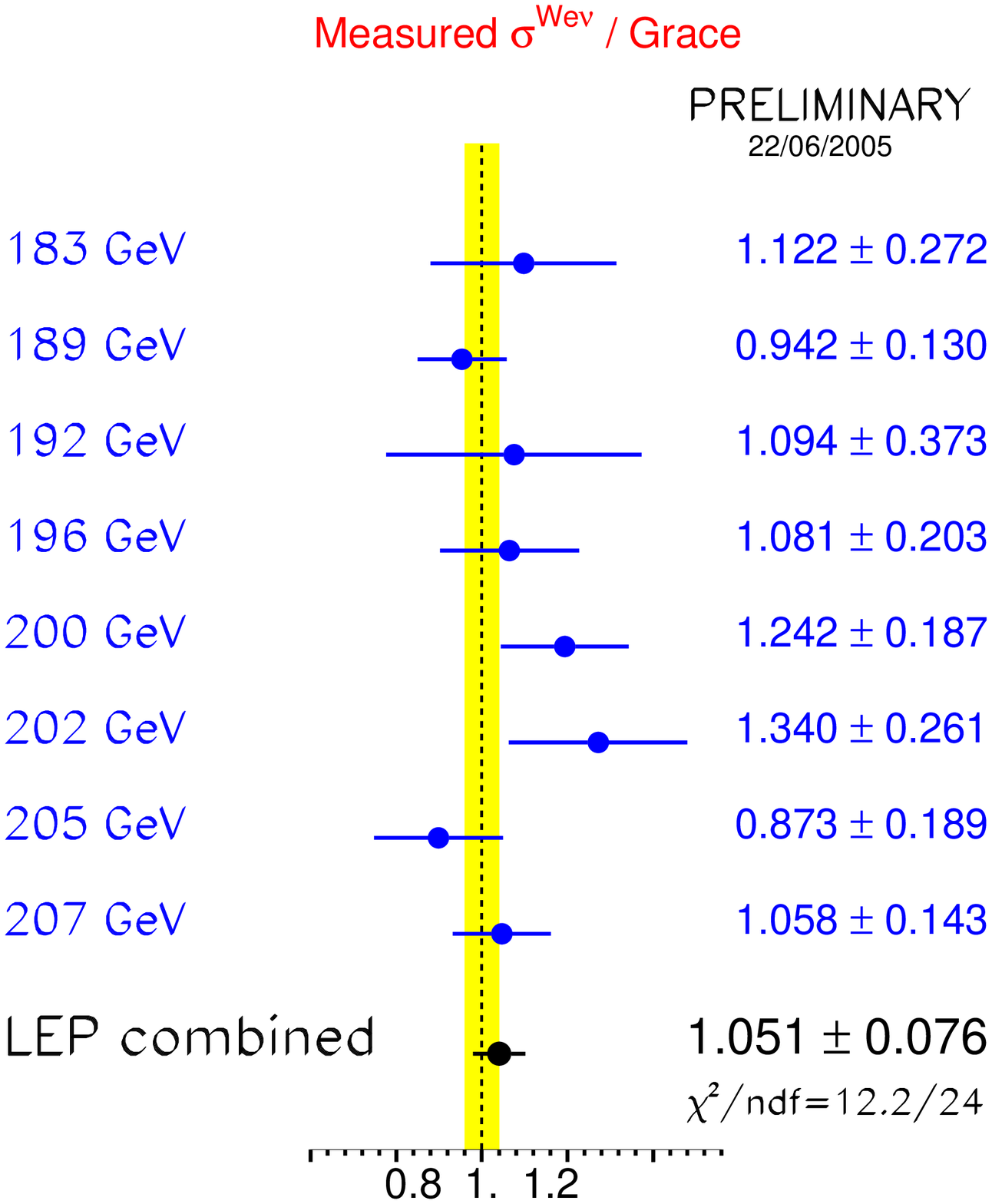,width=0.45\textwidth}}
  \hspace*{0.04\textwidth}
  \fbox{\epsfig{figure=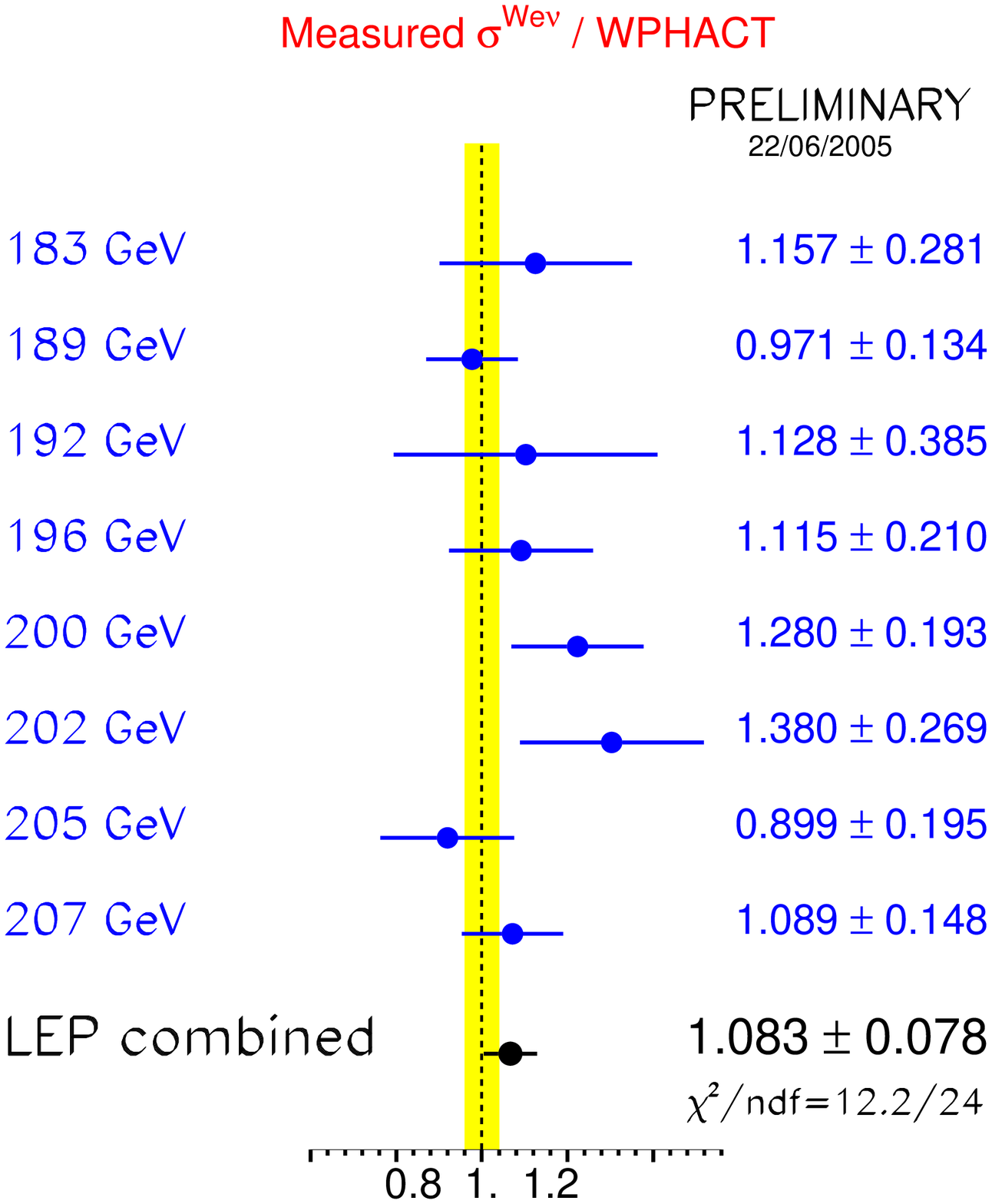,width=0.45\textwidth}}
  }
\vspace*{-0.5truecm}
\caption{%
  Ratios of LEP combined total single-W cross-section measurements
  to the expectations according to 
  \Grace~\protect\cite{4f_bib:grace} and 
  \WPHACT~\protect\cite{4f_bib:wphact}.
  The yellow bands represent constant relative errors 
  of 5\% on the two cross-section predictions.
}
\label{4f_fig:rwev}
\end{figure}

The agreement can also be appreciated in Table~\ref{4f_tab:wevratio},
where the values of the ratio between measured and expected cross-section 
values according to the computations by \Grace\ and \WPHACT\,
are reported. The combination is performed accounting for the energy 
and experiment correlations of the systematic sources. 
The results are also presented in Figure~\ref{4f_fig:rwev}.

\begin{table}[ht]
\begin{center}
\hspace*{-0.3cm}
\begin{tabular}{|c|c|c|} 
\hline 
\roots (GeV) & $\rwev^{\footnotesize\Grace}$ & $\rwev^{\footnotesize\WPHACT}$ \\
\hline
182.7             & $1.122\pm0.272$ & $1.157\pm0.281$  \\
188.6             & $0.942\pm0.130$ & $0.971\pm0.134$  \\
191.6             & $1.094\pm0.373$ & $1.128\pm0.385$  \\
195.5             & $1.081\pm0.203$ & $1.115\pm0.210$  \\
199.5             & $1.242\pm0.187$ & $1.280\pm0.193$  \\
201.6             & $1.340\pm0.261$ & $1.380\pm0.269$  \\
204.9             & $0.873\pm0.189$ & $0.899\pm0.195$  \\
206.6             & $1.058\pm0.143$ & $1.089\pm0.148$  \\
\hline
$\chi^2$/d.o.f    & 8.1/16         & 8.1/16         \\
\hline
\hline
Average           & $1.051\pm0.075$ & $1.083\pm0.078$  \\
\hline
$\chi^2$/d.o.f    & 12.2/24         & 12.2/24        \\
\hline
\end{tabular}
\caption{%
  Ratios of LEP combined total single-W cross-section measurements to
  the expectations according to \Grace~\protect\cite{4f_bib:grace} and
  \WPHACT~\protect\cite{4f_bib:wphact}.  The resulting averages over
  energies are also given.  The averages take into account
  inter-experiment as well as inter-energy correlations of systematic
  errors.  }
\label{4f_tab:wevratio}
\end{center}
\vspace*{-6mm}
\end{table}
\renewcommand{\arraystretch}{1.}

The theory predictions and the details of the experimental inputs
and the LEP combined values of the single-W cross-sections and the 
ratios to theory are reported in Appendix~\ref{4f_sec:appendix}. 

\clearpage

\section{Z-pair production cross-section}
\label{4f_sec:ZZxsec}

The Z-pair production cross-section is defined as the {\sc NC02}~\cite{4f_bib:fourfrep} 
contribution to four-fermion cross-section.
Final results from DELPHI, L3 and OPAL at all \CoM\ energies are 
available~\cite{4f_bib:delzz,4f_bib:ltrzz,4f_bib:opazz}. 
ALEPH published final results at 183 and 189 GeV~\cite{4f_bib:alezz189} and contributed
preliminary results for all other energies up to 207
GeV~\cite{4f_bib:alezzsc01}.

\renewcommand{\arraystretch}{1.2}
\begin{table}[bp]
\begin{center}
\begin{tabular}{|c|c|c|c|c|c|c|} 
\hline
\roots & \multicolumn{5}{|c|}{ZZ cross-section (pb)} & \\
\cline{2-6} 
(GeV) & \Aleph\ & \Delphi\ & \Ltre\ & \Opal\ & LEP & $\chi^2/\textrm{d.o.f.}$ \\ 
\hline
182.7 & 
$0.11^{\phz+\phz0.16\phz*}_{\phz-\phz0.12}$ & 
$0.35^{\phz+\phz0.20\phz*}_{\phz-\phz0.15}$ & 
$0.31\pm0.17^*$ & 
$0.12^{\phz+\phz0.20\phz*}_{\phz-\phz0.18}$ &
$0.22\pm0.08\phs^*$ & 
             \multirow{8}{20.3mm}{$
               \hspace*{-0.3mm}
               \left\}
                 \begin{array}[h]{rr}
                   &\multirow{8}{8mm}{\hspace*{-4.2mm}16.1/24}\\
                   &\\ &\\ &\\ &\\ &\\ &\\ &\\  
                 \end{array}
               \right.
               $}\\
188.6 & 
$0.67^{\phz+\phz0.14\phz*}_{\phz-\phz0.13}$ & 
$0.52^{\phz+\phz0.12\phz*}_{\phz-\phz0.11}$ & 
$0.73\pm0.15\phs^*$ & 
$0.80^{\phz+\phz0.15\phz*}_{\phz-\phz0.14}$ &
$0.66\pm0.07\phs^*$ &  \\
191.6 & 
$0.53^{\phz+\phz0.34}_{\phz-\phz0.27}\phzs$ &
$0.63^{\phz+\phz0.36*}_{\phz-\phz0.30}\phzs$ &
$0.29\pm0.22^*$ & 
$1.29^{\phz+\phz0.48*}_{\phz-\phz0.41}\phzs$ &
$0.65\pm0.17\phs$ &  \\
195.5 & 
$0.69^{\phz+\phz0.23}_{\phz-\phz0.20}\phzs$ & 
$1.05^{\phz+\phz0.25*}_{\phz-\phz0.22}\phzs$ & 
$1.18\pm0.26^*$ & 
$1.13^{\phz+\phz0.27*}_{\phz-\phz0.25}\phzs$ &
$0.99\pm0.12\phs$ &  \\
199.5 & 
$0.70^{\phz+\phz0.22}_{\phz-\phz0.20}\phzs$ & 
$0.75^{\phz+\phz0.20*}_{\phz-\phz0.18}\phzs$ & 
$1.25\pm0.27^*$ & 
$1.05^{\phz+\phz0.26*}_{\phz-\phz0.23}\phzs$ &
$0.90\pm0.12\phs$ &  \\
201.6 & 
$0.70^{\phz+\phz0.33}_{\phz-\phz0.28}\phzs$ & 
$0.85^{\phz+\phz0.33*}_{\phz-\phz0.28}\phzs$ & 
$0.95\pm0.39^*$ & 
$0.79^{\phz+\phz0.36*}_{\phz-\phz0.30}\phzs$ &
$0.81\pm0.17\phs$ &  \\
204.9 & 
$1.21^{\phz+\phz0.26}_{\phz-\phz0.23}\phzs$ & 
$1.03^{\phz+\phz0.23*}_{\phz-\phz0.20}\phzs$ & 
$0.77^{\phz+\phz0.21*}_{\phz-\phz0.19}\phzs$ & 
$1.07^{\phz+\phz0.28*}_{\phz-\phz0.25}\phzs$ &
$0.98\pm0.13\phs$ &  \\
206.6 & 
$1.01^{\phz+\phz0.19}_{\phz-\phz0.17}\phzs$ & 
$0.96^{\phz+\phz0.16*}_{\phz-\phz0.15}\phzs$ & 
$1.09^{\phz+\phz0.18*}_{\phz-\phz0.17}\phzs$ & 
$0.97^{\phz+\phz0.20*}_{\phz-\phz0.19}\phzs$ &
$0.99\pm0.09\phs$ & \\
\hline
\end{tabular}
\caption{%
Z-pair production cross-sections from the four LEP
experiments and combined values 
for the eight energies between 183 and 207~GeV.
All results are preliminary with the exception of those indicated by $^*$.}
\label{4f_tab:zzxsec}
\end{center}
\end{table}
\renewcommand{\arraystretch}{1.}

\begin{figure}[p]
\centering
\epsfig{figure=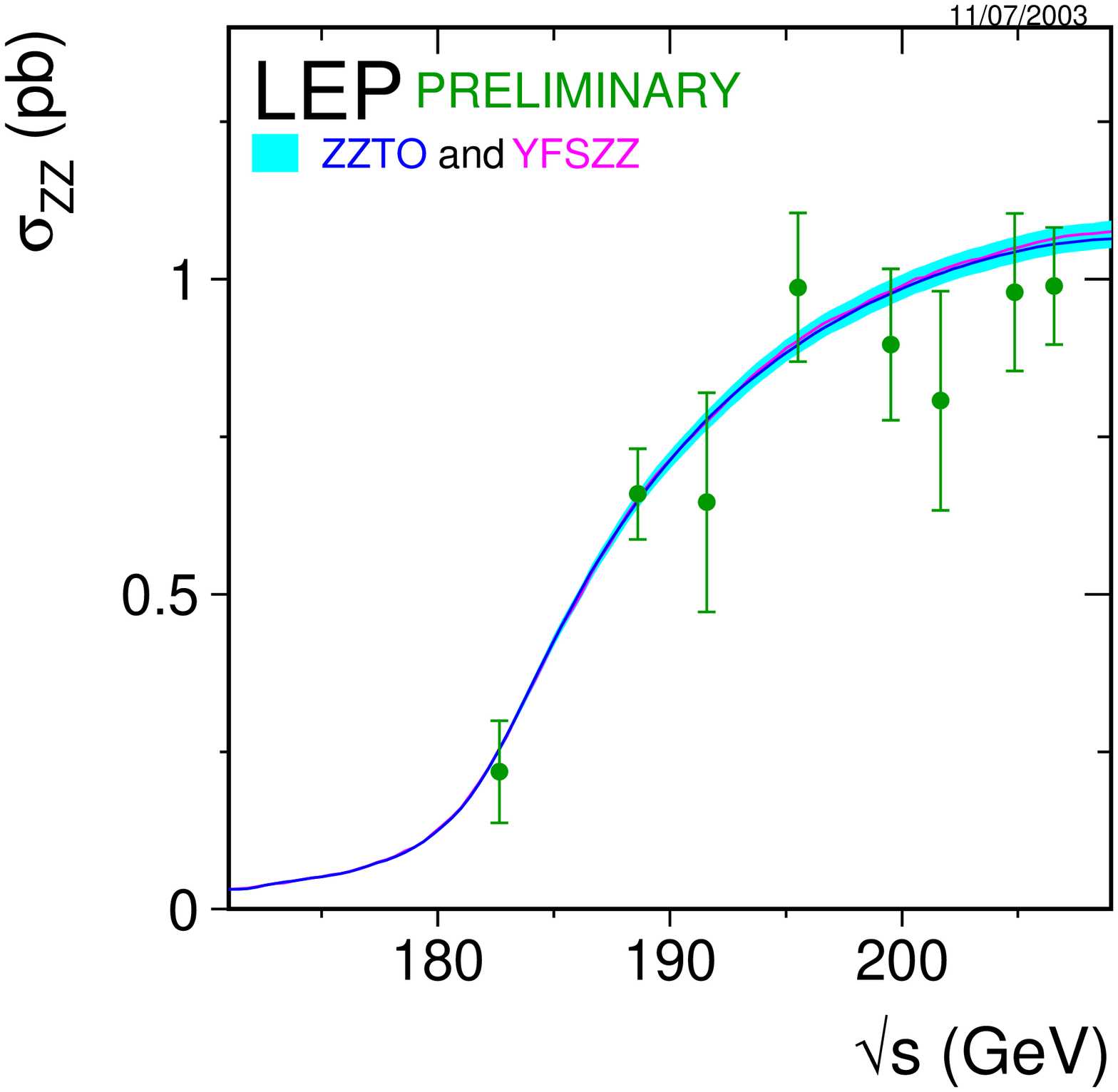,width=0.9\textwidth}
\caption{%
  Measurements of the Z-pair production cross-section,
  compared to the predictions 
  of \YFSZZ~\protect\cite{4f_bib:yfszz} and \ZZTO~\protect\cite{4f_bib:zzto}. 
  The shaded area represent the $\pm2$\% uncertainty 
  on the predictions.
}
\label{4f_fig:szz_vs_sqrts}
\end{figure}

\begin{table}[bhp]
\begin{center}
\begin{tabular}{|c|c|c|} 
\hline
\roots (GeV)      & $\rzz^{\footnotesize\ZZTO}$ 
           & $\rzz^{\footnotesize\YFSZZ}$ \\
\hline
182.7             & $0.857\pm0.320$ & $0.857\pm0.320$  \\
188.6             & $1.017\pm0.113$ & $1.007\pm0.111$  \\
191.6             & $0.831\pm0.225$ & $0.826\pm0.224$  \\
195.5             & $1.100\pm0.133$ & $1.100\pm0.133$  \\
199.5             & $0.915\pm0.125$ & $0.912\pm0.124$  \\
201.6             & $0.799\pm0.174$ & $0.795\pm0.173$  \\
204.9             & $0.937\pm0.121$ & $0.931\pm0.120$  \\
206.6             & $0.937\pm0.091$ & $0.928\pm0.090$  \\
\hline
$\chi^2$/d.o.f    & 16.1/24         & 16.1/24         \\
\hline
\hline
Average           & $0.952\pm0.052$ & $0.945\pm0.052$  \\
\hline
$\chi^2$/d.o.f    & 19.1/31         & 19.1/31        \\
\hline
\end{tabular}
\caption{%
Ratios of LEP combined Z-pair cross-section measurements
to the expectations according to 
\ZZTO~\protect\cite{4f_bib:zzto} and \YFSZZ~\protect\cite{4f_bib:yfszz}.
The results of the combined fits are given in the table together with 
the resulting $\chi^2$.
Both fits take into account inter-experiment 
as well as inter-energy correlations of systematic errors.
}
\label{4f_tab:zzratio}
\end{center}
\end{table}
\renewcommand{\arraystretch}{1.}

\begin{figure}[tp]
\centering
\vspace*{-0.5truecm}
\mbox{
  \fbox{\epsfig{figure=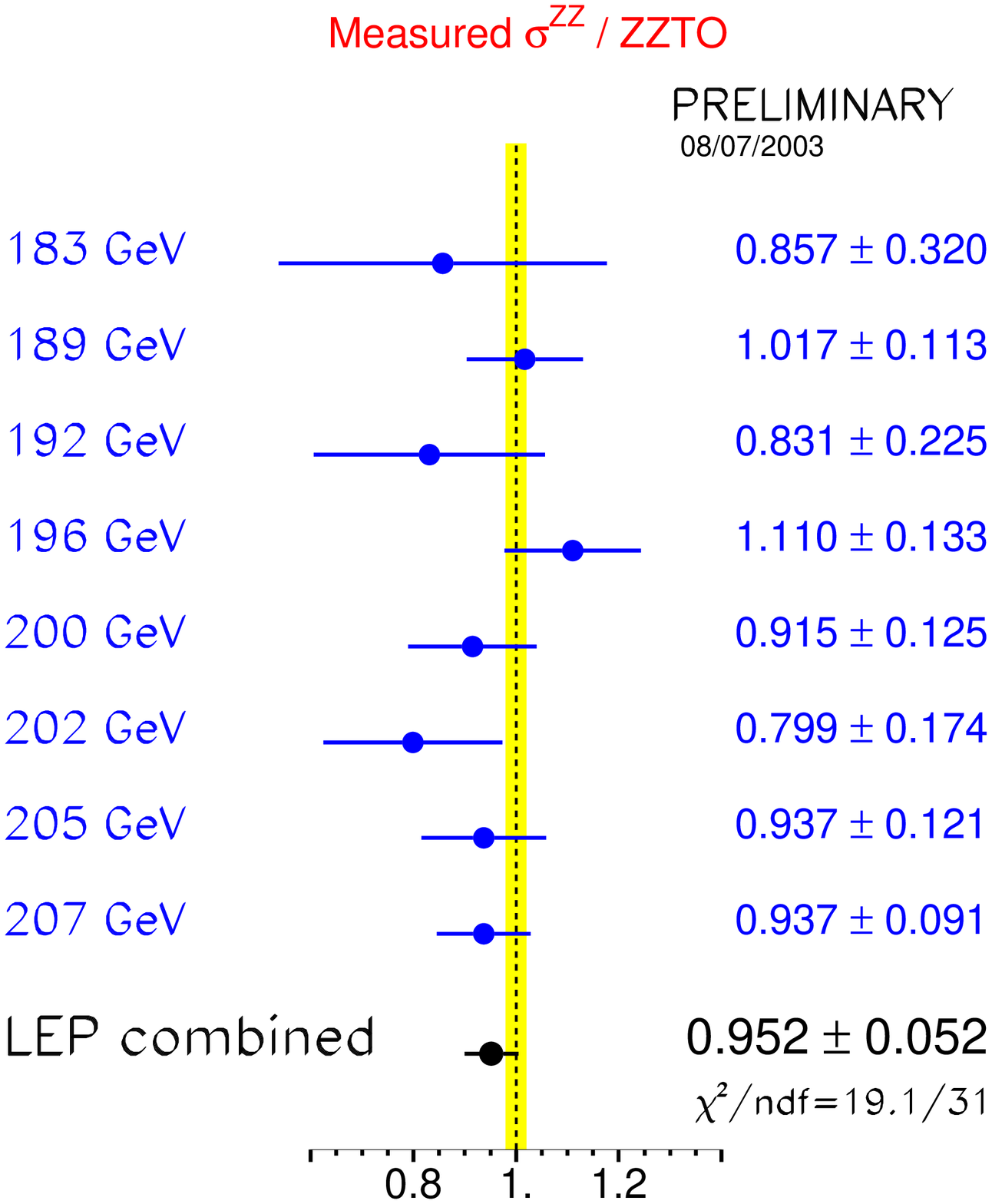,width=0.45\textwidth}}
  \hspace*{0.04\textwidth}
  \fbox{\epsfig{figure=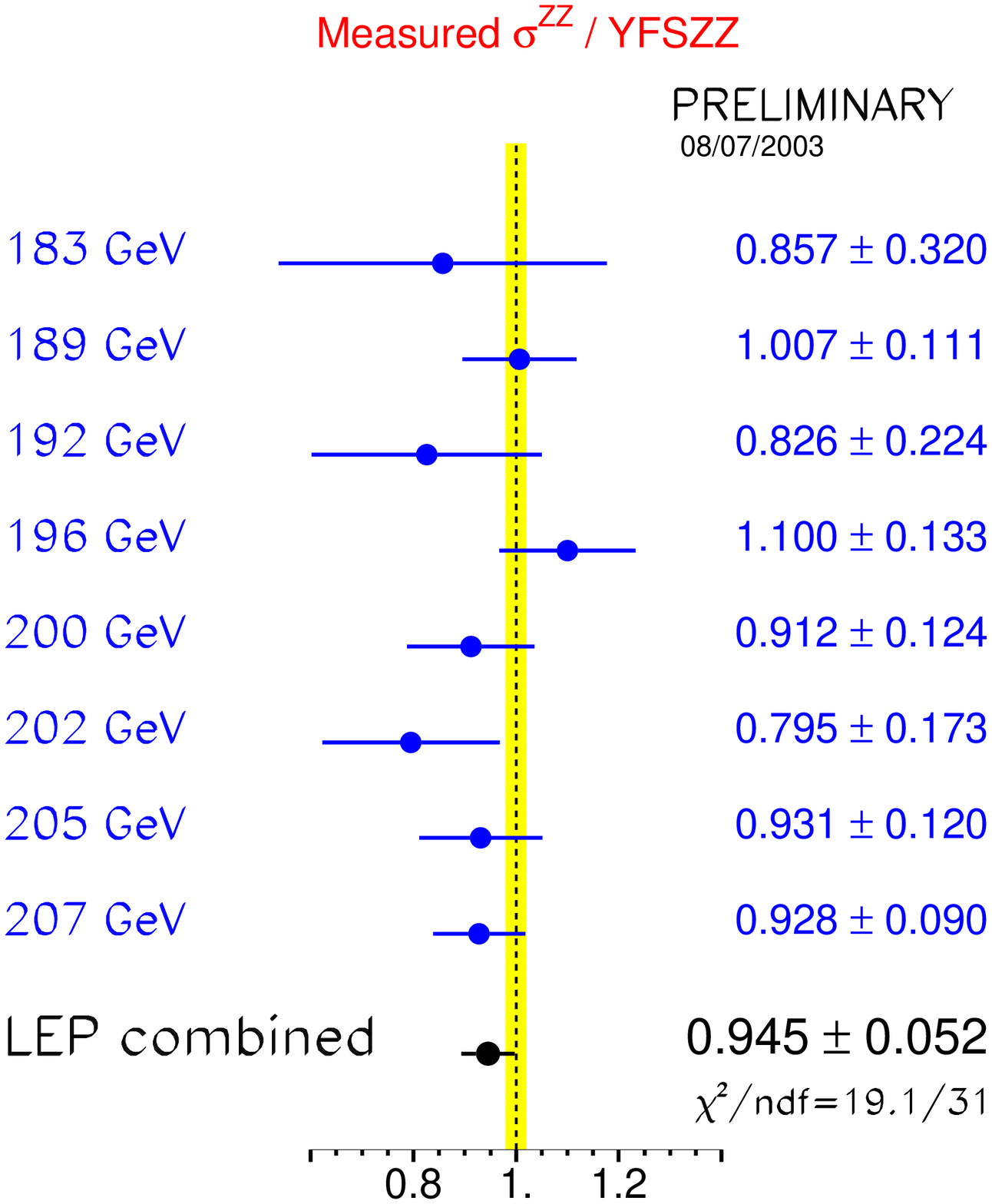,width=0.45\textwidth}}
  }
\vspace*{-0.5truecm}
\caption{%
  Ratios of LEP combined Z-pair cross-section measurements
  to the expectations according to 
  \ZZTO~\protect\cite{4f_bib:zzto} and 
  \YFSZZ~\protect\cite{4f_bib:yfszz}
  The yellow bands represent constant relative errors 
  of 2\% on the two cross-section predictions.
}
\label{4f_fig:rzz}
\end{figure} 

The combination of results is performed with the same technique used
for the WW cross-section. The symmetrized expected statistical error 
of each analysis is used, to avoid biases due to the limited number of 
selected events. 
All the cross-sections used for the combination and presented in 
Table~\ref{4f_tab:zzxsec} are determined by the experiments 
using the frequentist approach, i.e. without assuming any prior for 
the value of the cross-section itself.

The measurements are shown in Figure~\ref{4f_fig:szz_vs_sqrts} 
as a function of the LEP \CoM\ energy,
where they are compared to the \YFSZZ~\cite{4f_bib:yfszz} and
\ZZTO~\cite{4f_bib:zzto} predictions.
Both these calculations have an estimated 
uncertainty of $\pm2\%$~\cite{4f_bib:fourfrep}. 
The data do not show any significant deviation 
from the theoretical expectations.

In analogy with the W-pair cross-section, a value for $\rzz$ can 
also be determined: its definition and the procedure of the combination
follows the one described for $\rww$.
The data are compared with the \YFSZZ\ and \ZZTO\ predictions; 
Table~\ref{4f_tab:zzratio} reports the numerical values of $\rzz$ in energy 
and combined, whereas figure~\ref{4f_fig:rzz} show them in comparison
to unity, where the $\pm$2\% error on the theoretical ZZ cross-section is 
shown as a yellow band. The experimental accuracy on the combined 
value of $\rzz$ is about 5\%.

The theory predictions, the details of the experimental inputs
with the the breakdown of the error contributions and the LEP combined 
values of the total cross-sections and the ratios to theory are
reported in Appendix~\ref{4f_sec:appendix}. 

\clearpage

\section{Single-Z production cross-section}
\label{4f_sec:zeexsec}

Single-Z production at $\LEPII$ is studied considering only the 
$eeq\bar{q}$, $ee\mu\mu$ final states with the following phase space cuts 
and assuming one visible electron:
\mbox{m$_{q\bar{q}}(\mathrm{m}_{\mu\mu})>60$~GeV/c$^2$},
\mbox{$\theta_\mathrm{e^+}<12$~degrees},
\mbox{12 degrees$<\theta_\mathrm{e^-}<$120~degrees} and 
\mbox{E$_\mathrm{e^-}>$3~GeV}, with obvious notation and where the angle is
defined with respect to the beam pipe, with the positron direction
being along $+z$ 
and the electron direction being along $-z$. 
Corresponding cuts are imposed when the positron is visible:
\mbox{$\theta_\mathrm{e^-}>168$~degrees},
\mbox{60 degrees$<\theta_\mathrm{e^+}<$168~degrees} and 
\mbox{E$_\mathrm{e^+}>$3~GeV}.

The LEP combination of the single-Z production cross-section uses final results
by the ALEPH~\cite{4f_bib:alesw} and the L3~\cite{4f_bib:ltrzee} Collaborations and
has been updated with the final results by DELPHI~\cite{4f_bib:delswsc05}.

The results concern the hadronic and the leptonic channel and all the \CoM\ energies 
from 183 to 209~GeV.

\renewcommand{\arraystretch}{1.2}
\begin{table}[htb]
\begin{center}
\hspace*{-0.0cm}
\begin{tabular}{|c|c|c|c|c|c|c|} 
\hline
\roots & \multicolumn{5}{|c|}{Single-Z hadronic cross-section (pb)} 
       & \\
\cline{2-6} 
(GeV) & \Aleph\ & \Delphi\ & \Ltre\ & \Opal\ & LEP & $\chi^2/\textrm{d.o.f.}$ \\
\hline
182.7 & $0.27^{\phz+\phz0.21\phz*}_{\phz-\phz0.16}$ & $0.56^{\phz+\phz0.28\phz*}_{\phz-\phz0.23}$ &
        $0.51^{\phz+\phz0.19\phz*}_{\phz-\phz0.16}$ & --- &
$0.45\pm0.11\phs$ &  
             \multirow{8}{20.3mm}{$
               \hspace*{-0.3mm}
               \left\}
                 \begin{array}[h]{rr}
                   &\multirow{8}{8mm}{\hspace*{-4.2mm}13.0/16}\\
                   &\\ &\\ &\\ &\\ &\\ &\\ &\\  
                 \end{array}
               \right.
               $}\\

188.6 & $0.42^{\phz+\phz0.14*}_{\phz-\phz0.12}\phs$ & 
        $0.64^{\phz+\phz0.16*}_{\phz-\phz0.14}\phs$ &
        $0.55^{\phz+\phz0.11\phz*}_{\phz-\phz0.10}$ & --- &
$0.53\pm0.07\phs$ & \\

191.6 & $0.61^{\phz+\phz0.39*}_{\phz-\phz0.29}\phs$ & 
        $0.63^{\phz+\phz0.40*}_{\phz-\phz0.30}\phs$ &
        $0.60^{\phz+\phz0.26*}_{\phz-\phz0.21}\phs$ & --- &
$0.61\pm0.15\phs$ & \\

195.5 & $0.72^{\phz+\phz0.24*}_{\phz-\phz0.20}\phs$ & 
        $0.66^{\phz+\phz0.22*}_{\phz-\phz0.19}\phs$ &
        $0.40^{\phz+\phz0.13*}_{\phz-\phz0.11}\phs$ & --- &
$0.55\pm0.10\phs$ & \\

199.5 & $0.60^{\phz+\phz0.21*}_{\phz-\phz0.18}\phs$ & 
        $0.57^{\phz+\phz0.20*}_{\phz-\phz0.17}\phs$ &
        $0.33^{\phz+\phz0.13*}_{\phz-\phz0.11}\phs$ & --- &
$0.47\pm0.10\phs$ & \\

201.6 & $0.89^{\phz+\phz0.35*}_{\phz-\phz0.28}\phs$ & 
        $0.19^{\phz+\phz0.21*}_{\phz-\phz0.16}\phs$ &
        $0.81^{\phz+\phz0.27*}_{\phz-\phz0.23}\phs$ & --- &
$0.67\pm0.13\phs$ & \\

204.9 & $0.42^{\phz+\phz0.17*}_{\phz-\phz0.15}\phs$ & 
        $0.37^{\phz+\phz0.18*}_{\phz-\phz0.15}\phs$ & 
        $0.56^{\phz+\phz0.16*}_{\phz-\phz0.14}\phs$ & --- &
$0.47\pm0.10\phs$ & \\

206.6 & $0.70^{\phz+\phz0.17*}_{\phz-\phz0.15}\phs$ & 
        $0.69^{\phz+\phz0.16*}_{\phz-\phz0.14}\phs$ & 
        $0.59^{\phz+\phz0.12*}_{\phz-\phz0.11}\phs$ & --- &
$0.65\pm0.08\phs$ & \\
\hline
\end{tabular}
\end{center}
\caption{%
  Single-Z hadronic production cross-section from the four LEP
  experiments and combined values for the eight energies between 
  183 and 207~GeV. All results are preliminary with the exception 
  of those indicated by $^*$.}
\label{4f_tab:szxsecqq}
\vspace*{-0.1cm}
\end{table}
\renewcommand{\arraystretch}{1.}

\renewcommand{\arraystretch}{1.2}
\begin{table}[htb]
\begin{center}
\hspace*{-0.0cm}
\begin{tabular}{|c|c|c|c|c|c|} 
\hline
  & \multicolumn{5}{|c|}{Single-Z cross-section into muons(pb)} \\
\cline{2-6} 
    & \Aleph\ & \Delphi\ & \Ltre\ & \Opal\ & LEP  \\
\hline
Av. \roots (GeV) & 196.67 & 197.10 & 196.60 & --- & 196.79   \\
$\sigma_{\mathrm{Zee}\rightarrow \mu\mu \mathrm{ee}}$ 
& $0.055\pm0.016\phs^*$ &  $0.070^{\phz+\phz0.023\phz}_{\phz-\phz0.019}$ & 
  $0.043\pm0.013\phs^*$ & --- &
$0.057\pm0.009\phs$  \\
\hline
\end{tabular}
\end{center}
\caption{%
  Preliminary energy averaged single-Z production cross-section into muons 
  from the four LEP experiments and combined values. The results indicated
  with $^*$ are final.}
\label{4f_tab:szxsecmm}
\vspace*{-0.1cm}
\end{table}
\renewcommand{\arraystretch}{1.}

\begin{figure}[p]
\centering
\epsfig{figure=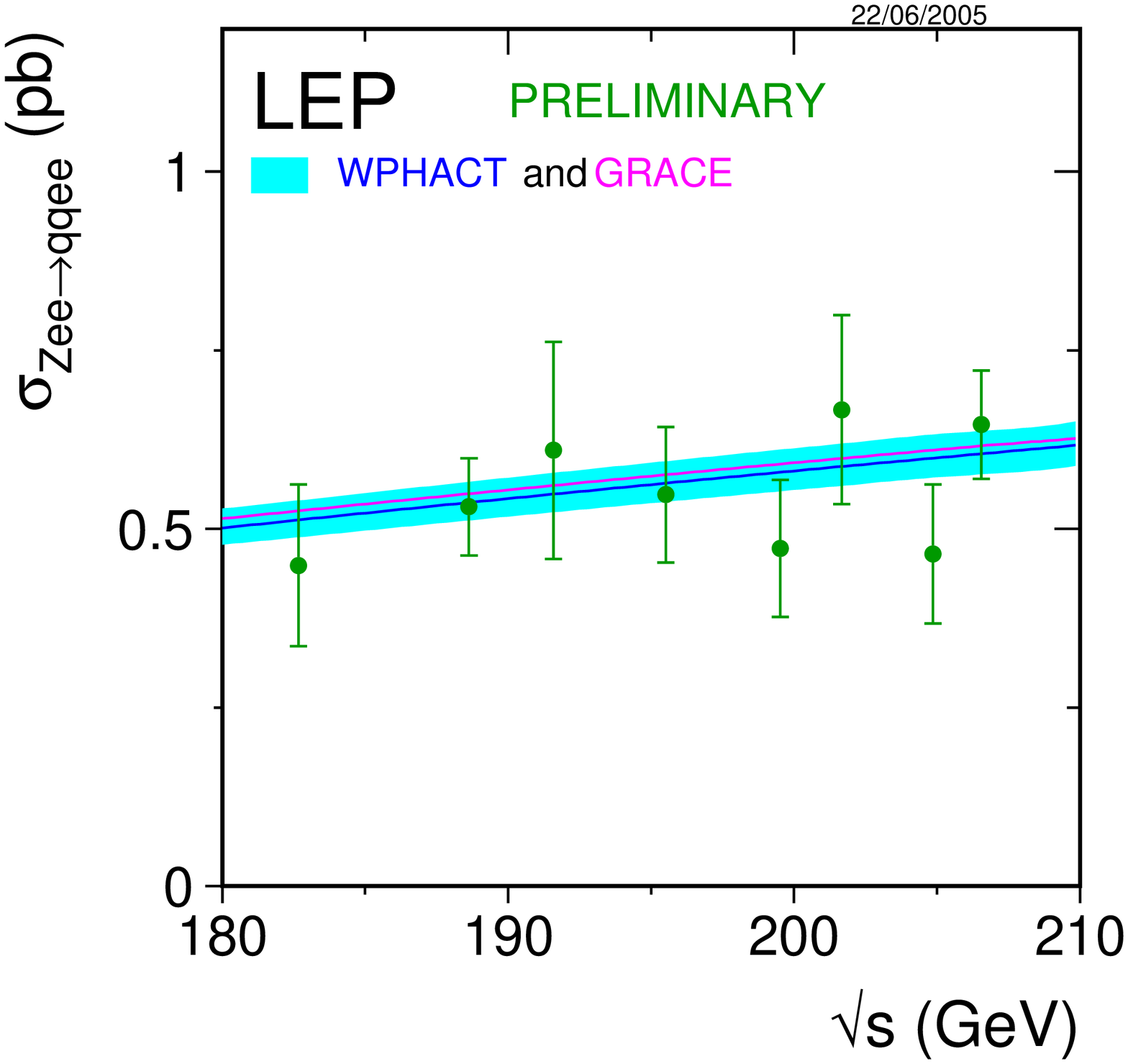,width=0.9\textwidth}
\caption{%
  Measurements of the single-Z hadronic production cross-section,
  compared to the predictions of \WPHACT\ and \Grace. 
  The shaded area represents the $\pm5$\% uncertainty 
  on the predictions.
}
\label{4f_fig:szee}
\end{figure}

\begin{figure}[tp]
\centering
\vspace*{-0.5truecm}
\mbox{
  \fbox{\epsfig{figure=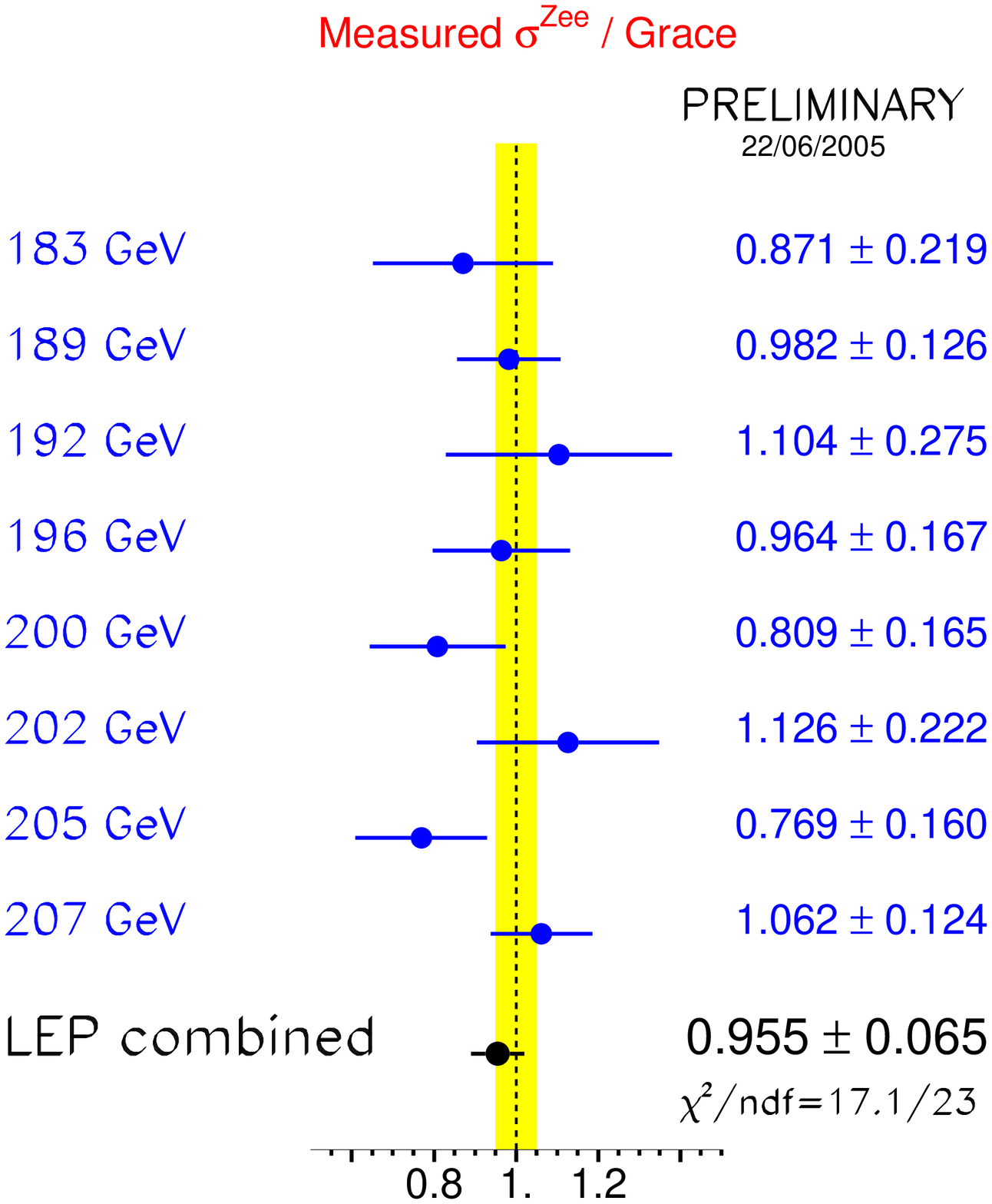,width=0.45\textwidth}}
  \hspace*{0.04\textwidth}
  \fbox{\epsfig{figure=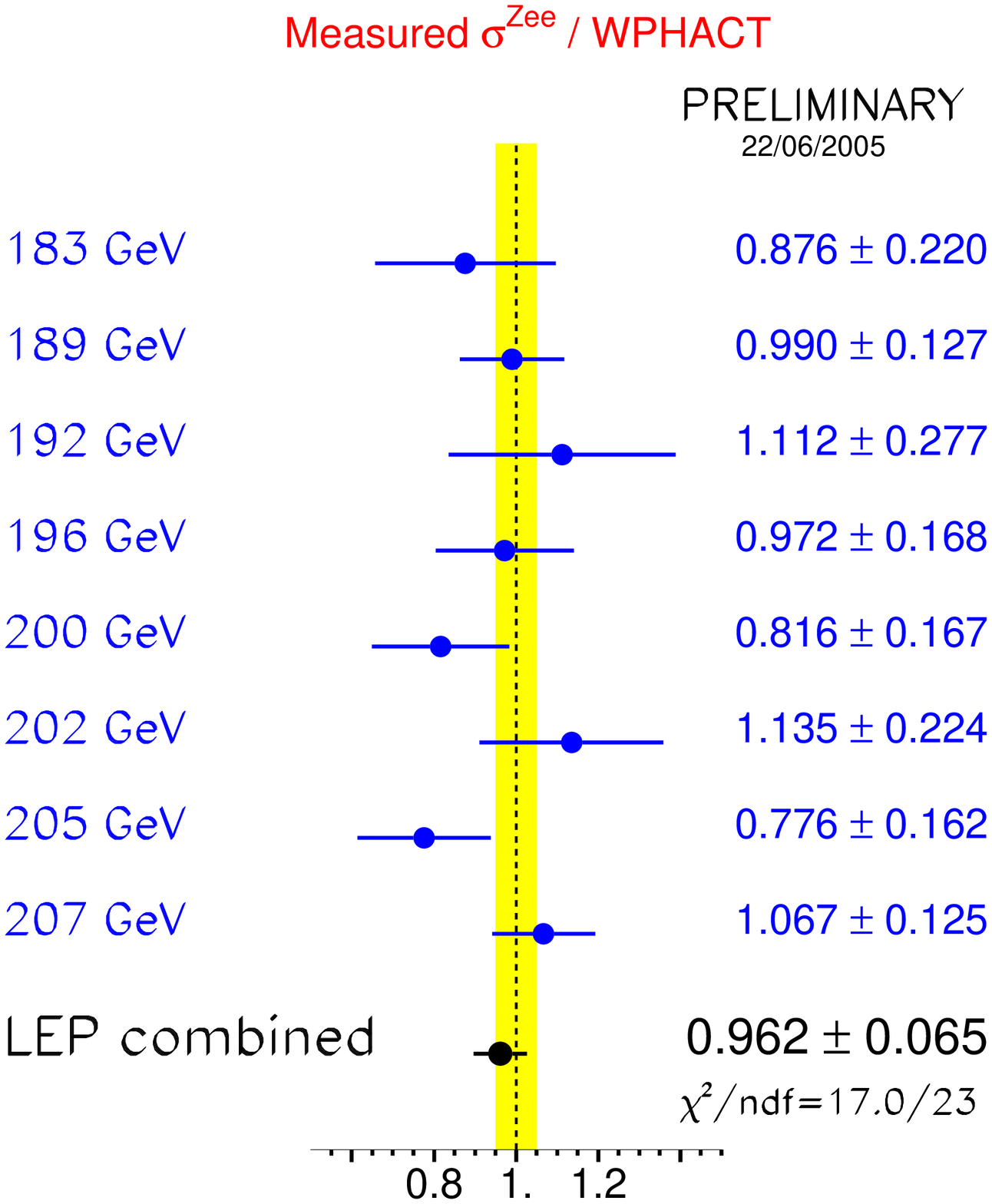,width=0.45\textwidth}}
  }
\vspace*{-0.5truecm}
\caption{%
  Ratios of LEP combined single-Z hadronic cross-section measurements
  to the expectations according to 
  \Grace~\protect\cite{4f_bib:grace} and 
  \WPHACT~\protect\cite{4f_bib:wphact}.
  The yellow bands represent constant relative errors 
  of 5\% on the two cross-section predictions.
}
\label{4f_fig:rzee}
\end{figure}

Tables~\ref{4f_tab:szxsecqq} and~\ref{4f_tab:szxsecmm} synthesize the inputs
by the experiments and the corresponding LEP combinations in the hadronic and
muon channel, respectively. 
The $ee\mu\mu$ cross-section is already combined in energy by the individual
experiments to increase the statistics of the data.
The combination accounts for energy and experiment
correlation of the systematic errors. 
The results in the hadronic channel are compared with the \WPHACT\ and \Grace\
predictions as a function of the \CoM\ energy and shown in figure~\ref{4f_fig:szee}.
Table~\ref{4f_tab:zeeratio} and figure~\ref{4f_fig:rzee} show the preliminary
values of the ratio between measured and expected cross-sections at the various
energy points and the combined value; the testing accuracy of the combined value 
is about 7\% with three experiments contributing in the average.

The detailed breakdown of the inputs of the experiments with the split up of the 
systematic contribution according to the correlations for the single-Z cross-section
and its ratio to theory can be found in Appendix~\ref{4f_sec:appendix}.

\begin{table}[bhtp]
\vspace*{-0mm}
\begin{center}
\hspace*{-0.3cm}
\begin{tabular}{|c|c|c|} 
\hline
\roots (GeV) & $\rzee^{\footnotesize\Grace}$ & $\rzee^{\footnotesize\WPHACT}$ \\
\hline
182.7             & $0.871\pm0.219$ & $0.876\pm0.220$  \\
188.6             & $0.982\pm0.126$ & $0.990\pm0.127$  \\
191.6             & $1.104\pm0.275$ & $1.112\pm0.277$  \\
195.5             & $0.964\pm0.167$ & $0.972\pm0.168$  \\
199.5             & $0.809\pm0.165$ & $0.816\pm0.167$  \\
201.6             & $1.126\pm0.222$ & $1.135\pm0.224$  \\
204.9             & $0.769\pm0.160$ & $0.776\pm0.162$  \\
206.6             & $1.062\pm0.124$ & $1.067\pm0.125$  \\
\hline
$\chi^2$/d.o.f    &  13.0/16          & 13.0/16         \\
\hline
\hline
Average           & $0.955\pm0.065$ & $0.962\pm0.065$  \\
\hline
$\chi^2$/d.o.f    & 17.1/23         & 17.0/23        \\
\hline
\end{tabular}
\caption{%
  Ratios of LEP combined single-Z hadronic cross-section measurements
  to the expectations according to \Grace~\protect\cite{4f_bib:grace}
  and \WPHACT~\protect\cite{4f_bib:wphact}.  The resulting averages
  over energies are also given.  The averages take into account
  inter-experiment as well as inter-energy correlations of systematic
  errors.  }
\label{4f_tab:zeeratio}
\end{center}
\vspace*{-6mm}
\end{table}
\renewcommand{\arraystretch}{1.}

\clearpage

\section{Z$\gamma^*$ production cross-section}
\label{4f_sec:zgstarxsec}

The Z$\gamma^*$ contribution to the four-fermion phase space is defined 
for the final states with two couples of same-kind, charge conjugate, leptons. 
It is required that one and only one of the invariant masses of the
couples satisfies:
\mbox{m$_{\mathrm{Z}}-2\Gamma_{\mathrm{Z}}<\mathrm{m}_{\mathrm{ff'}}<\mathrm{m}_{\mathrm{Z}}+2\Gamma_{\mathrm{Z}}$}
with m$_{\mathrm{ff'}}$ is the invariant mass of the two same-kind fermions.
In case of four identical leptons all oppositely charged couples have to be
considered.
Moreover the following cuts, final state dependent, have been introduced:
\begin{itemize}
\item{eeqq, $\mu\mu$qq: $|\cos{\theta_{\ell}}|<$0.95, m$_{\ell\ell}>$5GeV, m$_{\mathrm{qq}}>$10GeV, $\ell=$e,$\mu$}
\item{$\nu\nu$qq: m$_{\mathrm{qq}}>$10GeV}
\item{$\nu\nu\ell\ell$: m$_{\ell\ell}>$10GeV, m$_{\ell\nu}>$90GeV, m$_{\ell\nu}<$70GeV, $\ell=$e,$\mu$}
\item{$\ell_1\ell_1\ell_2\ell_2$: $|\cos{\theta_{\ell_1\ell_2}}|<$0.95, m$_{\ell_1\ell_1}>$5GeV, m$_{\ell_2\ell_2}>$5GeV, $\ell=$e,$\mu$}
\end{itemize}

The LEP collaborations did not provide a complete analysis of all possible 
Z$\gamma^*$ final states. Indeed, the \Delphi\ collaboration provided preliminary results
for the $\nu\nu$qq, $\ell\ell$qq final states~\cite{4f_bib:delzgstar}, whereas the 
\Ltre\ collaboration provided final results for 
the $\nu\nu$qq, $\ell\ell$qq, $\ell\ell\nu\nu$, $\ell\ell\ell\ell$ 
channels~\cite{4f_bib:ltrzgstar}.  
Final states with only quarks or containing $\tau$s were not studied.
\Aleph\ and \Opal\ did not present any result on Z$\gamma^*$.

To increase the statistics the cross-sections were determined using the full data sample
at an average $\LEPII$ centre-of-mass energy. Table~\ref{4f_tab:zgstarxsec} presents the measured
cross-sections, where the expected statistical errors were used for the combination.
The results agree well with the expectations. 

\renewcommand{\arraystretch}{1.2}
\begin{table}[hbt]
\begin{center}
\hspace*{-0.0cm}
\begin{tabular}{|c|c|c|c|c|c|c|c|} 
\hline

channel & $\sqrt{s}$ (GeV) & Lumi (pb$^{-1}$) & $\sigma$(pb) & $\delta\sigma_{\mathrm{stat}}$ (pb) & $\delta\sigma_{\mathrm{syst}}^{\mathrm{unc}}$ (pb) &  $\delta\sigma_{\mathrm{syst}}^{\mathrm{cor}}$ (pb) & $\delta\sigma_{\mathrm{MC}}$ (pb) \\

\hline
\hline
\multicolumn{8}{|c|}{\Delphi} \\
\hline
$\nu\nu$qq & 197.1 & 666.7 & 0.042 & $^{+0.022}_{-0.020}$ & 0.008 & 0.002 & 0.042 \\
$\mu\mu$qq & 197.1 & 666.7 & 0.031 & $^{+0.013}_{-0.011}$ & 0.004 & 0.001 & 0.016 \\
eeqq       & 197.1 & 666.7 & 0.063 & $^{+0.018}_{-0.016}$ & 0.009 & 0.001 & 0.016 \\

\hline
\hline
\multicolumn{8}{|c|}{\Ltre} \\
\hline
$\nu\nu$qq & 196.7 & 679.4 & 0.072 & $^{+0.047}_{-0.041}$ & 0.004 & 0.016 & 0.046 \\
$\mu\mu$qq & 196.7 & 681.9 & 0.040 & $^{+0.018}_{-0.016}$ & 0.002 & 0.003 & 0.017 \\
eeqq       & 196.7 & 681.9 & 0.100 & $^{+0.024}_{-0.022}$ & 0.004 & 0.007 & 0.020 \\

\hline

\hline
\hline
\multicolumn{8}{|c|}{LEP combined} \\
\hline
channel & $\sqrt{s}$ (GeV) & Lumi (pb$^{-1}$) & $\sigma$(pb) & $\delta\sigma_{\mathrm{stat}}$ (pb) & 
$\delta\sigma_{\mathrm{syst}}$ (pb) & $\delta\sigma_{\mathrm{tot}}$ (pb) & $\sigma_{\mathrm{theory}}$ (pb) \\
\hline
$\nu\nu$qq & 196.9 & 679.4 & 0.055 & 0.031 & 0.008 & 0.032 & 0.083 \\
$\mu\mu$qq & 196.9 & 681.9 & 0.035 & 0.012 & 0.003 & 0.012 & 0.042 \\
eeqq       & 196.9 & 681.9 & 0.079 & 0.012 & 0.005 & 0.013 & 0.059 \\

\hline

\end{tabular}
\end{center}
\caption[]{
  Z$\gamma^*$ input by the experiments and combined LEP measurements.
  In the columns are reported, respectively, the channel, the luminosity weighted
  centre-of-mass energy, the luminosity, the cross-section value, the measured statistical 
  error, the systematic contribution uncorrelated between experiments, the systematic contribution
  correlated between experiments and the expected statistical error from the simulation. 
  All results are final. For the LEP combination the full systematic error and the total error are 
  given and the last column presents the theory expectation with GRC4F.}
\label{4f_tab:zgstarxsec}
\end{table}
\renewcommand{\arraystretch}{1.}

\clearpage

\section{WW$\gamma$ production cross-section}
\label{4f_sec:wwgxsec}

A LEP combination of the WW$\gamma$ production cross-section has been
performed using final \Delphi~\cite{4f_bib:delwwg},
\Ltre~\cite{4f_bib:ltrwwg} and \Opal~\cite{4f_bib:opawwg} results
available since the Summer 2003 Conferences.  The signal is defined as
the part of the WW$\gamma$ process with the following cuts to the
photon: \mbox{E$_{\gamma}>$5~GeV},
\mbox{$|\cos\theta_{\gamma}|<$0.95},
\mbox{$|\cos\theta_{\gamma,\mathrm{f}}|<$0.90} and
\mbox{m$_{\mathrm{W}}-2\Gamma_{\mathrm{W}}<\mathrm{m}_{\mathrm{ff'}}<\mathrm{m}_{\mathrm{W}}+2\Gamma_{\mathrm{W}}$}
where $\theta_{\gamma,\mathrm{f}}$ is the angle between the photon and
the closest charged fermion and m$_{\mathrm{ff'}}$ is the invariant
mass of fermions from the Ws.

In order to increase the statistics the LEP combination is performed in energy
intervals rather than at each energy point; they are defined according to the 
$\LEPII$ running period where more statistics was accumulated.
The luminosity weighted \CoM\ per interval is determined in each experiment
and then combined to obtain the corresponding value in the combination.
Table~\ref{4f_tab:wwgxsec} reports those energies and the cross-sections
measured by the experiments, together with the combined LEP values.

\renewcommand{\arraystretch}{1.2}
\begin{table}[hbt]
\begin{center}
\hspace*{-0.0cm}
\begin{tabular}{|c|c|c|c|c|c|} 
\hline
\roots & \multicolumn{5}{|c|}{WW$\gamma$ cross-section (pb)}  \\
\cline{2-6} 
(GeV) & \Aleph\ & \Delphi\ & \Ltre\ & \Opal\ & LEP  \\
\hline
188.6 & --- & $0.05\pm0.08\phs$ & $0.20\pm0.09\phs$ & $0.16\pm0.04\phs$ & $0.15\pm0.03\phs$ \\
194.4 & --- & $0.17\pm0.12\phs$ & $0.17\pm0.10\phs$ & $0.17\pm0.06\phs$ & $0.17\pm0.05\phs$ \\
200.2 & --- & $0.34\pm0.12\phs$ & $0.43\pm0.13\phs$ & $0.21\pm0.06\phs$ & $0.27\pm0.05\phs$ \\
206.1 & --- & $0.18\pm0.08\phs$ & $0.13\pm0.08\phs$ & $0.30\pm0.05\phs$ & $0.24\pm0.04\phs$ \\
\hline
\end{tabular}
\end{center}
\vspace*{-0.3cm}
\caption{%
  WW$\gamma$ production cross-section from the four LEP experiments and combined values for 
  the four energy bins. All results are final.}
\label{4f_tab:wwgxsec}
\end{table}
\renewcommand{\arraystretch}{1.}

\begin{figure}[p]
\centering
\epsfig{figure=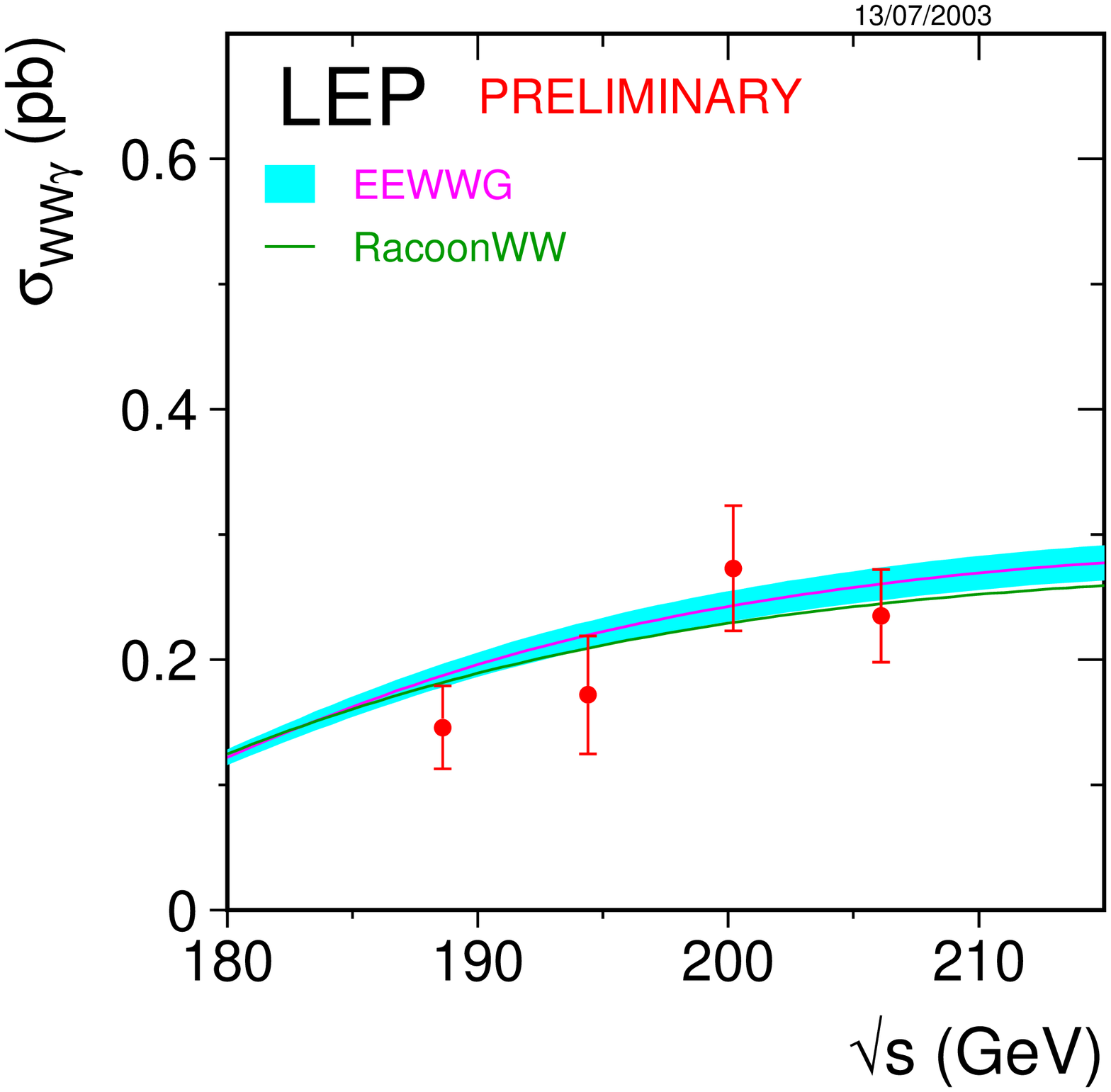,width=0.9\textwidth}
\caption{%
  Measurements of the WW$\gamma$ production cross-section,
  compared to the predictions of \EEWWG~\protect\cite{4f_bib:eewwg} and
  \RacoonWW~\protect\cite{4f_bib:racoonww}. 
  The shaded area in the \EEWWG\ curve represents the $\pm5$\% uncertainty 
  on the predictions.
}
\label{4f_fig:swwg}
\end{figure}

Figure~\ref{4f_fig:swwg} shows the combined data points compared with the
cross-section prediction by \EEWWG~\cite{4f_bib:eewwg} and by 
\RacoonWW. The \RacoonWW\ is shown in the figure without any theory error band.

\section{Summary}
\label{4f_sec:summary}
The updated LEP combinations of the W-pair and single boson production cross-section,
together with the W angular distributions, have been presented.
A first combination of some of the Z$\gamma^*$ final states has also been performed.
The combinations are based on data collected up to 209 GeV by the four LEP experiments.

All measurements agree with the expectations.
In the fit to the W branching fractions without the assumption of lepton universality
an excess of the W branching ratio into $\tau\nu_{\tau}$ with respect to the other
lapton families is observed in the data.
This excess is above two standard deviations from both the branching ratio into $e\nu_{e}$
and into $\mu\nu_{\mu}$.

This note still reflects a preliminary status of the analyses 
at the time of the Summer 2005 Conferences.
A definitive statement on these results and the ones not updated for these Conferences
must wait for publication by each collaboration.

\boldmath
\chapter{Electroweak Gauge Boson Self Couplings}
\label{sec-GC}
\unboldmath

\updates{ Unchanged w.r.t. summer 2003: Results are preliminary. 

  Note that some recent publications~\cite{ Heister:2004yd,
  Schael:2005am, 4f_bib:ltrzz, Achard:2004ji, Achard:2004ds,
  Abbiendi:2004bf} are not yet included in this combination.  }

\section{Introduction}
\label{sec:gc_introduction}

The measurement of  gauge boson couplings and the
search for possible anomalous contributions due to the effects of new
physics beyond the Standard Model are among the principal physics
aims at \LEPII~\cite{gc_bib:LEP2YR}.
Combined preliminary measurements of triple gauge boson
couplings are presented here. Results from W-pair production are
combined in single and two-parameter fits, including updated results from
ALEPH, L3 and OPAL as well as an improved treatment of the main systematic
effect in our previous combination, the uncertainty in the $O(\alpha_{em})$
correction.     
An updated combination of quartic gauge coupling (QGC) results for the \ZZgg\
vertex is also presented, including data from ALEPH, L3 and OPAL.
The combination of QGCs associated with the \WWgg\ vertex, including the sign
convention as reported 
in~\cite{gc_bib:Montagna:2001ej,gc_bib:denner} and the reweighting based
on~\cite{gc_bib:Montagna:2001ej} is foreseen for our next report. 
The combination of neutral TGCs measured in ZZ production (f-couplings) has
been updated, including new results from L3 and OPAL.
The combinations for neutral TGCs accessible through ${\rm Z} \gamma$
production (h-couplings) reported in 2001 still remain valid ~\cite{gc_bib:budapest01}.

The W-pair production process, $\mathrm{e^+e^-\rightarrow\WW}$,
involves charged triple gauge boson vertices between the $\WW$ and
the Z or photon.  During \LEPII\ operation, about 10,000 W-pair
events were collected by each experiment.  
Single W ($\enw$) and single photon ($\nng$) production at LEP are also
sensitive to the $\WWg$ vertex. Results from these channels are also included
in the combination for some experiments; the individual references should be
consulted for details.

For the charged TGCs, Monte Carlo calculations
(RacoonWW~\cite{common_bib:racoonww} and
YFSWW~\cite{common_bib:yfsww}) incorporating an improved treatment of
$O(\alpha_{em})$ corrections to the WW production have become our standard by
now. The corrections affect the measurements of the charged
TGCs in W-pair production. 
Results, some of them preliminary, including these
$O(\alpha_{em})$ corrections have been submitted from all four LEP
collaborations ALEPH~\cite{gc_bib:ALEPH-cTGC}, DELPHI~\cite{gc_bib:DELPHI-cTGC}, L3~\cite{gc_bib:L3-cTGC} and OPAL~\cite{gc_bib:OPAL-cTGC3}. 
LEP combinations are made for the charged TGC measurements in single- and
two-parameter fits. 

At centre-of-mass energies exceeding twice the Z boson mass, pair
production of Z bosons is kinematically allowed. Here, one searches
for the possible existence of triple vertices involving only neutral
electroweak gauge bosons. Such vertices could also contribute to
Z$\gamma$ production.  In contrast to triple gauge boson vertices with
two charged gauge bosons, purely neutral gauge boson vertices do not
occur in the Standard Model of electroweak interactions.

Within the Standard Model, quartic electroweak gauge boson vertices
with at least two charged gauge bosons exist. In $\ee$ collisions at
\LepII\ centre-of-mass energies, the $\WWZg$ and $\WWgg$ vertices
contribute to $\WWg$ and $\nngg$ production in $s$-channel and
$t$-channel, respectively.  The effect of the Standard Model quartic
electroweak vertices is below the sensitivity of \LepII.  
Quartic gauge boson vertices with only neutral bosons, like the
\ZZgg\ vertex, do not exist in the Standard Model. However, anomalous QGCs
associated with this vertex are studied at LEP.

Anomalous quartic vertices are searched for in the production of
$\WWg$, $\nngg$ and $\Zgg$ final states. The couplings related to the \ZZgg\
and \WWgg\ vertices are assumed to be different~\cite{gc_bib:QGC-Belanger},
and are therefore treated separately.
In this report, we only combine the results for the anomalous couplings
associated with the \ZZgg\ vertex. The combination of the \WWgg\ vertex
couplings is foreseen for the near future.

\subsection{Charged Triple Gauge Boson Couplings}
\label{sec:gc_cTGCs}

The parametrisation of the charged triple gauge boson vertices is
described in References~\cite{gc_bib:GAEMERS,gc_bib:Hagiwara1987vm,
gc_bib:HAGIWARA,gc_bib:BILENKY,gc_bib:KUSS,
gc_bib:PAPADOPOULOSCP,gc_bib:LEP2YR}.  
The most general Lorentz invariant
Lagrangian which describes the triple gauge boson interaction has
fourteen independent complex couplings, seven describing the
WW$\gamma$ vertex and seven describing the WWZ vertex.  Assuming
electromagnetic gauge invariance as well as C and P conservation, the
number of independent TGCs reduces to five.  A common set is \{$\gz,
\kz, \kg, \lz$, $\lg$\} where $\gz = \kz = \kg = 1$ and $\lz = \lg =
0$ in the \SM.  The parameters proposed in~\cite{gc_bib:LEP2YR} and used by
the LEP experiments are $\gz$, $\lg$ and $\kg$ with the gauge constraints:
\begin{eqnarray}
\kz & = & \gz - (\kg - 1) \twsq \,, \\
\lz & = & \lg \,,
\end{eqnarray}                               
where $\theta_W$ is the weak mixing angle.  The
couplings are considered as real, with the imaginary parts fixed to
zero. In contrast to previous LEP
combinations~\cite{gc_bib:moriond01,gc_bib:budapest01}, we are quoting
the measured coupling values themselves and not their deviation from the
Standard Model. 

Note that the photonic couplings $\lg$ and $\kg$ are related to the
magnetic and electric properties of the W-boson. One can write the
lowest order terms for a multipole expansion describing the W-$\gamma$
interaction as a function of $\lg$ and $\kg$. For the magnetic dipole
moment $\mu_{W}$ and the electric quadrupole moment $q_{W}$ one
obtains $e(1+\kappa_{\gamma}+\lambda_{\gamma})/2\MW$ and
$-e(\kappa_{\gamma}-\lambda_{\gamma})/\MW^2$, respectively.

The inclusion of $O(\alpha_{em})$ corrections in the Monte Carlo
calculations has a considerable effect on the charged TGC measurement. Both the
total cross-section and the differential distributions are affected. The  
cross-section is reduced by 1-2\% (depending on the energy). Amongst
the differential distributions, the effects are naturally more complex. The
polar W$^-$ production angle carries most of the information on the TGC
parameters; its 
shape is modified to be more forwardly peaked. In a fit to data, the
$O(\alpha_{em})$ effect
manifests itself as a negative shift of the obtained TGC values with a
magnitude of typically -0.015 for \lg\ and \gz\, and -0.04 for \kg.

\subsection{Neutral Triple Gauge Boson Couplings}
\label{sec:gc_nTGCs}

There are two classes of Lorentz invariant structures associated with
neutral TGC vertices which preserve $U(1)_{em}$ and Bose symmetry, as
described in~\cite{gc_bib:Hagiwara1987vm,gc_bib:Gounaris2000tb}.

The first class refers to anomalous Z$\gamma\gamma^*$ and $\rm Z\gamma
\rm Z^*$ couplings which are accessible at LEP in the process
$\mathrm{e^{+} e^{-}} \rightarrow {\rm Z} \gamma$. The parametrisation
contains eight couplings: $h_i^{V}$ with $i=1,...,4$ and $V=\gamma$,Z.
The superscript $\gamma$ refers to Z$\gamma\gamma^*$ couplings and
superscript Z refers to $\rm Z\gamma \rm Z^*$ couplings.  The photon
and the Z boson in the final state are considered as on-shell
particles, while the third boson at the vertex, the $s$-channel
internal propagator, is off shell.  The couplings $h_{1}^{V}$ and
$h_{2}^{V}$ are CP-odd while $h_{3}^{V}$ and $h_{4}^{V}$ are CP-even.

The second class refers to anomalous $\rm{ZZ}\gamma^*$ and
$\rm{ZZZ}^*$ couplings which are accessible at \LEPII\ in the process
$\mathrm{e^{+} e^{-}} \rightarrow$ ZZ.  This anomalous vertex is
parametrised in terms of four couplings: $f_{i}^{V}$ with $i=4,5$ and
$V=\gamma$,Z.  The superscript $\gamma$ refers to ZZ$\gamma^*$
couplings and the superscript Z refers to $\rm{ZZZ}^*$ couplings,
respectively.  Both Z bosons in the final state are assumed to be
on-shell, while the third boson at the triple vertex, the $s$-channel
internal propagator, is off-shell.
The couplings $f_{4}^{V}$ are CP-odd whereas $f_{5}^{V}$ are CP-even.

The $h_i^{V}$ and $f_{i}^{V}$ couplings are assumed to be real and they
vanish at tree level in the Standard Model.

\subsection{Quartic Gauge Boson Couplings}
\label{sec:gc_QGCs}
The couplings associated with the two QGC vertices \WWgg\ and \ZZgg\ are
assumed to be different, and are by convention treated as separate couplings
at LEP. In this report, we only combine QGCs related to the \ZZgg\ vertex.
The contribution of such anomalous quartic gauge boson couplings is described by two coupling parameters \acl\ and \azl,
which are zero in the Standard
Model~\cite{gc_bib:QGC-BelBou,4f_bib:eewwg}.  
Events from $\nngg$ and $\Zgg$ final states can originate from the \ZZgg\
vertex and are therefore used to study anomalous QGCs.

\section{Measurements}
\label{sec:gc_data}

The combined results presented here are obtained from charged and neutral
electroweak gauge boson coupling measurements, and from quartic gauge boson
couplings measurements as discussed above.  
The individual references should be consulted for details about the data
samples used. 

The charged TGC analyses of ALEPH, DELPHI, L3 and OPAL use data collected
at \LEPII\ up to centre-of-mass energies of 209~\GeV. These
analyses use different
channels, typically the semileptonic and fully hadronic W-pair
decays~\cite{gc_bib:ALEPH-cTGC,gc_bib:DELPHI-cTGC,
  gc_bib:L3-cTGC,gc_bib:OPAL-cTGC3}. 
The full data set is analysed by ALEPH, L3 and OPAL, whereas DELPHI presently
uses all data at 189 $\GeV$ and above. 
Anomalous
TGCs affect both the total production cross-section and the shape of
the differential cross-section as a function of the polar W$^-$
production angle.  The relative contributions of each helicity state
of the W bosons are also changed, which in turn affects the
distributions of their decay products.  The analyses presented by each
experiment make use of different combinations of each of these
quantities.  In general, however, all analyses use at least the
expected variations of the total production cross-section and the
W$^-$ production angle. Results from $\enw$ and $\nng$ production are
included by some 
experiments.  Single W production is particularly sensitive to \kg,
thus providing information complementary to that from W-pair
production.

The $h$-coupling analyses of ALEPH, DELPHI and L3 use data
collected up to centre-of-mass energies of 209~\GeV. The
OPAL measurements so far use the data at 189~\GeV.  The results of the
$f$-couplings are obtained from the whole data set above the
ZZ-production threshold by all of the experiments.  The experiments
already pre-combine different processes and final states for each of
the couplings.  For the neutral TGCs, the analyses use measurements of
the total cross sections of Z$\gamma$ and ZZ production and the
differential distributions: the $h_i^V$
couplings~\cite{gc_bib:ALEPH-nTGC,gc_bib:DELPHI-nTGC,gc_bib:L3-hTGC,
  gc_bib:OPAL-hTGC} and the $f_i^V$
couplings~\cite{gc_bib:ALEPH-nTGC,gc_bib:DELPHI-nTGC,gc_bib:L3-fTGC,
  gc_bib:OPAL-fTGC} are determined.

The combination of quartic gauge boson couplings associated with
the \ZZgg\ vertex is at present based on analyses of
ALEPH~\cite{gc_bib:ALEPH-QGC}, L3~\cite{gc_bib:L3-QGC} 
and OPAL~\cite{gc_bib:OPAL-QGC}.
The L3 analysis uses data from the \qqgg\
final state all at centre-of-mass energies above the Z resonance, from 130
GeV to 207 GeV. 
Both ALEPH and OPAL analyse the $\nngg$ final state, with ALEPH using data
from centre-of-mass energies ranging from 183 GeV to 209 GeV, and OPAL from
189 GeV to 209 GeV.  

\section{Combination Procedure}
\label{sec:gc_combination}

The combination is based on the individual likelihood functions from the four
LEP experiments.
Each experiment provides the negative log likelihood, $\LL$, as a
function of the coupling parameters to be combined.  
The single-parameter analyses are performed fixing 
all other parameters to their Standard Model values.  The
two-parameter analyses are performed setting the remaining
parameters to their Standard Model values. For the charged TGCs, the
gauge constraints listed in Section~\ref{sec:gc_cTGCs} are always
enforced.

The $\LL$ functions from each experiment include statistical as well
as those systematic uncertainties which are considered as
uncorrelated between experiments.  For both single- and
multi-parameter combinations, the individual $\LL$ functions are
combined.  It is necessary to use the $\LL$ functions directly in the
combination, since in some cases they are not parabolic, and hence it is not
possible to properly combine the results by simply taking weighted
averages of the measurements.

The main contributions to the systematic uncertainties that are
uncorrelated between experiments arise from detector effects,
background in the selected signal samples, limited Monte Carlo
statistics and the fitting method.  Their importance varies for each
experiment and the individual references should be consulted for
details.

In the neutral TGC sector, the systematic uncertainties arising from the
theoretical cross 
section prediction in Z$\gamma$-production ($\simeq 1\%$ in the
$\qq\gamma$- and $\simeq 2\%$ in the $\nng$ channel) are treated as
correlated.
For ZZ production, the uncertainty on the theoretical cross section
prediction is small compared to the statistical accuracy and therefore
is neglected.  Smaller sources of correlated systematic uncertainties,
such as those arising from the LEP beam energy, are for simplicity
treated as uncorrelated.

The combination procedure for neutral TGCs, where the relative
systematic uncertainties are small, is unchanged with respect to the
previous LEP combinations of electroweak gauge boson
couplings~\cite{gc_bib:moriond01,gc_bib:budapest01}. 
The correlated systematic uncertainties in the $h$-coupling analyses
are taken into account by scaling the combined log-likelihood
functions by the squared ratio of the sum of statistical and
uncorrelated systematic uncertainty over the total uncertainty
including all correlated uncertainties.  For the general case of
non-Gaussian probability density functions, this treatment of the
correlated errors is only an approximation; it also neglects
correlations in the systematic uncertainties between the parameters in
multi-parameter analyses.

In the charged TGC sector, systematic uncertainties considered
correlated between the experiments are the theoretical cross
section prediction ($0.5\%$ for W-pair production and $5\%$ for single W
production), hadronisation effects, the final
state interactions, namely Bose-Einstein correlations and colour reconnection,
and the uncertainty in the radiative corrections themselves. 
The latter was the dominant systematic error in our previous combination,
where we used a conservative estimate, the full effect from applying the
$O(\alpha_{em})$ corrections. 
New preliminary analyses on the subject are now available from several LEP
experiments~\cite{gc_bib:ALEPH-cTGC}, based on comparisons 
of fully simulated events using two different leading-pole approximation
schemes (LPA-A and LPA-B)~\cite{gc_bib:LPA_A-B}.
In addition, the availability of comparisons of both generators
incorporating $O(\alpha_{em})$ corrections (RacoonWW and
YFSWW~\cite{common_bib:racoonww,common_bib:yfsww}) makes 
it now possible to perform a more realistic estimation of this effect. 
In general, the TGC shift
measured in the comparison of the two generators is found to be larger than
the effect from the different LPA schemes.
This improved
estimation, whilst still being conservative, reduces the systematic
uncertainty from $O(\alpha_{em})$ corrections by about a
third for $\gz$ and $\lg$ and roughly halves it for $\kg$, compared to the
full $O(\alpha_{em})$ correction applied previously. The application of this
reduced systematic error renders the charged TGC measurements statistics
dominated. 

In case of the charged TGCs, the systematic uncertainties considered
correlated between the experiments amount to 58\% of the combined
statistical and uncorrelated uncertainties for $\lg$ and $\gz$, while for
$\kg$ it is 68\%.  This means that the measurements of $\lg$, $\gz$ and $\kg$
are now clearly limited by statistics.
An improved combination
procedure~\cite{gc_bib:Alcaraz} is used for the charged TGCs.  
This procedure allows the combination of statistical and correlated
systematic uncertainties, independently of the analysis method chosen by
the individual experiments. 

The combination of charged TGCs uses the likelihood curves and
correlated systematic errors submitted by each of the four experiments. 
The procedure is based on the introduction of an additional free parameter to
take into account the systematic uncertainties, which are treated as shifts on
the fitted TGC value, and are assumed to have a Gaussian distribution. 
A simultaneous minimisation of both parameters (TGC
and systematic error) is performed to the log-likelihood function. 

In detail, the combination proceeds in the following way: the set of
measurements from the LEP experiments
ALEPH, DELPHI, OPAL and L3 is given with statistical plus uncorrelated
systematic uncertainties in terms of likelihood curves:  
$-\log{\mathcal L}^A_{stat}(x)$,
$-\log{\mathcal L}^D_{stat}(x)$ 
$-\log{\mathcal L}^L_{stat}(x)$ 
and $-\log{\mathcal L}^O_{stat}(x)$, 
respectively, where $x$ is the coupling parameter in question. 
Also given are the shifts for each of the five totally correlated sources
of uncertainty mentioned above; each source $S$ is 
leading to systematic errors $\sigma^S_A$, $\sigma^S_D$, $\sigma^S_L$ and
$\sigma^S_O$.

Additional parameters $\Delta^S$ are included in order to take into 
account a Gaussian distribution for each of the systematic uncertainties.
The procedure then consists in minimising the function:
\noindent
\begin{eqnarray}
-\log {\mathcal L}_{total} = 
\sum_{E=A,D,L,O} \log {\mathcal L}^E_{stat} 
(x-\sum_{S=DPA,\sigma_{WW},HAD,BE,CR}(\sigma^S_E \Delta^S))
 + \sum_{S} {\displaystyle \frac{(\Delta^S)^{2}}{2}} \\ \nonumber
\end{eqnarray}

\noindent
where $x$ and $\Delta_S$ are the free parameters, and the sums run over the
four experiments and the five systematic errors.
The resulting uncertainty on $x$ will take into account all sources 
of uncertainty, yielding a measurement of the coupling with the error
representing statistical and systematic sources.
The projection of the minima of the log-likelihood as a function of $x$ 
gives the combined log-likelihood curve including statistical and
systematic uncertainties. 
The advantage over the scaling method used previously is that it
treats systematic uncertainties that are correlated between the experiments
correctly, while not forcing the averaging of these systematic
uncertainties into one global LEP systematics scaling factor. In other words,
the (statistical) precision of each experiment now gets reduced by its own
correlated systematic errors, instead of an averaged LEP systematic
error. 
The method has been cross-checked against the scaling method, and was found
to give comparable results. 
The inclusion of
the systematic uncertainties lead to small differences as expected by the
improved treatment of correlated systematic errors, a similar behaviour as
seen in Monte Carlo comparisons of these two combinations methods
~\cite{gc_bib:renaud}. Furthermore, it was shown that the minimisation-based
combination method used for the charged TGCs agrees with the method
based on optimal observables, where systematic effects are included directly
in the mean values of the optimal observables (see~\cite{gc_bib:renaud}), 
for any realistic ratio of statistical and systematic uncertainties. 
Further details on the improved combination method can be found
in~\cite{gc_bib:Alcaraz}. 

In the combination of the QGCs, the influence of correlated systematic
uncertainties is considered negligible compared to the statistical error,
arising from the small number of selected events. Therefore, the QGCs are
combined by adding the log-likelihood curves from the single experiments. 

For all single- and multi-parameter results quoted in numerical form,
the one standard deviation uncertainties (68\% confidence level) are
obtained by taking the coupling values for which $\Delta\LL=+0.5$
above the minimum.  The 95\% confidence level (C.L.)  limits are given
by the coupling values for which $\Delta\LL=+1.92$ above the minimum.
Note that in the case of the neutral TGCs, double minima structures
appear in the negative log-likelihood curves.  For multi-parameter
analyses, the two dimensional 68\%~C.L.~contour curves for any pair of
couplings are obtained by requiring $\Delta\LL=+1.15$, while for the
95\% C.L.~contour curves $\Delta\LL=+3.0$ is required.  Since the
results on the different parameters and parameter sets are obtained
from the same data sets, they cannot be combined.

\section{Results}

We present results from the four LEP experiments on the various
electroweak gauge boson couplings, and their combination.
The charged TGC combination has been updated with the inclusion of recent
results from ALEPH, L3 and OPAL.
The neutral TGC results include an update of the $f_i^V$ combinations,
whilst the $h_i^V$ combinations remain unchanged since
our last note~\cite{gc_bib:budapest01}.  The results
quoted for each individual experiment are calculated using the methods
described in Section~\ref{sec:gc_combination}.  Therefore they may differ
slightly from those reported in the individual references, as the experiments
in general use other methods to combine the data from different channels, and
to include systematic uncertainties. 
In particular for the charged couplings, experiments using a
combination method based on optimal observables (ALEPH, OPAL) 
obtain results with small differences
compared to the values given by our combination technique. These small
differences have been studied in Monte Carlo tests and are well
understood~\cite{gc_bib:renaud}.  
For the $h$-coupling result from OPAL and DELPHI, a slightly
modified estimate of the systematic uncertainty due to the theoretical
cross section prediction is responsible for slightly different limits
compared to the published results.

\subsection{Charged Triple Gauge Boson Couplings}

The individual analyses and results of the experiments for the charged couplings are described in~\cite{gc_bib:ALEPH-cTGC,gc_bib:DELPHI-cTGC,
gc_bib:L3-cTGC,gc_bib:OPAL-cTGC3}.

\subsubsection*{Single-Parameter Analyses}
\begin{figure}[htbp]
\begin{center}
\includegraphics[width=\linewidth]{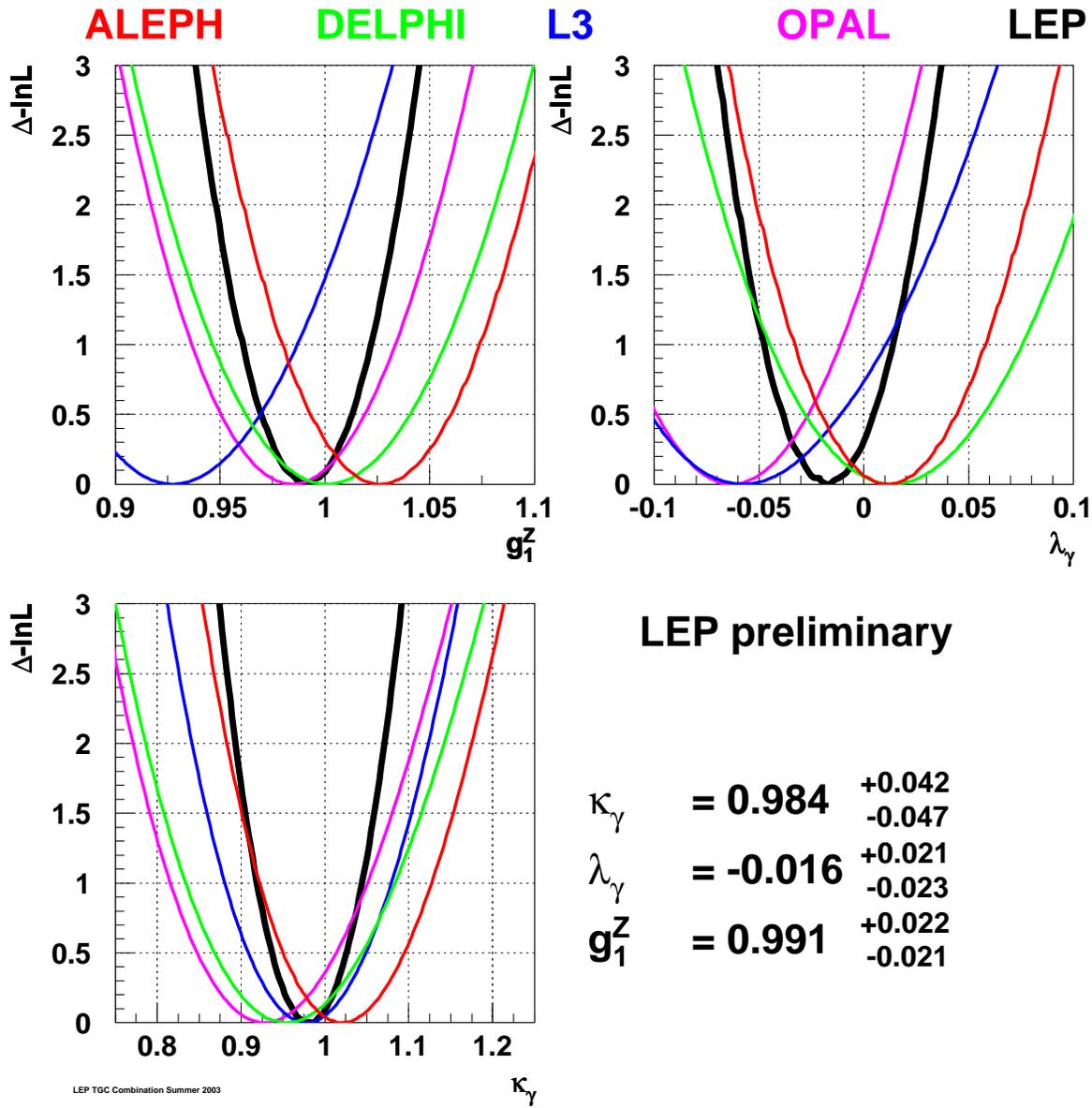}
\caption[]{
  The $\LL$ curves of the four experiments (thin lines) and the LEP
  combined curve (thick line) for the three charged TGCs $\gz$,
  $\kg$ and $\lg$.  In each case, the minimal value is subtracted.  }
\label{fig:cTGC-1}
\end{center}
\end{figure}

The results of single-parameter fits from each experiment are shown in
Table~\ref{tab:cTGC-1-ADLO}, where the errors include both statistical
and systematic effects. The individual $\LL$ curves and their sum are shown in
Figure~\ref{fig:cTGC-1}.  The results of the combination are given in
Table~\ref{tab:cTGC-1-LEP}. A list of the systematic errors treated as fully
correlated between the LEP experiments, and their shift on the combined fit
result are given in Table ~\ref{tab:cTGC-syst}.

\subsubsection*{Two-Parameter Analyses}
\begin{figure}[htbp]
\begin{center}
\includegraphics[width=\linewidth]{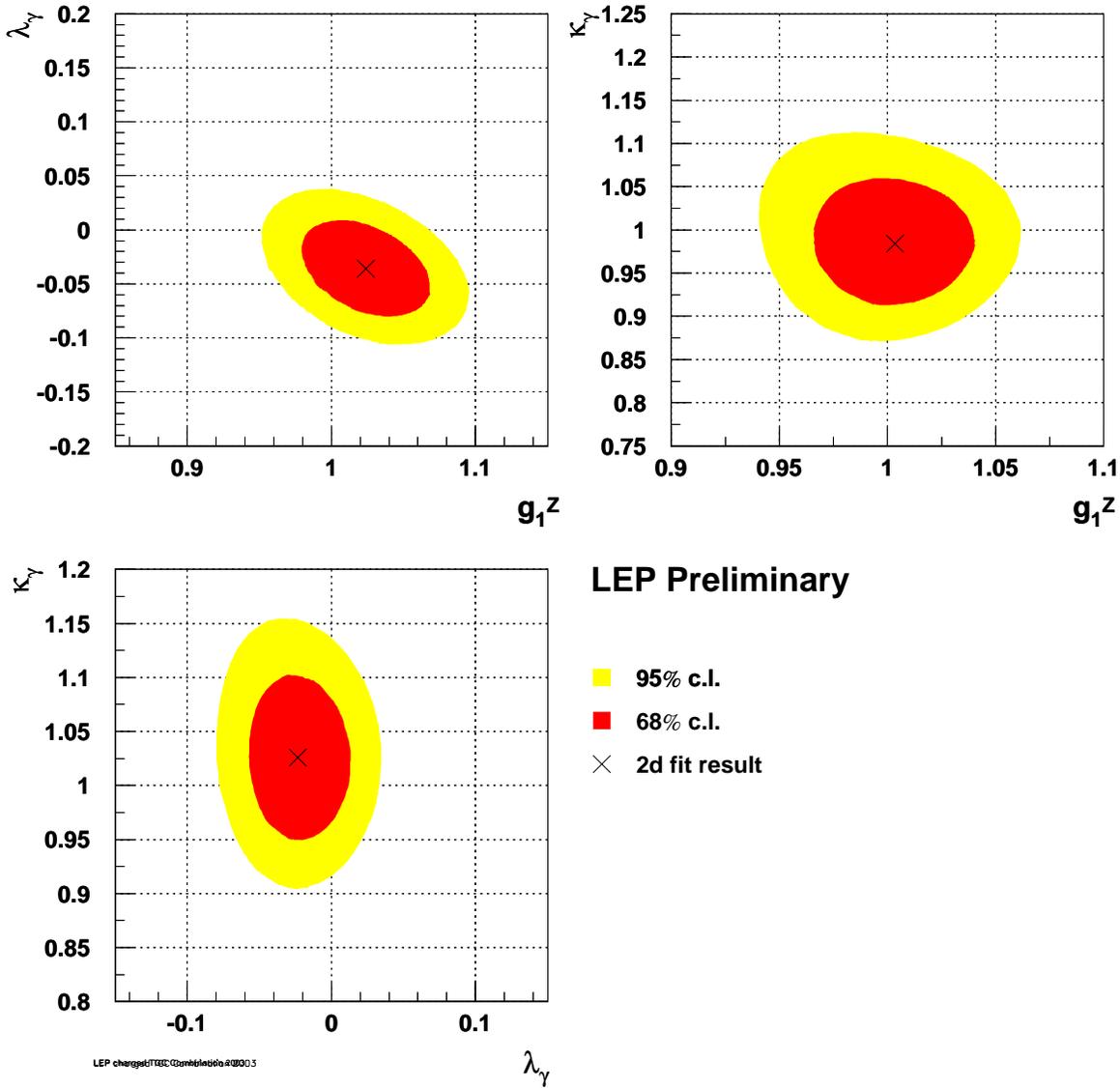}
\caption[]{
The 68\% and 95\% confidence level contours for the three two-parameter fits
to the charged TGCs $\gz$-$\lg$, $\gz$-$\kg$ and $\lg$-$\kg$. 
The fitted coupling
value is indicated with a cross; the Standard Model value for each fit is in
the centre of the grid. The contours include the contribution from systematic
uncertainties.}
 \label{fig:cTGC-2}
\end{center}
\end{figure}

Contours at 68\% and 95\% confidence level for the combined two-parameter
fits are shown in Figure~\ref{fig:cTGC-2}. The numerical results of the
combination are given in Table~\ref{tab:cTGC-2D-LEP}. The errors include both statistical and systematic effects. 

\begin{table}[htbp]
\begin{center}
\renewcommand{\arraystretch}{1.3}
\begin{tabular}{|l||r|r|r|r|} 
\hline
Parameter  & ALEPH   & DELPHI  &  L3   & OPAL  \\
\hline
\hline
\gz       & $1.026\apm{0.034}{0.033}$ & $1.002\apm{0.038}{0.040}$ 
           & $0.928\apm{0.042}{0.041}$ & $0.985\apm{0.035}{0.034}$ \\ 
\hline
\kg       & $1.022\apm{0.073}{0.072}$ & $0.955\apm{0.090}{0.086}$  
           & $0.922\apm{0.071}{0.069}$ & $0.929\apm{0.085}{0.081}$ \\  
\hline
\lg        & $0.012\apm{0.033}{0.032}$ & $0.014\apm{0.044}{0.042}$  
           & $-0.058\apm{0.047}{0.044}$ & $-0.063\apm{0.036}{0.036}$ \\
\hline
\end{tabular}
\caption[]{The measured central values and one standard deviation errors
  obtained by the four LEP experiments.  In each case the parameter
  listed is varied while the remaining two are fixed to their Standard Model
  values. Both
  statistical and systematic errors are included. The values given here
  differ slightly from the ones quoted in the individual contributions from
  the four LEP experiments, as a different combination method is used. See
  text in section \ref{sec:gc_combination} for details. }
\label{tab:cTGC-1-ADLO}
\end{center}
\end{table}

\begin{table}[htbp]
\begin{center}
\renewcommand{\arraystretch}{1.3}
\begin{tabular}{|l||r|c|} 
\hline
Parameter  & 68\% C.L.   & 95\% C.L.      \\
\hline
\hline
$\gz$     & $0.991\apm{0.022}{0.021} $  & [$0.949,~~1.034$]  \\ 
\hline
$\kg$     & $0.984\apm{0.042}{0.047}$  & [$0.895,~~1.069$]  \\ 
\hline
$\lg$     & $-0.016\apm{0.021}{0.023}$  & [$-0.059,~~0.026$]  \\ 
\hline
\end{tabular}
\caption[]{ The combined 68\% C.L. errors and 95\% C.L. intervals
  obtained combining the results from the four LEP experiments.  In
  each case the parameter listed is varied while the other two are
  fixed to their Standard Model values.  Both statistical and systematic
  errors are included.  }
 \label{tab:cTGC-1-LEP}
\end{center}
\end{table}

\begin{table}[htbp]
\begin{center}
\renewcommand{\arraystretch}{1.3}
\begin{tabular}{|l||r|r|r|} 
\hline
Source  & \gz  & \lg   & \kg  \\
\hline
\hline
$O(\alpha_{em})$ correction  & 0.010 & 0.010  & 0.020 \\ 
$\sigma_{WW}$ prediction   & 0.003  & 0.005 & 0.014 \\ 
Hadronisation   & 0.004 & 0.002 & 0.004 \\ 
Bose-Einstein Correlation   & 0.005  & 0.004  & 0.009 \\ 
Colour Reconnection   & 0.005 & 0.004 & 0.010 \\
$\sigma_{single W}$ prediction  & - & - &  0.011 \\ 
\hline
\end{tabular}
\caption[]{The systematic uncertainties considered correlated between the LEP
  experiments in the charged TGC combination and their effect on the combined
  fit results.}
\label{tab:cTGC-syst}
\end{center}
\end{table}

\begin{table}[htbp]
\begin{center}
\begin{tabular}{|l||r|r|rr|} \hline
Parameter   & 68\% C.L.  & 95\% C.L.   & \multicolumn{2}{|c|}{Correlations}  \\
\hline \hline
\gz       & $1.004\apm{0.024}{0.025}$ & [$+0.954,~~+1.050$]
  & 1.00 & +0.11 \\  
\kg       & $0.984\apm{0.049}{0.049}$ & [$+0.894,~~+1.084$]  
 & +0.11 & 1.00 \\ 
\hline
\gz       & $1.024\apm{0.029}{0.029}$ & [$+0.966,~~+1.081$] & 1.00
  & -0.40  \\  
\lg       & $-0.036\apm{0.029}{0.029}$ & [$-0.093,~~+0.022$] 
 & -0.40  & 1.00 \\ 
\hline
\kg        & $1.026\apm{0.048}{0.051}$ & [$+0.928,~~+1.127$]
 & 1.00  & +0.21 \\ 
\lg        & $-0.024\apm{0.025}{0.021}$ & [$-0.068,~~+0.023$] 
  & +0.21 & 1.00 \\ 
\hline
\end{tabular}
\end{center}
\caption{ The measured central values, one standard deviation errors and
  limits at 95\% confidence level, 
    obtained by combining the four LEP experiments for the 
    two-parameter fits of the charged TGC parameters.
    Since the shape of the log-likelihood
  is not parabolic, there is some ambiguity in the definition of the
  correlation coefficients and the values quoted here are approximate.
    The listed parameters are varied while the
    remaining one is fixed to its Standard Model value.
    Both statistical and systematic errors are included.
  }
\label{tab:cTGC-2D-LEP}
\end{table}

\clearpage

\clearpage

\subsection{Neutral Triple Gauge Boson Couplings in Z\boldmath$\gamma$ 
Production}

The individual analyses and results of the experiments for the $h$-couplings 
are described in~\cite{gc_bib:ALEPH-nTGC,gc_bib:DELPHI-nTGC,
gc_bib:L3-hTGC,gc_bib:OPAL-hTGC}.

\subsubsection*{Single-Parameter Analyses}
\begin{figure}[p]
\begin{center}
\includegraphics[width=\linewidth]{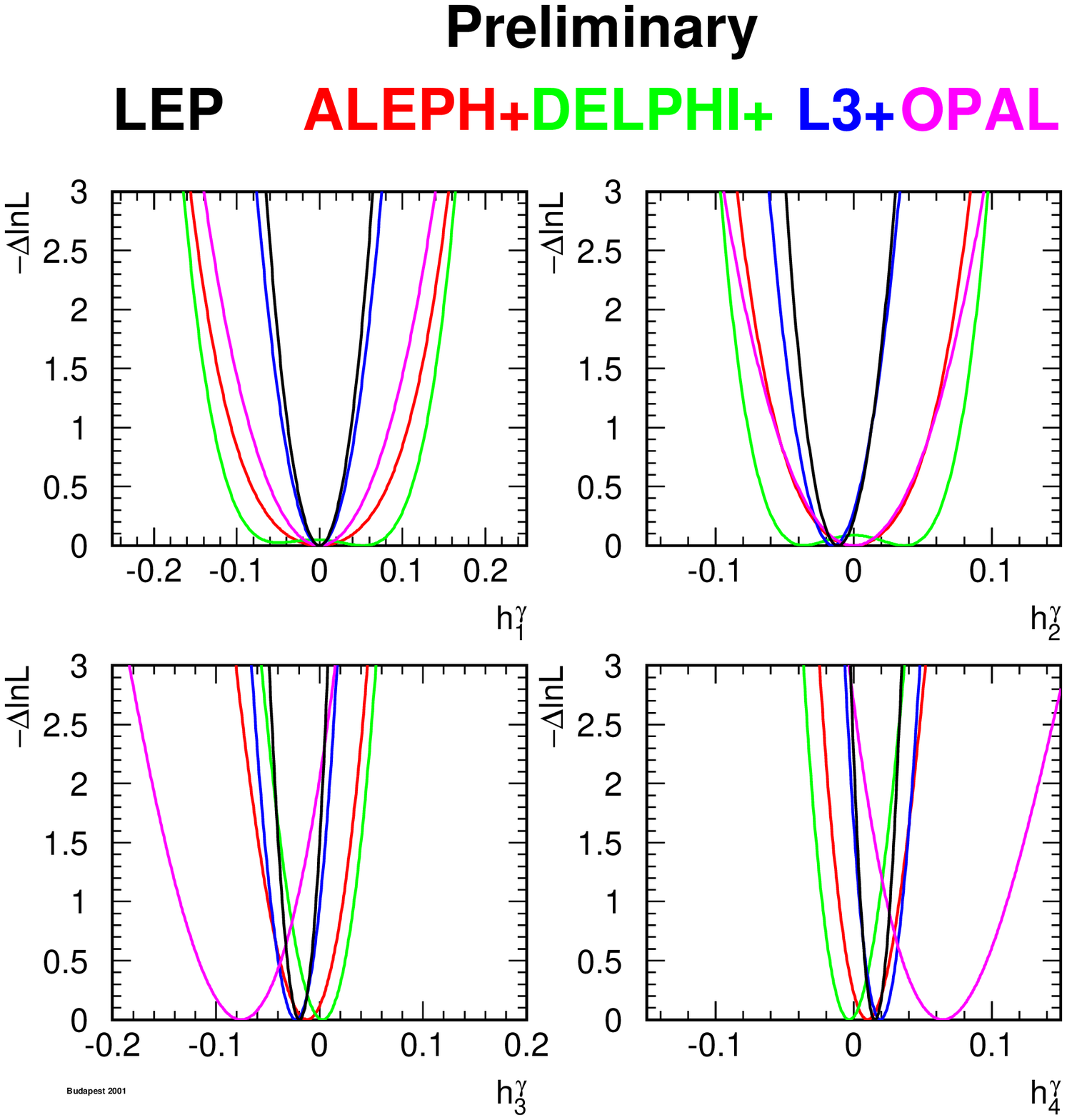}
\caption[]{
  The $\LL$ curves of the four experiments, and the LEP combined curve
  for the four neutral TGCs $h_i^\gamma,~i=1,2,3,4$. In each case, the
  minimal value is subtracted.  }
\label{fig:gc_hgTGC-1}
\end{center}
\end{figure}

\begin{figure}[p]
\begin{center}
\includegraphics[width=\linewidth]{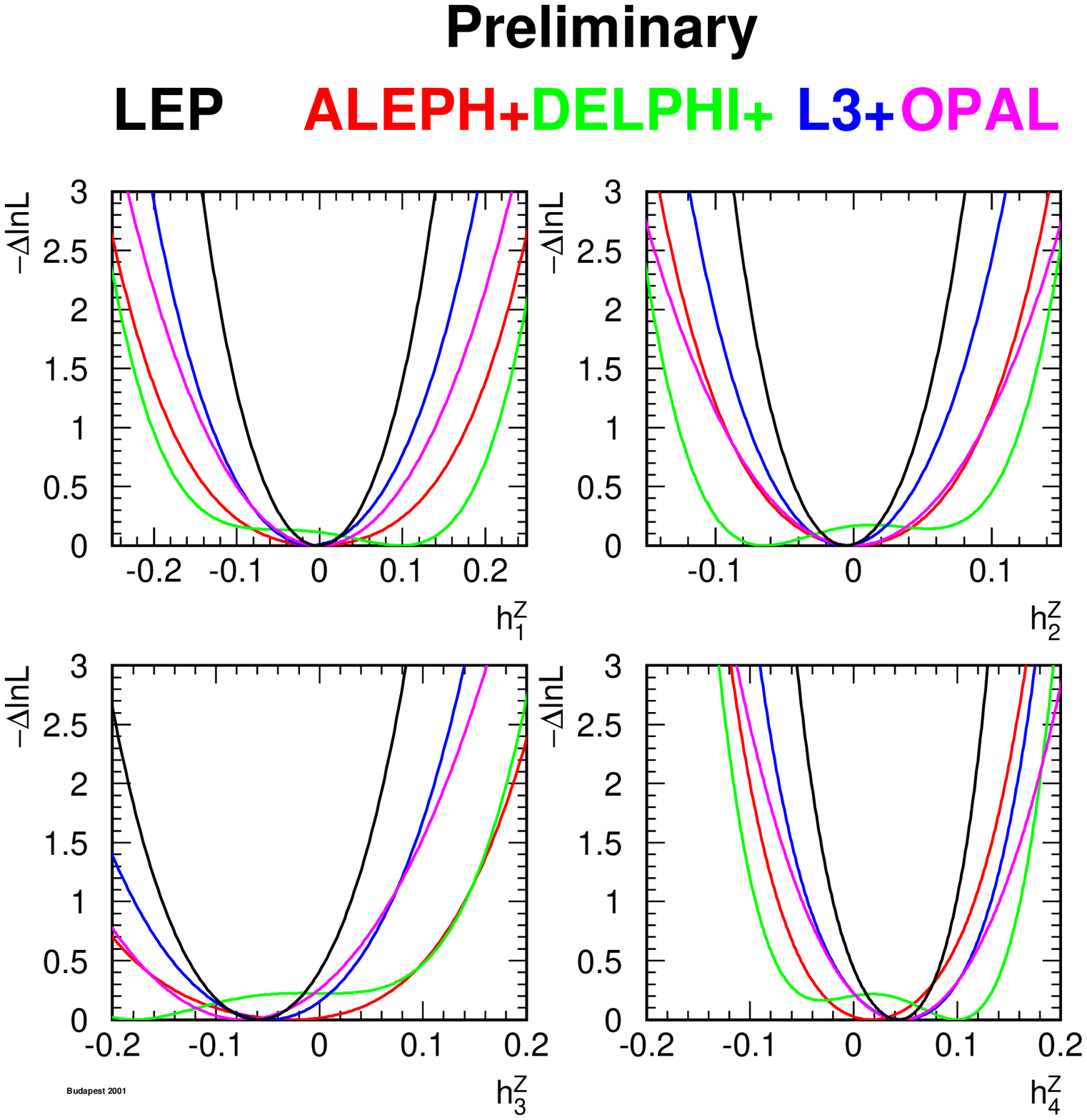}
\caption[]{
  The $\LL$ curves of the four experiments, and the LEP combined curve
  for the four neutral TGCs $h_i^Z,~i=1,2,3,4$.  In each case, the
  minimal value is subtracted.  }
\label{fig:gc_hzTGC-1}
\end{center}
\end{figure}

The results for each experiment are shown in
Table~\ref{tab:gc_hTGC-1-ADLO}, where the errors include both
statistical and systematic uncertainties.  The individual $\LL$ curves
and their sum are shown in Figures~\ref{fig:gc_hgTGC-1}
and~\ref{fig:gc_hzTGC-1}.  The results of the combination are given in
Table~\ref{tab:gc_hTGC-1-LEP}.  From
Figures~\ref{fig:gc_hgTGC-1} and \ref{fig:gc_hzTGC-1} it is clear that the
sensitivity of the L3 analysis~\cite{gc_bib:L3-hTGC} is the highest
amongst the LEP experiments. This is partially due to the use of a larger
phase space region, which increases the statistics by about a factor two,
and partially due to additional information from using an optimal-observable
technique.

\subsubsection*{Two-Parameter Analyses}

\begin{figure}[p]
\begin{center}
\includegraphics[width=0.49\linewidth]{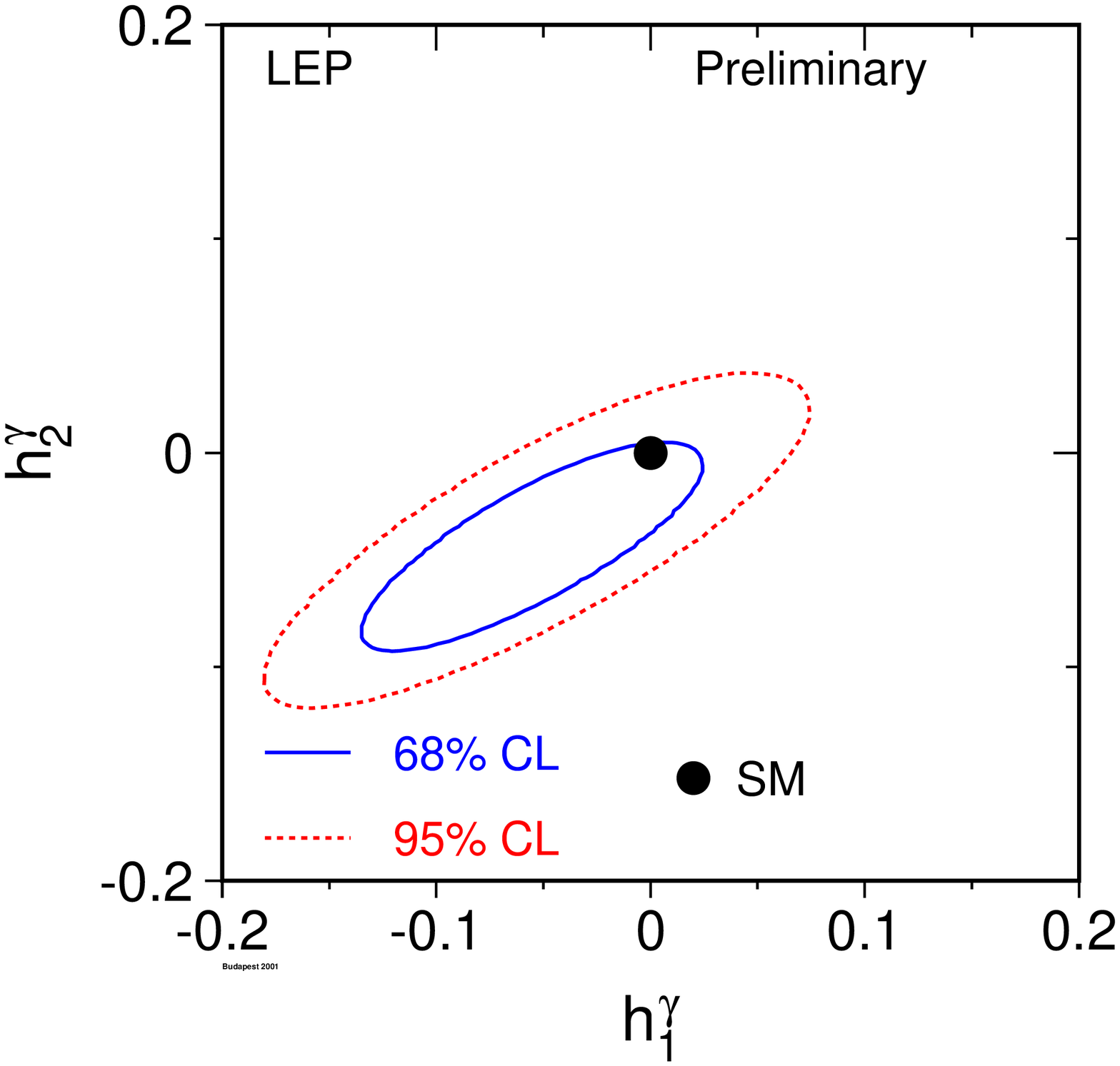}\hfill
\includegraphics[width=0.49\linewidth]{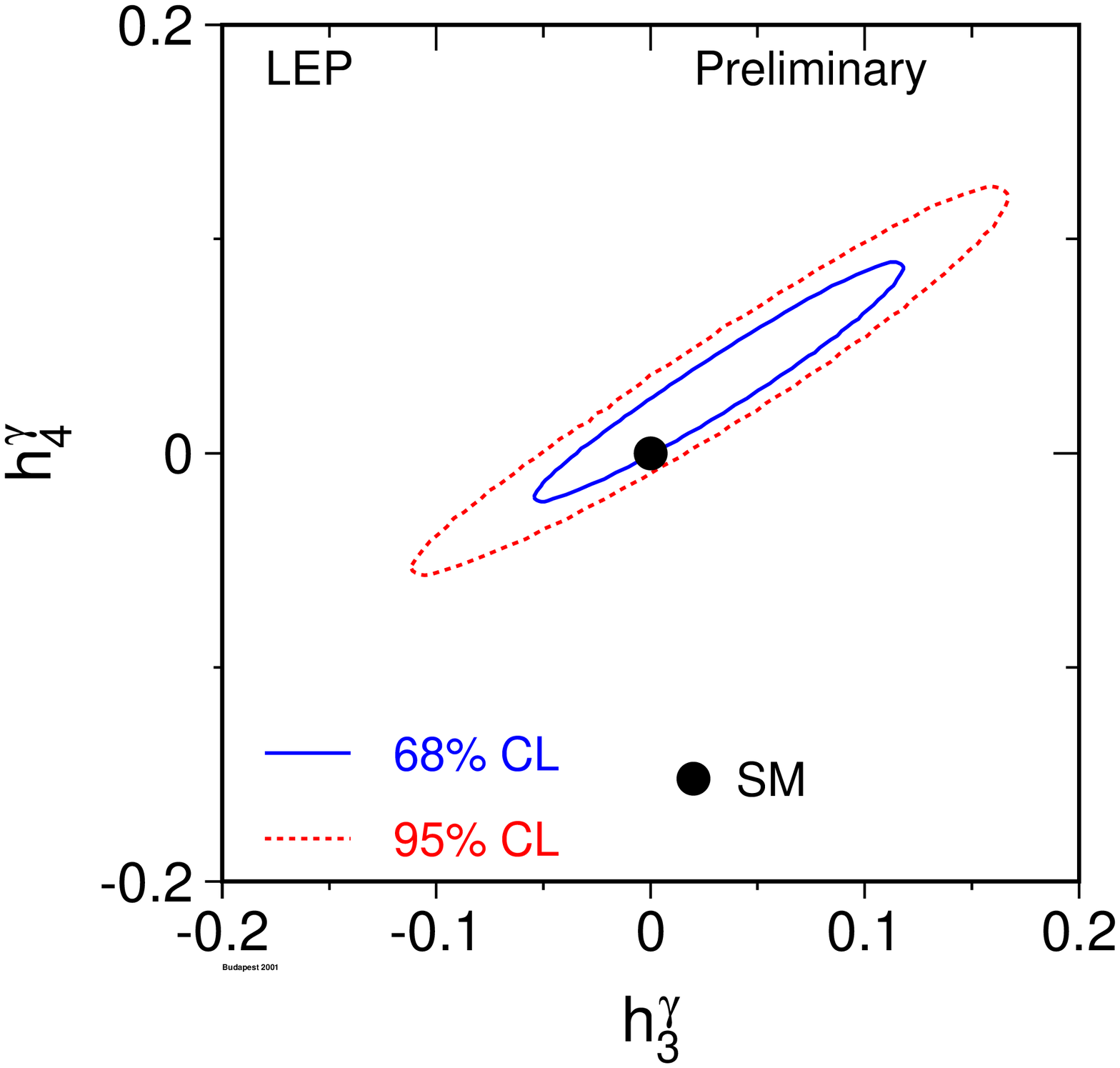}\\
\includegraphics[width=0.49\linewidth]{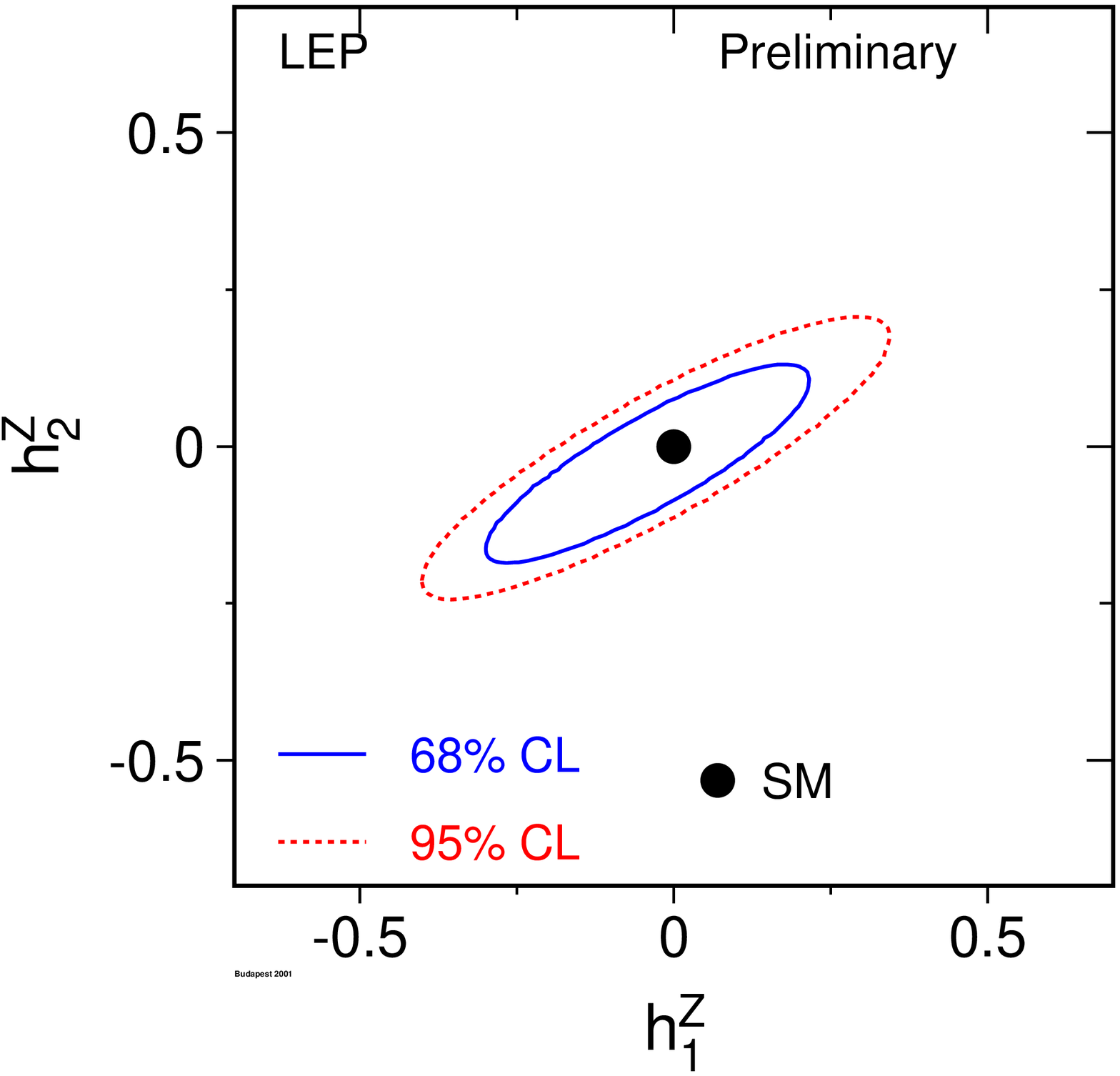}\hfill
\includegraphics[width=0.49\linewidth]{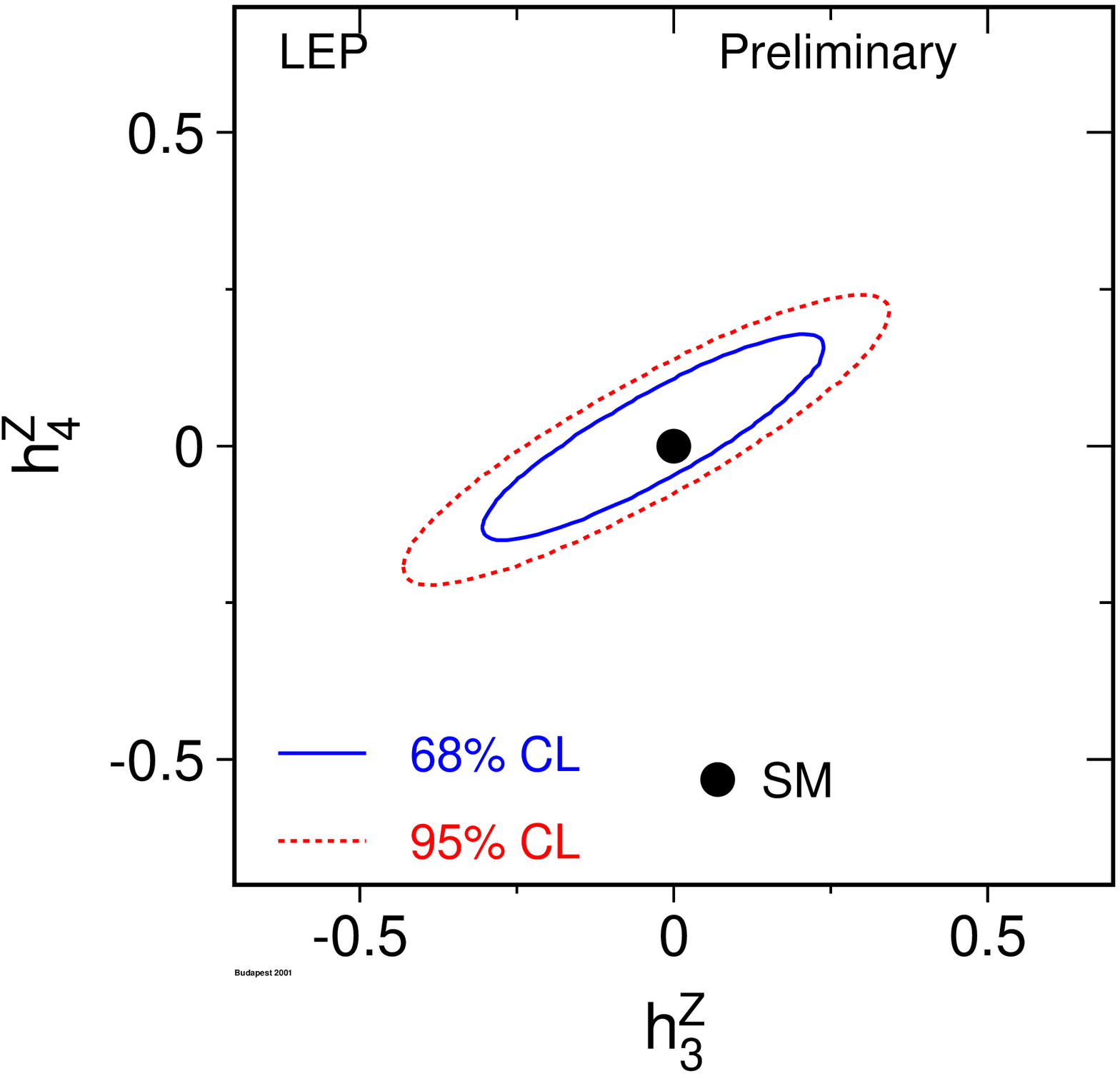}
\caption[]{
  Contour curves of 68\% C.L. and 95\% C.L. in the planes
  $(h_1^\gamma,h_2^\gamma)$, $(h_3^\gamma,h_4^\gamma)$,
  $(h_1^Z,h_2^Z)$ and $(h_3^Z,h_4^Z)$ showing the LEP combined result.
  }
\label{fig:gc_hTGC-2}
\end{center}
\end{figure}

The results for each experiment are shown in
Table~\ref{tab:gc_hTGC-2-ADLO}, where the errors include both
statistical and systematic uncertainties.  The 68\% C.L. and 95\% C.L.
contour curves resulting from the combinations of the two-dimensional
likelihood curves are shown in Figure~\ref{fig:gc_hTGC-2}.  The LEP
average values are given in Table~\ref{tab:gc_hTGC-2-LEP}.

\begin{table}[htbp]
\begin{center}
\renewcommand{\arraystretch}{1.3}
\begin{tabular}{|l||r|r|r|r|} 
\hline
Parameter  & ALEPH & DELPHI  &  L3  & OPAL \\
\hline
\hline
$h_1^\gamma$ & [$-0.14,~~+0.14$]  & [$-0.15,~~+0.15$]   & [$-0.06,~~+0.06$]   & [$-0.13,~~+0.13$] \\
\hline                             
$h_2^\gamma$ & [$-0.07,~~+0.07$]  & [$-0.09,~~+0.09$]   & [$-0.053,~~+0.024$] & [$-0.089,~~+0.089$] \\
\hline                             
$h_3^\gamma$ & [$-0.069,~~+0.037$]& [$-0.047,~~+0.047$] & [$-0.062,~~-0.014$] & [$-0.16,~~+0.00$] \\
\hline                             
$h_4^\gamma$ & [$-0.020,~~+0.045$]& [$-0.032,~~+0.030$] & [$-0.004,~~+0.045$] & [$+0.01,~~+0.13$] \\
\hline                             
$h_1^Z$      & [$-0.23,~~+0.23$]  & [$-0.24,~~+0.25$]   & [$-0.17,~~+0.16$]   & [$-0.22,~~+0.22$] \\
\hline                             
$h_2^Z$      & [$-0.12,~~+0.12$]  & [$-0.14,~~+0.14$]   & [$-0.10,~~+0.09$]   & [$-0.15,~~+0.15$] \\
\hline                             
$h_3^Z$      & [$-0.28,~~+0.19$]  & [$-0.32,~~+0.18$]   & [$-0.23,~~+0.11$]   & [$-0.29,~~+0.14$] \\
\hline                             
$h_4^Z$      & [$-0.10,~~+0.15$]  & [$-0.12,~~+0.18$]   & [$-0.08,~~+0.16$]   & [$-0.09,~~+0.19$] \\
\hline
\end{tabular}
\caption[]{The 95\% C.L. intervals ($\Delta\LL=1.92$) measured by
  the ALEPH, DELPHI, L3 and OPAL.  In each case the parameter listed is varied
  while the remaining ones are fixed to their Standard Model values.
  Both statistical and systematic uncertainties are included.  }
\label{tab:gc_hTGC-1-ADLO}
\end{center}
\end{table}

\begin{table}[htbp]
\begin{center}
\renewcommand{\arraystretch}{1.3}
\begin{tabular}{|l||c|} 
\hline
Parameter     & 95\% C.L.      \\
\hline
\hline
$h_1^\gamma$  & [$-0.056,~~+0.055$]  \\ 
\hline
$h_2^\gamma$  & [$-0.045,~~+0.025$]  \\ 
\hline
$h_3^\gamma$  & [$-0.049,~~-0.008$]  \\ 
\hline
$h_4^\gamma$  & [$-0.002,~~+0.034$]  \\ 
\hline
$h_1^Z$       & [$-0.13,~~+0.13$]  \\ 
\hline
$h_2^Z$       & [$-0.078,~~+0.071$]  \\ 
\hline
$h_3^Z$       & [$-0.20,~~+0.07$]  \\ 
\hline
$h_4^Z$       & [$-0.05,~~+0.12$]  \\ 
\hline
\end{tabular}
\caption[]{ The 95\% C.L. intervals ($\Delta\LL=1.92$) obtained
  combining the results from the four experiments.  In each case the
  parameter listed is varied while the remaining ones are fixed to
  their Standard Model values.  Both statistical and systematic
  uncertainties are included.  }
 \label{tab:gc_hTGC-1-LEP}
\end{center}
\end{table}

\begin{table}[htbp]
\begin{center}
\renewcommand{\arraystretch}{1.3}
\begin{tabular}{|l||r|r|r|} 
\hline
Parameter  & ALEPH & DELPHI  &  L3  \\
\hline
\hline
$h_1^\gamma$ & [$-0.32,~~+0.32$] & [$-0.28,~~+0.28$] & [$-0.17,~~+0.04$] \\
$h_2^\gamma$ & [$-0.18,~~+0.18$] & [$-0.17,~~+0.18$] & [$-0.12,~~+0.02$] \\
\hline                                               
$h_3^\gamma$ & [$-0.17,~~+0.38$] & [$-0.48,~~+0.20$] & [$-0.09,~~+0.13$] \\
$h_4^\gamma$ & [$-0.08,~~+0.29$] & [$-0.08,~~+0.15$] & [$-0.04,~~+0.11$] \\
\hline                                               
$h_1^Z$      & [$-0.54,~~+0.54$] & [$-0.45,~~+0.46$] & [$-0.48,~~+0.33$] \\
$h_2^Z$      & [$-0.29,~~+0.30$] & [$-0.29,~~+0.29$] & [$-0.30,~~+0.22$] \\
\hline
$h_3^Z$      & [$-0.58,~~+0.52$] & [$-0.57,~~+0.38$] & [$-0.43,~~+0.39$] \\
$h_4^Z$      & [$-0.29,~~+0.31$] & [$-0.31,~~+0.28$] & [$-0.23,~~+0.28$] \\
\hline
\end{tabular}
\caption[]{The 95\% C.L. intervals ($\Delta\LL=1.92$) measured  by
  ALEPH, DELPHI and L3.  In each case the two parameters listed are varied
  while the remaining ones are fixed to their Standard Model values.
  Both statistical and systematic uncertainties are included.  }
\label{tab:gc_hTGC-2-ADLO}
\end{center}
\end{table}

\begin{table}[htbp]
\begin{center}
\renewcommand{\arraystretch}{1.3}
\begin{tabular}{|l||c|rr|} 
\hline
Parameter  & 95\% C.L. & \multicolumn{2}{|c|}{Correlations} \\
\hline
\hline
$h_1^\gamma$  & [$-0.16,~~+0.05$]    & $ 1.00$ & $+0.79$ \\ 
$h_2^\gamma$  & [$-0.11,~~+0.02$]    & $+0.79$ & $ 1.00$ \\ 
\hline
$h_3^\gamma$  & [$-0.08,~~+0.14$]    & $ 1.00$ & $+0.97$ \\ 
$h_4^\gamma$  & [$-0.04,~~+0.11$]    & $+0.97$ & $ 1.00$ \\ 
\hline
$h_1^Z$       & [$-0.35,~~+0.28$]    & $ 1.00$ & $+0.77$ \\ 
$h_2^Z$       & [$-0.21,~~+0.17$]    & $+0.77$ & $ 1.00$ \\ 
\hline
$h_3^Z$       & [$-0.37,~~+0.29$]    & $ 1.00$ & $+0.76$ \\ 
$h_4^Z$       & [$-0.19,~~+0.21$]    & $+0.76$ & $ 1.00$ \\ 
\hline
\end{tabular}
\caption[]{ The 95\% C.L. intervals ($\Delta\LL=1.92$) obtained
  combining the results from ALEPH, DELPHI and L3.  In each case the two
  parameters listed are varied while the remaining ones are fixed to
  their Standard Model values.  Both statistical and systematic
  uncertainties are included.  Since the shape of the log-likelihood
  is not parabolic, there is some ambiguity in the definition of the
  correlation coefficients and the values quoted here are approximate.
  }
 \label{tab:gc_hTGC-2-LEP}
\end{center}
\end{table}

\clearpage

\clearpage

\subsection{Neutral Triple Gauge Boson Couplings in ZZ Production}

The individual analyses and results of the experiments for the $f$-couplings 
are described in~\cite{gc_bib:ALEPH-nTGC,gc_bib:DELPHI-nTGC,gc_bib:L3-fTGC,gc_bib:OPAL-fTGC}.

\subsubsection*{Single-Parameter Analyses}

\begin{figure}[p]
\begin{center}
\includegraphics[width=\linewidth]{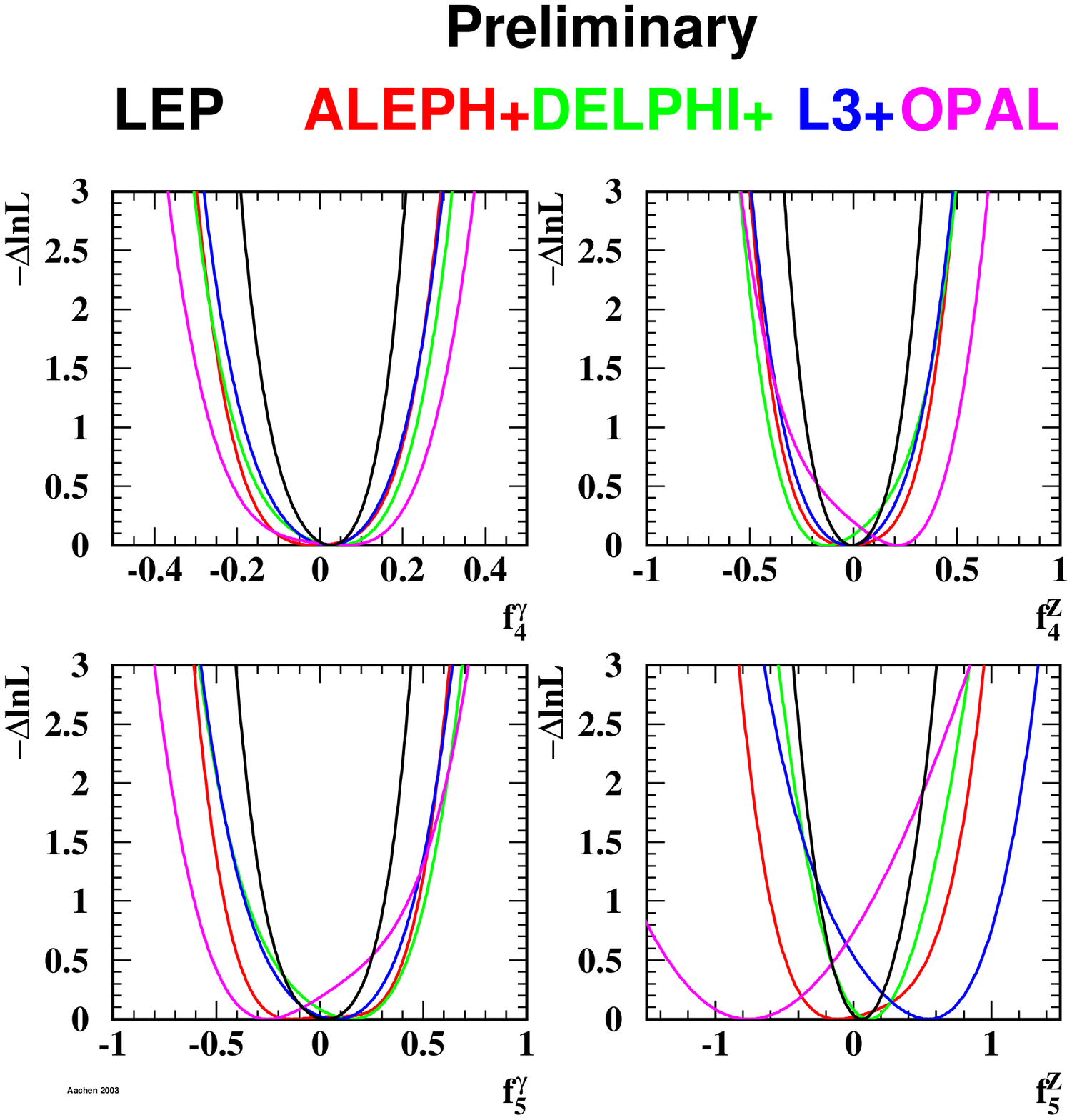}
\caption[]{
  The $\LL$ curves of the four experiments, and the LEP combined curve
  for the four neutral TGCs $f_i^V,~V=\gamma,Z,~i=4,5$.  In each case,
  the minimal value is subtracted.  }
\label{fig:gc_fTGC-1}
\end{center}
\end{figure}

The results for each experiment are shown in
Table~\ref{tab:gc_fTGC-1-ADLO}, where the errors include both statistical
and systematic uncertainties.  The individual $\LL$ curves and their sum are
shown in Figure~\ref{fig:gc_fTGC-1}.  The results of the combination are
given in Table~\ref{tab:gc_fTGC-1-LEP}.

\subsubsection*{Two-Parameter Analyses}

\begin{figure}[p]
\begin{center}
\includegraphics[width=0.55\linewidth]{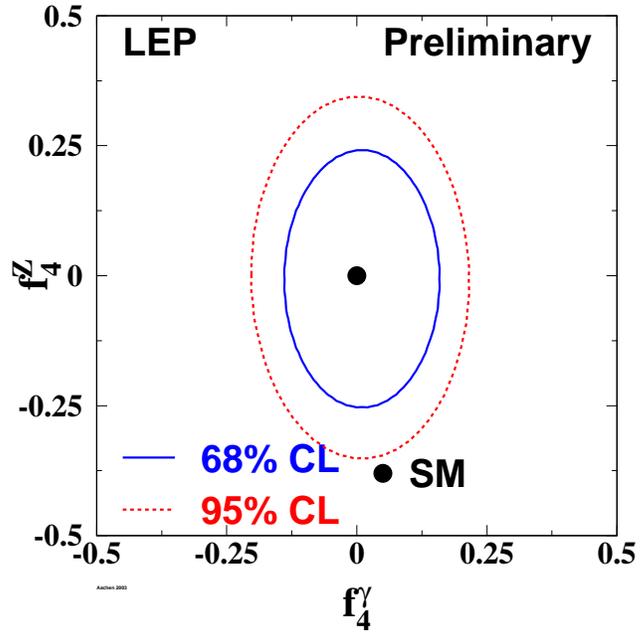}\\
\includegraphics[width=0.55\linewidth]{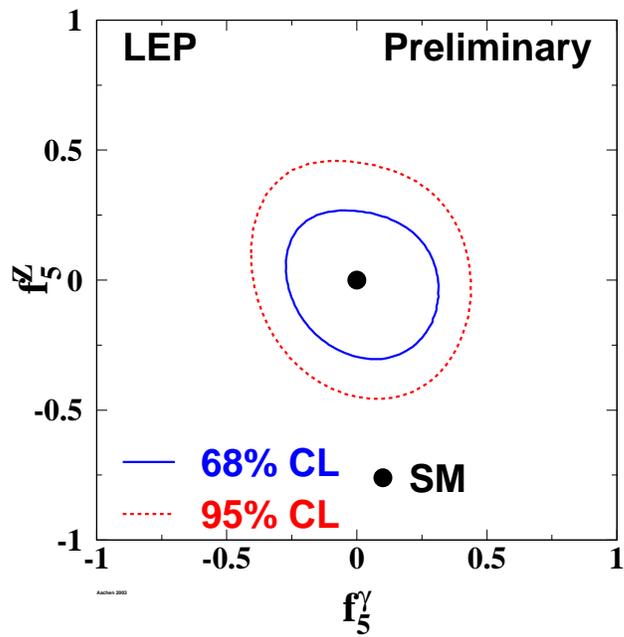}
\caption[]{
  Contour curves of 68\% C.L. and 95\% C.L. in the plane
  $(f_4^\gamma,f_4^Z)$ and $(f_5^\gamma,f_5^Z)$ showing the LEP
  combined result.  }
\label{fig:gc_fTGC-2}
\end{center}
\end{figure}

The results from each experiment are shown in
Table~\ref{tab:gc_fTGC-2-ADLO}, where the errors include both
statistical and systematic uncertainties.  The 68\% C.L. and 95\% C.L.
contour curves resulting from the combinations of the two-dimensional
likelihood curves are shown in Figure~\ref{fig:gc_fTGC-2}.  The LEP
average values are given in Table~\ref{tab:gc_fTGC-2-LEP}.

\begin{table}[htbp]
\begin{center}
\renewcommand{\arraystretch}{1.3}
\begin{tabular}{|l||r|r|r|r|} 
\hline
Parameter  & ALEPH & DELPHI  &  L3   & OPAL  \\
\hline
\hline
$f_4^\gamma$ & [$-0.26,~~+0.26$] & [$-0.26,~~+0.28$] & [$-0.28,~~+0.28$] & [$-0.32,~~+0.33$] \\ 
\hline
$f_4^Z$      & [$-0.44,~~+0.43$] & [$-0.49,~~+0.42$] & [$-0.48,~~+0.46$] & [$-0.45,~~+0.58$] \\ 
\hline
$f_5^\gamma$ & [$-0.54,~~+0.56$] & [$-0.48,~~+0.61$] & [$-0.39,~~+0.47$] & [$-0.71,~~+0.59$] \\ 
\hline
$f_5^Z$      & [$-0.73,~~+0.83$] & [$-0.42,~~+0.69$] & [$-0.35,~~+1.03$] & [$-0.94,~~+0.25$] \\ 
\hline
\end{tabular}
\caption[]{The 95\% C.L. intervals ($\Delta\LL=1.92$) measured by
  ALEPH, DELPHI, L3 and OPAL.  In each case the parameter listed is varied
  while the remaining ones are fixed to their Standard Model values.
  Both statistical and systematic uncertainties are included.  }
\label{tab:gc_fTGC-1-ADLO}
\end{center}
\end{table}

\begin{table}[htbp]
\begin{center}
\renewcommand{\arraystretch}{1.3}
\begin{tabular}{|l||c|} 
\hline
Parameter     & 95\% C.L.     \\
\hline
\hline
$f_4^\gamma$  & [$-0.17,~~+0.19$]  \\ 
\hline
$f_4^Z$       & [$-0.30,~~+0.30$]  \\ 
\hline
$f_5^\gamma$  & [$-0.32,~~+0.36$]  \\ 
\hline
$f_5^Z$       & [$-0.34,~~+0.38$]  \\ 
\hline
\end{tabular}
\caption[]{ The 95\% C.L. intervals ($\Delta\LL=1.92$) obtained
  combining the results from all four experiments.  In each case the
  parameter listed is varied while the remaining ones are fixed to
  their Standard Model values.  Both statistical and systematic
  uncertainties are included.  }
 \label{tab:gc_fTGC-1-LEP}
\end{center}
\end{table}

\begin{table}[htbp]
\begin{center}
\renewcommand{\arraystretch}{1.3}
\begin{tabular}{|l||r|r|r|r|} 
\hline
Parameter  & ALEPH & DELPHI  &  L3   & OPAL  \\

\hline
\hline
$f_4^\gamma$ & [$-0.26,~~+0.26$] & [$-0.26,~~+0.28$]& [$-0.28,~~+0.28$] & [$-0.32,~~+0.33$] \\ 
$f_4^Z$      & [$-0.44,~~+0.43$] & [$-0.49,~~+0.42$]& [$-0.48,~~+0.46$] & [$-0.47,~~+0.58$]  \\ 
\hline                                              
$f_5^\gamma$ & [$-0.52,~~+0.53$] & [$-0.52,~~+0.61$]& [$-0.52,~~+0.62$] & [$-0.67,~~+0.62$] \\ 
$f_5^Z$      & [$-0.77,~~+0.86$] & [$-0.44,~~+0.69$]& [$-0.47,~~+1.39$] & [$-0.95,~~+0.33$] \\ 
\hline
\end{tabular}
\caption[]{The 95\% C.L. intervals ($\Delta\LL=1.92$) measured by
  ALEPH, DELPHI, L3 and OPAL.  In each case the two parameters listed are
  varied while the remaining ones are fixed to their Standard Model
  values.  Both statistical and systematic uncertainties are included.
  }
\label{tab:gc_fTGC-2-ADLO}
\end{center}
\end{table}

\begin{table}[htbp]
\begin{center}
\renewcommand{\arraystretch}{1.3}
\begin{tabular}{|l||c|rr|} 
\hline
Parameter     & 95\% C.L. & \multicolumn{2}{|c|}{Correlations} \\
\hline
\hline
$f_4^\gamma$  &[$-0.17,~~+0.19$] & $ 1.00$ & $ 0.07$\\ 
$f_4^Z$       &[$-0.30,~~+0.29$] & $ 0.07$ & $ 1.00$\\ 
\hline
$f_5^\gamma$  &[$-0.34,~~+0.38$] & $ 1.00$ & $-0.17$\\ 
$f_5^Z$       &[$-0.38,~~+0.36$] & $-0.17$ & $ 1.00$\\ 
\hline
\end{tabular}
\caption[]{ The 95\% C.L. intervals ($\Delta\LL=1.92$) obtained
  combining the results from all four experiments.  In each case the
  two parameters listed are varied while the remaining ones are fixed
  to their Standard Model values.  Both statistical and systematic
  uncertainties are included. Since the shape of the log-likelihood is
  not parabolic, there is some ambiguity in the definition of the
  correlation coefficients and the values quoted here are approximate.
  }
 \label{tab:gc_fTGC-2-LEP}
\end{center}
\end{table}

\clearpage

\clearpage

\subsection{Quartic Gauge Boson Couplings}
\begin{figure}[htbp]
\begin{center}
\includegraphics[width=\linewidth]{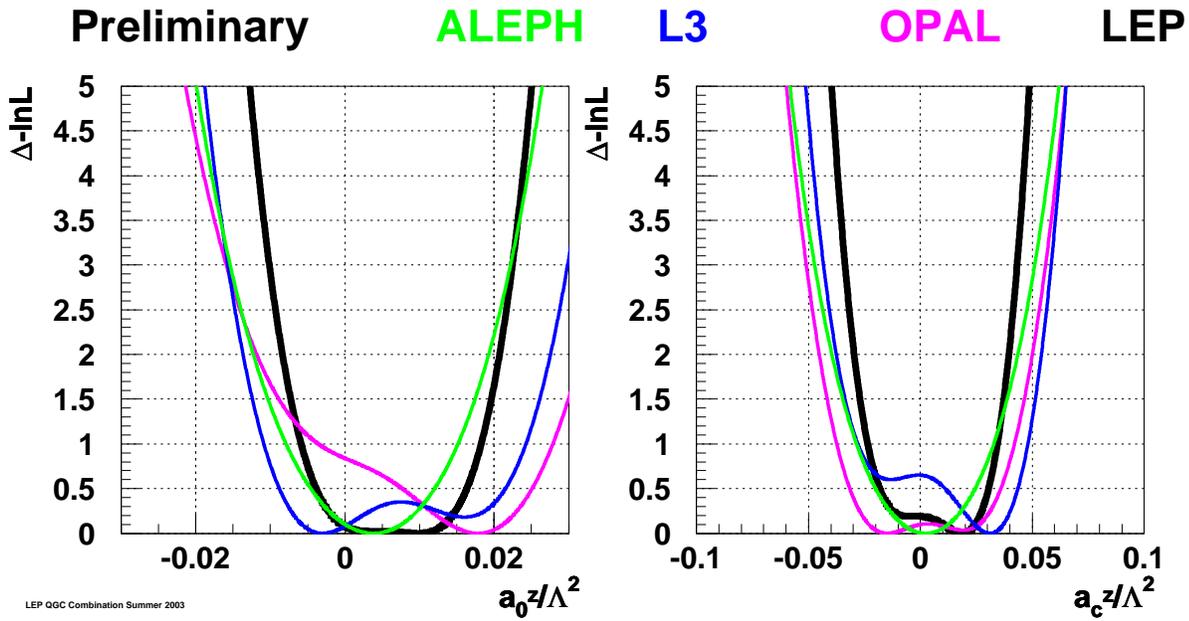}
\caption[]{
  The $\LL$ curves of L3 and OPAL (thin lines) and the 
  combined curve (thick line) for the QGCs \acl\ and \azl, associated with
  the \ZZgg\  vertex.  In each case, the minimal value is subtracted.  } 
\label{fig:gc_QGC-1}
\end{center}
\end{figure}

The individual numerical results from the experiments participating in the
combination, and the combined result are shown in Table~\ref{tab:gc_QGCs}. 
The corresponding $\LL$ curves are shown in
Figure~\ref{fig:gc_QGC-1}. The errors include both statistical and
systematic uncertainties. 

\begin{table}[ht]
\begin{center}
\begin{tabular}{|l||c|c|c|c|} \hline
Parameter & ALEPH  & L3  & OPAL   & Combined  \\
\hline \hline
\acl  &  [$-0.041,~~+0.044$]  &  [$-0.037,~~+0.054$] & 
         [$-0.045,~~+0.050$]  &  [$-0.029,~~+0.039$]  \\  
\azl  &  [$-0.012,~~+0.019$]  &  [$-0.014,~~+0.027$] & 
         [$-0.012,~~+0.031$]  &  [$-0.008,~~+0.021$] \\ 
\hline
\end{tabular}
\end{center}
\caption{ The limits for the QGCs \acl\ and \azl\ associated
    with the \ZZgg\ vertex at 95\% confidence level for ALEPH, L3 and OPAL, 
    and the LEP result obtained by combining them. 
    Both statistical and systematic errors are included.
  }
\label{tab:gc_QGCs}
\end{table}

\section*{Conclusions}
Combinations of charged and neutral triple gauge boson couplings, as well as
quartic gauge boson couplings associated with the \ZZgg\ vertex were made,
based on results from the four LEP experiments ALEPH, DELPHI, L3 and OPAL.
No significant deviation
from the Standard Model prediction is seen for any of the electroweak
gauge boson couplings studied. 
With the LEP-combined charged TGC results,
the existence of triple gauge boson couplings among the electroweak
gauge bosons is experimentally verified.
As an example, these data allow
the Kaluza-Klein theory~\cite{gc_bib:klein}, in which $\kg = -2$, to be
excluded completely~\cite{gc_bib:maiani}.

\boldmath
\chapter{Colour Reconnection in W-Pair Events}
\label{sec-CR}
\unboldmath

\updates{ Unchanged w.r.t. summer 2002: Results are preliminary.

  Note that some recent
  publications~\cite{Achard:2003pe,Abbiendi:2005es} are not yet
  included in this combination.  }

\section{Introduction}
 
 In \WWtoqqqq\ events, the products of the two (colour singlet) W
 decays in general have a significant space-time overlap as the
 separation of their decay vertices, $\tau_W \sim 1/\Gamma_W\approx
 0.1$~fm, is small compared to characteristic hadronic distance scales
 of $\sim 1$~fm.  Colour reconnection, also known as colour
 rearrangement (CR), was first introduced in\cite{bib:cr:GPZ} and
 refers to a reorganisation of the colour flow between the two W
 bosons.  A precedent is set for such effects by colour suppressed B
 meson decays, \eg\ $B \rightarrow J/\psi K$, where there is
 ``cross-talk'' between the two original colour singlets,
 $\bar{\mathrm c}$+s and
 c+spectator\cite{bib:cr:GPZ,bib:cr:SK_MODELS}.
 
 QCD interference effects between the colour singlets in \WW\ decays
 during the perturbative phase are expected to be small, affecting the
 W mass by $\sim (\frac{\alfas}{\pi N_{\mathrm{colours}}})^2 \Gamma_W$
 $\sim \cal{O}(\mathrm{1~\MeV})$ \cite{bib:cr:SK_MODELS}. In contrast,
 non-perturbative effects involving soft gluons with energies less
 than \GW\ may be significant, with effects on \MW\ $\sim
 \cal{O}(\mathrm{10~\MeV})$.  To estimate the impact of this phenomenon
 a variety of phenomenological models have been developed
 \cite{bib:cr:SK_MODELS,bib:cr:ARIADNECR_MODEL,HERWIG6,bib:cr:GH_MODEL,bib:cr:EG_MODEL,bib:cr:RATHSMAN},
 some of which are compared with data in this note.
 
 Many observables have been considered in the search for an
 experimental signature of colour reconnection.  The inclusive
 properties of events such as the mean charged particle multiplicity,
 distributions of thrust, rapidity, transverse momentum and
 $\ln(1/x_p)$ are found to have limited
 sensitivity\cite{bib:cr:OPAL_CR,bib:cr:DELPHI_CR,bib:cr:ALEPH_CR,bib:cr:L3_CR}.
 The effects of CR are predicted to be numerically larger in these
 observables when only higher mass hadrons such as kaons and protons
 are considered\cite{bib:cr:SK_HEAVYHAD}.  However, experimental
 investigations\cite{bib:cr:DELPHI_CR,bib:cr:OPAL_HEAVYHAD} find no
 significant gain in sensitivity due to the low production rate of
 such species in W decays and the finite size of the data sample.
 
 More recently, in analogy with the ``string effect'' analysis in
 3-jet $\eeqq g$
 events\cite{bib:cr:JADE_STRING2,*bib:cr:JADE_STRING3,%
 *bib:cr:JADE_STRING4,*bib:cr:JADE_STRING5,*bib:cr:TPC2GAM_STRING1,%
 *bib:cr:TPC2GAM_STRING2,*bib:cr:TASSO_STRING1}, the so-called
 ``particle flow'' method
 \cite{bib:cr:pflow1,bib:cr:pflow2,bib:cr:OXFORD_WS} has been
 investigated by all LEP collaborations
 \cite{bib:cr:ALEPH_PF,bib:cr:DELPHI_PF,bib:cr:L3_PF,bib:cr:OPAL_PF}.
 In this, pairs of jets in \WWtoqqqq\ events are associated with the
 decay of a W, after which four jet-jet regions are chosen: two corresponding
 to jets sharing the same W parent (intra-W), and two in which the
 parents differ (inter-W).  As there is a two-fold ambiguity in the
 assignment of inter-W regions, the configuration having the smaller sum
 of inter-W angles is chosen.
 
 Particles are projected onto the planes defined by these jet pairs
 and the particle density constructed as a function of $\phi$, the
 projected angle relative to one jet in each plane.  To account for
 the variation in the opening angles, $\phi_0$, of the jet-jet pairs
 defining each plane, the particle
 densities in $\phi$ are constructed as functions of normalised
 angles, $\phi_r=\phi/\phi_0$, by a simple rescaling of the projected
 angles for each particle, event by event.  Particles having
 projected angles $\phi$ smaller than $\phi_0$ in at least one of the
 four planes are considered further.  This gives particle densities,
 $\frac{1}{\Nevt}\dndphir$, in four regions with $\phi_r$ in the range
 0--1, and where $n$ and \Nevt\ are the number of particles and
 events, respectively.
 
 As particle density reflects the colour flow in an event, CR models
 predict a change in the relative particle densities between inter-W
 and intra-W regions.  On average, colour reconnection is expected to
 affect the particle densities of both inter-W regions in the same way
 and so they are added together, as are the two intra-W regions.  The
 observable used to quantify such changes, \Rn, is defined:
\begin{equation}
 \Rn =
  \frac{\frac{1}{\Nevt}\int^{0.8}_{0.2} \dndphir (\mathrm{intra-W}) \dphir}
       {\frac{1}{\Nevt}\int^{0.8}_{0.2} \dndphir (\mathrm{inter-W}) \dphir} \,.
 \label{fsi:cr:eq:Rn}
\end{equation}
 As the effects of CR are expected to be enhanced for low momentum
 particles far from the jet axis, the range of integration excludes
 jet cores ($\phi_r\approx 0$ and $\phi_r\approx 1$).  The precise
 upper and lower limits are optimised by model studies of predicted
 sensitivity.
 
 Each LEP experiment has developed its own variation on this analysis,
 differing primarily in the selection of \WWtoqqqq\ events.  In
 \Ltre\cite{bib:cr:L3_PF} and \Delphi\cite{bib:cr:DELPHI_PF}, events
 are selected in a very particular configuration (``topological
 selection'') by imposing restrictions on the jet-jet angles and on
 the jet resolution parameter for the three- to four-jet transition
 (Durham or LUCLUS schemes).  This selects events which are more planar
 than those in the inclusive \WWtoqqqq\ sample and the association between
 jet pairs and W's is given by the relative angular separation of the
 jets.  The overall efficiency for selecting events is $\sim15$\%.
 The \Aleph\cite{bib:cr:ALEPH_PF} and \Opal\cite{bib:cr:OPAL_PF} event
 selections are based on their W mass analyses.  Assignment of pairs
 of jets to W's also follows that used in measuring \MW, using either
 a 4-jet matrix element \cite{ALEPH-MW} or a multivariate algorithm
 \cite{OPAL-MW}.  These latter selections have much higher
 efficiencies, varying from 45\% to 90\%, but lead to samples of events
 having a less planar topology and hence a more complicated colour
 flow. ALEPH also uses the topological selection for consistency
 checks.
 
 The data are corrected bin-by-bin for background contamination in the
 inter-W and intra-W regions separately.  The possibility of CR
 effects existing in background processes, such as \ZZtoqqqq, is
 neglected.  Since the data are not corrected for the effects of event
 selection, momentum resolution and finite acceptance, the values of
 \Rn\ measured by the experiments cannot be compared directly with one
 another.  However, it is possible to perform a relative comparison by
 using a common sample of Monte Carlo events, processed using the
 detector simulation program of each experiment.

\section{Combination Procedure}
 
 The measured values of \Rn\ can be compared after they have been
 normalised using a common sample of events, processed using the
 detector simulation and particle flow analysis of each experiment.  A
 variable, $r$, is constructed:
\begin{equation}
          r = \frac{\Rndata}{\Rnnocr} \,,
\end{equation}
 where \Rndata\ and \Rnnocr\ are the values of \Rn\ measured by each
 experiment in data and in a common sample of events without CR.  In
 the absence of CR, all experiments should find $r$ consistent with
 unity.  The default no-CR sample used for this normalisation consists
 of \eeWW\ events produced using the \KoralW\cite{bib:fsi:KORALW} event
 generator and hadronised using either the \Jetset\cite{JETSET},
 \Ariadne\cite{ARIADNE} or \Herwig\cite{HERWIG6} model depending on
 the colour reconnection model being tested.  Input from experiments used to
 perform the combination is given in terms of \Rn\ and detailed in
 Appendix~\ref{fsi:cr:app:inputs}.

 \subsection{Weights}
 
 The statistical precision of \Rn\ measured by the experiments does
 not reflect directly the sensitivity to CR, for example the
 measurements of \Aleph\ and \Opal\ have efficiencies several times
 larger than the topological selections of \Ltre\ and \Delphi, yet
 only yield comparable sensitivity.  The relative sensitivity of the
 experiments may also be model dependent.  Therefore, results are
 averaged using model dependent weights, \ie\ 
\begin{equation}
  w_i = \frac{(\Rni - \Rninocr)^2}
             {\sigsqRn(\mathrm{stat.}) + \sigsqRn (\mathrm{syst.})}
             \,,
 \label{fsi:cr:eq:weight}
\end{equation}
 where \Rni\ and \Rninocr\ represent the \Rn\ values for CR model $i$
 and its corresponding no-CR scenario, and \sigsqRn\ are the total
 statistical and systematic uncertainties.  To test models, \Rn\ 
 values using common samples are provided by experiments for each of
 the following models:
 \begin{enumerate}
  \item \SKI, 100\% reconnected (\KoralW\ $+$ \Jetset),
  \item \Ariadne-II, inter-W reconnection rate about 22\% (\KoralW\ $+$ \Ariadne),
  \item \Herwig\ CR, reconnected fraction $\frac{1}{9}$ (\KoralW\ $+$
  \Herwig).
 \end{enumerate}
 Samples in parentheses are the corresponding no-CR scenarios used to
 define $w_i$.  In each case, $\KoralW$ is used to generate the events
 at least up to the four-fermion level. 
 These special Monte Carlo samples (called ``Cetraro''
 samples) have been generated with the ALEPH tuned parameters,
 obtained with hadronic Z decays, and have been processed through the
 detector simulation of each experiment.

 \subsection{Combination of centre-of-mass energies}
 
 The common files required to perform the combination are only
 available at a single centre-of-mass energy ($E_{\mathrm{cm}}$) of
 188.6~GeV.  The data from the experiments can only therefore be
 combined at this energy.  The procedure adopted to combine all LEP
 data is summarised below.
 
 \Rn\ is measured in each experiment at each centre-of-mass energy, in
 both data and Monte Carlo.
 The predicted variation of \Rn\ with centre-of-mass energy is
 determined separately by each experiment using its own samples of 
 simulated \eeWW\ events, with hadronisation performed using the no-CR
 \Jetset\ model.  This variation is parametrised by fitting
 a polynomial to these simulated \Rn.  The \Rn\ measured in data are
 subsequently extrapolated to the reference energy of 189~GeV using
 this function, and the weighted average of the rescaled values in
 each experiment is used as input to the combination.

\section{Systematics}

 The sources of potential systematic uncertainty identified are
 separated into those which are correlated between experiments and
 those which are not.  For correlated sources,
 the component correlated between all experiments is assigned as the
 smallest uncertainty found in any single experiment, with the
 quadrature remainder treated as an uncorrelated contribution.
 Preliminary estimates of the dominant systematics on \Rn\ are given
 in Appendix~\ref{fsi:cr:app:inputs} for each experiment, and
 described below.

 \subsection{Hadronisation}
 
 This is assigned by comparison of the single sample of \WW\ events
 generated using \KoralW, and hadronised with three different models,
 \ie\ \Jetset, \Herwig\ and \Ariadne.  The systematic is assigned as
 the spread of the \Rn\ values obtained when using the various models
 given in Appendix~\ref{fsi:cr:app:inputs}.  This is treated as a
 correlated uncertainty.

 \subsection{Bose-Einstein Correlations}
 
 Although a recent analysis by \Delphi\ reports the observation of
 inter-W Bose-Einstein correlation (BEC) in \WWtoqqqq\ events with a
 significance of 2.9 standard deviations for like-sign pairs and 1.9
 standard deviations for unlike-sign pairs~\cite{be:DELPHI03},
 analyses by other
 collaborations\cite{bib:cr:ALEPH_BEC,be:L302,bib:cr:OPAL_BEC} find no
 significant evidence for such effects, see also chapter~\ref{sec-BE}.
 Therefore, BEC effects are only considered within each W separately.
 The estimated uncertainty is assigned, using common MC samples, as
 the difference in \Rn\ between an intra-W BEC sample and the
 corresponding no-BEC sample.  This is treated as correlated between
 experiments.

 \subsection{Background}
 
 Background is dominated by the \eeqq\ process, with a smaller
 contribution from \ZZtoqqqq\ diagrams.  As no common background
 samples exist, apart from dedicated ones for BEC analyses, experiment
 specific samples are used.  The uncertainty is defined as the
 difference in the \Rn\ value relative to that obtained using the
 default background model and assumed cross-sections in each
 experiment.

  \subsubsection{\eeqq}
  
 The systematic is separated into two components, one accounting for
 the shape of the background, the other for the uncertainty in the
 value of the background cross-section, $\sigma(\eeqq)$.
 
 Uncertainty in the shape is estimated by comparing hadronisation
 models.  Experiments typically have large samples simulated using
 2-fermion event generators hadronised with various models.  This
 uncertainty is assigned as $\pm\frac{1}{2}$ of the largest difference
 between any pair of hadronisation models and treated as uncorrelated
 between experiments.
 
 The second uncertainty arises due to the accuracy of the
 experimentally measured cross-sections.  The systematic is assigned
 as the larger of the deviations in \Rn\ caused when $\sigma(\eeqq)$
 is varied by $\pm 10$\% from its default value.  This variation was
 based on the conclusions of a study comparing four-jet data
 with models\cite{bib:LEP2_MCWS}, and is significantly larger than the
 $\sim 1$\% uncertainty in the inclusive \eeqq\ ($\sqrt{s'/s}>0.85$)
 cross-section measured by the $\LEPII$ 2-fermion group.  It is treated as
 correlated between experiments.

 \subsubsection{\ZZtoqqqq}
 
 Similarly to the \eeqq\ case, this background cross-section is varied
 by $\pm 15$\%.  For comparison, the uncertainty on $\sigma(\ZZ)$
 measured by the $\LEPII$ 4-fermion group is $\sim 11$\% at
 $\sqrt{s}\simeq 189$~\GeV.  It is treated as correlated between
 experiments.

 \subsubsection{\WWtoqqlv}
 
 Semi-leptonic WW decays which are incorrectly identified as
 \WWtoqqqq\ events are the third main category of background, and its
 contribution is very small.  The fraction of \WWtoqqlv\ events
 present in the sample used for the particle flow analysis varies in
 the range 0.04--2.2\% between the experiments.  The uncertainty in
 this background consists of hadronisation effects and also
 uncertainty in the cross-section.  As this source is a very small
 background relative to those discussed above, and the effect of
 either varying the cross-section by its measured uncertainty or of
 changing the hadronisation model do not change the measured \Rn\ 
 significantly, this source is neglected.

 \subsection{Detector Effects}
 
 The data are not corrected for the effects of finite resolution or
 acceptance.  Various studies have been carried out, e.g. by analysing
 \WWtoqqlv\ events in the same way as \WWtoqqqq\ events in order to
 validate the method and the choice of energy flow objects used to
 measure the particle yields between jets~\cite{bib:cr:L3_PF}.  To
 take into account the effects of detector resolution and acceptance,
 ALEPH, L3 and OPAL have studied the impact of changing the object
 definition entering the particle flow distributions and have assigned
 a systematic error from the difference in the measured \Rn.

 \subsection{Centre-of-mass energy dependence}
 
 As there may be model dependence in the parametrised energy
 dependence, the second order polynomial used to perform the
 extrapolation to the reference energy of 189~GeV is usually
 determined using several different models, with and without colour
 reconnection.  DELPHI, L3 and OPAL use differences relative to the
 default no-CR model to assign a systematic uncertainty while ALEPH
 takes the spread of the results obtained with all the models with and
 without CR which have been used.  This error is assumed to be
 uncorrelated between experiments.

 \subsection{Weighting function}
 
 The weighting function of Equation~\ref{fsi:cr:eq:weight} could
 justifiably be modified such that only the uncorrelated components of
 the systematic uncertainty appear in the denominator.  To accommodate
 this, the average is performed using both variants of the weighting
 function.  This has an insignificant effect on the consistency
 between data and model under test, e.g. for \SKI\ the result is
 changed by 0.02 standard deviations, and this effect is therefore neglected.

\section{Combined Results}
 
 Experiments provide their results in the form of \Rn\ (or changes to
 \Rn) at a reference centre-of-mass energy of 189~\GeV\ by scaling
 results obtained at various energies using the predicted energy
 dependence of their own no-CR MC samples. This avoids having to
 generate common samples at multiple centre-of-mass energies.
 
 The detailed results from all experiments are included
 in Appendix~\ref{fsi:cr:app:inputs}.  These consist of preliminary
 results, taken from the publicly available
 notes\cite{bib:cr:ALEPH_PF,bib:cr:DELPHI_PF,bib:cr:L3_PF,bib:cr:OPAL_PF},
 and additional information from analysis of Monte Carlo samples.  The
 averaging procedure itself is carried out by each of the experiments
 and good agreement is obtained.
 
\begin{figure}[tbhp]
 \centerline{\epsfig{file=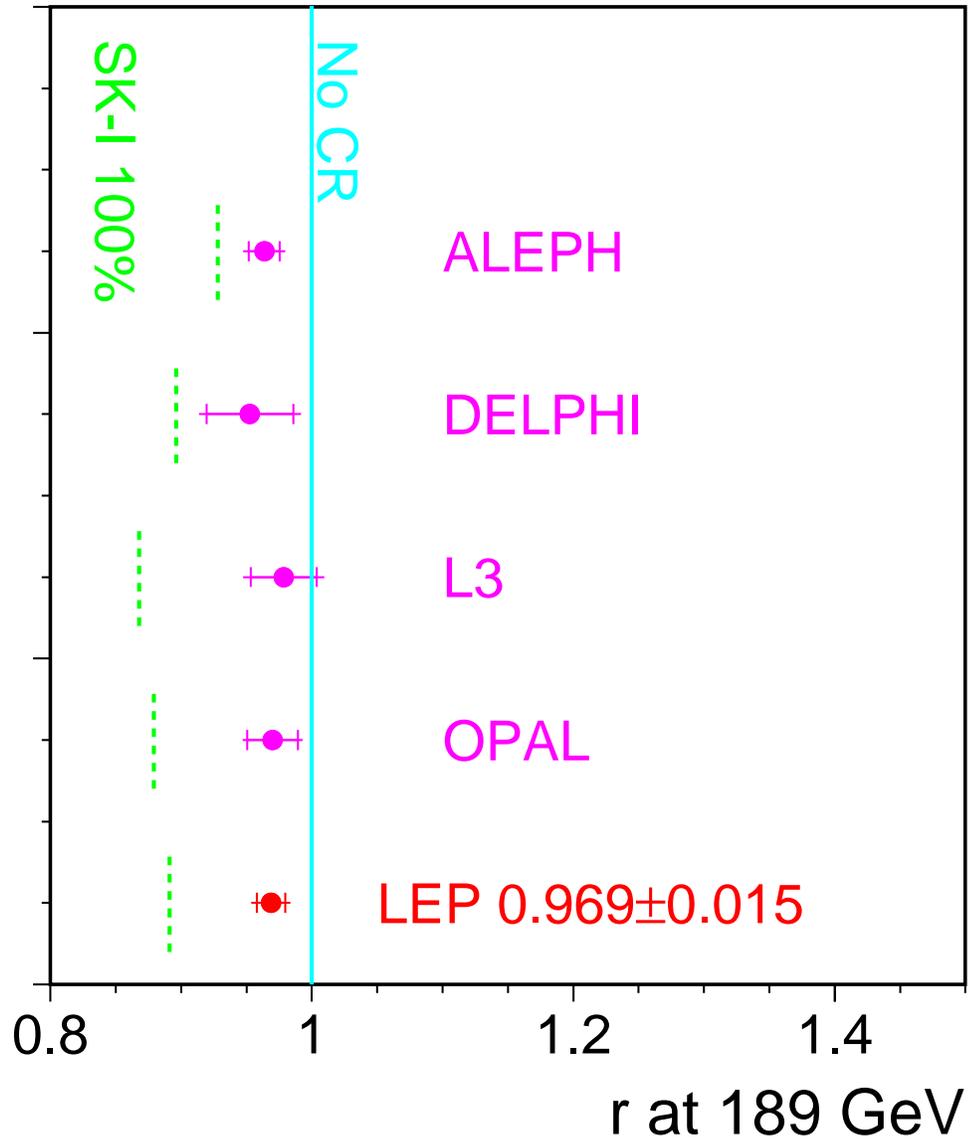,width=0.9\textwidth}}
 \caption[Particle flow combination, \SKI\ model.]  {Preliminary
   particle flow results using all data, combined to test the limiting
   case of the \SKI\ model in which more than 99.9\% of the events are
   colour reconnected. The error bars correspond to the total error
   with the inner part showing the statistical uncertainty.  The
   predicted values of $r$ for this CR model are indicated separately 
   for the analysis of each experiment by dashed lines.}
 \label{fsi:cr:fig:SKI_comb}
\end{figure}

 An example of this averaging to test an extreme scenario of the \SKI\ 
 CR model (full reconnection) is given in
 Appendix~\ref{fsi:cr:app:results}.  The average obtained in this case
 is:
\begin{eqnarray}
    r(data)    &  = & 0.969 \pm 0.011 (\mathrm{stat.})
                \pm 0.009 (\mathrm{syst.~corr.})
                \pm 0.006 (\mathrm{syst.~uncorr.}) \,, \\
   r (\mathrm{\SKI}\ 100\%) & = & 0.8909  \,.
 \end{eqnarray}
 The measurements of each experiment and this combined result are
 shown in Figure~\ref{fsi:cr:fig:SKI_comb}.  As the sensitivity of the
 analysis is different for each experiment, the value of $r$ predicted
 by the \SKI\ model is indicated separately for each experiment by a
 dashed line in the figure.  Thus the data disagree with the extreme
 scenario of this particular model at a level of 5.2 standard
 deviations. The data from the four experiments are consistent with
 each other and tend to prefer an intermediate colour reconnection
 scenario rather than the no colour reconnection one at the level of 2.2
 standard deviations in the \SKI\ framework.
 
 \subsection{Parameter space in \SKI\ model}
 
\begin{figure}[tbhp]
 \centerline{\epsfig{file=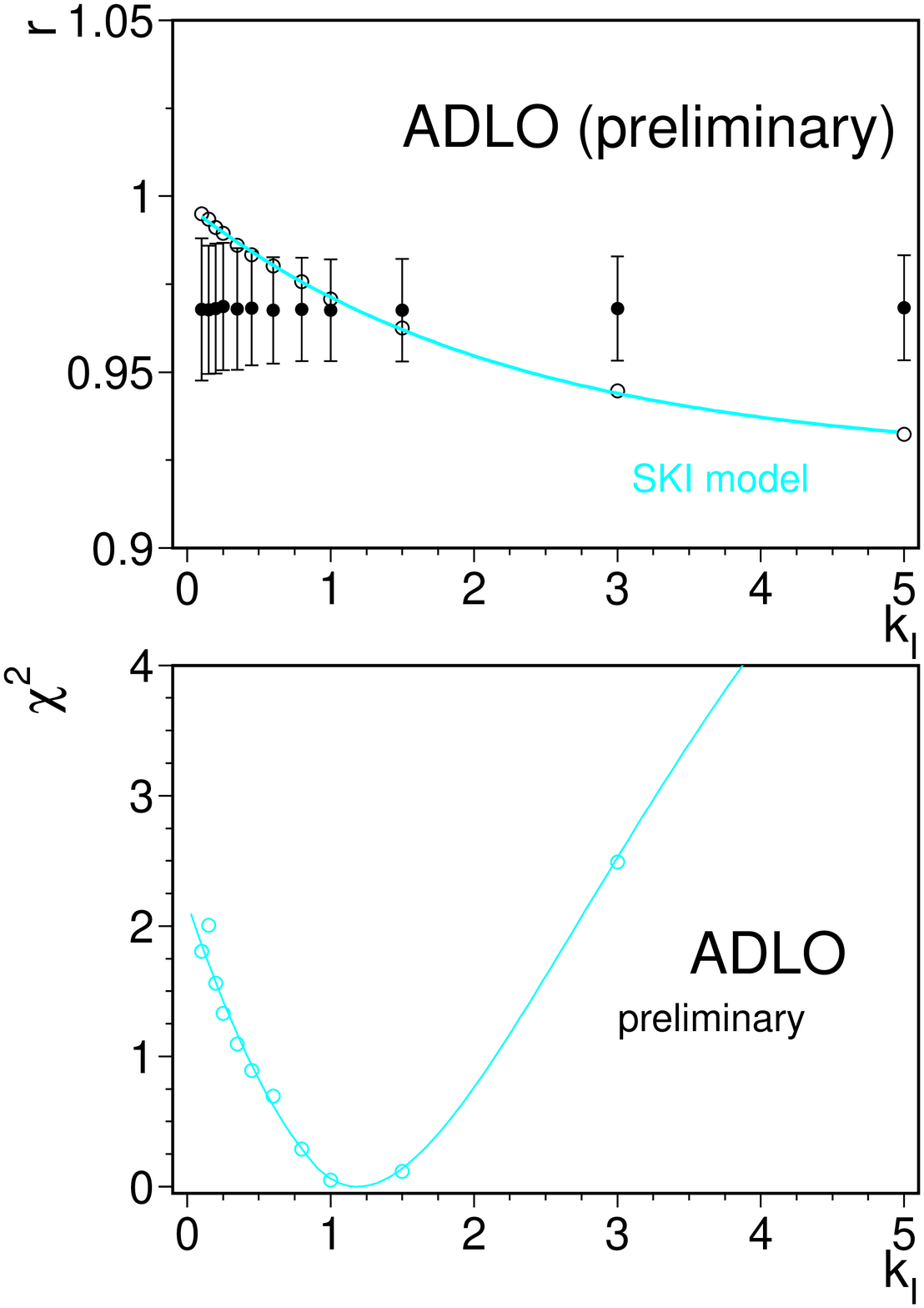,width=0.8\textwidth}}
 \caption[Constraints on \SKI\ model parameter, \kI.] {Comparison of the
   LEP average $r$ values with the \SKI\ model prediction obtained as
   a function of the \kI\ parameter.  The comparisons are performed
   after extrapolation of data to the reference centre-of-mass energy
   of 189~GeV.  In the upper plot, the solid line is the result of
   fitting a function of the form $r(k_{I}) = p_{1} (1-exp(-p_{2}
   k_{I}))+p_{3}$ to the MC predictions.  The lower plot shows the
   corresponding $\chi^{2}$ curve obtained from this comparison.  The
   best agreement between the model and the data is obtained when
   $\kI = 1.18$.}
 \label{fsi:cr:fig:ki_scan}
\end{figure}

\begin{figure}[tbhp]
 \centerline{\epsfig{file=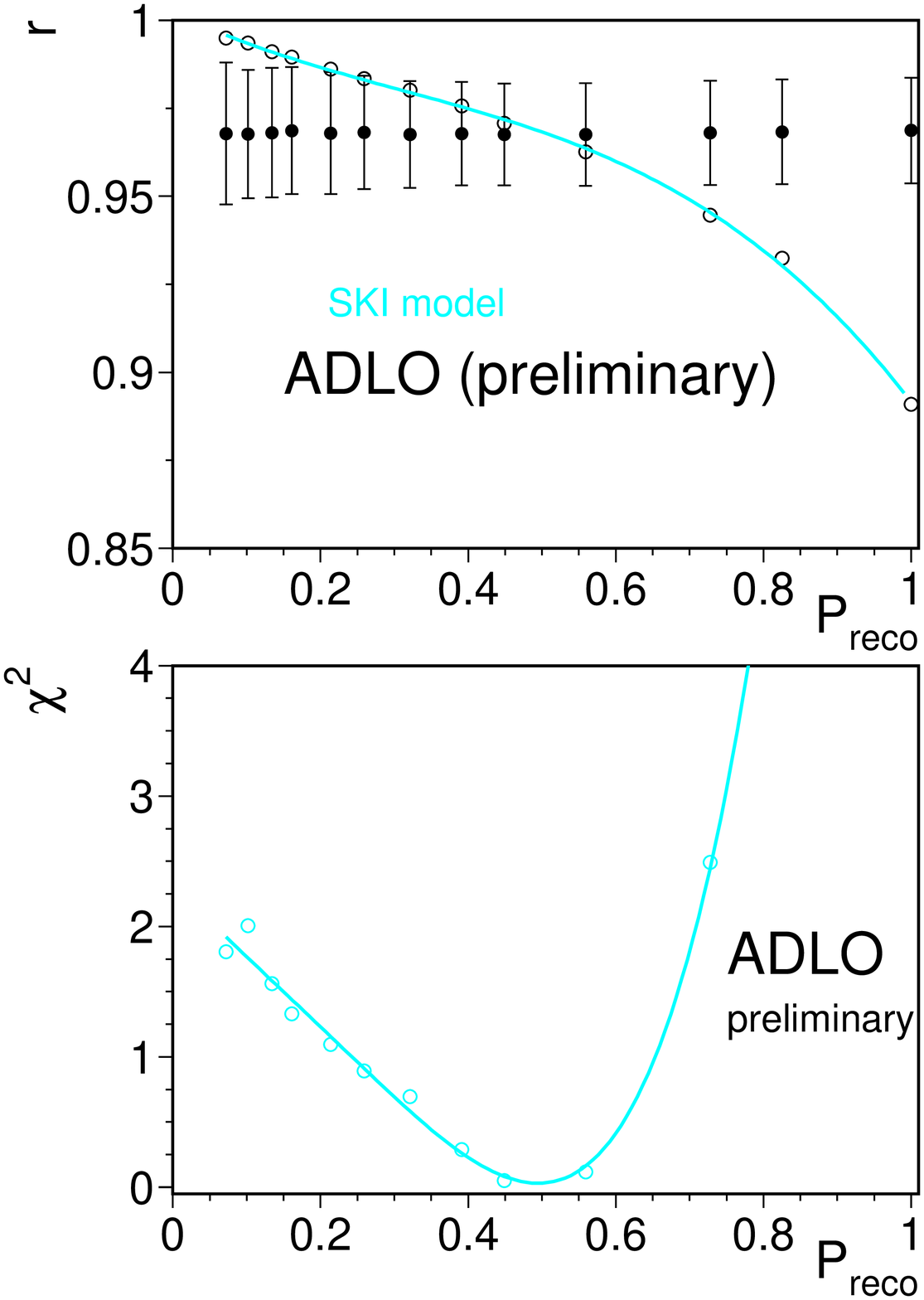,width=0.8\textwidth}}
 \caption[Constraints on \SKI\ reconnection probability.] {Comparison of the
   LEP average $r$ values with the \SKI\ model prediction obtained as
   a function of the reconnection probability.  In the upper plot, the
   solid line is the result of fitting a third order polynomial
   function to the MC predictions.  The lower plot shows a $\chi^{2}$
   curve obtained from this comparison using all LEP data at the
   reference centre-of-mass energy of 189 GeV.  The best agreement
   between the model and the data is obtained when 49\% of events are
   reconnected in this model.}
\label{fsi:cr:fig:preco_scan}
\end{figure}

 In the \SKI\ model, the reconnection probability is governed by an
 arbitrary, free parameter, \kI.  By comparing the data with model
 predictions evaluated at a variety of \kI\ values, it is possible to
 determine the reconnection probability that is most consistent with
 data, which can in turn be used to estimate the corresponding bias in
 the measured \MW.  By repeating the averaging procedure using model
 inputs for the set of \kI\ values given in
 Table~\ref{fsi:cr:tab:cetraro}, including a re-evaluation of the
 weights for each value of \kI, it is found that the data prefer a
 value of $\kI =1.18$ as shown in Figure~\ref{fsi:cr:fig:ki_scan}. The
 68\% confidence level lower and upper limits are 0.39 and 2.13
 respectively.  The LEP averages in $r$ obtained for the different \kI\
 values are summarised in Table~\ref{fsi:cr:tab:average}.  They
 correspond to a preferred reconnection probability of 49\% in this model at
 189 GeV as illustrated in Figure~\ref{fsi:cr:fig:preco_scan}.

 The small variations observed in the LEP average value of $r$ and its
 corresponding error as a function of $k_{I}$ (or $P_{reco}$) are
 essentially due to changes in the relative weighting of the
 experiments.

 \subsection{\Ariadne\ and \Herwig\ models}
\begin{figure}[tbhp]
   \centerline{\epsfig{file=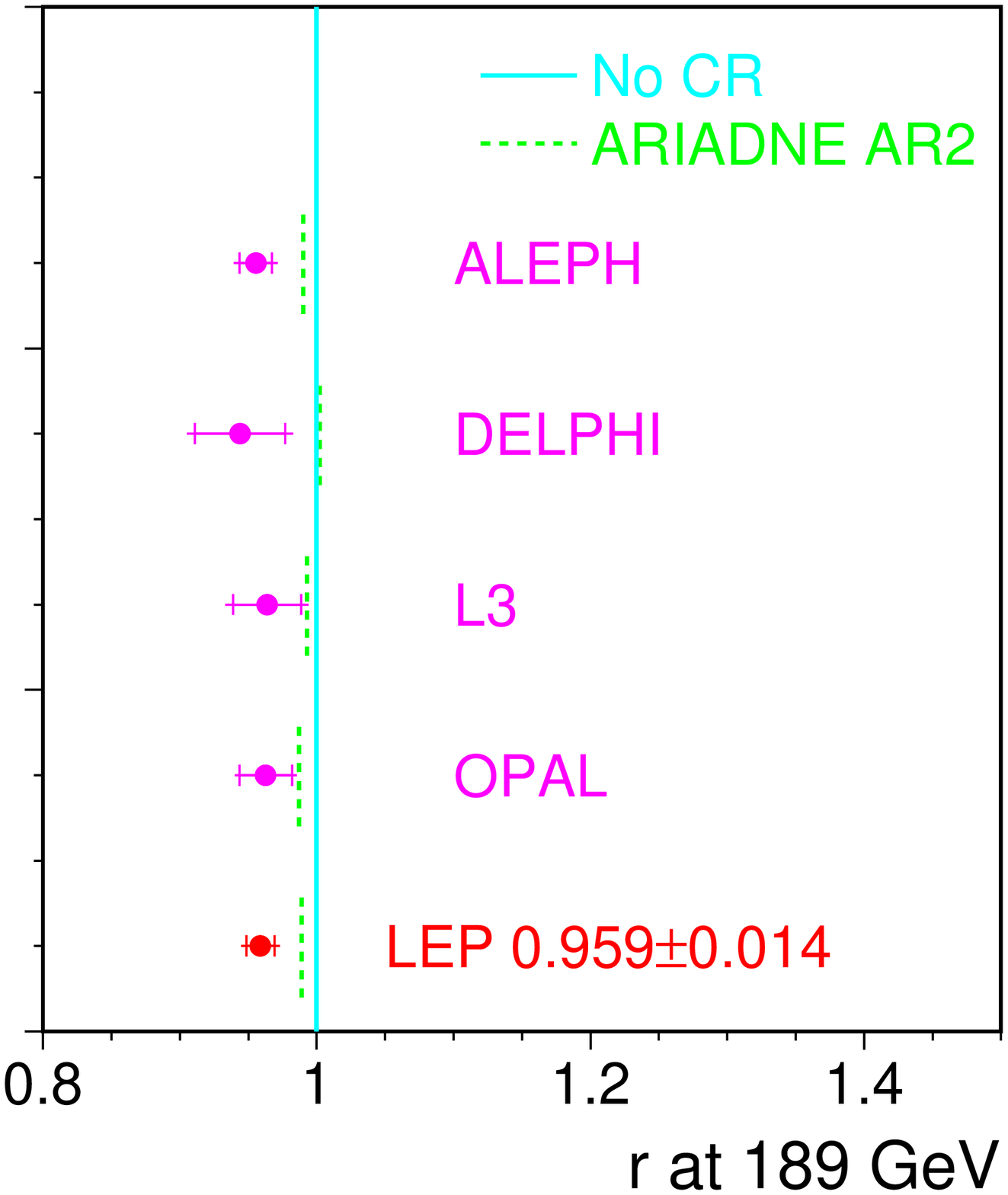,width=0.5\textwidth}
     \epsfig{file=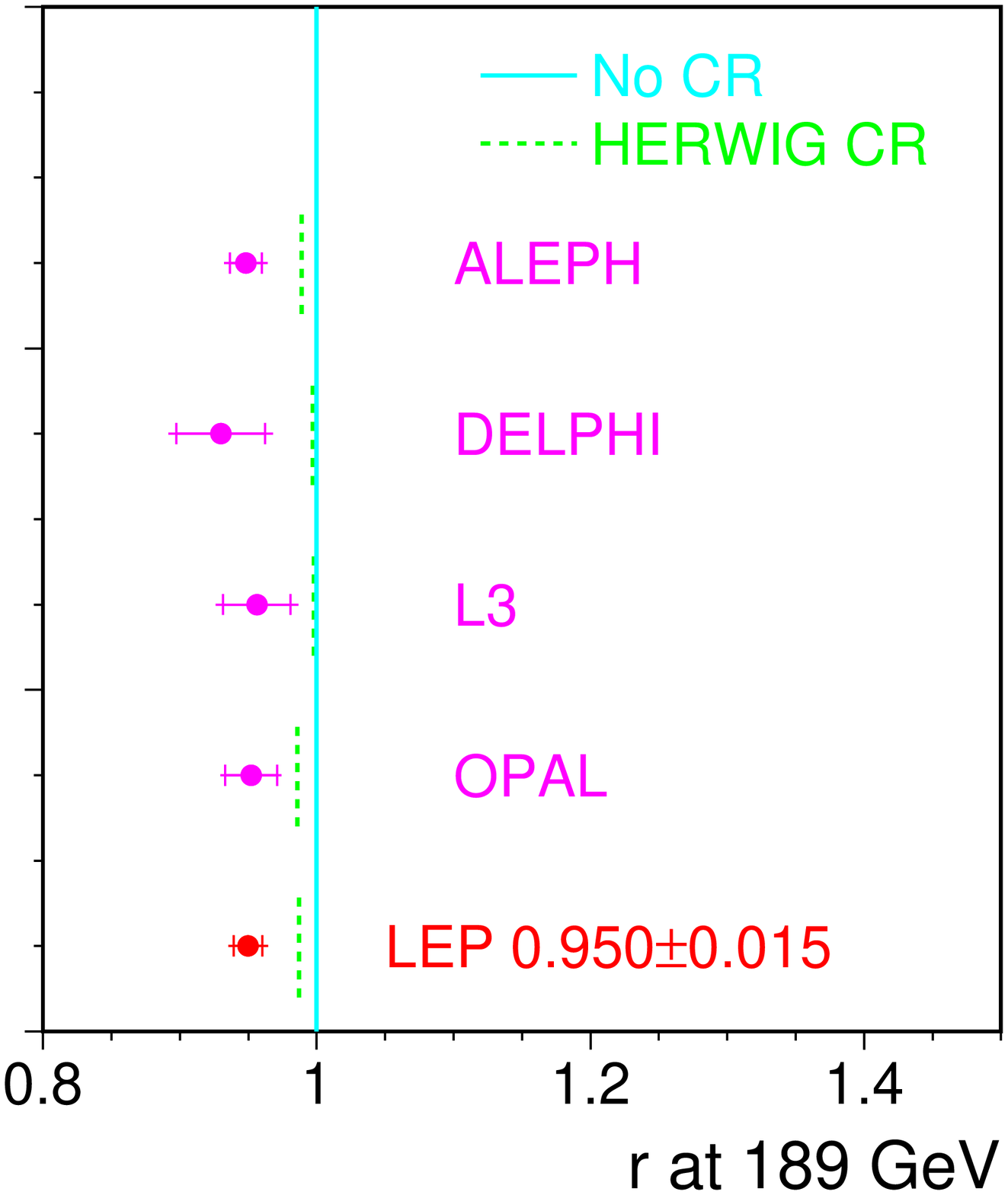,width=0.5\textwidth}}
  \caption[Particle flow combination, \ARII\ and \Herwig\ CR models.]
  {Preliminary particle flow results using all data, combined to test
    the \Ariadne\ and \Herwig\ colour reconnection models, based on
    the predicted sensitivity.  The predicted values of $r$ for this
    CR model are indicated separately for the analysis of each
    experiment by dashed lines.}
 \label{fsi:cr:fig:arhw}
 \end{figure}

 The combination procedure has been applied to common samples of
 \Ariadne\ and \Herwig\ Monte Carlo models. The \Rn\ average values
 obtained with these models based on their respective predicted
 sensitivity are summarised in Table~\ref{fsi:cr:tab:average2}.  The
 four experiments have observed a weak sensitivity to these colour
 reconnected samples with the particle flow analysis, as can be seen
 from Figure~\ref{fsi:cr:fig:arhw}.
 
\section{Summary}
 
 A first, preliminary combination of the LEP particle flow results is
 presented, using the entire $\LEPII$ data sample.  The data disfavour by
 5.2 standard deviations an extreme version of the \SKI\ model in
 which colour reconnection has been forced to occur in essentially all
 events.  The combination procedure has been generalised to the \SKI\ 
 model as a function of its variable reconnection probability. The
 combined data are described best by the model where 49\% of events
 at 189 GeV are reconnected, corresponding to $\kI =1.18$.  The LEP
 data, averaged using weights corresponding to $\kI=1.0$, \ie\ closest
 to the optimal fit, do not exclude the no colour reconnection
 hypothesis, deviating from it by 2.2 standard deviations.  A 68\%
 confidence level range has been determined for \kI\ and corresponds
 to [0.39,2.13].
 
 For both the \Ariadne\ and \Herwig\ models, which do not contain
 adjustable colour reconnection parameters, differences between the
 results of the colour reconnected and the no-CR scenarios are small
 and do not allow the particle flow analysis to discriminate between
 them.  To test consistency between data and the no-CR models, the
 data are averaged using weights where the factor accounting for
 predicted sensitivity to a given CR model has been set to unity.  The
 \Rn\ values obtained with the no colour reconnection \Herwig\ and
 \Ariadne\ models, using the common Cetraro samples, differ from the
 measured data value by 3.7 and 3.1 standard deviations.
 
 The observed deviations of the \Rn\ values from all no colour reconnection
 models may indicate a possible systematic effect in the description
 of particle flow for 4-jet events.  Independent studies of particle
 flow in WW semileptonic events as well as other CR-oriented analyses
 are required to investigate this.

\boldmath
\chapter{Bose-Einstein Correlations in W-Pair Events}
\label{sec-BE}
\unboldmath

\updates{ All experiments have now published final results which enter
in a new combination. }

\section{Introduction}

The LEP experiments have measured the strength of particle
correlations between two hadronic systems obtained from W-pair decay
occuring close in space-time at \LEPII.  The work presented in this
chapter is focused on so-called Bose-Einstein (BE) correlations, i.e.,
the enhanced probability of production of pairs (multiplets) of
identical mesons close together in phase space. The effect is readily
observed in particle physics, in particular in hadronic decays of the
Z boson, and is qualitatively understood as a result of
quantum-mechanical interference originating from the symmetry of the
amplitude of the particle production process under exchange of
identical mesons.

The presence of correlations between hadrons coming from the decay of
a $\WW$ pair, in particular those between hadrons originating from
different Ws, can affect the direct reconstruction of the mass of the
initial W bosons.  The measurement of the strength of these
correlations can be used for the estimation of the systematic
uncertainty of the W mass measurement.

\section{Method}

The principal method~\cite{be:chekanov}, called ``mixing method'',
used in this measurement is based on the direct comparison of
2-particle spectra of genuine hadronic WW events $\mathrm{WW
\rightarrow q\bar{q}q\bar{q}\,}$ and of mixed WW events.  The latter
are constructed by mixing the hadronic parts of two semileptonic WW
events $\mathrm{WW \rightarrow q\bar{q}\ell \nu\,}$ (first used
in~\cite{be:DELPHI97}). Such a reference sample has the advantage of
reproducing the correlations between particles belonging to the same
W, while the particles from different Ws are uncorrelated by
construction.

This method gives a model-independent estimate of the interplay
between the two hadronic systems, for which BE correlations and also
colour reconnection are considered as dominant sources. The
possibility of establishing the strength of inter-W correlations in a
model-independent way is rather unique; most correlations do carry an
inherent model dependence on the reference sample. In the present
measurement, the model dependence is limited to the background
subtraction.

\section{Distributions}

The two-particle correlations are evaluated using two-particle
densities defined in terms of the 4-momentum transfer
$Q=\sqrt{-(p_{1}-p_{2})^{2}}$, where $p_1,p_2$ are the 4-momenta of
the two particles:
\begin{equation}
\rho_{2}(Q)=\frac{1}{N_{ev}}\frac{dn_{pairs}}{dQ}
\end{equation}
Here $n_{pairs}$ stands for the number of like-sign (unlike-sign)
2-particle permutations.\footnote{For historical reasons, the number
of particle permutations rather than combinations is used in formulas.
For the same reason, a factor 2 appears in front of $\rho_{2}^{mix}$
in Eq.~\ref{eq:be:rho-mix}. The experimental statistical errors are,
however, based on the number of particle pairs, i.e., 2-particle
combinations.}  In the case of two stochastically independent
hadronically decaying W bosons the two-particle inclusive density is
given by:
\begin{equation}
 \rho_{2}^{WW}~=~\rho_{2}^{W^{+}}+\rho_{2}^{W^{-}}+2 \rho_{2}^{mix},
\label{eq:be:rho-mix}
\end{equation}
where $\rho_{2}^{mix}$ can be expressed via the single-particle
inclusive density $\rho_1(p)$ as:
\begin{equation}
\rho_2^{mix}(Q)~=~\int d^{4}p_{1}d^{4}p_{2}\rho^{W^{+}}(p_1)\rho^{W^{-}}(p_2)\delta(Q^{2}+(p_{1}-p_{2})^{2})\delta(p_1^2-m_{\pi}^2)\delta(p_2^2-m_{\pi}^2).
\end{equation}
Assuming further that:
\begin{equation}
\rho_{2}^{W^{+}}(Q)~=~\rho_{2}^{W^{-}}(Q)~=~\rho_{2}^{W}(Q),
\end{equation}
we obtain for the case of two stochastically independent hadronically
decaying W bosons:
\begin{equation}
\rho_{2}^{WW}(Q)~=~2\rho_{2}^{W}(Q)+2\rho_2^{mix}(Q).
\end{equation}
In the mixing method, we obtain $\rho_2^{mix}$ by combining two
hadronic W systems from two different semileptonic WW events.  The
direct search for inter-W BE correlations is done using the difference
of 2-particle densities:
\begin{equation}
\Delta \rho(Q) ~=~ \rho_{2}^{WW}(Q)-2\rho_{2}^{W}(Q)-2\rho_2^{mix}(Q),
\label{eq:be:dr}
\end{equation}
or, alternatively, their ratio:
\begin{equation}
    D(Q)~=~\frac{\rho_{2}^{WW}(Q)}{2\rho_{2}^{W}(Q)+2\rho^{mix}(Q)}
        ~=~ 1 + \frac{ \Delta \rho(Q) }{2\rho_{2}^{W}(Q)+2\rho^{mix}(Q)} .
\label{eq:be:D}
\end{equation}
Given the definition of the genuine inter-W correlations function
$\delta_I(Q)$ \cite{be:DeWolf}, it can be shown that
\begin{equation}
     \delta_I(Q)~=~\frac{\Delta \rho(Q)}{2\rho_2^{mix}(Q)}.
\end{equation}
To disentangle the BE correlation effects from other possible
correlation sources (such as energy-momentum conservation or color
reconnection), which are supposed to be the same for like-sign and
unlike-sign charge pairs, we analyze the double difference,
 \begin{equation}
 \delta \rho(Q) =  \Delta \rho^{like-sign}(Q)- \Delta \rho^{unlike-sign}(Q),
\end{equation}
or the double ratio,
 \begin{equation}
    d(Q) =  D^{like-sign}(Q)/D^{unlike-sign}(Q).
\end{equation}

The event mixing procedure may introduce artificial distortions, or
may not fully account for some detector effects or for correlations
other than BE correlations. Most of these possible effects are
simulated in the Monte Carlo without inter-W BE correlations.
Therefore they are reduced by using the double ratio or the double
difference:
\begin{equation}
    D'(Q)~=~\frac{D(Q)_{data}}{D(Q)_{MC,no inter}}\hspace{0.2cm} ,\hspace{1cm}
\Delta \rho'(Q) ~=~\Delta \rho(Q)_{data} - \Delta \rho(Q)_{MC,no inter}
\hspace{0.2cm} ,
\end{equation}
where $D(Q)_{MC,no inter}$ and $\Delta \rho(Q)_{MC,no inter}$ are
derived from a MC without inter-W BE correlations.

In addition to the mixing method, ALEPH \cite{be:ALEPH00} also uses
the double ratio of like-sign pairs ($N_{\pi}^{++,--}(Q)$) and
unlike-sign pairs $N_{\pi}^{+-}(Q)$ corrected with Monte-Carlo
simulations not including BE effects:
\begin{equation}
R^*(Q) ~=~ \left.\left(\frac{ N_{\pi}^{++,--}(Q) }
{ N_{\pi}^{+-}(Q) } \right)^{data}\right/
\left(\frac{ N_{\pi}^{++,--}(Q) }
{ N_{\pi}^{+-}(Q) }\right)^{MC}_{noBE}.
\end{equation}
In case of $\Delta\rho(Q)$, $\delta \rho(Q)$ or $\delta_I(Q)$, we look
for a deviation from 0, while in case of $D(Q)$, $D'(Q)$, $d(Q)$ or
$R^*(Q)$, inter-W BE correlations would manifest themselves by
deviation from 1.

\section{Results}

\begin{figure}[htbp]
\begin{center}
\epsfig{file=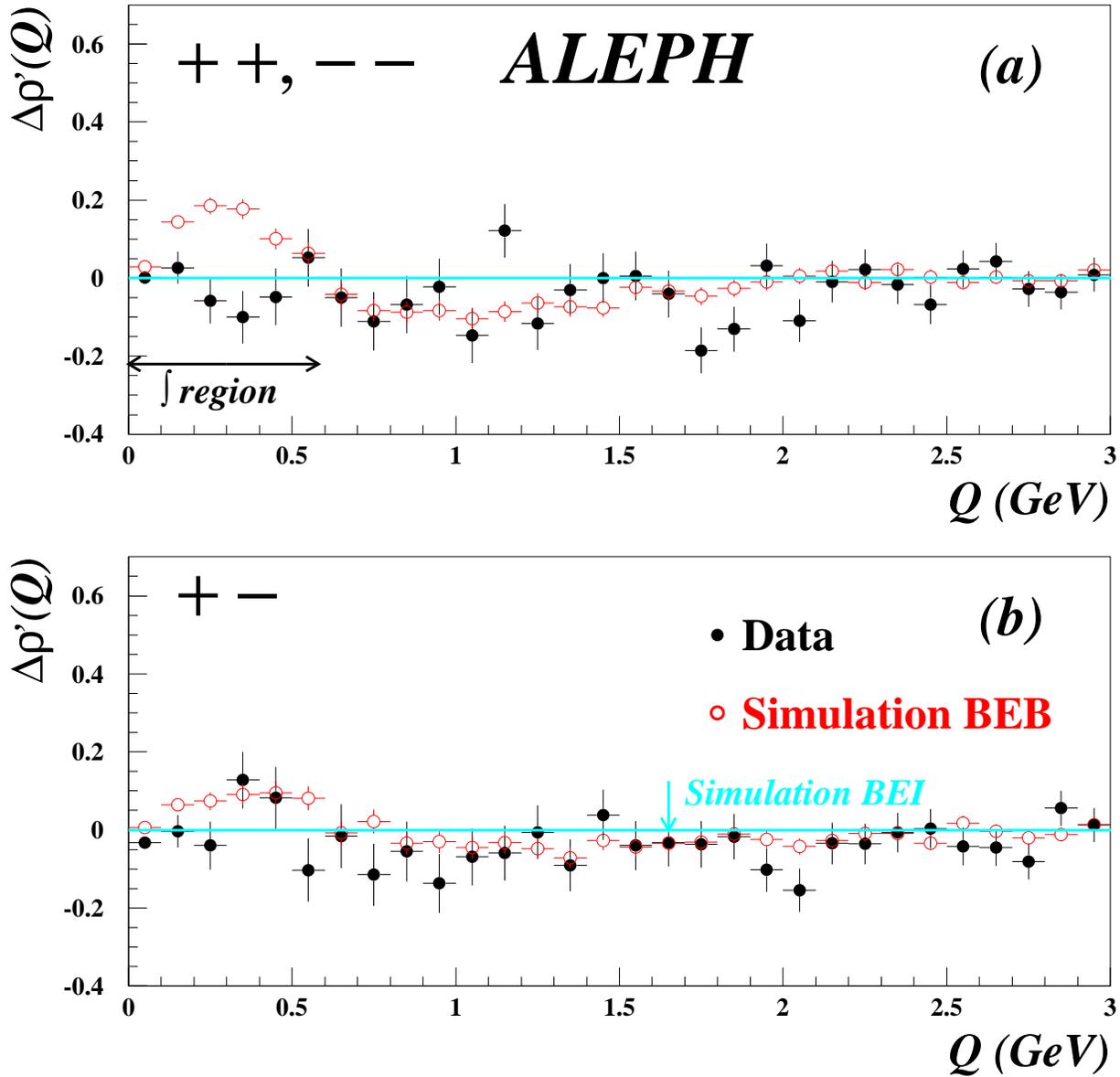,width=1.05\linewidth}
\end{center}
\caption[]{Distribution of the quantity $\Delta\rho'$ for like- and
  unlike-sign pairs as a function of
  $Q$ as measured by the ALEPH collaboration \cite{be:ALEPH05}. 
BEI stands for the case
in which Bose-Einstein correlations do not occur between decay products of
different W bosons, and BEB if they do.}
\label{be:fig:aleph}
\end{figure}

\begin{figure}[htbp]
\begin{center}
\epsfig{file=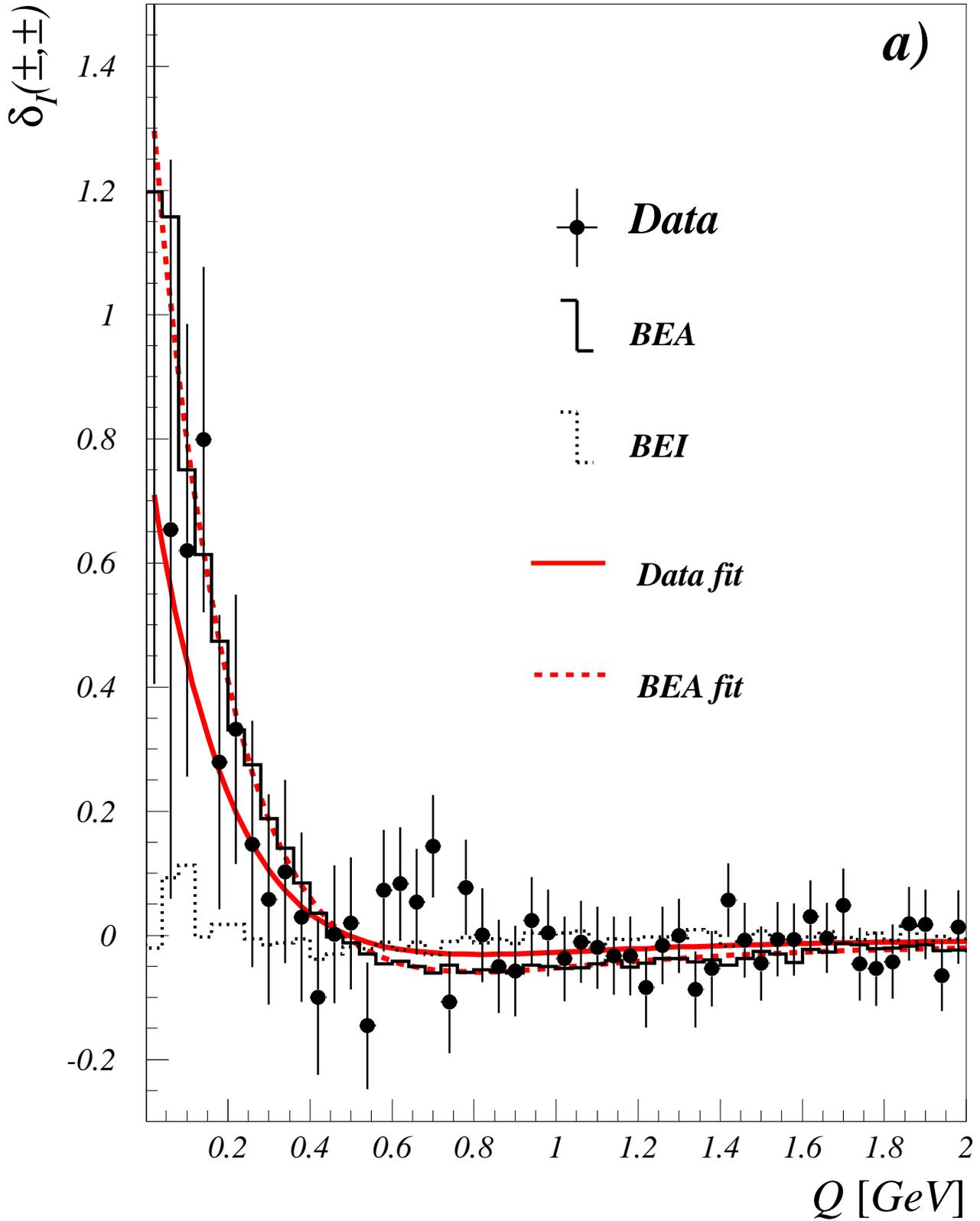,width=\linewidth}
\end{center}
\caption[]{Distributions of the quantity $\delta_I$  for like-sign pairs
as a function of $Q$ as measured by the DELPHI
collaboration \cite{be:DELPHI05}.
The solid line shows the fit results. BEI stands for the case
in which Bose-Einstein correlations do not occur between decay products of
different W bosons, and BEA if they do. }
\label{be:fig:delphi}
\end{figure}

\begin{figure}[htbp]
\begin{center}
\epsfig{file=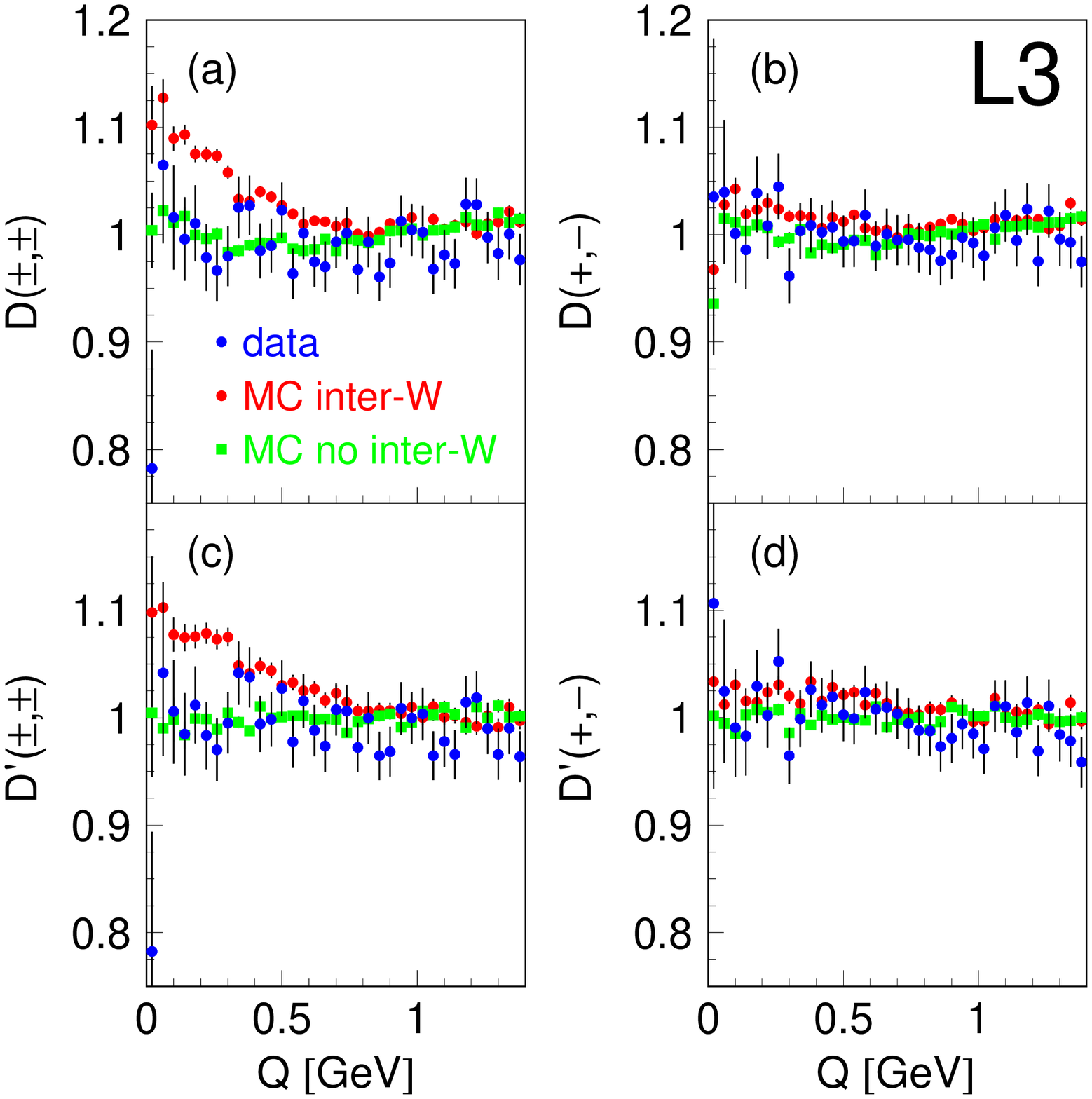,width=1.1\linewidth}
\end{center}
\caption[]{Distributions of the quantity $D$ and $D'$ for like- and
  unlike-sign pairs as a function of $Q$ as measured by the L3
  collaboration \cite{be:L302}.}
\label{be:fig:l3}
\end{figure}

\begin{figure}[htbp]
\begin{center}
\epsfig{file=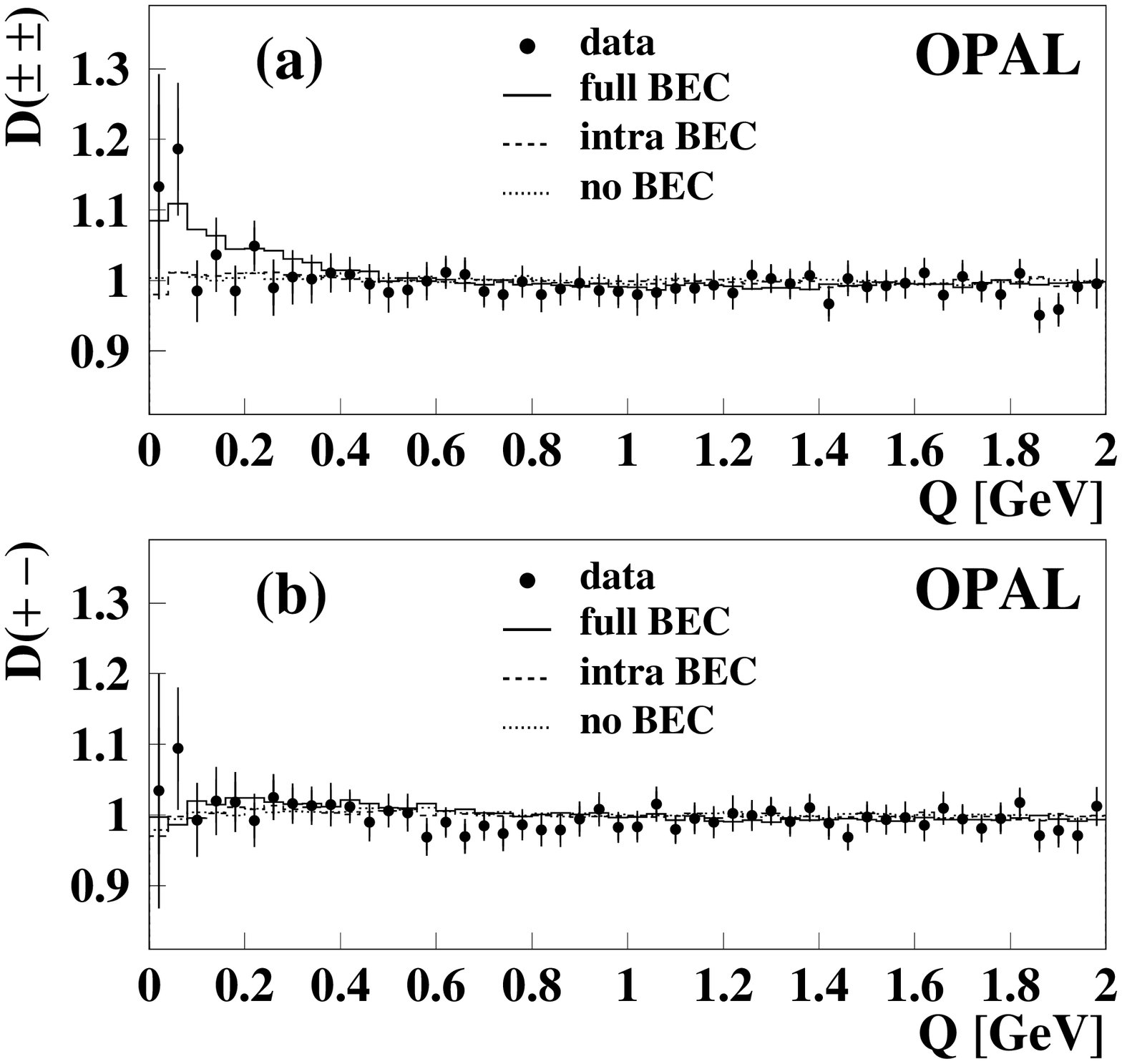,width=\linewidth}
\end{center}
\caption[]{Distribution of the quantity $D$ for like- and
  unlike-sign pairs as a function of
  $Q$ as measured by the OPAL collaboration \cite{be:OPAL05}. }
\label{be:fig:opal}
\end{figure}

The four LEP experiments have published results applying the mixing
method to the full \LEPII\ data sample.  As examples, the
distributions of $\Delta\rho'$ measured by ALEPH~\cite{be:ALEPH05},
$\delta_I$ measured by DELPHI~\cite{be:DELPHI05}, $D$ and $D'$
measured by L3~\cite{be:L302} and $D$ measured by
OPAL~\cite{be:OPAL05} are shown in Figures~\ref{be:fig:aleph},
\ref{be:fig:delphi}, \ref{be:fig:l3} and~\ref{be:fig:opal},
respectively.  In addition ALEPH have published results using $R^*$
variable \cite{be:ALEPH00}. The centre-of-mass energies, luminosities
and the number of events collected by different measurements are shown
in Table~\ref{be:table:lumi}.

\begin{table}[htb]   
\begin{center}
\vspace*{0.4cm}
\begin{tabular}{|l|c|c|r|r|}   \hline
                                                                                
& $\sqrt{s}$  & luminosity          & \multicolumn{2}{c|}{number of events} \\
&    [GeV]              & [pb$^{-1}$]        & $\mathrm{WW \rightarrow q\bar{q}q\bar{q}\,}$      & $\mathrm{WW \rightarrow q\bar{q}\ell \nu\,}$ \\ \hline
ALEPH&  183-209    & 683 &  6155  & 4849  \\ 
DELPHI& 189-209    & 550 &  3252  & 2567  \\  
L3    & 189-209    & 629 &  5100  & 3800  \\
OPAL  & 183-209    & 680 &  4470  & 4533  \\
ALEPH R$^*$ & 172-189 & 242 & 2021 & -    \\\hline
\end{tabular}
\caption[]{The centre-of-mass energies, 
luminosities and the number of events collected by different measurements. }
\label{be:table:lumi}
\end{center}   \end{table}

\begin{table}[htb]
\begin{center}
 \begin{tabular}{|l|l|l|l|l|l|c|}
 \hline
  & data--noBE &  stat. & syst. & corr. syst. &fullBE--noBE & Ref. \\ \hline
{\bf ALEPH (fit to $D'$)}  &{\bf  $-$0.004} &{\bf 0.011} &{\bf 0.014} & {\bf 0.003} & {\bf 0.081} &\cite{be:ALEPH05} \\
ALEPH (integral of $\Delta\rho$)  &  $-$0.127 & 0.143 & 0.199 & 0.044 &  0.699 &\cite{be:ALEPH05} \\
ALEPH (fit to $R^*$)  &  $-$0.004 & 0.0062 & 0.0036 & negligible &  0.0177 &\cite{be:ALEPH00} \\
{\bf DELPHI (fit to  $\delta_I$)} &{\bf $+$0.72} &{\bf 0.29} &{\bf 0.17} &{\bf 0.070} &{\bf 1.40} &\cite{be:DELPHI05} \\
{\bf L3 (fit to $D'$)}  &{\bf $+$0.008} &{\bf 0.018} &{\bf 0.012} &{\bf 0.0042} &{\bf  0.103} & \cite{be:L302} \\
L3 (integral of $\Delta\rho$) &   $+$0.03 & 0.33 &  0.15 &  0.055 & 1.38 & \cite{be:L302} \\
OPAL (integral of $\Delta\rho$) &   $-$0.01 & 0.27 &  0.23 &  0.06 & 0.77 & \cite{be:OPAL05} \\
{\bf OPAL (fit to $D$)} &{\bf $+$0.040} &{\bf 0.038} &{\bf  0.038} &{\bf  0.017} &{\bf 0.120} & \cite{be:OPAL05} \\
OPAL (fit to $D'$) &   $+$0.042 & 0.042 &  0.047 &  0.019 & 0.123 & \cite{be:OPAL05} \\
OPAL (fit to $d$) &   $-$0.017 & 0.055 &  0.050 &  0.003 & 0.133 & \cite{be:OPAL05} \\
\hline
\end{tabular}
\caption[]{
  An overview of the input values from different measurements: the
  difference between the measured correlations and the model without
  inter-W correlations (data--noBE), the corresponding statistical
  (stat.)~and total systematic (syst.)~errors, the correlated
  systematic error contribution (corr.~syst.), and the difference
  between ``fullBE'' and ``noBE'' scenario. The measurements used in 
  the combination are highlighted. }
\label{be:table:bei}
\end{center}
\end{table}

\begin{table}[htb]
\begin{center}
 \begin{tabular}{|l|c|c|c|}
 \hline
  & fraction of the model &  stat. & syst.  \\ \hline
{\bf ALEPH (fit to $D'$)} & {\bf $-$0.05} & {\bf 0.14} &{\bf 0.17}  \\
ALEPH (integral of $\Delta\rho$) &  $-$0.18 & 0.20 & 0.28  \\
ALEPH (fit to $R^*$) &  $-$0.23 & 0.35 & 0.20  \\
{\bf DELPHI (fit to $\delta_I$)} & {\bf  $+$0.51} &{\bf 0.21} &{\bf 0.12}  \\
{\bf L3  (fit to $D'$)} & {\bf $+$0.08} &{\bf 0.17} &{\bf 0.12}  \\
L3 (integral of $\Delta\rho$) & $+$0.02 & 0.24 & 0.11 \\
OPAL (integral of $\Delta\rho$) &   $-$0.01 & 0.35 & 0.30  \\
{\bf OPAL  (fit to $D$)} & {\bf  $+$0.33} &{\bf 0.32} &{\bf 0.32}  \\
OPAL  (fit to $D'$) &   $+$0.34 & 0.34 & 0.38  \\
OPAL  (fit to $d$) &   $-$0.13 & 0.41 & 0.38  \\
\hline
 \end{tabular}
\caption[]{
  The measured size of correlations expressed as the relative fraction
  of the model with inter-W correlations (see Eq. \ref{eq:be:model-fra}
and Table \ref{be:table:bei}). The measurements used in 
  the combination are highlighted.}
\label{be:table:rel}
\end{center}
\end{table}

A simple combination procedure is available through a $\chi^2$ average
of the numerical results of each
experiment~\cite{be:ALEPH00,be:ALEPH05,be:DELPHI05, be:L302,
be:OPAL05} with respect to a specific BE model under study, here based
on comparisons with various tuned versions of the LUBOEI
model~\cite{be:PYTHIA57,mw:bib:LUBOEI}.
The tuning is performed by adjusting the parameters of the model to
reproduce correlations in samples of Z and semileptonic W decays, and
applying identical parameters to the modelling of inter-W correlations
(so-called ``fullBE'' scenario).  In this way the tuning of each
experiment takes into account detector systematics in the track
measurements.

An important advantage of the combination procedure used here is that
it allows the combination of results obtained using different
analyses.  The combination procedure assumes a linear dependence of
the observed size of BE correlations on various estimators used to
analyse the different distributions.  It is also verified that there
is a linear dependence between the measured W mass shift and the
values of these estimators~\cite{be:lep-be}.  The estimators are: the
integral of the $\Delta\rho(Q)$ distribution (ALEPH, L3, OPAL); the
parameter $\Lambda$ when fitting the function $N(1+\delta
Q)(1+\Lambda\exp(-k^2Q^2))$ to the $D'(Q)$ distribution, with $N$
fixed to unity (L3), or $\delta$ fixed to zero and $k$ fixed to the
value obtained from a fit to the full BE sample (ALEPH); the parameter
$\Lambda$ when fitting the function $N(1+\delta
Q)(1+\Lambda\exp(-Q/R))$ to the $D(Q)$, $D(Q)'$ and $d$ distributions,
with $R$ fixed to the value obtained from a fit to the full BE sample
(OPAL); the parameter $\Lambda$ when fitting the function
$\Lambda\exp(-RQ)(1+\epsilon RQ)+\delta(1+\frac{\rho_{2}^{W}}{
\rho_{2}^{mix}})$ to the $\delta_I$ distribution, with $R$ and
$\epsilon$ fixed to the value obtained from a fit to the full BE
sample (DELPHI); and finally the integral of the term describing the
BE correlation part, $\int \lambda\exp(-\sigma^2Q^2)$, when fitting
the function $\kappa(1+\epsilon Q)(1+\lambda\exp(-\sigma^2Q^2))$ to
the $R^*(Q)$ distribution (ALEPH).

The size of the correlations for like-sign pairs of particles measured
in terms of these estimators is compared with the values expected in
the model with and without inter-W correlations in
Table~\ref{be:table:bei}.  Table~\ref{be:table:rel} summarizes the
normalized fractions of the model seen.  Note that DELPHI also finds a
1.4 standard deviation effect for pairs of unlike-sign particles from
different W bosons\cite{be:DELPHI05}, compatible with the prediction
of the LUBOEI model with full strength correlations.

\begin{figure}[htbp]
   \begin{center}
     \mbox{\hspace*{-1.0cm}\epsfig{file=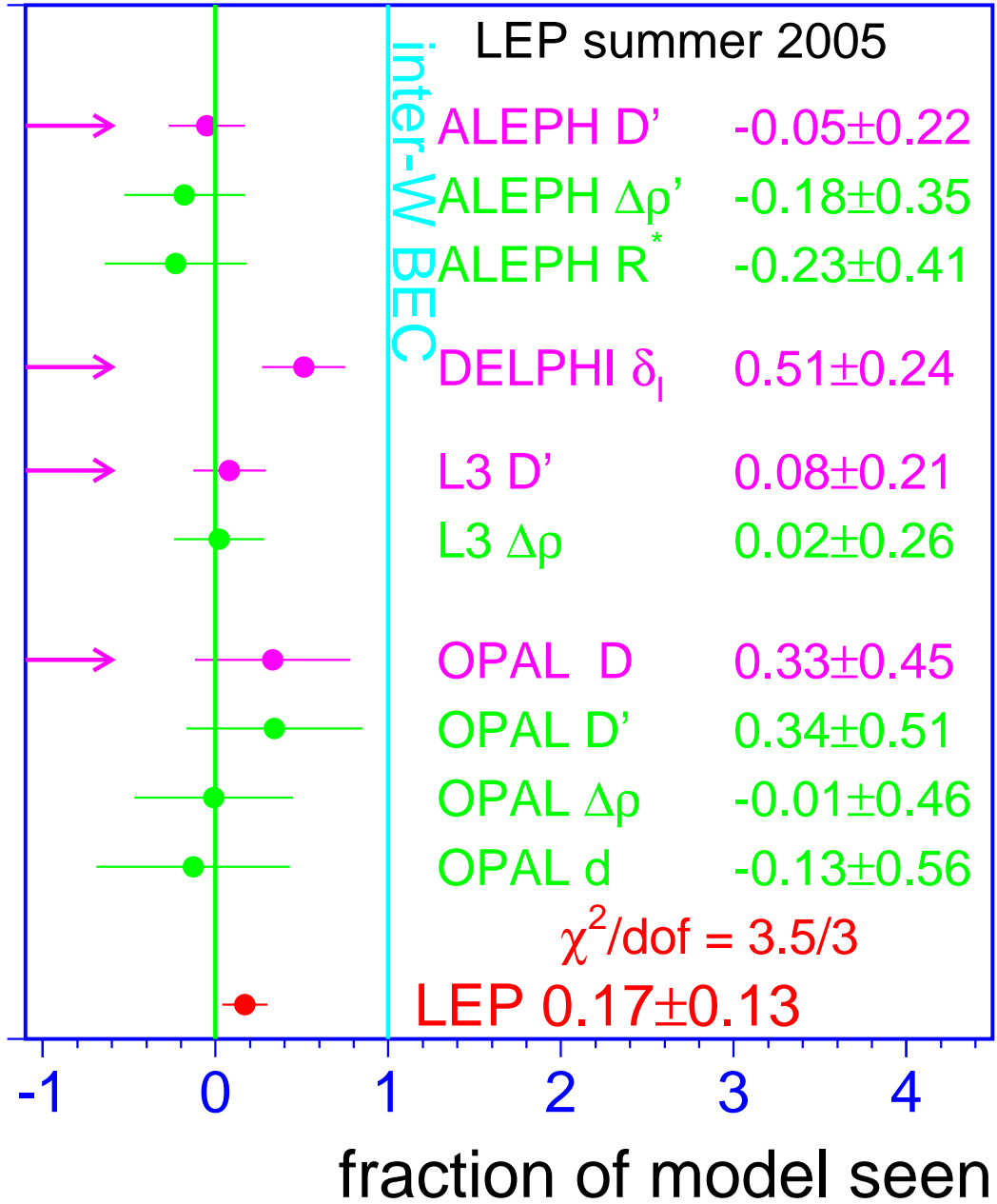,width=\linewidth}}
   \end{center}
 \caption[]{%
   Correlations expressed as the relative fraction of the model with
   inter-W correlations. The arrows indicate the measurements used in
   the combination. The LEP combination is shown at the bottom. }
 \label{be:chi-comb}
\end{figure}

For the combination of the above measurements one has to take into
account correlations between them. Correlations between results of the
same experiment are strong and are not available. It is however found,
for example, that taking reasonable value of these correlations and
combining three ALEPH measurements, one obtains normalized fractions
of the model seen very close to those of the most precise measurement.
Therefore, for simplicity, the combination of the most precise
measurements of each experiment is made here: $D'$ from ALEPH,
$\delta_I$ from DELPHI, $D'$ from L3 and $D$ from OPAL.  In this
combination only the uncertainties in the understanding of the
background contribution in the data are treated as correlated between
experiments (denoted as ``corr. syst.''  in Table~\ref{be:table:bei}).
The combination via a MINUIT fit gives:
\begin{equation}
\frac{\mathrm{data - model(noBE)}}{\mathrm{model(fullBE)-model(noBE)}}
 ~ = ~ 0.17 \pm 0.095(stat.)\pm 0.085(sys.) ~ = ~ 0.17 \pm 0.13~\,,~~
\label{eq:be:model-fra}
\end{equation}
where ``noBE'' includes correlations between decay products of each W,
but not the ones between decay products of different Ws and ``fullBE''
includes all the correlations.  A $\chi^2$/dof=3.5/3 of the fit is
observed.  The measurements and their average are shown in
Figure~\ref{be:chi-comb}. The measurements used in the combination are
marked with an arrow.

In conclusion, the results of LEP experiments are in good agreement
($\chi^2$/dof=3.5/3).  The LUBOEI model of BE correlations between
pions from different W bosons is disfavoured. The 68\% confidence
level (CL) upper limit on these correlations is
\begin{eqnarray}
0.17~+~0.13~=~0.30~.
\end{eqnarray}    
This result can be translated into a 68\% CL upper limit on the shift
of the W mass measurements due to the BE correlations between
particles from different Ws, $\Delta\MW$, assuming a linear dependence
of $\Delta\MW$ on the size of the correlation.  For the specific BE
model investigated, LUBOEI, a shift of $-35~\MeV$ in the W mass is
obtained at full BE correlation strength\cite{be:LEPW}. Thus the 68\%
CL upper limit on the magnitude of the mass shift within the LUBOEI
model is:
\begin{eqnarray}
 |\Delta\MW| & = & 0.30 \cdot 35~\MeV
             ~ = ~ 11~\MeV~~ .
\end{eqnarray}

\boldmath
\chapter{W-Boson Mass and Width at \LEPII}
\label{sec-MW}
\unboldmath

\updates{ New combinations using the final results from OPAL are
presented. The results from ALEPH, DELPHI and L3 used for the
combinations are preliminary, and thus the combinations are
preliminary.

Note that some recent publications~\cite{LEP2L3MWGW} are not yet
included in the combinations.  }

\section{Introduction}

The W boson mass and width results presented here are obtained from
data recorded over a range of centre-of-mass energies,
$\roots=161-209$~\GeV, during the 1996-2000 operation of the LEP
collider (\LEPII), and correspond to a luminosity of about $700~\ipb$
per experiment.  The results reported by the ALEPH, DELPHI and L3
collaborations are preliminary.  The ALEPH result does not include an
analysis of the small amount of data (about $10$~\ipb) collected in
1996 at a centre-of-mass energy of 172~\GeV.  The results presented in
this note differ from those in the previous
combination~\cite{bib-EWEP-04} due to the final results from the OPAL
collaboration, which are based on their complete $\LEPII$ data set.
The final W mass results in the $\WWqqqq$ channel from OPAL use an
analysis trading statistical accuracy for reduced FSI systematic
uncertainties in order to achieve a smaller total error.  The
sensitivity to FSI effects is reduced by excluding tracks of less than
$2.5~\GeV$ momentum in the determination of the jet directions in the
OPAL analysis.

The observables W mass and total width quoted here correspond to a
definition based on a Breit-Wigner denominator with a mass-dependent
width, $|(m^2-\Mw^2) + i m^2 \Gw/\Mw|$.

\section{W Mass Measurements}

Since 1996 until 2000 the LEP $\epem$ collider operated above the
threshold for $\WW$ pair production. Initially, 10~\ipb\ of data were
recorded close to the $\WW$ pair production threshold. At this energy
the $\WW$ cross section is sensitive to the W boson mass, $\Mw$.
Table \ref{mw:tab:wmass_threshold} summarises the W mass results from
the four LEP collaborations based on these
data~\cite{common_bib:adloww161}.

\begin{table}[htbp]
\begin{center}
\begin{tabular}{|r|c|}\hline
\multicolumn{2}{|c|}{THRESHOLD ANALYSIS~\cite{common_bib:adloww161}} \\
 Experiment &   \Mw(threshold)/\GeVm     \\ \hline
   ALEPH    & $80.14\pm0.35$             \\ 
   DELPHI   & $80.40\pm0.45$             \\
   L3       & $80.80^{+0.48}_{-0.42}$    \\
   OPAL     & $80.40^{+0.46}_{-0.43}$    \\ \hline
\end{tabular}
\caption{W mass measurements from the $\WW$ threshold cross section at
$\roots=161$~\GeV. The errors include statistical and systematic
contributions.}
\label{mw:tab:wmass_threshold}
\end{center}
\end{table} 

\begin{table}[htbp]
\begin{center}
\begin{tabular}{|r|c|c||c|}\hline
\multicolumn{1}{|c|}{ } & \multicolumn{3}{c|}{DIRECT RECONSTRUCTION } \\
           & \WWqqln         & \WWqqqq         & Combined        \\   
Experiment & \Mw/\GeVm        & \Mw/\GeVm        & \Mw/\GeVm        \\ \hline
     ALEPH \cite{mw:bib:ALEPHRevised}
           & $80.375 \pm 0.062$ & $80.431 \pm 0.117$ & $80.385 \pm 0.058$ \\ 
     DELPHI \cite{common_bib:delww172,mw:bib:D-mw183,mw:bib:D-mw189,mw:bib:D-mw20X}  
           & $80.414 \pm 0.089$ & $80.374 \pm 0.119$ & $80.402 \pm 0.075$ \\ 
     L3 \cite{common_bib:ltrww172,mw:bib:L-mw183,mw:bib:L-mw189,mw:bib:L-mw19X,mw:bib:L-mw20X}      
           & $80.314 \pm 0.087$ & $80.485 \pm 0.127$ & $80.367 \pm 0.078$ \\ 
     OPAL\cite{mw:bib:O-final}
           & $80.449 \pm 0.063$ & $80.353 \pm 0.083$ & $80.416 \pm 0.053$ \\ \hline
\end{tabular}
\caption{W mass measurements from direct reconstruction
         ($\roots=172-209$~\GeV). Results are given for the
         semi-leptonic, fully-hadronic channels and the combined
         value.  The $\WWqqln$ results from the OPAL collaboration
         include mass information from the $\WWlnln$ channel. The
         results given here differ from those in the publications of
         the individual experiments as they have been recalculated
         imposing FSI uncertainties evaluated in the same way.}
\label{mw:tab:wmass_experiments}
\end{center}
\end{table}

Subsequently LEP operated at energies significantly above the $\WW$
threshold, where the $\epem\rightarrow\WW$ cross section has little
sensitivity to $\Mw$. For these higher energy data $\Mw$ is measured
through the direct reconstruction of the W boson's invariant mass from
the observed jets and leptons.  Table~\ref{mw:tab:wmass_experiments}
summarises the W mass results presented by the four LEP experiments
using the direct reconstruction method.  The combined values of $\Mw$
from each collaboration take into account the correlated systematic
uncertainties between the decay channels and between the different
years of data taking. In addition to the combined numbers, each
experiment presents mass measurements from $\WWqqln$ and $\WWqqqq$
channels separately.  The DELPHI and OPAL collaborations provide
results from independent fits to the data in the $\qqln$ and $\qqqq$
decay channels separately and hence account for correlations between
years but do not need to include correlations between the two
channels. The $\qqln$ and $\qqqq$ results quoted by the ALEPH and L3
collaborations are obtained from a simultaneous fit to all data which,
in addition to other correlations, takes into account the correlated
systematic uncertainties between the two channels. The L3 result is
unchanged when determined through separate fits.  The systematic
uncertainties in the $\WWqqqq$ channel as quoted by the experiments
show a large variation between experiments; this is caused by
differing estimates of the possible effects of Colour Reconnection
(CR) and Bose-Einstein Correlations (BEC), discussed below.  The
systematic errors in the $\WWqqln$ channel are dominated by
uncertainties from hadronisation, with estimates ranging from
15~$\MeVm$ to 30~$\MeVm$.

\section{Combination Procedure}
 
A combined LEP W mass measurement is obtained from the results of the
four experiments. In order to perform a reliable combination of the
measurements, a more detailed input than that given in
Table~\ref{mw:tab:wmass_experiments} is required.  Each experiment
provided a W mass measurement for both the $\WWqqln$ and $\WWqqqq$
channels for each of the data taking years (1996-2000) that it had
analysed. In addition to the four threshold measurements a total of 36
direct reconstruction measurements are supplied: DELPHI provided 10
measurements (1996-2000), L3 gave 8 measurements (1996-2000) having
already combined the 1996 and 1997 results, ALEPH provided 8
measurements (1997-2000) and OPAL gave 10 measurements
(1996-2000). The $\WWlnln$ channel is also analysed by the OPAL
(1996-2000) collaboration; the lower precision results obtained from
this channel are combined with the $\WWqqln$ channel mass
determinations.

Subdividing the results by data-taking years enables a proper
treatment of the correlated systematic uncertainty from the LEP beam
energy and other dependences on the centre-of-mass energy or
data-taking period.  A detailed breakdown of the sources of systematic
uncertainty are provided for each result and the correlations
specified. The inter-year, inter-channel and inter-experiment
correlations are included in the combination. The main sources of
correlated systematic errors are: colour reconnection, Bose-Einstein
correlations, hadronisation, the LEP beam energy, and uncertainties
from initial and final state radiation. The full correlation matrix
for the LEP beam energy is employed\cite{mw:bib:energy}.  The
combination is performed and the evaluation of the components of the
total error assessed using the Best Linear Unbiased Estimate (BLUE)
technique, see Reference~\citen{common_bib:BLUE}.

\subsubsection{FSI Effects}
\label{mw:sec:cr}

A preliminary study of colour reconnection has been made by the LEP
experiments using the particle flow method \cite{mw:bib:CRcomb} on a
sample of fully-hadronic WW events, see
chapter~\ref{sec-CR}.\footnote{Note that this preliminary LEP analysis
does not take into account the final OPAL colour-reconnection analysis
published recently~\cite{cr:bib:O-final}.} These results are
interpreted in terms of the reconnection parameter $k_i$ of the SK-I
model \cite{mw:bib:ski} and yield a $68\%$ confidence level range of:
\begin{eqnarray}
 0.39 < k_i < 2.13 \,.
\end{eqnarray}
The method was found to be insensitive to the \HERWIG\ and \ARIADNE-II
models of colour reconnection.

\begin{figure}[p]
\begin{center}
\mbox{\epsfig{file=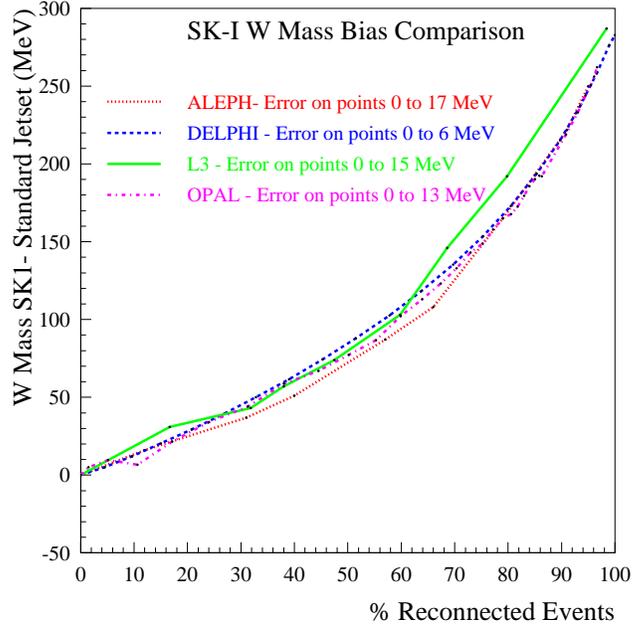,width=0.55\textwidth}}
\vskip -0.5cm
\caption{W mass bias obtained in the SK-I model of colour reconnection
  relative to a simulation without colour reconnection as a function
  of the fraction of events reconnected for the fully-hadronic decay
  channel at a centre of mass energy of 189 \GeV .  The preliminary
  W-mass analyses of all four LEP experiments show similar sensitivity
  to this effect. The points connected by the lines have correlated
  uncertainties increasing to the right in the range indicated.  The
  final OPAL analysis used in the combination has a lower sensitivity
  as discussed in the text.}
 \label{mw:fig:sk1}
\end{center}
\end{figure}

\begin{figure}[p]
\begin{center}
\mbox{\epsfig{file=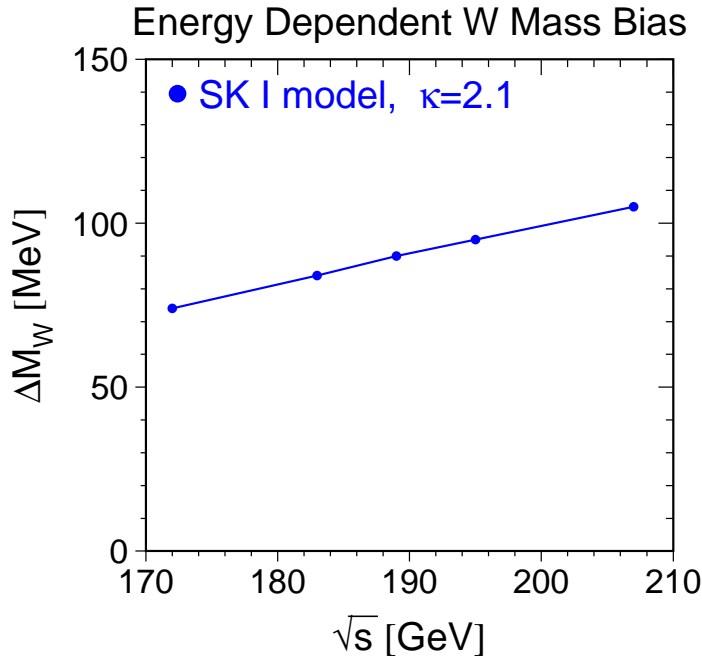,width=0.55\textwidth}}
\vskip -0.8cm
\caption{The values used in the W Mass combination for the uncertainty
 due to colour reconnection for the preliminary results from ALEPH,
 DELPHI and L3 are shown as a function of the centre of mass
 energy. These values were obtained from a linear fit to simulation
 results obtained with the SK1 model of colour reconnection at $k_i =
 2.13$. }
\label{mw:fig:sk1cm}
\end{center}
\end{figure}

Studies of simulation samples have demonstrated that the preliminary
results of all four experiments are equally sensitive to colour
reconnection effects, {\em i.e.} when looking at the same CR model
similar biases are seen by all experiments. This is shown in
Figure~\ref{mw:fig:sk1} for the SK-I model as a function of the
fraction of reconnected events.  For this reason a common value for
the preliminary ALEPH, DELPHI and L3 results of the CR systematic
uncertainty is used in the combination, which is set from a linear
extrapolation of simulation results obtained at $k_i = 2.13$.  The
values used in the combination are: $74~\MeVm$ uncertainty for the
1996 data at a centre-of-mass energy of $172~\GeV$, $84~\MeVm$ for
1997 at $183~\GeV$, $90~\MeV$ for 1998 at $189~\GeV$, $95~\MeV$ for
1999 at $195~\GeV$ and $105~\MeVm$ for 2000 at $207~\GeV$, they are
shown in Figure \ref{mw:fig:sk1cm}.  For the final OPAL results, an
energy-independent value of $39~\MeV$, corresponding to the same
$k_i=2.13$ value, is used.  No offset is applied to the central value
of \Mw\ due to colour reconnection effects and a symmetric systematic
error has been imposed.

Early \Mw\ combinations had relied upon theoretical expectations of
colour reconnection effects, in which there is considerable
uncertainty. This new data driven approach achieves a more robust
uncertainty estimate at the expense of a significantly increased
colour reconnection uncertainty. The \ARIADNE-II and \HERWIG\ models
of colour reconnection have also been studied and the W mass shift was
found to be lower than that from SK-I with $k_i = 2.13$ used for the
combination.

For Bose-Einstein correlations, a similar test has been made of the
respective experimental sensitivities with the LUBOEI
\cite{mw:bib:LUBOEI} model: the experiments observed compatible mass
shifts for their preliminary results. A common value for ALEPH, DELPHI
and L3 of the systematic uncertainty from BEC of 35~\MeVm\ is assumed
from studies of the LUBOEI model, while 19~\MeVm\ is used for the
final OPAL results.  These values may be compared with recent direct
measurements from LEP of this effect, Chapter~\ref{sec-BE}, where the
observed Bose-Einstein effect was of smaller magnitude than in the
LUBOEI model, see chapter~\ref{sec-BE}. Hence, the currently assigned
35~\MeVm\ and 19~\MeVm\ uncertainties are considered conservative
estimates.

\section{LEP Combined W Boson Mass }

The combined LEP W mass from direct reconstruction alone is
\begin{eqnarray}
   \Mw(\mathrm{direct}) = 80.391\pm0.027(\mathrm{stat.})\pm0.029(\mathrm{syst.})~\GeVm,
\end{eqnarray}
with a $\chi^2$/d.o.f. of 29.2/35, corresponding to a $\chi^2$
probability of 74\%.  The weight of the fully-hadronic channel in the
combined fit, previously 0.10~\cite{bib-EWEP-04} as a consequence of
the relatively large size of the estimates of the systematic errors
from CR and BEC, increased to 0.16 in this combination due to OPAL's
final results with reduced FSI sensitivity.  

Table \ref{mw:tab:errors} gives a breakdown of the contribution to the
total error of the various sources of systematic errors. Besides FSI
effects, the largest contribution to the systematic error comes from
hadronisation uncertainties, which are conservatively treated as
correlated between the two channels, between experiments and between
years. In the absence of systematic effects the current LEP
statistical precision on $\Mw$ would be $20$~$\MeVm$: the statistical
error contribution in the LEP combination is larger than this
(27~$\MeVm$) due to the reduced weight of the fully-hadronic channel.
Compared to the previous combination~\cite{bib-EWEP-04}, the final
OPAL results lead to an increased statistical error in the $\WWqqqq$
combination but reduced systematics due to CR and BEC, for an overall
smaller total error.

\begin{table}[htbp]
\begin{center}
\begin{tabular}{|l|r|r||r|}\hline
       Source  &  \multicolumn{3}{|c|}{Systematic Error on \Mw\ ($\MeVm$)}  \\  
                             &  \qqln & \qqqq  & Combined  \\ \hline   
 ISR/FSR                     & 10 &  9 & 10 \\
 Hadronisation               & 17 & 18 & 17 \\
 Detector Systematics        & 14 &  8 & 13 \\
 LEP Beam Energy             & 14 & 11 & 13 \\
 Colour Reconnection         & $-$& 49 &  7 \\
 Bose-Einstein Correlations  & $-$& 22 &  3 \\
 Other                       &  3 & 20 &  5 \\ \hline
 Total Systematic            & 28 & 63 & 29 \\ \hline
 Statistical                 & 31 & 48 & 27 \\ \hline\hline
 Total                       & 42 & 79 & 39 \\ \hline
 & & & \\
 Statistical in absence of Systematics  & 30 & 27 & 20 \\ \hline

\end{tabular}
 \caption{Error decomposition for the combined LEP W mass results.
          Detector systematics include uncertainties in the jet and
          lepton energy scales and resolution. The `Other' category
          refers to errors, all of which are uncorrelated between
          experiments, arising from: simulation statistics, background
          estimation, four-fermion treatment, fitting method and event
          selection. The error decomposition in the $\qqln$ and
          $\qqqq$ channels refers to the independent fits to the
          results from the two channels separately.}
 \label{mw:tab:errors}
\end{center}
\end{table}

In addition to the direct reconstruction results, the W boson mass is
measured at LEP from the 10~\ipb\ per experiment of data recorded at
threshold for W pair production:
\begin{eqnarray}
      {\Mw(\mathrm{threshold}) = 
  80.40\pm0.20(\mathrm{stat.})\pm
          0.07(\mathrm{syst.})\pm0.03(\mathrm{E_{beam}})~\GeVm}.
\end{eqnarray}
When the threshold measurements are combined with the much more
precise results obtained from direct reconstruction one achieves a W
mass measurement of
\begin{eqnarray}
           \Mw = 80.392\pm0.027(\mathrm{stat.})\pm0.028(\mathrm{syst.}) \GeVm.
\end{eqnarray}
The LEP beam energy uncertainty is the only correlated systematic
error source between the threshold and direct reconstruction
measurements.  The threshold measurements have a weight of only $0.02$
in the combined fit.  This LEP combined result is compared with the
results (threshold and direct reconstruction combined) of the four LEP
experiments in Figure \ref{mw:fig:mwgw}.

\section{Consistency Checks}

The difference between the combined W boson mass measurements obtained
from the fully-hadronic and semi-leptonic channels,
$\Delta\Mw(\qqqq-\qqln)$, is determined:
\begin{eqnarray}
 \Delta\Mw(\qqqq-\qqln) = +21\pm42~\MeVm.
\end{eqnarray}
A significant non-zero value for $\Delta\Mw$ could indicate that CR
and BEC effects are biasing the value of \Mw\ determined from \WWqqqq\
events.  Since $\Delta\Mw$ is primarily of interest as a check of the
possible effects of final state interactions, the errors from CR and
BEC are set to zero in its determination. The above result on the mass
difference is obtained from a fit where the imposed correlations are
the same as those for the results given in the previous sections. This
result is almost unchanged if the systematic part of the error on
$\Mw$ from hadronisation effects is considered as uncorrelated between
channels, although the uncertainty increases,
$\Delta\Mw=17\pm48~\MeVm$.

\begin{figure}[p]
\begin{center}
\mbox{\epsfig{file=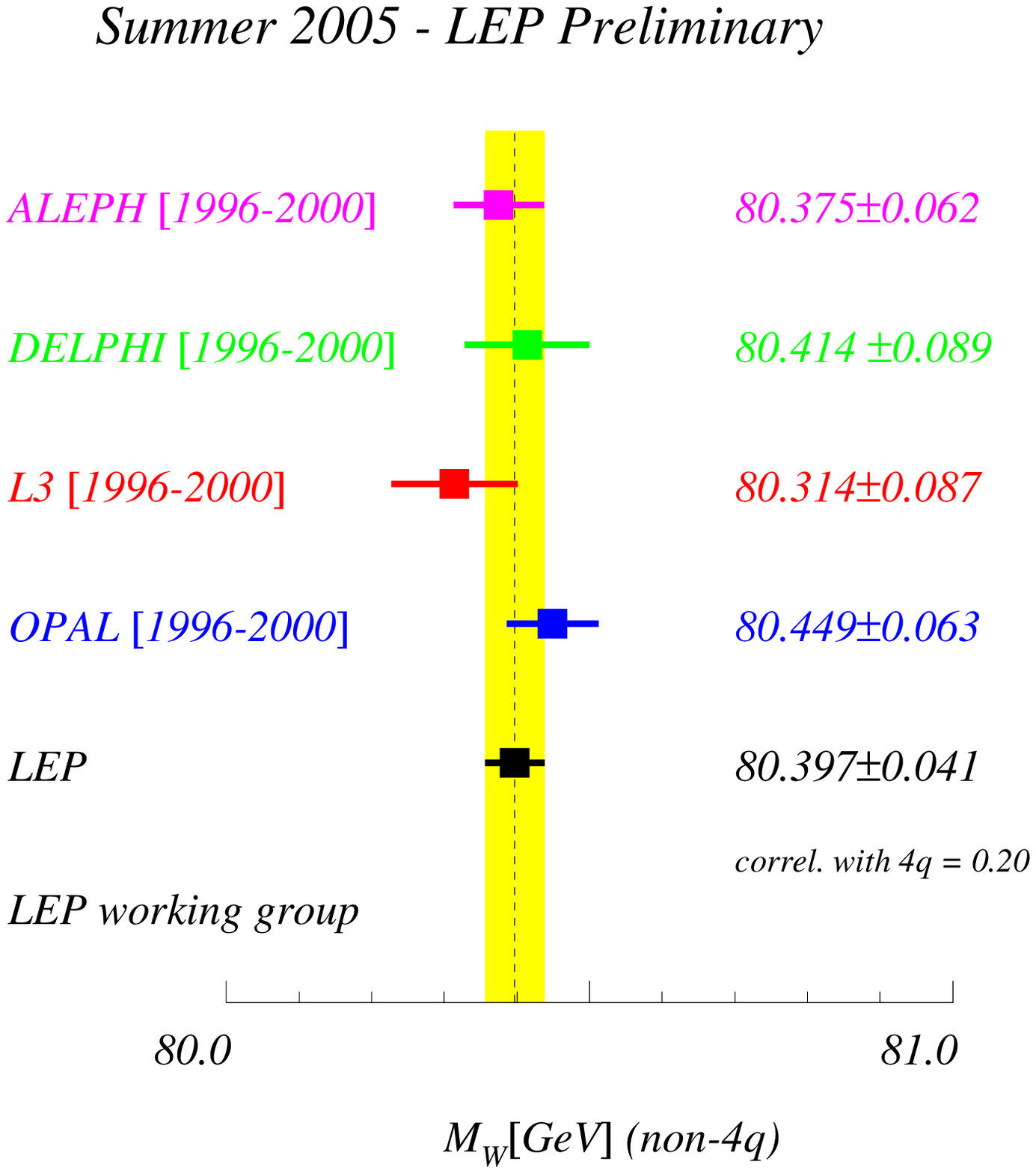,width=0.495\textwidth}}
\hfill
\mbox{\epsfig{file=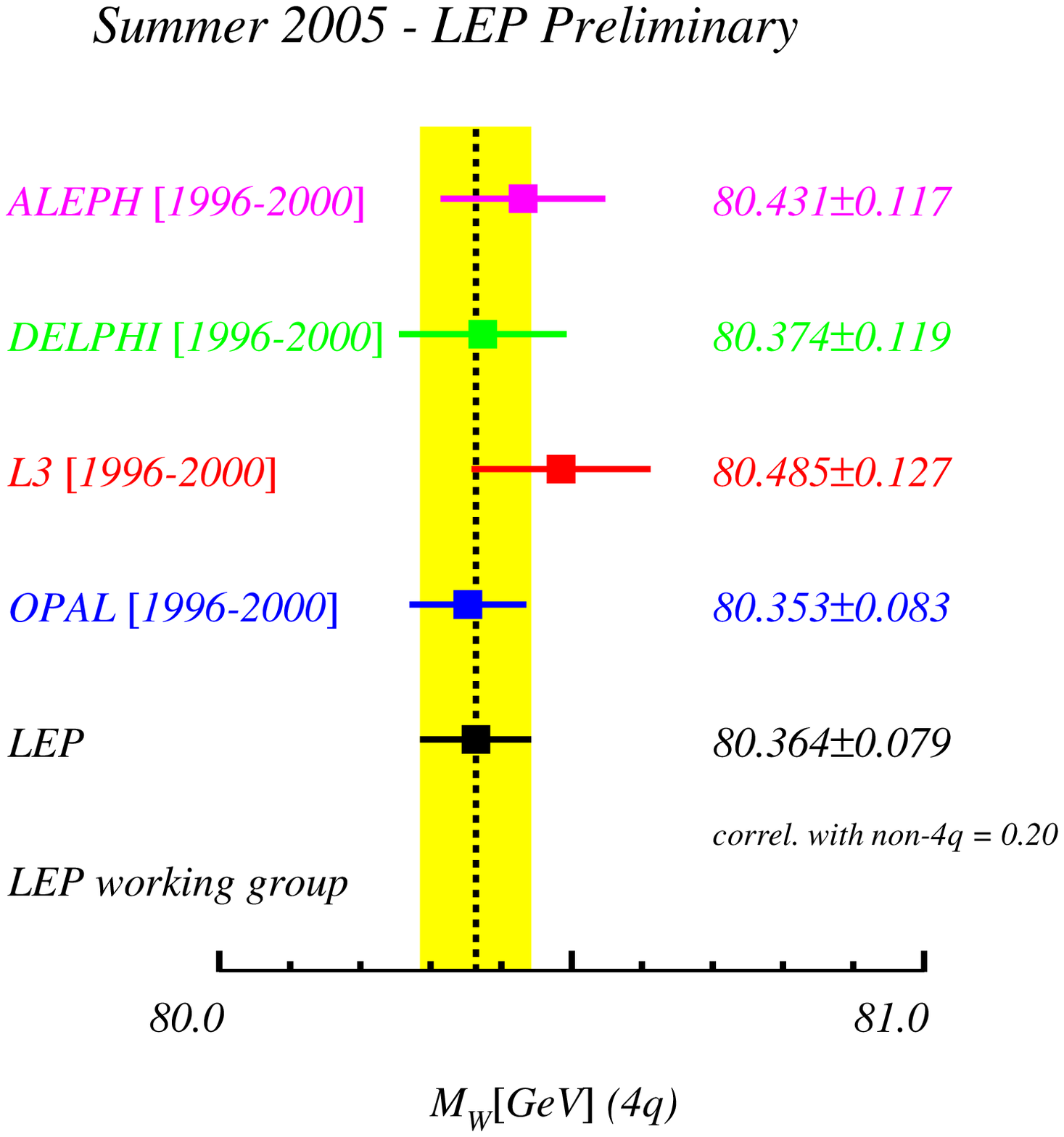,width=0.495\textwidth}}
\vskip -1cm
\caption{\label{mw:fig-qqlnqqqq} The W mass measurements from the
          $\WWqqln$ (left) and $\WWqqqq$ (right) channels obtained by
          the four LEP collaborations compared to the combined
          value. The combined values take into account correlations
          between experiments, years and the two channels.  In the LEP
          combination of the $\qqqq$ results common values for the CR
          and BEC errors of ALEPH, DELPHI and L3 are used (see text).
          The $\Mw$ values from ALEPH, DELPHI and L3 have been
          recalculated for this plot including these common CR and BEC
          errors. The ALEPH and L3 $\qqln$ and $\qqqq$ results are
          correlated since they are obtained from a fit to both
          channels taking into account inter-channel correlations. }
\end{center}
\end{figure}

The masses from the two channels with all errors and correlations
included are:
\begin{eqnarray}
\Mw(\WWqqln) & = & 80.397\pm0.031(\mathrm{stat.})\pm0.028(\mathrm{syst.})~\GeVm,\\
\Mw(\WWqqqq) & = & 80.364\pm0.048(\mathrm{stat.})\pm0.063(\mathrm{syst.})~\GeVm.  
\end{eqnarray}
The two results are correlated with a correlation coefficient of
0.20. The $\chi^2$/d.o.f is 29.0/34, corresponding to a $\chi^2$
probability of 75$\%$.  These results and the correlation between them
can be used to combine the two measurements or to form the mass
difference. The LEP combined results from the two channels are
compared with those quoted by the individual experiments in Figure
\ref{mw:fig-qqlnqqqq}, where the common CR and BEC errors have been
imposed.

Experimentally, separate $\Mw$ measurements are obtained from the
$\qqln$ and $\qqqq$ channels for each of the years of data.  The
combination using only the $\qqln$ measurements yields:
\begin{eqnarray*}
 \Mwindep(\WWqqln) = 80.400\pm0.031(\mathrm{stat.})\pm0.028(\mathrm{syst.})~\GeVm. 
\end{eqnarray*}
The largest contribution to the systematic error arises from the
hadronisation uncertainties, $\pm17$~$\MeVm$.  The combination using
only the $\qqqq$ measurements gives:
\begin{eqnarray*}
 \Mwindep(\WWqqqq) = 80.361\pm0.049(\mathrm{stat.})\pm0.062(\mathrm{syst.})~\GeVm.  
\end{eqnarray*}
where the dominant contributions to the systematic error are from CR
($\pm49$~$\MeVm$) and BEC ($\pm22$~$\MeVm$).

\section{LEP Combined W Boson Width}

\begin{figure}[p]
\begin{center}
\mbox{\epsfig{file=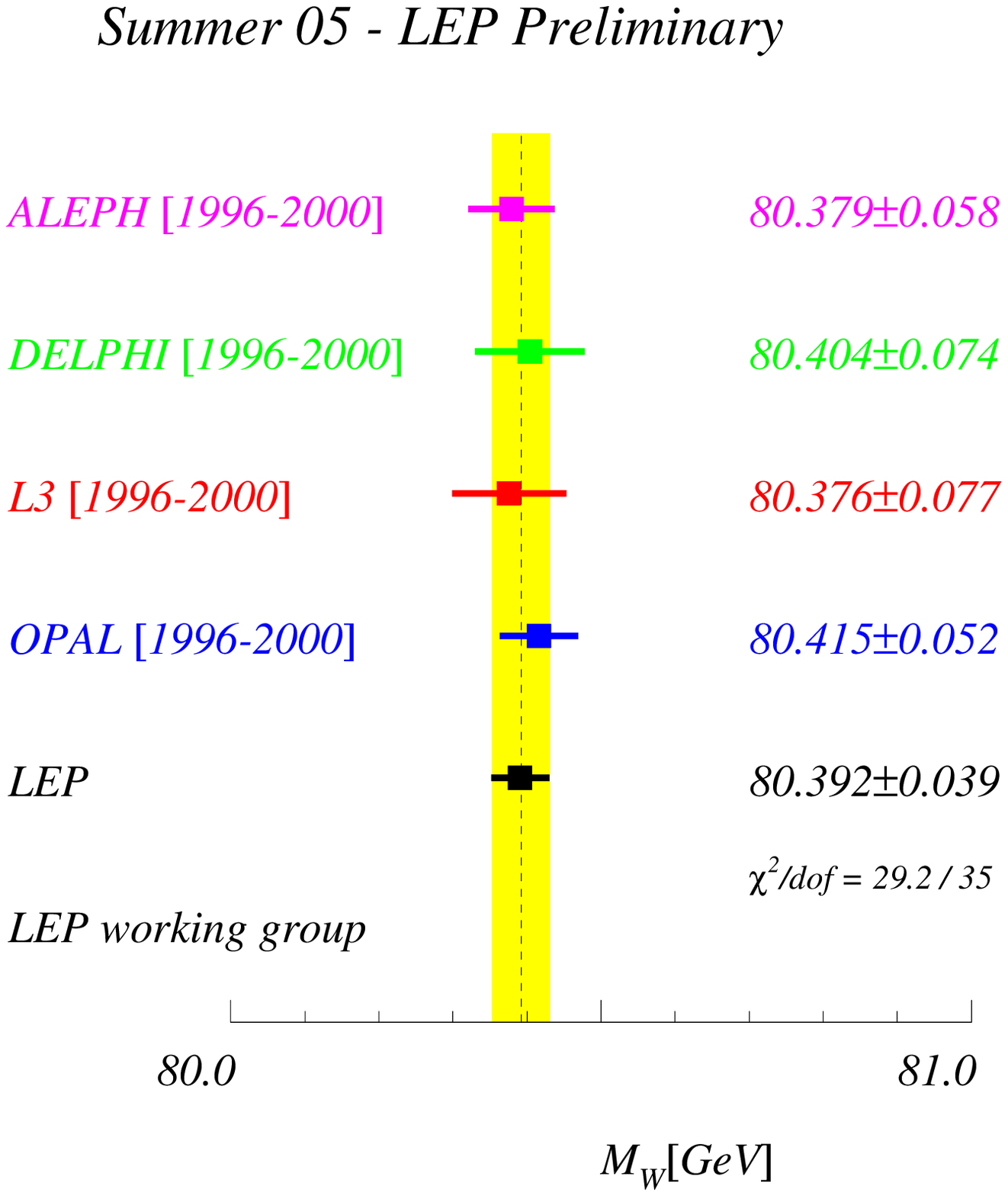,width=0.495\textwidth}}
\hfill
\mbox{\epsfig{file=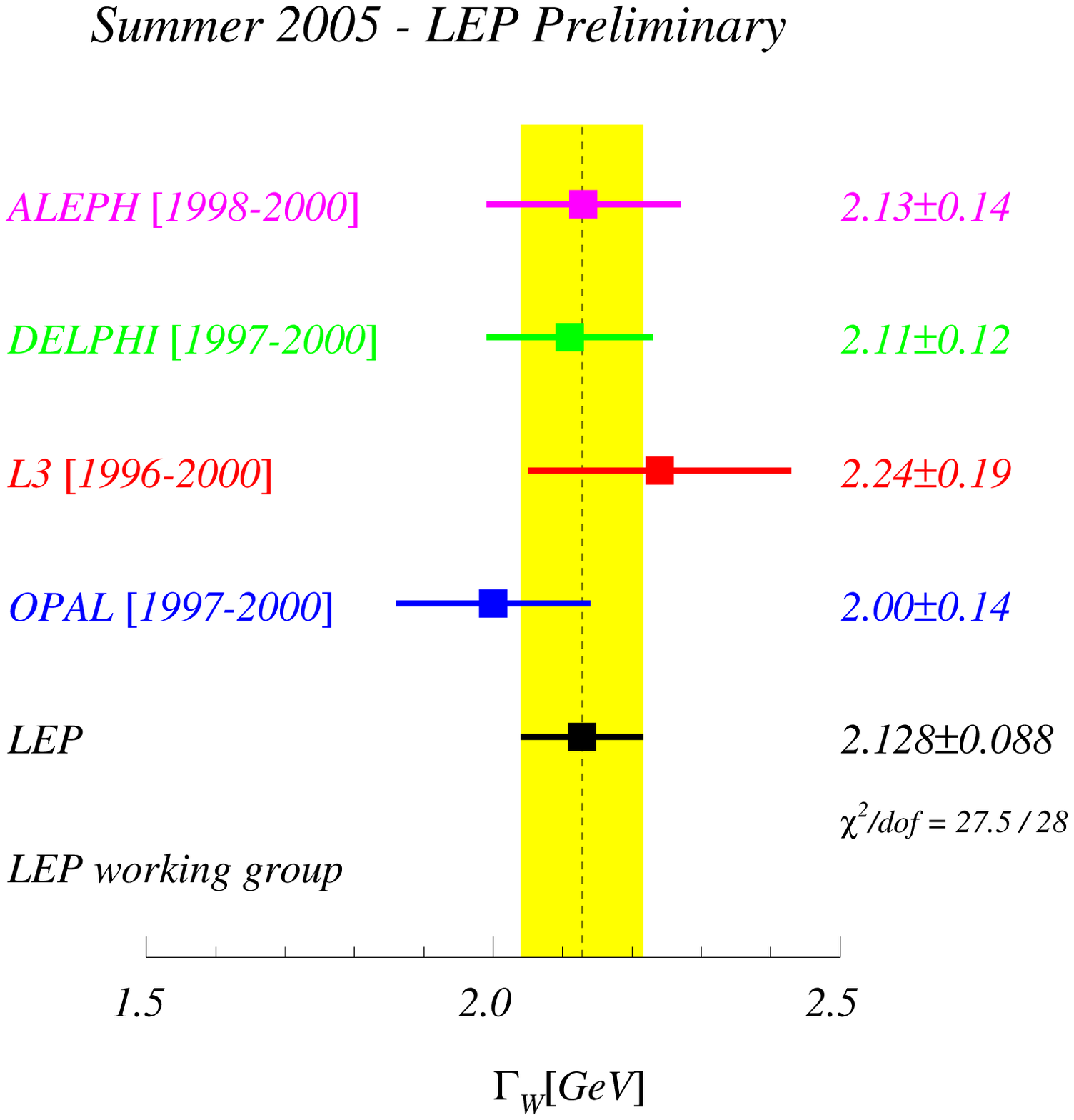,width=0.495\textwidth}}
\vskip -1cm
\caption{\label{mw:fig:mwgw} The combined results for the measurements
          of the W mass (left) and total W width (right) compared to
          the results obtained by the four LEP collaborations. The
          combined values take into account correlations between
          experiments and years and hence, in general, do not give the
          same central value as a simple average. In the combination
          of the $\qqqq$ results common values for the CR and BEC
          errors of ALEPH, DELPHI and L3 are used (see text). The
          $\Mw$ values from ALEPH, DELPHI and L3 have been
          recalculated for this plot including these common CR and BEC
          errors.  The individual and combined $\Mw$ results include
          the measurements from the threshold cross section. }
\end{center}
\end{figure}

The method of direct reconstruction is also well suited to the direct
measurement of the total width of the W boson. The results of the four
LEP experiments are shown in Table \ref{mw:tab:wwidth_experiments} and
in Figure \ref{mw:fig:mwgw}.
\begin{table}[htbp]
 \begin{center}
  \begin{tabular}{|c|c|}\hline
  Experiment & \Gw\ (\GeVm)        \\ \hline
   ALEPH    & $2.13\pm0.11\pm0.09$ \\ 
   DELPHI   & $2.11\pm0.10\pm0.07$ \\
   L3       & $2.24\pm0.11\pm0.15$ \\
   OPAL     & $2.00\pm0.10\pm0.10$ \\ \hline
\end{tabular}
 \caption{W width measurements ($\roots=172-209$~\GeV) from the
         individual experiments. The ALPEH, DELPHI and L3 results are
         preliminary and the OPAL result is final.  The first error is
         statistical and the second systematic.}
 \label{mw:tab:wwidth_experiments}
\end{center}
\end{table}

Each experiment provided a W width measurement for both $\WWqqln$ and
$\WWqqqq$ channels for each of the data taking years (1996-2000) that
it has analysed. A total of 29 measurements are supplied: ALEPH
provided 3 $\WWqqqq$ results (1998-2000) and 2 $\WWqqln$ results
(1998-1999), DELPHI 8 measurements (1997-2000), L3 8 measurements
(1996-2000) having already combined the 1996 and 1997 results, and
OPAL provided 8 measurements (1997-2000) not using the small 1996 data
set for the width analysis.

A common colour reconnection error of 80 $\MeVm$ and a common
Bose-Einstein correlation error of 35 $\MeVm$ are used in the
combination.  These common errors were determined such that the same
error was obtained on \Gw\ as when using the BEC/CR errors supplied by
the experiments. The change in the value of the width is less than
5~\MeVm.  The BEC and CR values supplied by the experiments were based
on studies of phenomenological models of these effects, the
uncertainty has not yet been determined from the particle flow
measurements of colour reconnection.  Note that the final OPAL W width
result does not use the momentum-cut technique introduced for the mass
analysis.

A simultaneous fit to the results of the four LEP collaborations is
performed in the same way as for the $\Mw$ measurement. Correlated
systematic uncertainties are taken into account and the combination
gives:
\begin{eqnarray}
 \Gw = 2.128\pm0.065(\mathrm{stat.})\pm0.059 (\mathrm{syst.})~\GeVm,
\end{eqnarray}
with a $\chi^2$/d.o.f. of 27.5/28, corresponding to a $\chi^2$
probability of 49$\%$.

\section{Summary}

The results of the four LEP experiments on the mass and width of the W
boson are combined taking into account correlated systematic
uncertainties, giving:
\begin{eqnarray*}
       \Mw & = & 80.392\pm0.039~\GeVm, \\
       \Gw & = &  2.128\pm0.088~\GeVm.
\end{eqnarray*}
The statistical correlation between mass and width is small and
neglected.  Their correlation due to common systematic effects is
under study.

\vskip -1cm

\boldmath
\chapter{Constraints on the Standard Model}
\label{sec-MSM}
\unboldmath

\updates{Updated preliminary and published measurements as discussed
  in the previous chapters are taken into account, including also (i)
  final published Z-pole results~\cite{bib-Z-pole}, (ii) an update of
  the hadronic vacuum polarisation, (iii) new preliminary results on
  the mass of the top-quark obtained at the Tevatron, as well as (iv)
  final results in atomic parity violation, Moller scattering and
  neutrino-nucleon interactions.  The new ZFITTER version 6.42 is used
  for the Standard-Model analyses.}

\section{Introduction}

The precise electroweak measurements performed at {\LEPII} and
elsewhere (\LEPI, SLC, Tevatron, etc.)  can be used to check the
validity of the Standard Model (SM) and, within its framework, to
infer valuable information about its fundamental parameters. The
accuracy of the measurements makes them sensitive to the mass of the
top quark $\Mt$, and to the mass of the Higgs boson $\MH$ through loop
corrections. While the leading $\Mt$ dependence is quadratic, the
leading $\MH$ dependence is logarithmic.  Therefore, the inferred
constraints on $\Mt$ are much stronger than those on $\MH$.

\section{Measurements}

The measurements considered here are reported in Table~\ref{tab-SMIN}.
Also shown are the results of the SM fit to these combined
measurements.  The measurements obtained at the Z pole by the LEP and
SLC experiments ALEPH, DELPHI, L3, OPAL and SLD, and their
combinations, reported in parts a), b) and c) of Table~\ref{tab-SMIN},
are final and published~\cite{bib-Z-pole}.

The results on the W-boson mass by UA2~\cite{UA2-MW},
CDF~\cite{CDF-MW-PRL90, *CDF-MW-PRD90, *CDF-MW-PRL95, *CDF-MW-PRD95,
*CDF-MW-2000} and D\O~\cite{D0-MW:central, *D0-MW:endcap, *D0-MW:edge,
*D0-MW:large} in Run-I, and the W-boson width by CDF\cite{CDF-GW} and
D\O\cite{D0-GW} in Run-I, are combined by the Tevatron Electroweak
Working Group based on a detailed treatment of common systematic
uncertainties. The results are\cite{PP-MW-GW:combination}: $\MW =
80452\pm59~\MeV$, $\GW = 2102\pm106~\MeV$, with a correlation of
$-17.4\%$.  Combining these results with the preliminary {\LEPII}
measurements as presented in Chapter~\ref{sec-MW}, the new preliminary
world averages used here are:
\begin{eqnarray}
\MW & = & 80.410 \pm 0.032~\GeV\\
\GW & = &  2.123 \pm 0.067~\GeV\,,
\end{eqnarray}
with a correlation of $-6.25\%$. 

For the mass of the top quark, $\Mt$, the published Run-I results from
CDF~\cite{Mtop1-CDF-di-l-PRLa, *Mtop1-CDF-di-l-PRLb,
*Mtop1-CDF-di-l-PRLb-E, *Mtop1-CDF-l+j-PRL, *Mtop1-CDF-l+j-PRD,
*Mtop1-CDF-all-j-PRL} and D\O~\cite{D0-top:prl-ll, *D0-top:prd-ll,
*D0-top:prl-lj, *D0-top:prd-lj, *Mtop1-D0-l+j-new1,
*Mtop1-D0-l+j-new2, *Mtop1-D0-all-j-PRL}, also including recent
preliminary results based on Run-II data, are combined by the Tevatron
Electroweak Working Group with the result:
$\Mt=172.7\pm2.9~\GeV$~\cite{TeVEWWGtop-0507}.

In addition, the following final results obtained in low-$Q^2$
interactions are considered: (i) the measurements of atomic parity
violation in caesium\cite{QWCs:exp:1, QWCs:exp:2}, with the numerical
result\cite{QWCs:theo:2003:new} taken from a recently published
revised analysis of QED radiative corrections applied to the raw
measurement; (ii) the result of the E-158 collaboration on the
electroweak mixing angle\footnote{ E-158 quotes in the MSbar scheme,
evolved to $Q^2=\MZ^2$.  We add 0.00029 to the quoted value in order
to obtain the effective electroweak mixing angle~\cite{PDG2004}.}
measured in Moller scattering~\cite{E158RunI, *E158RunI+II+III}; and
(iii) the final result of the NuTeV collaboration on neutrino-nucleon
neutral to charged current cross section
ratios~\cite{bib-NuTeV-final}.

Using both muon neutrino and muon anti-neutrino beams, the NuTeV
collaboration has published by far the most precise result in
neutrino-nucleon scattering~\cite{bib-NuTeV-final}, obtained at an
average $Q^2\simeq20~\GeV^2$.  Based on an analysis mainly exploiting
the Paschos-Wolfenstein quantity $R_-$~\cite{PaschosWolfenstein}, with
$R_\pm \equiv (\sigma_{NC}(\nu)\pm\sigma_{NC}(\bar\nu))/
(\sigma_{CC}(\nu)\pm\sigma_{CC}(\bar\nu)) = \gnlq^2\pm\gnrq^2$, where
$\gnlq^2=4\gln^2(\glu^2+\gld^2) =
[1/2-\swsqsqeff+(5/9)\swsqsqeff]\rhon\rho_{\mathrm{ud}}$ and
$\gnrq^2=4\gln^2(\gru^2+\grd^2) =
(5/9)\swsqsqeff\rhon\rho_{\mathrm{ud}}$, the NuTeV results for the
effective couplings defined above are: $\gnlq^2=0.30005\pm0.00137$ and
$\gnrq^2=0.03076\pm0.00110$, with a correlation of $-0.017$.  While
the result on $\gnrq$ agrees with the $\SM$ expectation, the result on
$\gnlq$, measured nearly eight times more precisely, shows a deficit
with respect to the expectation at the level of 3.0 standard
deviations.

An additional input parameter, not shown in the table, is the Fermi
constant $G_F$, determined from the $\mu$ lifetime, $G_F = 1.16637(1)
\cdot 10^{-5}~\GeV^{-2}$\cite{bib-Gmu-1, *bib-Gmu-2, *bib-Gmu-3}.  The
relative error of $G_F$ is comparable to that of $\MZ$; both errors
have negligible effects on the fit results.

\begin{table}[p]
\begin{center}
\renewcommand{\arraystretch}{1.10}
\begin{tabular}{|ll||r|r|r|r|}
\hline
 && \mcc{Measurement with}  &\mcc{Systematic} & \mcc{Standard} & \mcc{Pull} \\
 && \mcc{Total Error}       &\mcc{Error}      & \mcc{Model fit}&            \\
\hline
\hline
&&&&& \\[-3mm]
& $\Delta\alpha^{(5)}_{\mathrm{had}}(\MZ^2)$\cite{bib-BP05}
                & $0.02758 \pm 0.00035$ & 0.00034 &0.02767& $-0.3$ \\
&&&&& \\[-3mm]
\hline
a) & \underline{\LEPI}   &&&& \\
   & line-shape and      &&&& \\
   & lepton asymmetries: &&&& \\
&$\MZ$ [\GeV{}] & $91.1875\pm0.0021\pz$
                & ${}^{(a)}$0.0017$\pz$ &91.1874$\pz$ & $ 0.0$ \\
&$\GZ$ [\GeV{}] & $2.4952 \pm0.0023\pz$
                & ${}^{(a)}$0.0012$\pz$ & 2.4959$\pz$ & $-0.3$ \\
&$\shad$ [nb]   & $41.540 \pm0.037\pzz$ 
                & ${}^{(b)}$0.028$\pzz$ &41.478$\pzz$ & $ 1.7$ \\
&$\Rl$          & $20.767 \pm0.025\pzz$ 
                & ${}^{(b)}$0.007$\pzz$ &20.743$\pzz$ & $ 1.0$ \\
&$\Afbzl$       & $0.0171 \pm0.0010\pz$ 
                & ${}^{(b)}$0.0003\pz & 0.0164\pz     & $ 0.7$ \\
&+ correlation matrix~\cite{bib-Z-pole} &&&& \\
&                                             &&&& \\[-3mm]
&$\tau$ polarisation:                         &&&& \\
&$\cAl~(\ptau)$ & $0.1465\pm 0.0033\pz$ 
                & 0.0016$\pz$ & 0.1480$\pz$ & $-0.5$ \\
                      &                       &&&& \\[-3mm]
&$\qq$ charge asymmetry:                      &&&& \\
&$\swsqeffl(\Qfbhad)$
                & $0.2324\pm0.0012\pz$ 
                & 0.0010$\pz$ & 0.23140     & $ 0.8$ \\
&                                             &&&& \\[-3mm]
\hline
b) & \underline{SLD} &&&& \\
&$\cAl$ (SLD)   & $0.1513\pm 0.0021\pz$ 
                & 0.0010$\pz$ & 0.1480$\pz$ & $ 1.6$ \\
&&&&& \\[-3mm]
\hline
c) & \underline{{\LEPI}/SLD Heavy Flavour} &&&& \\
&$\Rbz{}$        & $0.21629\pm0.00066$  
                 & 0.00050     & 0.21579     & $ 0.8$ \\
&$\Rcz{}$        & $0.1721\pm0.0030\pz$
                 & 0.0019$\pz$ & 0.1723$\pz$ & $-0.1$ \\
&$\Afbzb{}$      & $0.0992\pm0.0016\pz$
                 & 0.0007$\pz$ & 0.1038$\pz$ & $-2.8$ \\
&$\Afbzc{}$      & $0.0707\pm0.0035\pz$
                 & 0.0017$\pz$ & 0.0742$\pz$ & $-1.0$ \\
&$\cAb$          & $0.923\pm 0.020\pzz$
                 & 0.013$\pzz$ & 0.935$\pzz$ & $-0.6$ \\
&$\cAc$          & $0.670\pm 0.027\pzz$
                 & 0.015$\pzz$ & 0.668$\pzz$ & $ 0.1$ \\
&+ correlation matrix~\cite{bib-Z-pole} &&&& \\
&                                              &&&& \\[-3mm]
\hline
d) & \underline{{\LEPII} and Tevatron} &&&& \\
&$\MW$ [\GeV{}] ({\LEPII}, Tevatron)
& $80.410 \pm 0.032\pzz$ &      $\pzz$   & 80.378$\pzz$ & $ 1.0$ \\
&$\GW$ [\GeV{}] ({\LEPII}, Tevatron)
& $ 2.123 \pm 0.067\pzz$ &      $\pzz$   &  2.092$\pzz$ & $ 0.5$ \\
&$\Mt$ [\GeV{}] (Tevatron~\cite{TeVEWWGtop-0507})
& $172.7\pm 2.9\pzz\pzz$ & 2.4$\pzz\pzz$ & 173.3$\pzz\pzz$ & $-0.2$ \\
\hline
\end{tabular}\end{center}
\caption[]{ Summary of high-$Q^2$ measurements included in the
  combined analysis of SM parameters. Section~a) summarises {\LEPI}
  averages, Section~b) SLD results ($\swsqeffl$ includes $\ALR$ and
  the polarised lepton asymmetries), Section~c) the {\LEPI} and SLD
  heavy flavour results, and Section~d) electroweak measurements from
  {\LEPII} and the Tevatron.  The total errors in column 2 include the
  systematic errors listed in column 3.  Although the systematic
  errors include both correlated and uncorrelated sources, the
  determination of the systematic part of each error is approximate.
  The $\SM$ results in column~4 and the pulls (difference between
  measurement and fit in units of the total measurement error) in
  column~5 are derived from the SM fit including all data
  (Table~\ref{tab-BIGFIT}, column~4).\\ $^{(a)}$\small{The systematic
  errors on $\MZ$ and $\GZ$ contain the errors arising from the
  uncertainties in the $\LEPI$ beam energy only.}\\
  $^{(b)}$\small{Only common systematic errors are indicated.}\\ }
\label{tab-SMIN}
\end{table}

\section{Theoretical and Parametric Uncertainties}

Detailed studies of the theoretical uncertainties in the SM
predictions due to missing higher-order electroweak corrections and
their interplay with QCD corrections had been carried out by the
working group on `Precision calculations for the $\Zzero$
resonance'\cite{bib-PCLI}, and later in~\cite{BP:98,PCP99}.
Theoretical uncertainties are evaluated by comparing different but,
within our present knowledge, equivalent treatments of aspects such as
resummation techniques, momentum transfer scales for vertex
corrections and factorisation schemes.  The effects of these
theoretical uncertainties are reduced by the inclusion of higher-order
corrections\cite{bib-twoloop,bib-QCDEW} in the electroweak libraries
TOPAZ0~\cite{Montagna:1993py, *Montagna:1993ai, *Montagna:1996ja,
*Montagna:1998kp} and ZFITTER~\cite{Bardin:1989di, *Bardin:1990tq,
*Bardin:1991fu, *Bardin:1991de, *Bardin:1992jc, *Bardin:1999yd,
*Kobel:2000aw, *Arbuzov:2005ma}.

The use of the QCD corrections\cite{bib-QCDEW} increases the value of
$\alfmz$ by 0.001, as expected.  The effects of missing higher-order
QCD corrections on $\alfmz$ covers missing higher-order electroweak
corrections and uncertainties in the interplay of electroweak and QCD
corrections. A discussion of theoretical uncertainties in the
determination of $\alfas$ can be found in References~\citen{bib-PCLI}
and~\citen{bib-SMALFAS}, with a recent analysis in
Reference~\citen{Stenzel:2005sg} where the theoretical uncertainty is
estimated to be about 0.001 for the analyses presented in the
following.

Recently, the complete (fermionic and bosonic) two-loop corrections
for the calculation of $\MW$~\cite{Twoloop-MW}, and the complete
fermionic two-loop corrections for the calculation of
$\swsqeffl$~\cite{Twoloop-sin2teff} have been calculated.  Including
three-loop top-quark contributions to the $\rho$ parameter in the
limit of large $\Mt$~\cite{Threeloop-rho}, efficient routines for
evaluating these corrections have been implemented since version 6.40
in the semi-analytical program ZFITTER.  The remaining theoretical
uncertainties are estimated to be $4~\MeV$ on $\MW$ and 0.000049 on
$\swsqeffl$.  The latter uncertainty dominates the theoretical
uncertainty in SM fits and the extraction of constraints on the mass
of the Higgs boson presented below. For a complete picture, the
complete two-loop calculation for the partial Z decay widths should be
calculated.

The determination of the size of remaining theoretical uncertainties
is under continued study.  The theoretical errors discussed above are
not included in the results presented in Table~\ref{tab-BIGFIT}.  At
present the impact of theoretical uncertainties on the determination
of $\SM$ parameters from the precise electroweak measurements is small
compared to the error due to the uncertainty in the value of
$\alpha(\MZ^2)$, which is included in the results.

The uncertainty in $\alpha(\MZ^2)$ arises from the contribution of
light quarks to the photon vacuum polarisation
($\Delta\alpha_{\mathrm{had}}^{(5)}(\MZ^2)$):
\begin{equation}
\alpha(\MZ^2) = \frac{\alpha(0)}%
   {1 - \Delta\alpha_\ell(\MZ^2) -
   \Delta\alpha_{\mathrm{had}}^{(5)}(\MZ^2) -
   \Delta\alpha_{\mathrm{top}}(\MZ^2)} \,,
\end{equation}
where $\alpha(0)=1/137.036$.  The top contribution, $-0.00007(1)$,
depends on the mass of the top quark, and is therefore determined
inside the electroweak libraries TOPAZ0 and ZFITTER.  The leptonic
contribution is calculated to third order\cite{bib-alphalept} to be
$0.03150$, with negligible uncertainty.

For the hadronic contribution, we no longer use the value $0.02804 \pm
0.00065$\cite{bib-JEG2,bib-Burk}, but rather the new evaluation
$0.02758\pm0.0035$~\cite{bib-BP05} which takes into account published
results on electron-positron annihilations into hadrons at low
centre-of-mass energies by the BES collaboration~\cite{BES_01}, as
well as the revised published results from CMD-2~\cite{CMD_03} and new
results from KLOE~\cite{KLOE_04}.  The reduced uncertainty still
causes an error of 0.00013 on the $\SM$ prediction of $\swsqeffl$, and
errors of 0.2~\GeV{} and 0.1 on the fitted values of $\Mt$ and
$\log(\MH)$, included in the results presented below.  The effect on
the $\SM$ prediction for $\Gll$ is negligible.  The $\alfmz$ values
for the $\SM$ fits presented here are stable against a variation of
$\alpha(\MZ^2)$ in the interval quoted.

There are also several evaluations of
$\Delta\alpha^{(5)}_{\mathrm{had}}(\MZ^2)$~\cite{bib-Swartz,
bib-Zeppe, bib-Alemany, bib-Davier, bib-alphaKuhn, bib-jeger99,
bib-Erler, bib-ADMartin, bib-Troconiz-Yndurain, bib-Hagiwara:2003,
bib-Troconiz-Yndurain-2004} which are more theory-driven.  The most
recent of these (Reference \citen{bib-Troconiz-Yndurain-2004}) also
includes the new results from BES, yielding $0.02749\pm0.00012$.  To
show the effects of the uncertainty of $\alpha(\MZ^2)$, we also use
this evaluation of the hadronic vacuum polarisation.  Note that all
these evaluations obtain values for
$\Delta\alpha^{(5)}_{\mathrm{had}}(\MZ^2)$ consistently lower than -
but in agreement with - the old value of $0.02804 \pm 0.00065$.

\section{Selected Results}

\begin{figure}[htbp]
\begin{center}
$ $ \vskip -1cm
  \mbox{\includegraphics[width=0.9\linewidth]{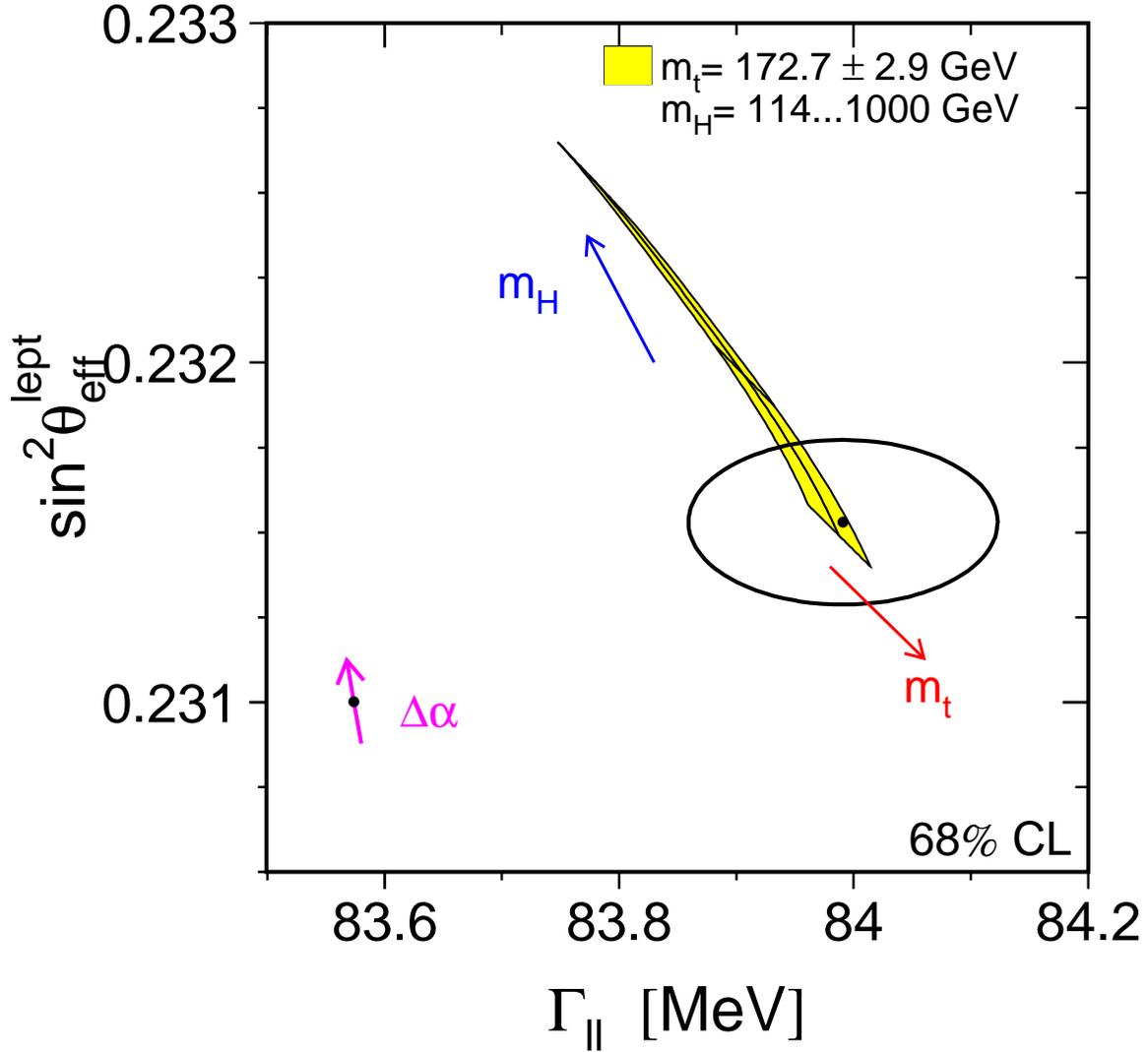}}
\end{center}
\vskip -1cm
\caption[]{ $\LEPI$+SLD measurements~\cite{bib-Z-pole} of $\swsqeffl$
  and $\Gll$ and the SM prediction.  The point shows the predictions
  if among the electroweak radiative corrections only the photon
  vacuum polarisation is included. The corresponding arrow shows
  variation of this prediction if $\alpha(\MZ^2)$ is changed by one
  standard deviation. This variation gives an additional uncertainty
  to the SM prediction shown in the figure.  }
\label{fig-gllsef}
\end{figure}

Figure~\ref{fig-gllsef} shows a comparison of the leptonic partial
width from {\LEPI}, $\Gll=83.985\pm0.086~\MeV$~\cite{bib-Z-pole}, and
the effective electroweak mixing angle from asymmetries measured at
{\LEPI} and SLD, $\swsqeffl=0.23153\pm0.00016$~\cite{bib-Z-pole}, with
the SM shown as a function of $\Mt$ and $\MH$.  Good agreement with
the $\SM$ prediction using the most recent result on $\Mt$ is
observed.  The point with the arrow indicates the prediction if among
the electroweak radiative corrections only the photon vacuum
polarisation is included, which shows that the precision electroweak
Z-pole data are sensitive to non-trivial electroweak corrections.
Note that the error due to the uncertainty on $\alpha(\MZ^2)$ (shown
as the length of the arrow) is not much smaller than the experimental
error on $\swsqeffl$ from {\LEPI} and SLD.  This underlines the
continued importance of a precise measurement of
$\sigma(\mathrm{e^+e^-\rightarrow hadrons})$ at low centre-of-mass
energies.

Of the measurements given in Table~\ref{tab-SMIN}, $\Rl$ is one of the
most sensitive to QCD corrections.  For $\MZ=91.1875$~\GeV{}, and
imposing $\Mt=172.7\pm2.9$~\GeV{} as a constraint,
$\alfas=0.1223\pm0.0038$ is obtained.  Alternatively,
$\slept\equiv\shad/Rl=2.0003\pm0.027~\nb$~\cite{bib-Z-pole} which has
higher sensitivity to QCD corrections and less dependence on $\MH$
yields: $\alfas=0.1179\pm0.0030$.  Typical errors arising from the
variation of $\MH$ between $100~\GeV$ and $200~\GeV$ are of the order
of $0.001$, somewhat smaller for $\slept$.  These results on $\alfas$,
as well as those reported in the next section, are in very good
agreement with recently determined world averages ($\alfmz=0.118 \pm
0.002$\cite{common_bib:pdg2000}, or $\alfmz=0.1178 \pm 0.0033$ based
solely on NNLO QCD results excluding the {\LEPI} lineshape results and
accounting for correlated errors~\cite{QCD:Bethke:2000,
*Bethke:2004uy}).

\section{Standard Model Analyses}

In the following, several different SM fits reported in
Table~\ref{tab-BIGFIT} are discussed.  The $\chi^2$ minimisation is
performed with the program MINUIT~\cite{MINUIT}, and the predictions
are calculated with ZFITTER as a function of the five SM input
parameters $\dalhad$, $\alfmz$, $\MZ$, $\Mt$ and $\LOGMH$ which are
all varied in the fits.  The somewhat increased $\chi^2$/d.o.f.{} for
all of these fits is caused by the large dispersion in the values of
the leptonic effective electroweak mixing angle measured through the
various asymmetries at {\LEPI} and SLD~\cite{bib-Z-pole}.
Following~\cite{bib-Z-pole} for the analyses presented here, this
dispersion is interpreted as a fluctuation in one or more of the input
measurements, and thus we neither modify nor exclude any of them.  A
further drastic increase in $\chi^2$/d.o.f.{} is observed when the
NuTeV results are included in the analysis.

\begin{figure}[htbp]
\begin{center}
  \leavevmode \includegraphics[width=0.9\linewidth]{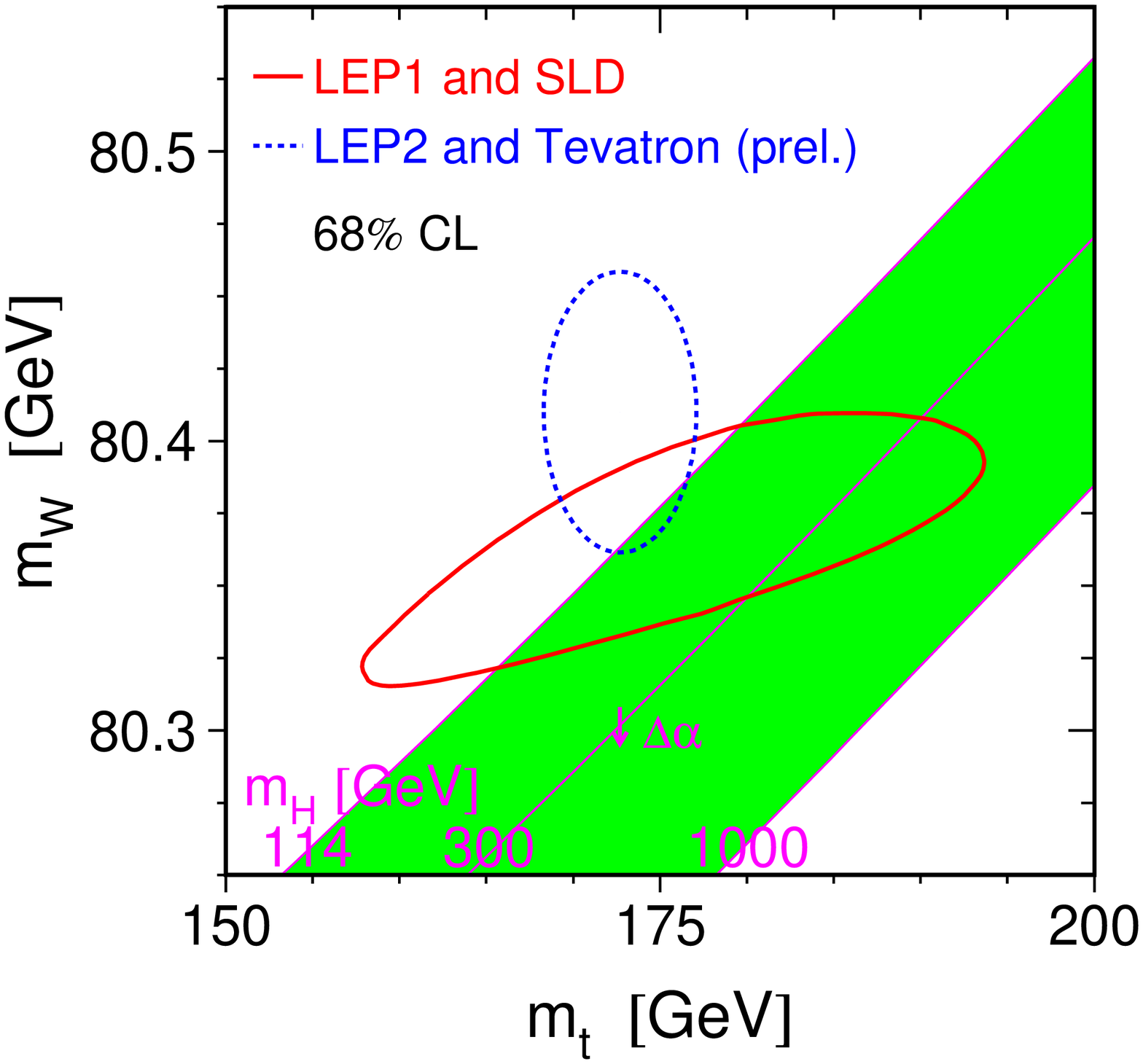}
\caption[]{ The comparison of the indirect measurements of $\MW$ and
  $\Mt$ ($\LEPI$+ SLD data) (solid contour) and the direct
  measurements ($\pp$ colliders and $\LEPII$ data) (dashed contour).
  In both cases the 68\% CL contours are plotted.  Also shown is the
  SM relationship for the masses as a function of the Higgs mass. The
  arrow labelled $\Delta\alpha$ shows the variation of this relation
  if $\alpha(\MZ^2)$ is changed by one standard deviation. This
  variation gives an additional uncertainty to the SM band shown in
  the figure.}
\label{fig:mtmW}
\end{center}
\end{figure}

To test the agreement between the Z-pole data~\cite{bib-Z-pole}
({\LEPI} and SLD) and the SM, a fit to this data is performed.  The
result is shown in Table~\ref{tab-BIGFIT}, column~1. The indirect
constraints on $\MW$ and $\Mt$ from this data sample are shown in
Figure~\ref{fig:mtmW}, compared with the direct measurements.  Also
shown are the SM predictions for Higgs masses between 114 and
1000~\GeV.  As can be seen in the figure, the indirect and direct
measurements of $\MW$ and $\Mt$ are in good agreement, and both sets
prefer a low value of the Higgs mass.

\begin{figure}[htbp]
\begin{center}
  \mbox{\includegraphics[width=0.9\linewidth]{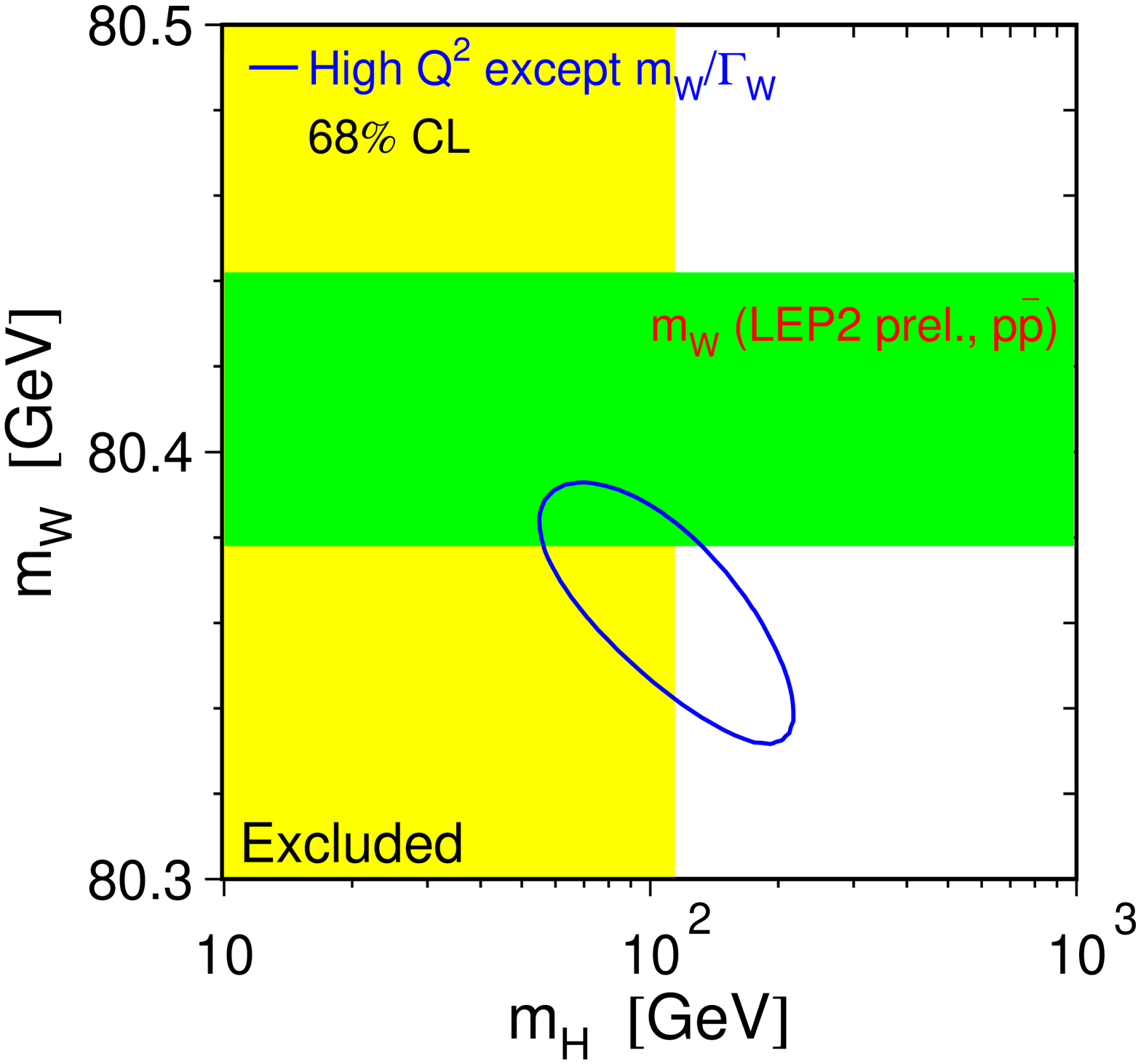}}
\end{center}
\vspace*{-0.6cm}
\caption[]{
  The 68\% confidence level contour in $\MW$ and $\MH$ for the fit to
  all data except the direct measurement of $\MW$, indicated by the
  shaded horizontal band of $\pm1$ sigma width.  The vertical band
  shows the 95\% CL exclusion limit on $\MH$ from the direct search.
  }
\label{fig-mhmw}
\end{figure}

\begin{figure}[htbp]
\begin{center}
  \mbox{\includegraphics[width=0.9\linewidth]{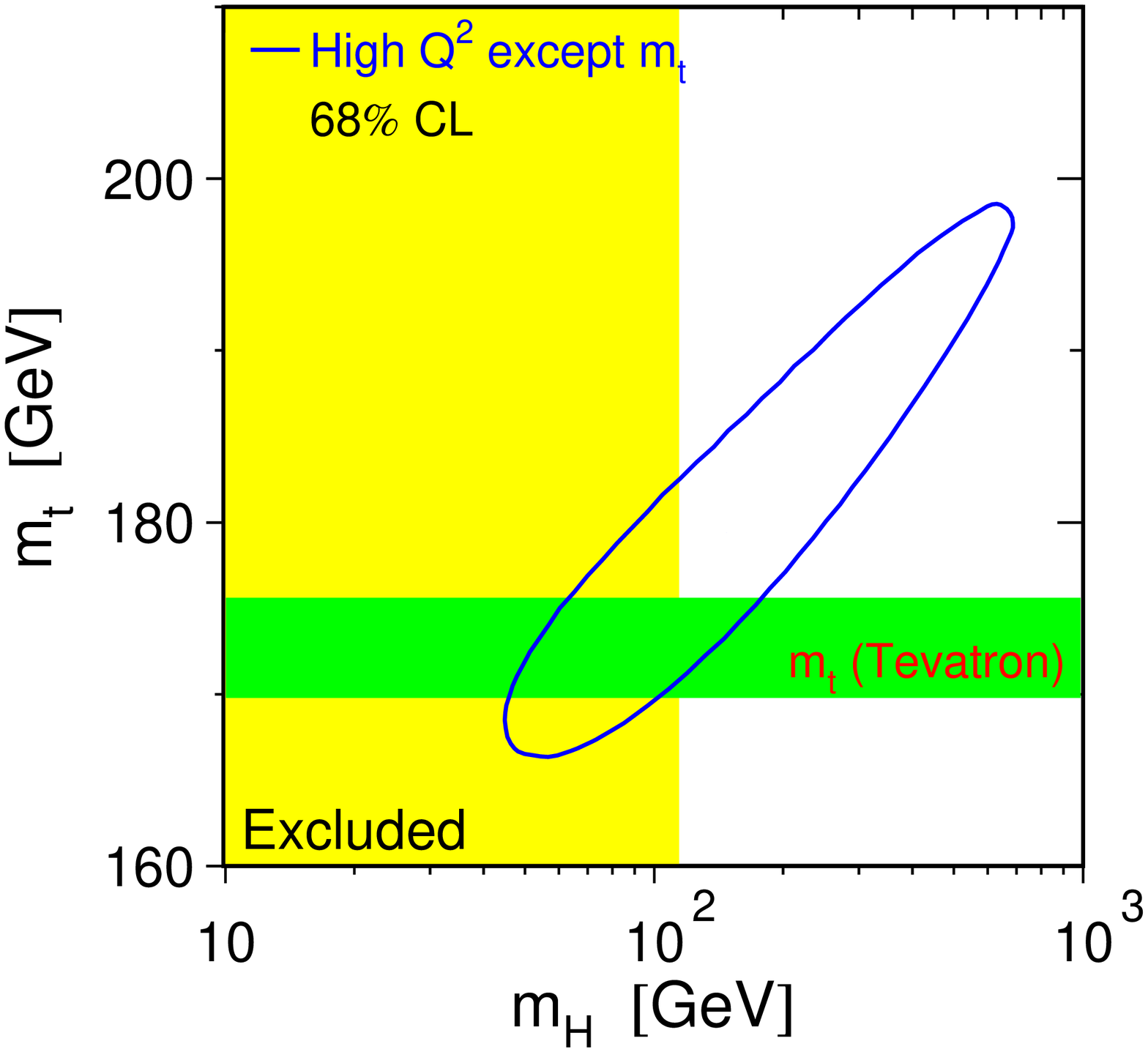}}
\end{center}
\vspace*{-0.6cm}
\caption[]{ The 68\% confidence level contour in $\Mt$ and $\MH$ for
  the fit to all data except the direct measurement of $\Mt$,
  indicated by the shaded horizontal band of $\pm1$ sigma width.  The
  vertical band shows the 95\% CL exclusion limit on $\MH$ from the
  direct search.  }
\label{fig-mhmt}
\end{figure}

For the fit shown in column~2 of Table~\ref{tab-BIGFIT}, the direct
$\Mt$ measurement is included to obtain the best indirect
determination of $\MW$.  The result is also shown in
Figure~\ref{fig-mhmw}.  Also in this case, the indirect determination
of W boson mass, $80.364\pm0.021~\GeV$, is in good agreement with the
direct measurements from {\LEPII} and the Tevatron, $\MW=
80.410\pm0.032~\GeV$.  For the fit shown in column~3 of
Table~\ref{tab-BIGFIT} and Figure~\ref{fig-mhmt}, the direct $\MW$ and
$\GW$ measurements from {\LEPII} and the Tevatron are included instead
of the direct $\Mt$ measurement in order to obtain the constraint
$\Mt= 179^{+12}_{-9}~\GeV$, in very good agreement with the direct
measurement of $\Mt = 172.7\pm2.9~\GeV$.

\begin{figure}[htbp]
\begin{center}
  \mbox{\includegraphics[width=0.9\linewidth]{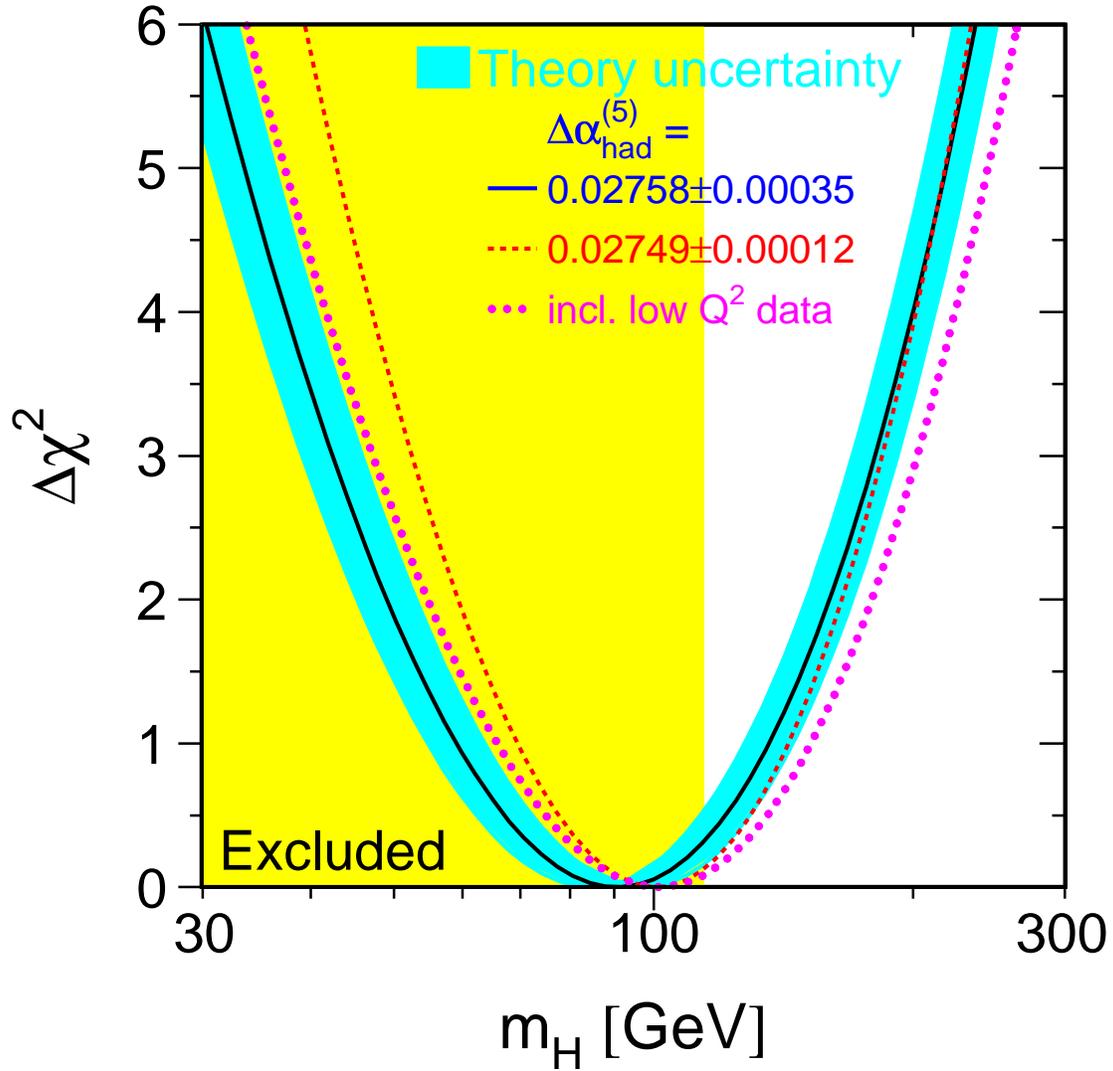}}
\end{center}
\vspace*{-0.6cm}
\caption[]{ $\Delta\chi^{2}=\chi^2-\chi^2_{min}$ {\it vs.} $\MH$
  curve.  The line is the result of the fit using all data (last
  column of Table~\protect\ref{tab-BIGFIT}); the band represents an
  estimate of the theoretical error due to missing higher order
  corrections.  The vertical band shows the 95\% CL exclusion limit on
  $\MH$ from the direct search.  The dashed curve is the result
  obtained using the evaluation of
  $\Delta\alpha^{(5)}_{\mathrm{had}}(\MZ^2)$ from
  Reference~\citen{bib-Troconiz-Yndurain-2004}. }
\label{fig-chiex}
\end{figure}

Finally, the best constraints on $\MH$ are obtained when all
high-$Q^2$ measurements are used in the fit.  The results of this fit
are shown in column~4 of Table~\ref{tab-BIGFIT}.  The predictions of
this fit for observables measured in high-$Q^2$ and low-$Q^2$
reactions are listed in Tables~\ref{tab-SMIN} and~\ref{tab-SMpred},
respectively.  In Figure~\ref{fig-chiex} the observed value of
$\Delta\chi^2 \equiv \chi^2 - \chi^2_{\mathrm{min}}$ as a function of
$\MH$ is plotted for this fit including all high-$Q^2$ results.  The
solid curve is the result using ZFITTER, and corresponds to the last
column of Table~\ref{tab-BIGFIT}.  The shaded band represents the
uncertainty due to uncalculated higher-order corrections, as estimated
by ZFITTER.

The 95\% confidence level upper limit on $\MH$ (taking the band into
account) is $186~\GeV$.  The 95\% C.L. lower limit on $\MH$ of
114.4~\GeV{} obtained from direct searches\cite{LEPSMHIGGS} is not
used in the determination of this limit.  Including it increases the
limit to $219~\GeV$.  Also shown is the result (dashed curve) obtained
when using $\Delta\alpha^{(5)}_{\mathrm{had}}(\MZ^2)$ of
Reference~\citen{bib-Troconiz-Yndurain-2004}.

\begin{table}[htbp]
\renewcommand{\arraystretch}{1.5}
  \begin{center}
\begin{tabular}{|c||c|c|c|c|c|}
\hline
&     - 1 -              &      - 2 -             &    - 3 -               &     - 4 -             \\
& all Z-pole             & all Z-pole data        & all Z-pole data        & all Z-pole data       \\[-3mm]
& data                   &    plus   $\Mt$        & plus $\MW$, $\GW$      & plus $\Mt,\MW,\GW$    \\
\hline
\hline
$\Mt$\hfill[\GeV] 
& $173^{+13 }_{-10}$     & $172.7^{+2.8}_{-2.8}$  & $179^{+12}_{- 9}$      & $173.3^{+2.7}_{-2.7}$  \\
$\MH$\hfill[\GeV] 
& $111^{+190}_{-60}$     & $112^{+62}_{-41}$      & $148^{+248}_{-83}$     & $ 91^{+45}_{-32}$      \\
$\log(\MH/\GeV)$  
& $2.05^{+0.43}_{-0.34}$ & $2.05^{+0.19}_{-0.20}$ & $2.17^{+0.43}_{-0.36}$ & $1.96^{+0.18}_{-0.19}$ \\
$\alfmz$          
& $0.1190\pm 0.0027$     & $0.1190\pm0.0027$      & $0.1190\pm 0.0028$     & $0.1186\pm 0.0027$     \\
\hline
$\chi^2$/d.o.f.{} ($P$)
& $16.0/10~(9.9\%)$      & $16.0/11~(14\%)$       & $17.3/12~(14\%)$       & $17.8/13~(17\%)$       \\
\hline
\hline
$\swsqeffl$
& $\pz0.23149$           & $\pz0.23149$           & $\pz0.23143$           & $\pz0.23140$ \\[-1mm]
& $\pm0.00016$           & $\pm0.00016$           & $\pm0.00014$           & $\pm0.00014$ \\
$\swsq$     
& $\pz0.22321$           & $\pz0.22331$           & $\pz0.22285$           & $\pz0.22304$ \\[-1mm]
& $\pm0.00062$           & $\pm0.00041$           & $\pm0.00043$           & $\pm0.00033$ \\
$\MW$\hfill[\GeV]
& $80.363\pm0.032$       & $80.364\pm0.021$       & $80.387\pm0.022$       & $80.377\pm0.017$   \\
\hline
\end{tabular}
\end{center}
\caption[]{ Results of the fits to: (1) all Z-pole data ({\LEPI} and
  SLD), (2) all Z-pole data plus direct $\Mt$ determination, (3) all
  Z-pole data plus direct $\MW$ and $\GW$ determinations, (4) all
  Z-pole data plus direct $\Mt,\MW,\GW$ determinations (i.e., all
  high-$Q^2$ results).  As the sensitivity to $\MH$ is logarithmic,
  both $\MH$ as well as $\log(\MH/\GeV)$ are quoted.  The bottom part
  of the table lists derived results for $\swsqeffl$, $\swsq$ and
  $\MW$.  See text for a discussion of theoretical errors not included
  in the errors above.  }
\label{tab-BIGFIT}
\renewcommand{\arraystretch}{1.0}
\end{table}

\begin{table}[htbp]
\begin{center}
  \renewcommand{\arraystretch}{1.30}
\begin{tabular}{|ll||r||r|r|l|}
\hline
 && {Measurement with}  & {Standard Model} & {Pull}  \\
 && {Total Error}       & {High-$Q^2$ Fit} & {    }  \\
\hline
\hline
&APV~\cite{QWCs:theo:2003:new}
                &                        &                      &       \\
\hline
&$\QWCs$        & $-72.74\pm0.46\pzz\pz$ & $-72.913\pm0.035\pzz$& $0.4$ \\
\hline
\hline
&M\o ller~\cite{E158RunI, *E158RunI+II+III}
                &                        &                      &        \\
\hline
&$\swsqMSb$     & $0.2330\pm0.0015\pz$   & $0.23111\pm0.00014$  & $1.3$  \\
\hline
\hline
&$\nu$N~\cite{bib-NuTeV-final}
                &                        &                      &        \\
\hline
&$\gnlq^2$      & $0.30005\pm0.00137$    & $0.30397\pm0.00020$  & $2.9$  \\
&$\gnrq^2$      & $0.03076\pm0.00110$    & $0.03012\pm0.00003$  & $0.6$  \\
\hline
\end{tabular}\end{center}
\caption[]{ Summary of measurements performed in low-$Q^2$ reactions,
  namely atomic parity violation, $e^-e^-$ Moller scattering and
  neutrino-nucleon scattering.  The SM results and the pulls
  (difference between measurement and fit in units of the total
  measurement error) are derived from the SM fit including all
  high-$Q^2$ data (Table~\ref{tab-BIGFIT}, column~4) with the Higgs
  mass treated as a free parameter.}
\label{tab-SMpred}
\end{table}

\begin{figure}[p]
\vspace*{-2.0cm}
\begin{center}
  \mbox{\includegraphics[height=21cm]{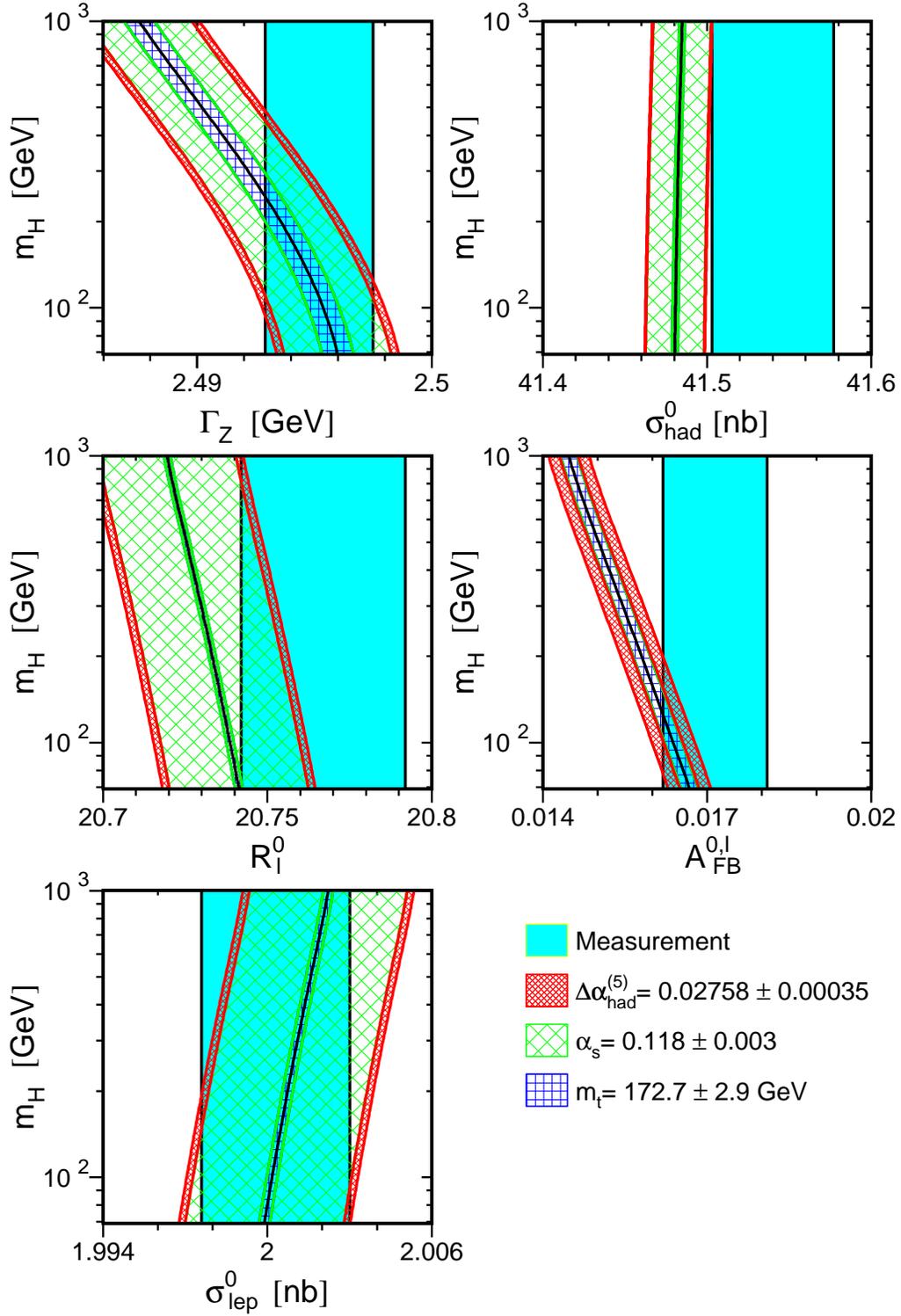}}
\end{center}
\vspace*{-0.6cm}
\caption[]{ Comparison of $\LEPI$ measurements with the SM prediction
  as a function of $\MH$.  The measurement with its error is shown as
  the vertical band.  The width of the SM band is due to the
  uncertainties in $\Delta\alpha^{(5)}_{\mathrm{had}}(\MZ^2)$,
  $\alfmz$ and $\Mt$.  The total width of the band is the linear sum
  of these effects.  }
\label{fig-higgs1}
\end{figure}

\begin{figure}[p]
\vspace*{-2.0cm}
\begin{center}
  \mbox{\includegraphics[height=21cm]{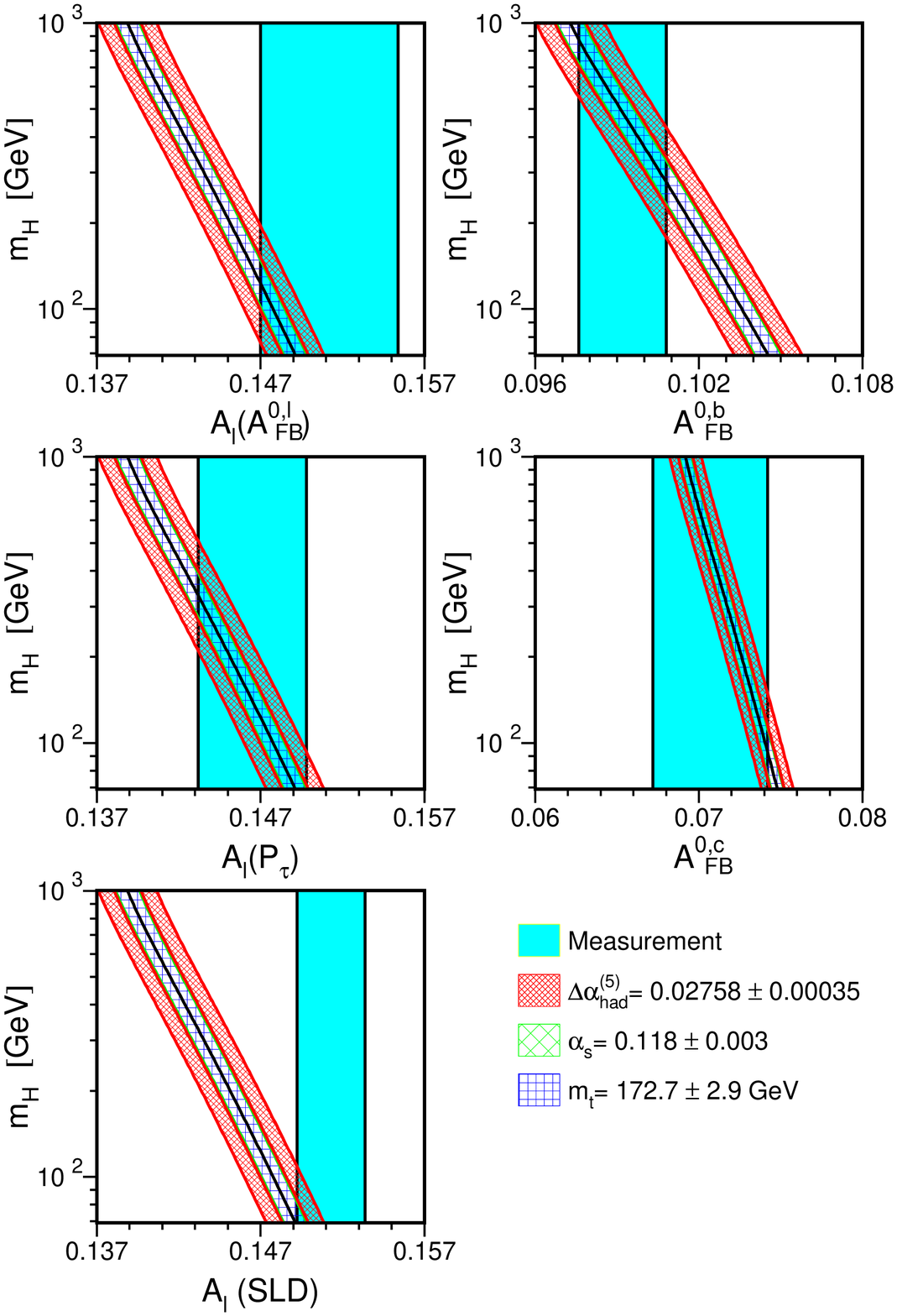}}
\end{center}
\vspace*{-0.6cm}
\caption[]{ Comparison of $\LEPI$ measurements with the SM prediction
  as a function of $\MH$.  The measurement with its error is shown as
  the vertical band.  The width of the SM band is due to the
  uncertainties in $\Delta\alpha^{(5)}_{\mathrm{had}}(\MZ^2)$,
  $\alfmz$ and $\Mt$.  The total width of the band is the linear sum
  of these effects.  Also shown is the comparison of the SLD
  measurement of $\cAl$, dominated by $\ALRz$, with the SM. }
\label{fig-higgs2}
\end{figure}

\begin{figure}[p]
\vspace*{-2.0cm}
\begin{center}
  \mbox{\includegraphics[height=21cm]{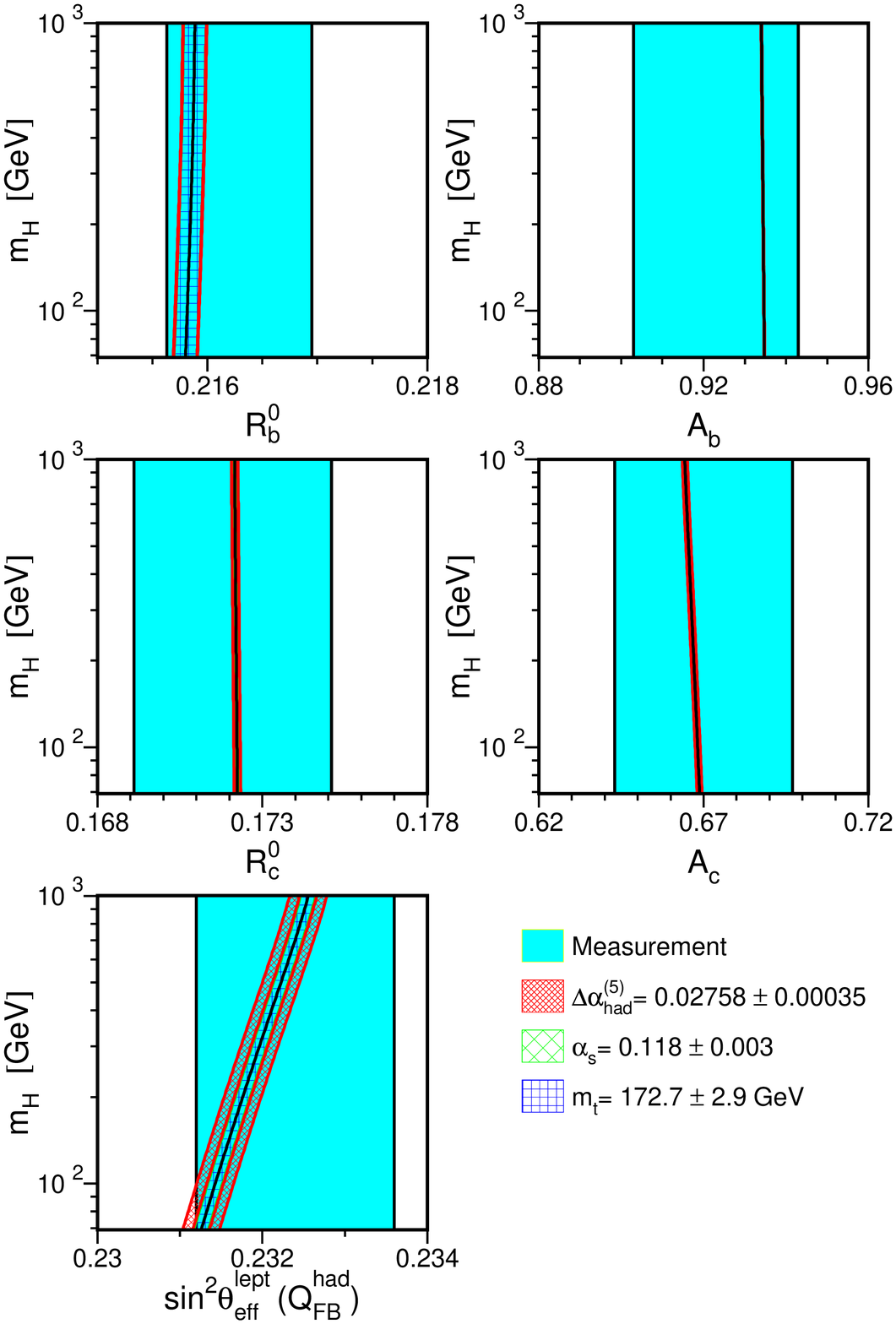}}
\end{center}
\vspace*{-0.6cm}
\caption[]{ Comparison of $\LEPI$ and SLD heavy-flavour measurements
  with the SM prediction as a function of $\MH$.  The measurement with
  its error is shown as the vertical band.  The width of the SM band
  is due to the uncertainties in
  $\Delta\alpha^{(5)}_{\mathrm{had}}(\MZ^2)$, $\alfmz$ and $\Mt$.  The
  total width of the band is the linear sum of these effects. Also
  shown is the comparison of the $\LEPI$ measurement of the inclusive
  hadronic charge asymmetry $\Qfbhad$ with the SM. }
\label{fig-higgs3}
\end{figure}

\begin{figure}[p]
\vspace*{-2.0cm}
\begin{center}
  \mbox{\includegraphics[height=21cm]{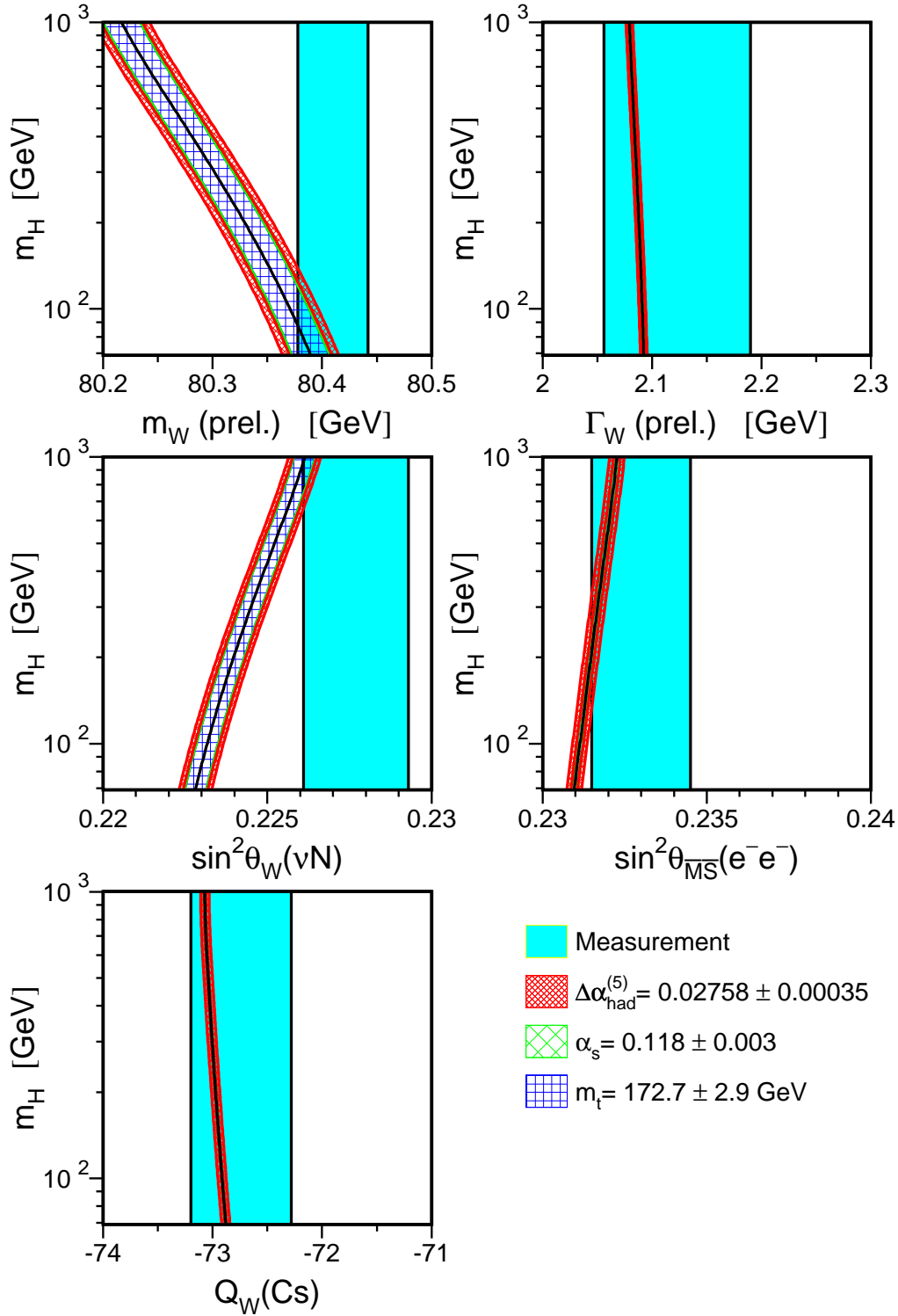}}
\end{center}
\vspace*{-0.6cm}
\caption[]{ Comparison of $\MW$ and $\GW$ measured at $\LEPII$ and
  $\pp$ colliders, of $\swsq$ measured by NuTeV and of APV in caesium
  with the SM prediction as a function of $\MH$.  The measurement with
  its error is shown as the vertical band.  The width of the SM band
  is due to the uncertainties in
  $\Delta\alpha^{(5)}_{\mathrm{had}}(\MZ^2)$, $\alfmz$ and $\Mt$.  The
  total width of the band is the linear sum of these effects.  }
\label{fig-higgs4}
\end{figure}

In Figures~\ref{fig-higgs1} to~\ref{fig-higgs4} the sensitivity of the
measurements to the Higgs mass is shown.  Besides the measurement of
the W mass, the most sensitive measurements are the asymmetries, \ie,
$\swsqeffl$.  A reduced uncertainty for the value of $\alpha(\MZ^2)$
would therefore result in an improved constraint on $\log\MH$ and thus
$\MH$, as already shown in Figures~\ref{fig-gllsef} and
\ref{fig-chiex}. 

\begin{figure}[p]
\vspace*{-2.0cm}
\begin{center}
  \mbox{\includegraphics[height=21cm]{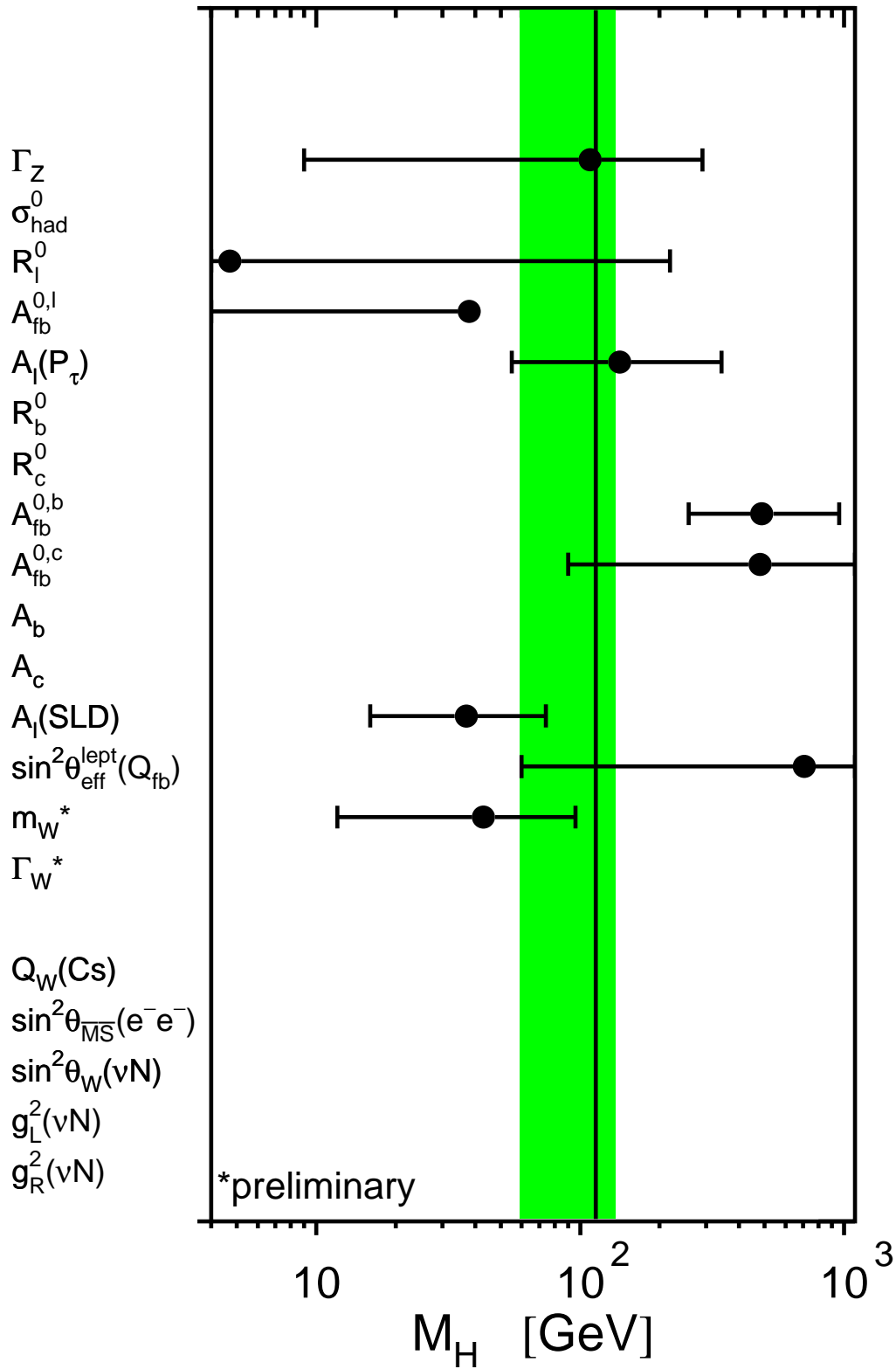}}
\end{center}
\vspace*{-0.6cm}
\caption[]{ Constraints on the mass of the Higgs boson from each
  pseudo-observable. The Higgs-boson mass and its 68\% CL uncertainty
  is obtained from a five-parameter SM fit to the observable,
  constraining
  $\Delta\alpha^{(5)}_{\mathrm{had}}(\MZ^2)=0.02761\pm0.00036$,
  $\alfmz=0.118\pm0.003$, $\MZ=91.1875\pm0.0021~\GeV$ and
  $\Mt=172.7\pm2.9~\GeV$.  Because of these four common constraints
  the resulting Higgs-boson mass values are highly correlated.  The
  shaded band denotes the overall constraint on the mass of the Higgs
  boson derived from all pseudo-observables including the above four
  SM parameters as reported in the last column of
  Table~\ref{tab-BIGFIT}.  }
\label{fig-higgs-obs}
\end{figure}

Given the constraints on the other four SM input parameters, each
observable is equivalent to a constraint on the mass of the SM Higgs
boson. The constraints on the mass of the SM Higgs boson resulting
from each observable are compared in Figure~\ref{fig-higgs-obs}.  For
vey low Higgs-masses, these constraints are qualitative only as the
effects of real Higgs-strahlung, neither included in the experimental
analyses nor in the SM calculations of expectations, may then become
sizeable~\cite{Kawamoto:2004pi}.


\boldmath
\chapter{Conclusions}
\label{sec-Conc}
\unboldmath

A combination of many electroweak measurements in electron-positron
collisions at centre-of-mass energies above the Z-pole is presented.
The LEP experiments ALEPH, DELPHI, L3 and OPAL wish to stress that
this report reflects for many of the results a preliminary status of
their analyses at the time of the 2005 summer conferences.  A
definitive statement on the results must wait for publication by each
collaboration.  Note that in some cases some experiments have already
published final results which are not yet included in the combinations
presented in this paper.

The preliminary and published results from the LEP experiments and
their combinations, test the Standard Model (SM) successfully at the
highest interaction energies.  The combination of the many precise
electroweak results, including those obtained at the
Z-pole~\cite{bib-Z-pole}, yields stringent constraints on the SM and
its free parameters.  Most measurements agree well with the
predictions.  The spread in values of the various determinations of
the effective electroweak mixing angle in asymmetry measurements at
the Z pole is somewhat larger than expected~\cite{bib-Z-pole}.  Within
the SM analysis, this seems to be caused by the measurement of the
forward-backward asymmetry in b-quark production, showing the largest
pull of all Z-pole measurements w.r.t. the SM expectation.  The final
result of the NuTeV collaboration on the low-$Q^2$ left-handed
couplings combination differs more, by three standard deviations from
the SM expectation calculated based on the high-$Q^2$ precision
electroweak measurements.

\boldmath
\section*{Prospects for the Future}
\unboldmath

The measurements from data taken at or near the Z resonance, both at
LEP as well as at SLC, are final and published~\cite{bib-Z-pole}.
Improvements in accuracy will therefore take place in the high energy
data (\LEPII), where each experiment has accumulated about
700~pb$^{-1}$ of data.  The measurements of $\MW$ are likely to reach
a precision not too far from the uncertainty on the prediction
obtained via the radiative corrections of the Z-pole data, providing
an important test of the Standard Model.  In the measurement of the
triple and quartic electroweak gauge boson self couplings, the
analysis of the complete $\LEPII$ statistics, together with the
increased sensitivity at higher beam energies, will lead to an
improvement in the current precision.

\section*{Acknowledgements}

We would like to thank the CERN accelerator divisions for the
efficient operation of the LEP accelerator, the precise information on
the absolute energy scale and their close cooperation with the four
experiments.  We would also like to thank members of the SLD, CDF,
D\O, E-158 and NuTeV collaborations for useful discussions concerning
their results.  Finally, the results of the section on Standard Model
constraints would not be possible without the close collaboration of
many theorists.

\clearpage

\begin{appendix}

\chapter{Detailed inputs and results on W-boson and four-fermion averages}
\label{4f_sec:appendix}

Tables~\ref{4f_tab:WWmeas}~-~\ref{4f_tab:rzeemeas}
give the details of the inputs and of the results
for the calculation of LEP averages
of the four-fermion cross-section and the corresponding cross-section ratios
For both inputs and results, whenever relevant,
the breakdown of the errors into their various components
is given in the table.

For each measurement, 
the Collaborations have privately 
provided
unpublished information which is necessary 
for the combination of LEP results,
such as the expected statistical error 
or the split up of the systematic uncertainty 
into its correlated and uncorrelated components.
Unless otherwise specified in the References,
all other inputs are taken from published papers 
and public notes submitted to conferences.

\begin{table}[hbtp]
\vspace*{-0.5cm}
\begin{center}
\begin{small}
\begin{tabular}{|c|ccccc|c|c|c|}
\cline{1-8}
\roots & & & {\scriptsize (LCEC)} & {\scriptsize (LUEU)} & 
{\scriptsize (LUEC)} & & &
\multicolumn{1}{|r}{$\quad$} \\
(GeV) & $\sww$ & 
$\Delta\sww^\mathrm{stat}$ &
$\Delta\sww^\mathrm{syst}$ &
$\Delta\sww^\mathrm{syst}$ &
$\Delta\sww^\mathrm{syst}$ &
$\Delta\sww^\mathrm{syst}$ &
$\Delta\sww$ & 
\multicolumn{1}{|r}{$\quad$} \\
\cline{1-8}
\multicolumn{8}{|c|}
{\Aleph~\cite{4f_bib:aleww}} &
\multicolumn{1}{|r}{$\quad$} \\
\cline{1-8}
182.7 & 15.86 & $\pm$0.61 & $\pm$0.08 & $\pm$0.08 & $\pm$0.09 & $\pm$0.14& $\pm$0.63 & \multicolumn{1}{|r}{$\quad$} \\
188.6 & 15.78 & $\pm$0.34 & $\pm$0.07 & $\pm$0.05 & $\pm$0.09 & $\pm$0.12& $\pm$0.36 & \multicolumn{1}{|r}{$\quad$} \\
191.6 & 17.10 & $\pm$0.90 & $\pm$0.07 & $\pm$0.07 & $\pm$0.09 & $\pm$0.14& $\pm$0.90 & \multicolumn{1}{|r}{$\quad$} \\
195.5 & 16.60 & $\pm$0.52 & $\pm$0.07 & $\pm$0.06 & $\pm$0.09 & $\pm$0.12& $\pm$0.54 & \multicolumn{1}{|r}{$\quad$} \\
199.5 & 16.93 & $\pm$0.50 & $\pm$0.07 & $\pm$0.06 & $\pm$0.09 & $\pm$0.12& $\pm$0.52 & \multicolumn{1}{|r}{$\quad$} \\
201.6 & 16.63 & $\pm$0.70 & $\pm$0.07 & $\pm$0.07 & $\pm$0.09 & $\pm$0.13& $\pm$0.71 & \multicolumn{1}{|r}{$\quad$} \\
204.9 & 16.84 & $\pm$0.53 & $\pm$0.07 & $\pm$0.06 & $\pm$0.09 & $\pm$0.13& $\pm$0.54 & \multicolumn{1}{|r}{$\quad$} \\
206.6 & 17.42 & $\pm$0.41 & $\pm$0.07 & $\pm$0.06 & $\pm$0.09 & $\pm$0.13& $\pm$0.43 & \multicolumn{1}{|r}{$\quad$} \\
\cline{1-8}
\multicolumn{8}{|c|}
{\Delphi~\cite{4f_bib:delww}} &
\multicolumn{1}{|r}{$\quad$} \\
\cline{1-8}
182.7 & 16.07 & $\pm$0.68 & $\pm$0.09 & $\pm$0.09 & $\pm$0.08 & $\pm$0.15& $\pm$0.70 & \multicolumn{1}{|r}{$\quad$} \\
188.6 & 16.09 & $\pm$0.39 & $\pm$0.08 & $\pm$0.09 & $\pm$0.09 & $\pm$0.15& $\pm$0.42 & \multicolumn{1}{|r}{$\quad$} \\
191.6 & 16.64 & $\pm$0.99 & $\pm$0.09 & $\pm$0.10 & $\pm$0.09 & $\pm$0.16& $\pm$1.00 & \multicolumn{1}{|r}{$\quad$} \\
195.5 & 17.04 & $\pm$0.58 & $\pm$0.09 & $\pm$0.10 & $\pm$0.09 & $\pm$0.16& $\pm$0.60 & \multicolumn{1}{|r}{$\quad$} \\
199.5 & 17.39 & $\pm$0.55 & $\pm$0.09 & $\pm$0.10 & $\pm$0.09 & $\pm$0.16& $\pm$0.57 & \multicolumn{1}{|r}{$\quad$} \\
201.6 & 17.37 & $\pm$0.80 & $\pm$0.10 & $\pm$0.10 & $\pm$0.09 & $\pm$0.17& $\pm$0.82 & \multicolumn{1}{|r}{$\quad$} \\
204.9 & 17.56 & $\pm$0.57 & $\pm$0.10 & $\pm$0.10 & $\pm$0.09 & $\pm$0.17& $\pm$0.59 & \multicolumn{1}{|r}{$\quad$} \\
206.6 & 16.35 & $\pm$0.44 & $\pm$0.10 & $\pm$0.10 & $\pm$0.09 & $\pm$0.17& $\pm$0.47 & \multicolumn{1}{|r}{$\quad$} \\
\cline{1-8}
\multicolumn{8}{|c|}
{\Ltre~\cite{4f_bib:ltrww}} &
\multicolumn{1}{|r}{$\quad$} \\
\cline{1-8}
182.7 & 16.53 & $\pm$0.67 & $\pm$0.19 & $\pm$0.13 & $\pm$0.12 & $\pm$0.26& $\pm$0.72 & \multicolumn{1}{|r}{$\quad$} \\
188.6 & 16.17 & $\pm$0.37 & $\pm$0.11 & $\pm$0.06 & $\pm$0.11 & $\pm$0.17& $\pm$0.41 & \multicolumn{1}{|r}{$\quad$} \\
191.6 & 16.11 & $\pm$0.90 & $\pm$0.11 & $\pm$0.07 & $\pm$0.11 & $\pm$0.17& $\pm$0.92 & \multicolumn{1}{|r}{$\quad$} \\
195.5 & 16.22 & $\pm$0.54 & $\pm$0.11 & $\pm$0.06 & $\pm$0.10 & $\pm$0.16& $\pm$0.57 & \multicolumn{1}{|r}{$\quad$} \\
199.5 & 16.49 & $\pm$0.56 & $\pm$0.11 & $\pm$0.07 & $\pm$0.11 & $\pm$0.17& $\pm$0.58 & \multicolumn{1}{|r}{$\quad$} \\
201.6 & 16.01 & $\pm$0.82 & $\pm$0.11 & $\pm$0.06 & $\pm$0.12 & $\pm$0.17& $\pm$0.84 & \multicolumn{1}{|r}{$\quad$} \\
204.9 & 17.00 & $\pm$0.58 & $\pm$0.12 & $\pm$0.06 & $\pm$0.11 & $\pm$0.17& $\pm$0.60 & \multicolumn{1}{|r}{$\quad$} \\
206.6 & 17.33 & $\pm$0.44 & $\pm$0.12 & $\pm$0.04 & $\pm$0.11 & $\pm$0.17& $\pm$0.47 & \multicolumn{1}{|r}{$\quad$} \\
\cline{1-8}
\multicolumn{8}{|c|}
{\Opal~\cite{4f_bib:opaww189,4f_bib:opawwsc01}} &
\multicolumn{1}{|r}{$\quad$} \\
\cline{1-8}
182.7 & 15.43 & $\pm$0.61 & $\pm$0.14 & $\pm$0.00 & $\pm$0.22 & $\pm$0.26& $\pm$0.66 & \multicolumn{1}{|r}{$\quad$} \\
188.6 & 16.30 & $\pm$0.35 & $\pm$0.11 & $\pm$0.12 & $\pm$0.07 & $\pm$0.18& $\pm$0.39 & \multicolumn{1}{|r}{$\quad$} \\
191.6 & 16.60 & $\pm$0.90 & $\pm$0.23 & $\pm$0.32 & $\pm$0.14 & $\pm$0.42& $\pm$0.99 & \multicolumn{1}{|r}{$\quad$} \\
195.5 & 18.59 & $\pm$0.61 & $\pm$0.23 & $\pm$0.34 & $\pm$0.14 & $\pm$0.43& $\pm$0.75 & \multicolumn{1}{|r}{$\quad$} \\
199.5 & 16.32 & $\pm$0.55 & $\pm$0.23 & $\pm$0.26 & $\pm$0.14 & $\pm$0.37& $\pm$0.67 & \multicolumn{1}{|r}{$\quad$} \\
201.6 & 18.48 & $\pm$0.82 & $\pm$0.23 & $\pm$0.33 & $\pm$0.14 & $\pm$0.42& $\pm$0.92 & \multicolumn{1}{|r}{$\quad$} \\
204.9 & 15.97 & $\pm$0.52 & $\pm$0.23 & $\pm$0.26 & $\pm$0.14 & $\pm$0.37& $\pm$0.64 & \multicolumn{1}{|r}{$\quad$} \\
206.6 & 17.77 & $\pm$0.42 & $\pm$0.23 & $\pm$0.28 & $\pm$0.14 & $\pm$0.38& $\pm$0.57 & \multicolumn{1}{|r}{$\quad$} \\
\cline{1-8}
\hline
\multicolumn{8}{|c|}{LEP Averages } & $\chi^2/\textrm{d.o.f.}$ \\
\hline
182.7 & 15.88 & $\pm$0.33 & $\pm$0.10 & $\pm$0.05 & $\pm$0.06 & $\pm$0.13& $\pm$0.35 & 
 \multirow{8}{20.3mm}{$
   \hspace*{-0.3mm}
   \left\}
     \begin{array}[h]{rr}
       &\multirow{8}{8mm}{\hspace*{-4.2mm}26.6/24}\\
       &\\ &\\ &\\ &\\ &\\ &\\ &\\  
     \end{array}
   \right.
   $}\\
188.6 & 16.03 & $\pm$0.18 & $\pm$0.08 & $\pm$0.04 & $\pm$0.05 & $\pm$0.10& $\pm$0.21 & \\
191.6 & 16.56 & $\pm$0.46 & $\pm$0.10 & $\pm$0.08 & $\pm$0.05 & $\pm$0.14& $\pm$0.48 & \\
195.5 & 16.90 & $\pm$0.29 & $\pm$0.09 & $\pm$0.06 & $\pm$0.05 & $\pm$0.12& $\pm$0.31 & \\
199.5 & 16.76 & $\pm$0.27 & $\pm$0.10 & $\pm$0.06 & $\pm$0.05 & $\pm$0.13& $\pm$0.30 & \\
201.6 & 16.99 & $\pm$0.39 & $\pm$0.10 & $\pm$0.07 & $\pm$0.05 & $\pm$0.13& $\pm$0.41 & \\
204.9 & 16.79 & $\pm$0.28 & $\pm$0.10 & $\pm$0.07 & $\pm$0.05 & $\pm$0.13& $\pm$0.31 & \\
206.6 & 17.15 & $\pm$0.22 & $\pm$0.10 & $\pm$0.06 & $\pm$0.05 & $\pm$0.13& $\pm$0.25 & \\
\hline
\end{tabular}
\end{small}
\caption[]{%
W-pair production cross-section (in pb) for different \CoM\ energies.
The first column contains the \CoM\ energy
and the second the measurements.
Observed statistical uncertainties are used in the fit
and are listed in the third column;
when asymmetric errors are quoted by the Collaborations,
the positive error is listed in the table and used in the fit.
The fourth, fifth and sixth columns contain
the components of the systematic errors,
as subdivided by the Collaborations into
LEP-correlated   energy-correlated   (LCEC),
LEP-uncorrelated energy-uncorrelated (LUEU),
LEP-uncorrelated energy-correlated   (LUEC).
The total systematic error is given in the seventh column,
the total error in the eighth.
For the LEP averages, the $\chi^2$ of the fit is also given
in the ninth column.}
\label{4f_tab:WWmeas} 
\end{center}
\end{table}

\begin{table}[hbtp]
\begin{center}
\hspace*{-0.3cm}
\renewcommand{\arraystretch}{1.2}
\begin{tabular}{|c|cccccccc|} 
\hline
\roots (GeV) 
      & 182.7 & 188.6 & 191.6 & 195.5 & 199.5 & 201.6 & 204.9 & 206.6 \\
\hline
182.7 & 1.000 & 0.161 & 0.090 & 0.128 & 0.138 & 0.101 & 0.141 & 0.165 \\
188.6 & 0.161 & 1.000 & 0.117 & 0.169 & 0.180 & 0.131 & 0.182 & 0.217 \\
191.6 & 0.090 & 0.117 & 1.000 & 0.093 & 0.099 & 0.072 & 0.101 & 0.119 \\
195.5 & 0.128 & 0.169 & 0.093 & 1.000 & 0.143 & 0.104 & 0.145 & 0.171 \\
199.5 & 0.138 & 0.180 & 0.099 & 0.143 & 1.000 & 0.111 & 0.155 & 0.183 \\
201.6 & 0.101 & 0.131 & 0.072 & 0.104 & 0.111 & 1.000 & 0.113 & 0.134 \\
204.9 & 0.141 & 0.182 & 0.101 & 0.145 & 0.155 & 0.113 & 1.000 & 0.186 \\
206.6 & 0.165 & 0.217 & 0.119 & 0.171 & 0.183 & 0.134 & 0.186 & 1.000 \\
\hline
\end{tabular}
\renewcommand{\arraystretch}{1.}
\caption[]{%
Correlation matrix for the LEP combined W-pair cross-sections
listed at the bottom of Table~\protect\ref{4f_tab:WWmeas}.
Correlations are all positive and range from 9\% to 22\%.}
\label{4f_tab:WWcorr} 
\end{center}
\end{table}

\begin{table}[hbtp]
\begin{center}
\hspace*{-0.3cm}
\renewcommand{\arraystretch}{1.2}
\begin{tabular}{|c|c|c|} 
\hline
\roots & \multicolumn{2}{|c|}{WW cross-section (pb)}                              \\
\cline{2-3} 
(GeV) & $\sww^{\footnotesize\YFSWW}$    
      & $\sww^{\footnotesize\RacoonWW}$ \\
\hline
182.7 & $15.361\pm0.005$ & $15.368\pm0.008$ \\
188.6 & $16.266\pm0.005$ & $16.249\pm0.011$ \\
191.6 & $16.568\pm0.006$ & $16.519\pm0.009$ \\
195.5 & $16.841\pm0.006$ & $16.801\pm0.009$ \\
199.5 & $17.017\pm0.007$ & $16.979\pm0.009$ \\
201.6 & $17.076\pm0.006$ & $17.032\pm0.009$ \\
204.9 & $17.128\pm0.006$ & $17.079\pm0.009$ \\
206.6 & $17.145\pm0.006$ & $17.087\pm0.009$ \\
\hline
\end{tabular}
\renewcommand{\arraystretch}{1.}
\caption[]{%
W-pair cross-section predictions (in pb) for different \CoM\ energies,
according to \YFSWW~\protect\cite{4f_bib:yfsww} and 
\RacoonWW~\protect\cite{4f_bib:racoonww},
for $\Mw=80.35$~GeV.
The errors listed in the table are only the statistical errors 
from the numerical integration of the cross-section.}
\label{4f_tab:WWtheo} 
\end{center}
\end{table}

\begin{table}[hbtp]
\begin{center}
\begin{small}
\begin{tabular}{|c|cccccc|c|c|}
\hline
\roots & & & {\scriptsize (LCEU)} & {\scriptsize (LCEC)} & 
{\scriptsize (LUEU)} & {\scriptsize (LUEC)} & & \\
(GeV) & $\rww$ & 
$\Delta\rww^\mathrm{stat}$ &
$\Delta\rww^\mathrm{syst}$ &
$\Delta\rww^\mathrm{syst}$ &
$\Delta\rww^\mathrm{syst}$ &
$\Delta\rww^\mathrm{syst}$ &
$\Delta\rww$ &
$\chi^2/\textrm{d.o.f.}$ \\
\hline
\hline

\multicolumn{9}{|c|}{\YFSWW~\cite{4f_bib:yfsww}}\\
\hline
182.7 & 1.034 & $\pm$0.021 & $\pm$0.000 & $\pm$0.006 & $\pm$0.003& $\pm$0.004 & $\pm$0.023&
\multirow{8}{20.3mm}{$
  \hspace*{-0.3mm}
  \left\}
    \begin{array}[h]{rr}
      &\multirow{8}{6mm}{\hspace*{-4.2mm}26.6/24}\\
      &\\ &\\ &\\ &\\ &\\ &\\ &\\  
    \end{array}
  \right.
  $}\\
188.6 & 0.986 & $\pm$0.011 & $\pm$0.000 & $\pm$0.005 & $\pm$0.003& $\pm$0.003 & $\pm$0.013&\\
191.6 & 1.000 & $\pm$0.028 & $\pm$0.000 & $\pm$0.006 & $\pm$0.005& $\pm$0.003 & $\pm$0.029&\\
195.5 & 1.003 & $\pm$0.017 & $\pm$0.000 & $\pm$0.006 & $\pm$0.004& $\pm$0.003 & $\pm$0.019&\\
199.5 & 0.985 & $\pm$0.016 & $\pm$0.000 & $\pm$0.006 & $\pm$0.004& $\pm$0.003 & $\pm$0.018&\\
201.6 & 0.995 & $\pm$0.023 & $\pm$0.000 & $\pm$0.006 & $\pm$0.004& $\pm$0.003 & $\pm$0.024&\\
204.9 & 0.980 & $\pm$0.016 & $\pm$0.000 & $\pm$0.006 & $\pm$0.004& $\pm$0.003 & $\pm$0.018&\\
206.6 & 1.000 & $\pm$0.013 & $\pm$0.000 & $\pm$0.006 & $\pm$0.003& $\pm$0.003 & $\pm$0.015&\\
\hline
Average & 
0.994 & $\pm$0.006 & $\pm$0.000 & $\pm$0.005 & $\pm$0.001& $\pm$0.003 & $\pm$0.009&
\hspace*{1.5mm}32.2/31\hspace*{-0.5mm}\\
\hline
\hline
\multicolumn{9}{|c|}{\RacoonWW~\cite{4f_bib:racoonww}}\\
\hline
182.7 & 1.033 & $\pm$0.021 & $\pm$0.001 & $\pm$0.006 & $\pm$0.003& $\pm$0.004 & $\pm$0.023&
\multirow{8}{20.3mm}{$
  \hspace*{-0.3mm}
  \left\}
    \begin{array}[h]{rr}
      &\multirow{8}{6mm}{\hspace*{-4.2mm}26.6/24}\\
      &\\ &\\ &\\ &\\ &\\ &\\ &\\  
    \end{array}
  \right.
  $}\\
188.6 & 0.987 & $\pm$0.011 & $\pm$0.001 & $\pm$0.005 & $\pm$0.003& $\pm$0.003 & $\pm$0.013&\\
191.6 & 1.003 & $\pm$0.028 & $\pm$0.001 & $\pm$0.006 & $\pm$0.005& $\pm$0.003 & $\pm$0.029&\\
195.5 & 1.006 & $\pm$0.017 & $\pm$0.001 & $\pm$0.006 & $\pm$0.004& $\pm$0.003 & $\pm$0.019&\\
199.5 & 0.987 & $\pm$0.016 & $\pm$0.001 & $\pm$0.006 & $\pm$0.004& $\pm$0.003 & $\pm$0.018&\\
201.6 & 0.998 & $\pm$0.023 & $\pm$0.001 & $\pm$0.006 & $\pm$0.004& $\pm$0.003 & $\pm$0.024&\\
204.9 & 0.983 & $\pm$0.016 & $\pm$0.001 & $\pm$0.006 & $\pm$0.004& $\pm$0.003 & $\pm$0.018&\\
206.6 & 1.004 & $\pm$0.013 & $\pm$0.001 & $\pm$0.006 & $\pm$0.003& $\pm$0.003 & $\pm$0.015&\\
\hline
Average & 
0.996 & $\pm$0.006 & $\pm$0.000 & $\pm$0.006 & $\pm$0.001& $\pm$0.003 & $\pm$0.009&
\hspace*{1.5mm}32.0/31\hspace*{-0.5mm}\\
\hline
\end{tabular}
\end{small}
\caption[]{%
Ratios of LEP combined W-pair cross-section measurements
to the expectations of the considered theoretical models,
for different \CoM\ energies and for all energies combined.
The first column contains the \CoM\ energy,
the second the combined ratios,
the third the statistical errors.
The fourth, fifth, sixth and seventh columns contain
the sources of systematic errors that are considered as 
LEP-correlated   energy-uncorrelated (LCEU),
LEP-correlated   energy-correlated   (LCEC),
LEP-uncorrelated energy-uncorrelated (LUEU),
LEP-uncorrelated energy-correlated   (LUEC).
The total error is given in the eighth column.
The only LCEU systematic sources considered 
are the statistical errors on the cross-section theoretical predictions,
while the LCEC, LUEU and LUEC sources are those coming from
the corresponding errors on the cross-section measurements.
For the LEP averages, the $\chi^2$ of the fit is also given
in the ninth column.}
\label{4f_tab:rWWmeas} 
\end{center}
\end{table}

 \renewcommand{\arraystretch}{1.2}
 \begin{table}[p]
 \begin{center}
 \begin{small}
 \hspace*{-0.0cm}
 \begin{tabular}{|l|cccc|c|c|c|}
 \cline{1-8}
 Decay & & & {\scriptsize (unc)} & {\scriptsize (cor)} & & & 
 3$\times$3 correlation \\
 channel & $\wwbr$ & 
 $\Delta\wwbr^\mathrm{stat}$ &
 $\Delta\wwbr^\mathrm{syst}$ &
 $\Delta\wwbr^\mathrm{syst}$ &
 $\Delta\wwbr^\mathrm{syst}$ &
 $\Delta\wwbr$ & 
 for $\Delta\wwbr$\\
 \cline{1-8}
 \multicolumn{8}{|c|}{\Aleph~\cite{4f_bib:aleww}}\\
 \hline
 \BWtoenu & 
 10.78 & $\pm$0.27 & $\pm$0.09 & $\pm$0.04 & $\pm$0.10 & $\pm$0.29 &
 \multirow{3}{47mm}{\mbox{$\Biggl(\negthickspace\negthickspace$
                      \begin{tabular}{ccc}
                       \phm1.000 &    -0.009 &    -0.332 \\
                          -0.009 & \phm1.000 &    -0.268 \\
                          -0.332 &    -0.268 & \phm1.000 \\
                      \end{tabular}
                      $\negthickspace\negthickspace\Biggr)$} } \\
 \BWtomnu & 
 10.87 & $\pm$0.25 & $\pm$0.07 & $\pm$0.04 & $\pm$0.08 & $\pm$0.26 & \\
 \BWtotnu & 
 11.25 & $\pm$0.32 & $\pm$0.19 & $\pm$0.05 & $\pm$0.20 & $\pm$0.38 & \\
 \hline
 \multicolumn{8}{c}{}\\
 
 \cline{1-8}
 \multicolumn{8}{|c|}{\Delphi~\cite{4f_bib:delww}}\\
 \hline
 \BWtoenu & 
 10.55 & $\pm$0.31 & $\pm$0.13 & $\pm$0.05 & $\pm$0.14 & $\pm$0.34 &
 \multirow{3}{47mm}{\mbox{$\Biggl(\negthickspace\negthickspace$
                      \begin{tabular}{ccc}
                       \phm1.000 &     0.030 &    -0.340 \\
                           0.030 & \phm1.000 &    -0.170 \\
                          -0.340 &    -0.170 & \phm1.000 \\
                      \end{tabular}
                      $\negthickspace\negthickspace\Biggr)$} } \\
 \BWtomnu & 
 10.65 & $\pm$0.26 & $\pm$0.06 & $\pm$0.05 & $\pm$0.08 & $\pm$0.27 & \\
 \BWtotnu & 
 11.46 & $\pm$0.39 & $\pm$0.17 & $\pm$0.09 & $\pm$0.19 & $\pm$0.43 & \\
 \hline
 \multicolumn{8}{c}{}\\
 
 \cline{1-8}
 \multicolumn{8}{|c|}{\Ltre~\cite{4f_bib:ltrww}}\\
 \hline
 \BWtoenu & 
 10.78 & $\pm$0.29 & $\pm$0.10 & $\pm$0.08 & $\pm$0.13 & $\pm$0.32 & 
 \multirow{3}{47mm}{\mbox{$\Biggl(\negthickspace\negthickspace$
                      \begin{tabular}{ccc}
                       \phm1.000 &    -0.016 &    -0.279 \\
                          -0.016 & \phm1.000 &    -0.295 \\
                          -0.279 &    -0.295 & \phm1.000 \\
                      \end{tabular}
                      $\negthickspace\negthickspace\Biggr)$} } \\
 \BWtomnu & 
 10.03 & $\pm$0.29 & $\pm$0.10 & $\pm$0.07 & $\pm$0.12 & $\pm$0.31 & \\
 \BWtotnu & 
 11.89 & $\pm$0.40 & $\pm$0.17 & $\pm$0.11 & $\pm$0.20 & $\pm$0.45 & \\
 \hline
 \multicolumn{8}{c}{}\\
 
 \cline{1-8}
 \multicolumn{8}{|c|}{\Opal~\cite{4f_bib:opaww189,4f_bib:opawwsc01}}\\
 \hline
 \BWtoenu & 
 10.40 & $\pm$0.25 & $\pm$0.24 & $\pm$0.05 & $\pm$0.25 & $\pm$0.35 & 
 \multirow{3}{47mm}{\mbox{$\Biggl(\negthickspace\negthickspace$
                      \begin{tabular}{ccc}
                       \phm1.000 &     0.141 &    -0.179 \\
                           0.141 & \phm1.000 &    -0.174 \\
                          -0.179 &    -0.174 & \phm1.000 \\
                      \end{tabular}
                      $\negthickspace\negthickspace\Biggr)$} } \\
 \BWtomnu & 
 10.61 & $\pm$0.25 & $\pm$0.23 & $\pm$0.06 & $\pm$0.24 & $\pm$0.35 & \\
 \BWtotnu & 
 11.18 & $\pm$0.31 & $\pm$0.37 & $\pm$0.05 & $\pm$0.37 & $\pm$0.48 & \\
 \hline
 \multicolumn{8}{c}{}\\
 
 \cline{1-8}
 \multicolumn{7}{|c}{LEP Average (without lepton universality assumption)}
 &\multicolumn{1}{c|}{}\\
 \hline
 \BWtoenu & 
 10.65 & $\pm$0.14 & $\pm$0.07 & $\pm$0.05 & $\pm$0.09 & $\pm$0.17 &
 \multirow{3}{47mm}{\mbox{$\Biggl(\negthickspace\negthickspace$
                      \begin{tabular}{ccc}
                        \phm1.000 & \phm0.110 &    -0.195 \\
                        \phm0.110 & \phm1.000 &    -0.132 \\
                           -0.195 &    -0.132 & \phm1.000 \\
                      \end{tabular}
                      $\negthickspace\negthickspace\Biggr)$} } \\
 \BWtomnu & 
 10.59 & $\pm$0.13 & $\pm$0.05 & $\pm$0.05 & $\pm$0.08 & $\pm$0.15 & \\
 \BWtotnu & 
 11.44 & $\pm$0.18 & $\pm$0.11 & $\pm$0.07 & $\pm$0.13 & $\pm$0.22 & \\
 \hline
 $\chi^2/\textrm{d.o.f.}$ & \multicolumn{1}{|c|}{6.3/9} & 
 \multicolumn{6}{c}{}\\
 \cline{1-2} 
 \multicolumn{8}{c}{}\\
 
 \cline{1-7} 
 \multicolumn{7}{|c|}{LEP Average (with lepton universality assumption)}
 &\multicolumn{1}{c}{}\\
 \cline{1-7} 
 \BWtolnu & 
 10.84 & $\pm$0.06 & $\pm$0.04 & $\pm$0.06 & $\pm$0.07 & $\pm$0.09 & 
 \multicolumn{1}{c}{}\\
 {\mbox{$\mathcal{B}(\mathrm{W}\rightarrow\mathrm{had.})$}}  & 
 67.48 & $\pm$0.19 & $\pm$0.12 & $\pm$0.18 & $\pm$0.21 & $\pm$0.28 & 
 \multicolumn{1}{c}{}\\
 \cline{1-7} 
 $\chi^2/\textrm{d.o.f.}$ & \multicolumn{1}{|c|}{15.4/11} &
 \multicolumn{6}{c}{}\\
 \cline{1-2} 
 \end{tabular}
 \vspace*{0.5cm}
 
 \end{small}
 \caption[]{%
 W branching fraction measurements (in \%).
 The first column contains the decay channel, 
 the second the measurements,
 the third the statistical uncertainty.
 The fourth and fifth column list 
 the uncorrelated and correlated components
 of the systematic errors,
 as provided by the Collaborations.
 The total systematic error is given in the sixth column and
 the total error in the seventh. 
 Correlation matrices 
 for the three leptonic branching fractions 
 are given in the last column.}
 \label{4f_tab:Wbrmeas} 
 \end{center}
 \end{table}
 \renewcommand{\arraystretch}{1.}

\begin{table}[hbtp]
\begin{center}
\begin{small}
\begin{tabular}{|c|}
\hline
ALEPH~\cite{4f_bib:aleww} \\
\hline 
\end{tabular}
\\
\begin{tabular}{|c|c|c|}
\hline
$\sqrt{s}$ interval (GeV) & Luminosity (pb$^{-1}$) & Lumi weighted $\sqrt{s}$ (GeV) \\
180-184 & 56.81 & 182.65 \\
\hline
\end{tabular}
\begin{tabular}{|c|c|c|c|c|c|c|c|c|c|c|}
\hline
cos$\theta_{\mathrm{W}-}$ bin $i$ & 1 & 2 & 3 & 4 & 5 & 6 & 7 & 8 & 9 & 10 \\
$\sigma_i$  (pb)                 & 0.216 & 0.498 & 0.696 & 1.568 & 1.293 & 1.954 & 2.486 & 2.228 & 4.536 & 6.088 \\
$\delta\sigma_i$(stat)  (pb)     & 0.053 & 0.137 & 0.185 & 0.517 & 0.319 & 0.481 & 0.552 & 0.363 & 0.785 & 0.874 \\
$\delta\sigma_i$(stat,exp.) (pb) & 0.263 & 0.276 & 0.309 & 0.341 & 0.376 & 0.415 & 0.459 & 0.523 & 0.597 & 0.714 \\
$\delta\sigma_i$(syst,unc)  (pb) & 0.012 & 0.018 & 0.017 & 0.025 & 0.023 & 0.021 & 0.036 & 0.047 & 0.047 & 0.066 \\
$\delta\sigma_i$(syst,cor)  (pb) & 0.004 & 0.003 & 0.003 & 0.003 & 0.003 & 0.004 & 0.004 & 0.003 & 0.004 & 0.006 \\
\hline
\end{tabular}

\begin{tabular}{|c|c|c|}
\hline
$\sqrt{s}$ interval (GeV) & Luminosity (pb$^{-1}$) & Lumi weighted $\sqrt{s}$ (GeV) \\
184-194 & 203.14 & 189.05 \\
\hline
\end{tabular}
\begin{tabular}{|c|c|c|c|c|c|c|c|c|c|c|}
\hline
cos$\theta_{\mathrm{W}-}$ bin $i$ & 1 & 2 & 3 & 4 & 5 & 6 & 7 & 8 & 9 & 10 \\
$\sigma_i$  (pb)                 & 0.665 & 0.743 & 0.919 & 0.990 & 1.156 & 2.133 & 2.795 & 3.070 & 3.851 & 5.772 \\
$\delta\sigma_i$(stat)  (pb)     & 0.148 & 0.140 & 0.158 & 0.142 & 0.144 & 0.287 & 0.337 & 0.297 & 0.300 & 0.366 \\
$\delta\sigma_i$(stat,exp.) (pb) & 0.132 & 0.147 & 0.157 & 0.175 & 0.196 & 0.223 & 0.246 & 0.282 & 0.332 & 0.408 \\
$\delta\sigma_i$(syst,unc)  (pb) & 0.010 & 0.016 & 0.015 & 0.024 & 0.021 & 0.020 & 0.035 & 0.047 & 0.049 & 0.075 \\
$\delta\sigma_i$(syst,cor)  (pb) & 0.003 & 0.003 & 0.003 & 0.002 & 0.002 & 0.003 & 0.003 & 0.003 & 0.005 & 0.005 \\
\hline
\end{tabular}

\begin{tabular}{|c|c|c|}
\hline
$\sqrt{s}$ interval (GeV) & Luminosity (pb$^{-1}$) & Lumi weighted $\sqrt{s}$ (GeV) \\
194-204 & 208.03 & 198.42 \\
\hline
\end{tabular}
\begin{tabular}{|c|c|c|c|c|c|c|c|c|c|c|}
\hline
cos$\theta_{\mathrm{W}-}$ bin $i$ & 1 & 2 & 3 & 4 & 5 & 6 & 7 & 8 & 9 & 10 \\

$\sigma_i$ (pb)                   & 0.802 & 0.475 & 0.886 & 0.972 & 1.325 & 1.889 & 2.229 & 3.581 & 4.428 & 6.380 \\
$\delta\sigma_i$(stat) (pb)       & 0.225 & 0.082 & 0.162 & 0.147 & 0.186 & 0.248 & 0.245 & 0.363 & 0.343 & 0.368 \\
$\delta\sigma_i$(stat,exp.) (pb)  & 0.124 & 0.134 & 0.149 & 0.167 & 0.188 & 0.214 & 0.241 & 0.281 & 0.338 & 0.433 \\
$\delta\sigma_i$(syst,unc) (pb)   & 0.007 & 0.013 & 0.012 & 0.021 & 0.018 & 0.016 & 0.032 & 0.046 & 0.049 & 0.082 \\
$\delta\sigma_i$(syst,cor) (pb)   & 0.003 & 0.002 & 0.002 & 0.002 & 0.002 & 0.002 & 0.002 & 0.003 & 0.003 & 0.004 \\
\hline
\end{tabular}

\begin{tabular}{|c|c|c|}
\hline
$\sqrt{s}$ interval (GeV) & Luminosity (pb$^{-1}$) & Lumi weighted $\sqrt{s}$ (GeV) \\
204-210 & 214.62 & 205.90 \\
\hline
\end{tabular}
\begin{tabular}{|c|c|c|c|c|c|c|c|c|c|c|}
\hline
cos$\theta_{\mathrm{W}-}$ bin $i$ & 1 & 2 & 3 & 4 & 5 & 6 & 7 & 8 & 9 & 10 \\

$\sigma_i$ (pb)                   & 0.334 & 0.637 & 0.800 & 1.229 & 1.229 & 1.789 & 2.810 & 2.740 & 4.192 & 8.005 \\
$\delta\sigma_i$(stat) (pb)       & 0.072 & 0.136 & 0.148 & 0.224 & 0.176 & 0.237 & 0.351 & 0.246 & 0.306 & 0.474 \\
$\delta\sigma_i$(stat,exp.) (pb)  & 0.114 & 0.126 & 0.143 & 0.155 & 0.180 & 0.206 & 0.234 & 0.273 & 0.338 & 0.443 \\
$\delta\sigma_i$(syst,unc) (pb)   & 0.008 & 0.013 & 0.013 & 0.020 & 0.018 & 0.017 & 0.033 & 0.046 & 0.052 & 0.089 \\
$\delta\sigma_i$(syst,cor) (pb)   & 0.003 & 0.003 & 0.003 & 0.002 & 0.002 & 0.003 & 0.003 & 0.003 & 0.004 & 0.005 \\
\hline
\end{tabular}
\end{small}
\caption[]{%
W$^{-}$ differential angular cross-section in the 10 angular bins for the four chosen energy intervals
for the \Aleph\ experiment. For each energy range, the measured integrated luminosity and the luminosity
weighted centre-of-mass energy is reported.
The results per angular bin in each of the energy interval are then presented: $\sigma_{i}$ indicates 
the average of d[$\sigma_{\mathrm{WW}}$(BR$_{e\nu}$+BR$_{\mu\nu}$)]/dcos$\theta_{\mathrm{W}^-}$ 
in the $i$-th bin of cos$\theta_{\mathrm{W}^-}$ with width 0.2.
The values, in each bin, of the measured and expected statistical error and of the systematic errors,
LEP uncorrelated and correlated, are reported as well. All values are expressed in pb
}
\label{4f_tab:dsdcost_aleph} 
\end{center}
\end{table}

\begin{table}[hbtp]
\begin{center}
\begin{small}
\begin{tabular}{|c|}
\hline
DELPHI~\cite{4f_bib:delww} \\
\hline 
\end{tabular}
\\
\begin{tabular}{|c|c|c|}
\hline
$\sqrt{s}$ interval (GeV) & Luminosity (pb$^{-1}$) & Lumi weighted $\sqrt{s}$ (GeV) \\
180-184 & 51.63 & 182.65 \\
\hline
\end{tabular}
\begin{tabular}{|c|c|c|c|c|c|c|c|c|c|c|}
\hline
cos$\theta_{\mathrm{W}-}$ bin $i$ & 1 & 2 & 3 & 4 & 5 & 6 & 7 & 8 & 9 & 10 \\
$\sigma_i$  (pb)                 & 0.715 & 0.795 & 1.175 & 1.365 & 1.350 & 1.745 & 1.995 & 2.150 & 4.750 & 6.040 \\
$\delta\sigma_i$(stat)  (pb)     & 0.320 & 0.315 & 0.380 & 0.400 & 0.400 & 0.450 & 0.485 & 0.510 & 0.775 & 0.895 \\
$\delta\sigma_i$(stat,exp.) (pb) & 0.320 & 0.315 & 0.350 & 0.370 & 0.405 & 0.450 & 0.505 & 0.580 & 0.695 & 0.850 \\ 
$\delta\sigma_i$(syst,unc)  (pb) & 0.020 & 0.025 & 0.035 & 0.035 & 0.040 & 0.085 & 0.050 & 0.065 & 0.095 & 0.075 \\
$\delta\sigma_i$(syst,cor)  (pb) & 0.045 & 0.025 & 0.020 & 0.015 & 0.015 & 0.025 & 0.015 & 0.015 & 0.030 & 0.035 \\
\hline
\end{tabular}

\begin{tabular}{|c|c|c|}
\hline
$\sqrt{s}$ interval (GeV) & Luminosity (pb$^{-1}$) & Lumi weighted $\sqrt{s}$ (GeV) \\
184-194 & 178.32 & 189.03 \\
\hline
\end{tabular}
\begin{tabular}{|c|c|c|c|c|c|c|c|c|c|c|}
\hline
cos$\theta_{\mathrm{W}-}$ bin $i$ & 1 & 2 & 3 & 4 & 5 & 6 & 7 & 8 & 9 & 10 \\
$\sigma_i$  (pb)                 & 0.865 & 0.760 & 0.990 & 0.930 & 1.330 & 1.460 & 1.675 & 2.630 & 4.635 & 5.4000 \\
$\delta\sigma_i$(stat)  (pb)     & 0.180 & 0.170 & 0.185 & 0.180 & 0.215 & 0.225 & 0.240 & 0.300 & 0.405 & 0.4550 \\
$\delta\sigma_i$(stat,exp.) (pb) & 0.165 & 0.170 & 0.180 & 0.200 & 0.215 & 0.240 & 0.270 & 0.320 & 0.385 & 0.4900 \\
$\delta\sigma_i$(syst,unc)  (pb) & 0.020 & 0.020 & 0.035 & 0.035 & 0.040 & 0.085 & 0.050 & 0.060 & 0.100 & 0.0850 \\
$\delta\sigma_i$(syst,cor)  (pb) & 0.040 & 0.020 & 0.020 & 0.015 & 0.015 & 0.020 & 0.015 & 0.015 & 0.025 & 0.0350 \\
\hline
\end{tabular}

\begin{tabular}{|c|c|c|}
\hline
$\sqrt{s}$ interval (GeV) & Luminosity (pb$^{-1}$) & Lumi weighted $\sqrt{s}$ (GeV) \\
194-204 & 193.52 & 198.46 \\
\hline
\end{tabular}
\begin{tabular}{|c|c|c|c|c|c|c|c|c|c|c|}
\hline
cos$\theta_{\mathrm{W}-}$ bin $i$ & 1 & 2 & 3 & 4 & 5 & 6 & 7 & 8 & 9 & 10 \\
$\sigma_i$ (pb)                  & 0.600 & 0.675 & 1.510 & 1.150 & 1.055 & 1.635 & 2.115 & 3.175 & 4.470 & 7.1400 \\
$\delta\sigma_i$(stat) (pb)      & 0.155 & 0.160 & 0.215 & 0.190 & 0.185 & 0.225 & 0.255 & 0.320 & 0.385 & 0.5000 \\
$\delta\sigma_i$(stat,exp.) (pb) & 0.150 & 0.160 & 0.170 & 0.180 & 0.200 & 0.230 & 0.260 & 0.310 & 0.380 & 0.5050 \\
$\delta\sigma_i$(syst,unc) (pb)  & 0.015 & 0.020 & 0.030 & 0.035 & 0.035 & 0.085 & 0.045 & 0.055 & 0.105 & 0.1000 \\
$\delta\sigma_i$(syst,cor) (pb)  & 0.025 & 0.015 & 0.015 & 0.015 & 0.015 & 0.015 & 0.010 & 0.015 & 0.025 & 0.0300 \\
\hline
\end{tabular}

\begin{tabular}{|c|c|c|}
\hline
$\sqrt{s}$ interval (GeV) & Luminosity (pb$^{-1}$) & Lumi weighted $\sqrt{s}$ (GeV) \\
204-210 & 198.59 & 205.91 \\
\hline
\end{tabular}
\begin{tabular}{|c|c|c|c|c|c|c|c|c|c|c|}
\hline
cos$\theta_{\mathrm{W}-}$ bin $i$ & 1 & 2 & 3 & 4 & 5 & 6 & 7 & 8 & 9 & 10 \\
$\sigma_i$ (pb)                  & 0.275 & 0.590 & 0.575 & 0.930 & 1.000 & 1.190 & 2.120 & 2.655 & 4.585 & 7.2900 \\
$\delta\sigma_i$(stat) (pb)      & 0.120 & 0.145 & 0.140 & 0.170 & 0.175 & 0.195 & 0.255 & 0.290 & 0.385 & 0.5050 \\
$\delta\sigma_i$(stat,exp.) (pb) & 0.145 & 0.150 & 0.160 & 0.175 & 0.195 & 0.220 & 0.250 & 0.300 & 0.380 & 0.5200 \\
$\delta\sigma_i$(syst,unc) (pb)  & 0.015 & 0.020 & 0.025 & 0.035 & 0.035 & 0.085 & 0.045 & 0.055 & 0.110 & 0.1100 \\
$\delta\sigma_i$(syst,cor) (pb)  & 0.020 & 0.015 & 0.010 & 0.010 & 0.015 & 0.010 & 0.010 & 0.010 & 0.020 & 0.0300 \\
\hline
\end{tabular}
\end{small}
\caption[]{%
W$^{-}$ differential angular cross-section in the 10 angular bins for the four chosen energy intervals
for the \Delphi\ experiment. For each energy range, the measured integrated luminosity and the luminosity
weighted centre-of-mass energy is reported.
The results per angular bin in each of the energy interval are then presented: $\sigma_{i}$ indicates 
the average of d[$\sigma_{\mathrm{WW}}$(BR$_{e\nu}$+BR$_{\mu\nu}$)]/dcos$\theta_{\mathrm{W}^-}$ 
in the $i$-th bin of cos$\theta_{\mathrm{W}^-}$ with width 0.2.
The values, in each bin, of the measured and expected statistical error and of the systematic errors,
LEP uncorrelated and correlated, are reported as well. All values are expressed in pb
}
\label{4f_tab:dsdcost_delphi} 
\end{center}
\end{table}

\begin{table}[hbtp]
\begin{center}
\begin{small}
\begin{tabular}{|c|}
\hline
L3~\cite{4f_bib:ltrww} \\
\hline 
\end{tabular}
\\
\begin{tabular}{|c|c|c|}
\hline
$\sqrt{s}$ interval (GeV) & Luminosity (pb$^{-1}$) & Lumi weighted $\sqrt{s}$ (GeV) \\
180-184 & 55.46 & 182.68 \\
\hline
\end{tabular}
\begin{tabular}{|c|c|c|c|c|c|c|c|c|c|c|}
\hline
cos$\theta_{\mathrm{W}-}$ bin $i$ & 1 & 2 & 3 & 4 & 5 & 6 & 7 & 8 & 9 & 10 \\
$\sigma_i$  (pb)                 & 0.691 & 0.646 & 0.508 & 0.919 & 1.477 & 2.587 & 3.541 & 3.167 & 3.879 & 4.467 \\
$\delta\sigma_i$(stat)  (pb)     & 0.270 & 0.265 & 0.243 & 0.322 & 0.407 & 0.539 & 0.640 & 0.619 & 0.708 & 0.801 \\
$\delta\sigma_i$(stat,exp.) (pb) & 0.269 & 0.290 & 0.329 & 0.364 & 0.404 & 0.453 & 0.508 & 0.591 & 0.704 & 0.877 \\
$\delta\sigma_i$(syst,unc)  (pb) & 0.016 & 0.009 & 0.007 & 0.011 & 0.018 & 0.031 & 0.043 & 0.039 & 0.048 & 0.058 \\
$\delta\sigma_i$(syst,cor)  (pb) & 0.002 & 0.002 & 0.002 & 0.003 & 0.005 & 0.009 & 0.012 & 0.011 & 0.013 & 0.015 \\
\hline
\end{tabular}

\begin{tabular}{|c|c|c|}
\hline
$\sqrt{s}$ interval (GeV) & Luminosity (pb$^{-1}$) & Lumi weighted $\sqrt{s}$ (GeV) \\
184-194 & 206.49 & 189.16 \\
\hline
\end{tabular}
\begin{tabular}{|c|c|c|c|c|c|c|c|c|c|c|}
\hline
cos$\theta_{\mathrm{W}-}$ bin $i$ & 1 & 2 & 3 & 4 & 5 & 6 & 7 & 8 & 9 & 10 \\
$\sigma_i$  (pb)                 & 0.759 & 0.902 & 1.125 & 1.320 & 1.472 & 1.544 & 2.085 & 2.870 & 4.144 & 6.022 \\
$\delta\sigma_i$(stat)  (pb)     & 0.128 & 0.151 & 0.173 & 0.190 & 0.209 & 0.213 & 0.254 & 0.303 & 0.370 & 0.459 \\
$\delta\sigma_i$(stat,exp.) (pb) & 0.115 & 0.137 & 0.160 & 0.180 & 0.205 & 0.223 & 0.262 & 0.304 & 0.367 & 0.461 \\
$\delta\sigma_i$(syst,unc)  (pb) & 0.017 & 0.013 & 0.015 & 0.015 & 0.017 & 0.018 & 0.024 & 0.034 & 0.048 & 0.074 \\
$\delta\sigma_i$(syst,cor)  (pb) & 0.003 & 0.003 & 0.004 & 0.005 & 0.005 & 0.005 & 0.007 & 0.010 & 0.014 & 0.021 \\
\hline
\end{tabular}

\begin{tabular}{|c|c|c|}
\hline
$\sqrt{s}$ interval (GeV) & Luminosity (pb$^{-1}$) & Lumi weighted $\sqrt{s}$ (GeV) \\
194-204 & 203.50 & 198.30 \\
\hline
\end{tabular}
\begin{tabular}{|c|c|c|c|c|c|c|c|c|c|c|}
\hline
cos$\theta_{\mathrm{W}-}$ bin $i$ & 1 & 2 & 3 & 4 & 5 & 6 & 7 & 8 & 9 & 10 \\
$\sigma_i$ (pb)                  & 0.652 & 0.709 & 0.880 & 0.859 & 1.140 & 1.295 & 2.114 & 2.334 & 3.395 & 5.773 \\
$\delta\sigma_i$(stat) (pb)      & 0.105 & 0.123 & 0.146 & 0.155 & 0.179 & 0.192 & 0.255 & 0.264 & 0.333 & 0.442 \\
$\delta\sigma_i$(stat,exp.) (pb) & 0.092 & 0.117 & 0.140 & 0.164 & 0.184 & 0.209 & 0.245 & 0.288 & 0.354 & 0.459 \\
$\delta\sigma_i$(syst,unc) (pb)  & 0.014 & 0.010 & 0.011 & 0.010 & 0.013 & 0.015 & 0.024 & 0.027 & 0.040 & 0.071 \\
$\delta\sigma_i$(syst,cor) (pb)  & 0.002 & 0.002 & 0.003 & 0.003 & 0.004 & 0.004 & 0.007 & 0.008 & 0.012 & 0.020 \\
\hline
\end{tabular}

\begin{tabular}{|c|c|c|}
\hline
$\sqrt{s}$ interval (GeV) & Luminosity (pb$^{-1}$) & Lumi weighted $\sqrt{s}$ (GeV) \\
204-210 & 217.30 & 205.96 \\
\hline
\end{tabular}
\begin{tabular}{|c|c|c|c|c|c|c|c|c|c|c|}
\hline
cos$\theta_{\mathrm{W}-}$ bin $i$ & 1 & 2 & 3 & 4 & 5 & 6 & 7 & 8 & 9 & 10 \\
$\sigma_i$ (pb)                  & 0.678 & 0.578 & 0.768 & 1.052 & 1.620 & 1.734 & 1.873 & 2.903 & 4.638 & 7.886 \\
$\delta\sigma_i$(stat) (pb)      & 0.111 & 0.114 & 0.140 & 0.168 & 0.212 & 0.226 & 0.238 & 0.302 & 0.394 & 0.534 \\
$\delta\sigma_i$(stat,exp.) (pb) & 0.089 & 0.117 & 0.141 & 0.164 & 0.186 & 0.216 & 0.251 & 0.303 & 0.387 & 0.528 \\
$\delta\sigma_i$(syst,unc) (pb)  & 0.015 & 0.008 & 0.010 & 0.012 & 0.019 & 0.020 & 0.021 & 0.034 & 0.054 & 0.097 \\
$\delta\sigma_i$(syst,cor) (pb)  & 0.002 & 0.002 & 0.003 & 0.004 & 0.006 & 0.006 & 0.006 & 0.010 & 0.016 & 0.027 \\
\hline
\end{tabular}
\end{small}
\caption[]{%
W$^{-}$ differential angular cross-section in the 10 angular bins for the four chosen energy intervals
for the \Ltre\ experiment. For each energy range, the measured integrated luminosity and the luminosity
weighted centre-of-mass energy is reported.
The results per angular bin in each of the energy interval are then presented: $\sigma_{i}$ indicates 
the average of d[$\sigma_{\mathrm{WW}}$(BR$_{e\nu}$+BR$_{\mu\nu}$)]/dcos$\theta_{\mathrm{W}^-}$ 
in the $i$-th bin of cos$\theta_{\mathrm{W}^-}$ with width 0.2.
The values, in each bin, of the measured and expected statistical error and of the systematic errors,
LEP uncorrelated and correlated, are reported as well. All values are expressed in pb
}
\label{4f_tab:dsdcost_l3} 
\end{center}
\end{table}

\begin{table}[p]
\renewcommand{\arraystretch}{1.2}
\vspace*{-0.7cm}
\begin{center}
\begin{small}
\begin{tabular}{|c|ccccc|c|c|}
\cline{1-8}
\roots & & & {\scriptsize (LCEC)} & {\scriptsize (LUEU)} & 
{\scriptsize (LUEC)} & & \\
(GeV) & $\swent$ & 
$\Delta\swent^\mathrm{stat}$ &
$\Delta\swent^\mathrm{syst}$ &
$\Delta\swent^\mathrm{syst}$ &
$\Delta\swent^\mathrm{syst}$ &
$\Delta\swent$ & 
$\Delta\swent^\mathrm{stat\,(exp)}$ \\
\hline
\multicolumn{8}{|c|}
{\Aleph~\cite{4f_bib:alesw}} \\
\hline
182.7 & 0.60 & $^{+0.32}_{-0.26}$ & $\pm$0.02 & $\pm$0.01 & $\pm$0.01 & $^{+0.32}_{-0.26}$ & $\pm$0.29 \\
188.6 & 0.55 & $^{+0.18}_{-0.16}$ & $\pm$0.02 & $\pm$0.01 & $\pm$0.01 & $^{+0.18}_{-0.16}$ & $\pm$0.18 \\
191.6 & 0.89 & $^{+0.58}_{-0.44}$ & $\pm$0.02 & $\pm$0.01 & $\pm$0.02 & $^{+0.58}_{-0.44}$ & $\pm$0.48 \\
195.5 & 0.87 & $^{+0.31}_{-0.27}$ & $\pm$0.03 & $\pm$0.01 & $\pm$0.02 & $^{+0.31}_{-0.27}$ & $\pm$0.28 \\
199.5 & 1.31 & $^{+0.32}_{-0.29}$ & $\pm$0.03 & $\pm$0.01 & $\pm$0.02 & $^{+0.32}_{-0.29}$ & $\pm$0.26 \\
201.6 & 0.80 & $^{+0.42}_{-0.35}$ & $\pm$0.03 & $\pm$0.01 & $\pm$0.02 & $^{+0.42}_{-0.35}$ & $\pm$0.38 \\
204.9 & 0.65 & $^{+0.27}_{-0.23}$ & $\pm$0.03 & $\pm$0.02 & $\pm$0.02 & $^{+0.27}_{-0.23}$ & $\pm$0.27 \\
206.6 & 0.81 & $^{+0.22}_{-0.20}$ & $\pm$0.03 & $\pm$0.02 & $\pm$0.02 & $^{+0.22}_{-0.20}$ & $\pm$0.22 \\
\hline
\multicolumn{8}{|c|}
{\Delphi~\cite{4f_bib:delsw2001,4f_bib:delswsc05}} \\
\hline
182.7 & 0.69 & $^{+0.41}_{-0.23}$ & $\pm$0.02 & $\pm$0.04 & $\pm$0.08 & $^{+0.42}_{-0.25}$ & $\pm$0.33 \\
188.6 & 0.75 & $^{+0.22}_{-0.20}$ & $\pm$0.02 & $\pm$0.04 & $\pm$0.08 & $^{+0.23}_{-0.22}$ & $\pm$0.20 \\
191.6 & 0.40 & $^{+0.54}_{-0.31}$ & $\pm$0.02 & $\pm$0.03 & $\pm$0.08 & $^{+0.55}_{-0.33}$ & $\pm$0.48 \\
195.5 & 0.68 & $^{+0.33}_{-0.28}$ & $\pm$0.02 & $\pm$0.03 & $\pm$0.08 & $^{+0.34}_{-0.38}$ & $\pm$0.30 \\
199.5 & 0.95 & $^{+0.33}_{-0.29}$ & $\pm$0.02 & $\pm$0.03 & $\pm$0.08 & $^{+0.34}_{-0.30}$ & $\pm$0.29 \\
201.6 & 1.24 & $^{+0.51}_{-0.42}$ & $\pm$0.02 & $\pm$0.04 & $\pm$0.08 & $^{+0.52}_{-0.43}$ & $\pm$0.41 \\
204.9 & 1.06 & $^{+0.36}_{-0.30}$ & $\pm$0.02 & $\pm$0.05 & $\pm$0.08 & $^{+0.37}_{-0.32}$ & $\pm$0.33 \\
206.6 & 1.14 & $^{+0.26}_{-0.23}$ & $\pm$0.02 & $\pm$0.04 & $\pm$0.08 & $^{+0.28}_{-0.25}$ & $\pm$0.23 \\
\hline
\multicolumn{8}{|c|}
{\Ltre~\cite{4f_bib:ltrsw2001,4f_bib:ltrsw}} \\
\hline
182.7 & 0.80 & $^{+0.28}_{-0.25}$ & $\pm$0.04 & $\pm$0.04 & $\pm$0.01 & $^{+0.28}_{-0.25}$ & $\pm$0.26 \\
188.6 & 0.69 & $^{+0.16}_{-0.14}$ & $\pm$0.03 & $\pm$0.03 & $\pm$0.01 & $^{+0.16}_{-0.15}$ & $\pm$0.15 \\
191.6 & 1.11 & $^{+0.48}_{-0.41}$ & $\pm$0.02 & $\pm$0.04 & $\pm$0.01 & $^{+0.48}_{-0.41}$ & $\pm$0.46 \\
195.5 & 0.97 & $^{+0.27}_{-0.25}$ & $\pm$0.02 & $\pm$0.02 & $\pm$0.01 & $^{+0.27}_{-0.25}$ & $\pm$0.25 \\
199.5 & 0.88 & $^{+0.26}_{-0.24}$ & $\pm$0.02 & $\pm$0.03 & $\pm$0.01 & $^{+0.26}_{-0.24}$ & $\pm$0.25 \\
201.6 & 1.50 & $^{+0.45}_{-0.40}$ & $\pm$0.03 & $\pm$0.04 & $\pm$0.02 & $^{+0.45}_{-0.40}$ & $\pm$0.38 \\
204.9 & 0.78 & $^{+0.29}_{-0.25}$ & $\pm$0.02 & $\pm$0.03 & $\pm$0.01 & $^{+0.29}_{-0.25}$ & $\pm$0.29 \\
206.6 & 1.08 & $^{+0.21}_{-0.20}$ & $\pm$0.02 & $\pm$0.03 & $\pm$0.01 & $^{+0.21}_{-0.20}$ & $\pm$0.23 \\
\hline
\multicolumn{8}{|c|}
{\Opal~\cite{4f_bib:opasw189}}  \\
\hline
188.6 GeV & 0.67 & $^{+0.16}_{-0.14}$ & $\pm$0.04 & $\pm$0.04 & $\pm$0.00 & $^{+0.17}_{-0.15}$ & $\pm$0.16 \\
\hline
\multicolumn{7}{|c|}
{LEP} & $\chi^2/\textrm{d.o.f.}$ \\
\hline
182.7 & 0.70 & $\pm$0.17 & $\pm$0.03 & $\pm$0.02 & $\pm$0.02 & $\pm$0.17 &
 \multirow{8}{20.3mm}{$
   \hspace*{-0.3mm}
   \left\}
     \begin{array}[h]{rr}
       &\multirow{8}{8mm}{\hspace*{-4.2mm}8.1/16}\\
       &\\ &\\ &\\ &\\ &\\ &\\ &\\  
     \end{array}
   \right.
   $}\\
188.6 & 0.66 & $\pm$0.08 & $\pm$0.03 & $\pm$0.02 & $\pm$0.01 & $\pm$0.09 & \\
191.6 & 0.81 & $\pm$0.27 & $\pm$0.02 & $\pm$0.02 & $\pm$0.02 & $\pm$0.28 & \\
195.5 & 0.85 & $\pm$0.16 & $\pm$0.02 & $\pm$0.01 & $\pm$0.02 & $\pm$0.16 & \\
199.5 & 1.05 & $\pm$0.15 & $\pm$0.02 & $\pm$0.01 & $\pm$0.02 & $\pm$0.16 & \\
201.6 & 1.17 & $\pm$0.23 & $\pm$0.03 & $\pm$0.02 & $\pm$0.02 & $\pm$0.23 & \\
204.9 & 0.80 & $\pm$0.17 & $\pm$0.02 & $\pm$0.02 & $\pm$0.02 & $\pm$0.17 & \\
206.6 & 1.00 & $\pm$0.13 & $\pm$0.03 & $\pm$0.02 & $\pm$0.02 & $\pm$0.14 & \\
\hline
\end{tabular}
\end{small}
\caption[]{%
Single-W total production cross-section (in pb) at different energies.
The first column contains the LEP \CoM\ energy,
and the second the measurements. 
The third column reports the statistical error, whereas in the fourth to the
sixth columns the different systematic uncertainties are listed.
The seventh column contains the total error and the eight lists,
for the four LEP measurements,
the symmetrized expected statistical error,
and for the LEP combined value,
the $\chi^2$ of the fit.}
\label{4f_tab:WevTOTmeas} 
\end{center}
\renewcommand{\arraystretch}{1.}
\end{table}

\begin{table}[p]
\renewcommand{\arraystretch}{1.2}
\vspace*{-0.7cm}
\begin{center}
\begin{small}
\begin{tabular}{|c|ccccc|c|c|}
\cline{1-8}
\roots & & & {\scriptsize (LCEC)} & {\scriptsize (LUEU)} & 
{\scriptsize (LUEC)} & & \\
(GeV) & $\swenh$ & 
$\Delta\swenh^\mathrm{stat}$ &
$\Delta\swenh^\mathrm{syst}$ &
$\Delta\swenh^\mathrm{syst}$ &
$\Delta\swenh^\mathrm{syst}$ &
$\Delta\swenh$ & 
$\Delta\swenh^\mathrm{stat\,(exp)}$ \\
\hline
\multicolumn{8}{|c|}
{\Aleph~\cite{4f_bib:alesw}} \\
\hline
182.7 & 0.44 & $^{+0.29}_{-0.24}$ & $\pm$0.01 & $\pm$0.01 & $\pm$0.01 & $^{+0.29}_{-0.24}$ & $\pm$0.26 \\
188.6 & 0.33 & $^{+0.16}_{-0.14}$ & $\pm$0.02 & $\pm$0.01 & $\pm$0.01 & $^{+0.16}_{-0.15}$ & $\pm$0.16 \\
191.6 & 0.52 & $^{+0.52}_{-0.40}$ & $\pm$0.02 & $\pm$0.01 & $\pm$0.01 & $^{+0.52}_{-0.40}$ & $\pm$0.45 \\
195.5 & 0.61 & $^{+0.28}_{-0.25}$ & $\pm$0.02 & $\pm$0.01 & $\pm$0.01 & $^{+0.28}_{-0.25}$ & $\pm$0.25 \\
199.5 & 1.06 & $^{+0.30}_{-0.27}$ & $\pm$0.02 & $\pm$0.01 & $\pm$0.01 & $^{+0.30}_{-0.27}$ & $\pm$0.24 \\
201.6 & 0.72 & $^{+0.39}_{-0.33}$ & $\pm$0.02 & $\pm$0.01 & $\pm$0.02 & $^{+0.39}_{-0.33}$ & $\pm$0.34 \\
204.9 & 0.34 & $^{+0.24}_{-0.21}$ & $\pm$0.02 & $\pm$0.01 & $\pm$0.02 & $^{+0.24}_{-0.21}$ & $\pm$0.25 \\
206.6 & 0.64 & $^{+0.21}_{-0.19}$ & $\pm$0.02 & $\pm$0.01 & $\pm$0.02 & $^{+0.21}_{-0.19}$ & $\pm$0.19 \\
\hline
\multicolumn{8}{|c|}
{\Delphi~\cite{4f_bib:delsw2001,4f_bib:delswsc05}} \\
\hline
182.7 & 0.11 & $^{+0.30}_{-0.11}$ & $\pm$0.02 & $\pm$0.03 & $\pm$0.08 & $^{+0.31}_{-0.14}$ & $\pm$0.30 \\
188.6 & 0.57 & $^{+0.19}_{-0.18}$ & $\pm$0.02 & $\pm$0.04 & $\pm$0.08 & $^{+0.21}_{-0.20}$ & $\pm$0.18 \\
191.6 & 0.30 & $^{+0.47}_{-0.30}$ & $\pm$0.02 & $\pm$0.03 & $\pm$0.08 & $^{+0.48}_{-0.31}$ & $\pm$0.43 \\
195.5 & 0.50 & $^{+0.29}_{-0.26}$ & $\pm$0.02 & $\pm$0.03 & $\pm$0.08 & $^{+0.30}_{-0.27}$ & $\pm$0.27 \\
199.5 & 0.57 & $^{+0.27}_{-0.25}$ & $\pm$0.02 & $\pm$0.02 & $\pm$0.08 & $^{+0.28}_{-0.26}$ & $\pm$0.25 \\
201.6 & 0.67 & $^{+0.39}_{-0.35}$ & $\pm$0.02 & $\pm$0.03 & $\pm$0.08 & $^{+0.40}_{-0.36}$ & $\pm$0.35 \\
204.9 & 0.99 & $^{+0.32}_{-0.30}$ & $\pm$0.02 & $\pm$0.05 & $\pm$0.08 & $^{+0.33}_{-0.31}$ & $\pm$0.28 \\
206.6 & 0.81 & $^{+0.22}_{-0.20}$ & $\pm$0.02 & $\pm$0.04 & $\pm$0.08 & $^{+0.23}_{-0.22}$ & $\pm$0.20 \\
\hline
\multicolumn{8}{|c|}
{\Ltre~\cite{4f_bib:ltrsw2001,4f_bib:ltrsw}} \\
\hline
182.7 & 0.58 & $^{+0.23}_{-0.20}$ & $\pm$0.03 & $\pm$0.03 & $\pm$0.00 & $^{+0.23}_{-0.20}$ & $\pm$0.21 \\
188.6 & 0.52 & $^{+0.14}_{-0.13}$ & $\pm$0.02 & $\pm$0.02 & $\pm$0.00 & $^{+0.14}_{-0.13}$ & $\pm$0.14 \\
191.6 & 0.84 & $^{+0.44}_{-0.37}$ & $\pm$0.03 & $\pm$0.03 & $\pm$0.00 & $^{+0.44}_{-0.37}$ & $\pm$0.41 \\
195.5 & 0.66 & $^{+0.24}_{-0.22}$ & $\pm$0.02 & $\pm$0.03 & $\pm$0.00 & $^{+0.25}_{-0.23}$ & $\pm$0.21 \\
199.5 & 0.37 & $^{+0.22}_{-0.20}$ & $\pm$0.01 & $\pm$0.02 & $\pm$0.00 & $^{+0.22}_{-0.20}$ & $\pm$0.22 \\
201.6 & 1.10 & $^{+0.40}_{-0.35}$ & $\pm$0.05 & $\pm$0.05 & $\pm$0.00 & $^{+0.40}_{-0.35}$ & $\pm$0.35 \\
204.9 & 0.42 & $^{+0.25}_{-0.21}$ & $\pm$0.02 & $\pm$0.03 & $\pm$0.00 & $^{+0.25}_{-0.21}$ & $\pm$0.25 \\
206.6 & 0.66 & $^{+0.19}_{-0.17}$ & $\pm$0.02 & $\pm$0.03 & $\pm$0.00 & $^{+0.20}_{-0.18}$ & $\pm$0.20 \\
\hline
\multicolumn{8}{|c|}
{\Opal~\cite{4f_bib:opasw189}}  \\
\hline
188.6 & 0.53 & $^{+0.13}_{-0.12}$ & $\pm$0.04 & $\pm$0.04 & $\pm$0.00 & $^{+0.14}_{-0.13}$ & $\pm$0.14 \\
\hline
\multicolumn{7}{|c|}
{LEP} & $\chi^2/\textrm{d.o.f.}$ \\
\hline
182.7 & 0.42 & $\pm$0.15 & $\pm$0.02 & $\pm$0.02 & $\pm$0.01 & $\pm$0.15 & 
 \multirow{8}{20.3mm}{$
   \hspace*{-0.3mm}
   \left\}
     \begin{array}[h]{rr}
       &\multirow{8}{8mm}{\hspace*{-4.2mm}13.3/16}\\
       &\\ &\\ &\\ &\\ &\\ &\\ &\\  
     \end{array}
   \right.
   $}\\
188.6 & 0.48 & $\pm$0.07 & $\pm$0.02 & $\pm$0.02 & $\pm$0.01 & $\pm$0.08 & \\
191.6 & 0.56 & $\pm$0.25 & $\pm$0.02 & $\pm$0.02 & $\pm$0.02 & $\pm$0.25 & \\
195.5 & 0.60 & $\pm$0.14 & $\pm$0.02 & $\pm$0.01 & $\pm$0.02 & $\pm$0.14 & \\
199.5 & 0.65 & $\pm$0.14 & $\pm$0.02 & $\pm$0.01 & $\pm$0.02 & $\pm$0.14 & \\
201.6 & 0.82 & $\pm$0.20 & $\pm$0.03 & $\pm$0.02 & $\pm$0.02 & $\pm$0.20 & \\
204.9 & 0.54 & $\pm$0.15 & $\pm$0.02 & $\pm$0.02 & $\pm$0.02 & $\pm$0.15 & \\
206.6 & 0.69 & $\pm$0.11 & $\pm$0.02 & $\pm$0.02 & $\pm$0.02 & $\pm$0.12 & \\
\hline
\end{tabular}
\end{small}
\caption[]{%
Single-W hadronic production cross-section (in pb) at different energies.
The first column contains the LEP \CoM\ energy,
and the second the measurements. 
The third column reports the statistical error, whereas in the fourth to the
sixth columns the different systematic uncertainties are listed.
The seventh column contains the total error and the eight lists,
for the four LEP measurements,
the symmetrized expected statistical error,
and for the LEP combined value,
the $\chi^2$ of the fit.}
\label{4f_tab:WevHADmeas} 
\end{center}
\renewcommand{\arraystretch}{1.}
\end{table}

\begin{table}[hbtp]
\begin{center}
\hspace*{-0.3cm}
\renewcommand{\arraystretch}{1.2}
\begin{tabular}{|c|c|c|c|c|c|} 
\hline
\roots & \multicolumn{3}{|c|}{We$\nu \rightarrow $qqe$\nu$ cross-section (pb)} 
& \multicolumn{2}{|c|}{We$\nu$ total cross-section (pb)} \\
\cline{2-6} 
(GeV) & $\swenh^{\footnotesize\Grace}$    
      & $\swenh^{\footnotesize\WPHACT}$ 
      & $\swenh^{\footnotesize\WTO}$ 
      & $\swent^{\footnotesize\Grace}$
      & $\swent^{\footnotesize\WPHACT}$  \\
\hline
182.7 & 0.4194[1] & 0.4070[2] & 0.40934[8] & 0.6254[1] & 0.6066[2] \\
188.6 & 0.4699[1] & 0.4560[2] & 0.45974[9] & 0.6999[1] & 0.6796[2] \\ 
191.6 & 0.4960[1] & 0.4810[2] & 0.4852[1] &  0.7381[2] & 0.7163[2] \\ 
195.5 & 0.5308[2] & 0.5152[2] & 0.5207[1] &  0.7896[2] & 0.7665[3] \\ 
199.5 & 0.5673[2] & 0.5509[3] & 0.5573[1] &  0.8431[2] & 0.8182[3] \\ 
201.6 & 0.5870[2] & 0.5704[4] & 0.5768[1] &  0.8718[2] & 0.8474[4] \\ 
204.9 & 0.6196[2] & 0.6021[4] & 0.6093[2] &  0.9185[3] & 0.8921[4] \\ 
206.6 & 0.6358[2] & 0.6179[4] & 0.6254[2] &  0.9423[3] & 0.9157[5] \\ 
\hline
\end{tabular}
\renewcommand{\arraystretch}{1.}
\caption[]{%
Single-W hadronic and total cross-section predictions (in pb) 
interpolated at the data \CoM\ energies,
according to the \Grace~\protect\cite{4f_bib:grace}, 
\WPHACT~\protect\cite{4f_bib:wphact} and 
\WTO~\protect\cite{4f_bib:wto} predictions.
The numbers in brackets are the errors on the last digit and are coming
from the numerical integration of the cross-section only.}
\label{4f_tab:Wentheo} 
\end{center}
\end{table}

\begin{table}[hbtp]
\begin{center}
\begin{small}
\begin{tabular}{|c|cccccc|c|c|}
\hline
\roots & & & {\scriptsize (LCEU)} & {\scriptsize (LCEC)} & 
{\scriptsize (LUEU)} & {\scriptsize (LUEC)} & & \\
(GeV) & $\rwev$ & 
$\Delta\rwev^\mathrm{stat}$ &
$\Delta\rwev^\mathrm{syst}$ &
$\Delta\rwev^\mathrm{syst}$ &
$\Delta\rwev^\mathrm{syst}$ &
$\Delta\rwev^\mathrm{syst}$ &
$\Delta\rwev$ &
$\chi^2/\textrm{d.o.f.}$ \\
\hline
\hline
\multicolumn{9}{|c|}{\Grace~\cite{4f_bib:grace}}\\
\hline
182.7 & 1.122 & $\pm$0.266 & $\pm$0.001 & $\pm$0.041 & $\pm$0.029 & $\pm$0.026 & $\pm$0.272 &
\multirow{8}{20.3mm}{$
  \hspace*{-0.3mm}
  \left\}
    \begin{array}[h]{rr}
      &\multirow{8}{6mm}{\hspace*{-4.2mm}8.1/16}\\
      &\\ &\\ &\\ &\\ &\\ &\\ &\\  
    \end{array}
  \right.
  $}\\
188.6 & 0.942 & $\pm$0.121 & $\pm$0.001 & $\pm$0.039 & $\pm$0.023 & $\pm$0.018 & $\pm$0.130 &\\
191.6 & 1.094 & $\pm$0.370 & $\pm$0.001 & $\pm$0.030 & $\pm$0.026 & $\pm$0.028 & $\pm$0.373 &\\
195.5 & 1.081 & $\pm$0.199 & $\pm$0.001 & $\pm$0.028 & $\pm$0.017 & $\pm$0.023 & $\pm$0.203 &\\
199.5 & 1.242 & $\pm$0.183 & $\pm$0.001 & $\pm$0.028 & $\pm$0.017 & $\pm$0.022 & $\pm$0.187 &\\
201.6 & 1.340 & $\pm$0.258 & $\pm$0.001 & $\pm$0.031 & $\pm$0.021 & $\pm$0.023 & $\pm$0.261 &\\
204.9 & 0.873 & $\pm$0.185 & $\pm$0.001 & $\pm$0.025 & $\pm$0.020 & $\pm$0.020 & $\pm$0.189 &\\
206.6 & 1.058 & $\pm$0.138 & $\pm$0.001 & $\pm$0.026 & $\pm$0.019 & $\pm$0.021 & $\pm$0.143 &\\
\hline
Average & 
1.051 & $\pm$0.065 & $\pm$0.000 & $\pm$0.031 & $\pm$0.009& $\pm$0.021 & $\pm$0.076&
\hspace*{1.5mm}12.2/24\hspace*{-0.5mm}\\
\hline
\hline
\multicolumn{9}{|c|}{\WPHACT~\cite{4f_bib:wphact}}\\
\hline
182.7 & 1.157 & $\pm$0.274 & $\pm$0.001 & $\pm$0.043 & $\pm$0.030 & $\pm$0.027 & $\pm$0.281 &
\multirow{8}{20.3mm}{$
  \hspace*{-0.3mm}
  \left\}
    \begin{array}[h]{rr}
      &\multirow{8}{6mm}{\hspace*{-4.2mm}8.1/16}\\
      &\\ &\\ &\\ &\\ &\\ &\\ &\\  
    \end{array}
  \right.
  $}\\
188.6 & 0.971 & $\pm$0.124 & $\pm$0.001 & $\pm$0.040 & $\pm$0.023 & $\pm$0.018 & $\pm$0.134 &\\
191.6 & 1.128 & $\pm$0.382 & $\pm$0.001 & $\pm$0.031 & $\pm$0.027 & $\pm$0.029 & $\pm$0.385 &\\
195.5 & 1.115 & $\pm$0.206 & $\pm$0.001 & $\pm$0.029 & $\pm$0.017 & $\pm$0.023 & $\pm$0.210 &\\
199.5 & 1.280 & $\pm$0.188 & $\pm$0.001 & $\pm$0.029 & $\pm$0.018 & $\pm$0.022 & $\pm$0.193 &\\
201.6 & 1.380 & $\pm$0.265 & $\pm$0.001 & $\pm$0.032 & $\pm$0.022 & $\pm$0.024 & $\pm$0.269 &\\
204.9 & 0.899 & $\pm$0.191 & $\pm$0.001 & $\pm$0.026 & $\pm$0.020 & $\pm$0.020 & $\pm$0.195 &\\
206.6 & 1.089 & $\pm$0.142 & $\pm$0.001 & $\pm$0.027 & $\pm$0.020 & $\pm$0.022 & $\pm$0.148 &\\
\hline
Average & 
1.083 & $\pm$0.067 & $\pm$0.000 & $\pm$0.032 & $\pm$0.009& $\pm$0.022 & $\pm$0.078&
\hspace*{1.5mm}12.2/24\hspace*{-0.5mm}\\
\hline
\end{tabular}
\end{small}
\caption[]{%
Ratios of LEP combined total single-W cross-section measurements
to the expectations, for different \CoM\ energies and for all energies combined.
The first column contains the \CoM\ energy,
the second the combined ratios,
the third the statistical errors.
The fourth, fifth, sixth and seventh columns contain
the sources of systematic errors that are considered as 
LEP-correlated   energy-uncorrelated (LCEU),
LEP-correlated   energy-correlated   (LCEC),
LEP-uncorrelated energy-uncorrelated (LUEU),
LEP-uncorrelated energy-correlated   (LUEC).
The total error is given in the eighth column.
The only LCEU systematic sources considered 
are the statistical errors on the cross-section theoretical predictions,
while the LCEC, LUEU and LUEC sources are those coming from
the corresponding errors on the cross-section measurements.}
\label{4f_tab:rwenmeas} 
\end{center}
\end{table}

\clearpage

\begin{table}[p]
\vspace*{-0.8cm}
\begin{center}
\begin{small}
\begin{tabular}{|c|ccccc|c|c|}
\cline{1-8}
\roots & & & {\scriptsize (LCEC)} & {\scriptsize (LUEU)} & 
{\scriptsize (LUEC)} & & 
\multicolumn{1}{|r}{$\quad$} \\
(GeV) & $\szz$ & 
$\Delta\szz^\mathrm{stat}$ &
$\Delta\szz^\mathrm{syst}$ &
$\Delta\szz^\mathrm{syst}$ &
$\Delta\szz^\mathrm{syst}$ &
$\Delta\szz$ & 
$\Delta\szz^\mathrm{stat\,(exp)}$ \\
\hline
\multicolumn{8}{|c|}
{\Aleph~\cite{4f_bib:alezz189,4f_bib:alezzsc01}} \\
\hline
182.7 & 0.11 & $^{+0.16}_{-0.11}$ & $\pm$0.01 & $\pm$0.03 & $\pm$0.03 & $^{+0.16}_{-0.12}$ & $\pm$0.14 \\
188.6 & 0.67 & $^{+0.13}_{-0.12}$ & $\pm$0.01 & $\pm$0.03 & $\pm$0.03 & $^{+0.14}_{-0.13}$ & $\pm$0.13 \\
191.6 & 0.53 & $^{+0.34}_{-0.27}$ & $\pm$0.01 & $\pm$0.01 & $\pm$0.01 & $^{+0.34}_{-0.27}$ & $\pm$0.33 \\
195.5 & 0.69 & $^{+0.23}_{-0.20}$ & $\pm$0.01 & $\pm$0.02 & $\pm$0.02 & $^{+0.23}_{-0.20}$ & $\pm$0.23 \\
199.5 & 0.70 & $^{+0.22}_{-0.20}$ & $\pm$0.01 & $\pm$0.02 & $\pm$0.02 & $^{+0.22}_{-0.20}$ & $\pm$0.23 \\
201.6 & 0.70 & $^{+0.33}_{-0.28}$ & $\pm$0.01 & $\pm$0.01 & $\pm$0.01 & $^{+0.33}_{-0.28}$ & $\pm$0.35 \\
204.9 & 1.21 & $^{+0.26}_{-0.23}$ & $\pm$0.01 & $\pm$0.02 & $\pm$0.02 & $^{+0.26}_{-0.23}$ & $\pm$0.27 \\
206.6 & 1.01 & $^{+0.19}_{-0.17}$ & $\pm$0.01 & $\pm$0.01 & $\pm$0.01 & $^{+0.19}_{-0.17}$ & $\pm$0.18 \\
\hline
\multicolumn{8}{|c|}
{\Delphi~\cite{4f_bib:delzz}} \\
\hline
182.7 & 0.35 & $^{+0.20}_{-0.15}$ & $\pm$0.01 & $\pm$0.00 & $\pm$0.02 & $^{+0.20}_{-0.15}$ & $\pm$0.16 \\
188.6 & 0.52 & $^{+0.12}_{-0.11}$ & $\pm$0.01 & $\pm$0.00 & $\pm$0.02 & $^{+0.12}_{-0.11}$ & $\pm$0.13 \\
191.6 & 0.63 & $^{+0.36}_{-0.30}$ & $\pm$0.01 & $\pm$0.01 & $\pm$0.02 & $^{+0.36}_{-0.30}$ & $\pm$0.35 \\
195.5 & 1.05 & $^{+0.25}_{-0.22}$ & $\pm$0.01 & $\pm$0.01 & $\pm$0.02 & $^{+0.25}_{-0.22}$ & $\pm$0.21 \\
199.5 & 0.75 & $^{+0.20}_{-0.18}$ & $\pm$0.01 & $\pm$0.01 & $\pm$0.01 & $^{+0.20}_{-0.18}$ & $\pm$0.21 \\
201.6 & 0.85 & $^{+0.33}_{-0.28}$ & $\pm$0.01 & $\pm$0.01 & $\pm$0.01 & $^{+0.33}_{-0.28}$ & $\pm$0.32 \\
204.9 & 1.03 & $^{+0.23}_{-0.20}$ & $\pm$0.02 & $\pm$0.01 & $\pm$0.01 & $^{+0.23}_{-0.20}$ & $\pm$0.23 \\
206.6 & 0.96 & $^{+0.16}_{-0.15}$ & $\pm$0.02 & $\pm$0.01 & $\pm$0.01 & $^{+0.16}_{-0.15}$ & $\pm$0.17 \\
\hline
\multicolumn{8}{|c|}
{\Ltre~\cite{4f_bib:ltrzz}} \\
\hline
182.7 & 0.31 & $\pm$0.16 & $\pm$0.05 & $\pm$0.00 & $\pm$0.01 & $\pm$0.17 & $\pm$0.16 \\
188.6 & 0.73 & $\pm$0.15 & $\pm$0.02 & $\pm$0.02 & $\pm$0.02 & $\pm$0.15 & $\pm$0.15 \\
191.6 & 0.29 & $\pm$0.22 & $\pm$0.01 & $\pm$0.01 & $\pm$0.02 & $\pm$0.22 & $\pm$0.34 \\
195.5 & 1.18 & $\pm$0.24 & $\pm$0.04 & $\pm$0.05 & $\pm$0.06 & $\pm$0.26 & $\pm$0.22 \\
199.5 & 1.25 & $\pm$0.25 & $\pm$0.04 & $\pm$0.05 & $\pm$0.07 & $\pm$0.27 & $\pm$0.24 \\
201.6 & 0.95 & $\pm$0.38 & $\pm$0.03 & $\pm$0.04 & $\pm$0.05 & $\pm$0.39 & $\pm$0.35 \\
204.9 & 0.77 & $^{+0.21}_{-0.19}$ & $\pm$0.01 & $\pm$0.01 & $\pm$0.04 & $^{+0.21}_{-0.19}$ & $\pm$0.22 \\
206.6 & 1.09 & $^{+0.17}_{-0.16}$ & $\pm$0.02 & $\pm$0.02 & $\pm$0.06 & $^{+0.18}_{-0.17}$ & $\pm$0.17 \\
\hline
\multicolumn{8}{|c|}
{\Opal~\cite{4f_bib:opazz}}  \\
\hline
182.7 & 0.12 & $^{+0.20}_{-0.18}$ & $\pm$0.00 & $\pm$0.03 & $\pm$0.00 & $^{+0.20}_{-0.18}$ & $\pm$0.19 \\
188.6 & 0.80 & $^{+0.14}_{-0.13}$ & $\pm$0.01 & $\pm$0.05 & $\pm$0.03 & $^{+0.15}_{-0.14}$ & $\pm$0.14 \\
191.6 & 1.29 & $^{+0.47}_{-0.40}$ & $\pm$0.02 & $\pm$0.09 & $\pm$0.05 & $^{+0.48}_{-0.41}$ & $\pm$0.36 \\
195.5 & 1.13 & $^{+0.26}_{-0.24}$ & $\pm$0.02 & $\pm$0.06 & $\pm$0.05 & $^{+0.27}_{-0.25}$ & $\pm$0.25 \\
199.5 & 1.05 & $^{+0.25}_{-0.22}$ & $\pm$0.02 & $\pm$0.05 & $\pm$0.04 & $^{+0.26}_{-0.23}$ & $\pm$0.25 \\
201.6 & 0.79 & $^{+0.35}_{-0.29}$ & $\pm$0.02 & $\pm$0.05 & $\pm$0.03 & $^{+0.36}_{-0.30}$ & $\pm$0.37 \\
204.9 & 1.07 & $^{+0.27}_{-0.24}$ & $\pm$0.02 & $\pm$0.06 & $\pm$0.04 & $^{+0.28}_{-0.25}$ & $\pm$0.26 \\
206.6 & 0.97 & $^{+0.19}_{-0.18}$ & $\pm$0.02 & $\pm$0.05 & $\pm$0.04 & $^{+0.20}_{-0.19}$ & $\pm$0.20 \\
\hline
\multicolumn{7}{|c|}
{LEP} & $\chi^2/\textrm{d.o.f.}$ \\
\hline
182.7 & 0.22 & $\pm$0.08 & $\pm$0.02 & $\pm$0.01 & $\pm$0.01 & $\pm$0.08 & 
 \multirow{8}{20.3mm}{$
   \hspace*{-0.3mm}
   \left\}
     \begin{array}[h]{rr}
       &\multirow{8}{8mm}{\hspace*{-4.2mm}16.1/24}\\
       &\\ &\\ &\\ &\\ &\\ &\\ &\\  
     \end{array}
   \right.
   $}\\
188.6 & 0.66 & $\pm$0.07 & $\pm$0.01 & $\pm$0.01 & $\pm$0.01 & $\pm$0.07 & \\
191.6 & 0.65 & $\pm$0.17 & $\pm$0.01 & $\pm$0.02 & $\pm$0.01 & $\pm$0.17 & \\
195.5 & 0.99 & $\pm$0.11 & $\pm$0.02 & $\pm$0.02 & $\pm$0.02 & $\pm$0.12 & \\
199.5 & 0.90 & $\pm$0.12 & $\pm$0.02 & $\pm$0.02 & $\pm$0.02 & $\pm$0.12 & \\
201.6 & 0.81 & $\pm$0.17 & $\pm$0.02 & $\pm$0.02 & $\pm$0.01 & $\pm$0.17 & \\
204.9 & 0.98 & $\pm$0.12 & $\pm$0.01 & $\pm$0.01 & $\pm$0.02 & $\pm$0.13 & \\
206.6 & 0.99 & $\pm$0.09 & $\pm$0.02 & $\pm$0.01 & $\pm$0.02 & $\pm$0.09 & \\
\hline
\end{tabular}
\end{small}
\caption[]{%
Z-pair production cross-section (in pb) at different energies.
The first column contains the LEP \CoM\ energy,
the second the measurements and
the third the statistical uncertainty. 
The fourth, the fifth and the sixth columns list 
the different components of the systematic errors, 
as provided by the Collaborations.
The total error is given in the seventh column,
whereas the eighth column lists, for the four LEP measurements,
the symmetrized expected statistical error,
and for the LEP combined value,
the $\chi^2$ of the fit.}
\label{4f_tab:ZZmeas} 
\end{center}
\end{table}

\begin{table}[hbtp]
\begin{center}
\hspace*{-0.3cm}
\renewcommand{\arraystretch}{1.2}
\begin{tabular}{|c|c|c|} 
\hline
\roots & \multicolumn{2}{|c|}{ZZ cross-section (pb)}  \\
\cline{2-3} 
(GeV) & $\szz^{\footnotesize\YFSZZ}$    
      & $\szz^{\footnotesize\ZZTO}$ \\
\hline
182.7 & 0.254[1] & 0.25425[2] \\
188.6 & 0.655[2] & 0.64823[1] \\
191.6 & 0.782[2] & 0.77670[1] \\
195.5 & 0.897[3] & 0.89622[1] \\
199.5 & 0.981[2] & 0.97765[1] \\
201.6 & 1.015[1] & 1.00937[1] \\
204.9 & 1.050[1] & 1.04335[1] \\
206.6 & 1.066[1] & 1.05535[1] \\
\hline
\end{tabular}
\renewcommand{\arraystretch}{1.}
\caption[]{%
Z-pair cross-section predictions (in pb) interpolated at the data 
\CoM\ energies,according to the \YFSZZ~\protect\cite{4f_bib:yfszz} and 
\ZZTO~\protect\cite{4f_bib:zzto} predictions.
The numbers in brackets are the errors on the last digit and are coming
from the numerical integration of the cross-section only.}
\label{4f_tab:ZZtheo} 
\end{center}
\end{table}

\begin{table}[hbtp]
\begin{center}
\begin{small}
\begin{tabular}{|c|cccccc|c|c|}
\hline
\roots & & & {\scriptsize (LCEU)} & {\scriptsize (LCEC)} & 
{\scriptsize (LUEU)} & {\scriptsize (LUEC)} & & \\
(GeV) & $\rzz$ & 
$\Delta\rzz^\mathrm{stat}$ &
$\Delta\rzz^\mathrm{syst}$ &
$\Delta\rzz^\mathrm{syst}$ &
$\Delta\rzz^\mathrm{syst}$ &
$\Delta\rzz^\mathrm{syst}$ &
$\Delta\rzz$ &
$\chi^2/\textrm{d.o.f.}$ \\
\hline
\hline
\multicolumn{9}{|c|}{\YFSZZ~\cite{4f_bib:yfszz}}\\
\hline
182.7 & 0.857 & $\pm$0.307 & $\pm$0.018 & $\pm$0.068 & $\pm$0.041 & $\pm$0.040 & $\pm$0.320 &
\multirow{8}{20.3mm}{$
  \hspace*{-0.3mm}
  \left\}
    \begin{array}[h]{rr}
      &\multirow{8}{6mm}{\hspace*{-4.2mm}16.1/24}\\
      &\\ &\\ &\\ &\\ &\\ &\\ &\\  
    \end{array}
  \right.
  $}\\
188.6 & 1.007 & $\pm$0.104 & $\pm$0.020 & $\pm$0.019 & $\pm$0.022& $\pm$0.018 & $\pm$0.111&\\
191.6 & 0.826 & $\pm$0.220 & $\pm$0.017 & $\pm$0.014 & $\pm$0.025& $\pm$0.017 & $\pm$0.224&\\
195.5 & 1.100 & $\pm$0.127 & $\pm$0.022 & $\pm$0.021 & $\pm$0.019& $\pm$0.020 & $\pm$0.133&\\
199.5 & 0.912 & $\pm$0.119 & $\pm$0.019 & $\pm$0.018 & $\pm$0.016& $\pm$0.017 & $\pm$0.124&\\
201.6 & 0.795 & $\pm$0.170 & $\pm$0.016 & $\pm$0.017 & $\pm$0.015& $\pm$0.013 & $\pm$0.173&\\
204.9 & 0.931 & $\pm$0.116 & $\pm$0.019 & $\pm$0.014 & $\pm$0.013& $\pm$0.014 & $\pm$0.120&\\
206.6 & 0.928 & $\pm$0.085 & $\pm$0.019 & $\pm$0.014 & $\pm$0.010& $\pm$0.015 & $\pm$0.090&\\
\hline
Average & 
0.945 & $\pm$0.045 & $\pm$0.008 & $\pm$0.017 & $\pm$0.006& $\pm$0.016 & $\pm$0.052&
\hspace*{1.5mm}19.1/31\hspace*{-0.5mm}\\
\hline
\hline
\multicolumn{9}{|c|}{\ZZTO~\cite{4f_bib:zzto}}\\
\hline
182.7 & 0.857 & $\pm$0.307 & $\pm$0.018 & $\pm$0.068 & $\pm$0.041 & $\pm$0.040 & $\pm$0.320 &
\multirow{8}{20.3mm}{$
  \hspace*{-0.3mm}
  \left\}
    \begin{array}[h]{rr}
      &\multirow{8}{6mm}{\hspace*{-4.2mm}16.1/24}\\
      &\\ &\\ &\\ &\\ &\\ &\\ &\\  
    \end{array}
  \right.
  $}\\
188.6 & 1.017 & $\pm$0.105 & $\pm$0.021 & $\pm$0.019 & $\pm$0.022& $\pm$0.019 & $\pm$0.113&\\
191.6 & 0.831 & $\pm$0.222 & $\pm$0.017 & $\pm$0.014 & $\pm$0.025& $\pm$0.017 & $\pm$0.225&\\
195.5 & 1.100 & $\pm$0.127 & $\pm$0.022 & $\pm$0.021 & $\pm$0.019& $\pm$0.020 & $\pm$0.133&\\
199.5 & 0.915 & $\pm$0.120 & $\pm$0.019 & $\pm$0.018 & $\pm$0.016& $\pm$0.017 & $\pm$0.125&\\
201.6 & 0.799 & $\pm$0.171 & $\pm$0.016 & $\pm$0.017 & $\pm$0.015& $\pm$0.013 & $\pm$0.174&\\
204.9 & 0.937 & $\pm$0.117 & $\pm$0.019 & $\pm$0.014 & $\pm$0.013& $\pm$0.014 & $\pm$0.121&\\
206.6 & 0.937 & $\pm$0.085 & $\pm$0.019 & $\pm$0.014 & $\pm$0.011& $\pm$0.015 & $\pm$0.091&\\
\hline
Average & 
0.952 & $\pm$0.046 & $\pm$0.008 & $\pm$0.017 & $\pm$0.006& $\pm$0.016 & $\pm$0.052&
\hspace*{1.5mm}19.1/31\hspace*{-0.5mm}\\
\hline
\end{tabular}
\end{small}
\caption[]{%
Ratios of LEP combined Z-pair cross-section measurements
to the expectations, for different \CoM\ energies and for all energies combined.
The first column contains the \CoM\ energy,
the second the combined ratios,
the third the statistical errors.
The fourth, fifth, sixth and seventh columns contain
the sources of systematic errors that are considered as 
LEP-correlated   energy-uncorrelated (LCEU),
LEP-correlated   energy-correlated   (LCEC),
LEP-uncorrelated energy-uncorrelated (LUEU),
LEP-uncorrelated energy-correlated   (LUEC).
The total error is given in the eighth column.
The only LCEU systematic sources considered 
are the statistical errors on the cross-section theoretical predictions,
while the LCEC, LUEU and LUEC sources are those coming from
the corresponding errors on the cross-section measurements.
For the LEP averages, the $\chi^2$ of the fit is also given
in the ninth column.}
\label{4f_tab:rZZmeas} 
\end{center}
\end{table}

\begin{table}[p]
\renewcommand{\arraystretch}{1.2}
\vspace*{-0.0cm}
\begin{center}
\begin{small}
\begin{tabular}{|c|ccccc|c|c|}
\cline{1-8}
\roots & & & {\scriptsize (LCEC)} & {\scriptsize (LUEU)} & 
{\scriptsize (LUEC)} & & \\
(GeV) & $\szee$ & 
$\Delta\szee^\mathrm{stat}$ &
$\Delta\szee^\mathrm{syst}$ &
$\Delta\szee^\mathrm{syst}$ &
$\Delta\szee^\mathrm{syst}$ &
$\Delta\szee$ & 
$\Delta\szee^\mathrm{stat\,(exp)}$ \\
\hline
\multicolumn{8}{|c|}
{\Aleph~\cite{4f_bib:alesw}} \\
\hline
182.7 & 0.27 & $^{+0.21}_{-0.16}$ & $\pm$0.01 & $\pm$0.02 & $\pm$0.01 & $^{+0.21}_{-0.16}$ & $\pm$0.20 \\
188.6 & 0.42 & $^{+0.14}_{-0.12}$ & $\pm$0.01 & $\pm$0.03 & $\pm$0.01 & $^{+0.14}_{-0.12}$ & $\pm$0.12 \\
191.6 & 0.61 & $^{+0.39}_{-0.29}$ & $\pm$0.01 & $\pm$0.03 & $\pm$0.01 & $^{+0.39}_{-0.29}$ & $\pm$0.29 \\
195.5 & 0.72 & $^{+0.24}_{-0.20}$ & $\pm$0.01 & $\pm$0.03 & $\pm$0.01 & $^{+0.24}_{-0.20}$ & $\pm$0.18 \\
199.5 & 0.60 & $^{+0.21}_{-0.18}$ & $\pm$0.01 & $\pm$0.03 & $\pm$0.01 & $^{+0.21}_{-0.18}$ & $\pm$0.17 \\
201.6 & 0.89 & $^{+0.35}_{-0.28}$ & $\pm$0.01 & $\pm$0.03 & $\pm$0.01 & $^{+0.35}_{-0.28}$ & $\pm$0.24 \\
204.9 & 0.42 & $^{+0.17}_{-0.14}$ & $\pm$0.01 & $\pm$0.03 & $\pm$0.01 & $^{+0.17}_{-0.15}$ & $\pm$0.17 \\
206.6 & 0.70 & $^{+0.17}_{-0.15}$ & $\pm$0.01 & $\pm$0.03 & $\pm$0.01 & $^{+0.17}_{-0.15}$ & $\pm$0.14 \\
\hline
\multicolumn{8}{|c|}
{\Delphi~\cite{4f_bib:delswsc05}} \\
\hline
182.7 & 0.56 & $^{+0.27}_{-0.22}$ & $\pm$0.01 & $\pm$0.06 & $\pm$0.02 & $^{+0.28}_{-0.23}$ & $\pm$0.24 \\
188.6 & 0.64 & $^{+0.15}_{-0.14}$ & $\pm$0.01 & $\pm$0.03 & $\pm$0.02 & $^{+0.16}_{-0.14}$ & $\pm$0.14 \\
191.6 & 0.63 & $^{+0.40}_{-0.30}$ & $\pm$0.01 & $\pm$0.03 & $\pm$0.03 & $^{+0.40}_{-0.30}$ & $\pm$0.32 \\
195.5 & 0.66 & $^{+0.22}_{-0.18}$ & $\pm$0.01 & $\pm$0.02 & $\pm$0.03 & $^{+0.22}_{-0.19}$ & $\pm$0.19 \\
199.5 & 0.57 & $^{+0.20}_{-0.17}$ & $\pm$0.01 & $\pm$0.02 & $\pm$0.02 & $^{+0.20}_{-0.17}$ & $\pm$0.18 \\
201.6 & 0.19 & $^{+0.21}_{-0.16}$ & $\pm$0.01 & $\pm$0.02 & $\pm$0.01 & $^{+0.21}_{-0.16}$ & $\pm$0.25 \\
204.9 & 0.37 & $^{+0.18}_{-0.15}$ & $\pm$0.01 & $\pm$0.02 & $\pm$0.02 & $^{+0.18}_{-0.15}$ & $\pm$0.19 \\
206.6 & 0.69 & $^{+0.16}_{-0.14}$ & $\pm$0.01 & $\pm$0.01 & $\pm$0.03 & $^{+0.16}_{-0.14}$ & $\pm$0.14 \\
\hline
\multicolumn{8}{|c|}
{\Ltre~\cite{4f_bib:ltrzee}} \\
\hline
182.7 & 0.51 & $^{+0.19}_{-0.16}$ & $\pm$0.02 & $\pm$0.01 & $\pm$0.03 & $^{+0.19}_{-0.16}$ & $\pm$0.16 \\
188.6 & 0.55 & $^{+0.10}_{-0.09}$ & $\pm$0.02 & $\pm$0.01 & $\pm$0.03 & $^{+0.11}_{-0.10}$ & $\pm$0.09 \\
191.6 & 0.60 & $^{+0.26}_{-0.21}$ & $\pm$0.01 & $\pm$0.01 & $\pm$0.03 & $^{+0.26}_{-0.21}$ & $\pm$0.21 \\
195.5 & 0.40 & $^{+0.13}_{-0.11}$ & $\pm$0.01 & $\pm$0.01 & $\pm$0.03 & $^{+0.13}_{-0.11}$ & $\pm$0.13 \\
199.5 & 0.33 & $^{+0.12}_{-0.10}$ & $\pm$0.01 & $\pm$0.01 & $\pm$0.03 & $^{+0.13}_{-0.11}$ & $\pm$0.14 \\
201.6 & 0.81 & $^{+0.27}_{-0.23}$ & $\pm$0.02 & $\pm$0.02 & $\pm$0.03 & $^{+0.27}_{-0.23}$ & $\pm$0.19 \\
204.9 & 0.56 & $^{+0.16}_{-0.14}$ & $\pm$0.01 & $\pm$0.01 & $\pm$0.03 & $^{+0.16}_{-0.14}$ & $\pm$0.14 \\
206.6 & 0.59 & $^{+0.12}_{-0.10}$ & $\pm$0.01 & $\pm$0.01 & $\pm$0.03 & $^{+0.12}_{-0.11}$ & $\pm$0.11 \\
\hline
\multicolumn{7}{|c|}
{LEP} & $\chi^2/\textrm{d.o.f.}$ \\
\hline
182.7 & 0.45 & $\pm$0.11 & $\pm$0.01 & $\pm$0.02 & $\pm$0.01 & $\pm$0.11 & 
 \multirow{8}{20.3mm}{$
   \hspace*{-0.3mm}
   \left\}
     \begin{array}[h]{rr}
       &\multirow{8}{8mm}{\hspace*{-4.2mm}13.0/16}\\
       &\\ &\\ &\\ &\\ &\\ &\\ &\\  
     \end{array}
   \right.
   $}\\
188.6 & 0.53 & $\pm$0.07 & $\pm$0.01 & $\pm$0.01 & $\pm$0.01 & $\pm$0.07 &  \\
191.6 & 0.61 & $\pm$0.15 & $\pm$0.01 & $\pm$0.02 & $\pm$0.01 & $\pm$0.15 &  \\
195.5 & 0.55 & $\pm$0.09 & $\pm$0.01 & $\pm$0.01 & $\pm$0.01 & $\pm$0.10 &  \\
199.5 & 0.47 & $\pm$0.09 & $\pm$0.01 & $\pm$0.02 & $\pm$0.01 & $\pm$0.10 &  \\
201.6 & 0.67 & $\pm$0.13 & $\pm$0.01 & $\pm$0.01 & $\pm$0.01 & $\pm$0.13 &  \\
204.9 & 0.47 & $\pm$0.10 & $\pm$0.01 & $\pm$0.01 & $\pm$0.01 & $\pm$0.10 &  \\
206.6 & 0.65 & $\pm$0.07 & $\pm$0.01 & $\pm$0.01 & $\pm$0.01 & $\pm$0.08 &  \\
\hline
\end{tabular}
\end{small}
\caption[]{%
Single-Z hadronic production cross-section (in pb) at different energies.
The first column contains the LEP \CoM\ energy,
and the second the measurements. 
The third column reports the statistical error, whereas in the fourth to the
sixth columns the different systematic uncertainties are listed.
The seventh column contains the total error and the eight lists,
for the four LEP measurements,
the symmetrized expected statistical error,
and for the LEP combined value,
the $\chi^2$ of the fit.}
\label{4f_tab:Zeemeas} 
\end{center}
\renewcommand{\arraystretch}{1.}
\end{table}

\begin{table}[hbtp]
\begin{center}
\hspace*{-0.3cm}
\renewcommand{\arraystretch}{1.2}
\begin{tabular}{|c|c|c|} 
\hline
\roots & \multicolumn{2}{|c|}{Zee cross-section (pb)}  \\
\cline{2-3} 
(GeV) & $\szee^{\footnotesize\WPHACT}$    
      & $\szee^{\footnotesize\Grace}$ \\
\hline
182.7 & 0.51275[4] & 0.51573[4] \\
188.6 & 0.53686[4] & 0.54095[5] \\
191.6 & 0.54883[4] & 0.55314[5] \\
195.5 & 0.56399[5] & 0.56891[4] \\
199.5 & 0.57935[5] & 0.58439[4] \\
201.6 & 0.58708[4] & 0.59243[4] \\
204.9 & 0.59905[4] & 0.60487[4] \\
206.6 & 0.61752[4] & 0.60819[4] \\
\hline
\end{tabular}
\renewcommand{\arraystretch}{1.}
\caption[]{%
Zee cross-section predictions (in pb) interpolated at the data 
\CoM\ energies,according to the \WPHACT~\protect\cite{4f_bib:wphact} and 
\Grace~\protect\cite{4f_bib:grace} predictions.
The numbers in brackets are the errors on the last digit and are coming
from the numerical integration of the cross-section only.}
\label{4f_tab:Zeetheo} 
\end{center}
\end{table}

\begin{table}[hbtp]
\begin{center}
\begin{small}
\begin{tabular}{|c|cccccc|c|c|}
\hline
\roots & & & {\scriptsize (LCEU)} & {\scriptsize (LCEC)} & 
{\scriptsize (LUEU)} & {\scriptsize (LUEC)} & & \\
(GeV) & $\rzee$ & 
$\Delta\rzee^\mathrm{stat}$ &
$\Delta\rzee^\mathrm{syst}$ &
$\Delta\rzee^\mathrm{syst}$ &
$\Delta\rzee^\mathrm{syst}$ &
$\Delta\rzee^\mathrm{syst}$ &
$\Delta\rzee$ &
$\chi^2/\textrm{d.o.f.}$ \\
\hline
\hline
\multicolumn{9}{|c|}{\Grace~\cite{4f_bib:grace}}\\
\hline
182.7 & 0.871 & $\pm$0.214 & $\pm$0.000 & $\pm$0.020 & $\pm$0.035 & $\pm$0.025 & $\pm$0.219 &
\multirow{8}{20.3mm}{$
  \hspace*{-0.3mm}
  \left\}
    \begin{array}[h]{rr}
      &\multirow{8}{6mm}{\hspace*{-4.2mm}13.0/16}\\
      &\\ &\\ &\\ &\\ &\\ &\\ &\\  
    \end{array}
  \right.
  $}\\
188.6 & 0.982 & $\pm$0.120 & $\pm$0.000 & $\pm$0.022 & $\pm$0.023 & $\pm$0.024 & $\pm$0.126 &\\
191.6 & 1.104 & $\pm$0.272 & $\pm$0.000 & $\pm$0.019 & $\pm$0.027 & $\pm$0.025 & $\pm$0.276 &\\
195.5 & 0.964 & $\pm$0.163 & $\pm$0.000 & $\pm$0.016 & $\pm$0.024 & $\pm$0.025 & $\pm$0.167 &\\
199.5 & 0.809 & $\pm$0.160 & $\pm$0.000 & $\pm$0.018 & $\pm$0.030 & $\pm$0.023 & $\pm$0.165 &\\
201.6 & 1.126 & $\pm$0.219 & $\pm$0.000 & $\pm$0.023 & $\pm$0.024 & $\pm$0.021 & $\pm$0.222 &\\
204.9 & 0.769 & $\pm$0.157 & $\pm$0.000 & $\pm$0.019 & $\pm$0.019 & $\pm$0.021 & $\pm$0.160 &\\
206.6 & 1.062 & $\pm$0.119 & $\pm$0.000 & $\pm$0.018 & $\pm$0.018 & $\pm$0.024 & $\pm$0.124 &\\
\hline
Average & 
0.955 & $\pm$0.057 & $\pm$0.000 & $\pm$0.019 & $\pm$0.009 & $\pm$0.023 & $\pm$0.065&
\hspace*{1.5mm}17.1/23\hspace*{-0.5mm}\\
\hline
\hline
\multicolumn{9}{|c|}{\WPHACT~\cite{4f_bib:wphact}}\\
\hline
182.7 & 0.876 & $\pm$0.215 & $\pm$0.000 & $\pm$0.020 & $\pm$0.035 & $\pm$0.025 & $\pm$0.220 &
\multirow{8}{20.3mm}{$
  \hspace*{-0.3mm}
  \left\}
    \begin{array}[h]{rr}
      &\multirow{8}{6mm}{\hspace*{-4.2mm}13.0/16}\\
      &\\ &\\ &\\ &\\ &\\ &\\ &\\  
    \end{array}
  \right.
  $}\\
188.6 & 0.990 & $\pm$0.120 & $\pm$0.000 & $\pm$0.022 & $\pm$0.023 & $\pm$0.025 & $\pm$0.127 &\\
191.6 & 1.112 & $\pm$0.274 & $\pm$0.000 & $\pm$0.020 & $\pm$0.027 & $\pm$0.026 & $\pm$0.277 &\\
195.5 & 0.972 & $\pm$0.164 & $\pm$0.000 & $\pm$0.016 & $\pm$0.025 & $\pm$0.025 & $\pm$0.168 &\\
199.5 & 0.816 & $\pm$0.161 & $\pm$0.000 & $\pm$0.019 & $\pm$0.030 & $\pm$0.023 & $\pm$0.167 &\\
201.6 & 1.135 & $\pm$0.221 & $\pm$0.000 & $\pm$0.023 & $\pm$0.024 & $\pm$0.021 & $\pm$0.224 &\\
204.9 & 0.776 & $\pm$0.158 & $\pm$0.000 & $\pm$0.019 & $\pm$0.019 & $\pm$0.021 & $\pm$0.162 &\\
206.6 & 1.067 & $\pm$0.120 & $\pm$0.000 & $\pm$0.018 & $\pm$0.018 & $\pm$0.024 & $\pm$0.125 &\\
\hline
Average & 
0.962 & $\pm$0.057 & $\pm$0.000 & $\pm$0.020 & $\pm$0.009 & $\pm$0.024 & $\pm$0.065&
\hspace*{1.5mm}17.0/23\hspace*{-0.5mm}\\
\hline
\end{tabular}
\end{small}
\caption[]{%
Ratios of LEP combined single-Z cross-section measurements
to the expectations, for different \CoM\ energies and for all energies combined.
The first column contains the \CoM\ energy,
the second the combined ratios,
the third the statistical errors.
The fourth, fifth, sixth and seventh columns contain
the sources of systematic errors that are considered as 
LEP-correlated   energy-uncorrelated (LCEU),
LEP-correlated   energy-correlated   (LCEC),
LEP-uncorrelated energy-uncorrelated (LUEU),
LEP-uncorrelated energy-correlated   (LUEC).
The total error is given in the eighth column.
The only LCEU systematic sources considered 
are the statistical errors on the cross-section theoretical predictions,
while the LCEC, LUEU and LUEC sources are those coming from
the corresponding errors on the cross-section measurements.
For the LEP averages, the $\chi^2$ of the fit is also given
in the ninth column.}
\label{4f_tab:rzeemeas} 
\end{center}
\end{table}

\chapter{Colour Reconnection Combination}

 \section{Inputs}
  \label{fsi:cr:app:inputs}
 \begin{table}[hbt]
  \center
  \begin{tabular}{|c||c|c|c|c|} \hline
               & \multicolumn{4}{c|}{Experiment} \\
   \Rn         & \multicolumn{1}{c}{\Aleph} & \multicolumn{1}{c}{\Delphi} &
                 \multicolumn{1}{c}{\Ltre}  & \multicolumn{1}{c|}{\Opal} \\ \hline\hline

   Data &   $1.0951\pm0.0135$   & $0.8996\pm0.0314$    & $0.8436\pm0.0217$ & $1.2570\pm0.0251$ \\
   \SKI\ (100\%)
        &   $1.0548\pm0.0012$   & $0.8463\pm0.0036$    & $0.7482\pm0.0033$    &     $1.1386\pm0.0027$   \\
  \Jetset
        &   $1.1365\pm0.0013$   & $0.9444\pm0.0039$    & $0.8622\pm0.0037$    &     $1.2958\pm0.0028$  \\
  \ARII
        &   $1.1341\pm0.0013$   & $0.9552\pm0.0041$    & $0.8696\pm0.0037$    &     $1.2887\pm0.0028$   \\
  \Ariadne
        &   $1.1461\pm0.0013$   & $0.9530\pm0.0039$    & $0.8754\pm0.0037$    &     $1.3057\pm0.0028$    \\
  \Herwig\ CR
        &   $1.1416\pm0.0013$   & $0.9649\pm0.0039$    & $0.8805\pm0.0037$    &     $1.3016\pm0.0029$    \\
  \Herwig
        &   $1.1548\pm0.0013$   & $0.9675\pm0.0040$    & $0.8822\pm0.0038$    &     $1.3204\pm0.0029$     \\
                                                                         \hline\hline
Systematics    &             &                 &                 &       \\ \hline\hline
 Intra-W BEC
              &  $\pm0.0020$ & $\pm0.0094$     & $\pm0.0017$     & $\pm0.0015$            \\
  \eeqq\ shape
              &  $\pm0.0012$ & $\pm0.0013$     & $\pm0.0086$     & $\pm0.0035$            \\
  $\pm10$\% $\sigma(\eeqq)$
              &  $\pm0.0036$ & $\pm0.0042$     & $\pm0.0071$     & $\pm0.0040$            \\
  $\pm15$\% $\sigma(\ZZtoqqqq)$
              &  $\pm0.0004$ & $\pm0.0001$     & $\pm0.0020$     & $\pm0.0013$            \\
 Detector effects
              &  $0.0040$    &         $-$     & $\pm0.0016$     & $\pm0.0072$            \\
 $E_{\mathrm{cm}}$ dependence
              &  $\pm0.0062$ & $\pm0.0012$     & $\pm0.0020$     & $\pm0.0030$            \\ \hline
  \end{tabular}
  \center
 \caption[Experimental inputs in particle flow.]{Inputs provided by the experiments for the combination.}
 \label{fsi:cr:tab:inputs}
 \end{table}

\begin{table}[bht]
\begin{center}
\begin{tabular}{||c|c|c|c|c|c||}
\hline
$k_{i}$  & $P_{reco}$ (\%) & ALEPH & DELPHI & L3 & OPAL \\
\hline
 \hline
 0.10 & \phantom{0}7.2& $1.1357\pm0.0057$  & $0.9410\pm0.0034$  & $0.8613\pm0.0037$ & $1.2887\pm0.0028$  \\
\hline
  0.15 & 10.2 & $1.1341\pm0.0057$ &  $0.9393\pm0.0032$  & 0.8598 $\pm$  0.0037 & $1.2859\pm0.0028$ \\
\hline
 0.20  & 13.4 & $1.1336\pm0.0057$ &  $0.9378\pm0.0031$  & 0.8585 $\pm$  0.0037 & $1.2823\pm0.0028$ \\
\hline
 0.25 & 16.1  & $1.1336\pm0.0057$ &  $0.9363\pm0.0030$  & 0.8561 $\pm$  0.0037 & $1.2800\pm0.0028$ \\
\hline
 0.35  & 21.4 & $1.1303\pm0.0057$ &  $0.9334\pm0.0028$  & 0.8551 $\pm$  0.0037 & $1.2741\pm0.0028$ \\
\hline
 0.45  & 25.9 & $1.1269\pm0.0057$ &  $0.9307\pm0.0027$  & 0.8509 $\pm$  0.0036 & $1.2693\pm0.0028$ \\
\hline
 0.60 & 32.1  & $1.1216\pm0.0057$ &  $0.9271\pm0.0025$  & 0.8482 $\pm$  0.0036 & $1.2639\pm0.0028$ \\
\hline
 0.80  & 39.1 & $1.1166\pm0.0056$ &  $0.9227\pm0.0024$  & 0.8414 $\pm$  0.0037 & $1.2576\pm0.0028$  \\
\hline
 1.00  & 44.9 & $1.1109\pm0.0056$ &  $0.9189\pm0.0024$  & 0.8381 $\pm$  0.0036 & $1.2499\pm0.0028$ \\
\hline
 1.50  & 55.9 & $1.1048\pm0.0056$ &  $0.9110\pm0.0025$  & 0.8318 $\pm$  0.0036 & $1.2368\pm0.0028$ \\
\hline
 3.00 &  72.8 & $1.0929\pm0.0056$ &  $0.8959\pm0.0028$  & 0.8135 $\pm$  0.0036 & $1.2093\pm0.0027$  \\
\hline
 5.00 &  82.5 & $1.0852\pm0.0056$ &  $0.8846\pm0.0030$  & 0.7989 $\pm$  0.0035 & $1.1920\pm0.0022$ \\
\hline
\end{tabular}
\caption[\SKI\ model predictions for \Rn.]
{\SKI\ Model predictions for $R_{N}$ obtained with the common LEP
  samples at 189 GeV. The second column gives the fraction of
  reconnected events in the common samples obtained for the different
  choice of $k_{I}$ values.}
\label{fsi:cr:tab:cetraro}
\end{center}
\end{table}

\clearpage

 \section{Example Average}
 \begin{table}[hbt]
  \begin{center}
  \begin{tabular}{|c||c|c|c|c|} \hline
     Model tested  & \multicolumn{4}{c|}{Experiment} \\
     \SKI\ (100\%)        & \multicolumn{1}{c}{\Aleph} & \multicolumn{1}{c}{\Delphi} &
                 \multicolumn{1}{c}{\Ltre}  & \multicolumn{1}{c|}{\Opal} \\ \hline\hline

\Rn\ (no-CR)      
    &   $1.1365\pm0.0013$   & $0.9444\pm0.0039$    & $0.8622\pm0.0037$    &     $1.2958\pm0.0028$  \\
\Rn\ (with CR)   
     &   $1.0548\pm0.0012$   & $0.8463\pm0.0036$    & $0.7482\pm0.0033$    &     $1.1386\pm0.0027$   \\
  weight &  19.688     &  7.054           &  18.250       &         28.202          \\ \hline\hline
  $r$ ($\equiv$\Rn(data)/\Rnnocr)
               &             &                 &                 &       \\
  Data                         
               &   0.9636    &   0.9526   &   0.9784   &  0.9701  \\
  Stat.\ error &  0.0119     &   0.0332   &   0.0252   &  0.0194  \\
  Syst.\ error &  0.0110     &   0.0206   &   0.0180   &  0.0121  \\ \hline\hline

Uncorrel.\ syst.
               &             &             &             &        \\
 Background
               &   0.0013    &   0.0035    &     0.0128  & 0.0029 \\
 Hadronisation
               &   0.0000    &   0.0094    &   0.0086    & 0.0051 \\
 Intra-W BEC
               &   0.0013    &   0.0099    &    0.0016   & 0.0000 \\
 Detector effects
               &   0.0035    &  $-$        &    0.0019   & 0.0056 \\
 $E_{\mathrm{cm}}$ dependence
               &    0.0055   &   0.0123    &    0.0023   & 0.0023 \\ \hline\hline
 Total uncorr. error
               &   0.0068    &  0.0187     &    0.0158   & 0.0084 \\ \hline\hline
 Correl.\ syst.
               &             &             &             &        \\
 Background
               &  0.0031     &  0.0031     &  0.0031     & 0.0031 \\
 Hadronisation
               & 0.0081      &  0.0081     &  0.0081     & 0.0081 \\
 Intra-W BEC
               &  0.0012     &  0.0012     &  0.0012     & 0.0012 \\ \hline\hline
 Total correl. error
               &  0.0087     &  0.0087     &  0.0087     & 0.0087 \\ \hline
 \hline
  \end{tabular}
  \end{center}
 \caption[Normalised results of particle flow to \SKI\ model.]
 {Normalised results of particle flow analysis, based on the
  predicted \SKI\ 100\% sensitivity.}
 \label{fsi:cr:tab:outputs}
 \end{table}

  \label{fsi:cr:app:results}

\begin{table}[hbt]
\begin{center}
\begin{tabular}{||c|c|c|c|c|c||}
\hline
$k_{i}$  & $P_{reco}$ (\%) &$\langle r\rangle^{MC}$  &
                           $\langle r\rangle^{ADLO}$ & data-MC ($\sigma$) \\\hline \hline
 0.10 & \phantom{0}7.2 & 0.9950& $ 0.9679 \pm 0.0167\pm 0.0087\pm 0.0076$   & -1.34  \\
\hline
  0.15 & 10.2 & 0.9935& $ 0.9677 \pm 0.0146\pm 0.0087\pm 0.0065$   & -1.42  \\
\hline
 0.20  & 13.4 & 0.9911& $ 0.9681 \pm 0.0148\pm 0.0087\pm 0.0066$   & -1.25  \\
\hline
 0.25 & 16.1  & 0.9895& $ 0.9687 \pm 0.0144\pm 0.0087\pm 0.0066$   & -1.15  \\
\hline
 0.35  & 21.4 & 0.9861& $ 0.9680 \pm 0.0136\pm 0.0087\pm 0.0062$   & -1.05  \\
\hline
 0.45  & 25.9 & 0.9834& $ 0.9681 \pm 0.0123\pm 0.0087\pm 0.0057$   & -0.98  \\
\hline
 0.60 & 32.1  & 0.9802& $ 0.9676 \pm 0.0112\pm 0.0087\pm 0.0053$   & -0.84  \\
\hline
 0.80  & 39.1 & 0.9757& $ 0.9678 \pm 0.0106\pm 0.0087\pm 0.0052$   & -0.54   \\
\hline
 1.00  & 44.9 &  0.9708& $ 0.9676 \pm 0.0103\pm 0.0087\pm 0.0051$   & -0.22 \\
\hline
 1.50  & 55.9 & 0.9626& $ 0.9676 \pm 0.0105\pm 0.0087\pm 0.0051$   & +0.34  \\
\hline
 3.00 &  72.8 &  0.9447& $ 0.9680 \pm 0.0107\pm 0.0087\pm 0.0053$   & +1.58   \\
\hline
 5.00 &  82.5 & 0.9324& $ 0.9683 \pm 0.0108\pm 0.0087\pm 0.0054$   & +2.42  \\
\hline
 10000 & 100 & 0.8909& $ 0.9687 \pm 0.0108\pm 0.0087\pm 0.0057$   & +5.20 \\
\hline
\end{tabular}
\caption[LEP Averages in $r$ for \SKI\ model.]{LEP Average values of $r$ in Monte
  Carlo, $\langle r\rangle^{MC}$
  ($\equiv\langle\Rn/\Rnnocr\rangle^{MC}$), and data, $\langle
  r\rangle^{ADLO}$ ($\equiv\langle\Rn/\Rnnocr\rangle^{ADLO}$), for
  various $k_I$ values in \SKI\ model.  The first uncertainty is
  statistical, the second corresponds to the correlated systematic
  error and the third corresponds to the uncorrelated systematic
  error.}
\label{fsi:cr:tab:average}
\end{center}
\end{table}

\begin{table}[hbt]
\begin{center}
\begin{tabular}{||c|c|c|c|c|c||}
\hline
Model & $ \langle r \rangle^{MC}$  & $\langle r\rangle^{ADLO}$ & data-MC ($\sigma$) \\
\hline
 \hline
 AR2 & 0.9888& $ 0.9589 \pm 0.0101\pm 0.0086\pm 0.0050$   & -2.10  \\
\hline
 \Herwig\ CR & 0.9874& $ 0.9498 \pm 0.0105\pm 0.0086\pm 0.0052$   & -2.59  \\
\hline
\hline
\end{tabular}
\caption[LEP Averages in $r$ for \ARII\ and \Herwig\ CR models.]
{LEP Average values of $r$ in Monte Carlo, $\langle r\rangle^{MC}$
  ($\equiv\langle\Rn/\Rnnocr\rangle^{MC}$), and data, $\langle r
  \rangle^{ADLO}$ ($\equiv\langle\Rn/\Rnnocr\rangle^{ADLO}$), for
  \Ariadne\ and \Herwig\ models with colour reconnection.  The first
  uncertainty is statistical, the second corresponds to the correlated
  systematic error and the third corresponds to the uncorrelated
  systematic error.}
\label{fsi:cr:tab:average2}
\end{center}
\end{table}

\end{appendix}

\clearpage

\bibliographystyle{PhysRep}
\bibliography{s05_ew,common,gg,ff,smat,fsi,be,4f_s05,gc,mw}

\begin{mcbibliography}{100}

\bibitem{bib-EWEP-04}
{The LEP Collaborations ALEPH, DELPHI, L3, OPAL, the LEP Electroweak Working
  Group, the SLD Electroweak and Heavy Flavour Groups},
\newblock  {\em A Combination of Preliminary Electroweak Measurements and
  Constraints on the Standard Model},
\newblock  Eprint hep-ex/0412015, CERN, 2004\relax
\relax
\bibitem{bib-Z-pole}
{The ALEPH, DELPHI, L3, OPAL, SLD Collaborations, the LEP Electroweak Working
  Group, the SLD Electroweak and Heavy Flavour Groups},
\newblock  {\em Precision Electroweak Measurements on the Z Resonance},
\newblock  Eprint hep-ex/0509008, CERN, 2005\relax
\relax
\bibitem{Abdallah:2004rc}
DELPHI Collaboration, J. Abdallah {\it et~al.},
\newblock  Eur. Phys. J. {\bf C37}  (2004) 405--419\relax
\relax
\bibitem{ref:QED}
W. Heitler,
\newblock  Quantum Theory of Radiation,
\newblock  (Oxford University Press, second edition, 1944),
\newblock  pages 204--207\relax
\relax
\bibitem{ref:radcor}
F.~A. Berends and R. Kleiss,
\newblock  Nucl. Phys. {\bf B186}  (1981) 22\relax
\relax
\bibitem{gg:ref:LEPGG}
ALEPH Collab., Eur. Phys. J., C 28 (2003) 1 and ref. therein;\\ DELPHI Collab.,
  DELPHI 2001-093 CONF 521 and ref. therein; \\ L3 Collab., \PL {\bf B531}
  (2002) 28 and ref. therein;\\ OPAL Collab., Eur.Phys.J.C26 (2003) 331 and
  ref. therein\relax
\relax
\bibitem{gg:ref:drell}
S.~D. Drell,
\newblock  Ann. Phys. {\bf 4}  (1958) 75\relax
\relax
\bibitem{gg:ref:low}
F.~E. Low,
\newblock  \PRL {\bf 14}  (1965) 238\relax
\relax
\bibitem{gg:ref:eboli}
O.~J.~P. Eboli, A.~A. Natale, and S.~F. Novaes,
\newblock  \PL {\bf B271}  (1991) 274\relax
\relax
\bibitem{gg:ref:estar}
P. Mery, M. Perrottet, and F.~M. Renard,
\newblock  Z. Phys. {\bf C38}  (1988) 579\relax
\relax
\bibitem{gg:ref:g2_brodsky}
S.~J. Brodsky and S.~D. Drell,
\newblock  Phys. Rev. {\bf D22}  (1980) 2236\relax
\relax
\bibitem{gg:ref:boudjema:1993}
F. Boudjema, A. Djouadi, and J.~L. Kneur,
\newblock  Z. Phys. {\bf C57}  (1993) 425--450\relax
\relax
\bibitem{gg:ref:vachon}
B. Vachon,
\newblock  {\em Excited electron contribution to the $\eegg$ cross-section},
\newblock  hep-ph/0103132, 2001\relax
\relax
\bibitem{gg:ref:ad}
K. Agashe and N.~G. Deshpande,
\newblock  \PL {\bf B456}  (1999) 60\relax
\relax
\bibitem{Abbiendi:2003dh}
OPAL Collaboration, G. Abbiendi {\it et~al.},
\newblock  Eur. Phys. J. {\bf C33}  (2004) 173--212\relax
\relax
\bibitem{Abbiendi:2004vw}
OPAL Collaboration, G. Abbiendi {\it et~al.},
\newblock  Phys. Lett. {\bf B609}  (2005) 212--225\relax
\relax
\bibitem{LEP2L3ffbar}
L3 Collaboration, P. Achard {\it et~al.},
\newblock  {\em Measurement of Hadron and Lepton-Pair Production in $e^+e^-$
  Collisions at $\sqrt{s}=192-208$~GeV at LEP},
\newblock  Eprint CERN-PH-EP/2005-044, 2005\relax
\relax
\bibitem{bib-EWEP-02}
{The LEP Collaborations ALEPH, DELPHI, L3, OPAL, the LEP Electroweak Working
  Group, the SLD Electroweak and Heavy Flavour Groups},
\newblock  {\em A Combination of Preliminary Electroweak Measurements and
  Constraints on the Standard Model},
\newblock  Eprint hep-ex/0212036, CERN, 2002\relax
\relax
\bibitem{ff:ref:ffbar_web}
LEPEWWG $\ff$ subgroup: http://www.cern.ch/LEPEWWG/lep2/~\relax
\relax
\bibitem{ff:ref:expts}
ALEPH Collaboration ``Study of Fermion Pair Production in $\ee$ Collisions at
  130-183 GeV'', Eur. Phys. J. {\bf{C12}} (2000) 183; \\ ALEPH Collaboration,
  ``Fermion Pair Production in $\ee$ Collisions at 189 GeV and Kimits on
  Physics Beyond the Standard Model'', ALEPH 99-018 CONF 99-013; \\ ALEPH
  Collaboration ``Fermion Pair Production in $\ee$ Collisions from 192 to 202
  GeV and Limits on Physics beyond the Standard Model'', ALEPH 2000-047 CONF
  2000-030; \\ ALEPH Collaboration ``Fermion Pair Production in $\ee$
  Collisions at high energy and Limits on Physics beyond the Standard Model'',
  ALEPH 2002-032 CONF 2002-021; \\ ALEPH Collaboration, ``Fermion Pair
  Production in $\ee$ Collisions at high energy and Limits on Physics beyond
  the Standard Model '', ALEPH 2001-019 CONF 2001-016; \\ DELPHI
  Collaboration,``Measurement and Interpretation of Fermion-Pair Production at
  LEP energies from 130 to 172 GeV'', Eur. Phys. J. {\bf{C11}} (1999), 383; \\
  DELPHI Collaboration, ``Measurement and Interpretation of Fermion-Pair
  Production at LEP Energies from 183 to 189 GeV'', Phys.Lett. {\bf{B485}}
  (2000), 45; \\ DELPHI Collaboration, ``Results on Fermion-Pair Production at
  LEP running from 192 to 202 GeV'', DELPHI 2000-128 OSAKA CONF 427; \\ DELPHI
  Collaboration, ``Results on Fermion-Pair Production at LEP running in 2000'',
  DELPHI 2001-094 CONF 522; \\ L3 Collaboration, ``Measurement of Hadron and
  Lepton-Pair Production at 161 GeV $< \sqrt{s} <$ 172 GeV at LEP'', Phys.
  Lett. {\bf{B407}} (1997) 361; \\ L3 Collaboration, ``Measurement of Hadron
  and Lepton-Pair Production at 130 GeV $< \sqrt{s} <$ 189 GeV at LEP'', Phys.
  Lett. {\bf{B479}} (2000), 101. \\ L3 Collaboration, ``Preliminary L3 Results
  on Fermion-Pair Production in 1999'', L3 note 2563; \\ L3
  Collaboration,``Preliminary L3 Results on Fermion-Pair Production in 2000'',
  L3 note 2648; \\ OPAL Collaboration, ``Tests of the Standard Model and
  Constraints on New Physics from Measurements of Fermion Pair Production at
  130 - 172 GeV at LEP'', Euro. Phys. J. {\bf C2} (1998) 441; \\ OPAL
  Collaboration, ``Tests of the Standard Model and Constraints on New Physics
  from Measurements of Fermion Pair Production at 183 GeV at LEP'', Euro. Phys.
  J. {\bf C6} (1999) 1; \\ OPAL Collaboration, ``Tests of the Standard Model
  and Constraints on New Physics from Measurements of Fermion Pair Production
  at 189 GeV at LEP'', Euro. Phys. J. {\bf C13} (2000) 553; \\ OPAL
  Collaboration, ``Tests of the Standard Model and Constraints on New Physics
  from Measurements of Fermion Pair Production at 192-202 GeV at LEP'', OPAL
  PN424 (2000); \\ OPAL Collaboration, ``Measurement of Standard Model
  Processes in e+e- Collisions at sqrt{s}~203-209 GeV'', OPAL PN469
  (2001)\relax
\relax
\bibitem{ff:ref:lepffwrkshp}
M. Kobel {\it et al.}, ``Two-Fermion Production in Electron Positron
  Collisions'' in S. Jadach {\it et al.} [eds] , ``Reports of the Working
  Groups on Precision Calculations for LEP2 Physics: proceedings'' CERN
  2000-009, hep-ph/0007180\relax
\relax
\bibitem{ff:ref:ZFITTER}
D.~Bardin {\it et al.}, CERN-TH 6443/92;
  http://www.ifh.de/$\sim$riemann/Zfitter/zf.html~.\\ Predictions are from
  ZFITTER versions 6.04 or later.\\ Definition 1 corresponds to the ZFITTER
  flags FINR=0 and INTF=0; definition 2 corresponds to FINR=0 and INTF=1 for
  hadrons, FINR=1 and INTF=1 for leptons\relax
\relax
\bibitem{common_bib:BLUE}
L. Lyons \etal, NIM A270 (1988) 110; A. Valassi, NIM A500 (2003) 391.\relax
\relax
\bibitem{ff:ref:BHWIDE}
S.~Jadach, W.~Placzek and B.Ward, Phys. Lett. {\bf B390} (1997) 298\relax
\relax
\bibitem{ff:ref:lepff-osaka}
LEPEWWG $\ff$ Subgroup, C.~Geweniger {\it et. al.}, LEP2FF/00-03,ALEPH 2000-088
  PHYSIC 2000-034, DELPHI 2000-168 PHYS 881, L3 note 2624, OPAL TN673\relax
\relax
\bibitem{ff:ref:hfpublished}
ALEPH Collaboration, Euro. Phys J. {\bf{C12}} (2000) 183; \\ DELPHI
  Collaboration, P.Abreu {\it et al.}, Euro. Phys J. {\bf{C11}}(1999); \\ L3
  Collaboration, M.Acciarri {\it et al.}, Phys.\ Lett.\ {\bf B485} (2000) 71;
  \\ OPAL Collaboration, G.Abbiendi {\it et al.}, Euro.\ Phys.\ J.\ {\bf C16}
  (2000) 41\relax
\relax
\bibitem{ff:ref:hfpreliminary}
ALEPH Collaboration, ALEPH 99-018 CONF 99-013; \\ ALEPH Collaboration, ALEPH
  2000-046 CONF 2000-029; \\ ALEPH Collaboration, ICHEP2002, ABS388; \\ DELPHI
  Collaboration, DELPHI 2001-095 CONF 523; \\ L3 Collaboration, L3 Internal
  note 2640, 1 March 2001\relax
\relax
\bibitem{ff:ref:hfconfnote}
LEPEWWG Heavy Flavour at LEP2 Subgroup, ``Combination of Heavy Flavour
  Measurements at LEP2'', LEP2FF/00-02\relax
\relax
\bibitem{ff:ref:hfzfit}
ZFITTER V6.23 is used.\\ D. Bardin {\it et al.}, Preprint hep-ph/9908433. \\
  Relevant ZFITTER settings used are FINR=0 and INTF=1\relax
\relax
\bibitem{ff:ref:hflep1-99}
DELPHI Collaboration, P.Abreu {\it et al.}, Euro Phys J. {\bf{C10}}(1999) 415.
  \\ The LEP collaborations {\it et al.}, CERN-EP/2000-016\relax
\relax
\bibitem{ff:ref:ELPthr}
E. Eichten, K. Lane, and M. Peskin, Phys. Rev. Lett. {\bf 50} (1983) 811\relax
\relax
\bibitem{ff:ref:Kroha}
H. Kroha, Phys. Rev. {\bf D46} (1992) 58\relax
\relax
\bibitem{ff:ref:zprime-thry}
P. Langacker, R.W. Robinett and J.L. Rosner, Phys. Rev. {\bf D30} (1984) 1470;
  \\ D. London and J.L. Rosner, Phys. Rev. {\bf D34} (1986) 1530; \\ J.C. Pati
  and A. Salam, Phys. Rev. {\bf D10} (1974) 275; \\ R.N. Mohapatra and J.C.
  Pati, Phys. Rev. {\bf D11} (1975) 566\relax
\relax
\bibitem{ff:ref:sqsm}
G. Altarelli \etal, Z. Phys. {\bf{C45}} (1989) 109; \\ erratum Z. Phys.
  {\bf{C47}} (1990) 676\relax
\relax
\bibitem{ff:ref:lep1zprime}
DELPHI Collaboration, P. Abreu {\it et al.}, Zeit. Phys. {\bf{C65}} (1995)
  603\relax
\relax
\bibitem{ff:ref:lq-thry}
W.~Buchm{\"{u}}ller, R.~R{\"{u}}ckl, D.~Wyler, Phys. Lett. {\bf B191} (1987)
  442; Erratum-ibid. {\bf B448} (1999) 320.\relax
\relax
\bibitem{ff:ref:lq-squ}
J.~Kalinowski {\it et al.}, Z. Phys. {\bf C74} (1997) 595\relax
\relax
\bibitem{ff:ref:lq-h1}
H1 Collab., C.~Adloff {\it et al.}, Phys. Lett. {\bf B523} (2001) 234\relax
\relax
\bibitem{ff:ref:lq-zeus}
ZEUS Collaboration; J.~Breitweg {\it et al.}, Phys. Rev. {\bf D63} (2001)
  052002\relax
\relax
\bibitem{ff:ref:lq-tevatron}
CDF Collab., F. Abe {\it et al.}, Phys. Rev. Lett. {\bf 79} (1997) 4327;\\
  D{\O} Collab., B.~Abbott {\it et al.}, Phys. Rev. Lett. {\bf 80} (1998)
  2051\relax
\relax
\bibitem{ff:ref:ADD}
N.~Arkani-Hamed, S.~Dimopoulos and G.~Dvali, Phys. Lett. {\bf B429} (1998)
  263\relax
\relax
\bibitem{ff:ref:ADD2}
I.~Antoniadis, N.~Arkani-Hamed, S.~Dimopoulos and G.~Dvali, Phys. Lett. {\bf
  B436} (1998) 257\relax
\relax
\bibitem{ff:ref:ADD3}
N.~Arkani-Hamed, S.~Dimopoulos and G.~Dvali, Phys. Rev. {\bf D59} (1999)
  086004\relax
\relax
\bibitem{ff:ref:Hewett}
J.~Hewett, Phys. Rev. Lett. {\bf 82} (1999) 4765\relax
\relax
\bibitem{ff:ref:Rizzo}
T.~Rizzo, Phys. Rev. {\bf D59} (1999) 115010\relax
\relax
\bibitem{ff:ref:Giudice}
G.~Giudice, R.~Rattazi and J.~Wells, Nucl. Phys. {\bf B544} (1999) 3\relax
\relax
\bibitem{ff:ref:Lykken}
T.~Han, J.D.~Lykken and R-J.~ Zhang, Phys. Rev. {\bf D59} (1999) 105006\relax
\relax
\bibitem{ff:ref:Shrock}
S.~Nussinov and R.~Shrock, Phys. Rev. {\bf D59} (1999) 105002\relax
\relax
\bibitem{ff:ref:Bourilkov:1999}
D.~Bourilkov, J. High Energy Phys. {\bf 08} (1999) 006\relax
\relax
\bibitem{ff:ref:Bourilkov:2000}
D.~Bourilkov, Phys. Rev. {\bf D62} (2000) 076005\relax
\relax
\bibitem{ff:ref:Bourilkov:2001}
D.~Bourilkov, Phys. Rev. {\bf D64} (2001) 071701\relax
\relax
\bibitem{smat:ref:5and9}
Lineshape subgroup of the LEP EWWG,``Combination procedure for the precise
  determination of Z boson parameters from results of the LEP experiments'',
  CERN-EP/2000-153, hep-ex/0101027 (2000)\relax
\relax
\bibitem{smat:ref:smat1997}
S-Matrix subgroup of the LEP EWWG,``An investigation of the interference
  between photon and Z boson exchange'', LEPEWWG/LS/97-02, ALEPH 97-92 PHYSICS
  97-82, DELPHI 97-153 PHYS 732, L3 Note 2164, OPAL TN510 (1997)\relax
\relax
\bibitem{smat:ref:bcms}
A. Borelli {\it et~al.},
\newblock  Nucl. Phys. {\bf B333}  (1990) 357\relax
\relax
\bibitem{smat:ref:rgs}
R.~G. Stuart,
\newblock  Phys.~Lett. {\bf B272}  (1991) 353\relax
\relax
\bibitem{smat:ref:arr}
A. Leike, T. Riemann, and J. Rose,
\newblock  Phys.~Lett. {\bf B273}  (1991) 513\relax
\relax
\bibitem{smat:ref:tr}
T. Riemann,
\newblock  Phys.~Lett. {\bf B293}  (1992) 451\relax
\relax
\bibitem{smat:ref:smatasy}
S. Kirsch and T. Riemann,
\newblock  Comput.~Phys.~Commun. {\bf 88}  (1995) 89\relax
\relax
\bibitem{smat:ref:lep2xsafbave}
LEP EWWG ``A combination of preliminary electroweak measurements and
  constriants on the Standard Model'', CERN-EP/2001-098, hep-ex/0112021
  (2001)\relax
\relax
\bibitem{smat:ref:aleph}
``S-Matrix fits to cross-section and forward-backward asymmetry measurements at
  LEP 1 and 2'', ALEPH 2000-042 PHYSICS 2000-015 (2002)\relax
\relax
\bibitem{smat:ref:delphi}
DELPHI collaboration, ``S-Matrix fits to LEP 1 and LEP 2 data'', DELPHI 2002-04
  CONF 545 (2002)\relax
\relax
\bibitem{smat:ref:l3lep1}
L3 Collaboration, M. Acciarri {\it et~al.},
\newblock  Euro. Phys. J. {\bf C16}  (2000) 1\relax
\relax
\bibitem{smat:ref:opallep1}
OPAL Collaboration, G. Abbiendi {\it et~al.},
\newblock  Euro. Phys. J. {\bf C19}  (2001) 587\relax
\relax
\bibitem{smat:ref:l3lep2}
L3 collaboration, ``Preliminary results on fermion-pair production in 2000'',
  L3 Note 2648 (2001)\relax
\relax
\bibitem{smat:ref:opallep2}
OPAL Collaboration, ``Determination of S-Matrix parameters from fermion pair
  production at OPAL'', OPAL PN474 (2001)\relax
\relax
\bibitem{smat:ref:venus}
VENUS Collaboration, K. Yusa {\it et~al.},
\newblock  Phys. Lett. {\bf B447}  (1999) 167\relax
\relax
\bibitem{smat:ref:topaz}
TOPAZ Collaboration, K. Miyabuyaski {\it et~al.},
\newblock  Phys. Lett. {\bf B347}  (1995) 171\relax
\relax
\bibitem{4f_bib:4f_s04}
The LEP WW Working Group, hep-ex/0412015 prepared for the Summer 2004
  conferences, {\tt http://lepewwg.web.cern.ch/LEPEWWG/stanmod/}\relax
\relax
\bibitem{4f_bib:fourfrep}
M.W.~Gr{\"{u}}newald, G.~Passarino \etal, {\it ``Four fermion production in
  electron positron collisions''}, Four fermion working group report of the
  LEP2 \MC\ Workshop 1999/2000, in {\it ``Reports of the working groups on
  precision calculations for LEP2 Physics''} CERN 2000--009, {\tt
  http://arXiv.org/abs/hep-ph/0005309}\relax
\relax
\bibitem{4f_bib:adloww161}
\Aleph\ Collaboration, Phys.~Lett.~{\bf B401} (1997) 347.\\ \Delphi\
  Collaboration, Phys.~Lett.~{\bf B397} (1997) 158.\\ \Ltre\ Collaboration,
  Phys.~Lett. {\bf B398} (1997) 223.\\ \Opal\ Collaboration, Phys.~Lett. {\bf
  B389} (1996) 416\relax
\relax
\bibitem{4f_bib:adloww172}
\Aleph\ Collaboration, Phys.~Lett.~{\bf B415} (1997) 435.\\ \Delphi\
  Collaboration, Eur.~Phys.~J.~{\bf C2} (1998) 581.\\ \Ltre\ Collaboration,
  Phys.~Lett.~{\bf B407} (1997) 419;\\ \Opal\ Collaboration, Eur.~Phys.~J.~{\bf
  C1} (1998) 395\relax
\relax
\bibitem{4f_bib:aleww}
\Aleph\ Collaboration, Eur. Phys. Jour. {\bf C38} (2004) 147\relax
\relax
\bibitem{4f_bib:delww}
\Delphi\ Collaboration, Eur. Phys. Jour. {\bf C34} (2004) 127\relax
\relax
\bibitem{4f_bib:ltrww}
\Ltre\ Collaboration, Phys.~Lett. {\bf B600} (2004) 22\relax
\relax
\bibitem{4f_bib:opaww189}
\Opal\ Collaboration, \PLB{493}{2000}{249}\relax
\relax
\bibitem{4f_bib:opawwsc01}
\Opal\ Collaboration, \Opal\ Physics Note PN469, submitted to the Summer 2001
  Conferences, and PN437 and PN420\relax
\relax
\bibitem{4f_bib:valassi}
A. Valassi,
\newblock  Nucl. Instrum. Meth. {\bf A500}  (2003) 391\relax
\relax
\bibitem{MINUIT}
F. James and M. Roos,
\newblock  Comput. Phys. Commun. {\bf 10}  (1975) 343\relax
\relax
\bibitem{4f_bib:lepewwg97}
The LEP Collaborations \Aleph, \Delphi, \L3, \Opal, the LEP Electroweak Working
  Group and the SLD Heavy Flavour Working Group, {\it ``A Combination of
  Preliminary Electroweak Measurements and Constraints on the Standard
  Model''}, CERN--PPE/97--154\relax
\relax
\bibitem{4f_bib:lepewwg98}
The LEP Collaborations \Aleph, \Delphi, \L3, \Opal, the LEP Electroweak Working
  Group and the SLD Heavy Flavour Working Group, {\it ``A Combination of
  Preliminary Electroweak Measurements and Constraints on the Standard
  Model''}, CERN--EP/99--15\relax
\relax
\bibitem{4f_bib:yfsww}
S. Jadach, W. P{\l}aczek, M. Skrzypek, B.F.L. Ward, Phys. Rev. {\bf D54} (1996)
  5434. \\ S. Jadach, W. P{\l}aczek, M. Skrzypek, B.F.L. Ward, Z. W\c{a}s,
  Phys. Lett. {\bf B417} (1998) 326. \\ S. Jadach, W. P{\l}aczek, M. Skrzypek,
  B.F.L. Ward, Z. W\c{a}s, Phys. Rev. {\bf D61} (2000) 113010; preprint
  CERN-TH-99-222, hep-ph/9907346.\\ S. Jadach, W. P{\l}aczek, M. Skrzypek,
  B.F.L. Ward, Z. W\c{a}s, preprint CERN-TH/2000-337, hep-ph/0007012; submitted
  to Phys. Lett. B.\\ S. Jadach, W. P{\l}aczek, M. Skrzypek, B.F.L. Ward, Z.
  W\c{a}s, {\it The Monte Carlo Event Generator {\tt YFSWW3} version {\tt 1.16}
  for W-Pair Production and Decay at LEP2/LC Energies}, preprint
  CERN-TH/2001-017, UTHEP-01-0101, hep-ph/0103163, accepted for publication by
  Comput. Phys. Commun.\\ The \YFSWW\ cross-sections at 155--215 GeV have been
  kindly provided by the authors\relax
\relax
\bibitem{4f_bib:racoonww}
A.~Denner, S.~Dittmaier, M.~Roth and D.~Wackeroth, Nucl. Phys. {\bf B560}
  (1999) 33.\\ A.~Denner, S.~Dittmaier, M.~Roth and D.~Wackeroth, Nucl. Phys.
  {\bf B587} (2000) 67.\\ A.~Denner, S.~Dittmaier, M.~Roth and D.~Wackeroth,
  \PLB{475}{2000}{127}.\\ A.~Denner, S.~Dittmaier, M.~Roth and D.~Wackeroth,
  hep-ph/0101257.\\ The \RacoonWW\ cross-sections at 155--215 GeV have been
  kindly provided by the authors\relax
\relax
\bibitem{4f_bib:dpa}
See~\cite{4f_bib:fourfrep} and references therein for a discussion of complete
  $\oa$ radiative corrections to W-pair production in the LPA/DPA
  approximations\relax
\relax
\bibitem{4f_bib:dpaerr}
The theoretical uncertainty $\Delta\sigma/\sigma$ on the W-pair production
  cross section calculated in the LPA/DPA above 170 GeV can be parametrised as
  $\Delta\sigma/\sigma=0.4\oplus0.072\cdot t_1\cdot t_2$, where
  $t_1=(200-2\cdot\Mw)/(\roots-2\cdot\Mw)$ and $t_2=(1-(\frac{2\cdot
  M_{\mathrm{W}}}{200})^2)/ (1-(\frac{2\cdot M_{\mathrm{W}}}{\sqrt{s}})^2)$. In
  the threshold region, a 2\% uncertainty is assigned.\relax
\relax
\bibitem{4f_bib:gentle}
D.~Bardin, J.~Biebel, D.~Lehner, A.~Leike, A.~Olchevski and T.~Riemann,
  \CPC{104}{1997}{161}. See also~\cite{4f_bib:fourfrep}.\\ The \Gentle\
  cross-sections at 155--215 GeV have been kindly provided by Eric Lan\c{c}on
  and Anne Ealet\relax
\relax
\bibitem{4f_bib:koralw}
M. Skrzypek, S. Jadach, M.~Martinez, W.~P{\l}aczek, Z.~W\c{a}s, Phys. Lett.
  {\bf B372} (1996) 289. \\ S. Jadach, W. P{\l}aczek, M. Skrzypek, Z. W\c{a}s,
  Comput. Phys. Commun. {\bf 94} (1996) 216. \\ S. Jadach, W. P{\l}aczek, M.
  Skrzypek, B.F.L. Ward, Z. W\c{a}s, Comput. Phys. Commun. {\bf 119} (1999)
  272. \\ S. Jadach, W. P{\l}aczek, M. Skrzypek, B.F.L. Ward, Z. W\c{a}s,
  preprint hep-ph/0104049, submitted to Comput. Phys. Commun.\\ The ``\KoralW''
  cross-sections at 155--215 GeV have been kindly provided by the authors. They
  have actually been computed using \YFSWW~\cite{4f_bib:yfsww}, switching off
  non-leading $\oa$ radiative corrections and the screening of the Coulomb
  correction, to reproduce the calculation from \KoralW\relax
\relax
\bibitem{4f_bib:pdg02}
K. Hagiwara et al., Phys. Rev. {\bf D66}, (2002) 010001\relax
\relax
\bibitem{4f_bib:kandy}
S. Jadach, W. P{\l}aczek, M. Skrzypek, B.F.L. Ward, Z. W\c{a}s, Comput. Phys.
  Commun. {\bf 140} (2001) 475\relax
\relax
\bibitem{4f_bib:delswsc05}
\Delphi\ Collaboration, \Delphi\ 2005-005 \Conf\ 725, EPS 2005 abstract
  1-84\relax
\relax
\bibitem{4f_bib:wto}
G.~Passarino, \NPB{578}{2000}{3}.\\ G.~Passarino, \NPB{574}{2000}{451}.\\ The
  \WTO\ cross-sections at 160--210 GeV have been kindly provided by the
  author\relax
\relax
\bibitem{4f_bib:wphact}
E.~Accomando and A.~Ballestrero, \CPC{99}{1997}{270}.\\ E.~Accomando,
  A.~Ballestrero and E. Maina, \CPC{150}{2003}{166}.\\ The \WPHACT\
  cross-sections at 160--210 GeV have been kindly provided by
  A.~Ballestrero\relax
\relax
\bibitem{4f_bib:grace}
J.~Fujimoto \etal, \CPC{100}{1997}{74}.\\ Y.~Kurihara \etal,
  Prog.~Theor.~Phys.~{\bf 103}~(2000)~1199.\\ The \Grace\ cross-sections at
  160--210 GeV have been kindly computed by R.~Tanaka\relax
\relax
\bibitem{4f_bib:wwichep00}
The LEP WW Working Group, LEPEWWG/XSEC/2000-01, note prepared for the summer
  2000 conferences, {\tt
  http://lepewwg.web.cern.ch/LEPEWWG/lepww/4f/Summer00}\relax
\relax
\bibitem{4f_bib:swap}
G.~Montagna \etal, {\it ``Higher--order QED corrections to single W production
  in electron--positron collisions''}, FNT/T--2000/08, {\tt
  http://arXiv.org/abs/hep-ph/0006307}\relax
\relax
\bibitem{4f_bib:delzz}
\Delphi\ Collaboration, Eur. Phys. J {\bf C30} (2003) 447\relax
\relax
\bibitem{4f_bib:ltrzz}
\Ltre\ Collaboration, \PLB{572}{2003}{133}\relax
\relax
\bibitem{4f_bib:opazz}
\Opal\ Collaboration, Eur. Phys. J {\bf C32} (2004) 303\relax
\relax
\bibitem{4f_bib:alezz189}
\Aleph\ Collaboration, \PLB{469}{1999}{287}\relax
\relax
\bibitem{4f_bib:alezzsc01}
\Aleph\ Collaboration, \Aleph\ 2001--006 \Conf\ 2001--003, submitted to the
  Winter 2001 Conferences\relax
\relax
\bibitem{4f_bib:yfszz}
S. Jadach, W. P{\l}aczek, B.F.L. Ward, \PRD{56}{1997}{6939}\relax
\relax
\bibitem{4f_bib:zzto}
G.~Passarino, in~\cite{4f_bib:fourfrep}\relax
\relax
\bibitem{4f_bib:alesw}
\Aleph\ Collaboration, CERN-PH-EP/2004-034, submitted to Phys. Lett. B.\relax
\relax
\bibitem{4f_bib:ltrzee}
\Ltre\ Collaboration, \PLB{561}{2003}{73}\relax
\relax
\bibitem{4f_bib:delzgstar}
\Delphi\ Collaboration, DELPHI 2004-019 CONF 694, contributed to Summert 2004
  conferences\relax
\relax
\bibitem{4f_bib:ltrzgstar}
\Ltre\ Collaboration, \PLB{616}{2005}{159}\relax
\relax
\bibitem{4f_bib:delwwg}
\Delphi\ Collaboration, Eur. Phys. Jour. {\bf C31} (2003) 139\relax
\relax
\bibitem{4f_bib:ltrwwg}
\Ltre\ Collaboration, \PLB{527}{2002}{39}\relax
\relax
\bibitem{4f_bib:opawwg}
\Opal\ Collaboration, \PLB{580}{2004}{17}\relax
\relax
\bibitem{4f_bib:eewwg}
J.~W.~Stirling and A.~Werthenbach, \EPJC{14}{2000}{103}\relax
\relax
\bibitem{Heister:2004yd}
ALEPH Collaboration, A. Heister {\it et~al.},
\newblock  Phys. Lett. {\bf B602}  (2004) 31--40\relax
\relax
\bibitem{Schael:2005am}
ALEPH Collaboration, S. Schael {\it et~al.},
\newblock  {\em Branching ratios and spectral functions of tau decays: Final
  ALEPH measurements and physics implications},
\newblock  Eprint hep-ex/0506072, 2005\relax
\relax
\bibitem{Achard:2004ji}
L3 Collaboration, P. Achard {\it et~al.},
\newblock  Phys. Lett. {\bf B586}  (2004) 151--166\relax
\relax
\bibitem{Achard:2004ds}
L3 Collaboration, P. Achard {\it et~al.},
\newblock  Phys. Lett. {\bf B597}  (2004) 119--130\relax
\relax
\bibitem{Abbiendi:2004bf}
OPAL Collaboration, G. Abbiendi {\it et~al.},
\newblock  Phys. Rev. {\bf D70}  (2004) 032005\relax
\relax
\bibitem{gc_bib:LEP2YR}
G. Gounaris \etal, in {\em Physics at LEP 2}, Report CERN 96-01 (1996), eds G.
  Altarelli, T. Sj{\"o}strand, F. Zwirner, Vol. 1, p. 525\relax
\relax
\bibitem{gc_bib:Montagna:2001ej}
G. Montagna {\it et~al.},
\newblock  Phys. Lett. {\bf B515}  (2001) 197--205\relax
\relax
\bibitem{gc_bib:denner}
A. Denner \etal, Eur. Phys. J. {\bf C 20} (2001) 201\relax
\relax
\bibitem{gc_bib:budapest01}
The LEP-TGC combination group, LEPEWWG/TGC/2001-03, September 2001\relax
\relax
\bibitem{common_bib:racoonww}
A.~Denner, S.~Dittmaier, M.~Roth and D.~Wackeroth, Nucl. Phys. {\bf B560}
  (1999) 33.\\ A.~Denner, S.~Dittmaier, M.~Roth and D.~Wackeroth, Nucl. Phys.
  {\bf B587} (2000) 67.\\ A.~Denner, S.~Dittmaier, M.~Roth and D.~Wackeroth,
  \PLB{475}{2000}{127}.\\ A.~Denner, S.~Dittmaier, M.~Roth and D.~Wackeroth,
  hep-ph/0101257.\\ The \RacoonWW\ cross-sections at 155--215 GeV have been
  kindly provided by the authors\relax
\relax
\bibitem{common_bib:yfsww}
S. Jadach, W. P{\l}aczek, M. Skrzypek, B.F.L. Ward, Phys. Rev. {\bf D54} (1996)
  5434. \\ S. Jadach, W. P{\l}aczek, M. Skrzypek, B.F.L. Ward, Z. W\c{a}s,
  Phys. Lett. {\bf B417} (1998) 326. \\ S. Jadach, W. P{\l}aczek, M. Skrzypek,
  B.F.L. Ward, Z. W\c{a}s, Phys. Rev. {\bf D61} (2000) 113010. \\ S. Jadach, W.
  P{\l}aczek, M. Skrzypek, B.F.L. Ward, Z. W\c{a}s, preprint CERN-TH/2000-337,
  hep-ph/0007012; submitted to Phys. Lett. B.\\ S. Jadach, W. P{\l}aczek, M.
  Skrzypek, B.F.L. Ward, Z. W\c{a}s, Comput. Phys. Commun. {\bf 140} (2001)
  432.\\ The \YFSWW\ cross-sections at 155--215 GeV have been kindly provided
  by the authors\relax
\relax
\bibitem{gc_bib:ALEPH-cTGC}
ALEPH Collaboration, {\em Measurement of Triple Gauge-Boson Couplings in
  $e^+e^-$ collisions from 183 to 209GeV}, ALEPH 2003-015 CONF 2003-011\relax
\relax
\bibitem{gc_bib:DELPHI-cTGC}
DELPHI Collaboration, {\em Measurement of Charged Trilinear Gauge Boson
  Couplings }, DELPHI 2003-051 (July 2003) CONF-671\relax
\relax
\bibitem{gc_bib:L3-cTGC}
L3 Collaboration, {\em Preliminary Results on the Measurement of
  Triple-Gauge-Boson Couplings of the W Boson at LEP}, L3 Note 2820 (September
  2003)\relax
\relax
\bibitem{gc_bib:OPAL-cTGC3}
\Opal\ Collaboration, {\em Measurement of charged current triple gauge boson
  couplings using W pairs at LEP}, submitted to Eur. Phys. J. C., CERN-EP
  2003-042, hep-ex/0308067\relax
\relax
\bibitem{gc_bib:QGC-Belanger}
G. B\'elanger \etal, Eur. Phys. J. {\bf C 13} (2000) 283\relax
\relax
\bibitem{gc_bib:GAEMERS}
K. Gaemers and G. Gounaris, Z. Phys. {\bf C 1} (1979) 259\relax
\relax
\bibitem{gc_bib:Hagiwara1987vm}
K. Hagiwara {\it et~al.},
\newblock  Nucl. Phys. {\bf B282}  (1987) 253\relax
\relax
\bibitem{gc_bib:HAGIWARA}
K. Hagiwara, S. Ishihara, R. Szalapski, and D. Zeppenfeld, Phys. Lett. {\bf B
  283} (1992) 353; \\ K. Hagiwara, S. Ishihara, R. Szalapski, and D.
  Zeppenfeld, Phys. Rev. {\bf D 48} (1993) 2182; \\ K.~Hagiwara, T.~Hatsukano,
  S.~Ishihara and R.~Szalapski, Nucl. Phys. {\bf B 496} (1997) 66\relax
\relax
\bibitem{gc_bib:BILENKY}
M. Bilenky, J.L. Kneur, F.M. Renard and D. Schildknecht, Nucl. Phys. {\bf B
  409} (1993) 22;\\ M. Bilenky, J.L. Kneur, F.M. Renard and D. Schildknecht,
  Nucl. Phys. {\bf B 419} (1994) 240\relax
\relax
\bibitem{gc_bib:KUSS}
I. Kuss and D. Schildknecht, Phys. Lett. {\bf B 383} (1996) 470\relax
\relax
\bibitem{gc_bib:PAPADOPOULOSCP}
G. Gounaris and C.G. Papadopoulos, DEMO-HEP-96/04, THES-TP 96/11,
  hep-ph/9612378\relax
\relax
\bibitem{gc_bib:moriond01}
The LEP-TGC combination group, LEPEWWG/TGC/2001-01, March 2001\relax
\relax
\bibitem{gc_bib:Gounaris2000tb}
G.~J. Gounaris, J. Layssac, and F.~M. Renard,
\newblock  Phys. Rev. {\bf D62}  (2000) 073013\relax
\relax
\bibitem{gc_bib:QGC-BelBou}
G. B\'elanger and F. Boudjema, Phys. Lett. {\bf B 288} (1992) 201\relax
\relax
\bibitem{gc_bib:ALEPH-nTGC}
ALEPH Collaboration, {\em Limits on anomalous neutral gauge couplings using
  data from ZZ and Z$\gamma$ production between 183-208 GeV}, ALEPH 2001-061
  (July 2001) CONF 2001-041\relax
\relax
\bibitem{gc_bib:DELPHI-nTGC}
DELPHI Collaboration, {\em Study of Trilinear Gauge Boson Couplings ZZZ,
  $ZZ\gamma$ and $Z\gamma\gamma$}, DELPHI 2001-097 (July 2001) CONF 525\relax
\relax
\bibitem{gc_bib:L3-hTGC}
L3 Collaboration, M. Acciarri \etal, Phys. Lett. {\bf B 436} (1999) 187;\\ L3
  Collaboration, M. Acciarri \etal, Phys. Lett. {\bf B 489} (2000) 55;\\ L3
  Collaboration, {\em Search for anomalous ZZg and Zgg couplings in the process
  ee$\rightarrow$Zg at LEP}, L3 Note 2672 (July 2001)\relax
\relax
\bibitem{gc_bib:OPAL-hTGC}
OPAL Collaboration, G. Abbiendi \etal, Eur. Phys. J. {\bf C 17} (2000) 13\relax
\relax
\bibitem{gc_bib:L3-fTGC}
\Ltre\ Collaboration, M.~Acciarri \etal, \PLB{450}{1999}{281}. The Z-pair
  cross-section at 183 GeV therein follows the \Ltre\ definition: the
  corresponding {\sc NC02} cross-section is given in \L3\ Collaboration, \L3\
  Note 2366, submitted to the Winter 1999 Conferences.\\ \Ltre\ Collaboration,
  M.~Acciarri \etal, \PLB{465}{1999}{363}.\\ \Ltre\ Collaboration, \L3\ Note
  2805, EPS 2003 abstract 228\relax
\relax
\bibitem{gc_bib:OPAL-fTGC}
See \Opal\ Collaboration, G.~Abbiendi \etal, \PLB{476}{2000}{256} and
  reference~\cite{4f_bib:opazz}\relax
\relax
\bibitem{gc_bib:ALEPH-QGC}
ALEPH Collaboration, {\em Constraints on Anomalous Quartic Gauge Boson
  Couplings}, ALEPH 2003-009 CONF 2003-006\relax
\relax
\bibitem{gc_bib:L3-QGC}
L3 Collaboration, {\em The e$^+$e$^- \rightarrow {\mathrm{Z} \gamma \gamma}
  \rightarrow {\mathrm{q} \bar{\mathrm{q}} \gamma \gamma} $ Reaction at LEP and
  Constraints on Anomalous Quartic Gauge Boson Couplings}, Phys. Lett. {\bf B
  540} (2002) 43\relax
\relax
\bibitem{gc_bib:OPAL-QGC}
\Opal\ Collaboration, {\em Constraints on Anomalous Quartic Gauge Boson
  Couplings using Acoplanar Photon Pairs et LEP-2}, \Opal\ Physics Note
  PN510\relax
\relax
\bibitem{gc_bib:LPA_A-B}
R. Bruneli{\`e}re \etal, Phys. Lett. {\bf B 533} (2002) 75 and references
  therein\relax
\relax
\bibitem{gc_bib:Alcaraz}
J. Alcaraz, {\em A proposal for the combination of TGC measurements}, L3 Note
  2718\relax
\relax
\bibitem{gc_bib:renaud}
R. Bruneli{\`e}re, {\em Tests on the LEP TGC combination procedures}, ALEPH
  2002-008 PHYS-2002-007 (2002)\relax
\relax
\bibitem{gc_bib:klein}
O.Klein, {\em On the Theory of Charged Fields}, New Theories in Physics,
  Proceedings, Warsaw, 1938; reprinted in: Surveys of High Energ. Phys. {\bf 5}
  (1986) 269\relax
\relax
\bibitem{gc_bib:maiani}
L.Maiani and P.M.Zerwas, {\em W Static ELM Parameters}, Memorandum to the TGC
  Combination Group (1998)\relax
\relax
\bibitem{Achard:2003pe}
L3 Collaboration, P. Achard {\it et~al.},
\newblock  Phys. Lett. {\bf B561}  (2003) 202--212\relax
\relax
\bibitem{Abbiendi:2005es}
OPAL Collaboration, G. Abbiendi {\it et~al.},
\newblock  {\em Colour reconnection in e+ e- --> W+ W- at s**(1/2) = 189- GeV -
  209-GeV},
\newblock  Eprint hep-ex/0508062, 2005\relax
\relax
\bibitem{bib:cr:GPZ}
G. Gustafson, U. Pettersson, and P.~M. Zerwas,
\newblock  Phys. Lett. {\bf B209}  (1988) 90\relax
\relax
\bibitem{bib:cr:SK_MODELS}
T. Sjostrand and V.~A. Khoze,
\newblock  Z. Phys. {\bf C62}  (1994) 281--310\relax
\relax
\bibitem{bib:cr:ARIADNECR_MODEL}
L. Lonnblad,
\newblock  Z. Phys. {\bf C70}  (1996) 107--114\relax
\relax
\bibitem{HERWIG6}
G. Corcella {\it et~al.},
\newblock  JHEP {\bf 01}  (2001) 010\relax
\relax
\bibitem{bib:cr:GH_MODEL}
G. Gustafson and J. Hakkinen,
\newblock  Z. Phys. {\bf C64}  (1994) 659--664\relax
\relax
\bibitem{bib:cr:EG_MODEL}
J.~R. Ellis and K. Geiger,
\newblock  Phys. Rev. {\bf D54}  (1996) 1967--1990\relax
\relax
\bibitem{bib:cr:RATHSMAN}
J. Rathsman,
\newblock  Phys. Lett. {\bf B452}  (1999) 364--371\relax
\relax
\bibitem{bib:cr:OPAL_CR}
OPAL Collaboration, {\it Colour Reconnection Studies in \eeWW\ at
  $\roots=$189~GeV}, OPAL PN417\relax
\relax
\bibitem{bib:cr:DELPHI_CR}
DELPHI Collaboration, P. Abreu {\it et~al.},
\newblock  Eur. Phys. J. {\bf C18}  (2000) 203--228\relax
\relax
\bibitem{bib:cr:ALEPH_CR}
ALEPH Collaboration, {\it Preliminary Charged Particle Multiplicity in
  $\ee\rightarrow$ W-pairs}, ALEPH 2000-058 CONF 2000-038\relax
\relax
\bibitem{bib:cr:L3_CR}
L3 Collaboration, {\it Search for Colour Reconnection Effects in
  $\eeWW\rightarrow\textrm{hadrons}$}, L3 Note 2560\relax
\relax
\bibitem{bib:cr:SK_HEAVYHAD}
V.~A. Khoze and T. Sjostrand,
\newblock  Eur. Phys. J. {\bf C6}  (1999) 271--284\relax
\relax
\bibitem{bib:cr:OPAL_HEAVYHAD}
OPAL Collaboration, {\it Investigation of Colour Reconnection via Heavy
  Particle Production in \eeWW}, OPAL PN412\relax
\relax
\bibitem{bib:cr:JADE_STRING2}
JADE Collaboration, W. Bartel {\it et~al.},
\newblock  Phys. Lett. {\bf B101}  (1981) 129\relax
\relax
\bibitem{bib:cr:JADE_STRING3}
JADE Collaboration, W. Bartel {\it et~al.},
\newblock  Z. Phys. {\bf C21}  (1983) 37\relax
\relax
\bibitem{bib:cr:JADE_STRING4}
JADE Collaboration, W. Bartel {\it et~al.},
\newblock  Phys. Lett. {\bf B134}  (1984) 275\relax
\relax
\bibitem{bib:cr:JADE_STRING5}
JADE Collaboration, W. Bartel {\it et~al.},
\newblock  Phys. Lett. {\bf B157}  (1985) 340\relax
\relax
\bibitem{bib:cr:TPC2GAM_STRING1}
TPC/Two Gamma Collaboration, H. Aihara {\it et~al.},
\newblock  Z. Phys. {\bf C28}  (1985) 31\relax
\relax
\bibitem{bib:cr:TPC2GAM_STRING2}
TPC/Two Gamma Collaboration, H. Aihara {\it et~al.},
\newblock  Phys. Rev. Lett. {\bf 57}  (1986) 945\relax
\relax
\bibitem{bib:cr:TASSO_STRING1}
TASSO Collaboration, M. Althoff {\it et~al.},
\newblock  Z. Phys. {\bf C29}  (1985) 29\relax
\relax
\bibitem{bib:cr:pflow1}
L3 Collaboration, {\it Colour Reconnection Studies in \eeWW\ Events at
  $\roots=189$~GeV}, L3 Note 2406\relax
\relax
\bibitem{bib:cr:pflow2}
D.\ Duchesneau, {\it New Method Based on Energy and Particle Flow in
  $\eeWW\rightarrow$ hadron Events for Colour Reconnection Studies},
  LAPP-EXP-2000-02\relax
\relax
\bibitem{bib:cr:OXFORD_WS}
A. Ballestrero {\it et~al.},
\newblock  J. Phys. {\bf G24}  (1998) 365--403\relax
\relax
\bibitem{bib:cr:ALEPH_PF}
ALEPH Collaboration, {\it Colour Reconnection Studies Using Particle Flow
  Between W Bosons at $\roots=$189--208~GeV}, ALEPH 2002-020 CONF
  2002-009\relax
\relax
\bibitem{bib:cr:DELPHI_PF}
DELPHI Collaboration, {\it Update of the Investigation of Colour Reconnection
  in WW Pairs Using Particle Flow}, DELPHI 2002-047 CONF 581\relax
\relax
\bibitem{bib:cr:L3_PF}
L3 Collaboration, {\it Search for Colour Reconnection Effects in
  $\eeWW\rightarrow\textrm{hadrons}$ Through Particle-Flow Studies at
  $\roots=$189--208~GeV}, L3 Note 2748\relax
\relax
\bibitem{bib:cr:OPAL_PF}
OPAL Collaboration, {\it Colour Reconnection Studies in \eeWW\ at
  $\roots=$189--208~GeV Using Particle Flow}, OPAL PN506\relax
\relax
\bibitem{ALEPH-MW}
ALEPH Collaboration, R. Barate {\it et~al.},
\newblock  Eur. Phys. J. {\bf C17}  (2000) 241--261\relax
\relax
\bibitem{OPAL-MW}
OPAL Collaboration, G. Abbiendi {\it et~al.},
\newblock  Phys. Lett. {\bf B507}  (2001) 29--46\relax
\relax
\bibitem{bib:fsi:KORALW}
S. Jadach {\it et~al.},
\newblock  Comput. Phys. Commun. {\bf 140}  (2001) 475--512\relax
\relax
\bibitem{JETSET}
T. Sjostrand,
\newblock  Comput. Phys. Commun. {\bf 82}  (1994) 74--90,
\newblock  (JETSET)\relax
\relax
\bibitem{ARIADNE}
L.\,Lonnblad, Comp. Phys. Comm. {\bf 71} (1992) 15\relax
\relax
\bibitem{be:DELPHI03}
  DELPHI 2003-020-CONF-640 
\relax
\bibitem{bib:cr:ALEPH_BEC}
ALEPH Collaboration, {\it Further Studies on Bose-Einstein Correlations in
  W-pair Decays}, ALEPH 2001-064 CONF 2001-044\relax
\relax
\bibitem{be:L302}
L3 Coll., {\em Measurement of Bose-Einstein Correlations in $e^+ e^-
  \rightarrow W^+ W^-$ Events at LEP}, Phys. Lett. B547 (2002) 139\relax
\relax
\bibitem{bib:cr:OPAL_BEC}
OPAL Collaboration, {\it Bose-Einstein Correlations in \eeWW\ Events at 172,
  183 and 189 GeV}, OPAL PN393\relax
\relax
\bibitem{bib:LEP2_MCWS}
A. Ballestrero {\it et~al.},
\newblock  hep-ph/0006259 (2000)\relax
\relax
\bibitem{be:chekanov}
S.V. Chekanov, E.A. De Wolf, W. Kittel, Eur. Phys. J. C6 (1999) 403\relax
\relax
\bibitem{be:DELPHI97}
DELPHI Coll., P. Abreu et al., Phys. Lett. B401 (1997) 181\relax
\relax
\bibitem{be:DeWolf}
E.A. De Wolf, {\em Correlations in $e^+ e^- \rightarrow W^+ W^-$ hadronic
  decays}, hep-ph/0101243\relax
\relax
\bibitem{be:ALEPH00}
ALEPH Coll., {\em Bose-Einstein Correlations in W-pair decays}, Phys. Lett.
  B478 (2000) 50\relax
\relax
\bibitem{be:ALEPH05}
ALEPH Coll., {\em Bose-Einstein correlations in W-pair decays with an
  event-mixing technique}, Phys. Lett. B606 (2005) 265\relax
\relax
\bibitem{be:DELPHI05}
DELPHI Coll., {\em Bose-Einstein Correlations in $W^+ W^-$ events at LEP2},
  Eur. Phys. J. C44 (2005) 161\relax
\relax
\bibitem{be:OPAL05}
OPAL Coll., {\em Study of Bose-Einstein Correlations in $e^+ e^- \rightarrow
  W^+ W^-$ Events at LEP}, Eur. Phys. J. C36 (2004) 297\relax
\relax
\bibitem{be:PYTHIA57}
T. Sj{\"o}strand: PYTHIA 5.7 and JETSET 7.4, Computer Physics Commun. 82 (1994)
  74\relax
\relax
\bibitem{mw:bib:LUBOEI}
L. L{\"o}nnblad and T. Sj{\"o}strand,
\newblock  Eur. Phys. J. {\bf C2}  (1998) 165\relax
\relax
\bibitem{be:lep-be}
LEPWW FSI group http://lepewwg.web.cern.ch/LEPEWWG/lepww/fsi.html\relax
\relax
\bibitem{be:LEPW}
This report, section {\em W-Boson Mass and Width at LEP-II}\relax
\relax
\bibitem{LEP2L3MWGW}
L3 Collaboration, P. Achard {\it et~al.},
\newblock  {\em Measurement of the Mass and Width of the W Boson at LEP},
\newblock  Eprint hep-ex/0511049, 2005\relax
\relax
\bibitem{common_bib:adloww161}
\Aleph\ Collaboration, R.~Barate \etal, Phys.~Lett.~{\bf B401} (1997) 347.\\
  \Delphi\ Collaboration, P.~Abreu \etal, Phys.~Lett.~{\bf B397} (1997) 158.\\
  \Ltre\ Collaboration, M.~Acciarri \etal, Phys.~Lett. {\bf B398} (1997) 223.\\
  \Opal\ Collaboration, K.~Ackerstaff \etal, Phys.~Lett. {\bf B389} (1996)
  416\relax
\relax
\bibitem{mw:bib:ALEPHRevised}
ALEPH Collaboration, {\it Measurement of the W Mass in $\epem$ Collisions at
  $\roots$ between 183 and 209~$\GeV$}, ALEPH note 2003-005 CONF 2003-003\relax
\relax
\bibitem{common_bib:delww172}
DELPHI Collaboration, P.~A. \etal,
\newblock  Eur.~Phys.~J. {\bf C2}  (1998) 581\relax
\relax
\bibitem{mw:bib:D-mw183}
DELPHI Collaboration, P. Abreu {\it et~al.},
\newblock  Phys. Lett. {\bf B462}  (1999) 410--424\relax
\relax
\bibitem{mw:bib:D-mw189}
DELPHI Collaboration, P. Abreu {\it et~al.},
\newblock  Phys. Lett. {\bf B511}  (2001) 159--177\relax
\relax
\bibitem{mw:bib:D-mw20X}
DELPHI Collaboration, {\it{Measurement of the mass and width of the W Boson in
  $\epem$ collisions at $\roots = 192-209~\GeV$}}, DELPHI 2001-103 CONF
  531\relax
\relax
\bibitem{common_bib:ltrww172}
L3 Collaboration, M.~A. \etal,
\newblock  Phys. Lett. {\bf B407}  (1997) 419\relax
\relax
\bibitem{mw:bib:L-mw183}
L3 Collaboration, M. Acciarri {\it et~al.},
\newblock  Phys. Lett. {\bf B454}  (1999) 386--398\relax
\relax
\bibitem{mw:bib:L-mw189}
L3 Collaboration, {\it{Preliminary Results on the Measurement of Mass and Width
  of the W Boson at LEP}}, L3 Note 2377, March 1999\relax
\relax
\bibitem{mw:bib:L-mw19X}
L3 Collaboration, {\it{Preliminary Results on the Measurement of Mass and Width
  of the W Boson at LEP}}, L3 Note 2575, July 2000\relax
\relax
\bibitem{mw:bib:L-mw20X}
L3 Collaboration, {\it{Preliminary Results on the Measurement of Mass and Width
  of the W Boson at LEP}}, L3 Note 2637, February 2001\relax
\relax
\bibitem{mw:bib:O-final}
OPAL Collaboration, G. Abbiendi,
\newblock  hep-ex/0508060 (2005)\relax
\relax
\bibitem{mw:bib:energy}
LEP Energy Working Group, LEPEWG 01/01, March 2001\relax
\relax
\bibitem{mw:bib:CRcomb}
LEP W Working Group, LEPEWWG/FSI/2002-01, July 2002\relax
\relax
\bibitem{cr:bib:O-final}
OPAL Collaboration, G. Abbiendi,
\newblock   (2005)\relax
\relax
\bibitem{mw:bib:ski}
T. Sj{\"o}strand and V.~A. Khoze,
\newblock  Z. Phys. {\bf C62}  (1994) 281--310\relax
\relax
\bibitem{UA2-MW}
UA2 Collaboration, J. Alitti {\it et~al.},
\newblock  Phys. Lett. {\bf B276}  (1992) 354--364\relax
\relax
\bibitem{CDF-MW-PRL90}
CDF Collaboration, F. Abe {\it et~al.},
\newblock  Phys. Rev. Lett. {\bf 65}  (1990) 2243--2246\relax
\relax
\bibitem{CDF-MW-PRD90}
CDF Collaboration, F. Abe {\it et~al.},
\newblock  Phys. Rev. {\bf D43}  (1991) 2070--2093\relax
\relax
\bibitem{CDF-MW-PRL95}
CDF Collaboration, F. Abe {\it et~al.},
\newblock  Phys. Rev. Lett. {\bf 75}  (1995) 11--16\relax
\relax
\bibitem{CDF-MW-PRD95}
CDF Collaboration, F. Abe {\it et~al.},
\newblock  Phys. Rev. {\bf D52}  (1995) 4784--4827\relax
\relax
\bibitem{CDF-MW-2000}
CDF Collaboration, T. Affolder {\it et~al.},
\newblock  Phys. Rev. {\bf D64}  (2001) 052001\relax
\relax
\bibitem{D0-MW:central}
{D\O} Collaboration, B. Abbott {\it et~al.},
\newblock  Phys. Rev. Lett. {\bf 80}  (1998) 3008\relax
\relax
\bibitem{D0-MW:endcap}
{D\O} Collaboration, B. Abbott {\it et~al.},
\newblock  Phys. Rev. Lett. {\bf 84}  (2000) 222--227\relax
\relax
\bibitem{D0-MW:edge}
{D\O} Collaboration, V.~M. Abazov {\it et~al.},
\newblock  Phys. Rev. {\bf D66}  (2002) 012001\relax
\relax
\bibitem{D0-MW:large}
{D\O} Collaboration, B. Abbott {\it et~al.},
\newblock  Phys. Rev. {\bf D62}  (2000) 092006\relax
\relax
\bibitem{CDF-GW}
CDF Collaboration, T. Affolder {\it et~al.},
\newblock  Phys. Rev. Lett. {\bf 85}  (2000) 3347--3352\relax
\relax
\bibitem{D0-GW}
{D\O} Collaboration, V.~M. Abazov {\it et~al.},
\newblock  Phys. Rev. {\bf D66}  (2002) 032008\relax
\relax
\bibitem{PP-MW-GW:combination}
{The CDF Collaboration, the D\O\ Collaboration, and the Tevatron Electroweak
  Working Group},
\newblock  Phys. Rev. {\bf D70}  (2004) 092008\relax
\relax
\bibitem{Mtop1-CDF-di-l-PRLa}
CDF Collaboration, F. Abe {\it et~al.},
\newblock  Phys. Rev. Lett. {\bf 80}  (1998) 2779--2784\relax
\relax
\bibitem{Mtop1-CDF-di-l-PRLb}
CDF Collaboration, F. Abe {\it et~al.},
\newblock  Phys. Rev. Lett. {\bf 82}  (1999) 271--276\relax
\relax
\bibitem{Mtop1-CDF-di-l-PRLb-E}
CDF Collaboration, F. Abe {\it et~al.},
\newblock  Erratum: Phys. Rev. Lett. {\bf 82}  (1999) 2808--2809\relax
\relax
\bibitem{Mtop1-CDF-l+j-PRL}
CDF Collaboration, F. Abe {\it et~al.},
\newblock  Phys. Rev. Lett. {\bf 80}  (1998) 2767--2772\relax
\relax
\bibitem{Mtop1-CDF-l+j-PRD}
CDF Collaboration, T. Affolder {\it et~al.},
\newblock  Phys. Rev. {\bf D63}  (2001) 032003\relax
\relax
\bibitem{Mtop1-CDF-all-j-PRL}
CDF Collaboration, F. Abe {\it et~al.},
\newblock  Phys. Rev. Lett. {\bf 79}  (1997) 1992--1997\relax
\relax
\bibitem{D0-top:prl-ll}
{D\O} Collaboration, B. Abbott {\it et~al.},
\newblock  Phys. Rev. Lett. {\bf 80}  (1998) 2063--2068\relax
\relax
\bibitem{D0-top:prd-ll}
{D\O} Collaboration, B. Abbott {\it et~al.},
\newblock  Phys. Rev. {\bf D60}  (1999) 052001\relax
\relax
\bibitem{D0-top:prl-lj}
{D\O} Collaboration, S. Abachi {\it et~al.},
\newblock  Phys. Rev. Lett. {\bf 79}  (1997) 1197--1202\relax
\relax
\bibitem{D0-top:prd-lj}
{D\O} Collaboration, B. Abbott {\it et~al.},
\newblock  Phys. Rev. {\bf D58}  (1998) 052001\relax
\relax
\bibitem{Mtop1-D0-l+j-new1}
{D\O} Collaboration, V.~M. Abazov {\it et~al.},
\newblock  Nature {\bf 429}  (2004) 638--642\relax
\relax
\bibitem{Mtop1-D0-l+j-new2}
{D\O} Collaboration, V.~M. Abazov {\it et~al.},
\newblock  {\em New measurement of the top quark mass in lepton + jets t anti-t
  events at {D\O}},
\newblock  Eprint hep-ex/0407005, 2004\relax
\relax
\bibitem{Mtop1-D0-all-j-PRL}
{D\O} Collaboration, V.~M. Abazov {\it et~al.},
\newblock  Phys. Lett. {\bf B606}  (2005) 25--33\relax
\relax
\bibitem{TeVEWWGtop-0507}
{The CDF Collaboration, the D\O\ Collaboration, and the Tevatron Electroweak
  Working Group},
\newblock  {\em Combination of CDF and {D\O} results on the top-quark mass},
\newblock  Eprint hep-ex/0507091, Fermilab, 2005\relax
\relax
\bibitem{QWCs:exp:1}
C.~S. Wood {\it et~al.},
\newblock  Science {\bf 275}  (1997) 1759\relax
\relax
\bibitem{QWCs:exp:2}
S.~C. Bennett and C.~E. Wieman,
\newblock  Phys. Rev. Lett. {\bf 82}  (1999) 2484--2487\relax
\relax
\bibitem{QWCs:theo:2003:new}
J.~S.~M. Ginges and V.~V. Flambaum,
\newblock  Phys. Rept. {\bf 397}  (2004) 63--154\relax
\relax
\bibitem{PDG2004}
Particle Data Group Collaboration, S. Eidelman {\it et~al.},
\newblock  Phys. Lett. {\bf B592}  (2004) 1\relax
\relax
\bibitem{E158RunI}
SLAC E158 Collaboration, P. Anthony {\it et~al.},
\newblock  Phys. Rev. Lett. {\bf 92}  (2004) 181602\relax
\relax
\bibitem{E158RunI+II+III}
SLAC E158 Collaboration, P.~L. Anthony {\it et~al.},
\newblock  Phys. Rev. Lett. {\bf 95}  (2005) 081601\relax
\relax
\bibitem{bib-NuTeV-final}
NuTeV Collaboration, G.~P. Zeller {\it et~al.},
\newblock  Phys. Rev. Lett. {\bf 88}  (2002) 091802,
\newblock  erratum: 90 (2003) 239902\relax
\relax
\bibitem{PaschosWolfenstein}
E.~A. Paschos and L. Wolfenstein,
\newblock  Phys. Rev. {\bf D7}  (1973) 91--95\relax
\relax
\bibitem{bib-Gmu-1}
T. van Ritbergen and R.~G. Stuart,
\newblock  Phys. Rev. Lett. {\bf 82}  (1999) 488--491\relax
\relax
\bibitem{bib-Gmu-2}
T. van Ritbergen and R.~G. Stuart,
\newblock  Nucl. Phys. {\bf B564}  (2000) 343--390\relax
\relax
\bibitem{bib-Gmu-3}
M. Steinhauser and T. Seidensticker,
\newblock  Phys. Lett. {\bf B467}  (1999) 271--278\relax
\relax
\bibitem{bib-BP05}
H. Burkhardt and B. Pietrzyk,
\newblock  Phys. Rev. {\bf D72}  (2005) 057501\relax
\relax
\bibitem{bib-PCLI}
D. Bardin {\it et~al.}, in {\em Reports of the working group on precision
  calculations for the Z resonance}, CERN 95-03, ed. D. Bardin, W. Hollik, and
  G. Passarino,  (CERN, Geneva, Switzerland, 1995), pp. 7--162\relax
\relax
\bibitem{BP:98}
D.~Y. Bardin and G. Passarino,
\newblock  {\em Upgrading of precision calculations for electroweak
  observables},
\newblock  Eprint hep-ph/9803425, 1998\relax
\relax
\bibitem{PCP99}
D.~Y. Bardin, M. Gr{\"u}newald, and G. Passarino,
\newblock  {\em Precision calculation project report},
\newblock  Eprint hep-ph/9902452, 1999\relax
\relax
\bibitem{bib-twoloop}
G.~Degrassi, S.~Fanchiotti and A.~Sirlin, Nucl.~Phys.~{\bf B351} (1991) 49;\\
  G.~Degrassi and A.~Sirlin, Nucl.~Phys.~{\bf B352} (1991) 342;\\ G.~Degrassi,
  P.~Gambino and A.~Vicini, Phys.~Lett.~{\bf B383} (1996) 219;\\ G.~Degrassi,
  P.~Gambino and A.~Sirlin, Phys.~Lett.~{\bf B394} (1997) 188;\\ G.~Degrassi
  and P.~Gambino, Nucl.~Phys.~{\bf B567} (2000) 3\relax
\relax
\bibitem{bib-QCDEW}
A.~Czarnecki and J.~K{\"u}hn, Phys.~Rev.~Lett.~{\bf 77} (1996) 3955;\\
  R.~Harlander, T.~Seidensticker and M.~Steinhauser, Phys.~Lett.~{\bf B426}
  (1998) 125\relax
\relax
\bibitem{Montagna:1993py}
G. Montagna {\it et~al.},
\newblock  Nucl. Phys. {\bf B401}  (1993) 3--66\relax
\relax
\bibitem{Montagna:1993ai}
G. Montagna {\it et~al.},
\newblock  Comput. Phys. Commun. {\bf 76}  (1993) 328--360\relax
\relax
\bibitem{Montagna:1996ja}
G. Montagna {\it et~al.},
\newblock  Comput. Phys. Commun. {\bf 93}  (1996) 120--126\relax
\relax
\bibitem{Montagna:1998kp}
G. Montagna {\it et~al.},
\newblock  Comput. Phys. Commun. {\bf 117}  (1999) 278--289,
\newblock  updated to include initial state pair radiation (G. Passarino, priv.
  comm.)\relax
\relax
\bibitem{Bardin:1989di}
D.~Y. Bardin {\it et~al.},
\newblock  Z. Phys. {\bf C44}  (1989) 493\relax
\relax
\bibitem{Bardin:1990tq}
D.~Y. Bardin {\it et~al.},
\newblock  Comput. Phys. Commun. {\bf 59}  (1990) 303--312\relax
\relax
\bibitem{Bardin:1991fu}
D.~Y. Bardin {\it et~al.},
\newblock  Nucl. Phys. {\bf B351}  (1991) 1--48\relax
\relax
\bibitem{Bardin:1991de}
D.~Y. Bardin {\it et~al.},
\newblock  Phys. Lett. {\bf B255}  (1991) 290--296\relax
\relax
\bibitem{Bardin:1992jc}
D.~Y. Bardin {\it et~al.},
\newblock  {\em ZFITTER: An Analytical program for fermion pair production in
  ${\ee}$ annihilation},
\newblock  Eprint arXiv:hep-ph/9412201, 1992\relax
\relax
\bibitem{Bardin:1999yd}
D.~Y. Bardin {\it et~al.},
\newblock  Comput. Phys. Commun. {\bf 133}  (2001) 229--395,
\newblock  updated with results from~\cite{Arbuzov}\relax
\relax
\bibitem{Kobel:2000aw}
Two Fermion Working Group Collaboration, M. Kobel {\it et~al.},
\newblock  {\em Two-fermion production in electron positron collisions},
\newblock  Eprint hep-ph/0007180, 2000\relax
\relax
\bibitem{Arbuzov:2005ma}
A.~B. Arbuzov {\it et~al.},
\newblock  {\em ZFITTER: a semi-analytical program for fermion pair production
  in e+e- annihilation, from version 6.21 to version 6.42},
\newblock  Eprint hep-ph/0507146, 2005\relax
\relax
\bibitem{bib-SMALFAS}
T.~Hebbeker, M.~Martinez, G.~Passarino and G.~Quast, Phys.~Lett. {\bf B331}
  (1994) 165;\\ P.A.~Raczka and A.~Szymacha, Phys. Rev. {\bf D54} (1996)
  3073;\\ D.E.~Soper and L.R.~Surguladze, Phys. Rev. {\bf D54} (1996)
  4566\relax
\relax
\bibitem{Stenzel:2005sg}
H. Stenzel,
\newblock  JHEP {\bf 07}  (2005) 0132\relax
\relax
\bibitem{Twoloop-MW}
M. Awramik {\it et~al.},
\newblock  Phys. Rev. {\bf D69}  (2004) 053006\relax
\relax
\bibitem{Twoloop-sin2teff}
M. Awramik {\it et~al.},
\newblock  Phys. Rev. Lett. {\bf 93}  (2004) 201805\relax
\relax
\bibitem{Threeloop-rho}
M. Faisst {\it et~al.},
\newblock  Nucl. Phys. {\bf B665}  (2003) 649--662\relax
\relax
\bibitem{bib-alphalept}
M. Steinhauser,
\newblock  Phys. Lett. {\bf B429}  (1998) 158--161\relax
\relax
\bibitem{bib-JEG2}
S. Eidelman and F. Jegerlehner,
\newblock  Z. Phys. {\bf C67}  (1995) 585--602\relax
\relax
\bibitem{bib-Burk}
H. Burkhardt and B. Pietrzyk,
\newblock  Phys. Lett. {\bf B356}  (1995) 398--403\relax
\relax
\bibitem{BES_01}
BES Collaboration, J.~Z. Bai {\it et~al.},
\newblock  Phys. Rev. Lett. {\bf 88}  (2002) 101802\relax
\relax
\bibitem{CMD_03}
CMD-2 Collaboration, R.~R. Akhmetshin {\it et~al.},
\newblock  Phys. Lett. {\bf B578}  (2004) 285--289\relax
\relax
\bibitem{KLOE_04}
KLOE Collaboration, A. Aloisio {\it et~al.},
\newblock  Phys. Lett. {\bf B606}  (2005) 12--24\relax
\relax
\bibitem{bib-Swartz}
M.~L. Swartz,
\newblock  Phys. Rev. {\bf D53}  (1996) 5268--5282\relax
\relax
\bibitem{bib-Zeppe}
A.~D. Martin and D. Zeppenfeld,
\newblock  Phys. Lett. {\bf B345}  (1995) 558--563\relax
\relax
\bibitem{bib-Alemany}
R. Alemany, M. Davier, and A. Hocker,
\newblock  Eur. Phys. J. {\bf C2}  (1998) 123--135\relax
\relax
\bibitem{bib-Davier}
M. Davier and A. Hocker,
\newblock  Phys. Lett. {\bf B419}  (1998) 419--431\relax
\relax
\bibitem{bib-alphaKuhn}
J.~H. Kuhn and M. Steinhauser,
\newblock  Phys. Lett. {\bf B437}  (1998) 425--431\relax
\relax
\bibitem{bib-jeger99}
F. Jegerlehner, in Proceedings, 4th International Symposium, RADCOR'98, ed. J.
  Sola,  (World Scientific, Singapore, Sep 1999), p.~75\relax
\relax
\bibitem{bib-Erler}
J. Erler,
\newblock  Phys. Rev. {\bf D59}  (1999) 054008\relax
\relax
\bibitem{bib-ADMartin}
A.~D. Martin, J. Outhwaite, and M.~G. Ryskin,
\newblock  Phys. Lett. {\bf B492}  (2000) 69--73\relax
\relax
\bibitem{bib-Troconiz-Yndurain}
J.~F. de~Troconiz and F.~J. Yndurain,
\newblock  Phys. Rev. {\bf D65}  (2002) 093002\relax
\relax
\bibitem{bib-Hagiwara:2003}
K. Hagiwara {\it et~al.},
\newblock  Phys. Rev. {\bf D69}  (2004) 093003\relax
\relax
\bibitem{bib-Troconiz-Yndurain-2004}
J.~F. de~Troconiz and F.~J. Yndurain,
\newblock  Phys. Rev. {\bf D71}  (2005) 073008\relax
\relax
\bibitem{common_bib:pdg2000}
Particle Data Group, D.E.~Groom \etal, \EPJ{15} {(2000)} {1}\relax
\relax
\bibitem{QCD:Bethke:2000}
S. Bethke,
\newblock  J. Phys. {\bf G26}  (2000) R27\relax
\relax
\bibitem{Bethke:2004uy}
S. Bethke,
\newblock  Nucl. Phys. Proc. Suppl. {\bf 135}  (2004) 345--352\relax
\relax
\bibitem{LEPSMHIGGS}
ALEPH, DELPHI, L3, and OPAL Collaboration,
\newblock  Phys. Lett. {\bf B565}  (2003) 61--75\relax
\relax
\bibitem{Kawamoto:2004pi}
T. Kawamoto and R.~G. Kellogg,
\newblock  Phys. Rev. {\bf D69}  (2004) 113008\relax
\relax
\bibitem{4f_bib:delsw2001}
\Delphi\ Collaboration, \PLB{515}{2001}{238}\relax
\relax
\bibitem{4f_bib:ltrsw2001}
\Ltre\ Collaboration, \PLB{436}{1998}{417};\\ \Ltre\ Collaboration,
  \PLB{487}{2000}{229}\relax
\relax
\bibitem{4f_bib:ltrsw}
\Ltre\ Collaboration, \PLB{547}{2002}{151}\relax
\relax
\bibitem{4f_bib:opasw189}
\Opal\ Collaboration, \Opal\ Physics Note PN427, March 2000\relax
\relax
\bibitem{Arbuzov}
A.~B. Arbuzov,
\newblock  {\em Light pair corrections to electron positron annihilation at
  LEP/SLC},
\newblock  Eprint arXiv:hep-ph/9907500, Turin U. and INFN, Turin, 1999\relax
\relax
\end{mcbibliography}

\vfill

\section*{Links to LEP results on the World Wide Web}
The physics notes describing the preliminary results of the four LEP
experiments submitted to the 2005 summer conferences, as well as
additional documentation from the LEP electroweak working
group are available on the World Wide Web at: \\[2mm]
\begin{tabular}{ll}
  ALEPH:   &
  {\tt http://aleph.web.cern.ch/aleph/alpub/oldconf/conferences.html}\\
  DELPHI:  &
  {\tt http://delphiwww.cern.ch/pubxx/conferences/summer05/}\\
  L3:      &
  {\tt http://l3.web.cern.ch/l3/conferences/Lisboa2005/}\\
  OPAL:    &
  {\tt http://opal.web.cern.ch/Opal/pubs/eps2005/abstr.html}\\     
  LEP-EWWG: &
  {\tt http://www.cern.ch/LEPEWWG/}\\
\end{tabular}

\end{document}